%% file: master_thesis_Buchholz_Benjamin_00893529.tex
\documentclass[
headsepline, 
12pt, 
a4paper, 
parskip=on,
liststotoc] 
{scrreprt}	
\usepackage[utf8]{inputenc} 
	\usepackage{cite}						
\usepackage{footmisc}
\usepackage{mathrsfs}						
\usepackage[automark]{scrlayer-scrpage} 
\usepackage[latin,english]{babel}
\usepackage[T1]{fontenc}
\usepackage{amssymb}
\usepackage{mathtools}
\usepackage{tikz}
\usetikzlibrary{arrows}
\usetikzlibrary{calc}
\usetikzlibrary{angles, quotes, babel}
\usepackage{commath}
\usepackage{extarrows}
\usepackage{esvect}
\usepackage{lmodern} 
\usepackage{textcmds}
\usepackage[export]{adjustbox}
\usepackage{dsfont}
\usepackage[left=30mm,right=25mm,bottom=25mm,top=30mm]{geometry}
\renewcommand{\arraystretch}{1.2} 
\usepackage{subcaption}
\usepackage[onehalfspacing]{setspace} 

\usepackage{graphicx}	
\usepackage{xspace}

\usepackage{textpos} 

\usepackage[colorlinks,bookmarks]{hyperref} 

\usepackage{tabularx}	

\usepackage{supertabular} 

\usepackage{longtable} 

\usepackage{ragged2e}	
	\newcolumntype{L}[1]{>{\RaggedRight\arraybackslash\hspace{0pt}}p{#1}} 	
	\newcolumntype{R}[1]{>{\RaggedLeft\arraybackslash\hspace{0pt}}p{#1}} 		
	\newcolumntype{C}[1]{>{\centering\arraybackslash\hspace{0pt}}p{#1}} 		

\usepackage{endnotes} 

\usepackage{hyperref}	
 
\hypersetup{colorlinks = true, linkcolor  = black, urlcolor=black, linkcolor=black}

\usepackage{booktabs} 
\usepackage{multirow} 

\usepackage[binary-units = true]{siunitx}
\DeclareSIUnit\px{px}
\usepackage{amsmath} 

\usepackage{float} 
\usepackage{placeins} 
\usepackage{makecell} 

\usepackage{caption} 
\usepackage{appendix} 

\usepackage{pdfpages} 

\usepackage[version=4]{mhchem}

\usepackage{lscape} 

\usepackage{rotating} 
   
\usepackage{upgreek} 

\usepackage{wrapfig} 

\usepackage{icomma}

\makeatletter
\begingroup
\catcode`\_=\active
\protected\gdef_{\@ifnextchar|\subtextit\subtextup }
\endgroup
\def\subtextit|#1|{\sb{#1}}
\def\subtextup#1{\sb{\mathrm{#1}}}
\AtBeginDocument{\catcode`\_=12 \mathcode`\_=32768}
\makeatother

\RequirePackage{ifthen}

\newcommand{\nobarfrac}{\genfrac{}{}{0pt}{}}
\usepackage{courier}
\usepackage{cancel}
\usepackage{chngcntr}
\counterwithout{footnote}{chapter}
\usepackage{caption}
\usepackage{pgfplots}
\usepackage{listings} 
\lstdefinestyle{Matlab}{
  frame=L,
  language=Matlab,
  showstringspaces=false,
}
\lstdefinestyle{c}{
  frame=L,
  language=C++,
  showstringspaces=false,
}

\usepackage{cleveref}
\crefname{lstlisting}{listing}{listings}
\Crefname{lstlisting}{Listing}{Listings}

\Crefname{listing}{}{}
\begin{document}

\setcounter{tocdepth}{3}	
\setcounter{secnumdepth}{3}	

\hypersetup{
pdftitle={Calculation of the runaway electron current in tokamak disruptions},
pdfsubject={master thesis},
pdfauthor={Benjamin Buchholz},
citecolor=black}
\addtolength{\footskip}{-5mm}
\pagestyle{scrheadings} 
\clearscrheadings	
\clearscrplain		
\cfoot{\pagemark}
\lohead{\headmark}	

\setlength\abovedisplayshortskip{0.4cm}
\setlength\belowdisplayshortskip{0.4cm}
\setlength\abovedisplayskip{0.4cm}
\setlength\belowdisplayskip{0.4cm}

\clubpenalty = 10000
\widowpenalty = 10000
\displaywidowpenalty = 10000

\thispagestyle{empty}
\begin{titlepage}
	
\begin{textblock}{65}(106,-16)
\includegraphics[width=0.5\textwidth]{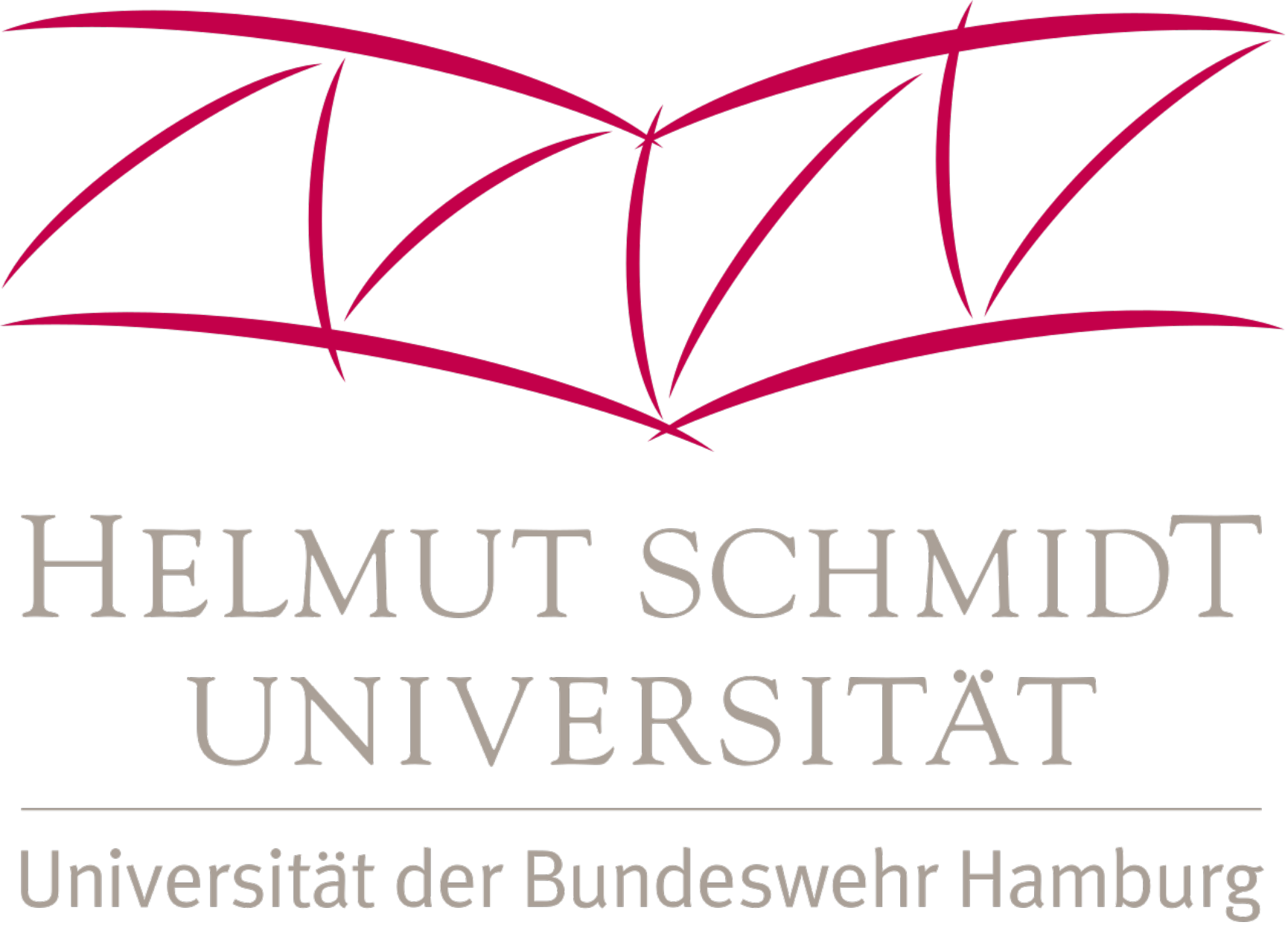}
\end{textblock}
					
\begin{textblock}{65}(80,-1.5)
\includegraphics[width=1.33\textwidth]{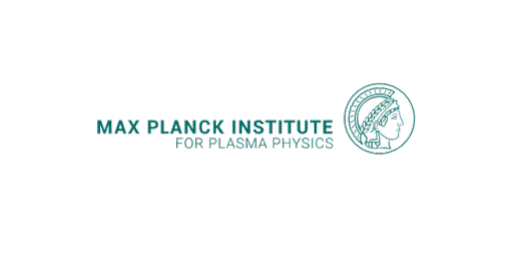}
\end{textblock}

\textcolor[RGB]{140,130,121}{	
\begin{textblock}{80}(0,-23)
\renewcommand{\baselinestretch}{1.1}\footnotesize
  \textsf{Helmut-Schmidt-University/ 
University of the \\German Federal Armed Forces Hamburg\\
  	Faculty of Mechanical and Civil Engineering\\ 
 	Institute for Applied Mathematics\\
	Prof. \hspace{-0.75mm}Dr. \hspace{-0.75mm}rer. \hspace{-0.75mm}nat. \hspace{-0.35mm}Thomas Carraro}
\end{textblock}}

\textcolor[RGB]{140,130,121}{	
\begin{textblock}{80}(0,-1)
\renewcommand{\baselinestretch}{1.1}\footnotesize
\textsf{Max Planck Institute for Plasma Physics, Garching\\
Tokamak Theory Division\\Dr. \hspace{-0.45mm}Gergely Papp}
\end{textblock}}

\textcolor[RGB]{140,130,121}{
\begin{textblock}{65}(0,6)										{\large\textsf{\underline{~~~~~~~~~~~~~~~~~~~~~~~~~~~~~~~~~~~~~~~~~~~~~~~~~~~~~~~~~~~~~~~~~~~~~~~~~~~~~~~~~~~~~~~~~~~~~}}}            
\end{textblock}}
	
\begin{textblock}{65}(44,29.5)
\begin{color}{black}											{\Huge \textbf{\textsf {Master Thesis}}}            
\end{color}
\end{textblock}																	
\begin{textblock}{120}(0,60.5)									
\textcolor[RGB]{165,0,52}{				
\textbf{{\Large Benjamin Buchholz}}}
\end{textblock}	

\begin{textblock}{130}(0,83.5)
\textcolor[RGB]{140,130,121}{
\textbf{{\Large Calculation of the runaway electron current in tokamak disruptions \\[1.2cm] 
Berechnung des Runaway-Elektronen-Stromes in Tokamak-Disruptionen}} \\ [1.75cm]
\begin{tabular}{@{}cl@{}}
{\Large Supervisors:} & {\Large Prof. \hspace{-0.75mm}Dr. \hspace{-0.75mm}rer. \hspace{-0.75mm}nat. \hspace{-0.35mm}Thomas Carraro}\\
&{\Large Dr. \hspace{-0.45mm}Gergely Papp}\\
\end{tabular}\\[1.75cm]
{\Large Hamburg, July 31st, 2023}}

\end{textblock}

\begin{textblock}{100}(-30,135)												
 \includegraphics{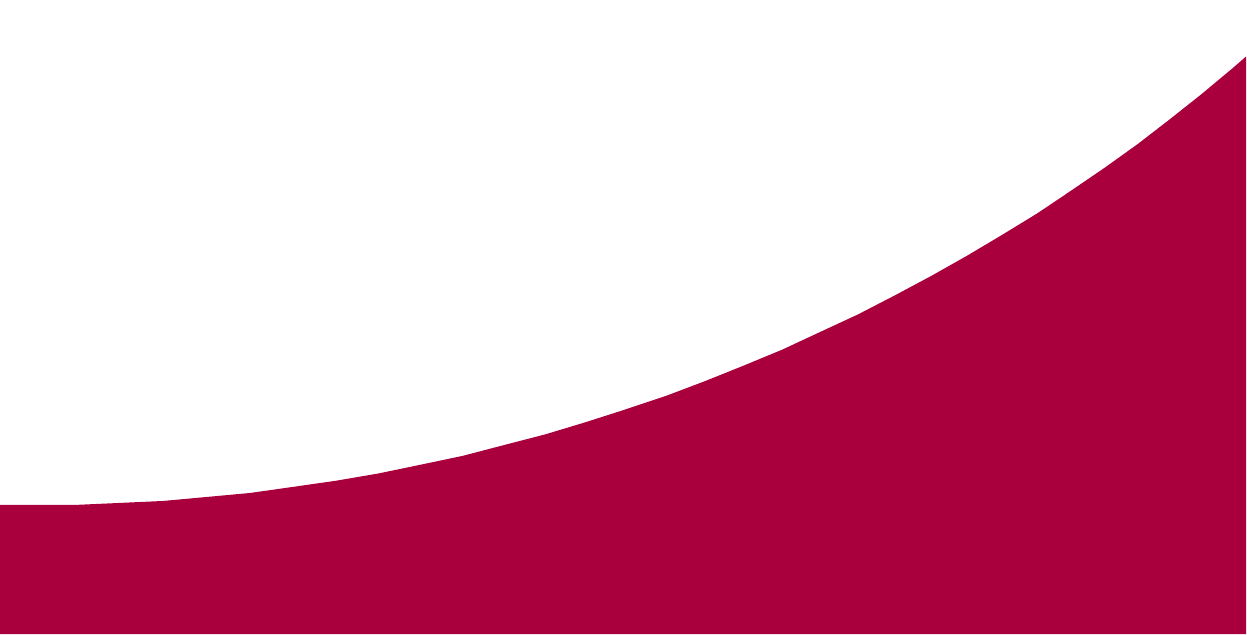}
\end{textblock}

\end{titlepage}
\newpage\thispagestyle{empty}\mbox{}\newpage 
\begin{titlepage}
\newgeometry{left=30mm,right=25mm,bottom=20mm,top=15mm}

\begin{center}
\begin{tabular}{p{\textwidth}}
\vspace{-2cm}
\centering
\begin{minipage}[c]{0.4\textwidth}
\hspace{-2.1cm}\includegraphics[scale=1.3]{IPP_eng.pdf}
\end{minipage}
\hspace{2.8cm}
\begin{minipage}[c]{0.2\textwidth}
\includegraphics[width=\textwidth]{Intellus-Partner-HSU.pdf}
\end{minipage}

\vspace{-0.76cm}

\large{Helmut-Schmidt-University/ \\
University of the German Federal Armed Forces Hamburg}\\[3pt]
\normalsize{Faculty of Mechanical and Civil Engineering\\
Institute for Applied Mathematics}

\vspace{0.83cm}

\LARGE{\textsc{
Calculation of the runaway electron current in tokamak disruptions}}

\vspace{0.84cm}

\textbf{\LARGE{Master Thesis}}

\vspace{1.07cm}

\small{for the acquisition of the academic degree} 

\large{Master of Science (M.Sc.)}

\vspace{0.63cm}

\normalsize{submitted by}

\vspace{0.53cm}

\large{\textbf{Benjamin Buchholz}} \\[1pt]
\normalsize{born on June 24th, 1998 in Dresden }

\vspace{0.64cm}

\normalsize{supervised by}

\vspace{0.53cm}

\large{\textbf{Dr. \hspace{-0.45mm}Gergely Papp}} 

\vspace{0.2cm}

Max Planck Institute for Plasma Physics, Garching\\[1pt]
\normalsize{Tokamak Theory Division }

\vspace{1.5cm}

\begin{tabular}{lllll}
\textbf{Student ID Number:} & & & 00893529\\ 
\textbf{Date of submission:} & & & July 31st, 2023 \\
\textbf{First examiner:} & & & Prof. \hspace{-0.75mm}Dr. \hspace{-0.75mm}rer. \hspace{-0.75mm}nat. \hspace{-0.35mm}Thomas Carraro\\
\textbf{Second examiner:} & & & Dr. \hspace{-0.45mm}Gergely Papp \\
\end{tabular}

\end{tabular}
\end{center}
\end{titlepage}

\thispagestyle{empty}

\begin{titlepage}
\newgeometry{left=30mm,right=25mm,bottom=20mm,top=17mm}

\begin{center}
\begin{tabular}{p{\textwidth}}

\centering
\begin{minipage}[c]{0.4\textwidth}
\includegraphics[scale=0.4]{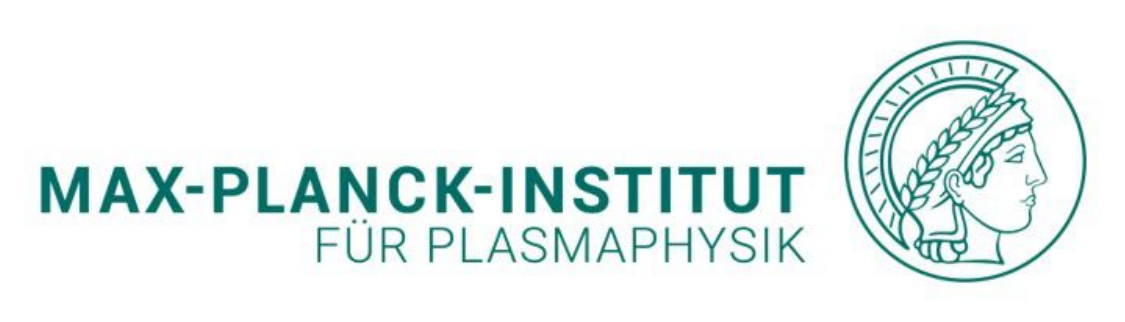}
\end{minipage}
\hspace{3cm}
\begin{minipage}[c]{0.2\textwidth}
\includegraphics[width=\textwidth]{Intellus-Partner-HSU.pdf}
\end{minipage}

\vspace{0.8cm}

\large{Helmut-Schmidt-Universit\"at/ \\
Universit\"at der Bundeswehr Hamburg}\\[3pt]
\normalsize{Fakult\"at f\"ur Maschinenbau und Bauingenieurwesen \\
Institut f\"ur Angewandte Mathematik }

\vspace{0.69cm}

\LARGE{\textsc{
Berechnung des Runaway-Elektronen-Stromes in Tokamak-Disruptionen}}

\vspace{0.70cm}

\textbf{\LARGE{Masterarbeit}}

\vspace{0.90cm}

\small{zur Erlangung des akademischen Grades} 

\large{Master of Science (M.Sc.)}

\vspace{0.57cm}

\normalsize{vorgelegt von}

\vspace{0.48cm}

\large{\textbf{Benjamin Buchholz}} \\[1pt]
\normalsize{geboren am 24.\hspace{1.4mm}Juni\hspace{1.5mm}1998 in Dresden}

\vspace{0.57cm}

betreut von

\vspace{0.48cm}

\large{\textbf{Dr. \hspace{-0.45mm}Gergely Papp}} 

\vspace{0.2cm}

Max-Planck-Institut f\"ur Plasmaphysik, Garching\\[1pt]
\normalsize{Bereich Tokamaktheorie}

\vspace{1.38cm}

\begin{tabular}{lllll}
\textbf{Martikelnummer:} & & & 00893529\\ 
\textbf{Abgabedatum:} & & & 31.\hspace{1.15mm}Juli 2023\\
\textbf{Erstgutachter:} & & & Prof. \hspace{-0.75mm}Dr. \hspace{-0.75mm}rer. \hspace{-0.75mm}nat. \hspace{-0.35mm}Thomas Carraro\\
\textbf{Zweitgutachter:} & & & Dr. \hspace{-0.45mm}Gergely Papp\\
\end{tabular}

\end{tabular}
\end{center}
\end{titlepage}

\thispagestyle{empty}

\addchap*{Eidesstattliche Erklärung}
\label{cap:Erklaerung}

Hiermit erkläre ich, Benjamin Buchholz, die vorliegende Arbeit selbstständig angefertigt zu haben. Die Erstellung erfolgte ohne das unerlaubte Zutun Dritter. Alle Hilfs-mittel, die für die Erstellung der vorliegenden Arbeit benutzt wurden, befinden sich ausschließlich im Literaturverzeichnis. Alles, was aus anderen Arbeiten unverändert oder mit Abänderungen übernommen wurde, ist kenntlich gemacht.

Die Arbeit wurde bisher keiner anderen Prüfungsbehörde vorgelegt. 

\vspace{2em}
Hamburg, 31.\hspace{1.15mm}Juli\hspace{1.5mm}2023

\vspace{1em}

\rule[-1cm]{5cm}{0.1mm}\\[1em]
Benjamin Buchholz

\clearpage

\pagenumbering{Roman} 

\phantomsection 

\addcontentsline{toc}{chapter}{Abstract}

\setcounter{page}{6}

\begin{abstract}
\vspace*{8.5mm}  
{\Large \textbf{Abstract}}
\noindent \\ \vspace*{-4mm} \\ \noindent 
The research on \textit{tokamak} fusion reactors, which have an intrinsic toroidal plasma current \cite{wesson}, is funded \cite{Euratom}, because it is expected to contribute to the fulfillment of the increasing future energy demand \cite{IEA_report,EPSpostitionpaper}. At this, the prediction, avoidance and mitigation of a \textit{disruption} \cite{Hoppe_2022}, which defines the abrupt loss of the magnetic confinement and rapid plasma cooling, i.a.\ due to plasma instabilities \cite{Hender_2007,REsimulation}, is relevant. Tokamak disruptions can give rise to the \textit{runaway phenomenon} \cite{Smith_2009}, which is typical in plasma physics \cite{Gurevich1994,Holman1985} and describes here the almost unbound acceleration of electrons to relativistic velocities and can lead to the formation of a \textit{runaway electron beam} \cite{stahl,REdistfuncderivation}. In tokamak reactors like ITER \cite{Hoppe_2021,REsimulation,Hoppe_2022}, impacts of such a beam can damage the reactor wall \cite{Bazylev_2011,Breizman_2019,Matthews2016,Reux2015}. This motivates the development of computationally efficient and accurate simulation methods for the runaway electron current \cite{Hoppe_2022}.
 
In present simulation software \cite{Hoppe_2022}, the \textit{reduced kinetic modeling} approach is used, which can be extended by using physically relevant moments of analytical runaway electron distribution functions. Because of this, calculation schemes for moments in connection with the density, the average velocity and the average kinetic energy of runaway\linebreak electrons are deduced in this work and analysed with the help of \textsc{MATLAB}-imple-\linebreak mentations. At that, the screening effects of partially ionized impurities and different representations of the runaway electron generation region in momentum space are taken into account.
\\
First, numerical calculation rules for the primary \textit{hot-tail} generation mechanism for isotropic and anisotropic two-dimensional descriptions of the runaway region are stated, which allow computations with standard quadrature formulas. The derived calculation schemes are then evaluated using the results of an ITER disruption simulation \cite{Smith_2009}.
\\
After that, calculation concepts for said moments, related to the secondary \textit{avalanche} generation mechanism, are derived. Different lower momentum boundaries for the runaway region and the influence of the partial screening of the nucleus by bound electrons are discussed on the basis of results calculated for different density combinations of a singly ionized deuterium-neon plasma.

It is proven, that the calculation of certain moments of distribution functions allows for the rapid investigation of physical quantities and is therefore suitable for parameter studies, assessing the applicability of assumptions or approximations, and expanding understanding. Finally, the validity and applicability of the analysed calculation schemes is examined.
\end{abstract}
\begin{abstract}
\vspace*{-11.0mm} 
{\Large \textbf{Kurzzusammenfassung}}
\noindent \\ \vspace*{-4mm} \\ \noindent 
Die Forschung an \textit{Tokamak}-Fusionsreaktoren, welchen ein toroidalen Plasmastrom intrinsisch ist \cite{wesson}, wird gefördert \cite{Euratom}, da erwartet werden kann, dass sie zur Deckung des steigenden Energiebedarfes der Zukunft beitragen \cite{IEA_report,EPSpostitionpaper}. Relevant ist dabei die Vorhersage, Vermeidung und Abschwächung einer \textit{Disruption} \cite{Hoppe_2022}, welche den abrupten Verlust des magnetischen Einschlusses und eine schnelle Plasma-Abkühlung u.\ a.\ durch Plasmainstabilitäten definiert \cite{Hender_2007,REsimulation}. Tokamak-Disruptionen können das, in der Plasmaphysik typische \cite{Gurevich1994,Holman1985}, \textit{Runaway-Phänomen} \cite{Smith_2009} hervorrufen, welches hier die nahezu ungebundene Beschleunigung von Elektronen auf relativistische Geschwindigkeiten be-\linebreak schreibt und zur Bildung eines \textit{Runaway-Elektronen-Strahles} führen kann \cite{stahl,REdistfuncderivation}. In Tokamak-Reaktoren wie ITER \cite{Hoppe_2021,REsimulation,Hoppe_2022}, können Einschläge eines solchen Strahles die Reaktorwand beschädigen \cite{Bazylev_2011,Breizman_2019,Matthews2016,Reux2015}. Dies motiviert die Entwicklung recheneffizienter und genauer Simulationsmethoden des Runaway-Elektronen-Stromes \cite{Hoppe_2022}.
 
In aktueller Simulationssoftware \cite{Hoppe_2022} findet der \textit{Reduced Kinetic Modeling}-Ansatz Anwendung, welcher durch die Verwendung physikalisch relevanter Momente analytischer Runaway-Elektronen-Verteilungsfunktionen erweitert werden kann. Aufgrund dessen werden in dieser Arbeit Berechnungsschemata für Momente im Zusammenhang mit der Dichte, der mittleren Geschwindigkeit und der mittleren kinetischen Energie von Runaway-Elektronen aufgestellt und mit Hilfe von \textsc{MATLAB}-Implementierungen ana-\linebreak lysiert. Dabei werden die Abschirmeffekte teilweise ionisierter Verunreinigungen sowie unterschiedliche Darstellungen der Runaway-Elektronen-Erzeugungsregion im Impuls-\linebreak raum berücksichtigt.
\\
Zunächst werden numerische Berechnungsregeln für den primären \textit{Hot-Tail}-Generie-\linebreak rungsmechanismus für isotrope und anisotrope zweidimensionale Beschreibungen der Runaway-Region angegeben, die eine Berechnung mittels Standardquadraturformeln ermöglichen. Die Auswertung der abgeleiteten Berechnungsschemata erfolgt daraufhin unter Verwendung der Ergebnisse einer ITER-Disruptionssimulation \cite{Smith_2009}.
\\
Anschließend werden Rechenkonzepte für die besagten Momente abgeleitet, die sich auf den sekundären \textit{Avalanche}-Erzeugungsmechanismus beziehen. Basierend auf Rechenergebnissen für verschiedene Dichtekombinationen eines einfach ionisierten Deuterium-Neon-Plasmas werden mehrere untere Impulsgrenzen für die Runaway-Region sowie die Einflüsse der Teilabschirmung des Kerns durch gebundene Elektronen diskutiert.

Es wird bewiesen, dass die Berechnung bestimmter Momente von Verteilungsfunktionen die schnelle Untersuchung physikalischer Größen ermöglicht und sie somit für Parameterstudien, die Beurteilung der Anwendbarkeit von Annahmen oder Näherungen sowie zur Verständniserweiterung geeignet ist. Schlussendlich wird die Gültigkeit und Anwendbarkeit der analysierten Berechnungsschemata untersucht.
\end{abstract}

\phantomsection

\addcontentsline{toc}{chapter}{Acknowledgement}

\addchap*{Acknowledgements}

In preparation and throughout the process of writing this thesis I have received support, assistance and motivation, without which I would not have been able to cope with a subject beyond the study of mechanical engineering.
\\[11pt]
First, I would like to thank my supervisor, Dr.\hspace{1.2mm}Gergely Papp, whose expertise and experience was crucial within the process of analyzing and understanding. His insightful feedback and support helped me in dealing with my complex subject and allowed me to make progress easily. Furthermore, I am grateful for the several discussions and meetings which helped me to extend my knowledge and were able to motivate me.
\\[11pt]
Moreover, I want to acknowledge my first examiner \mbox{Prof. Dr. rer. nat. Thomas} Carraro, who supported the cooperation with the Max Planck Institute for Plasma Physics, was always open for questions and very interested and open-minded about the subject.
\\[11pt]
As well, I would like to thank \mbox{Prof. Dr. Frank Jenko} for making this collaboration with Max Planck Institute for Plasma Physics possible, and for his continuous encouragement of the project.
\\[11pt]
I would like to thank \mbox{Dr. Mathias Hoppe} for the original idea behind the research project, and both him and Peter Halldestam for their help with the work and in particular the ongoing DREAM implementation.
\\[11pt]
Finally, I am grateful for the motivation and confirmation, that I received from my family, helping me to never lose sight of my aims. Therefore, I dedicate this work to my dear grandparents. 

\pdfbookmark[chapter]{Contents}{contents}

\tableofcontents

\clearpage

\input{Formelzeichen}

\clearpage

\pagenumbering{arabic}	

\input{Introduction}

\input{Kinetic_theory_of_electrons_in_tokamak_plasmas}

\input{Calculation_of_the_moments_of_the_hot_tail_runaway_electron_distribution_function}

\input{Calculation_of_the_moments_of_the_avalanche_runaway_electron_distribution_function}

\input{outlook}

\phantomsection 

\addcontentsline{toc}{chapter}{Bibliography}

\interlinepenalty=10000 

\bibliography{Master_Ref}

\interlinepenalty=4000

\clearpage

\listoffigures

\listoftables

\appendix

\input{Appendix}

\newpage\thispagestyle{empty}\mbox{}\newpage

\end{document}

%% file: Formelzeichen.tex
\addchap{Nomenclature}

\textbf{Latin letters}
\begin{table}[H]
\renewcommand{\arraystretch}{1.05}
\newcolumntype{s}{>{\hsize=.27\hsize}X}
\newcolumntype{S}{>{\hsize=.52\hsize}X}
\begin{tabularx}{\textwidth}{sSX}
\toprule
Symbol				&				Unit				& 				Denotation \\    
\midrule
$a$ &\si{\meter}& minor radius \\

$B$\,=\,$\vert\mathbf{B}\vert$		&				\si{\tesla}\,=\,\si{\kilogram\per\ampere\per\square\second}		&	magnetic flux density \\

$\mathcal{E}$&				\si{\joule}\,=\,\si{\kilogram\meter\squared\per\second\squared}		&	energy\\

$E$\,=\,$\vert\mathbf{E}\vert$		&					\si{\volt\per\meter}\,=\,\si{\kilogram\meter\per\second\per\cubed\ampere}		&	electric field strength \\

$f_{\alpha}$ &				-		&	\mbox{distribution function of the particle species $\alpha$} \\

$F$\,=\,$\vert\mathbf{F}\vert$&					\si{\kilogram\meter\square\per\second} & force\\

$I$		&				\si{\ampere}		&	current strength \\

$j$\,=\,$\vert\mathbf{j}\vert$		&				\si{\ampere\per\square\meter}		&	current density \\

$k$		&				\si{\joule\per\kilogram}\,=\,\si{\meter\squared\per\second\squared}		&	mass-related kinetic energy density \\

$K$		&				\si{\joule}\,=\,\si{\kilogram\meter\squared\per\second\squared}		&	kinetic energy\\

$\ln{\hspace{-0.45mm}\Lambda}$& -&\textit{Coulomb} logarithm\\

$m_{\alpha0}$ &				\si{\kilogram}		&	rest mass of the particle species $\alpha$ \\


$M_{\alpha}^{m}$ & -& $m$th moment of a distribution function $f_{\alpha}$\\

$n_{\alpha}$		&				\si{\per\cubic\meter}		&	particle density of particle species $\alpha$ \\

$N_{\alpha}$		&				-	&	number of particles of a species $\alpha$ \\

$N_{p}$		&				-	&	number of plasma components \\

$p$\,=\,$\vert\mathbf{p}\vert$&					\si{\kilogram\meter\per\second} &  relativistic momentum\\

$q_{\alpha}$		&				\si{\ampere\second}		&	electric charge of particle species $\alpha$ \\

$\mathbf{r}$		&					\si{\meter} &position vector\\

$r_{\perp}$		&					\si{\meter} &cylindrical radius\\

$R_{0}$ &\si{\meter}& major radius \\

$t$		&				\si{\second}		&	time \\

$T$		&				\si{\kelvin}		&	temperature \\

$u$\,=\,$\abs{\mathbf{u}}$		&					\si{\meter\per\second} & mean bulk velocity\\
 
$v$\,=\,$\abs{\mathbf{v}}$		&					\si{\meter\per\second} &velocity\\

$\mathbf{z}=(\mathbf{r},\,\mathbf{p})$&					- & phase space state vector\\

$Z$ &				-		&	nuclear charge number \\

$Z_{eff}$ &				-		&	effective ion charge  \\
\bottomrule						
\end{tabularx}
\end{table}

\clearpage
\textbf{Greek letters}
\begin{table}[H]
\renewcommand{\arraystretch}{1.05}
\newcolumntype{s}{>{\hsize=.26\hsize}X}
\newcolumntype{S}{>{\hsize=0.95\hsize}X}
\newcolumntype{P}{>{\hsize=0.93\hsize}X}
\begin{tabularx}{\textwidth}{sSP}
\toprule
Symbol				&				Unit				& 				Denotation \\      
\midrule
$\alpha,\,\beta$		&			-	&	particle species index  \\
$\chi$		&			-	&	numerical constant  \\
$\delta$ & - & absolute difference \\
$\Delta$ & - & relative deviation \\
$\eta$		&		\si{\volt\meter\per\ampere}$\;=\;$\si{\kilogram\meter\cubed\per\second\cubed\per\square\ampere}	&	electrical resistivity \cite{Stroth_2018}\\
$\nu$		&			\si{\per\second}	&	characteristic frequency \\
$\tau$		&			\si{\second}	&	characteristic time scale \\
$\gamma$		&			-	&	\textit{Lorentz} or gamma factor \\
$\Gamma$		&			\si{\per\second\per\meter\cubed}	&	growth rate \\
$\kappa$ & - & numerical parameter\\
$\sigma$ & \si{\ampere\per\volt\per\meter}$\;=\;$\si{\square\ampere\second\cubed\per\kilogram\per\meter\cubed} & electrical conductivity\\
$\theta$		&			radiant	&	pitch/polar angle \\
$\Theta$		&			-	&	normalized temperature \\
$\phi$ &			-	&	physical/numerical parameter  \\
$\varphi$		&			radiant	&	azimuthal angle  \\
$\xi$		&			-	&	pitch coordinate \\
$\zeta$		&			-	&	abbreviatory constant \\
\bottomrule						
\end{tabularx}
\end{table}

\textbf{Physical and mathematical constants}
\begin{table}[H]
\renewcommand{\arraystretch}{1.2}
\newcolumntype{s}{>{\hsize=.24\hsize}X}
\newcolumntype{S}{>{\hsize=0.94\hsize}X}
\newcolumntype{P}{>{\hsize=0.93\hsize}X}
\begin{tabularx}{\textwidth}{sSP}
\toprule
Symbol		&				Value and Unit				& 				Denotation \\    
\midrule

$\pi$ & $3.14159265358979323846264$&	\mbox{ratio of a circle's circumference to} its diameter \cite{WolframPI} \\

e & $2.71828182845904523536028$&	\textit{Euler's} number \cite{OEISe} \\

$c$  &				$2.99792458 \cdot 10^{8}$ \si{\meter\per\second}		&	speed of light in vacuum \cite{NISTc}\\

$e $ &				$1.602176634 \cdot 10^{-19}$ \si{\ampere\second}		&	elementary charge \cite{NISTe} \\

$m_{e0}$  &				$9.1093837015 \cdot 10^{-31}$ \si{\kilogram}		&	electron rest mass \cite{NISTme}\\

$\mu_{0} $&				$1.25663706212\cdot 10^{-6}$ \si{\newton\per\square\ampere}		& \mbox{vacuum magnetic permeability \cite{NISTmue0}}  \\

$k_{B} $&				$8.617333262\cdot 10^{-5}$ \si{\electronvolt\per\kelvin}		& \textit{Boltzmann} constant \cite{NISTkB}  \\

$\varepsilon_{0}$  &				\mbox{$8.8541878128\cdot10^{-12}$  \si{\ampere\second\per\volt\per\meter}}		&	
vacuum electric permittivity \cite{NISTeps0}\\

$\textup{eV}$  &	$e\cdot 1\textup{V}=1.602176634\cdot10^{-19}$ \si{\ampere\volt\second}			 	&	
electron volt \cite{NISTeV}\\

\bottomrule						
\end{tabularx}
\end{table}
\clearpage

\textbf{Mathematical and physical symbols as well as operators}
\begin{table}[H]
\renewcommand{\arraystretch}{1.25}
\newcolumntype{s}{>{\hsize=.17\hsize}X}
\newcolumntype{S}{>{\hsize=.84\hsize}X}
\begin{tabularx}{\textwidth}{sXS}
\toprule
Symbol				&				representation				& 				Denotation \\      
\midrule

$ \mathbf{e}_{i}$			&	 $ \mathbf{e}_{i}\coloneqq( a_{j})_{j=\left\lbrace 1,2,...,d\right\rbrace};  $		$\,\vert\mathbf{e}_{i}\vert=1$&	 $i$th unit vector in $d$ dimensions \\[11pt]

%
$ \mathcal{H}_{b}(x)	$			&		$  \mathcal{H}_{b}(x) \coloneqq  \mathcal{H}(x-b)=\biggl\lbrace\begin{aligned}
\,0\,\,;\,\, x<b \\[-7pt]
\,1 \,\,;\,\, x\geq b
\end{aligned} $			&	\mbox{\textit{Heaviside} function \cite{WolframHeaviside}}\\[15pt]

$ \textup{exp}(x)	$			&		$ \textup{exp}(x)\equiv\textup{e}^{x} $			&	exponential function \cite{WolframEXP}\\[11pt]

$ \textup{erf}(z)	$			&		$ \textup{erf}(z)=\dfrac{2}{\sqrt{\pi}}\,\displaystyle{\int\limits^{z}_{x=0}} \,\textup{e}^{-x^2}\, \mathrm{d}x$			&	error function \cite{helander}\label{erf_label}\\[20pt] 

$ \textup{erfc}(z)	$			&		$ \textup{erfc}(z)=1-\textup{erf}(z)  $			&	\mbox{complementary error function \cite{WolframERFC}} \\[10pt]

\vspace{1.4mm}$ \mathcal{K}_{2}(x)	$			&	\vspace{1.4mm}	$ \mathcal{K}_{2}(x)\approx \sqrt{\dfrac{\pi}{2x}}\cdot\textup{e}^{-x}   \;\;  ; \;\; x \gg 3.75$			&	approximation for the second-order modified \textit{Bessel} function of the second
kind \cite{Stroth_2018}\\[24pt]

%
%
%

$ \gamma	$			&		$ \gamma=\gamma(v)=\dfrac{1}{\sqrt{1-\left(\dfrac{v}{c}\right)^2}}$			&	\vspace{-4.6mm}\mbox{\textit{Lorentz} or gamma factor \cite{Stroth_2018}} \\[28pt] 

$\ln{\hspace{-0.45mm}\Lambda_{th}}	$	&	 $\ln{\hspace{-0.45mm}\Lambda_{th}}\approx 14.9-0.5\cdot\ln{\hspace{-0.6mm}\left(\hspace{-0.35mm}10^{-20}\hspace{-0.3mm}\cdot\hspace{-0.05mm} n_{\mathrm{e}}\left[\mathrm{m}^{-3}\right] \hspace{-0.15mm}\right)} $			&	\mbox{relation for the thermal} \\[4pt]
 &	 \hspace{1.55cm}$  +\; \ln{\hspace{-0.5mm}\left(\hspace{-0.35mm}10^{-3}\hspace{-0.3mm}\cdot\hspace{-0.05mm}k_{B}T_e\left[\mathrm{eV}\right]  \hspace{-0.15mm}\right)}$			& \mbox{\textit{Coulomb} logarithm \cite{wesson,Hesslow_2018II}} \\[7pt] 

\vspace*{-2.5mm}$\ln{\hspace{-0.45mm}\Lambda_{rel}}	$	&	\vspace*{-3.2mm} $ \ln{\hspace{-0.45mm}\Lambda_{rel}} \approx 14.6+\dfrac{1}{2}\cdot\ln{\hspace{-0.5mm}\left( \hspace{-0.25mm}\dfrac{ k_{\mathrm{B}}T_{\mathrm{e}}\left[\mathrm{eV}\right] }{10^{-20}\hspace{-0.3mm}\cdot\hspace{-0.05mm} n_{\mathrm{e}}\left[\mathrm{m}^{-3}\right]}\hspace{-0.15mm}\right)}$	 	&	 \mbox{relation for the relativistic} \textit{Coulomb} logarithm \cite{Svensson_2021,Hesslow_2018II}\\[24pt]

\vspace{-3.0mm} $\ln{\hspace{-0.45mm}\Lambda}	$	& \vspace{-4.9mm}$\ln{\hspace{-0.45mm}\Lambda} = \ln{\hspace{-0.45mm}\Lambda_{th}} + \dfrac{1}{\kappa}\cdot\ln{\hspace{-0.6mm}\left(\hspace{-0.25mm}1+\left(\dfrac{2\hspace{0.2mm}p}{p_{th}}\right)^{\hspace{-1.3mm}\kappa}  \right)}$&	 \mbox{relation for the energy-dependent} \textit{Coulomb} logarithm \cite{Hesslow_2018II}\\[20pt]
 
\bottomrule						
\end{tabularx}
\end{table}

\clearpage

\textbf{Spherical position and momentum space coordinates}
\begin{table}[H]
\renewcommand{\arraystretch}{0.75}
\newcolumntype{s}{>{\hsize=.2\hsize}X}
\newcolumntype{D}{>{\hsize=0.872\hsize}X}
\begin{tabularx}{\textwidth}{sDX}
\toprule
Symbol				&				representation				& 				Denotation \\      
\midrule\\[-2pt]
$ r$			&	 $ r \in  \mathbb{R}^{+} \coloneqq	\left[0,\,\infty\right)	$		&	spherical radius \cite{Bartelmann_2015} \\[9pt]

$ p$			&	 $ p \in  \mathbb{R}^{+} \coloneqq	\left[0,\,\infty\right)	$		&	radial momentum coordinate \cite{Bartelmann_2015} \\[9pt]

$ \theta$			&	 $ \theta \in  \left[0,\,\pi\right]	$		&	polar angle \cite{Bartelmann_2015} \\[9pt]

$ \varphi$			&	 $ \varphi \in  \left[0,\,2\pi\right)		$		&	 azimuthal angle \cite{Bartelmann_2015} \\[8pt]

$ \mathbf{e}_{r}$			&	 $  \mathbf{e}_{r}\coloneqq  \begin{pmatrix} 
\sin{\left(\theta\right)}\cos{\left(\varphi\right)}   \\ 
\sin{\left(\theta\right)}\sin{\left(\varphi\right) }   \\ 
 \cos{\left(\theta\right)}    \\
\end{pmatrix} 	$		&	radial basis vector \cite{Bartelmann_2015} \\[21pt]

$ \mathbf{e}_{\theta}$			&	 $  \mathbf{e}_{\theta}\coloneqq  \begin{pmatrix}
\cos{\left(\theta\right)}\cos{\left(\varphi\right) }  \\ 
\cos{\left(\theta\right)}\sin{\left(\varphi\right)  }  \\ 
 -\sin{\left(\theta\right)}    \\
\end{pmatrix} 	$		&	polar basis vector \cite{Bartelmann_2015}\\[21pt]

$ \mathbf{e}_{\varphi}$		&	 $  \mathbf{e}_{\varphi}\coloneqq  \begin{pmatrix}
-\sin{\left(\varphi\right)}   \\ 
\cos{\left(\varphi\right) }   \\ 
 0    \\
\end{pmatrix} 	$		&	azimuthal basis vector \cite{Bartelmann_2015}\\[23pt]

$ \mathbf{r}$			&	 $ \mathbf{r}\coloneqq r\cdot\mathbf{e}_{r}\;;\;\;r\coloneqq\vert\mathbf{r} \vert	$		&	position vector  \\[10pt]

$ \mathbf{p}$			&	 $ \mathbf{p}\coloneqq p\cdot\mathbf{e}_{r}\;;\;\;p\coloneqq\vert\mathbf{p} \vert	$		&	momentum vector  
\end{tabularx}
\end{table}
\vspace*{-5mm}
\begin{table}[H]
\renewcommand{\arraystretch}{0.75}
\newcolumntype{s}{>{\hsize=.10\hsize}X}
\newcolumntype{S}{>{\hsize=.65\hsize}X}
\begin{tabularx}{\textwidth}{sX}

 	$ \mathrm{d}^3 r$			&	  	 position space volume element with \textit{Jacobian} determinant \cite{Bartelmann_2015}\\[8pt]

 	&	 $ \mathrm{d}^3 r=r^{2}\sin{\left(\theta\right)}\,\mathrm{d}r\,\mathrm{d}\theta \,\mathrm{d}\varphi	$		  \\[10pt]

$ \mathrm{d}^3 p$			&	  	 momentum space volume element with \textit{Jacobian} determinant \cite{Bartelmann_2015}\\[8pt]

 	&	 $ \mathrm{d}^3 p=p^{2}\sin{\left(\theta\right)}\,\mathrm{d}p\,\mathrm{d}\theta \,\mathrm{d}\varphi	$		  \\[13pt]
 	
%
%

$ \Delta_{\mathbf{r}}	$			 	&	\textit{Laplace}-Operator in spherical position space coordinates \cite{Bartelmann_2015} \\[7pt]

 	&		$\Delta_{\mathbf{r}}\coloneqq\dfrac{1}{r^{2}}\dfrac{\partial}{\partial r}\left(r^{2}\,\dfrac{\partial}{\partial r}\right)+ \dfrac{1}{r^{2}\sin{\left(\theta\right)}}\dfrac{\partial}{\partial \theta}\left(\sin{\left(\theta\right)}\,\dfrac{\partial}{\partial \theta}\right)+ \dfrac{1}{r^{2}\sin^{2}{\left(\theta\right)}}\dfrac{\partial^{2}}{\partial \varphi^{2}} $\\[20pt]
 	
 	$ \Delta_{\mathbf{p}}	$			 	&	\textit{Laplace}-Operator in spherical momentum space coordinates \cite{Bartelmann_2015} \\[7pt]

 	&		$\Delta_{\mathbf{p}}\coloneqq\dfrac{1}{p^{2}}\dfrac{\partial}{\partial p}\left(p^{2}\,\dfrac{\partial}{\partial p}\right)+ \dfrac{1}{p^{2}\sin{\left(\theta\right)}}\dfrac{\partial}{\partial \theta}\left(\sin{\left(\theta\right)}\,\dfrac{\partial}{\partial \theta}\right)+ \dfrac{1}{p^{2}\sin^{2}{\left(\theta\right)}}\dfrac{\partial^{2}}{\partial \varphi^{2}} $\\[17pt]

\bottomrule						
\end{tabularx}
\end{table}

\clearpage

\textbf{Cylindrical position and momentum space coordinates}
\begin{table}[H]
\renewcommand{\arraystretch}{0.75}
\newcolumntype{s}{>{\hsize=.2\hsize}X}
\newcolumntype{D}{>{\hsize=0.872\hsize}X}
\begin{tabularx}{\textwidth}{sDX}
\toprule
Symbol				&				representation				& 				Denotation \\      
\midrule
$ r_{\perp}$			&	 $ r_{\perp} \in  \mathbb{R}^{+} \coloneqq 	\left[0,\,\infty\right)	$		&	\mbox{cylindrical radius \cite{Bartelmann_2015}} \\[9pt]

$ p_{\perp}$			&	 $ p_{\perp} \in  \mathbb{R}^{+} \coloneqq 	\left[0,\,\infty\right)	$		&	\mbox{orthogonal momentum coordinate \cite{Bartelmann_2015}} \\[9pt]

$  r_{\|}$			&	 $ r_{\|} \in  \left(-\infty,\,\infty\right)	$		&	\mbox{parallel position coordinate \cite{Bartelmann_2015}} \\[9pt]

$  p_{\|}$			&	 $ p_{\|} \in  \left(-\infty,\,\infty\right)	$		&	\mbox{parallel momentum coordinate \cite{Bartelmann_2015}} \\[9pt]

$ \varphi$			&	 $ \varphi \in  \left[0,\,2\pi\right)		$		&	 \mbox{azimuthal angle \cite{Bartelmann_2015}} \\[7pt]

$ \mathbf{e}_{\perp}$			&	 $ \mathbf{e}_{\perp}\hspace{-0.2mm} \coloneqq \hspace{-0.2mm}\begin{pmatrix} 
\cos{\left(\varphi\right) } \\ 
\sin{\left(\varphi\right) } \\ 
 0   \\
\end{pmatrix} \;;\; \mathbf{e}_{\perp}\cdot\mathbf{B} =0$		&	\mbox{orthogonal basis vector \cite{Bartelmann_2015}} \\[22pt]

$ \mathbf{e}_{\|}$			&	 $ \mathbf{e}_{\|}\hspace{-0.2mm} \coloneqq \hspace{-0.2mm} \begin{pmatrix}
0 \\ 
0  \\ 
 1   \\
\end{pmatrix} \;;\; \mathbf{e}_{\|}\cdot\mathbf{B} =\vert\mathbf{B}\vert =B$		&	\mbox{basis vector parallel to vector $\mathbf{B}$ \cite{Bartelmann_2015}} \\[22pt]

$ \mathbf{e}_{\varphi}$		&	 $  \mathbf{e}_{\varphi}\coloneqq   \begin{pmatrix}
-\sin{\left(\varphi\right)}   \\ 
\cos{\left(\varphi\right) }   \\ 
 0    \\
\end{pmatrix} 	$		&	\mbox{azimuthal basis vector \cite{Bartelmann_2015}}\\[24pt]

$ \mathbf{r}$			&	 $ \mathbf{r}=\mathbf{r}_{\|}+\mathbf{r}_{\perp}=r_{\|}\cdot\mathbf{e}_{\|}+r_{\perp}\cdot\mathbf{e}_{\perp}	$		&	\mbox{position vector}\\[10pt]  

$ \mathbf{p}$			&	 $ \mathbf{p}=\mathbf{p}_{\|}+\mathbf{p}_{\perp}=p_{\|}\cdot\mathbf{e}_{\|}+p_{\perp}\cdot\mathbf{e}_{\perp}	$		&	\mbox{momentum vector}
\end{tabularx}
\end{table}
\vspace*{-5mm}
\begin{table}[H]
\renewcommand{\arraystretch}{0.75}
\newcolumntype{s}{>{\hsize=.10\hsize}X}
\newcolumntype{S}{>{\hsize=.65\hsize}X}
\begin{tabularx}{\textwidth}{sX}

 	$ \mathrm{d}^{3}r$			&	  	 position space volume element with \textit{Jacobian} determinant \cite{Bartelmann_2015}\\[8pt]

&	 $  \mathrm{d}^{3}r=r_{\perp} \,\mathrm{d}r_{\perp}\,\mathrm{d}r_{\|} \,\mathrm{d}\varphi	$ \\[10pt] 

 	$ \mathrm{d}^{3}p$			&	  	 momentum space volume element with \textit{Jacobian} determinant \cite{Bartelmann_2015}\\[8pt]
 	
 	&	 $  \mathrm{d}^{3}p=p_{\perp} \,\mathrm{d}p_{\perp}\,\mathrm{d}p_{\|} \,\mathrm{d}\varphi	$ \\[13pt] 
 	
$ \Delta_{\mathbf{r}}	$			 	&	\textit{Laplace}-Operator in cylindrical position space coordinates \cite{Bartelmann_2015} \\[7pt]

 	&		$\Delta_{\mathbf{r}}\coloneqq \dfrac{1}{r_{\perp}}\,\dfrac{\partial }{\partial\hspace{0.15mm}r_{\perp}\hspace{-0.9mm}}\hspace{-0.15mm}\left(\hspace{-0.2mm}r_{\perp}\,\dfrac{\partial }{\partial\hspace{0.15mm}r_{\perp}\hspace{-0.9mm}} \right) + \dfrac{1}{r_{\perp}^{2}}\dfrac{\partial^{2}}{\partial \varphi^{2}}+  \dfrac{\partial^{2}}{\partial r_{\|}^{2}} $\\[20pt]
 	
 	$ \Delta_{\mathbf{p}}	$			 	&	\textit{Laplace}-Operator in cylindrical momentum space coordinates \cite{Bartelmann_2015} \\[7pt]

 	&		$\Delta_{\mathbf{p}}\coloneqq \dfrac{1}{p_{\perp}}\,\dfrac{\partial }{\partial\hspace{0.15mm}p_{\perp}\hspace{-0.9mm}}\hspace{-0.15mm}\left(\hspace{-0.2mm}p_{\perp}\,\dfrac{\partial }{\partial\hspace{0.15mm}p_{\perp}\hspace{-0.9mm}} \right) + \dfrac{1}{p_{\perp}^{2}}\dfrac{\partial^{2}}{\partial \varphi^{2}}+  \dfrac{\partial^{2}}{\partial p_{\|}^{2}} $\\[17pt]
 	
%
%
\bottomrule						
\end{tabularx}
\end{table}

\clearpage

\textbf{Indices and abbreviations}
\begin{table}[H]
\renewcommand{\arraystretch}{0.9}
\newcolumntype{S}{>{\hsize=0.35\hsize}X}
\begin{tabularx}{\textwidth}{SX}
\toprule
Symbol				&				meaning \\      
\midrule
acc & acceleration \\

analyt & analytical \\

Ar & argon \\

ava & \textit{avalanch}e generation mechanism \\

B & \textit{Boltzmann} \\

bd & bound \\

br & bremsstrahlung \\

c & critical \\

C & \textit{Coulomb}\\

comp. scr. & complete screening limit \\

CQ & current quench \\

d & deflection \\

D & \textit{Dreicer}\\

DREAM & Disruption Runaway Electron Analysis Model \cite{Hoppe_2021}\\

eff & effective \\

E & electric field\\

e & electron \\

ee & electron-electron \\

\mbox{e.\hspace{0.9mm}g.} & exempli gratia \\

\mbox{et.\hspace{0.9mm}al.} &  et alia  \\

Euratom & Europ\"aische Atomgemeinschaft \\

exp & exponential function \\

fr & friction \\

g & gyro \\

H & hydrogen \\

He & helium \\

ht & \textit{hot-tail} generation mechanism \\

\mbox{i.\hspace{0.9mm}a.} & inter alia \\

I.b.P. & integration by parts \\

i & ion \\

IEA & International Energy Agency \cite{IEA_report}\\

IPCC & Intergovernmental Panel on Climate Change \\

ITER & International Thermonuclear Experimental Reactor \cite{ITER_old_2002}\\

JET & Joint European Torus\\

L & \textit{Lorentz}\\

M &\textit{Maxwell} \\

max & maximal \\

MGI & Massive Gas Injection \cite{PappSlides}\\

min & minimal \\

\bottomrule						
\end{tabularx}
\end{table}
\newpage
\begin{table}[H]
\renewcommand{\arraystretch}{0.9}
\newcolumntype{S}{>{\hsize=0.35\hsize}X}
\begin{tabularx}{\textwidth}{SX}     
\midrule
MMI & Massive Material Injection \cite{Vallhagen_2020_MMI}\\

n & neutron\\

Ne & neon \\

norm & normalized\\

num & numerical \\

$\Omega$ & ohmic \\

p & plasma \\

rel & relativistic \\

RE & runaway electrons \\

s & slowing-down \\

sa & slide-away\\

sep & separatrix\\

scr & partially screened\\

SI &  International System of Units \\

SPI & Shattered Pellet Injection \cite{PappSlides}\\

syn & synchrotron \\

th & thermal \\

tot & total \\

tp & test particle \\

TQ & thermal quench \\

Z & effective ion charge $Z_{eff}$\\

$\|$ & parallel \\

$\perp$ & orthogonal \\

\bottomrule						
\end{tabularx}
\end{table}

%% file: Introduction.tex
\chapter{Introduction}

\begin{figure}[H]
\begin{center}
\includegraphics[trim=53 405 51 74,width=0.67\textwidth,clip]{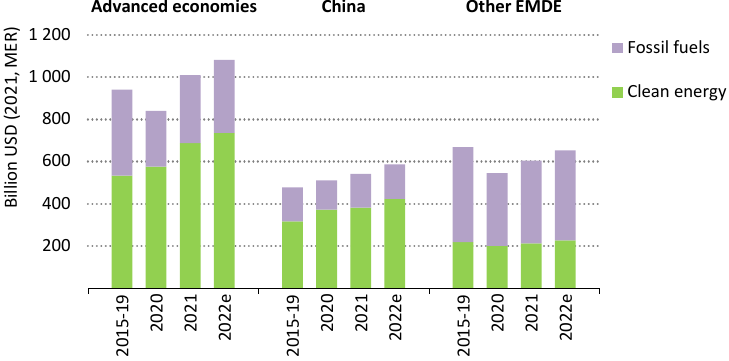}
\caption[\qq{Global energy investment by region}\hspace{0.7mm}(World Energy Outlook 2022\linebreak\cite{IEA_report}).]{\qq{Global energy investment by region}\protect\footnotemark{}\hspace{0.7mm}(World Energy Outlook 2022 \cite{IEA_report}).}
\label{green_invest_fig}
\end{center}
\end{figure}

Humanity is facing an overall rising energy demand \cite{IEA_report}. Moreover, it is forced to develop a sustainable energy production, in order to cope with the effects of the human-caused climate change as it is discussed in the most recent report by the \textit{Intergovernmental Panel on Climate Change} (IPCC) \cite{IPCC}. Furthermore, global energy security and resilience is needed for each country, which wants to preserve its independence or wants to reduce international dependencies connected to the import of fossil fuels. This is i.a.\ covered in the \textit{\qq{World Energy Outlook 2022}}\cite{IEA_report} published by the \textit{International Energy Agency} (IEA) and motivated by the current Ukrainian-Russian conflict. As a consequence of the finite reserves of fossil fuels and the emission of greenhouse gases like carbon dioxide accompanying their combustion, the future energy production has to focus on renewable energy technologies with low carbon footprints. This trend is already apparent and can for instance be verified on the basis of the development of green energy investments, as it is depicted in figure \ref{green_invest_fig} \cite{IEA_report}.\vspace{0.5mm} 
\footnotetext{\hspace{1mm}EMDE = emerging market and developing economies; MER = market exchange rate;\\ \hspace*{8.7mm}2015-19 indicates
average annual figure; 2022e = estimated values for 2022 \cite{IEA_report}.}
\vspace{-9.0mm}
Hence, the focus of industrial countries and strong emerging economies in Europe, Asia and America on green energy research is comprehensible and is e.g.\ expressed by the \textit{\qq{Euratom Research and Training Programme}}, which funds nuclear fusion research with 1.38 billion euros from 2021 until 2025 \cite{Euratom}. In this context, one important field of research are nuclear fusion reactors, because they promise an environmentally friendly and secure alternative to nuclear fission power plants, while providing a similar base load capability in energy supply. Especially, the last characteristic is vital in coping with the rising energy demand, since renewable energy technologies like photovoltaic, wind or tidal power intrinsically have a highly volatile energy production, which in the most unfavourable case cannot satisfy the energy base load demand \cite{IEA_report,EPSpostitionpaper}. 

\clearpage

\section{Nuclear fusion}\label{nuclear_fusion_section}

Today different concepts of nuclear fusion reactors exist. At that, the most known are the inertial and magnetic confinement fusion. Both rely on the ignition of a plasma, in which light nuclei unify to a heavier nucleus, while releasing excess nuclear binding energy. The most promising fusion reaction, due to its higher cross-section, and thus increased reactivity, in comparison to other fusion reactions, uses the hydrogen isotopes deuterium $_{1}^{2}\mathrm{H}$ and tritium $_{1}^{3}\mathrm{H}$ in their fully ionized forms \cite{wesson}:\vspace*{-3.0mm}
\begin{equation}\label{nuc_fus_reac}
_{1}^{2}\mathrm{H}^{+} \,+\;\, _{1}^{3}\mathrm{H}^{+} \;\longrightarrow  \;\; _{2}^{4}\mathrm{He}^{2+}\, +\;\, _{0}^{1}\mathrm{n}\;+\; 17.6\,\mathrm{MeV}\,.
\end{equation}
\vspace*{-10.5mm}\\The reaction products are a helium nucleus and a neutron carrying a kinetic energy of $3.5\,\mathrm{MeV}$ and $14.1\,\mathrm{MeV}$ \cite{wesson}. 
 
The most advanced research area in fusion plasma physics is the magnetic confinement of the plasma with current-carrying coils, whose arrangement and shape define the two main reactor types. At this, the so-called \textit{tokamak}-type reactors have toroidally symmetric coils, while a twisted magnetic torus consisting of three-dimensionally shaped coils is one of the main characteristics of the reactor type of the \textit{stellarator}.  
\\ 
Both systems rely on the basic procedure of a \textit{plasma discharge}, which starts with the heating of a neutral gas. The initial neutral gas fueling always contains a trace number of ions and free electrons, and through the application of heating, ionization rates close to one are reached, which leads to a large number of charged particles. Those approximately move along the magnetic field lines and are hence confined in the vacuum vessel of the fusion device. The consecutive heating deposits thermal energy and increases the degree of ionization within the plasma, which in the case of the tokamak enhances the magnetic confinement by itself. Consequently, an increasing fraction of the plasma particles reaches the necessary parameters, so that the desired nuclear fusion reaction takes place. The energy gained in this fusion reaction then adds to the plasma heating. If this self-heating, makes the external heating unnecessary, one speaks of a \textit{plasma ignition}. This means, that excess energy of the fusion reaction is partly self-heating the plasma, while the rest might be converted into electricity. Here it is important to understand that, not the whole amount of the released fusion energy can be used in further energy conversion, because for instance perturbations in the magnetic confinement, plasma impurities like the reaction products, neutron-caused plasma-wall interactions and radiative loss processes require the ongoing fueling and self-heating. From this one can understand, that a fusion plasma consists of ions, neutrals and electrons with different energies, momenta and trajectories, leading to the high complexity of the plasma physics in nuclear fusion reactors. In addition, photons of various energies and plasma influencing abilities are present inside the vacuum vessel of the reactor. 
 
A description of the plasma composition is possible, by means of the density vector $\mathbf{n}$, containing the densities of the individual plasma components defined in the vector $\boldsymbol{\alpha}$. For a research plasma with only singly-charged deuterium (\mbox{$\alpha_{1}\coloneqq\,_{1}^{2}\mathrm{H}^{+}$}) and neon (\mbox{$\alpha_{2}\coloneqq\, _{10}^{20}\mathrm{Ne}^{+}$}) ions, one can for example consider $N_{p}=2$ plasma components with:\vspace*{-3.5mm}
\begin{equation}\label{alpha_vec_def}
\boldsymbol{\alpha}\coloneqq \left[ \alpha_{k} \right]_{k=1}^{\mathrm{k}=  N_{p}}\overset{\scriptsize{\underbrace{N_{p}=2}}{}}{=}\left( _{1}^{2}\mathrm{H}^{+},\,_{10}^{20}\mathrm{Ne}^{+}\right)\,,
\end{equation}
\vspace*{-9.5mm}\\which leads to the following density vector with the component assigned unit $\mathrm{m}^{-3}$:\vspace*{-3.5mm}
\begin{equation}\label{n_vec_def}
\mathbf{n} \coloneqq \left[ n_{\alpha_{k}} \right]_{k=1}^{\mathrm{k}= N_{p}}\overset{\scriptsize{\underbrace{N_{p}=2}}{}}{=} \left(n_{_{1}^{2}\mathrm{H}^{+}},\,n_{_{10}^{20}\mathrm{Ne}^{+}}\right)\,. 
\end{equation}
\vspace*{-9.0mm}\\Each plasma species in the density vector, for instance neutral atoms or ions with different charge states, carries a certain charge, which is stored in the charge vector in units of the elementary charge $e$:\vspace*{-5.5mm} 
\begin{equation}\label{q_vec_def}
\mathbf{q} \coloneqq \left[ q_{\alpha_{k}} \right]_{k=1}^{\mathrm{k}= N_{p}} \overset{\scriptsize{\underbrace{N_{p}=2}}{}}{=} \left(q_{_{1}^{2}\mathrm{H}^{+}},\,q_{_{10}^{20}\mathrm{Ne}^{+}}\right) = \left(1,\,1\right) \,.
\end{equation}
\vspace*{-9.0mm}\\The free electrons from the ionization of the neutral atoms towards the charged species in the density vector define the free electron density:\vspace*{-3.5mm}
\begin{equation}\label{n_e_def}
n_{\mathrm{e}}= \mathbf{n}\cdot\mathbf{q}=\displaystyle{\sum\limits_{k=1}^{N_{p}}} \,n_{\alpha_{k}}q_{\alpha_{k}}\,.
\end{equation}
\vspace*{-8.5mm}\\In contrast, one finds the total electron density, respectively the sum of the free electron density $n_{\mathrm{e}}$ and the bound electron density $n_{e}^{\mathrm{bd}}$, from the number density of all electrons within the plasma. Therefore, the subsequently stated equation is suggested:\vspace*{-3.0mm}
\begin{equation}\label{n_tot_def}
\begin{split}
\begin{gathered}
n_{e}^{\mathrm{tot}}= n_{\mathrm{e}}+n_{e}^{\mathrm{bd}}= \mathbf{n}\cdot\mathbf{Z}= \displaystyle{\sum\limits_{k=1}^{N_{p}}} \,n_{\alpha_{k}} Z_{\alpha_{k}}\,.
\end{gathered}
\end{split}
\end{equation}
\vspace*{-8.5mm}\\Note, that in $(\ref{n_tot_def})$ the vector of the nuclear charge numbers: \vspace*{-3.5mm}
\begin{equation}\label{Z_vec_def}
\begin{split}
\begin{gathered}
\mathbf{Z}\coloneqq \left[ Z_{\alpha_{k}} \right]_{k=1}^{\mathrm{k}= N_{p}} \overset{\scriptsize{\underbrace{N_{p}=2}}{}}{=} \left(Z_{_{1}^{2}\mathrm{H}^{+}},\,Z_{_{10}^{20}\mathrm{Ne}^{+}}\right)  = \left(1,\,10\right) 
\end{gathered}
\end{split}
\end{equation}
\vspace*{-8.5mm}\\was introduced.
\\
Another common quantity, especially if ion species with different charges and densities are present, is the effective ion charge of the plasma. It is defined as a density-weighted charge square, in accordance with the book by \textit{U.\hspace{0.9mm}Stroth} \cite{Stroth_2018}:\vspace*{-3.5mm}
\begin{equation}\label{Z_eff_def}
\begin{split}
\begin{gathered}
Z_{\mathrm{eff}}= \dfrac{1}{n_{e}}\;\displaystyle{\sum\limits_{k=1}^{N_{p}}} \,n_{\alpha_{k}}\hspace{-1mm}\left(q_{\alpha_{k}}\right)^{2}\,.
\end{gathered}
\end{split}
\end{equation}

\clearpage

\section{Tokamak}\label{tokamak_section}

The most advanced thermonuclear fusion reactor concept is the magnetic plasma confinement in a toroidally symmetric \textit{tokamak} \cite{wesson}, which also occupies the widest area in nuclear fusion and plasma physics research. Hereinafter, the functional principles of such fusion devices, like the \textit{Joint European Torus} (JET) reactor or the under construction \textit{International Thermonuclear Experimental Reactor} (original meaning: abbreviation \qq{ITER} \cite{ITER_old_2002}, current meaning: latin word \qq{\textit{iter}} = \qq{the way} \cite{ITER_website}), shall be explored based on the book from \textit{J.\hspace{0.9mm}Wesson} \cite{wesson}.
 
\begin{figure} 
\centering
\begin{subfloat} 
 \centering
 \includegraphics[trim=150 526 151 73,width=0.591\linewidth,valign=c,clip]{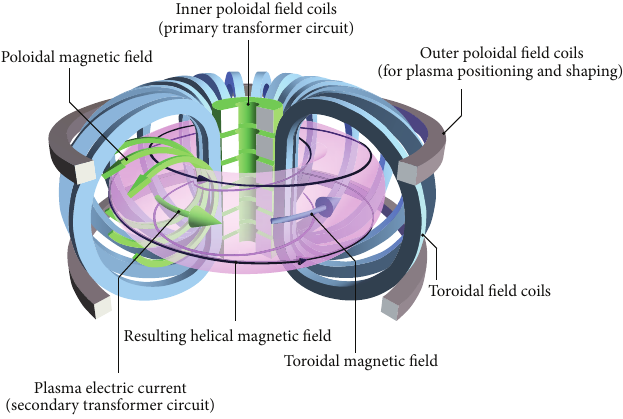} 
\end{subfloat} 
 \hspace{-3mm}
\begin{subfloat} 
  \centering
  \includegraphics[trim=130 425 90 81,width=0.39\linewidth,valign=c,clip]{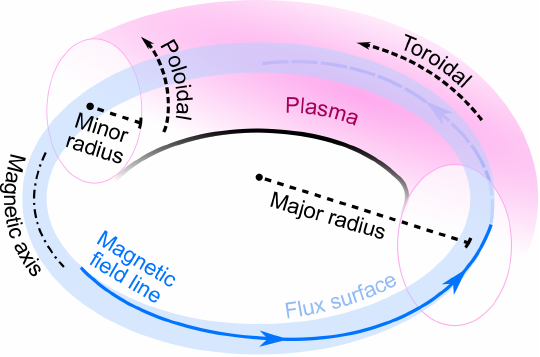}
\end{subfloat} 
\caption[Schematic depiction of the magnetic field and coil configuration of a tokamak (left graphic, \cite{Li_2014}) as well as principles of plasma description with position space coordinates (right graphic, \cite{stahl}).]{Schematic depiction of the magnetic field and coil configuration of a tokamak (left graphic, \cite{Li_2014}) as well as principles of plasma description with position space coordinates (right graphic, \cite{stahl}).}
\label{tokamak_fig}
\end{figure}

The main idea of magnetic confinement is, that the dynamics of charged particles of the species $\alpha$ with electric charge $q_{\alpha}$ will be governed by the \textit{Lorentz} force \cite{wesson}:\vspace{-3.2mm} 
\begin{equation}\label{F_Lorentz_def}
\mathbf{F}_{L}=q_{\alpha}\left(\mathbf{E}+\mathbf{v}\times\mathbf{B}\,\right)\,,\vspace{-1.5mm}
\end{equation}
\vspace{-7.95mm}\\resulting from a macroscopic electric field $\mathbf{E}$ and a magnetic field $\mathbf{B}$ for a particle velocity $\mathbf{v}$. In the case of tokamaks, the main magnetic field component is the toroidal field, which is produced by the toroidal field coils. The maximum toroidal field strength is reached at the inner minor radius of the torus, where the highest coil density is prevalent and it decreases towards the major radius, due to the same reason. The current coil and conductor technology allows values up to $16$ T \cite{wesson}, while the typical magnetic field strength at the center of the torus' toroidal cross-section, often referred to as the \textit{magnetic axis}, is between $6$ and $8$ T. In consequence, the plasma particles approximately perform a gyration motion around the toroidal field lines as a consequence of the \textit{Lorentz} force. At this, the gyration motion is a superposition of a motion along the toroidal magnetic field lines with a circular motion orthogonal to those field lines. Further details are considered in \mbox{section \ref{mom_space_coord_section}.}
\\
However, the plasma pressure has to be sufficiently compensated by a magnetic pressure, in order to truly confine the plasma within the magnetic torus-shaped cage. Therefore, one utilizes a central solenoid to generate a toroidal plasma current, which is not needed in a stellarator. This can be thought of as a transformer, where the plasma itself is the secondary coil. The driven plasma current, which for example has a maximum current strength of $7$ MA in JET \cite{wesson}, surrounds itself with a magnetic field, generating a poloidal magnetic field component, which has a smaller field strength than the toroidal component. Hence, the resulting magnetic field lines in this fusion reactor are helically wound and the plasma can be confined in a stable equilibrium apart from magnetic perturbations. For the purpose of visualization of the typical coil and magnetic field arrangement, one is referred to figure \ref{tokamak_fig} \cite{Li_2014,stahl}, which shows a schematic depiction of a tokamak alongside a position space description of a plasma.
\\
Since, the induction of a plasma current requires a changing magnetic field, the electric\linebreak 
\vspace*{-10.0mm}\\current within the first transformer circuit has to be varied. This is not possible unlimitedly, which is why \textit{pulsed} plasma discharges are intrinsic to a tokamak. In order to present a scenario for a larger tokamak in comparison to the JET reactor, one might choose an ITER-like research scenario \cite{ITER_scenario}. It is defined on the basis of fundamental parameters in table \ref{ITER_table}.\vspace*{-0.3mm}
\begin{table}[H]
\renewcommand{\arraystretch}{0.92}
\newcolumntype{s}{>{\hsize=.4\hsize}X}
\newcolumntype{S}{>{\hsize=.6\hsize}X}
\centering
\begin{tabularx}{0.7\textwidth}{Ss}
\toprule
Parameter	&	Value 	 \\    
\midrule
minor radius &	$a=2.0\,\mathrm{m}$	 \\ 
major radius &	$R_{0}=6.2\,\mathrm{m}$		 \\ 
plasma current strength &	$I_{\mathrm{p}}=15\,\mathrm{MA}$ \\ 
toroidal magnetic field at $R_{0}$ &	$B=5.3\,\mathrm{T}$\\ 
average electron density  &	$\langle n_{e}\rangle=1.01\cdot10^{20}\,\mathrm{m}^{-3}$ \\ 
average electron temperature &	$\langle k_{B}T_{e}\rangle=8.8 \,\mathrm{keV}$  \\ 
burn time &	$t_{\mathrm{burn}} =400 \,\mathrm{s}$ \\ 
\bottomrule						
\end{tabularx}
\caption[Main parameters for an ITER-like research scenario \cite{ITER_scenario}.]{\label{ITER_table}Main parameters for an ITER-like research scenario \cite{ITER_scenario}.}
\end{table}
\vspace*{-5.8mm}
With regard to the plasma physics research case, it should be mentioned, that often a pure deuterium instead of a deuterium-tritium plasma is used. For this plasma, two possible fusion reactions channels are conceivable:\vspace{-3.3mm}
\begin{equation}\label{deut_nuc_fus_reac}
\begin{split}
& _{1}^{2}\mathrm{H}^{+} \,+\;\, _{1}^{2}\mathrm{H}^{+} \;\longrightarrow  \;\; _{2}^{3}\mathrm{He}^{2+}\, +\;\, _{0}^{1}\mathrm{n}\;+\; 3.27\,\mathrm{MeV} \;\;\mathrm{and}
\\
& _{1}^{2}\mathrm{H}^{+} \,+\;\, _{1}^{2}\mathrm{H}^{+} \;\longrightarrow  \;\; _{1}^{3}\mathrm{H}^{+}\,  +\;\, _{1}^{0}\mathrm{H}^{+}\;+\; 4.03\,\mathrm{MeV}\;, 
\end{split}
\end{equation}
\vspace*{-6.9mm}\\which are further elaborated in \textit{J.\hspace{0.9mm}Wesson's} book \cite{wesson}. The utilization of such a research plasma is reasonable, because the main plasma behaviour can be studied without the additional fueling with the radioactive and expensive gas tritium. Consequently, no additional licensing is required and less radioactive radiation is produced within the plasma, which reduces the neutron-activation of the reactor wall. 


\clearpage


\section{Tokamak plasma disruptions}\label{tokamak_disruptions_section}

\begin{figure}[H]
\begin{center}
\includegraphics[trim=0 0 0 0,width=0.9\textwidth,clip]
{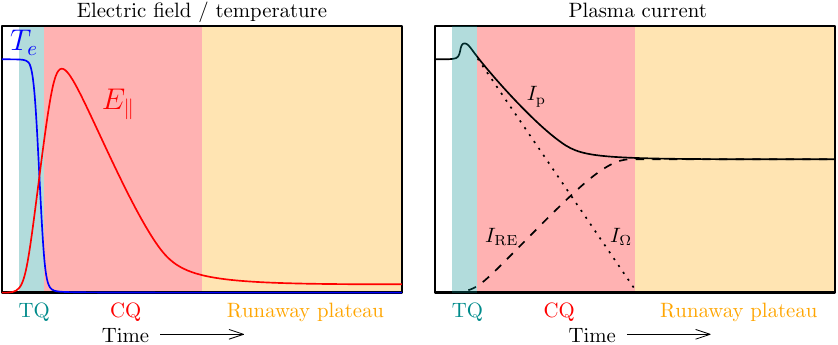}
\caption[Schematic time evolution of significant plasma parameters during a disruption and typical disruption phases \cite{Hoppe_PHD}.]{Schematic time evolution of significant plasma parameters during a disruption and typical disruption phases \cite{Hoppe_PHD} (graphic provided by \textit{M.\hspace{0.6mm}Hoppe}).}
\label{disruption_fig}
\end{center}
\end{figure}

The plasma within the tokamak is confined in a stable magneto-hydrodynamic equilibrium. However, for instance a large plasma current, high plasma densities or too many heavy impurity atoms can trigger an abrupt loss of energy and magnetic confinement \cite{Kowslowski,Hender_2007}. This phenomenon, which is connected to a sudden plasma cool down and an eventual termination of the plasma discharge, is known as a \textit{disruption} and often occurs in several successive phases, which shall be explained in the following with the help of the reference \cite{wesson}. At this, the different phases of a disruption are visualized in figure \ref{disruption_fig} \cite{Hoppe_PHD}, by means of the time evolution of some selected parameters. 
\vspace{-8.5mm}
During the \textit{pre-disruption phase} a local plasma current instability grows within less than $10$ ms for tokamaks of medium size and becomes a global instability. Hence, the quality of the energy confinement and thus also the magnetic confinement decreases. 
\\
This leads to a collapse of the plasma or electron temperature $T_{e}$ at the center of the tokamak cross-section. This phase happens within the order of \mbox{$1\,\mathrm{ms}$} and is called the \textit{thermal quench} (TQ). 
\\
Caused by the decreasing temperature of the thermal quench, the resistivity of the plasma increases with \mbox{$\eta_{p}\propto T_{e}^{-1.5}$} \cite{wesson}. This perturbs the ability of the plasma to carry a strong toroidal current, so that the radial plasma current profile flattens and the ohmic part $I_{\Omega}$ of the plasma current $I_{p}$ decays. At that, decay rates of 100 MA per second are possible, which is equivalent to a duration of this \textit{current quench} (CQ) in the magnitude of \mbox{$10\,\mathrm{ms}$}. Note, that the plasma inductance prevents the parallel plasma current density $j_{p,\|}$ from changing this quickly, which leads to an increase of the component of the electric field parallel to the magnetic field lines with \mbox{$E_{\|}=\eta_{p}\,j_{p,\|}$} as long as the runaway current $I_{\mathrm{RE}}$ is negligible, due to the mentioned evolution of the resistivity \cite{Smith_2005}.
\\
The rapid changing plasma current produces a changing magnetic field, which instantaneously induces an electric field. If the component of the electric field parallel to the magnetic field lines $E_{\|}$, which is often used as a model parameter, exceeds a certain critical electric field strength, it is able to accelerate electrons within the plasma to relativistic velocities, generating a beam of \textit{runaway electrons} \cite{Dreicer_1959}. This runaway phenomenon, which is discussed in detail in section \ref{RE_phenom_section}, is therefore the reason why only the ohmic plasma current decays to zero, while a fraction of the initial plasma current is transformed into a runaway current $I_{\mathrm{RE}}$. Such a runaway beam has the potential to damage components of the reactor vessel \cite{Hoppe_2021} and is a major concern for future reactors such as ITER \cite{Hoppe_2021,REsimulation,Hoppe_2022}, since it stays nearly constant in time and is highly energetic with particle energies in the magnitude of \mbox{$10\,\mathrm{MeV}$}. Thus, this last phase of a disruption is characterized and named by the \textit{runaway plateau}. 
 
Disruptions and disruption-related phenomena provide potentially dangerous physics phenomena with the capability of damaging the reactor \cite{Bazylev_2011,Breizman_2019}. Based on this, the research area of \textit{disruption prediction}, \textit{avoidance} and \textit{mitigation} \cite{Hoppe_2022}, especially in connection with the simulation of the evolution of a runaway current in tokamaks, is motivated. Subsequently, those three notions shall be explained briefly on the basis of the lectures slides \cite{PappSlides}. The \textit{disruption prediction} requires a broad knowledge of the prevalent plasma physics, in order to e.g.\ train machine learning models for the purpose of providing real-time computable disruption triggers, so that a reaction time scale is achieved, which allows to avoid disruption-related phenomena. From this follows, that \textit{disruption avoidance} represents a control challenge, which has the goal to use actuators, like additional fueling or changes in the external plasma heating, which are triggered by certain plasma configurations, to avoid an uncontrolled disruption. However, a different approach is the procedure of \textit{disruption mitigation}, which can be thought of as a controlled plasma shut down, by means of the initiation of a controlled, externally-triggered disruption, which is based on the cooling of the plasma through homogeneous electromagnetic radiation.
\\
One of the major disruption mitigation schemes is the principle of \textit{\qq{Massive\linebreak Material Injection}} (MMI) \cite{Vallhagen_2020_MMI}, which relies on the isotropic and broad dissipation of the energy on the components facing the plasma, before an uncontrolled disruption and a runaway electron population occurs and causes damage, due to high local energy deposition. This shall be achieved by material injection in the form of additional deuterium and/or impurities like neon or argon atoms, which are heated up and become ionized by interacting with the plasma \cite{REsimulation}. At this, a lower ionization of the impurity atoms is intended, in order to keep some of their bound electrons, because they are responsible for a large fraction of the radiative dissipation of energy. In consequence, the stored energy within the plasma should be transformed into electromagnetic radiation, which needs to be distributed in such a way, that it does not damage any reactor components. This is called \textit{thermal load mitigation}. However, the addition of hydrogen or deuterium is necessary to facilitate dilution cooling and an increase of the free electron density, both of which are aimed at reducing the primary generation of runaways \cite{Vallhagen2022xx}. In general, one distincts the \textit{\qq{Shattered Pellet Injection}} (SPI) and \textit{\qq{Massive Gas Injection}} (MGI) as the two most common types of MMI \cite{REsimulation}. More precisely, the MGI-procedure is understood to by an injection of cold gas into the vacuum vessel of the tokamak through a valve. At that, atoms with a high nuclear charge number, typically neon or argon, are used and about ten to one hundred times the number of particles in the reactor is injected \cite{PappSlides}. In contrast, the material is shot into the plasma as a frozen pellet at high speed in case of SPI-approach. This allows the thermal mitigation to take place near the magnetic axis, where it is most effective. For this, however, it is necessary that the pellet shatters, because otherwise the required ablation of the material into the surrounding plasma cannot take place \cite{PappSlides}.

\clearpage

\section{Thesis outline and motivation}

In the previous section \ref{tokamak_disruptions_section}, it has been elaborated, that during a disruption within the a tokamak plasma a runaway beam can be created, which has can potentially damage the plasma-facing reactor components \cite{Hoppe_2021}. Due to the fact, that this is a major concern for future reactors such as ITER \cite{Hoppe_2021,REsimulation,Hoppe_2022}, research on the runaway phenomenon and in particular a computational efficient simulation of the runaway current, which provides sufficient physical accuracy, is motivated. Hence, the \textit{reduced kinetic modeling} approach is often used as an intermediate option between the computation-intensive and highly accurate calculations based on the complete solution of the kinetic equation and the simplified simulation on the basis of the so-called \textit{fluid description} of a plasma. Conceivably, the utilization of certain moments of analytic runaway electron distribution functions might provide improved accuracy alongside a tolerable increase in the computational effort, which is why in this work corresponding calculation rules shall be established, analysed and evaluated. 

For this purpose, an overview of the kinetic theory of runaway electrons in a tokamak plasma, the runaway electron phenomenon and its generation and loss mechanisms will be provided in chapter \ref{theory_chapter}. In addition, the effects of partially ionized impurities are explained and connected to the runaway electron generation region in momentum space. On this occasion, a two-dimensional momentum space coordinate system is introduced, which will be used throughout the thesis.
\\
Based on this theoretical framework, the calculation of the moments of a \textit{hot-tail} electron distribution function is going to be explained in chapter \ref{hot_tail_chapter}. At this, calculation schemes for the runaway electron density, the mean velocity and the mean kinetic energy of a \textit{hot-tail} runaway electron population are deduced and evaluated based on an ITER-like disruption simulation.
\\
In chapter \ref{avalanche_chapter}, the \textit{avalanche} generation of runaway electrons is treated similarly to the \textit{hot-tail} generation mechanism. Here, two different models are considered, so that computation rules will be defined and discussed separately for each model. At this, the \textit{Rosenbluth-Putvinski} model and the \textit{Hesslow} model, which accounts for the effects of partial screening, will be compared on the basis of computational results for the mean velocity and the mean kinetic energy of avalanche runaway electrons corresponding to a wide range of density combinations for a singly-ionized deuterium-neon plasma.

Eventually, a conclusion about the applicability of the analysed calculation schemes for the moments connected to the \textit{hot-tail} and \textit{avalanche} runaway generation mechanism is drawn and a related outlook is set out. Finally, it shall be remarked, that precise data and high-resolution contour plots of the performed computer-aided calculations are displayed in the digital appendix.

\clearpage


%% file: Kinetic_theory_of_electrons_in_tokamak_plasmas.tex
\chapter[Kinetic theory of electrons in tokamak plasmas]{Kinetic theory of electrons in tokamak plasmas}\label{theory_chapter}

The idea of \textit{kinetic theory} is the description of a population of $N_{\alpha}$ particles of the species $\alpha$, which does not allow the solution of single-particle equations of motion, in order to investigate their dynamics and their trajectories. This is reasoned by the fact the resulting system of coupled partial differential equations is not solvable numerically with the current affordable computational capabilities for \mbox{$N_{\alpha}\gtrsim 20$} \cite{RichterLecture}, if all $N$-particle interactions are taken into account. 
\\
Hence, kinetic theory utilizes a \textit{statistical} approach based on the probability density for a single particle in a seven-dimensional phase space with a time coordinate as well as three position and momentum coordinates. The most general function describing this probability density, under the distinction of $N_{\alpha}$ different particles, is the \textit{many-particle distribution function} \mbox{$f_{\alpha}(\mathbf{r}_{1},\,\mathbf{p}_{1},\dots,\,\mathbf{r}_{N_{\alpha}},\,\mathbf{p}_{N_{\alpha}},\,t)$} \cite{Stroth_2018,wesson}. With regard to a plasma, this function would contain all physical information including pair correlations and collisions. However, a more efficient description is possible, if collisions are treated by a yet to be introduced collision term and one drops the distinction between individual particles \cite{wesson}. Thus, one can introduce the \textit{single-particle distribution function} \mbox{$f_{\alpha}(\mathbf{r},\,\mathbf{p},\,t)$}. Through the expression \mbox{$f_{\alpha}(\mathbf{r},\,\mathbf{p},\,t)\,\mathrm{d}^3r\,\mathrm{d}^3p$} it represents the number of particles of one kind per unit phase space volume \mbox{$\mathrm{d}^3r\,\mathrm{d}^3p$} at the time $t$, whose state vectors are near the state $\mathbf{z}$ in phase space \cite{helander}. Here \mbox{$\mathbf{z}\coloneqq(\mathbf{r},\,\mathbf{p})$} is the phase space state vector, which consists of the position vector $\mathbf{r}$ and the momentum vector $\mathbf{p}$. 

\section{The kinetic equation}\label{kin_equa_section}

The distribution function connects a given configuration of particles, consisting of point-like contributions from individual particles, to a state distribution in the phase space and allows to draw inferences from the dynamics of the phase space points about the dynamics of the real particle population. In particular the single-particle distribution function is a probability density denoting the ensemble average over a significantly large number of macroscopically equivalent particle configurations \cite{HesslowPHD}.
 
In order to deduce a governing equation for this phase space dynamics, one follows the references \cite{HesslowPHD,wesson,Hoppe_PHD,helander} and imagines a closed system without external particle sources. In such a system the total derivative of the distribution function or its rate of change is equal to the divergence of the phase space flow. At that, one interprets the total time derivative $\dot{\mathbf{z}}$ of the state vector $\mathbf{z}$ as the velocity of a fluid of phase space particles. For the considered system the number of particles is conserved in the absence of collisions \cite{slides}, which means that the single-particle distribution function has to be constant along phase space trajectories and its divergence of the phase space flow is zero. An expression for this relation is provided by the \textit{Liouville} equation \cite{Liouville1838,Lifshitz}:
\noindent\\\vspace{-7.5mm}\noindent  
\begin{equation}\label{Liouville_equation}
\dfrac{\mathrm{d} f_{\alpha}}{\mathrm{d} t}\equiv\dfrac{\partial f_{\alpha}}{\partial t} + \dot{\mathbf{r}}\cdot\dfrac{\partial f_{\alpha}}{\partial \mathbf{r}}+\dot{\mathbf{p}}\cdot\dfrac{\partial f_{\alpha}}{\partial \mathbf{p}} =0 
\end{equation}
\vspace{-8.5mm}\\with the total time derivatives of the position and momentum vectors $\dot{\mathbf{r}}$ and $\dot{\mathbf{p}}$. In a plasma the force $\dot{\mathbf{p}}$ can be replaced by a macroscopic equation of motion, like for instance the \textit{Lorentz} force as defined in ($\ref{F_Lorentz_def}$), so that the equation ($\ref{Liouville_equation}$) becomes the \textit{Vlasov} equation. 
\\
Since the single-particle distribution function is used, the microscopic fields determining collisions are separated from macroscopic fields. Hence, a \textit{collision operator} $C_{\alpha}\lbrace f_{\alpha}\rbrace$ is required as a non-zero right-hand side in ($\ref{Liouville_equation}$), in order to take e.g.\ \textit{Coulomb} interactions into account. Here, the operator indicates the time rate of change in the distribution function, due to collisions with all species $\beta$ in a plasma with $N_{p}$ different plasma components. On this occasion, the following notation, in accordance with \textit{P.\hspace{0.9mm}Helander's} book \cite{helander}, is introduced:\vspace{-4.8mm}
\begin{equation}\label{coll_op_short}
C_{\alpha}\lbrace f_{\alpha}\rbrace \;\equiv\; \left.\dfrac{\partial f_{\alpha}}{\partial t}\right|_{coll}\coloneqq \; \displaystyle{\sum\limits_{\beta=1}^{N_{p}}} \,C_{\alpha\beta}\;\equiv\; \displaystyle{\sum\limits_{\beta=1}^{N_{p}}} \,C_{\alpha\beta} \lbrace f_{\alpha},\,f_{\beta}  \rbrace \,.
\end{equation}
\vspace{-8.0mm}\\Furthermore, it is as well possible to further add an operator $S$ on the right-hand side in ($\ref{Liouville_equation}$), which represents the effects of sources or sinks of particles, due to ionization, recombination, fueling or loss processes \cite{stahl}. Hence, a general form of a \textit{kinetic equation} could be written as stated in the PhD thesis of \textit{A.\hspace{0.9mm}Stahl} \cite{stahl}:\vspace{-2.8mm}
\begin{equation}\label{kinetic_equation}
\dfrac{\partial f_{\alpha}}{\partial t} + \dot{\mathbf{r}}\cdot\dfrac{\partial f_{\alpha}}{\partial \mathbf{r}}+\dot{\mathbf{p}}\cdot\dfrac{\partial f_{\alpha}}{\partial \mathbf{p}} = C_{\alpha}\lbrace f_{\alpha}\rbrace +S\,.
\end{equation}
\vspace*{-8.5mm}\\Nevertheless, it should be remarked that the kinetic equation and especially the collision operator is expressed differently for each combination of approximations, assumptions and modeling principles \cite{helander}.

In general, an analytic solution for the kinetic equation ($\ref{kinetic_equation}$) can only be obtained for simplified cases. Moreover, a numerical solution is runtime- or computation-power-expensive, if the distribution function is to be resolved in the full six-dimensional phase space, while including several sources and collision operators. Note, that here the time dimension of the phase space is treated as an evolving parameter. Therefore, approaches like the \textit{fluid description} of a plasma, are used to overcome those difficulties \cite{helander}.
\\
The fluid description of a plasma utilizes physically motivated moments of the kinetic equation, in order to deduce balance equations for the number of particles, momentum and energy. The definition of the $m$th moment, according to the lecture slides from \mbox{\textit{Y.\hspace{0.9mm}Mizuno}} \cite{slides}, reads:\vspace{-7mm}
\begin{equation}\label{mth_moment}
M_{\alpha}^{m}(\mathbf{r},\,t)\coloneqq \displaystyle{\iiint\limits_{\mathbf{p}\,\in\,\mathbb{R}^3}}\hspace{0.5mm} \mathbf{v}^m\hspace{-0.8mm}\cdot\hspace{-0.5mm} f_{\alpha}(\mathbf{r},\,\mathbf{p},\,t)\;\mathrm{d}^3p\,.
\end{equation}
\vspace{-7.5mm}\\For example the balance equation for the particle density follows from the integration of the kinetic equation over the whole momentum space, because the zeroth moment can be interpreted as the particle density \cite{slides,Bellan_2006}:\vspace{-2.7mm} 
\begin{equation}\label{zeroth_moment}
n_{\alpha}(\mathbf{r},\,t)\coloneqq M_{\alpha}^{0}(\mathbf{r},\,t)= \displaystyle{\iiint\limits_{\mathbf{p}\,\in\,\mathbb{R}^3}}\hspace{0.5mm} f_{\alpha}(\mathbf{r},\,\mathbf{p},\,t)\;\mathrm{d}^3p\,.
\end{equation} 
\vspace{-7.5mm}\\The set of coupled partial differential equations containing the  balance equations is solvable by existing solvers in a finite runtime. This is also possible, if the set of equations is extended by the \textit{Maxwell} equations and the plasma is described as a single fluid, which leads to the research area of \textit{magnetohydrodynamics} \cite{slidesMHD,wesson}. At this, it should be remarked, that both of the mentioned simulation methods require further modeling, since they simplify the kinetic equation and are therefore not closed, so that for instance turbulence models are necessary. However, the runtime of fluid or magnetohydrodynamic simulations is usually still too large for rapid predictions, for which simpler models are required. 
 
Those simple models often further reduce the dimensionality of the problem and e.g.\ assume a certain spatial distribution in combination with a simulation with respect to time and momentum. Additionally, a homogeneous distribution of some simulated quantities in one or more phase space dimensions is introduced, which further improves the efficiency of the simulation. Furthermore, analytic and numeric models might be used for the fast computation of certain moments of the distribution function, which are then used to evolve physical quantities in time, while others are computed self-consistently. Thus, one can simulate and analyse certain plasma phenomena. 
\\
Such a simulation model is the runaway electron generation computation with \textit{self-consistent} electric field \cite{Papp2013,Hoppe_2021}, which calculates the time evolution of the radial runaway electron current density profile. Its governing equations are a differential equation for the growth rate of the runaway electron density and a diffusion equation for the electric field deduced from the parallel component of the induction equation \cite{pappPHD}. In addition, simplified balance equations are solved i.a.\ for the electron temperature and the effective ion charge. The applied approximations and assumptions as well as additional understanding for this model can be provided, by the PhD-thesis of \mbox{\textit{G.\hspace{0.9mm}Papp}} \cite{pappPHD} and the publication \cite{Papp2013}, which explains the \qq{GO-Code} as a possible implementation of said simulation approach. At this, the induction equation for the parallel electric field component in cylindrical coordinates of the form \cite{Eriksson_2004,pappPHD}:\vspace{-3.2mm} 
\begin{equation}\label{induction_eq}
\dfrac{1}{r_{\perp}}\,\dfrac{\partial }{\partial\hspace{0.15mm}r_{\perp}\hspace{-0.9mm}}\hspace{-0.15mm}\left(\hspace{-0.2mm}r_{\perp}\,\dfrac{\partial\hspace{0.05mm}E_{\|}}{\partial\hspace{0.15mm}r_{\perp}\hspace{-0.9mm}} \right)= \mu_{0}\,\dfrac{\partial\hspace{0.15mm}j_{\|} }{\partial\hspace{0.15mm}t}= \mu_{0}\,\dfrac{\partial\hspace{0.15mm} }{\partial\hspace{0.15mm}t}\left[j_{\|,\Omega}+j_{\|,RE}\right]
\end{equation} 
\vspace{-8.0mm}\\is used. Note, that the left-hand side of the partial differential equation is the\linebreak\textit{Laplace} operator in cylindrical position space coordinates acting on \mbox{$E_{\|}=E_{\|}(r_{\perp},\,t)$} and\linebreak\mbox{$\mu_{0}=1.25663706212\cdot 10^{-6} \;\si{\newton\per\square\ampere}$} is the vacuum magnetic permeability \cite{NISTmue0}. Because the spatial resolution of the self-consistently calculated electric field is only expressed by the cylindrical radius $r_{\perp}$, measured orthogonally from the magnetic axis, one classifies this model as one-dimensional. In the equation $(\ref{induction_eq})$, the \textit{ohmic} part of the \textit{current density}, which can be written in the following form:\vspace{-3.5mm}
\begin{equation}\label{Ohmic_curr_dens_def}
j_{\|,\Omega}=\dfrac{1}{\eta_{th}}\cdot E_{\|}=\sigma_{th} \cdot E_{\|} \,,
\end{equation}
\vspace*{-9.4mm}\\appears. It is given by \textit{Ohm's} law, as the product of the electric field and the conductivity $\sigma_{th}$, thus the inverse of the resistivity $\eta_{th}$, of the thermal electron population, which is often expressed through the \textit{Spitzer} conductivity \cite{Spitzer_1953} for fully ionized plasmas \cite{Stahl_2016}. As a second fraction of the electron current density, the \textit{runaway electron current density}, expressed by means of the relation:\vspace{-4.0mm} 
\begin{equation}\label{RE_curr_dens_def}
j_{\|,\mathrm{RE}}= q_{e}\cdot n_{\mathrm{RE}}\cdot u_{\|,\mathrm{RE}}= -\,e\cdot n_{\mathrm{RE}}\cdot u_{\|,\mathrm{RE}}\,,
\end{equation} 
\vspace{-10.5mm}\\appears in equation $(\ref{induction_eq})$, which depends on the  \textit{runaway electron density} $n_{\mathrm{RE}}$ and the \textit{mean velocity} of the associated runaway electron population. This runaway electron density is determined by the growth rates of different generation mechanisms and is hence the solution of the partial differential equation \cite{Eriksson_2004,pappPHD}:\vspace{-2.8mm} 
\begin{equation}\label{n_RE_eq}
\dfrac{\partial\hspace{0.05mm}n_{RE}}{\partial\hspace{0.15mm}t}  = \Gamma_{D}(E_{\|},\,t)+  \Gamma_{ht}(E_{\|},\,t)+\Gamma_{ava}(E_{\|},\,n_{RE},\,t)\,.
\end{equation} 
\vspace{-8.5mm}\\Here, only the \textit{Dreicer} (D), the \textit{hot-tail} (ht) and the \textit{avalanche} (ava) growth rate are considered, although typically the right-hand side of the equation $(\ref{n_RE_eq})$ is tailored to the regarded simulation scenario in terms of the dominant generation and loss mechanism. A further explanation of the named generation mechanism as well as an analysis of their growth rates can be found in section \ref{mechanisms_section}. Both equations $(\ref{induction_eq})$ and $(\ref{n_RE_eq})$ are coupled and have to be solved numerically and in particular together, explaining the description as a self-consistent simulation. 
\\
In addition, a comment concerning the \textit{runaway current strength} $I_{RE}$, which was firstly mentioned in section \ref{tokamak_disruptions_section}, can now be made. It is defined as the integral of the runaway current density over the toroidal cross-section area of the plasma tube, which is bounded by a characteristic magnetic field line, the \textit{separatrix}. It represents the rim of the plasma cross-section, which is ideally not transcended by plasma particles, in case of a perfect magnetic confinement. Consequently, the strength of a current in the plasma volume, for example of a runaway electron population, is defined as the charge per unit time, which passes orthogonally through the cross-section area \cite{Bartelmann_2015}. The main component of the runaway current density is parallel to the local magnetic field, which will be explained in the subsequent section, and thus approximately oriented in toroidal direction and therefore perpendicular to the toroidal plasma cross-section. In section \ref{tokamak_disruptions_section}, it was already discussed, that the runaway electron populations are highly energetic and can have the ability to damage the reactor wall \cite{Hoppe_2021}. In order to avoid this, the research area of the disruption mitigation deduced for instance the criterion \mbox{$I_{RE}<150\,\mathrm{kA}$} for the tolerable runaway current strength in ITER \cite{Lehnen_ITER}, due to the fact that for larger current strengths the deposited thermal load to the wall, leads to surface melting.

The definition of a \textit{zeroth moment} of a distribution function from $(\ref{zeroth_moment})$ motivates the idea of connecting the runaway electron density to a given distribution function, for a runaway electron population. Here it is possible to investigate contributions from different generation or loss mechanisms. Furthermore, the \textit{mean velocity} $u_{\|,\mathrm{RE}}$ from $(\ref{RE_curr_dens_def})$ is related to the \textit{first moment} of a runaway electron distribution function. This relation follows from the definition \cite{slides,Bellan_2006}:\vspace{-2.5mm} 
\begin{equation}\label{first_moment}
\mathbf{u}_{\alpha}(\mathbf{r},\,t)\coloneqq\dfrac{M_{\alpha}^{1}(\mathbf{r},\,t)}{n_{\alpha}(\mathbf{r},\,t)}=\dfrac{1}{n_{\alpha}(\mathbf{r},\,t)}\; \displaystyle{\iiint\limits_{\mathbf{p}\,\in\,\mathbb{R}^3}}\hspace{0.3mm} \mathbf{v}\hspace{-0.4mm}\cdot\hspace{-0.5mm}f_{\alpha}(\mathbf{r},\,\mathbf{p},\,t)\;\mathrm{d}^3p\,,
\end{equation}
\vspace*{-8.0mm}\\so that one might compute the magnitude of the mean velocity from the integral of a numerically or analytically given distribution function, weighted with the absolute value of the velocity vector. This could improve existing simulations like the GO-code, which assumes that all runaway electrons move with \mbox{$u_{\mathrm{RE}}\equiv c$}. Moreover, this could allow a higher efficiency of simulation tools like the \qq{Disruption Runaway Electron Analysis Model (DREAM)-Code} \cite{Hoppe_2021}. This is because the \textit{reduced kinetic modeling} approach of the DREAM-implementation utilizes a split into a partly solve, based on the computation-intensive full kinetic description, where the runaway velocity is accurate and a solve in the less runtime consuming fluid description, which makes the mentioned assumption. A known runaway velocity from an efficient computation of the first moment could lead to a possibly more accurate solution in the fluid description without increasing the runtime too much.
\\
In addition, one should mention, that further applications exist for the mean velocity magnitude \mbox{$u=\vert\mathbf{u}\vert$} and its components parallel or orthogonal to the local magnetic field vector $u_{\|}$ and $u_{\perp}$. For instance, the radial transport model for runaway electrons from \mbox{\textit{A.\hspace{0.7mm}B.\hspace{0.9mm}Rechester}} and \mbox{\textit{M.\hspace{0.7mm}N.\hspace{0.9mm}Rosenbluth}} \cite{Rechester_1978,Entrop_1998}, which accounts for the \textit{radial diffusion} due to magnetic perturbations, makes use of the parallel velocity component $u_{\|}$ and usually approximates it with the speed of light. At that, a more accurate value for $u_{\|}$ would improve said model. Additionally, it is imaginable, that advective transport velocities and diffusion coefficients in general include velocity components computed from a moment of a distribution function and might find applications in existing or future simulations.
\\
Finally, a motivation for the computation and utilization of the moments of distribution functions is the ability to analyse certain characteristic physical quantities and their behaviour within certain parameter regions. An example for this would be the \textit{mean rest mass-related kinetic energy density} of a confined runaway electron population $k_{\mathrm{RE}}$, which has the ability to influence e.g.\ the equilibrium confinement \cite{Ficker2019,Ficker2021}, the evolution of atomic physics processes \cite{Breizman_2019} or the electron impact ablation of mitigation pellet injections \cite{James_2011,Hollmann2016}. It can be normalized to the square of the speed of light in vacuum $c^2$ and is related to the subsequently defined moment of a runaway electron distribution function \mbox{$f_{\mathrm{RE}}(\mathbf{r},\,\mathbf{p},\,t)$} \cite{Bartelmann_2015,Svenningsson_2021,stahl}:\vspace{-2.3mm} 
\begin{equation}\label{kin_dens_ava_def}
\begin{split}
\dfrac{k_{\mathrm{RE}}}{c^2}\hspace{0.3mm} \coloneqq&\hspace{1.5mm}\dfrac{\langle K_{\mathrm{RE}}\rangle }{m_{e0}\hspace{0.25mm}c^2}\hspace{0.2mm}=\hspace{0.2mm} \langle\gamma-1\rangle \hspace{0.2mm}= \hspace{0.2mm}\dfrac{1}{n_{\mathrm{RE}}}\, \iiint\limits_{\mathbf{p}\,\in\,\mathbb{R}^3} (\gamma-1)\hspace{-0.4mm}\cdot\hspace{-0.5mm}f_{\mathrm{RE}}(\mathbf{r},\,\mathbf{p},\,t)\;\mathrm{d}^3p 
\\
=&\hspace{1.5mm}\dfrac{1}{n_{\mathrm{RE}}}\,\iiint\limits_{\mathbf{p}\,\in\,\mathbb{R}^3}  \gamma \hspace{-0.4mm}\cdot\hspace{-0.5mm}f_{\mathrm{RE}}(\mathbf{r},\,\mathbf{p},\,t)\;\mathrm{d}^3p 
\,-\,\dfrac{1}{n_{\mathrm{RE}}}\, \underbrace{ \displaystyle{\iiint\limits_{\mathbf{p}\,\in\,\mathbb{R}^3}}  f_{\mathrm{RE}}(\mathbf{r},\;\mathbf{p},\,t)\;\mathrm{d}^3p }_{=\,n_{\mathrm{RE}}}
\\[-10pt]
=&\hspace{1.5mm}\dfrac{1}{n_{\mathrm{RE}}}\,\iiint\limits_{\mathbf{p}\,\in\,\mathbb{R}^3}  \gamma \hspace{-0.5mm}\cdot\hspace{-0.4mm}f_{\mathrm{RE}}(\mathbf{r},\,\mathbf{p},\,t)\;\mathrm{d}^3 p \,-\,1\,.
\end{split}
\end{equation}
\vspace*{-6.8mm}\\Note, that the last equality only holds, if the chosen distribution function is normalized as written in $(\ref{zeroth_moment})$. As well, it is shown, that the normalized mean rest mass-related kinetic energy density is equivalent to the mean kinetic energy $\langle K_{\mathrm{RE}}\rangle$ divided by the electron rest mass energy $m_{e0}\hspace{0.25mm}c^2$.
\\
The calculation of such quantities, could enable more efficient and accurate simulations, which might include distribution functions, based on experimental data. Furthermore, one can imagine the computation of certain moments for a wide parameter space and their usage as training sets for neural networks with the goal of improved simulations on the basis of machine learning. Eventually, it is also thinkable, that certain moments are used as a criterion, in order to decide, when certain assumptions are useful or in which extend physical phenomena have to be simulated. This might also increase the efficiency and applicability of existing and future simulations.

\clearpage

\section{Gyro-radius-averaged two-dimensional momentum space}\label{mom_space_coord_section}

As discussed in the previous section \ref{kin_equa_section} reduced models rely on a kinetic description of the plasma with reduced dimensionality, in order to increase the computation efficiency of simulations. Therefore one often utilizes a two-dimensional momentum-space instead of the whole seven-dimensional phase space with one time dimension as well as three position and momentum dimensions for the parameterisation of particle orbits. This is achieved, by assuming a known spatial resolution and averaging over the momentum coordinate associated with the \textit{gyro motion} of the charged particles. In consequence, the plasma physics, determining the motion of particles, only depends on two momentum space coordinates and the evolution of time.
 
The gyro-radius-averaging is valid, if the \textit{gyro radius} \cite{wesson}:\vspace*{-1.5mm}
\begin{equation}\label{gyroradius}
r_{g}=\frac{m_{\alpha0}v_{\perp}}{\vert q_{\alpha}\vert \vert\mathbf{B}\vert} 
\end{equation}
\vspace*{-6.5mm}\\for moving particles of a species $\alpha$ with electric charge $q_{\alpha}$ and $v_{\perp}$ is the velocity component perpendicular to the toroidal magnetic field $\mathbf{B}$ is negligibly small compared to the typical length scale of the gradients and the typical time scale of the gyration is much smaller than those of other processes within the plasma \cite{stahl,helander}.
\\
In the following, drifts of particle orbits are neglected and a small \textit{Larmor} radius is assumed, which holds approximately for the considered fusion plasma scenarios. Thus, one can average over the gyro motion, implying that the particles mainly move along the magnetic field lines. Thereby, the set of spatial coordinates describing the approximated torus geometry of the reactor also defines the magnetic field lines. At this, the radius measured perpendicular from the magnetic axis, a poloidal and a toroidal angle represent suitable spatial parameters for a circular cross-section torus.

The three-dimensional velocity and momentum space is usually described by a local spherical or cylindrical coordinate system moving with the particle along the magnetic field lines. In order to provide a visualization of the following explanations, the reader is referred to the figure \ref{fig_mom_coord}. 
\\
Typically, a normalized relativistic momentum vector $\mathbf{p}$, involving the \textit{Lorentz} factor $\gamma$ either depend on the magnitude of the velocity vector \mbox{$v=\vert\mathbf{v}\vert$} or the momentum vector \mbox{$p=\vert\mathbf{p}\vert$}, is applied:\vspace{-2mm}
\begin{equation}\label{p_norm_gamma_def}
\mathbf{p} = \dfrac{m_{e0} \gamma\hspace{0.1mm}\mathbf{v}}{m_{e0}c} = \dfrac{\gamma\hspace{0.1mm}\mathbf{v}}{c} \;\;\;;\;\;\; \gamma= \gamma (v)=\dfrac{1}{\sqrt{1-\left(\dfrac{v}{c}\right)^2}}=\sqrt{1+p^2}\,.
\end{equation}
\vspace*{-6.5mm}\\The spherical momentum coordinate system makes use the triplet \mbox{$(p,\,\varphi,\,\theta)$}, consisting of the length of the normalized momentum vector \mbox{$p \in[0,\,\infty)$}, the azimuthal angle \mbox{$\varphi\in[0,\,2\pi]$} measured orthogonally to the magnetic field and the polar or pitch angle \mbox{$\theta\in[0,\,\pi]$} measured from the local magnetic field direction. The corresponding volume element is \mbox{$\mathrm{d}^{3}p=p^2\sin{(\theta)}\mathrm{d}p\,\mathrm{d}\varphi\,\mathrm{d}\theta$}. Due to the neglection of the gyro motion the dependence on the azimuthal angle $\varphi$ is suppressed by integration, leading to the following two-dimensional momentum space volume or area element:\vspace{-2.5mm}
\begin{equation}\label{volelem_sphere_2D}
\mathrm{d}^{2}p=\int\limits_{\varphi}\mathrm{d}^{3}p=\int\limits_{0}^{2\pi}d\varphi\,p^2\sin{(\theta)} \,\mathrm{d}p\,\mathrm{d}\theta=2\pi\,p^2\sin{(\theta)}\,\mathrm{d}p\,\mathrm{d}\theta\,,
\end{equation}
\vspace*{-5.5mm}\\where a point is characterized by the coordinate pair \mbox{$(p,\,\theta)$}. This can be rewritten, by means of the substitution of the pitch coordinate \mbox{$\xi:=\cos{(\theta)}$}, by which one finds \mbox{$\mathrm{d}^{3}p=-2\pi\,p^2\,\mathrm{d}p\,\mathrm{d}\xi$} for \mbox{$\xi\in[1,-1]$}. Interchanging the boundaries to \mbox{$\xi\in[-1,\,1]$} and dropping the negative sign from the pitch-coordinate substitution yields:\vspace{-1.5mm}
\begin{equation}\label{pitch_angle_volelem_sphere_2D}
\mathrm{d}^{2}p=2\pi\,p^2\,\mathrm{d}p\,\mathrm{d}\xi\,.
\end{equation} 
\vspace*{-8.5mm}\\In this case, the duplet \mbox{$(p,\,\xi)$} for \mbox{$p\in[0,\,\infty)$} and \mbox{$\xi\in[-1,\,1]$} defines a momentum space point. Note, that a pitch angle of \mbox{$\theta=0$} with a pith coordinate \mbox{$\xi=1$} represents the direction parallel to the magnetic field, while \mbox{$\theta=-\pi$} with a pitch coordinate \mbox{$\xi=-1$} describes the antiparallel direction.

As well, it is possible to describe the three-dimensional momentum space with the cylindrical coordinates triplet \mbox{$(p_{\|},\,p_{\perp},\,\varphi)$}, using the component of the momentum parallel to the magnetic field \mbox{$p_{\|}=\vert\mathbf{p}_{\|}\vert=(\gamma\,\vert\mathbf{v}_{\|}\vert)/c \in(-\infty,\,\infty)$}, the component of the momentum perpendicular to the magnetic field direction \mbox{$p_{\perp}=\vert\mathbf{p}_{\perp}\vert$} with \mbox{$p_{\perp}=( \gamma \,\vert\mathbf{v}_{\perp}\vert)/c\in[0,\,\infty)$} and the azimuthal angle \mbox{$\varphi\in[0,\,2\pi]$} measured in the plane orthogonal to the magnetic field. This is graphically displayed in figure \ref{fig_mom_coord}. \vspace{0.5mm}
\begin{figure}[H]
\centering
\begin{tikzpicture}
\begin{axis}[hide axis,x=1cm,y=1cm, ymin=-1.3,ymax=2.4, xmin=-3.5,xmax=9.5]
\addplot+ [solid,->,mark=none,line width=1.5,black] coordinates { (0,0) (-1.6*0.1+5.5*0.9949,1.6*0.9949+5.5*0.1)};
\addplot+ [solid,->,mark=none,line width=1.5pt,blue] coordinates { (0,0) (9*0.9949,9*0.1)}; 
\addplot+ [dashed,->,mark=none,line width=1.5pt, black] coordinates { (0,0) (-1.6*0.1,1.6*0.9949)}; 
\addplot+ [dashed,->,mark=none,line width=1.5pt, black] coordinates { (0,0) (5.5*0.9949,5.5*0.1)}; 
\addplot+ [solid,->,mark=none,line width=1.5pt,red] coordinates { (0,0) (0.9949,0.1)}; 
\addplot+ [solid,->,mark=none,line width=1.5pt,red] coordinates { (0,0) (-0.1,0.9949)}; 
\addplot+ [dotted,mark=none,line width=1.5,black!25] coordinates { (-1.6*0.1,1.6*0.9949) (-1.6*0.1+5.5*0.9949,1.6*0.9949+5.5*0.1)};
\addplot+ [dotted,mark=none,line width=1.5,black!25] coordinates { (5.5*0.9949,5.5*0.1) (-1.6*0.1+5.5*0.9949,1.6*0.9949+5.5*0.1)};
\node at (axis cs:9.5,9*0.1) [anchor=east] {$\mathbf{B}$};
\node at (axis cs:5.85,1.6*0.9949+5.5*0.1) [anchor=east] {$\mathbf{p}$};
\node at (axis cs:-3.3,2.5) [anchor=north east] {$\mathrlap{\mathbf{p}_{\perp}=p\overbrace{\sin{(\theta)}}^{=\,\sqrt{1-\xi^2}}\mathbf{e}_{\perp}}$};
\node at (axis cs:5.3,0.45) [anchor=north east] {$\mathrlap{\mathbf{p}_{\|}=p\underbrace{\cos{(\theta)}}_{\coloneqq\,\xi}\,\mathbf{e}_{\|}}$};
\node at (axis cs:-0.16,1) [anchor=north east] {\textcolor{red}{$\mathbf{e}_{\perp}=\dfrac{\mathbf{p}_{\perp}}{\vert\mathbf{p}_{\perp}\vert}=\dfrac{\mathbf{p}_{\perp}}{p_{\perp}}$}};
\node at (axis cs:2.8,0.06) [anchor=north east] {\textcolor{red}{$\mathbf{e}_{\|}=\dfrac{\mathbf{p}_{\|}}{\vert\mathbf{p}_{\|}\vert}=\dfrac{\mathbf{p}_{\|}}{p_{\|}}$}};
\node at (axis cs:2.40,0.83) [anchor=north east] {$\theta$};
\coordinate[] (A) at (axis cs:0.9949,0.1); 
\coordinate[] (B) at (axis cs:0,0); 
\coordinate[] (C) at (axis cs:-0.1,0.9949); 
\draw pic [draw,angle radius=4mm,line width=1pt]{angle =A--B--C};
\coordinate[] (D) at (axis cs:-1.6*0.1+5.5*0.9949,1.6*0.9949+5.5*0.1); 
\coordinate[] (E) at (axis cs:5.5*0.9949,5.5*0.1); 
\coordinate[] (F) at (axis cs:-0.9949,-0.1); 
\draw pic [draw,angle radius=4mm,line width=1pt]{angle =D--E--F};
\coordinate[] (G) at (axis cs:0,0); 
\coordinate[] (H) at (axis cs:-1.6*0.1,1.6*0.9949); 
\coordinate[] (I) at (axis cs:-1.6*0.1+5.5*0.9949,1.6*0.9949+5.5*0.1); 
\draw pic [draw,angle radius=4mm,line width=1pt]{angle =G--H--I};
\node at (axis cs:0.155,0.165)[circle,fill,inner sep=1pt]{};
\node at (axis cs:0.03,1.42*0.9949)[circle,fill,inner sep=1pt]{};
\node at (axis cs:5.31*0.9949,0.7)[circle,fill,inner sep=1pt]{};
\node at (axis cs:0,0)[circle,fill,inner sep=1.4pt,green]{};
\draw[->,line width=1] (axis cs:3.6,0.80) arc [start angle=-225,end angle=130,x radius=0.39cm,y radius =0.59cm];
\draw[->,line width=1] (axis cs:2.43,0.99) arc [start angle=60,end angle=39,x radius=0.65cm,y radius =3cm];
\node at (axis cs:4.15,0.65) [anchor=east] {$\varphi$};
\end{axis}
\end{tikzpicture}
\captionsetup{format=hang,indention=0cm}
\caption[Two-dimensional moving momentum coordinate system for a particle (light green) with respect to the local magnetic field $\mathbf{B}$ and depiction of the relations to the pitch-coordinate \mbox{$\xi\hspace{-0.2mm}\coloneqq\hspace{-0.2mm}\cos(\theta)\hspace{-0.2mm}\in\hspace{-0.2mm}\lbrack -1,\,1 \rbrack$} for \mbox{$\theta\hspace{-0.2mm}\in\hspace{-0.2mm}\lbrack -\pi,\,0 \rbrack$}.]{Two-dimensional moving momentum coordinate system\protect\footnotemark{} for a particle (light green) with respect to the local magnetic field $\mathbf{B}$ and depiction of the relations to the pitch-coordinate \mbox{$\xi:=\cos{(\theta)}\in[-1,\,1]$} for \mbox{$\theta\in[-\pi,\,0]\,$}.}
\label{fig_mom_coord}
\end{figure}
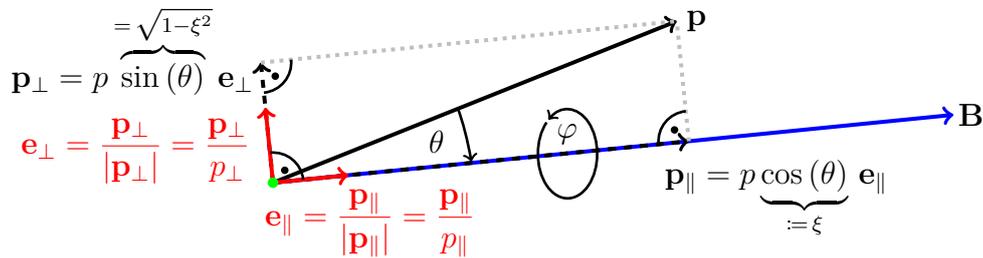\footnotetext{\label{coord_sys_fig_footnote} The graphic in figure \ref{fig_mom_coord} was created with \LaTeX-internal routines.}
\vspace*{-4.0mm}Therewith, the volume element in the three-dimensional momentum space reads\linebreak\mbox{$\mathrm{d}^{3}p=p_{\perp}\mathrm{d}p_{\perp}\mathrm{d}p_{\|}\,\mathrm{d}\varphi$}. By eliminating the dependence on the azimuthal angle it becomes\linebreak the following two-dimensional volume or area element: \vspace*{-7.5mm}\\
\begin{equation}\label{volelem_cyl_2D}
\mathrm{d}^{2}p=\int\limits_{\varphi}\mathrm{d}^{3}p=\int\limits_{0}^{2\pi}\mathrm{d}\varphi\,p_{\perp}\mathrm{d}p_{\perp}\mathrm{d}p_{\|}=2\pi\,p_{\perp}\mathrm{d}p_{\perp}\mathrm{d}p_{\|}\,.
\end{equation}
\vspace*{-5.5mm}\\The equivalence of the volume elements $(\ref{pitch_angle_volelem_sphere_2D})$ and $(\ref{volelem_cyl_2D})$ in the two discussed coordinate descriptions can de shown, with the help of the relations:\vspace*{-7.5mm}\\
\begin{equation}\label{p_relations}
p_{\|}=p\,\xi \;\;\;;\;\;\;p_{\perp}=p\,\sqrt{1-\xi^2}\,
\end{equation}
\vspace*{-8.0mm}\\which can be derived from figure \ref{fig_mom_coord}, using the trigonometric pythagoras and the fact that $\mathbf{e}_{\|}$ and $\mathbf{e}_{\perp}$ are orthonormal basis vectors. Said verifying calculation also involves the \textit{Jacobian} determinant, in order to perform the transformation between the coordinate systems and is carried out explicitly in the appendix of the reference \cite{study_thesis}.

\clearpage

\section{The runaway electron phenomenon}\label{RE_phenom_section}

In section \ref{kin_equa_section}, the kinetic equation was introduced as the governing equation for the behavior of plasma particles through the distribution function $f_{\alpha}$. However, to fully define the kinetic equation an adequate right-hand side, which describes collisions, energy sources and sinks, is required. For the purpose of the understanding of the runaway phenomenon, as mentioned in section \ref{tokamak_disruptions_section}, the following \textit{test-particle collision operator} $C^{\mathrm{tp}}_{\alpha\beta}$ for the particle species $\alpha$, which is connected to $C_{\alpha}\lbrace f_{\alpha}\rbrace$ through the relation $(\ref{coll_op_short})$, is suitable \cite{helander,HesslowPHD}:\vspace{-3.0mm}
\begin{equation}\label{coll_operator}
C_{\alpha\beta}^{\mathrm{tp}} =\nu_{d}^{\alpha\beta}\mathscr{L}\lbrace f_{\alpha}\rbrace+\dfrac{1}{p^2}\dfrac{\partial }{\partial p}\hspace{-0.4mm}\left[p^3\hspace{-0.4mm}\left(\nu_{s}^{\alpha\beta} f_{\alpha}+\nu_{\|}^{\alpha\beta}\dfrac{1}{2}p\dfrac{\partial f_{\alpha}}{\partial p}\right)\right]\,.
\end{equation}
\vspace*{-6.0mm}\\On this occasion, the three parameterizing frequencies and the \textit{Lorentz} scattering operator:\vspace{-3.5mm}
\begin{equation}\label{Lorentz_scatter_operator}
\mathscr{L}\lbrace f_{\alpha}\rbrace= \dfrac{p^2}{2}\cdot \Delta_{\mathbf{p}}\hspace{-0.4mm}\left[f_{\alpha}(\theta,\,\varphi)\right]= \dfrac{1}{2}\cdot\left(\dfrac{1}{\sin{\hspace{-0.8mm}\left(\hspace{-0.1mm}\theta\hspace{-0.1mm}\right)}}\dfrac{\partial}{\partial\theta}\hspace{-0.6mm}\left(\sin{\hspace{-0.8mm}\left(\hspace{-0.1mm}\theta\hspace{-0.1mm}\right)}\dfrac{\partial f_{\alpha}}{\partial\theta}\right)+\dfrac{1}{\left(\sin{\hspace{-0.8mm}\left(\hspace{-0.1mm}\theta\hspace{-0.1mm}\right)}\right)^2}\dfrac{\partial^2 f_{\alpha}}{\partial\varphi^2}\right)\,,
\end{equation}
\vspace*{-5.5mm}\\are introduced \cite{HesslowPHD}. The \textit{Lorentz} scattering operator is connected to the \textit{Laplace} operator $\Delta_{\mathbf{p}}$ in spherical momentum space coordinates $(p,\,\theta,\,\varphi)$ acting on a function with constant momentum magnitude $p$. Hence, one can comprehend, that it describes deflections at constant energy, which are often referred to as \textit{pitch-angle scattering}\cite{HesslowPHD,helander}. The associated \textit{deflection frequency} $\nu_{d}^{\alpha\beta}$, consequently leads to a more isotropic distribution function in momentum space. The second term in the operator from $(\ref{coll_operator})$, defines the distribution of the kinetic energy, since it depends on the magnitude of the momentum vector. The term contains, the
\textit{slowing-down frequency} $\nu_{s}^{\alpha\beta}$, determining collision induced deceleration of particles thus friction and the \textit{parallel momentum diffusion frequency} $\nu_{\|}^{\alpha\beta}$, responsible for the smoothing of energy distribution gradients \cite{HesslowPHD}. 

The slowing-down frequency for electron-electron interactions $\nu_{s}^{ee}$ defines the momentum transfer between electrons and thus the average \textit{dynamical friction force} $F_{fr}$ acting on an electron. Its approximative velocity dependence, under neglection of the energy-dependency of the \textit{Coulomb} logarithm, in the non-relativistic temperature limit, expressed by means of the normalized temperature:\vspace{-1.5mm}
\begin{equation}\label{normal_temp}
\Theta\coloneqq \dfrac{k_{B}T_{e}}{m_{e0}c^2} \ll 1
\end{equation}
\vspace*{-7.0mm}\\was deduced by \textit{L.\hspace{0.9mm}Hesslow} in \cite{HesslowPHD} together with the functions representing the asymptotic behaviour in the superthermal respectively relativistic limit \mbox{$\gamma-1\gg\Theta$} and the non-relativistic velocity limit \mbox{$p\ll 1$}. Expressed with respect to the gamma factor and the normalized momentum, which are functions of the velocity as given in equation $(\ref{p_norm_gamma_def})$, it reads \cite{HesslowPHD}\vspace{-4.5mm}
\begin{equation}\label{F_fr}
\begin{split}
\begin{gathered}
F_{fr}(v) = - p \hspace*{0.1mm}\nu_{s}^{ee}\sim  \dfrac{\gamma^2}{p^2}\cdot\textup{erf}\hspace{-0.5mm}\left(\hspace{-0.3mm}\dfrac{p}{\sqrt{2\Theta}}\hspace{-0.3mm}\right) -\dfrac{1}{p }\sqrt{\dfrac{2}{\pi\Theta}}\cdot\textup{exp}\hspace{-0.5mm}\left(\hspace{-0.3mm}-\dfrac{\gamma-1}{\Theta}\hspace{-0.3mm}\right) 
\\[6pt]
= \begin{cases}\longrightarrow \;\dfrac{1}{\Theta}\cdot G\hspace{-0.7mm}\left(\hspace{-0.5mm}\dfrac{p}{\sqrt{2\Theta}}\hspace{-0.5mm}\right)\;\;;\;\;p \ll 1 \\ \longrightarrow \;\dfrac{c^2}{v^2}\;\;;\;\; \gamma-1 \gg \Theta
\end{cases} .
\end{gathered}
\end{split}
\end{equation}
\vspace*{-5.5mm}\\Here, the \textit{error function} $\textup{erf}(x)$ and the  \mbox{\textit{Chandrasekhar}} function \cite{helander}:
\begin{equation}\label{Chadrasekhar}
G(x) = \dfrac{1}{2 x^2}\cdot\left( \textup{erf}(x)-x\cdot\dfrac{\mathrm{d}}{\mathrm{d}x}\left[ \textup{erf}(x)\right] \right)= \dfrac{1}{2x^2}\cdot\left( \textup{erf}(x)-x\cdot\dfrac{2}{\sqrt{\pi}}\cdot e^{-x^2} \right) 
\end{equation}
appear, where $G(x)$ itself depends on the error function and its first derivative. 

\begin{figure}[H]
\begin{center}
\includegraphics[trim=127 42 71 17,width=0.98\textwidth,clip]{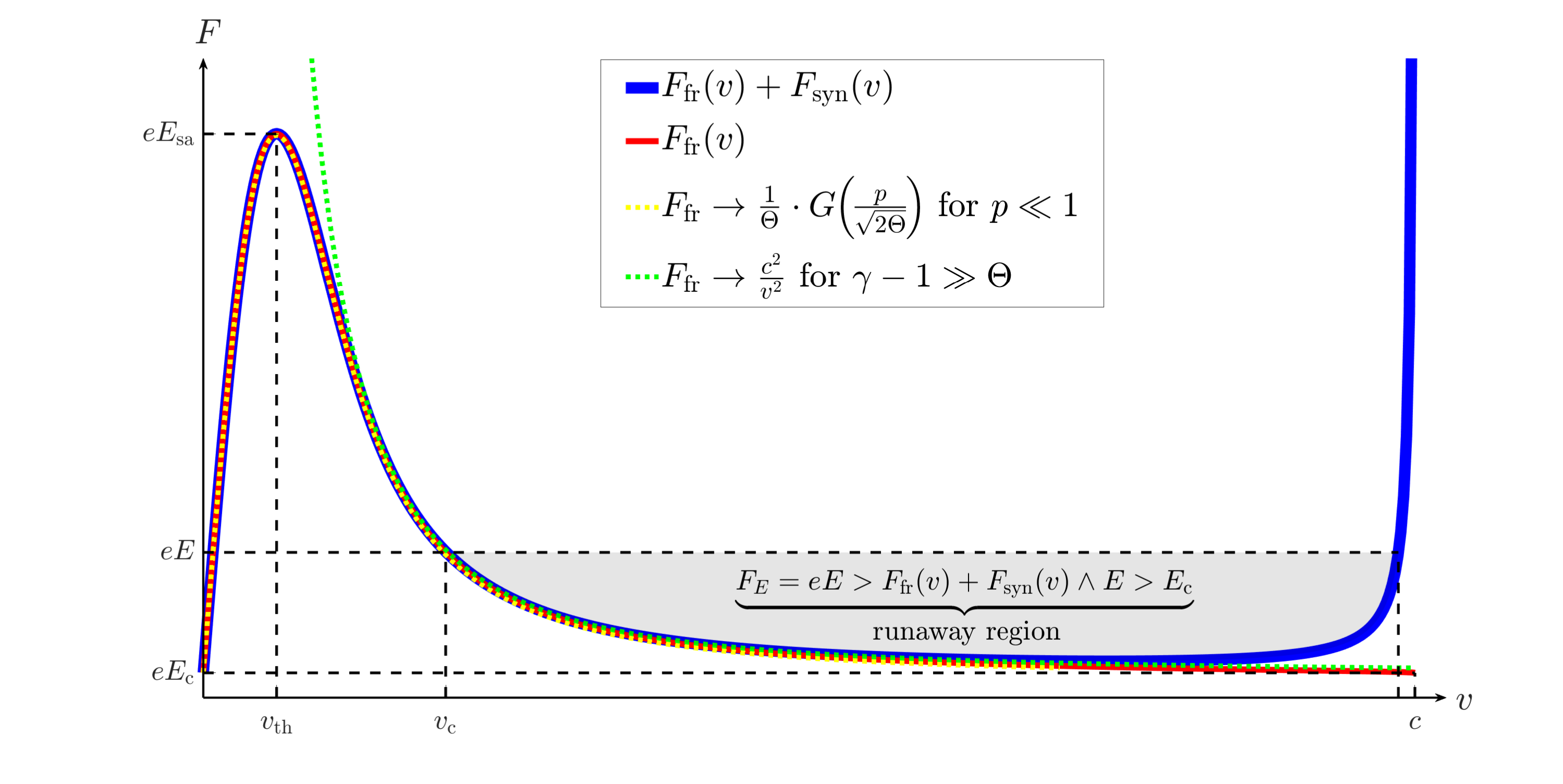}
\caption[Qualitative depiction of the forces effecting the electron dynamics for an induced electric field with \mbox{$E_{c} < E \ll E_{sa}$} and schematic representation of the region for runaway-electron generation (see also\linebreak\cite{stahl,pappPHD,Hoppe_PHD,HesslowPHD}).]{Qualitative depiction\protect\footnotemark{} of the forces effecting the electron dynamics for an induced electric field with \mbox{$E_{c} < E \ll E_{sa}$} and schematic representation of the region for runaway-electron generation (see also \cite{stahl,pappPHD,Hoppe_PHD,HesslowPHD}).}
\label{fig_RE_region}
\end{center}
\end{figure}

As described in section \ref{tokamak_disruptions_section}, the plasma current in tokamak reactors can abruptly change, due to the occurrence of plasma instabilities \cite{Hender_2007}. During such a disruption \cite{Hoppe_2021} or while the disruption mitigation takes place \cite{REsimulation} inductive toroidal electric fields up to $100\,\mathrm{V}\hspace{0.25mm}\mathrm{m}^{-1}$ can be produced \cite{REsimulation,Hender_2007}. In addition, smaller toroidal electric fields are also induced in the start-up phase of a tokamak reactor \cite{Hoppe_2022}, where they are necessary to fastly create a sufficiently high plasma current and temperature, in order to achieve a fully ionized plasma and magnetic confinement.
\\
In both scenarios the electrons with the negative elementary charge \mbox{$q_{e}=-e$} within the plasma are accelerated by the force:\vspace*{-3.5mm}
\begin{equation}\label{F_e}
F_{E}=q_{e}E=-eE\,,
\end{equation}
\vspace*{-10.0mm}\\originating from the induced electric field $E$. The resulting force from the dynamical friction force and the accelerating force:\vspace*{-2.5mm}
\begin{equation}\label{F_res1}
F_{res}=\dfrac{\mathrm{d}p_{res}}{\mathrm{d}t} \coloneqq \vert F_{E} \vert -  \vert F_{fr}(v) \vert 
\end{equation}
\vspace*{-9.0mm}\\defines an equation of motion for the electrons, which allows oneself to determine, if the electrons are accelerated or slowed down. The analysis of the graphs of the asymptotic functions of $F_{fr}(v)$, defined through the expressions in $(\ref{F_fr})$, are displayed in figure \ref{fig_RE_region} on page \pageref{fig_RE_region}. They show, that the dynamical friction force decreases for velocities greater than the thermal velocity \cite{wesson,stahl}, respectively the thermal momentum, for \mbox{$\gamma(v_{th})\approx 1$}:\vspace*{-2.5mm}
\begin{equation}\label{v_th}
v_{th}=\sqrt{\dfrac{2\hspace{0.25mm}e\hspace{0.25mm}k_{B}T_{e}}{m_{e0}}} \;\;;\;\; p_{th}=\dfrac{m_{e0}\hspace{0.25mm}\gamma(v_{th})\hspace{0.25mm}v_{th}}{m_{e0}\hspace{0.25mm}c}\approx\sqrt{\dfrac{2\hspace{0.25mm}e\hspace{0.25mm}k_{B}T_{e}}{m_{e0}\hspace{0.25mm}c^2}}
\end{equation}
\vspace*{-8.5mm}\\and reaches a minimum value in the limit \mbox{$v \rightarrow c$}, if radiation losses are neglected. The electric field associated to this minimum of the friction force follows from the collision frequency for a highly relativistic electron and was found by \mbox{\textit{J.\hspace{0.7mm}W.\hspace{0.9mm}Connor}} and \mbox{\textit{R.\hspace{0.7mm}J.\hspace{0.9mm}Hastie}} in $1975$ \cite{stahl,Connor_1975}. It is known as the \textit{critical} electric field $E_{c}$, which is connected to the characteristic relativistic collision time $\tau_{rel}$. It can be written as:\vspace*{-2mm}
\begin{equation}\label{E_crit}
E_{c}= \dfrac{m_{e0}\,c}{e\,\tau_{rel}} = \dfrac{n_{e}\,e^3\ln{\hspace{-0.45mm}\Lambda_{rel}}}{4\pi\,\varepsilon_{0}^2 \,m_{e0}\,c^2}\,,
\end{equation}
\vspace*{-7.0mm}\\where the \textit{Coulomb} logarithm was used, which is the factor by which small-angle collisions are more effective than large-angle collisions. It can be calculated, for instance for collisions between relativistic and thermal
electrons, from the relation \cite{Svensson_2021,Hesslow_2018II}:\vspace{-2.0mm}
\begin{equation}\label{CoulombLogrel}
\ln{\hspace{-0.45mm}\Lambda_{rel}} = \ln{\hspace{-0.45mm}\Lambda_{th}}+\dfrac{1}{2}\cdot\ln{\hspace{-0.5mm}\left( \dfrac{ m_{\mathrm{e}0}\hspace{0.3mm}c^2}{ k_{\mathrm{B}}T_{\mathrm{e}}  }\right)}\approx 14.6+\dfrac{1}{2}\cdot\ln{\hspace{-0.5mm}\left( \dfrac{ k_{\mathrm{B}}T_{\mathrm{e}}\left[\mathrm{eV}\right] }{10^{-20}\cdot n_{\mathrm{e}}\left[\mathrm{m}^{-3}\right]}\right)} 
\end{equation}
\vspace*{-6.5mm}\\with the \textit{Boltzmann} constant expressed as \mbox{$k_{B}=8.617333262\cdot 10^{-5} \,\si{\electronvolt\per\kelvin}$} \cite{NISTkB} and the thermal \textit{Coulomb} logarithm $\ln{\hspace{-0.45mm}\Lambda_{th}}$. It should be remarked, that the neglected energy dependence of the \textit{Coulomb} logarithm would increase the critical electric field in the limit \mbox{$v \rightarrow c$}, proceeding from the \textit{Coulomb} logarithm for the collisions of thermal electrons, which can be expressed as \cite{wesson}:\vspace{-2.0mm}
\begin{equation}\label{CoulombLogth}
\ln{\hspace{-0.45mm}\Lambda_{th}} \approx 14.9-0.5\cdot\ln{\hspace{-0.6mm}\left(\hspace{-0.35mm}10^{-20}\hspace{-0.3mm}\cdot\hspace{-0.05mm} n_{\mathrm{e}}\left[\mathrm{m}^{-3}\right] \hspace{-0.15mm}\right)}+  \ln{\hspace{-0.5mm}\left(\hspace{-0.35mm}10^{-3}\hspace{-0.3mm}\cdot\hspace{-0.05mm}k_{B}T_e\left[\mathrm{eV}\right]  \hspace{-0.15mm}\right)}
\end{equation}
\vspace*{-8.5mm}\\Therefore a relation might be used, which includes the transition between a thermal and a relativistic or superthermal expression for the energy-dependent $\ln{\hspace{-0.4mm}\Lambda}$, as defined in the reference \cite{Hesslow_2018II} through:\vspace{-5.0mm}
\begin{equation}\label{CoulombLogHesslow}
\ln{\hspace{-0.45mm}\Lambda} =  \ln{\hspace{-0.45mm}\Lambda_{th}} + \dfrac{1}{\kappa}\cdot\ln{\hspace{-0.6mm}\left(\hspace{-0.25mm}1+\left(\dfrac{2\hspace{0.2mm}p}{p_{th}}\right)^{\hspace{-1.3mm}\kappa}\right)}
\end{equation}
\vspace*{-8.0mm}\\with \mbox{$\kappa=5$} and $p_{th}$ from $(\ref{v_th})$.  
\\
The critical velocity and thus the critical relativistic momentum is related to the critical electric field strength from (\ref{E_crit}) by the following expressions \cite{stahl}:\vspace*{-2.0mm}
\begin{equation}\label{p_crit}
p_{c} = \dfrac{\gamma_{c}\hspace{-0.2mm}(v_{c})\hspace{0.25mm}v_{c}}{c}= \left(\dfrac{E}{E_{c}}-1 \right)^{-\frac{1}{2}} \;\;;\;\;  \gamma(v_{c})=\left(1-\left(\dfrac{v_{c}}{c}\right)^2\right)^{-\frac{1}{2}}=\left(1-\dfrac{E_{c}}{E}\right)^{-\frac{1}{2}}\,.
\end{equation} 
\vspace*{-6.5mm}\\In conclusion, one discovers that electrons with a velocity $v>v_{c}$ experience a continuous net acceleration, if an induced electric field $E$ is larger than the critical electric field $E_{c}$. Those electrons with \mbox{$v\gg v_{th}$} \cite{wesson,stahl} are superthermal in relation to the thermal electrons of a distribution of electrons in the phase space with \mbox{$v\approx v_{th}$}. Moreover, all electrons with \mbox{$v\gg v_{c}$} reach relativistic velocities and are referred to as \textit{runaway electrons}. 

For electric fields with $E>E_{sa}\cong0.214E_{D}$ \cite{Dreicer_1959} the whole electron population becomes a runaway electron population, where the \textit{Dreicer} field is given as \cite{stahl}:\vspace*{-2.0mm}
\begin{equation}\label{Dreicerfield}
E_{D} = \dfrac{m_{e0}\,c^2}{e\,k_{B}T_{e}}\cdot E_{c} = \dfrac{n_{e}\,e^3\ln{\hspace{-0.45mm}\Lambda_{rel}}}{4\pi\,\varepsilon_{0}^{2}\,e\,k_{B}T_{e}}\,.
\end{equation}
\vspace*{-7.0mm}\\However, this so-called \textit{slide-away} phenomenon \cite{Coppi_1976} does almost never appear in tokamak reactors, because it often holds \mbox{$E\ll E_{sa}$} for fusion plasmas (compare \cite{stahl}). Hence, the important interval for the study of runaway electron dynamics is \mbox{$E_{c} < E \ll E_{sa}$}. 

In the case of a tokamak disruption, the total energy from the plasma current bounds the acceleration of the electrons. Nevertheless, one might consider other reaction forces like \textit{synchrotron radiation}, which results from the gyro motion of the electrons around the curved magnetic field lines \cite{stahl} and is further discussed in subsection \ref{RE_loss_subsection}. Note, that also \textit{Bremsstrahlung}, as explained in more detail in subsection \ref{RE_loss_subsection}, leads to a reaction force, which counteracts the acceleration of the electrons. However, radiation losses will be represented by synchrotron radiation for the rest of this section. 
\\
In order to visualize the runaway region by means of the resulting force under consideration of synchrotron radiation the following condition, using $(\ref{F_e})$, is applicable: \vspace*{-8.5mm} 
\begin{equation}\label{F_res2}
F_{res}(v)=  \vert F_{E}  \vert -  \vert F_{fr}(v) \vert -  \vert F_{syn}(v) \vert \overset{!}{>} 0 \;\longleftrightarrow\; eE \overset{!}{>}   F_{fr}    +    F_{syn}   \;\wedge \; v>v_{c}\,.
\end{equation}
\vspace*{-9.0mm}\\For this purpose, the \textsc{MATLAB}-script$^{\ref{fig_RE_region_footnote}}$, which generated the figure \ref{fig_RE_region}, computed the pitch-averaged magnitude of the synchrotron radiation reaction force vector. At this, the utilized computation rule is based on a relation for $\mathbf{F}_{syn}$ with respect to the parallel and orthogonal momentum as depicted in figure \ref{fig_mom_coord} from section \ref{mom_space_coord_section} and stated redundantly in the references \cite{Hoppe_PHD} and \cite{Papp_2011}:
\vspace*{-2.5mm}
\begin{equation}\label{F_syn0}
\mathbf{F}_{syn} =  -\dfrac{\nu_{syn}}{\gamma}\left(\mathbf{p}_{\perp}+p_{\perp}^2\mathbf{p} \right)\;\;;\;\; \nu_{syn}=\dfrac{1}{\tau_{syn}} =\dfrac{e^4 B^2}{6\pi\hspace{0.15mm}\varepsilon_{0}\hspace{0.15mm} m_{e0}^3\hspace{0.15mm} c^3}\,,
\end{equation}
\vspace*{-7.5mm}\\introducing the characteristic synchrotron radiation timescale $\tau_{syn}$ and the associated frequency $\nu_{syn}$. With the help of $(\ref{p_relations})$ and $(\ref{F_syn0})$, the calculation equation with respect to the normalized momentum $p$, where $p$ was defined as $p(v)$ in the expression $(\ref{p_norm_gamma_def})$, reads:\vspace*{-4.5mm} 
\begin{equation}\label{F_syn}
F_{syn}(v)\hspace{-0.5mm}\coloneqq \hspace{-0.3mm}  \left\langle \vert  \mathbf{F}_{syn}(p,\,\xi) \vert  \right\rangle_{\xi} \hspace{-1mm}\sim\hspace{-0.1mm} \dfrac{1}{2}\hspace{-0.1mm}\int\limits^{1}_{\xi=-1} \hspace{-2.5mm}p\hspace{0.5mm} \sqrt{\dfrac{ 1-\xi^2}{1+p^2}\hspace{-1.0mm}\left(\hspace{-0.1mm}\xi^2 p^4\hspace{-0.2mm}-\hspace{-0.1mm}\xi^4 p^4\hspace{-0.15mm}+\hspace{-0.2mm} \left[1\hspace{-0.1mm}+\hspace{-0.1mm}p^2\hspace{-0.7mm}\left(1\hspace{-0.15mm}-\hspace{-0.1mm}\xi^2\right)\right]^2\right)} \; \mathrm{d}\xi\,.
\end{equation}
\vspace*{-6.0mm}\\The generation region for runaway electrons based on the velocity dependence of the dynamical friction force under consideration of synchrotron radiation can then be looked upon schematically in figure \ref{fig_RE_region}. 

On this occasion, the \textit{runaway region} is emphasized, where the magnitude of the force $F_{E}$, which accelerates the electrons due to the prevalent electric field, is larger than the sum of the absolute values of the decelerating forces $F_{fr}+F_{syn}$, caused by collisions and synchrotron radiation. The intersections of the magnitude of the force $F_{E}$ with said decelerating forces defines an electron population with velocities \mbox{$v_{c} < v < v _{max}\approx c$} as the fraction of all electrons, which will run away. At this, the height of the runaway region for a given velocity \mbox{$v\in[v_{c} ,\, v _{max}]$} represents the runaway net acceleration, of all electrons with this initial velocity. In addition, one has to remark, that the runaway condition \mbox{$E>E_{c}\wedge v>v_{c}$} only holds, if no radiation reaction forces and a fully ionized plasma are considered. This \textit{Connor-Hastie} runaway region is \textit{not} colored in figure \ref{fig_RE_region} and would postulate that electrons with a velocity \mbox{$v\in[v_{c} ,\, c]$} will run away, since the corresponding critical electric field $E_{c}$ was deduced solely from the friction force in the limit \mbox{$v\rightarrow c$}. The runaway region in reality is narrower, because the effective critical electric field and momentum are higher. Therefore more elaborate runaway modeling, as it will be explained in the following section \ref{part_screen_section}, include radiation losses and partial screening effects, due to the fact that real plasmas are not fully ionized.  
\vspace{1.0mm}
\footnotetext{\label{fig_RE_region_footnote} The diagram in figure \ref{fig_RE_region} was produced with the \textsc{MATLAB}-file \qq{\texttt{plot_runaway_region.m}}.\\ \hspace*{8.7mm}The script and its output \qq{\texttt{output_plot_runaway_region.txt}} can be found in the\\ \hspace*{8.7mm}digital appendix.}
\vspace{-9.5mm}
Typically, one finds the character of the runaway electron population in a tokamak disruption to be of a highly energetic beam along the magnetic field lines or flux surfaces with a radially centered localization \cite{REdistfuncderivation}. If this beam comes into contact with the inner wall of the vacuum vessel of the reactor, this localised energy input can lead to severe damages \cite{Matthews2016,Reux2015}. In case of the JET experiment a runaway electron current of \mbox{$I_{RE}\approx 1\,\mathrm{MA}$} was able to damage the wall and for ITER current strength of up to \mbox{$10\,\mathrm{MA}$} are expected \cite{Linder_2020}. Hence, runaway electron current simulations are required to be as efficient as possible, in order to develop optimal mechanisms to avoid damage to the reactor. With such efficient computation models, it is possible to run more simulations per unit time without sacrificing accuracy, so that the research progress can be supported. Therefore, the \textit{reduced kinetic modeling} approach, which relies on the self-consistent simulation of the plasma evolution in combination with the application of approximations, simplified assumptions and lower-dimensional plasma descriptions, conduces as a basis of typical simulation software. A typical simulation software that adopts this approach is the DREAM-code \cite{Hoppe_2022}, which is also capable of performing fully kinetic calculations. Similarly, a self-consistent evolution of the runaway current is possible \cite{pappPHD}. For this, the calculation of the moments of analytically or numerically defined distribution functions for the runaway electrons, where e.g.\ the runaway electron current density is determined by the first moment, can be an enrichment for fast simulation software.   
 
Finally, it should be remarked, that the runaway electron phenomenon is not restricted to a thermonuclear fusion plasma in tokamak reactors. In fact, runaway electrons also form e.g.\ in atmospheric plasmas during lightning discharges \cite{Gurevich1994} or in astrophysical plasmas like solar flares \cite{Holman1985}, where one is referred to the PhD thesis of \textit{A.\hspace{0.9mm}Stahl} \cite{stahl} for additional examples.

\clearpage

\section{The effect of partially ionized impurities}\label{part_screen_section}

Realistic plasmas are never in a completely ionized state and in addition a continuous entry and appearance of impurities is inevitable. In a tokamak, those impurities might originate from the wall or are already present in the material flow of the nuclear fusion fuel. Moreover, the injection of material is often used for disruption mitigation as described in section \ref{tokamak_disruptions_section}. In the following, the effects of partially ionized impurities on the plasma physics within a nuclear fusion reactor shall be discussed, by means of the references \cite{HesslowPHD,Hesslow_2018,Svensson_2021}.
 
For such partially ionized plasmas it holds, that electrons with relativistic speeds close to the speed of light in vacuum can penetrate the cloud of the bound electrons of the partially ionized atoms. Those electrons therefore experience an only partially screened nuclear charge and hence a stronger interaction with this positive charge, e.g.\ in the form of an intensified \textit{Coulomb} interaction. Since avalanche runaway electrons intrinsically reach ultra-relativistic velocities, their slowing-down and deflection frequencies, related to inelastic and elastic \textit{Coulomb} collisions \cite{Hesslow_2017}, are influenced by the effect of partial screening. Consequently, the dynamical friction force at large momenta is enhanced, which will add to the already mentioned mechanisms, providing an additional limit to the energies that runaway electrons can reach.

A further effect of partial screening is a up to tens of percent larger effective critical electric field $E_{c}^{\mathrm{eff}}$ in comparison to the \textit{Connor-Hastie} critical electric field $E_{c}$ from $(\ref{E_crit})$ \cite{Hesslow_2018}. This is caused, by higher collision rates thus stronger deflection and pitch-angle scattering. Furthermore, \textit{Bremsstrahlung} is enhanced directly, while the emission of synchrotron radiation increases as a consequence of the larger pitch-angle scattering rate \cite{HesslowPHD}. The larger effective critical electric field then leads to a higher critical momentum as the lower bound in momentum space for the runaway region.  

\begin{figure}[H]
  \centering
  \subfloat{\label{fig_E_c_ava_p_c_scr_E100_main} 
    \includegraphics[trim=320 27 381 17,width=0.32438\textwidth,clip]
    {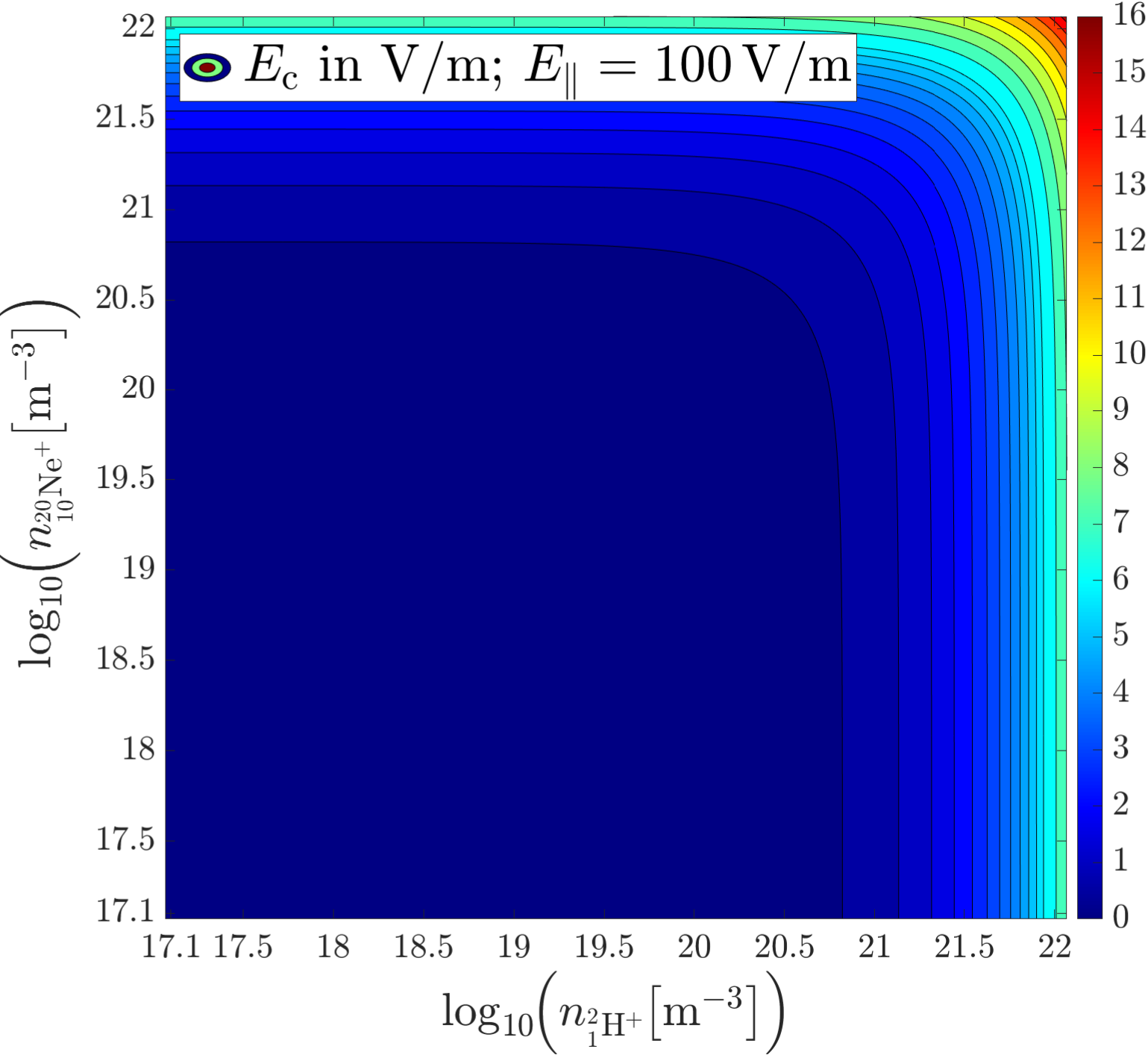}}\hfill
  \subfloat{\label{fig_E_c_tot_ava_p_c_scr_E10_main}
    \includegraphics[trim=320 27 381 17,width=0.324455\textwidth,clip]
    {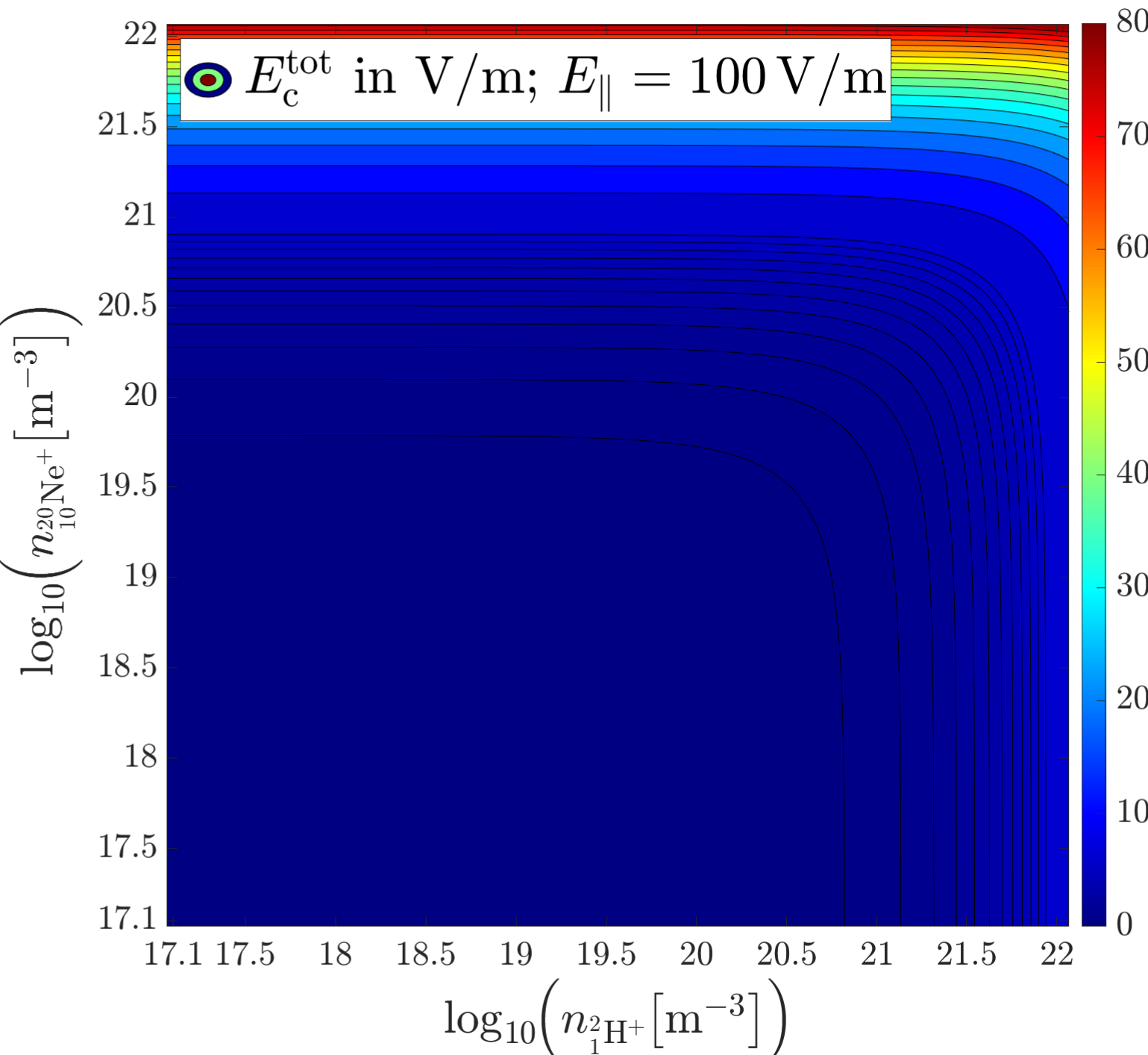}}\hfill
  \subfloat{\label{fig_E_c_eff_ava_p_c_scr_E100_main} 
    \includegraphics[trim=320 22 366 23,width=0.324455\textwidth,clip]
    {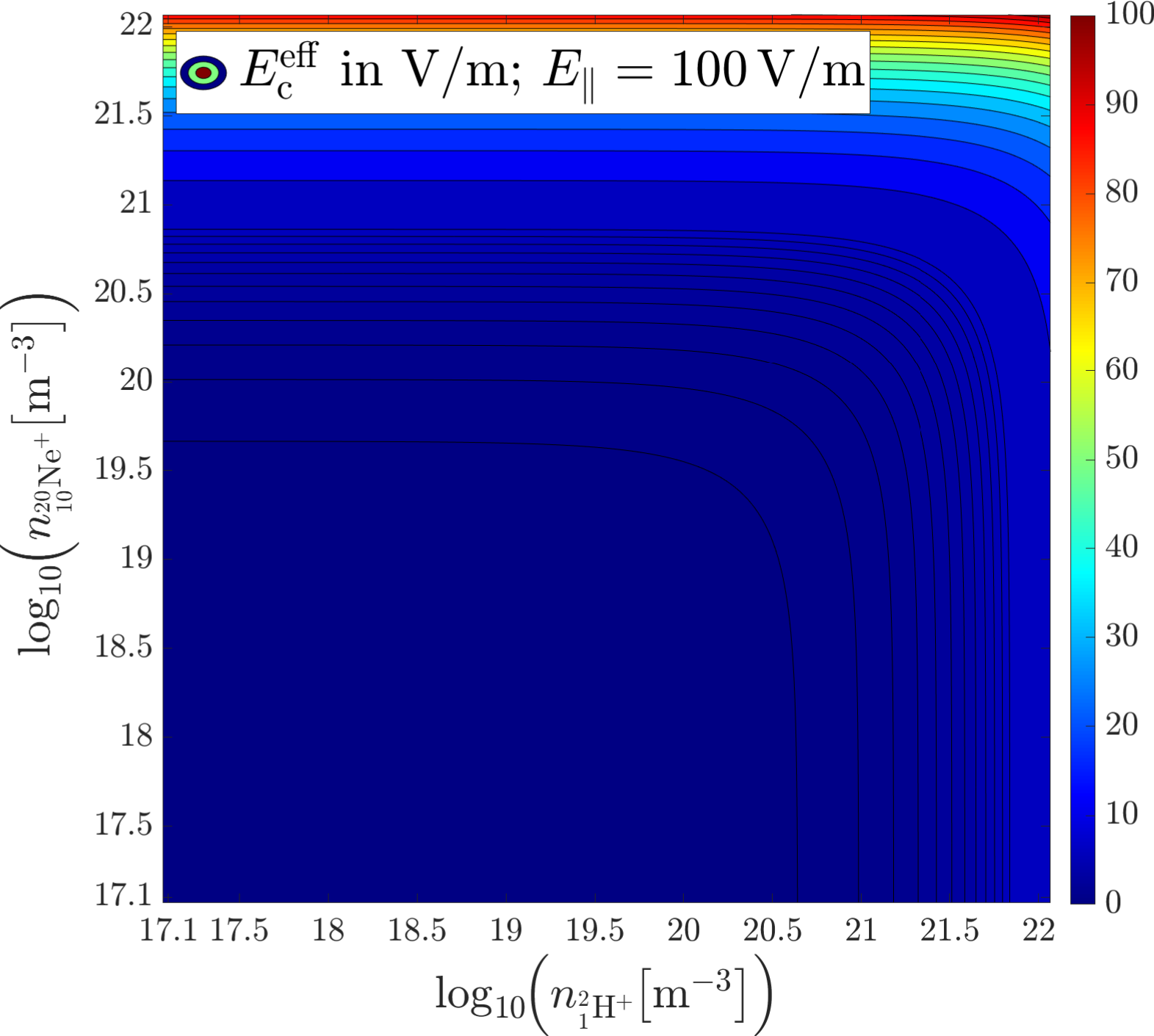}}\\[7pt]
  \subfloat{\label{fig_rel_E_c_tot_ava_p_c_scr_E100_main}
    \includegraphics[trim=320 28 348 14,width=0.32446\textwidth,clip]
    {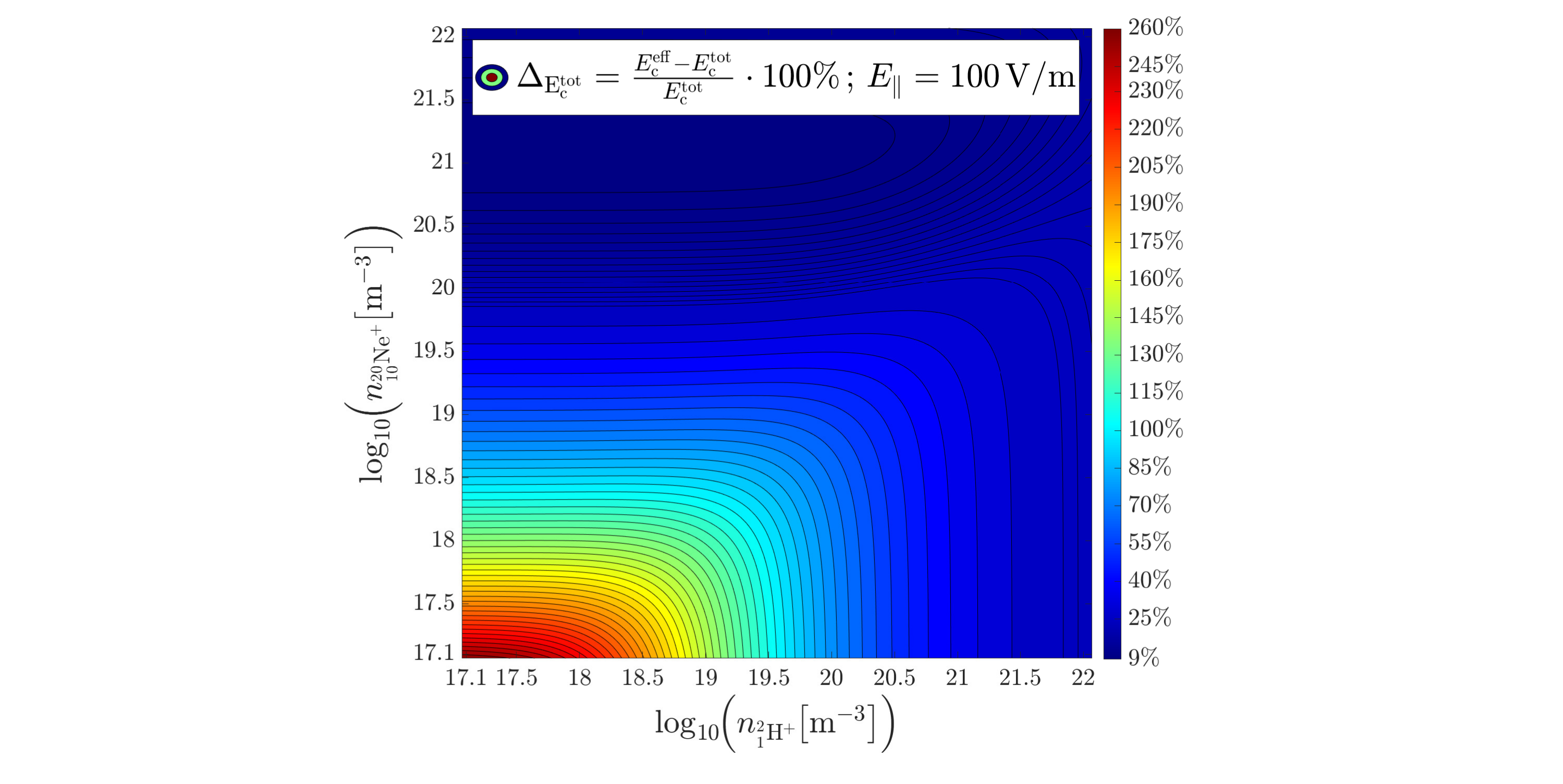}}\quad
  \subfloat{\label{fig_rel_E_c_eff_ava_p_c_scr_E100_main} 
    \includegraphics[trim=319 22 342 16,width=0.32448\textwidth,clip]
    {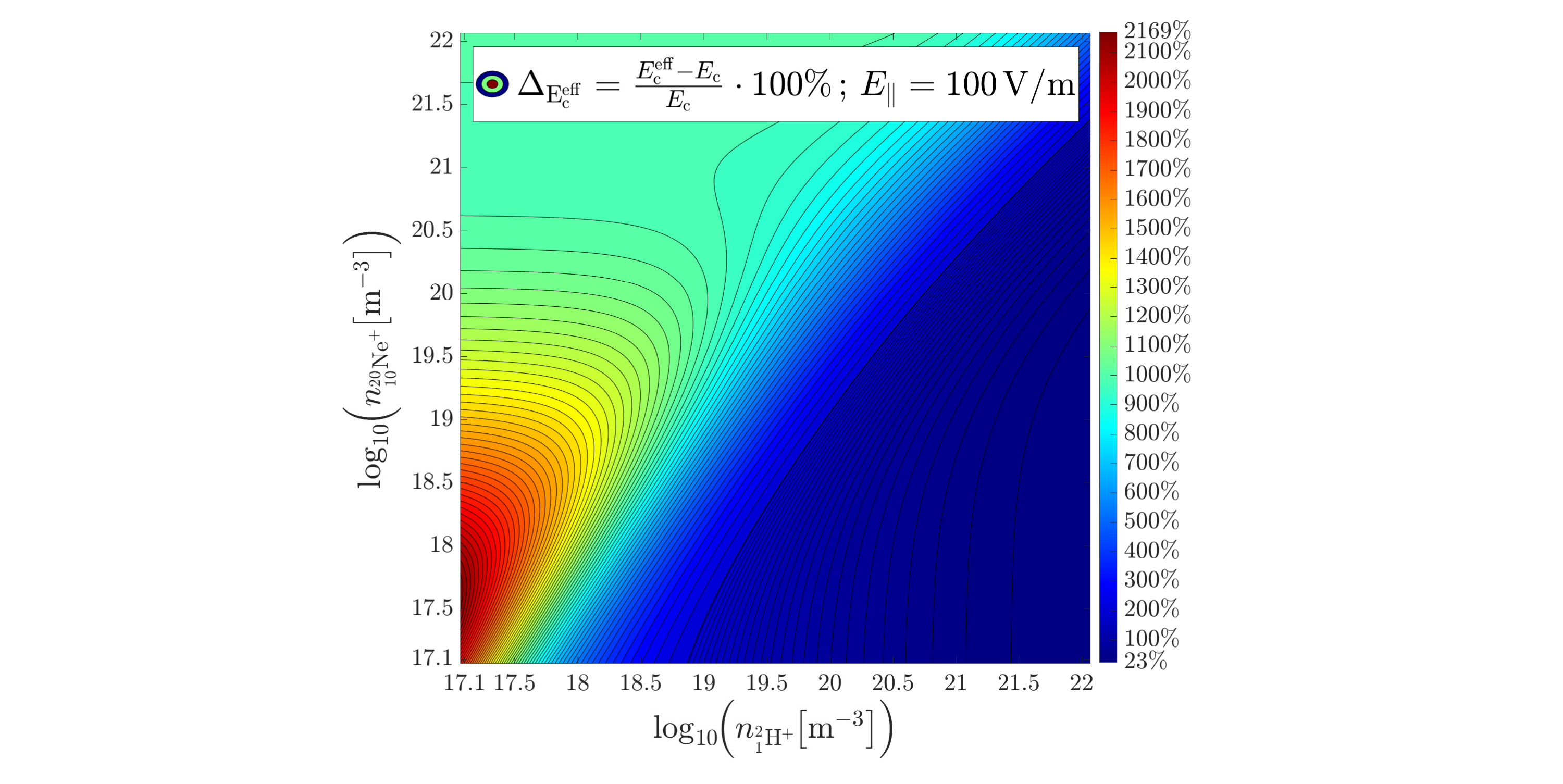}}
  \caption[Contour plots of the \textit{Connor-Hastie} critical electric field from the free electron density $E_{\mathrm{c}}$, from the total electron density $E_{\mathrm{c}}^{\mathrm{tot}}$, the effective critical electric field $E_{\mathrm{c}}^{\mathrm{eff}}$ and the relative deviations $\Delta_{E_{\mathrm{c}}^{\mathrm{tot}}}$ and $\Delta_{E_{\mathrm{c}}^{\mathrm{eff}}}$ for \mbox{$k_{\mathrm{B}}T_{\mathrm{e}}=10\,\textup{eV}$}, \mbox{$B=5.25\,\textup{T}$}, \mbox{$Z_{\mathrm{eff}}=1$} and \mbox{$E_{\|}=100\,\mathrm{V/m}$} (larger view in figure \ref{fig_E_C_ava_p_c_scr} of the appendix).]{Contour plots\protect\footnotemark{} of the \textit{Connor-Hastie} critical electric field from the free electron density $E_{\mathrm{c}}$, from the total electron density $E_{\mathrm{c}}^{\mathrm{tot}}$, the effective critical electric field $E_{\mathrm{c}}^{\mathrm{eff}}$ and the relative deviations $\Delta_{E_{\mathrm{c}}^{\mathrm{tot}}}$ and $\Delta_{E_{\mathrm{c}}^{\mathrm{eff}}}$ for \mbox{$k_{\mathrm{B}}T_{\mathrm{e}}=10\,\textup{eV}$}, \mbox{$B=5.25\,\textup{T}$}, \mbox{$Z_{\mathrm{eff}}=1$} and \mbox{$E_{\|}=100\,\mathrm{V/m}$} (larger view in figure \ref{fig_E_C_ava_p_c_scr} of the appendix).}
\label{fig_E_C_ava_p_c_scr_main}
\end{figure}

\begin{figure}[H]
\begin{center}
\includegraphics[trim=15 155 9 46,width=0.97\textwidth,clip]{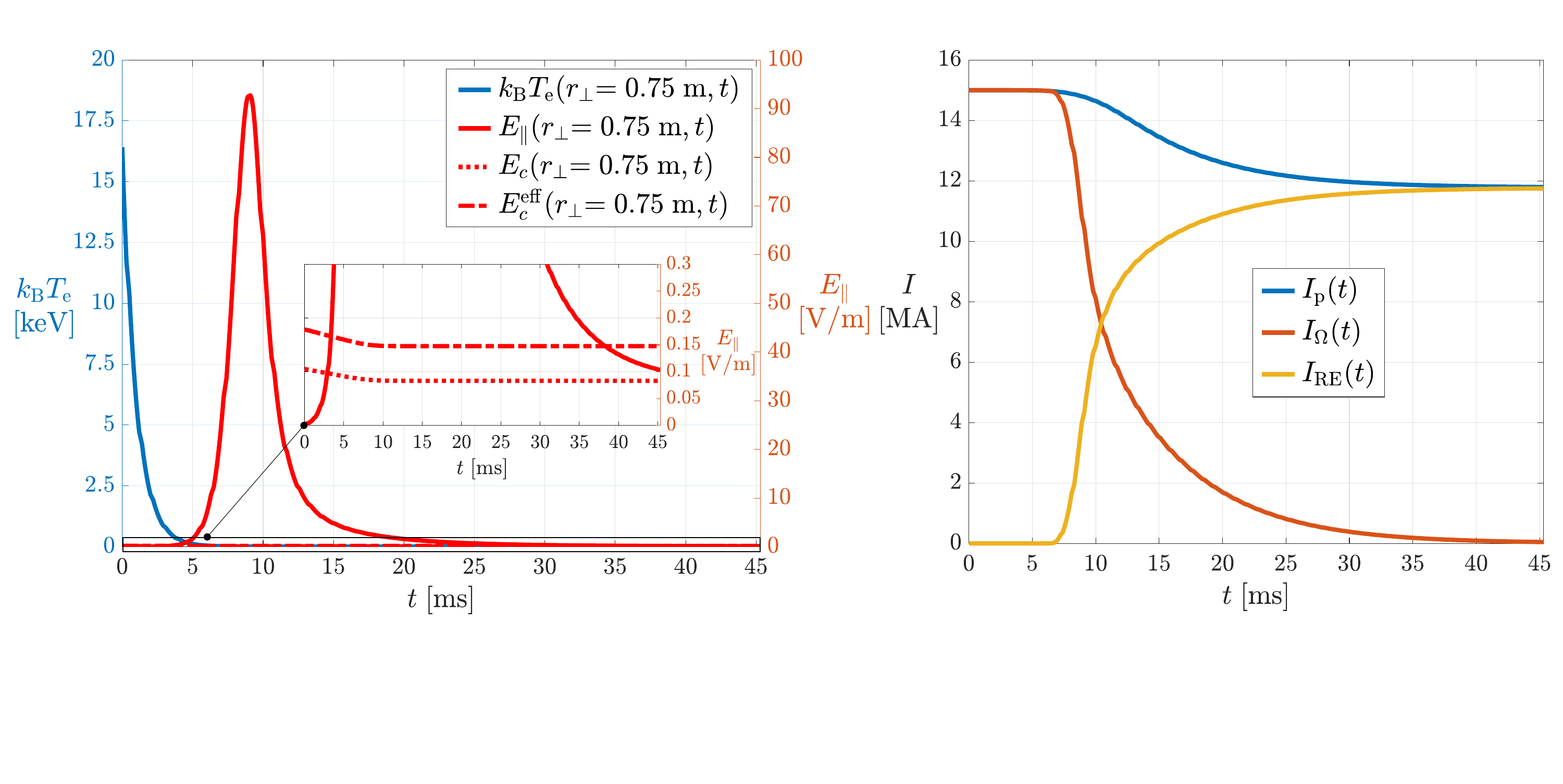}  
\caption[Time evolution of characteristic plasma quantities during a disruption for an ITER-scenario \cite{Smith_2009}.]{Time evolution\protect\footnotemark{} of characterisitc plasma quantities during a disruption for an ITER-scenario \cite{Smith_2009}.}
\label{fig_RE_ITER_simu}
\end{center}
\end{figure}

\begin{figure}[H]
\begin{center}
\includegraphics[trim=70 15 130 27,width=0.9\textwidth,clip]{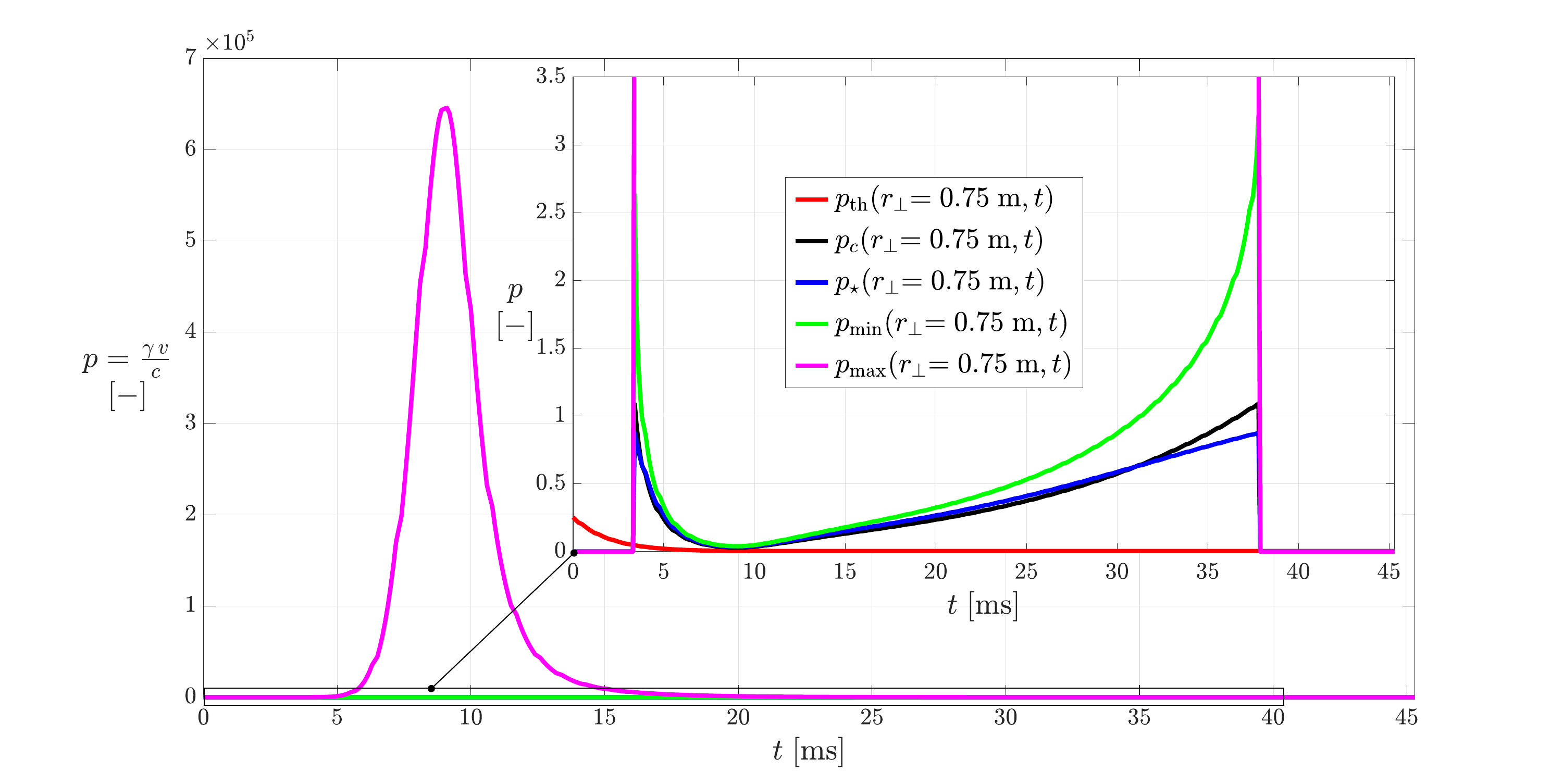}
\caption[Time evolution of different representations of the momentum boundaries of the runaway region, during a disruption for an ITER-scenario \cite{Smith_2009}.]{Time evolution$^{\text{\ref{fig_RE_ITER_simu_footnote}}}$ of different representations of the momentum boundaries of the runaway region, during a disruption for an ITER-scenario \cite{Smith_2009}.}
\label{fig_RE_mom_bound}
\end{center}
\end{figure}

The effective critical electric field $E^{\mathrm{eff}}_{\mathrm{c}}$ can be computed in $\mathrm{V}\hspace{0.25mm}\mathrm{m}^{-1}$ for a given magnetic field strength $B$, an electron temperature $k_{B}T_{e}$ in electron volts, a density vector $\mathbf{n}$ in $\mathrm{m}^{-3}$, the related charge vector $\mathbf{q}$ in units of the elementary charge and the corresponding vector of the nuclear charge numbers $\mathbf{Z}$. At this, one is referred to the plasma description defined in section \ref{nuclear_fusion_section}. The calculation itself is then carried out with the \textsc{MATLAB}-script \qq{\texttt{calculate_E_c_eff.m}} from \textit{L.\hspace{0.9mm}Hesslow} \cite{Hesslow_2018}. In general, the script iteratively finds the minimum electric field $E= E^{\mathrm{eff}}_{\mathrm{c}}$ satisfying the pitch averaged force balance equation \cite{Hesslow_2018}:\vspace{-3.0mm}
\begin{equation}\label{force_balance}
 \left\langle e\hspace{0.3mm}\xi E  - F_{fr,\|} - F_{syn,\|} - F_{br,\|} \right\rangle_{\xi} = 0\,.
\end{equation}
\vspace*{-9.0mm}\\Note, that this equation considers the influences of the dynamical friction force $F_{fr,\|}$ and the radiation reaction forces $F_{syn,\|}$ and $F_{br,\|}$ related to \textit{Bremsstrahlung} and synchrotron radiation losses, which are further evaluated in subsection \ref{RE_loss_subsection}. As well, it should be mentioned, that equation $(\ref{force_balance})$ pitch-averages the balance of the components of the force vectors parallel to the magnetic field, which can be deduced from the accelerating force $F_{E}=eE$, since its parallel component is $F_{E,\|}=e\hspace{0.3mm}\xi E$. In addition, it shall be remarked, that the output of the mentioned \textsc{MATLAB}-script is the effective critical electric field normalized to the total \textit{Connor-Hastie} critical electric field in $\mathrm{V}\hspace{0.25mm}\mathrm{m}^{-1}$:\vspace{-2.0mm}
\begin{equation}\label{E_c_tot_def}
E^{\mathrm{tot}}_{\mathrm{c}} =\dfrac{n_{e}^{\mathrm{tot}}}{n_{\mathrm{e}}}\cdot E_{\mathrm{c}} =\left(1+\dfrac{n_{e}^{\mathrm{bd}}}{n_{\mathrm{e}}}\right)\cdot E_{\mathrm{c}}= \dfrac{n_{e}^{\mathrm{tot}}\hspace{0.3mm}e^{3}\hspace{0.2mm}\ln{\hspace{-0.45mm}\Lambda_{rel}}}{4\pi\hspace{0.4mm}\varepsilon_{0}^{2}\hspace{0.6mm} m_{e0}\hspace{0.3mm}c^{2}}\,
\end{equation}   
\vspace{-8.4mm}\\which is always larger than $E_{\mathrm{c}}$ and equivalent to the expression from $(\ref{E_crit})$, if the free electron density $n_{\mathrm{e}}$ is exchanged with the total electron density $n_{e}^{\mathrm{tot}}$. Here, the definitions of the electron densities can be found in the equations $(\ref{n_e_def})$  and $(\ref{n_tot_def})$ from section \ref{nuclear_fusion_section}.
\\ 
A comparison of the effective critical electric field $E^{\mathrm{eff}}_{\mathrm{c}}$ with the \textit{Connor-Hastie} critical electric fields $E_{\mathrm{c}}$ and $E_{\mathrm{c}}^{\mathrm{tot}}$, calculated from the free and the total electron density and is possible, on the basis of figure \ref{fig_E_C_ava_p_c_scr_main}. At this, a deuterium-neon research plasma, as\linebreak\vspace{-4.6mm}
\vspace*{-0.5mm} 
\footnotetext{\label{fig_plot_footnote_E_c} The contour plots were generated, with the help of the \textsc{MATLAB}-scripts\\ \hspace*{8.7mm}\qq{\texttt{generate_num_data_densities_p_c_scr_E100.m}} and\\ \hspace*{8.7mm}\qq{\texttt{plot_num_data_densities_p_c_scr_E100.m}}, which can be found in the digital\\ \hspace*{8.7mm}appendix.}
\noindent\newpage\noindent
presented in section \ref{nuclear_fusion_section}, was investigated for different densities of the singly-charged deuterium and neon ions, whereat the \textit{Coulomb} logarithm was calculated, by means of the expression $(\ref{CoulombLogrel})$. For that, a fixed electric field strength was used, because all expressions of the critical electric field do not depend on this plasma parameter. However, it should be noticed, that for the increasing densities the critical electric field grows and reaches the chosen value of the parallel component of the electric field, so that the electron density parameter region is bound by the electron density, which is related to this point within the parameter space. Moreover, one observes, that the \textit{Connor-Hastie} model predicts electric field values, whose minimum, for a given deuterium and neon density parameter point \mbox{$\left(n_{_{1}^{2}\mathrm{H}^{+}},\,n_{_{10}^{20}\mathrm{Ne}^{+}}\right)$}, corresponds to the consideration of the free electron density and $E_{\mathrm{c}}$, whilst the maximum value for this parameter tuple result from the total electron density and thus $E_{\mathrm{c}}^{\mathrm{tot}}$. In comparison with the effective critical electric field $E_{\mathrm{c}}^{\mathrm{eff}}$ one finds, that this model always underestimates the true critical electric field, since it does not include the effects of the only partially ionized plasma. This is particularly evident by the contour plots of the relative deviations in figure \ref{fig_E_C_ava_p_c_scr}.  
\\
As well the critical momentum is affected, if partial screening is considered. Hence, the \textit{Connor-Hastie} critical momentum $p_{c}$ from $(\ref{p_crit})$ has to be replaced by the always larger normalized effective critical momentum $p_{\star}$. However, this lower momentum boundary for the generation of runaway electrons can only be evaluated iteratively as the root of a non-polynomial function stated by \textit{L.\hspace{0.9mm}Hesslow} \cite{Hesslow_2019} for \mbox{$\vert E_{\|}\vert\gg E_{\mathrm{c}}$}:\vspace{-3.0mm}
\begin{equation}\label{func_p_c_eff_def}
f_{p_{\star}}(p) = \dfrac{E_{\mathrm{c}}}{\vert E_{\|}\vert}\cdot\sqrt{\left( \nu_{\mathrm{d}0}+\nu_{\mathrm{d}1}\cdot\ln{\hspace{-0.5mm}(p)} \right)\cdot\left( \nu_{\mathrm{s}0}+\nu_{\mathrm{s}1}\cdot\ln{\hspace{-0.5mm}(p)} \right) }-p^2 \,.
\end{equation}
\vspace{-8.5mm}\\This function involves the normalized deflection frequency $\nu_{\mathrm{d}}(p)$ and the normalized slowing-down
frequency $\nu_{\mathrm{s}}(p)$ from \cite{Hesslow_2018,Hesslow_2018II}, which are expressed through $\tilde{\nu}_{\mathrm{s}}(p)$ and $\tilde{\nu}_{\mathrm{s}}(p)$ in their approximated ultra-relativistic limit $p\gg1$:\vspace{-3.0mm}
\begin{equation}\label{nue_s_nue_d_def}
\begin{split}
\nu_{\mathrm{d}}(p)\, =\,  \dfrac{\gamma}{p^3}\cdot\tilde{\nu}_{\mathrm{d}}(p)  \;\;;\;\; &\tilde{\nu}_{\mathrm{d}}(p)\,\approx \,\nu_{\mathrm{d}0}+\hspace{0.6mm}\nu_{\mathrm{d}1}\cdot\ln{\hspace{-0.5mm}(p)}    \;\; \xrightarrow{\text{ comp. scr. }}\; \; 1+Z_{eff}
\\[0pt]
\nu_{\mathrm{s}}(p) \, = \, \dfrac{\gamma^2}{p^3}\cdot\tilde{\nu}_{\mathrm{s}}(p)  \;\;;\;\; & \tilde{\nu}_{\mathrm{s}}(p)\,\hspace{0.6mm}\approx \,\nu_{\mathrm{s}0}\hspace{0.6mm}+\hspace{0.6mm}\nu_{\mathrm{s}1}\cdot\ln{\hspace{-0.5mm}(p)} \hspace{0.30mm} \; \;   \xrightarrow{\text{ comp. scr. }}  \;\; 1 \,.
\end{split}
\end{equation}
\vspace{-7.5mm}\\Here, the constants $\nu_{\mathrm{d}0}$, $\nu_{\mathrm{d}1}$, $\nu_{\mathrm{s}0}$ and $\nu_{\mathrm{s}1}$ are also computable by the \textsc{MATLAB}-script\footnote{\label{MatlabEcefffootnote} \qq{\texttt{calculate_E_c_eff.m}} (available at \url{https://github.com/hesslow/Eceff})} from \textit{L.\hspace{0.9mm}Hesslow}. The full generalized expressions for $\tilde{\nu}_{\mathrm{s}}(p)$ and $\tilde{\nu}_{\mathrm{s}}(p)$ are to be read in the equations $(5)$ and $(9)$ of the publication \cite{Hesslow_2018}. Moreover, the \textit{complete screening limit} \mbox{(comp.\ scr.)} of the frequencies, for a fully ionized plasma under neglection of energy variation of the \textit{Coulomb} logarithm, can be regarded in $(\ref{nue_s_nue_d_def})$ \cite{Hesslow_2019}. Furthermore, one remarks, that both frequencies reproduce the effects of partial screening and are normalized to the inverse of the relativistic collision time $\tau^{-1}_{rel}$ \cite{Hesslow_2018}:\vspace{-2.5mm}
\begin{equation}\label{tau_rel}
\tau_{rel}=\dfrac{4\pi\hspace{0.2mm}\epsilon_{0}^{2}\hspace{0.2mm}m_{e0}^{2}\hspace{0.2mm}c^{3}}{n_{e}\hspace{0.2mm}e^{4}\hspace{0.2mm}\ln{\hspace{-0.45mm}\Lambda_{rel}}} \,,
\end{equation}
\vspace{-7.5mm}\\which was already implicitly defined in equation $(\ref{E_crit})$. Finally, one should note, that the computation of $p_{\star}$ as the root of the function from $(\ref{func_p_c_eff_def})$ requires an initial value \mbox{$p_{\star,0}=p^{\mathrm{scr}}_{\mathrm{c}}$}, in order to carry out a numerical calculation, as it is e.g.\ done by the \textsc{MATLAB}-routine \qq{\texttt{fzero}}, which uses an iterative algorithm and a combination of bisection, secant, and inverse quadratic interpolation method \cite{fzero}. For instance, one can choose the initial value to have a similar form as the \textit{Connor-Hastie} critical momentum $p_{c}$ from $(\ref{p_crit})$:\vspace{-7.5mm}
\begin{equation}\label{p_c_scr_def}
p^{\mathrm{scr}}_{\mathrm{c}} =\left(\dfrac{\vert E_{\|}\vert}{E^{\mathrm{eff}}_{\mathrm{c}}}-1\right)^{-\frac{1}{2}} \,.
\end{equation}
\vspace{-7.5mm}\\In conclusion, an improved description of the runaway phenomenon is, that all electrons with a momentum \mbox{$p_{\star} < p < p_{max}$} will runaway, if the component of an electric field parallel to the magnetic field is present and satisfies the inequality \mbox{$E_{\|}\gg E_{\mathrm{c}}$} \cite{Hesslow_2019}. Here it should be remarked, that this is an optimistic condition and one might use \mbox{$E_{\|}\gg E_{\mathrm{c}}^{\mathrm{eff}}$} as a more conservative applicability threshold. At this, the magnitude of the parallel component of their momentum-dependent net acceleration force, thus the height of the improved runaway region with regard to figure \ref{fig_RE_region}, can be expressed through the pitch-average \cite{Hesslow_2018}:\vspace{-3mm}
\begin{equation}\label{F_acc}
\begin{split}
F_{acc,\|}(p) =\;&   \left\langle e E_{\|}  - F_{fr,\|} - F_{br,\|} - F_{syn,\|} \right\rangle_{\xi}
\\[3pt]
 = \;& \dfrac{E_{\|}}{E_{c}}\hspace{-0.8mm}\cdot\hspace{-0.4mm}\textup{coth}\hspace{-0.5mm}\left(\hspace{-0.3mm} \dfrac{2\hspace{0.2mm}E_{\|}}{p\hspace{0.3mm}\nu_{D}\hspace{0.2mm}E_{c}}\hspace{-0.3mm}\right)-p\hspace{0.3mm}\nu_{s}-\dfrac{p\hspace{0.3mm}\nu_{D}}{2}-p\hspace{-0.3mm}\cdot\hspace{-0.5mm}\left(\phi_{br0}+\phi_{br1}\hspace{-0.3mm}\ln{\hspace{-0.45mm}(p)}\right)
\\[3pt]
\;&  -\dfrac{\tau_{rel}\hspace{0.6mm}p^2\hspace{0.4mm}\gamma\hspace{0.4mm}\nu_{D}\hspace{0.3mm}E_{c}}{\tau_{syn}\hspace{0.3mm}E_{\|}}\hspace{-0.3mm}\cdot\hspace{-0.5mm}\left(\hspace{-0.45mm}\textup{coth}\hspace{-0.5mm}\left(\hspace{-0.3mm} \dfrac{2\hspace{0.2mm}E_{\|}}{p\hspace{0.2mm}\nu_{D}\hspace{0.2mm}E_{c}}\hspace{-0.3mm}\right)-\dfrac{p\hspace{0.3mm}\nu_{D}\hspace{0.2mm}E_{c}}{2\hspace{0.2mm}E_{\|}}\hspace{-0.45mm}\right)\,,
\end{split}
\end{equation}
\vspace{-6mm}\\using the parameters $\phi_{br0}$ and $\phi_{br1}$ from the previously mentioned \textsc{MATLAB}-script$^{\text{\ref{MatlabEcefffootnote}}}$ and the relations $(\ref{p_norm_gamma_def})$, $(\ref{E_crit})$, $(\ref{CoulombLogHesslow})$, $(\ref{F_res2})$, $(\ref{nue_s_nue_d_def})$ and $(\ref{tau_rel})$. 
\\ 
The runaway region momentum boundaries $p_{min}$ and $p_{max}$ can be computed, similarly to $p_{\star}$, as the roots of the function in $(\ref{F_acc})$ with the help of a \textsc{MATLAB}-script\footnote{\label{MatlabITERfootnote} \qq{\texttt{RE_region_momentum_boundaries.m}}}. At this, the calculation is based on a self-consistent one-dimensional model of the electric field evolution, governed by an induction equation. Here the ITER-scenario from the paper \cite{Smith_2009}, covering the thermal and current quench phase of a disruption with a plasma current of \mbox{$I_{p}=15\;\mathrm{MA}$}, a magnetic field of \mbox{$B=5.3\,\mathrm{T}$} and a time-independent electron density of \mbox{$n_{e}=1.06\cdot10^{20}\;\mathrm{m}^{-3}$}, is considered. In \mbox{figure \ref{fig_RE_ITER_simu}}, one can observe the time evolution of the parallel component of the electric field, the \textit{Connor-Hastie} and the effective critical electric field, the electron temperature and the plasma current strength at the radius \mbox{$r_{\perp}=0.75\,\mathrm{m}$}, because the simulation is based on a one-dimensional cylindrical plasma model \cite{Smith_2009}. Thus, one is able to verify the general character of the behaviour of said quantities, like for instance the decomposition of the plasma current into an ohmic and a runaway electron part, during the first two phases of a disruption, by comparing the figure \ref{fig_RE_ITER_simu} with figure \ref{disruption_fig} from section \ref{tokamak_disruptions_section}. 
\vspace{0.6mm}
\footnotetext{\label{fig_RE_ITER_simu_footnote} The diagrams in figure \ref{fig_RE_ITER_simu} and \ref{fig_RE_mom_bound} were generated with the \textsc{MATLAB}-file$^{\textup{\ref{MatlabITERfootnote}}}$. The script and its\\ \hspace*{8.7mm}output \qq{\texttt{output_RE_region_mom_bound.txt}} can be found in the digital appendix.}
\vspace{-9.5mm}
By means of the data from this disruption simulation, it is possible to analyse the momentum boundaries of the runaway region. Therefore, the behaviour of the lower boundaries respectively the representations of the effective critical momentum $p_{\star}$ and $p_{min}$, determined by the functions $(\ref{func_p_c_eff_def})$ and $(\ref{F_acc})$ is of interest. Consequently, those momenta are plotted in figure \ref{fig_RE_mom_bound}.\vspace{-0.2mm}
\vspace{-9.5mm}
In addition, the thermal momentum $p_{th}$ from $(\ref{v_th})$ and the upper momentum boundary $p_{max}$ derived from $(\ref{F_acc})$ are displayed, in order to facilitate the comparability with the figure \ref{fig_RE_region}. In this context, the thermal momentum is connected to the thermal electrons, the main part of the distribution function at the beginning of the disruption and to the ohmic current. The maximum momentum however shows the threshold above which the radiation losses become dominant. In the interval \mbox{$0.3\;\mathrm{ms}<t<3.8\;\mathrm{ms}$} the creation, the growth and the decay of the runaway region \mbox{$p_{min}<p<p_{max}$} or \mbox{$p_{\star}<p<p_{max}$} is clearly visible. This shows that a significant runaway current can be produced within a short time duration of approximately $3.5\;\mathrm{ms}$, which is in accordance with the runaway current evolution shown in figure \ref{fig_RE_ITER_simu}. 
\\
As well, the large order of magnitude of the maximum momentum $p_{max}$ is apparent in figure \ref{fig_RE_mom_bound}. Furthermore, one can receive estimations for the $p_{max}$ of the runaway electrons during a tokamak disruption, based on the maximum change of the poloidal magnetic flux, originating from the plasma current decay \cite{Papp_2011}. At this, the magnetic flux through a surface parallel to the magnetic axis is defined as the poloidal flux \cite{pappPHD}. For example, a maximum reachable energy of \mbox{$\mathcal{E}_{p,max}\approx 340\,\mathrm{MeV}$}, equivalent to the total energy stored in the whole plasma, and the highest possible energy of the electrons \mbox{$\mathcal{E}_{e,max}\approx 100\,\mathrm{MeV}$}, including loss effects and electric field diffusion, was calculated with the help of a self-consistent simulation for an ITER-scenario, as stated in the publication \cite{Papp_2011}. The relativistic momentum associated with those two upper limits might be obtained, if the given energy limits are normalized to the rest mass energy and set equal to the rest mass-related kinetic energy density as defined in equation $(\ref{kin_dens_ava_def})$:
\vspace{-3.0mm}
\begin{equation}\label{p_max_ITER_I_p}
\dfrac{k}{c^2}\hspace{-0.1mm}=\hspace{-0.1mm}\dfrac{ K}{m_{e0}\hspace{0.3mm}c^2}\hspace{-0.1mm}=\gamma(p)\hspace{0.2mm}-\hspace{0.2mm}1\hspace{-0.35mm}\overset{(\ref{p_norm_gamma_def})}{=}\hspace{-0.75mm} \sqrt{1\hspace{-0.3mm}+\hspace{-0.2mm}p^2}\hspace{0.2mm}-\hspace{0.2mm}1 \overset{!}{=} \dfrac{ e\hspace{0.3mm}\mathcal{E} }{m_{e0}\hspace{0.3mm}c^2} \;\longleftrightarrow\;  p = \sqrt{\hspace{-0.55mm}\left(\hspace{-0.35mm}\dfrac{e\hspace{0.35mm}\mathcal{E} }{m_{e0}\hspace{0.3mm}c^2}\hspace{-0.35mm}+\hspace{-0.25mm}1\hspace{-0.3mm}\right)^{\hspace{-0.95mm}2}\hspace{-0.85mm}-\hspace{-0.1mm}1}\,.
\end{equation}
\vspace{-8.0mm}\\If this relation and the limiting energies $\mathcal{E}_{p,max}$ and $\mathcal{E}_{e,max}$, expressed in electron volts, are used to calculated upper limits for the maximum runaway electron momentum, one receives \mbox{$p_{max}\lesssim 196.7 \lesssim 666.4$}. For the ITER-disruption used for the figure \ref{fig_RE_mom_bound}, it can be asserted, that between $0.6\,\mathrm{ms}$ and $1.4\,\mathrm{ms}$ the upper limit for the runaway momentum is determined through the largest change in the poloidal magnetic flux. This leads to the understanding, that $p_{max}$ does not have to be computed as the second root of a defining function as written in equation $(\ref{F_acc})$ for the mentioned time interval, which is connected to large values of the electric field with \mbox{$E_{\|}\gg E_{c}$}.
\\
Furthermore and due to the analysis of figure \ref{fig_RE_mom_bound}, one can see that $p_{c}$, $p_{\star}$ and $p_{min}$ represent a similar approximation of the critical effective momentum during the thermal quench and the beginning of the current quench. In contrast, they predict different lower momentum boundaries for the runaway region in the last part of the current quench for \mbox{$t>3\;\mathrm{ms}$}, if electric field and electron temperature are small. This seems to be plausible, since the governing equation $(\ref{func_p_c_eff_def})$ for $p_{\star}$ only holds for \mbox{$E_{\|}\gg E_{\mathrm{c}}$} \cite{Hesslow_2019} and for \mbox{$t>3\;\mathrm{ms}$} one finds that \mbox{$E_{\|}\rightarrow E_{\mathrm{c}}^{\mathrm{eff}}$}. Hence, the more complicated equation $(\ref{F_acc})$ can generally be replaced by the expression $(\ref{func_p_c_eff_def})$ for \mbox{$E_{\|} \gtrsim E_{\mathrm{c}}^{\mathrm{eff}},\,E_{\mathrm{c}}$}, if the interesting quantities within a simulation are not too sensitive to the lower momentum bound. This could as well be regarded as an enhancement of the calculation efficiency, since a simpler function could allow a faster computation of their roots. With regard to the \textit{Connor-Hastie} critical momentum $p_{c}$ as defined in $(\ref{p_crit})$, one can observe, that is as well a suitable approximation of the critical effective momentum. However, it has to be considered, that the displayed values in figure \ref{fig_RE_mom_bound} do not include a time-evolving electron density, which will further influence the deviation between the different lower momentum boundary approximations. In conclusion, one should consider $p_{\star}$ as the most effective and accurate approximation, since it accounts for partial screening effects in contrast to $p_{c}$ and allows a better understood and presumably faster computation in terms of required iterations than $p_{min}$. 

\begin{figure}[H]
  \centering
  \subfloat{\label{p_c_ava_p_c_scr_E100_main} 
   \includegraphics[trim=310 23 360 24,width=0.322\textwidth,clip]
    {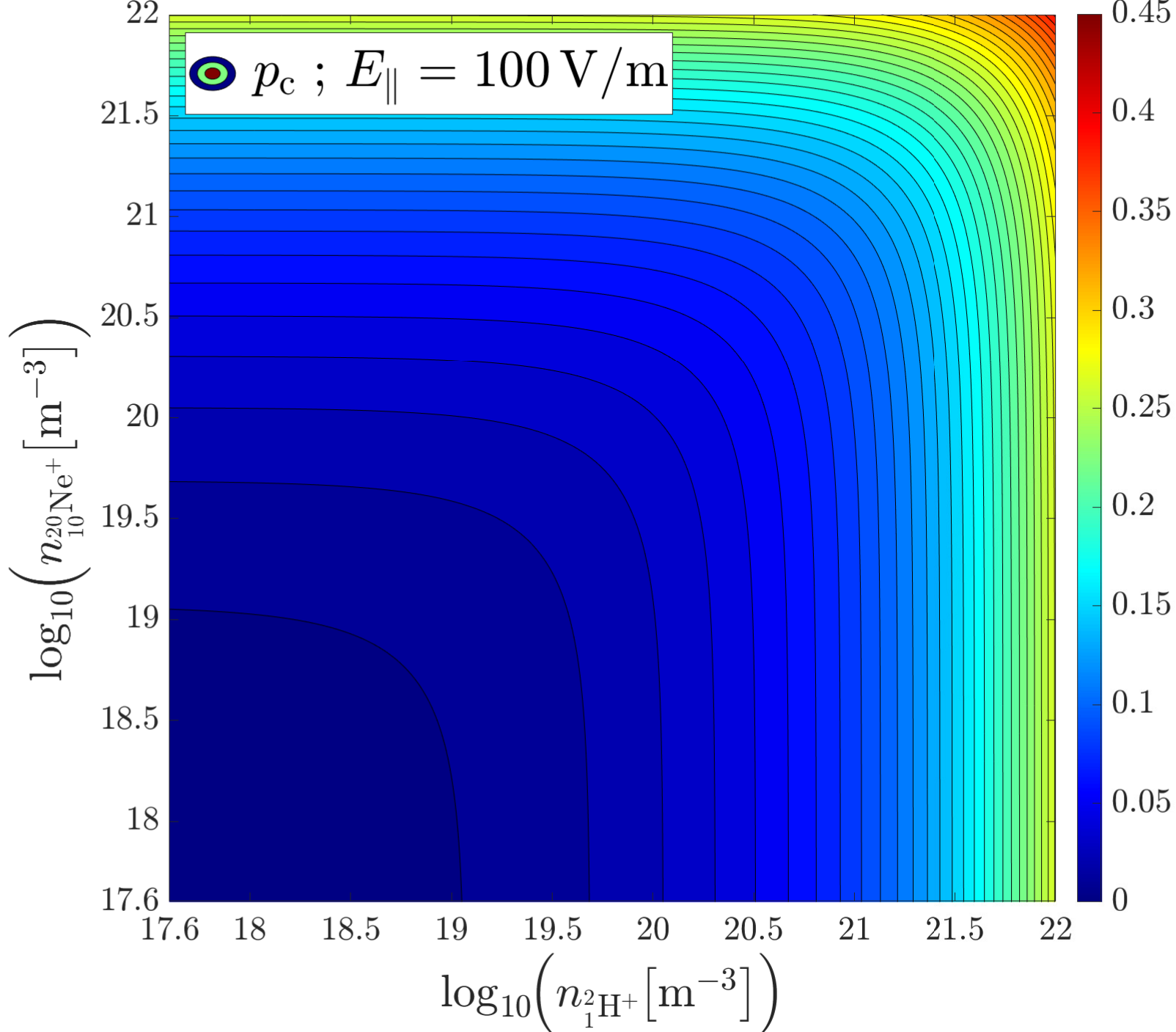}}\hfill
  \subfloat{\label{fig_p_c_scr_ava_p_star_E100_main}
    \includegraphics[trim=320 24 354 20,width=0.322\textwidth,clip]
    {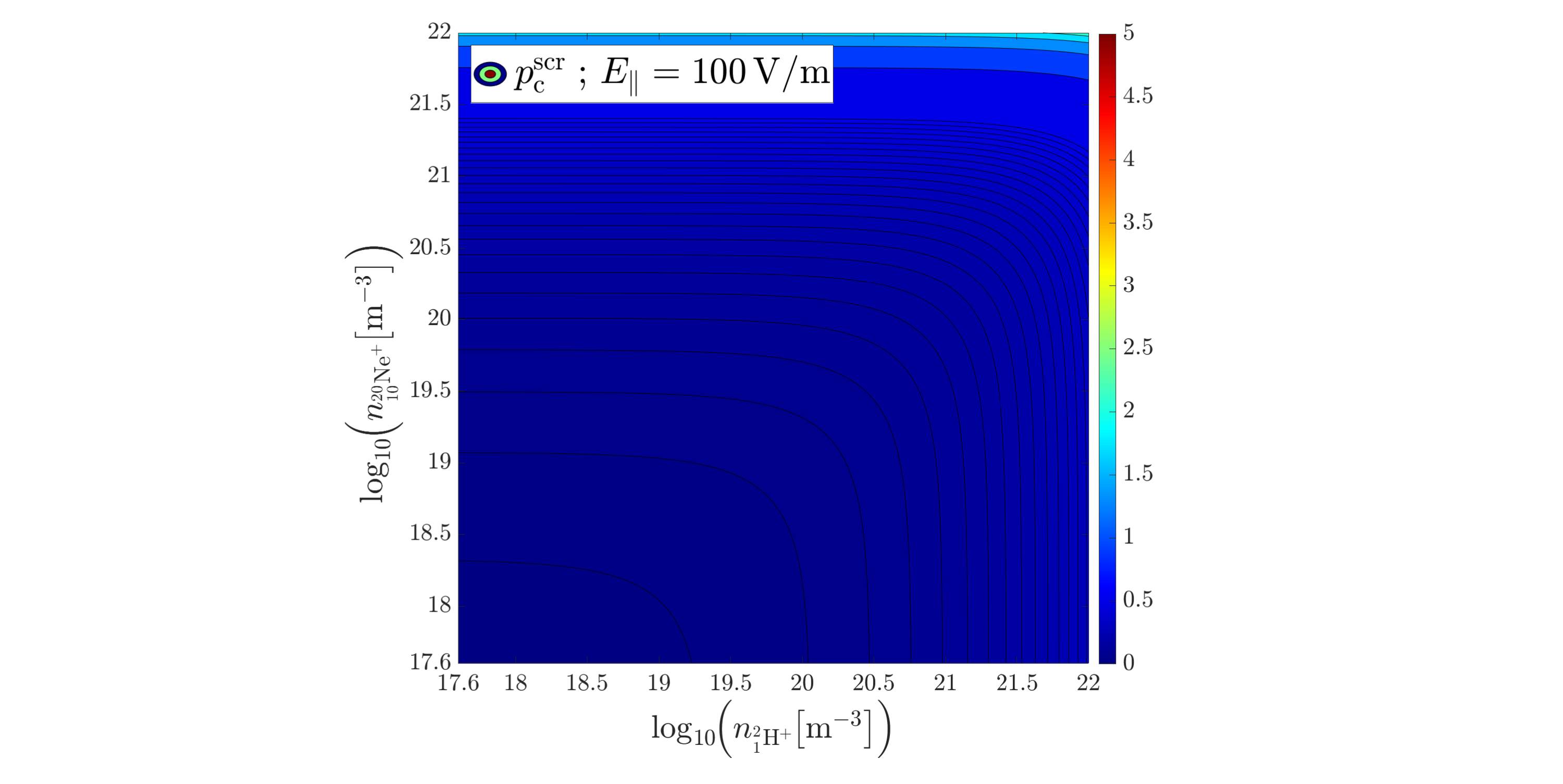}}\hfill
  \subfloat{\label{fig_p_star_ava_p_star_E100_main} 
    \includegraphics[trim=321 23 351 18,width=0.329\textwidth,clip]
    {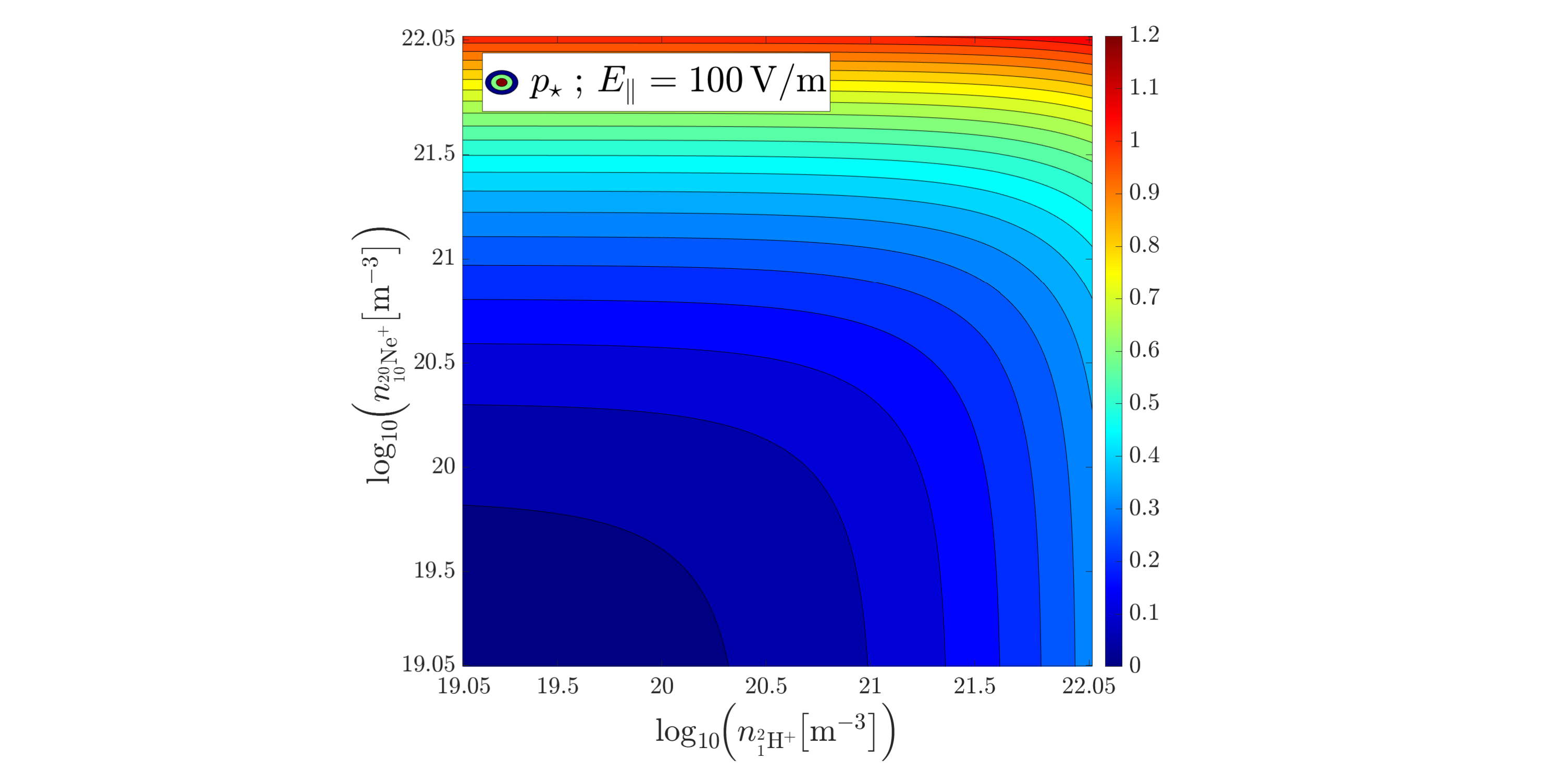}} \\[5pt]
  \subfloat{\label{fig_rel_p_c_scr_ava_E100_main} 
   \includegraphics[trim=310 23 360 21,width=0.322\textwidth,clip]
    {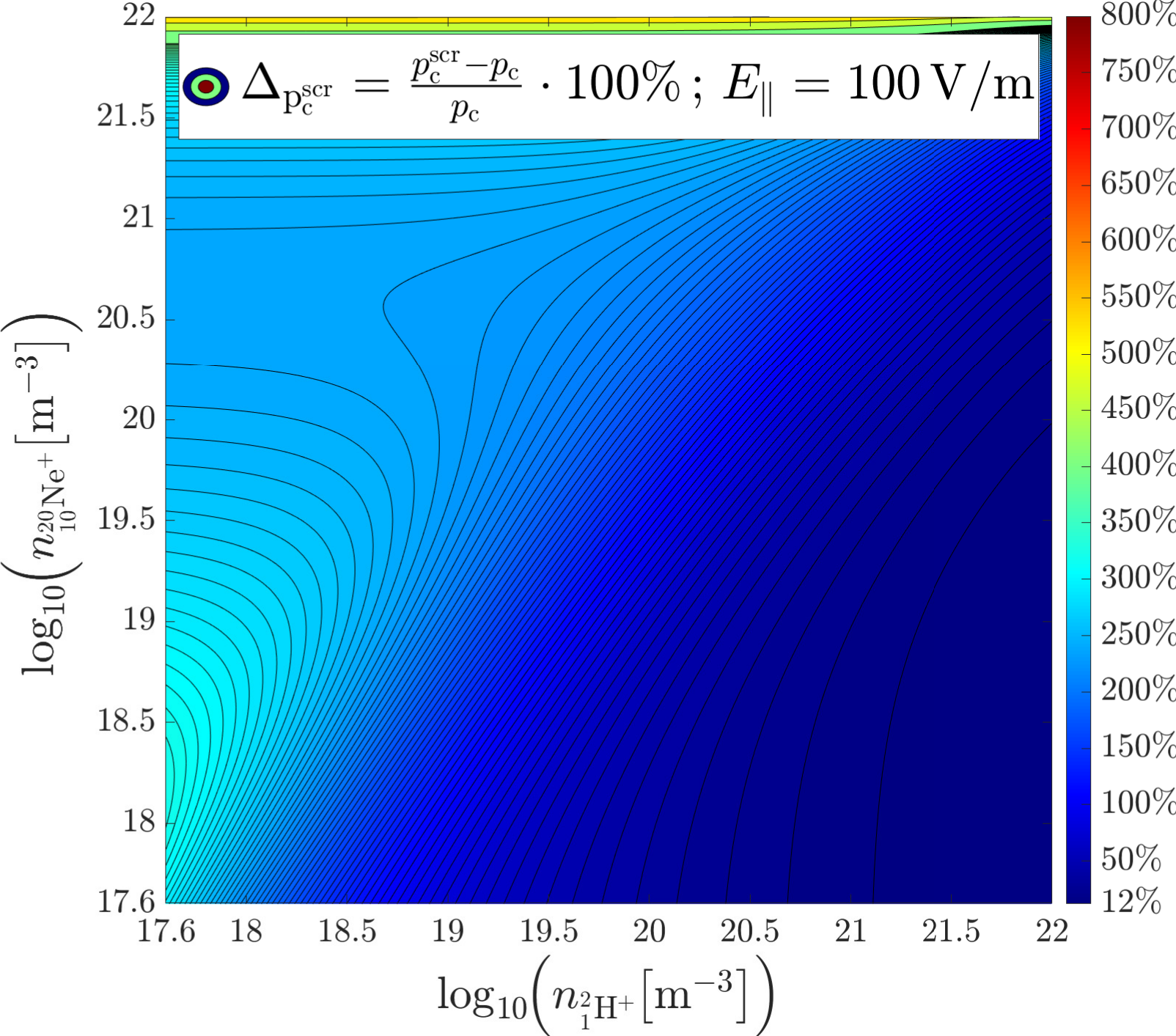}}\hfill
  \subfloat{\label{fig_rel_p_star_ava_E100_main}
    \includegraphics[trim=321 24 355 18,width=0.322\textwidth,clip]
    {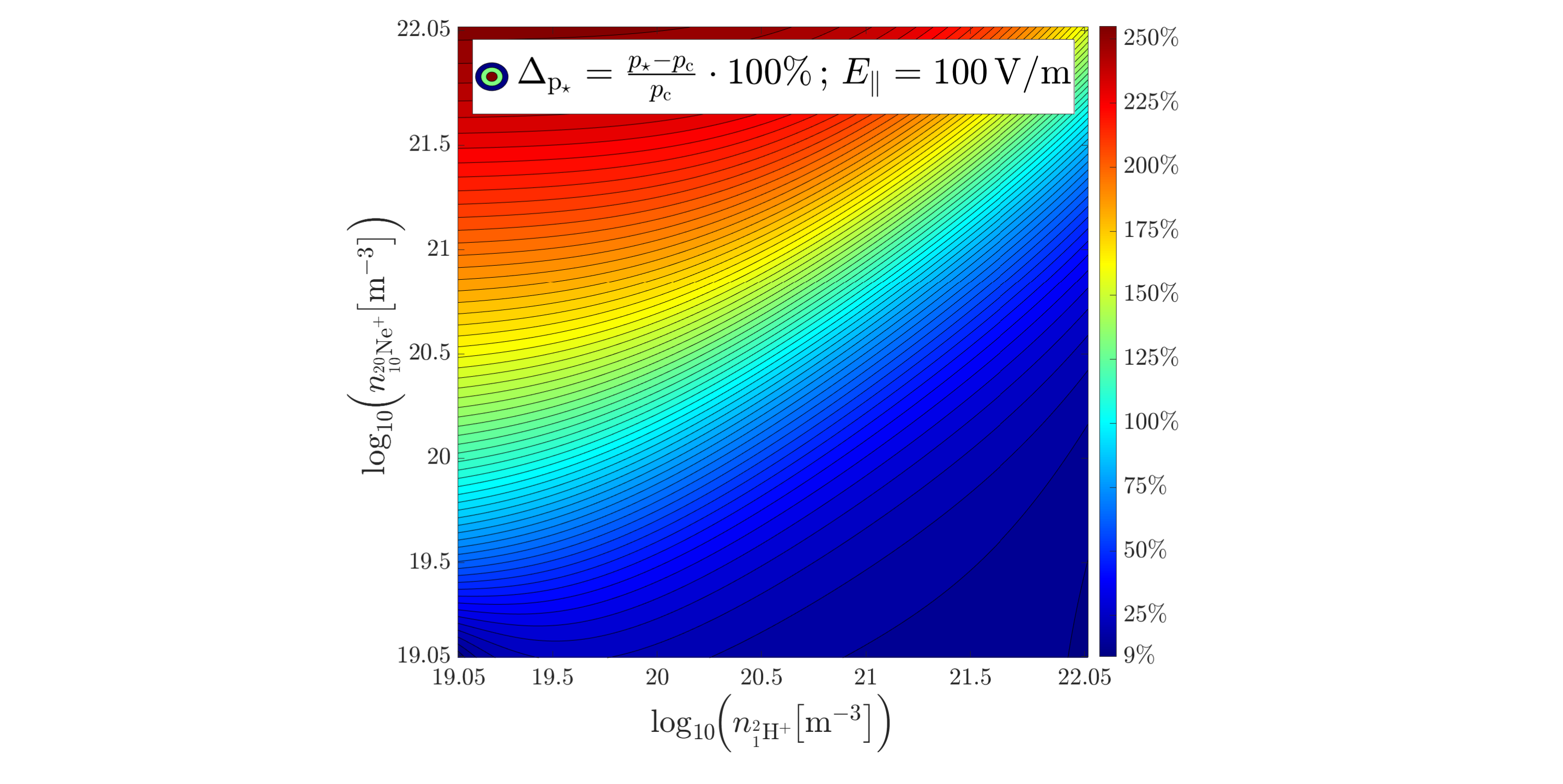}}\hfill
  \subfloat{\label{fig_tilde_rel_p_c_scr_ava_E100_main} 
    \includegraphics[trim=320 24 355 19,width=0.329\textwidth,clip]
    {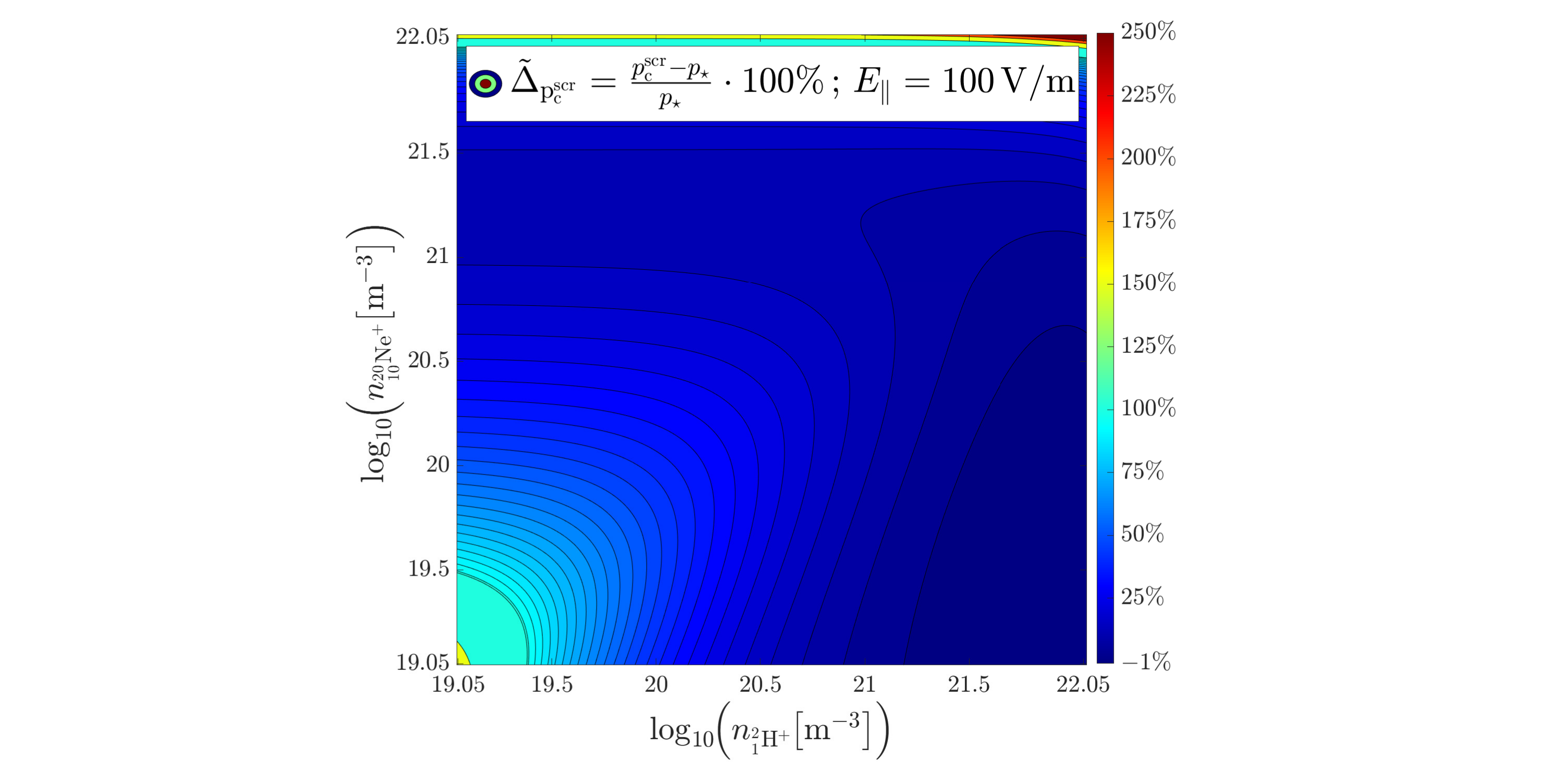}} 
 \caption[Contour plots of the approximations $p_{c}$, $p^{\mathrm{scr}}_{\mathrm{c}}$ and $p_{\star}$ for the effective critical momentum $p^{\mathrm{eff}}_{\mathrm{c}}$ as well as their relative deviations $\Delta_{p^{\mathrm{scr}}_{c}}$, $\Delta_{p_{\star}}$ and $\tilde{\Delta}_{p^{\mathrm{scr}}_{\mathrm{c}}}$ from each other, plotted over a density parameter space for a singly-ionized deuterium-neon plasma with \mbox{$k_{\mathrm{B}}T_{\mathrm{e}}=10\,\textup{eV}$}, \mbox{$B=5.25\,\textup{T}$}, \mbox{$Z_{\mathrm{eff}}=1$} and \mbox{$E_{\|}\coloneqq\vert E_{\|}\vert=100\,\mathrm{V/m}$} (see also figures \labelcref{fig_p_c_ava_p_c_scr,fig_p_c_ava_screen_p_c_scr,fig_p_star_ava,fig_rel_p_c_scr_ava,fig_rel_p_star_ava,fig_tilde_rel_p_c_scr_ava}).]{Contour plots\protect\footnotemark{} of the approximations $p_{c}$, $p^{\mathrm{scr}}_{\mathrm{c}}$ and $p_{\star}$ for the effective critical momentum $p^{\mathrm{eff}}_{\mathrm{c}}$ as well as their relative deviations $\Delta_{p^{\mathrm{scr}}_{c}}$, $\Delta_{p_{\star}}$ and $\tilde{\Delta}_{p^{\mathrm{scr}}_{\mathrm{c}}}$ from each other, plotted over a density parameter space for a singly-ionized deuterium-neon plasma with \mbox{$k_{\mathrm{B}}T_{\mathrm{e}}=10\,\textup{eV}$}, \mbox{$B=5.25\,\textup{T}$}, \mbox{$Z_{\mathrm{eff}}=1$} and \mbox{$E_{\|}\coloneqq\vert E_{\|}\vert=100\,\mathrm{V/m}$} (see also figures \labelcref{fig_p_c_ava_p_c_scr,fig_p_c_ava_screen_p_c_scr,fig_p_star_ava,fig_rel_p_c_scr_ava,fig_rel_p_star_ava,fig_tilde_rel_p_c_scr_ava}).}
\label{fig_p_comparison}
\end{figure}

Finally, a comparison and an improved understanding of the different relations $p_{c}$, $p^{\mathrm{scr}}_{\mathrm{c}}$ and $p_{\star}$, which approximate the true value the effective critical momentum $p^{\mathrm{eff}}_{\mathrm{c}}$ might be supported by the figure \ref{fig_p_comparison}. \vspace*{-1.0mm}
\vspace*{-2.5mm}
\footnotetext{\label{fig_p_relations_footnote} The contour plots were generated by means of the \textsc{MATLAB}-scripts\\ \hspace*{8.7mm}\qq{\texttt{plot_num_data_densities_p_c_scr_E100.m}} and\\ \hspace*{8.7mm}\qq{\texttt{plot_num_data_densities_p_star_E100.m}}, utilizing the results calculated with the\\ \hspace*{8.7mm}implementations \qq{\texttt{generate_num_data_densities_p_c_scr_E100.m}} and\\ \hspace*{8.7mm}\qq{\texttt{generate_num_data_densities_p_star_E100.m}}, which can be found in the digital\\ \hspace*{8.7mm}appendix.}
At this, the deuterium-neon research plasma with only singly-charged ions, as presented in section \ref{nuclear_fusion_section}, was considered and the displayed data from the \textsc{MATLAB}-implementations$^{\ref{fig_p_relations_footnote}}$ corresponds to different combinations of the ion densities and the utilization of the relation $(\ref{CoulombLogrel})$ for the \textit{Coulomb} logarithm. Due to the fact, that all of the computed expressions depend on the electric field strength, one can analyse the growth of the critical momentum within the density parameter space for four nearly logarithmically increasing values of the electric field strength in the figures \labelcref{fig_p_c_ava_p_c_scr,fig_p_c_ava_screen_p_c_scr,fig_p_star_ava,fig_rel_p_c_scr_ava,fig_rel_p_star_ava,fig_tilde_rel_p_c_scr_ava} and from subsection \ref{contour_plots_p_c_eff_ava_appendix_subsection} of the appendix. In addition, the outputs of the \textsc{MATLAB}-implementations$^{\ref{fig_p_relations_footnote}}$ can be viewed in the listings \cref{MATLABoutput_plot_p_scr_E3,MATLABoutput_plot_p_scr_E10,MATLABoutput_plot_p_scr_E30,MATLABoutput_plot_p_scr_E100,MATLABoutput_plot_p_star_E3,MATLABoutput_plot_p_star_E10,MATLABoutput_plot_p_star_E30,MATLABoutput_plot_p_star_E100} from subsection \ref{output_matlab_appendix_subsection} of the appendix. 
\\
However, for a comparison with respect to an ion density combination \mbox{$\left(n_{_{1}^{2}\mathrm{H}^{+}},\,n_{_{10}^{20}\mathrm{Ne}^{+}}\right)$} only the results for an electric field strength of \mbox{$E=100\,\mathrm{V/m}$} are shown in figure \ref{fig_p_comparison}. Note, that the increase of $p^{\mathrm{scr}}_{\mathrm{c}}$ and $p_{\star}$ is stronger in the direction of an increasing neon ion density than in the direction of an increasing deuterium ion density. This is reasoned, by the fact, that the influence of the higher nuclear charge of neon, is only modeled appropriately, if partial screening effects are considered. Furthermore, one can confirm, that the effects of partially ionized impurities lead to a higher critical momentum. Whereat this and the fact that the analytic expression $p^{\mathrm{scr}}_{\mathrm{c}}$ generally overestimates the more physically accurate $p_{\star}$ follows from the analysis of the relative deviations. A further discussion with a focus on the applicability of the presented relations for the lower momentum boundary of the runaway region will be held in \mbox{section \ref{Hesslow_avalanche_dist_subsection}}.

\clearpage

\section{Generation mechanisms for runaway electrons}\label{mechanisms_section}

In general, two kinds of runaway electron generation are possible. The first kind, which is referred to as the \textit{primary generation}, describes how runaway electrons are initially produced through acceleration towards relativistic speeds. This means, that no runaway electrons are present before a primary generation mechanism is triggered. In the following, the \textit{Dreicer} and the \textit{hot-tail} generation mechanism are further explained, because they are dominant in the runaway seed generation, which mainly takes place during the thermal quench and the early phase of the current quench \cite{HesslowPHD}. However, tritium decay and \textit{Compton} scattering of photons emitted by the activated reactor wall are also contributing to primarily generated runaway electrons \cite{REsimulation}. However, they are modeled by means of two different source terms in the kinetic equation $(\ref{kinetic_equation})$, where for the \textit{Compton} scattering source term the highly energetic photon spectrum from the reactor wall and the plasma composition is needed \cite{Mart_n_Sol_s_2017}. In terms of the tritium decay source, the determining factors are the half-life and the energy spectrum of tritium beta-decay \cite{Mart_n_Sol_s_2017}. As a matter of fact, a seed runaway electron population is often mainly explainable with the \textit{Dreicer} and the \textit{hot-tail} mechanism for tokamak scenarios with a low activation wall as it can be assumed for research deuterium plasmas. Eventually, it shall be remarked, that within the framework of this thesis, exclusively the \textit{Dreicer} and the \textit{hot-tail} mechanism are considered as primary generation mechanisms, due to the fact that they are, in contrast to the tritium decay and the \textit{Compton} scattering, described by kinetic theory.  

The second kind of runaway electron generation, requires an existing seed population of runaway electrons and is therefore called a \textit{secondary generation}. At that, the production of runaway electrons caused by the interaction of already existing runaway electrons with thermal electrons, leading to an additional population of runaway electrons. The seed runaway current can increase exponentially, due to secondary generation mechanisms \cite{Rosenbluth_1997}, while overpowering primary generation mechanisms \cite{Hender_2007,Rosenbluth_1997}. This reasons the name \textit{avalanche} generation mechanism and the more detailed discussion in subsection \ref{Avalanche_subsection}, since this mechanism is mainly responsible for damages to the reactor walls, in particular in tokamaks with large plasma currents such as ITER \cite{REsimulation}.

\subsection{\textit{Dreicer} generation mechanism}\label{Dreicer_subsection}

The collisions between electrons lead to a momentum diffusion, which will balance out the friction force acting on them, if no electric field is present \cite{EmbreusPHD}. In consequence, the existing electron population experiences an equilibrium between diffusion and friction and is described by the \textit{Maxwell} distribution function with respect to velocity in a non-relativistic regard \cite{Stroth_2018,helander}:\vspace{-3.5mm}
\begin{equation}\label{f_Maxwell}
f^{\mathrm{M}}_{e}(v) = n_{e}\cdot \left(\hspace{-0.3mm}\dfrac{m_{e0}}{\pi\hspace{0.4mm} 2 \hspace{0.25mm}e \hspace{0.25mm}k_{B}T_{e}}\hspace{-0.3mm}\right)^{\frac{3}{2}}\hspace{-0.4mm}\cdot\textup{exp}\hspace{-0.5mm}\left(\hspace{-0.3mm}-\dfrac{m_{e0}\hspace{0.15mm}v^2}{2\hspace{0.25mm}e\hspace{0.25mm}k_{B}T_{e}}\hspace{-0.3mm}\right) \overset{(\ref{v_th})}{=} \dfrac{n_{e}} { \left(\sqrt{\pi }\hspace{0.25mm}v_{th}\right)^{3}} \cdot\textup{exp}\hspace{-0.5mm}\left(\hspace{-0.3mm}-\dfrac{v^2}{v^{2}_{th}}\hspace{-0.3mm}\right) 
\end{equation}
\vspace{-7.5mm}\\and by the \textit{Maxwell-J\"uttner} distribution function with respect to the normalized momentum from $(\ref{p_norm_gamma_def})$ for a relativistic consideration, using the normalized temperature $\Theta$ from $(\ref{normal_temp})$ and the second-order modified \textit{Bessel} function of the second
kind $\mathcal{K}_{2}(x)$ \cite{Hoppe_2021}:\vspace{-5.0mm}
\begin{equation}\label{f_Maxwell_Juettner}
f^{\mathrm{MJ}}_{e}(p) = \dfrac{n_{e}}{4\pi\hspace{0.35mm}\Theta\hspace{0.35mm} \mathcal{K}_{2}\hspace{-0.2mm}\left(\Theta^{-1}\right)}  \cdot\textup{exp}\hspace{-0.5mm}\left(\hspace{-0.3mm}-\dfrac{\gamma(p)}{\Theta}\hspace{-0.3mm}\right) \overset{(\ref{p_norm_gamma_def})}{=} \dfrac{n_{e}}{4\pi\hspace{0.35mm}\Theta\hspace{0.35mm} \mathcal{K}_{2}\hspace{-0.2mm}\left(\Theta^{-1}\right)}  \cdot\textup{exp}\hspace{-0.5mm}\left(\hspace{-0.3mm}-\dfrac{\sqrt{1+p^2}}{\Theta}\hspace{-0.3mm}\right) ,
\end{equation}
\vspace{-8.0mm}\\where the approximation \cite{Stroth_2018}:\vspace{-3.5mm}
\begin{equation}\label{K_2_approxiamtion}
\mathcal{K}_{2}(x)\approx \sqrt{\dfrac{\pi}{2x}}\cdot\textup{e}^{-x}   \;\;\; ; \;\;\; x \gg 3.75
\end{equation}
\vspace{-8.5mm}\\shows the connection to the \textit{Maxwell} distribution, since \mbox{$\Theta\ll 1$} and thus \mbox{$\Theta^{-1}\gg 1$} holds in the non-relativistic limit \cite{HesslowPHD}.
\\
If however an electric field is induced, which exceeds the critical electric field, a quasi-steady state can be created. Electrons with a velocity above the critical velocity are accelerated towards the speed of light and therefore the tail of the thermal distribution with \mbox{$v>v_{c}$} is transferred to the runaway region \cite{pappPHD}. Consequently, the \textit{Maxwellian} distribution function experiences a non-equilibrium gradient with respect to momentum or velocity, causing a diffusive flux of electrons across the threshold \mbox{$v>v_{c}$} into the velocity space due to small-angle collisions, where those electrons become runaway electrons \cite{stahl,pappPHD}. This primary generation of runaway electrons as a consequence of momentum-space diffusion is commonly called the \textit{Dreicer} generation mechanism \cite{stahl}, due to the fundamental understanding provided by the work \cite{Dreicer_1959,Dreicer_1960} of \textit{H.\hspace{0.9mm}Dreicer} in 1959/60. 
\\
The characteristic electric field for this mechanism is the \textit{Dreicer} field as denoted in equation $(\ref{Dreicerfield})$. This is because, the steady-state growth
rate of the runaway population originating from this generation mechanism mainly depends on an exponential factor of the inverse of \mbox{$\hat{E}_{D}\coloneqq E/E_{D}$} \cite{Connor_1975,stahl}, representing the electric field normalized to the \textit{Dreicer} field. This can be seen from the analytic expression for said growth rate proposed by \textit{J.\hspace{0.7mm}W.\hspace{0.9mm}Connor} and \textit{R.\hspace{0.7mm}J.\hspace{0.9mm}Hastie} in 1975 \cite{Connor_1975,Hesslow_2019Dreicer}:\vspace{-2.5mm}
\begin{equation}\label{growth_rate_Dreicer}
\Gamma_{D}\coloneqq\left(\hspace{-0.85mm}\dfrac{\mathrm{d}\hspace{0.15mm}n_{RE}}{\mathrm{d}\hspace{0.15mm}t}\hspace{-0.85mm}\right)_{\hspace{-0.85mm}D}\hspace{-0.45mm}= \;\hspace{0.65mm} C_{D}\cdot\dfrac{n_{e}}{\tau_{ee}}\cdot \hat{E}_{D}^{\hspace{0.3mm}-\chi_{1}}\hspace{-0.5mm}\cdot\textup{exp}\hspace{-0.5mm}\left(\hspace{-0.3mm}-\dfrac{\chi_{2} }{\hat{E}_{D}}\hspace{-0.3mm}-\hspace{-0.3mm}\sqrt{\dfrac{\chi_{3} }{\hat{E}_{D}}} \right) \;\hspace{0.55mm}\propto \,\;\hspace{0.2mm}\textup{exp}\hspace{-0.5mm}\left(\hspace{-0.5mm}-\dfrac{1}{\hat{E}_{D}} \right) \,.
\end{equation}
\vspace{-6mm}\\In equation $(\ref{growth_rate_Dreicer})$ the constant \mbox{$C_{D}\approx 1.0$} \cite{Kruskal,Jayakumar_1993}, the abbreviation \mbox{$\hat{E}_{c}\coloneqq E/E_{c}$} and the collision time between thermal electrons:\vspace{-1mm}
\begin{equation}\label{tau_ee}
\tau_{ee}=\dfrac{4\pi\hspace{0.25mm}\varepsilon_{0}^2\hspace{0.25mm}m_{e0}^2\hspace{0.25mm}v_{th}^3}{n_{e}\hspace{0.25mm}e^4\hspace{0.25mm}\ln{\hspace{-0.45mm}\Lambda}}\,,
\end{equation}
\vspace{-8mm}\\depending on the energy-dependent \textit{Coulomb} logarithm, were used, while the numerical quantities are given by \cite{Connor_1975,Hesslow_2019Dreicer}:\vspace{-3mm}
\begin{equation}\label{growth_rate_Dreicer_num_quantities}
\begin{split}
\begin{gathered}
\chi_{1}\coloneqq  \dfrac{1}{16}\hspace{-0.6mm}\cdot\hspace{-0.7mm}\dfrac{Z_{eff}+1}{ \hat{E}_{c}-1 }\hspace{-0.4mm}\cdot\hspace{-0.9mm}\left[\hat{E}_{c}+2\hspace{-0.6mm}\cdot\hspace{-0.7mm}\left(\hspace{-0.5mm}\hat{E}_{c}-2\right)\hspace{-0.7mm}\cdot\hspace{-0.7mm}\sqrt{\dfrac{\hat{E}_{c} }{\hat{E}_{c}-1}}-\dfrac{Z_{eff}-7}{Z_{eff}+1}\right]\hspace{-0.5mm},
\\[0pt]
\chi_{2}\coloneqq   2\hspace{-0.4mm}\cdot\hspace{-0.7mm} \hat{E}_{c}^{\hspace{0.3mm}2}\hspace{-0.4mm}\cdot\hspace{-0.9mm}\left[ 1\hspace{-0.3mm}-\hspace{-0.3mm}\dfrac{1}{2\hat{E}_{c}}\hspace{-0.3mm}-\hspace{-0.3mm}\sqrt{1\hspace{-0.3mm}-\hspace{-0.3mm}\dfrac{1}{\hat{E}_{c}}}\right] \hspace{-0.5mm},
\;
\chi_{3}\coloneqq   \dfrac{ \hat{E}_{c}^{\hspace{0.3mm}2}}{4}\hspace{-0.6mm}\cdot\hspace{-0.7mm}\dfrac{ Z_{eff}+1 }{ \hat{E}_{c}-1  }\hspace{-0.4mm}\cdot\hspace{-0.9mm}\left[ \dfrac{\pi}{2}\hspace{-0.3mm}-\hspace{-0.3mm}\textup{arcsin}\hspace{-0.7mm}\left(\hspace{-0.6mm}1\hspace{-0.3mm}-\hspace{-0.3mm}\dfrac{2}{\hat{E}_{c}}\hspace{-0.5mm}\right)\hspace{-0.3mm}\right]^{\hspace{-0.5mm}2}\hspace{-0.5mm}.
\end{gathered}
\end{split}
\end{equation}
\vspace{-6mm}\\In $(\ref{growth_rate_Dreicer})$ it was found, in accordance with the references \cite{EmbreusPHD,stahl}, that the
\textit{Dreicer} runaway growth rate is exponentially sensitive to the electric field, based on its most dominant proportionality, expressed by an electric field dependent function. Consequentially, the normalized electric field $\hat{E}_{D}$ has to be large, in order to increase the \textit{Dreicer} growth rate, implying that the electric field $E$ is a non-negligible fraction of the \textit{Dreicer} field. Further analysis concerning the magnitude of the growth rate entails, that the \mbox{\textit{Dreicer}} runaway production rate is not significant for \mbox{$E_{c}\hspace{-0.25mm}<\hspace{-0.25mm}E \lesssim 0.03\hspace{0.45mm}E_{D}$} in comparison to other generation mechanisms \cite{EmbreusPHD,stahl}. However, for a tokamak disruption occurring in a hydrogen plasma with an argon ion \mbox{$\bigl({}^{40}_{18}\mathrm{Ar}^{4+} \bigr)$} impurity density of $1\%$ of the hydrogen density and electric fields between $3\%$ and $6\%$ of the \textit{Dreicer} field, the \textit{Dreicer} runaway current generation rate can be as large as $1\hspace{1.2mm}\mathrm{MA}\hspace{0.45mm}\mathrm{m}^{-2}\hspace{0.45mm}\mathrm{ms}^{-1}$ as it was calculated by \mbox{\textit{O.\hspace{0.9mm}Linder}} \cite{LinderPHD}. Nevertheless, the \mbox{\textit{Dreicer}} mechanism is expected to cause only negligible contributions to the total runaway electron population, if massive material injections with impurity densities of several multiples of the deuterium content in the vacuum vessel of the tokamak are applied \cite{Insulander_Bj_rk_2020}.

The \textit{Dreicer} growth rate is exponentially sensitive to the plasma properties, as it can be observed in $(\ref{growth_rate_Dreicer})$ and $(\ref{growth_rate_Dreicer_num_quantities})$. This finding thus promises a complex behavior at near-thermal energies, due to the complicated energy dependence of the collision frequencies \cite{Hesslow_2019Dreicer}. In particular, this is the case in partially ionized plasmas produced by the injection of cold impurities amplifying the effects of partial screening, partial and full ionization and various electron populations with different temperatures. Based on those insights, a neural network for the \textit{Dreicer} growth rate was trained by \mbox{\textit{L.\hspace{0.9mm}Hesslow\hspace{1.5mm}et\hspace{1.1mm}al.}\hspace{1.2mm}\cite{Hesslow_2019Dreicer}} with the results of kinetic simulations for plasmas consisting of hydrogen isotopes, neon and argon. The results indicate a smaller steady-state \mbox{\textit{Dreicer}} runaway generation rate, if one also accounts for collisions with partially ionized atoms \cite{HesslowPHD,Hesslow_2019Dreicer}. 

\subsection{\textit{Hot-tail} generation mechanism}\label{Hot_tail_subsection}

\begin{figure}[H]
\begin{center}
\includegraphics[trim=0 0 0 0,width=0.7\textwidth,clip]{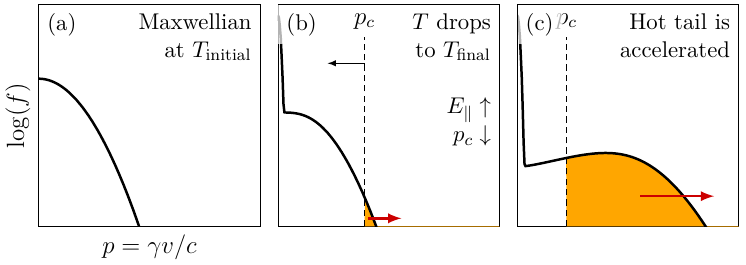} 
\caption[Illustration of the \textit{hot-tail} runaway electron generation, during a disruption in a tokamak (depiction from the master thesis by \textit{I.\hspace{0.9mm}Svenningsson} \cite{Svenningsson2020}).]{Illustration of the \textit{hot-tail} runaway electron generation, during a disruption in a tokamak (depiction from the master thesis by \textit{I.\hspace{0.9mm}Svenningsson} \cite{Svenningsson2020}).}
\label{fig_RE_hot_tail}
\end{center}
\end{figure}

Another primary generation mechanism can be understood, by recapitulating the phase of the \textit{thermal quench} from section \ref{tokamak_disruptions_section}, which appears after the sudden loss of magnetic confinement initiates a disruption. Hereby, the thermal quench describes a rapid cooling of the plasma, thus a drop of the electron temperature in electron volts, which is often modeled as an exponential decay from an initial temperature $T_{\mathrm{e,0}}$ to a final temperature $T_{\mathrm{e,fin}}$ \cite{hottailREdistfunc,Svenningsson2020}:\vspace{-4mm}
\begin{equation}\label{T_e_func}
k_{\mathrm{B}}T_{e} (t)=k_{\mathrm{B}}T_{\mathrm{e,fin}}+(k_{\mathrm{B}}T_{e,0}-k_{\mathrm{B}}T_{\mathrm{e,fin}})\cdot  \exp{\hspace{-0.4mm}\left(\frac{t}{t_{\mathrm{TQ}}} \right)} \,,
\end{equation}
\vspace{-8.0mm}\\where $t_{\mathrm{TQ}}$ is the characteristic cooling time or equivalently the timescale of the thermal quench. If the thermal quench time is significantly smaller than the characteristic collision time $\tau_{rel}$ from $(\ref{tau_rel})$, which is typical for tokamak disruptions during the thermal quench, one can not assume a quasi-steady state distribution \cite{hottailREdistfunc} as in subsection \ref{Dreicer_subsection} for the \textit{Dreicer} mechanism. This means that the slowing-down process of the initially \textit{Maxwellian} distribution occurs slower than the cooling. However, the dynamical friction force still decreases for higher velocities as discussed in \mbox{section \ref{RE_phenom_section}}. Therefore the energetic electrons, thus the \textit{hot tail} of the distribution, equilibrates slower to smaller velocities than the cold tail. This hot tail remains in the post-thermal quench electron distribution, while the induced electric field grows as shown in figure \ref{disruption_fig}. Due to this, the critical velocity and respectively the critical momentum $p_{c}$ is decreasing, which can be verified by means of the relation $(\ref{p_crit})$ in section \ref{RE_phenom_section}. In consequence, the runaway region as depicted in \mbox{figure \ref{fig_RE_region}} expands towards lower velocities and lower critical electric fields, so that a part of the \textit{hot-tail} electrons accelerates and becomes a burst of runaway electrons \cite{hottailREdistfunc,stahl,pappPHD}. A visualization of the hot-tail runaway electron generation can be seen in figure \ref{fig_RE_hot_tail}, where three snapshots of the graph of the momentum-dependent electron distribution function are shown. \vspace{-1.0mm}
\vspace{-9.5mm}
They correspond to the beginning of the disruption, where a \textit{Maxwell}-distribution function describes the electrons at a moment in time after the thermal quench and at a point in time after the current quench. In addition, one should mention, that all electrons, which are present in the runaway region before the disruption, will create an additional seed runaway electron population, whose net acceleration enhances during the thermal quench and as long as the electric field increases during the current quench.
\\ 
The explained primary mechanism is the \textit{hot-tail} generation mechanism, which is responsible for a significant conversion of the highly energetic tail of a previously thermal electron distribution into a runaway electron population, even if for the present electric field the inequality \mbox{$E \lesssim 0.01\hspace{0.45mm}E_{D}$} holds at all times \cite{EmbreusPHD}. \\
Further, one should notice that, in contrast to the other primary generation mechanisms, the volatile behaviour of the different fractions of the electron distribution during the thermal and also the current quench aggravate the development of accurate models for the hot-tail
mechanism, which are preferably analytic or at least numerically efficient in their evaluation \cite{Hoppe_PHD}. Hence, the first analytic models from 2004/5, as proposed in the publications \cite{Helander_2004,Smith_2005}, stated that the bulk of the electron distribution remains in thermal equilibrium while the cooling process takes place, under the assumption, that the collisions are faster than the plasma cooling. However, this is not necessarily true for a tokamak disruption, because the thermal quench time can be notably smaller than the collision timescale, as a consequence of the injection of large amounts of cold impurities, for the purpose of disruption mitigation through MMI in the form of SPI or MGI as explained in section \ref{tokamak_disruptions_section}. This was elaborated on in the above paragraphs as well as in further research by \textit{H.\hspace{0.7mm}M.\hspace{0.9mm}Smith} and \textit{E.\hspace{0.9mm}Verwichte} in their paper \cite{hottailREdistfunc} from 2008.
\\
Moreover, an expression for the \textit{hot-tail} growth rate under the assumption, that no electrons escape from the runaway region, can be found in said publication \cite{hottailREdistfunc}: \vspace{-4.0mm}
\begin{equation}\label{growth_rate_hot_tail}
\Gamma_{ht}\coloneqq\left(\hspace{-0.85mm}\dfrac{\mathrm{d}\hspace{0.15mm}n_{RE}}{\mathrm{d}\hspace{0.15mm}t}\hspace{-0.85mm}\right)_{\hspace{-0.85mm}ht} \hspace{-1mm}= \,\hspace{0.65mm} 4\pi\hspace{-0.6mm}\cdot\hspace{-0.6mm}\dfrac{\partial }{\partial\hspace{0.15mm}t}\hspace{-0.8mm}\left[\,\int\limits^{\infty}_{v_{c}} \hspace{-0.8mm}\left(v^2-v_{c}^2\right)\hspace{-0.6mm}\cdot\hspace{-0.6mm}f^{\mathrm{ht}}_{RE}(v,\,t)\hspace{0.6mm}\mathrm{d}t\,\right]\,.
\end{equation}
\vspace{-7.1mm}\\Here, the following distribution function, utilizing the free electron density\linebreak\mbox{$n_{e,0}=n_{e}(t=t_{0})$}, the thermal velocity \mbox{$v_{th,0}=v_{th}(T_{e}(t=t_{0}))$} according to $(\ref{v_th})$ and the thermal collision time \mbox{$\tau_{ee,0}=\tau_{ee}(t=t_{0})$} as defined in $(\ref{tau_ee})$ at the starting time $t_{0}$ of the disruption, was introduced \cite{Svenningsson2020}:
\vspace{-3.0mm}
\begin{equation}\label{f_RE_hot_tail}
f_{RE}^{\mathrm{ht}}(v,\,t)=\dfrac{n_{e,0}}{\left(\hspace{-0.2mm}\sqrt{\pi}\hspace{-0.55mm}\cdot\hspace{-0.55mm}v_{th,0}\right)^3}\hspace{-0.25mm}\cdot \exp{\hspace{-0.5mm}\left(\hspace{-0.4mm}-\left(\left(\dfrac{v}{v_{\mathrm{th}0}} \right)^{\hspace{-1.2mm}3}\hspace{-0.3mm}+\hspace{0.2mm}\dfrac{1}{\tau_{ee0}}\hspace{-0.05mm}\cdot\hspace{-0.99mm}\int\limits_{\tilde{t}=t_{0}}^{t}\hspace{-1mm}\dfrac{n_{e}(\tilde{t})}{n_{e}}\hspace{1mm}\mathrm{d}\tilde{t} \right)^{\hspace{-1.2mm}\frac{2}{3}}\right)}\,.
\end{equation}
\vspace{-6.5mm}\\The critical velocity was previously defined implicitly through the critical momentum from equation $(\ref{p_crit})$ and can be calculated from the expression:\vspace{-3.0mm}
\begin{equation}\label{v_c_hot_tail}
v_{c}=c\cdot\sqrt{\dfrac{E_{c}}{E_{\|}}}\,,
\end{equation}
\vspace{-6.5mm}\\where one can use the \textit{Connor-Hastie} field $E_{c}$ from $(\ref{E_crit})$ or perhaps the effective critical electric field $E_{c}^{\mathrm{eff}}$ as it was introduced in section \ref{part_screen_section}.

\subsection{\textit{Avalanche} generation mechanism}\label{Avalanche_subsection}

As mentioned at the beginning of the section, runaway electrons are also generated, due to the interactions of a seed runaway electron population with the thermal electrons. This \textit{avalanche} generation mechanism was first mentioned as a \qq{\textit{Multiplication} of accelerated electrons in a tokamak} by \textit{Y.\hspace{0.7mm}A.\hspace{0.9mm}Sokolov} \cite{Sokolov} in 1979. In detail, the mechanism describes \textit{knock-on}, \textit{large-angle} or \textit{close-range} collisions of seed runaway electrons with thermal electrons, which transfer the thermal electrons into the runaway region, while the post-collision momentum of the seed runaway electrons is greater than the critical momentum, implying their stay in the runaway region \cite{Rosenbluth_1997,Jayakumar_1993}.  
\\
Normally, large-angle \textit{Coulomb} collisions are less relevant for the dynamics in a fusion plasma, because the factor, by which they are less effective than small-angle collisions, is the \textit{Coulomb} logarithm with typical values between $10$ and $20$ for laboratory plasmas \cite{helander} and around $17$ for tokamak plasmas \cite{wesson}. Nevertheless, they play a vital role in the context of secondary runaway generation \cite{Stahl_2016}, since a runaway electron can quickly gain enough kinetic energy to cause the momentum of a thermal electron to become larger than the critical momentum. This means, that the minimum kinetic energy transferred during the knock-on collision has to be approximately of the order of the critical kinetic energy. Therefore, one can estimate, that the seed runaway electrons are required to have energies above two times the critical energy \cite{stahl}. With regard to the fact, that the \textit{avalanche} generation mechanism becomes dominant, if the fastest electron in a plasma reaches a kinetic energy of:\vspace{-3.0mm}
\begin{equation}\label{K_min_Avalanche}
K_{min}^{\mathrm{ava}}\approx \dfrac{\ln{\hspace{-0.45mm}\Lambda}}{2} \hspace{-0.45mm}\cdot\hspace{-0.45mm} \sqrt{5+Z_{eff}}\;\,\textup{MeV} \approx 12 ...  30\;\, \textup{MeV}\;\;;\;\;\ln{\hspace{-0.45mm}\Lambda}\in[10,\,20]\;;\;Z_{eff}\in[1,\,4]\,,
\end{equation}
\vspace{-9.0mm}\\according to an estimation of \textit{O.\hspace{0.9mm}Embréus}.

The first analytic avalanche growth rate was proposed by \mbox{\textit{M.\hspace{0.9mm}Rosenbluth}} and\linebreak\mbox{\textit{S.\hspace{0.9mm}Putvinski}} in $1997$ \cite{Rosenbluth_1997}, which reads in the limit \mbox{$E/E_{c} \gg \sqrt{1+Z_{eff}}$} and for large aspect ratio \cite{EmbreusPHD}:\vspace{-6.0mm}
\begin{equation}\label{growth_rate_Avalanche}
\Gamma_{ava}\coloneqq\left(\hspace{-0.85mm}\dfrac{\mathrm{d}\hspace{0.15mm}n_{RE}}{\mathrm{d}\hspace{0.15mm}t}\hspace{-0.85mm}\right)_{\hspace{-0.85mm}ava} \hspace{-1mm}\approx \,\hspace{0.65mm} n_{RE}\cdot \dfrac{e}{m_{e0}\hspace{0.25mm}c\hspace{0.25mm}\ln{\hspace{-0.45mm}\Lambda_{rel}}}\cdot\dfrac{E -E_{c}}{\sqrt{5+Z_{eff}}} \,.
\end{equation}
\vspace{-7.5mm}\\Here, the limit of a large aspect ratio, which is the quotient of the major and minor radius of the tokamak, as introduced in section \ref{tokamak_section}, was used. Runaway generation is expected to be strongest at the center of the plasma, therefore one can use the large aspect ratio approximation, which is always valid near the magnetic axis \cite{Breizman_2019}.
\\
From the growth rate $\Gamma_{ava}$, one can understand the sudden creation of a large number of runaway electrons. Since the acceleration of the electrons in a tokamak mainly results from the parallel component of the electric field, because the charged particles in magnetically confined fusion plasmas predominantly move parallel to the magnetic field lines. In addition, one should notice, that this growth rate is linear in the electric field $E$, which corresponds to a stronger electric field dependency than for the \textit{Dreicer} mechanism, where the growth rate was mainly proportional to the exponential of the reciprocal electric field.
\\
Furthermore, one can understand the description as an \textit{avalanche-like} generation with the help of the observation, that the growth rate in equation $(\ref{growth_rate_Avalanche})$ is linear in the runaway electron density $n_{RE}$. For this purpose, one applies the technique of the \textit{separation of variables}, which eventually reveals, that the seed runaway electron density \mbox{$n_{RE,0}\hspace{-0.25mm}\coloneqq n_{RE}(t\hspace{-0.25mm}=\hspace{-0.15mm}t_{0})$} gets multiplied by a large exponential factor:\vspace{-3.0mm}
\begin{equation}\label{sep_var_Avalanche}
n_{RE}(t) \hspace{-0.5mm}\approx\hspace{-0.5mm} n_{RE,0}\hspace{-0.5mm}\cdot\hspace{-0.3mm} \textup{exp}\hspace{-0.85mm}\left(\hspace{0.3mm}\int\limits^{t}_{\tilde{t}=t_{0}} \hspace{-1mm}\dfrac{e\hspace{-0.5mm}\cdot\hspace{-0.5mm}\left(E -E_{c}\right)}{m_{e0}\hspace{0.25mm}c\hspace{0.25mm}\ln{\hspace{-0.45mm}\Lambda_{rel}}\sqrt{5+Z_{eff}}}  \hspace{0.5mm}\mathrm{d}\tilde{t} \hspace{-0.45mm}\right)\eqqcolon n_{RE,0}\hspace{-0.5mm}\cdot\hspace{-0.3mm} \textup{e}^{\hspace{0.35mm}\mathcal{M}_{ava}}\,.
\end{equation}
\vspace{-7.0mm}\\A formula for the purpose of an estimation of the intensity of the avalanche generation for an fully ionized plasma, based on the avalanche multiplication factor \mbox{$ \textup{e}^{\hspace{0.35mm}\mathcal{M}_{ava}}$} under the assumption of a purely toroidal electric field \mbox{$E\gg E_{c}$} and in limit of large aspect ratio \mbox{($R_{0}\gg a$) }was derived by \textit{L.\hspace{0.9mm}Hesslow} \cite{HesslowPHD}:\vspace{-3.0mm}
\begin{equation}\label{exp_M_Avalanche}
\textup{e}^{\hspace{0.35mm}\mathcal{M}_{ava}}\, \approx \, \textup{exp}\hspace{-0.85mm}\left( \hspace{-0.45mm}\dfrac{1.6\hspace{-0.5mm}\cdot\hspace{-0.6mm}I_{p}}{ \ln{\hspace{-0.45mm}\Lambda_{rel}}\hspace{-0.5mm}\cdot\hspace{-0.6mm}\sqrt{5+Z_{eff}}\hspace{-0.5mm}\cdot\hspace{-0.6mm}I_{A}}    \hspace{-0.45mm}\right) \;\;;\;\; I_{A} = \dfrac{4\pi\hspace{0.25mm}m_{e0}\hspace{0.25mm}c}{\mu_{0}\hspace{0.25mm}e}\approx 17.05\;\textup{kA}\,.
\end{equation}
\vspace{-7.0mm}\\At this, the \textit{Alfvén} current $I_{A}$ was defined and for an ITER scenario as displayed in table \ref{ITER_table}, with \mbox{$\langle k_{B}T_{e}\rangle=8.8\,\textup{eV}$}, \mbox{$\langle n_{e}\rangle=10^{20}\,\textup{m}^{-3}$} and a plasma current of \mbox{$I_{p}=15\,\textup{MA}$}, one can calculate an avalanche multiplication factor of \mbox{$\textup{e}^{\hspace{0.35mm}\mathcal{M}_{ava}}\approx \textup{e}^{\hspace{0.35mm}36.64 }\approx 8\cdot10^{15}$}. A different approximation by \textit{Rosenbluth} \& \textit{Putvinski} \cite{Rosenbluth_1997} prognosticates an even larger multiplication factor for the case of ITER of \mbox{$ \textup{e}^{\hspace{0.35mm}\mathcal{M}_{ava}}\approx \textup{e}^{\hspace{0.35mm}50}\approx 5\cdot10^{21}$}. As a result of this, the runaway generation, especially in larger tokamaks such as ITER \cite{HesslowPHD}, can be substantial and might convert a significant fraction of the plasma current into a runaway current, even if the seed runaway density produced by primary generation mechanisms is only a single seed runaway electron \cite{Hoppe_PHD}.  

Typically, one finds the character of a runaway electron population in a tokamak disruption to be of a beam-like shape along magnetic field lines or flux surfaces. But it does not have to be centered at the magnetic axis, since its magnetic confinement is not necessarily undisturbed and the generation of the runaway beam can start everywhere within the plasma. This was analysed in the references \cite{Besedin_1986} and \cite{REdistfuncderivation}, which discovered, that the beam has a strongly anisotropic non-equilibrium momentum distribution function with respect to the orthogonal momenta and is able to excite various instabilities such as electromagnetic waves.

Recent research by \mbox{\textit{L.\hspace{0.9mm}Hesslow\hspace{1.4mm}et\hspace{1.1mm}al.}\hspace{1.2mm}\cite{Hesslow_2019}} discovered, that the avalanche multiplication factor might exceed \mbox{$\textup{e}^{\hspace{0.35mm}\mathcal{M}_{ava}}\approx 10^{35}$} for an ITER-like deuterium density and impurity densities near \mbox{$10^{20}\,\textup{m}^{-3}$}, due to the effect of partial screening and an only partially ionized plasma. As discussed in section \ref{part_screen_section}, one finds enhanced collision frequencies and therefore stronger friction and pitch-angle scattering influencing the electrons \cite{Hesslow_2018,HesslowPHD}. However, the partial or full ionization of the impurities increases the total free electron density, which leads to a higher probability for large-angle collisions and hence increases the avalanche growth rate, in contrast to the decreasing \textit{Dreicer} generation rate \cite{Hesslow_2019,HesslowPHD}. The avalanche growth rate proposed by \mbox{\textit{Rosenbluth} \& \textit{Putvinski}} is consequently unsuitable for plasmas with partially ionized impurities and is suggested to be replaced by a form-invariant, interpolated formula published by \mbox{\textit{L.\hspace{0.9mm}Hesslow}} in 2019 \cite{Hesslow_2019}:\vspace{-3.0mm}
\begin{equation}\label{growth_rate_Avalanche_Hesslow}
\Gamma^{\hspace{0.3mm}\mathrm{scr}}_{ava}\coloneqq\left(\hspace{-0.85mm}\dfrac{\mathrm{d}\hspace{0.15mm}n_{RE}}{\mathrm{d}\hspace{0.15mm}t}\hspace{-0.85mm}\right)_{\hspace{-0.85mm}ava}^{\hspace{-0.85mm}\mathrm{scr}} \hspace{-0.85mm}=\,\hspace{0.65mm} n_{RE}\cdot \dfrac{e}{m_{e0}\hspace{0.25mm}c\hspace{0.25mm}\ln{\hspace{-0.45mm}\Lambda_{rel}}}\cdot\dfrac{n_{e}^{\mathrm{tot}}}{n_{e}}\cdot\dfrac{\vert E_{\|}\vert -E^{\mathrm{eff}}_{c}}{\sqrt{4+ \tilde{\nu}_{s}\hspace{-0.35mm}\left(p_{\star}\right)\tilde{\nu}_{d}\hspace{-0.35mm}\left(p_{\star}\right)}} \,,
\end{equation}
\vspace{-6.5mm}\\which holds for \mbox{$\vert E_{\|}\vert \gtrsim E^{\mathrm{eff}}_{c}$} and uses the effective critical electric field $E^{\mathrm{eff}}_{c}$ as well as the ultra-relativistic approximations \mbox{$\tilde{\nu}_{d}\hspace{-0.35mm}\left(p_{\star}\right)$} and \mbox{$\tilde{\nu}_{s}\hspace{-0.35mm}\left(p_{\star}\right)$} for \mbox{$p\gg 1$} of the normalized deflections and the slowing-down frequency, to evaluate the effective critical momentum $p_{\star}$. Note, that all of the mentioned quantities were discussed alongside their computation in section \ref{part_screen_section}. Furthermore, it should be remarked, that the growth rate $(\ref{growth_rate_Avalanche_Hesslow})$ transfers to the growth rate from $(\ref{growth_rate_Avalanche})$ in the complete screening limit, respectively in the limit of a fully ionized plasma, so that \mbox{$\Gamma^{\hspace{0.3mm}\mathrm{scr}}_{ava}\rightarrow\Gamma_{ava}$} holds for \mbox{$\vert E_{\|}\vert=E$}, due to \mbox{$\tilde{\nu}_{d}\hspace{-0.35mm}\left(p_{\star}\right)\rightarrow 1+Z_{eff}$} and \mbox{$\tilde{\nu}_{s}\hspace{-0.35mm}\left(p_{\star}\right)\rightarrow 1$} as given in \mbox{$(\ref{nue_s_nue_d_def})$}, while \mbox{$n_{e}\rightarrow n_{e}^{\mathrm{tot}}$} and \mbox{$E^{\mathrm{eff}}_{c}\rightarrow E_{c}$} follows from the full ionization of all plasma components.
\\
By comparing the growth rates $\Gamma^{\hspace{0.3mm}\mathrm{scr}}_{ava}$ and $\Gamma_{ava}$, one eventually notices a stronger dependency on the electric field with \mbox{$\Gamma^{\hspace{0.3mm}\mathrm{scr}}_{ava}\propto E^{\hspace{0.2mm}1.5}_{\|}$} \cite{Hesslow_2019}, in contrast to \mbox{$\Gamma_{ava}\propto E$} for the expression from \mbox{\textit{Rosenbluth} \& \textit{Putvinski}}. The reason therefore is the consideration of the partial screening phenomena in the growth rate $\Gamma^{\hspace{0.3mm}\mathrm{scr}}_{ava}$, which yields an additional influence of variations in the electric field through the deflection and the slowing-down frequency.

\subsection{Runaway electron loss mechanisms}\label{RE_loss_subsection}

The zeroth loss mechanism originates from the \textit{Coulomb} collisions in-between electrons and between electrons and ions as discussed in section \ref{RE_phenom_section}. Here, it should be remarked, that mainly the mentioned slowing-down part of the dynamical friction force is described as a loss mechanism, since elastic collisions and pitch-angle scattering as well as parallel momentum diffusion contribute less to the transport of runaway electrons out of the runaway region \cite{stahl,HesslowPHD}.    

The first main energy loss channel is by the so-called \textit{Bremsstrahlung}, resulting from the acceleration of electrons as a consequence of inelastic scattering with heavier ions \cite{stahl,Stroth_2018}. \textit{Bremsstrahlung} especially increases for collisions with ions of a high nuclear charge number \cite{Stroth_2018} and has a dominant influence on the electron dynamics for electron energies above hundred MeV \cite{Papp_2011}.

The second energy loss channel is the radiation, which is emitted by highly relativistic particles experiencing an acceleration perpendicular to their main direction of motion towards their own trajectory. This \textit{synchrotron radiation} as described in detail in the work of \textit{A. Stahl} \cite{stahl} is therefore also produced by runaway electrons, since they perform a gyro motion around their main direction of motion. In general, the synchrotron radiation is not isotropic and concentrates in the direction parallel to the magnetic field in tokamaks, because this is the major direction of motion for the electrons. In this regard, the angle of opening of the emission cone was found to be roughly proportional to $\gamma^{-1}$ \cite{Hoppe_PHD}.

Both radiation losses are basically photon populations, which carry away a \mbox{non-neglec-}\linebreak\mbox{table} amount of momentum from the electrons, since the total momentum has to be conserved. In consequence, this leads to an additional force which can counteract the acceleration of those electrons for instance due to an electric field \cite{stahl}. 

In addition, the presence of impurities or more generally of a not fully ionized plasma is a contributing loss mechanism. It is caused by the partial screening of the nuclear charges as discussed in section \ref{part_screen_section}. In detail, it leads to enhanced collision rates causing a larger dynamical friction force and an increased synchrotron radiation, while also directly enlarging \textit{Bremsstrahlung} losses \cite{HesslowPHD}. In addition, runaway electrons might lose energy in ionizing collisions \cite{stahl}.

Furthermore, the high kinetic energy of the runaway electrons can cause spontaneous electron-positron pair production in collisions with thermal ions \cite{pappPHD}. This happens, if the runaway energy exceeds approximately three times the rest mass energy \mbox{$m_{e0}\hspace{0.25mm}c^2$}, so that \mbox{$k_{RE}/c^2\gtrsim 3$} holds for the rest mass energy-related kinetic energy density, which corresponds to an energy of roughly \mbox{$1.53\,\mathrm{MeV}$} \cite{Helander_2003}. Pair production can also originate from the collisions with thermal electrons, which however requires runaway energies above \mbox{$3.58\,\mathrm{MeV}$} or equivalently with \mbox{$k_{RE}/c^2\gtrsim 7$} \cite{Helander_2003}. In post-disruption plasmas within large tokamaks, the mean energy of the runaway electrons is known to be between $10$ and \mbox{$20\,\mathrm{MeV}$} \cite{RunawayPositrons}. Therefore, electron-positron pair production, is as well present as a runaway loss mechanism in nuclear fusion devices like ITER, since those collisions between energetic runaway electrons and thermal ions or electrons are accompanied, by a momentum and energy transfer. However, it has to be remarked, that this loss mechanism is only responsible for minor loss contributions in ITER \cite{RunawayPositrons}. On this occasion, it can be remarked that positrons and ions in a plasma can also run away during tokamak discharges \cite{Furth_1972,Helander_2003,RunawayPositrons}, although the electron runaway phenomenon is the most significant in the context of tokamak plasmas \cite{Dreicer_1960,Connor_1975}.  

Finally, also perturbations in the magnetic confinement play a role as a loss mechanism for runaway electrons through several channels. Those are for instance magnetic trapping effects, radial transport as a consequence of magnetic field turbulences \cite{Rechester_1978} and collisions with the wall for high-enough electron energies, allowing a drift-induced escape from the magnetic confinement \cite{Hoppe_PHD,stahl}. In order to further understand the named as well as additional loss mechanisms, like the interaction of runaway electrons with different kinds of waves within the plasma, one is referred to the literature \cite{Hoppe_PHD,stahl}. 

\clearpage

%% file: Calculation_of_the_moments_of_the_hot_tail_runaway_electron_distribution_function.tex
\chapter{Calculation of the moments of a \textit{hot-tail} runaway electron distribution function}\label{hot_tail_chapter}

The potential generation of a runaway electron beam, during a disruption, in future tokamak reactors, such as ITER, carries the risk of damaging plasma-facing components of the vacuum vessel \cite{Hoppe_2021,REsimulation,Hoppe_2022}. This motivates i.a.\ the research area of \textit{disruption mitigation} as introduced in the section \ref{tokamak_disruptions_section}. At this the method of \textit{massive material injection} was described, which relies on injected impurity atoms with high nuclear charge, which can convert the thermal and magnetic energy contained in the plasma into isotropic electromagnetic radiation, so that high local energy deposition into the reactor wall is avoided. In addition, the electron density of the plasma increases as a consequence of the material injection, which raises the critical electric field and therefore the threshold for acceleration of thermal electrons into the runaway region. This mitigation procedure is also the currently proposed disruption mitigation method for the ITER fusion device \cite{Hollmann2014,Nardon2020}. 
\\
In section \ref{mechanisms_section}, different mechanisms for the generation of runaway electrons were discussed. In detail, it was explained in subsection \ref{Avalanche_subsection}, that the primary production of a seed runaway electron population is crucial in the development of a large number of runaway electrons, because it triggers the \textit{avalanche} generation mechanisms. For an ITER-like deuterium density and impurity densities near \mbox{$10^{20}\,\textup{m}^{-3}$}, it was found, that even small primary runaway electron densities might be enhanced by a multiplicative avalanche amplification factor of approximately $10^{35}$ \cite{Hesslow_2019}. Hence, it is important to simulate and analyse physical quantities related to primary runaway electron generation mechanisms.

During the first phase of a disruption, the \textit{thermal quench}, one observes a rapid cooling of the plasma, which is followed by an abrupt ramp up of the electric field in the subsequent phase named the \textit{current quench}. This can be recapitulated in section \ref{tokamak_disruptions_section} and in particular by means of the figure \ref{disruption_fig}. The thermal quench or cooling time is significantly smaller than the characteristic collision time for the electron interactions, so that the tail of the momentum space distribution of the electrons with high momenta equilibrates slower towards smaller velocities then the electrons with lower momentum. Meanwhile, the critical momentum decreases during the first phase of the current quench due to the growing electric field, so that the runaway region grows and electrons of the hot tail of the distribution accelerate and become runaway electrons \cite{hottailREdistfunc,stahl,pappPHD}. This \textit{hot-tail} generation mechanism is further analysed in subsection \ref{Hot_tail_subsection} and can be the dominant primary source of seed runaway electrons at the start of a disruption \cite{Smith_2005,hottailREdistfunc,REsimulation,Mart_n_Sol_s_2017,Svenningsson_2021}. Therefore, it is required to do research on the \textit{hot-tail} mechanism. 

In this chapter one focuses on the evaluation of the moments of a \textit{hot-tail} runaway electron distribution function. Especially, the moments connected to the \textit{hot-tail} runaway electron density and the mean velocity are interesting. This is because, they lead to the current density:\vspace*{-7.0mm} 
\begin{equation}\label{RE_HT_curr_dens_def}
j_{\mathrm{RE}}^{\mathrm{ht}}= q_{e}\cdot n^{\mathrm{ht}}_{\mathrm{RE}}\cdot u_{\mathrm{RE}}^{\mathrm{ht}}= -\,e\cdot n^{\mathrm{ht}}_{\mathrm{RE}}\cdot u_{\mathrm{RE}}^{\mathrm{ht}}\,,
\end{equation} 
\vspace*{-11.0mm}\\which contributes to the total runaway current density. At this, the magnitude of the mean velocity of \textit{hot-tail} runaway electrons can be calculated from a distribution function, with the help of the integral:\vspace*{-2.9mm} 
\begin{equation}\label{u_RE_HT_def}
u_{\mathrm{RE}}^{\mathrm{ht}}(\mathbf{r},\,t)  =\dfrac{1}{n_{\mathrm{RE}}^{\mathrm{ht}}(\mathbf{r},\,t)}\, \displaystyle{\iiint\limits_{\mathbb{R}^3}}\;\hspace{-0.8mm} v \cdot f_{\mathrm{RE}}^{\mathrm{ht}}(\mathbf{r},\,\mathbf{p},\,t)\;\mathrm{d}^3p\,,
\end{equation} 
\vspace*{-7.5mm}\\as stated in relation $(\ref{first_moment})$. Furthermore, the \textit{hot-tail} runaway electron density:\vspace*{-2.7mm} 
\begin{equation}\label{n_RE_HT_def}
n_{\mathrm{RE}}^{\mathrm{ht}}(\mathbf{r},\,t) \,= \displaystyle{\iiint\limits_{\mathbb{R}^3}} f_{\mathrm{RE}}^{\mathrm{ht}}(\mathbf{r},\,\mathbf{p},\,t)\;\mathrm{d}^3p\,,
\end{equation} 
\vspace*{-7.5mm}\\according to the definition $(\ref{zeroth_moment})$ has to be computed, due to the fact, that it is i.a.\ needed to determine the avalanche runaway density after it is multiplied by the avalanche amplification factor, so that it subsequently allows the simulation of the major contribution to the runaway current. Moreover, the moment of the mean kinetic energy normalized to the electron rest mass energy:\vspace*{-2.8mm} 
\begin{equation}\label{k_RE_HT_def}
\begin{split}
\dfrac{k_{\mathrm{RE}}^{\mathrm{ht}}}{c^2} \,\coloneqq \,\dfrac{\langle K_{\mathrm{RE}}^{\mathrm{ht}}\rangle }{m_{e0}\hspace{0.25mm}c^2}\,=\, \langle\gamma-1\rangle =  \dfrac{1}{n^{\mathrm{ht}}_{\mathrm{RE}}}\,\iiint\limits_{\mathbb{R}^3}  \hspace{-0.5mm}\gamma \cdot  f^{\mathrm{ht}}_{\mathrm{RE}}(\mathbf{r},\,\mathbf{p},\,t)\;\mathrm{d}^3 p \,-\,1\,,
\end{split}
\end{equation} 
\vspace*{-8.0mm}\\from equation $(\ref{kin_dens_ava_def})$, might be used in energy balance equations within simulations or for the purpose of further characterization of a \textit{hot-tail} runaway electron population.
 
In this thesis, the isotropic electron distribution function $f^{\mathrm{ht}}_{\mathrm{RE}}(p,\,t)$ by \textit{H.\hspace{0.7mm}M.\hspace{0.9mm}Smith} and \textit{E.\hspace{0.9mm}Verwichte} \cite{hottailREdistfunc} is used for the modeling of the \textit{hot-tail} generation mechanism. However, different representations of the runaway region in momentum space are applied, in order to analyse general tendencies, which result from the pitch-angle dependency of this region and/or the effects of partial screening. The different fractions of the electron distribution show a volatile behaviour in time during the thermal and also the current quench of a disruption \cite{Hoppe_PHD}, as discussed in subsection \ref{Hot_tail_subsection}. Therefore, the simulation of a disruption for an ITER-scenario, covering the thermal and current quench phase, from the paper \cite{Smith_2009} with a plasma current of \mbox{$I_{p}=15\;\mathrm{MA}$}, a magnetic field of \mbox{$B=5.3\,\mathrm{T}$} and a time-independent electron density of \mbox{$n_{e}=1.06\cdot10^{20}\;\mathrm{m}^{-3}$} is recapitulated from section \ref{part_screen_section}. The time evolution of the main physical quantities was displayed in figure \ref{fig_RE_ITER_simu} and in addition the results for different models of the boundaries of the runaway region with respect to the momentum magnitude can be seen in figure \ref{fig_RE_mom_bound}. For the purpose of a physics validation of the derived calculation schemes of the moments of \textit{hot-tail} runaway electrons in the \textit{Smith-Verwichte} model, one makes use of the mentioned ITER-disruption simulation for the subsequent analysis. The effects of partial screening are then discussed, by means of the utilization of the same data for a calculation with a time-independent singly-ionized neon impurity density of \mbox{$n_{_{10}^{20}\mathrm{Ne}^{+}}=2.40\cdot 10^{20}\,\mathrm{m}^{-3}$}. For this case, it has to be addressed, that the electric field and temperature evolution of the regarded simulation is not self-consistent anymore, because the original data corresponds to a pure deuterium plasma. 






\section[\textit{Smith-Verwichte} approach and pitch angle-dependent runaway region]{\textit{Smith-Verwichte} approach and pitch angle- dependent runaway region}\label{SmithVerwichte_ht_dist_section}

In the following subsections, the \textit{Smith-Verwichte} approach is presented, which models a cooling electron population and is based on an isotropic distribution function. It is the typically used approach in disruption and more precise runaway current simulations. For example, it is also applied implicitly in the DREAM-code \cite{Hoppe_2021}. The derivation and the associate framework of the distribution function were carried out by \textit{H.\hspace{0.7mm}M.\hspace{0.9mm}Smith} and \textit{E.\hspace{0.9mm}Verwichte} in the publication \cite{hottailREdistfunc} from 2008. They also stated integral calculation rules for the \textit{hot-tail} runaway electron density with and without a consideration of the pitch-angle dependency of the runaway region. In this chapter, the results for the density shall be reproduced and evaluated for different representations of the runaway region. Although, the derivation of the connection of the \textit{hot-tail} runaway density calculation rules by \textit{H.\hspace{0.7mm}M.\hspace{0.9mm}Smith} and \textit{E.\hspace{0.9mm}Verwichte} to the pitch-dependent runaway region is not shown explicitly, since it can be found for example in the master thesis of \mbox{\textit{I.\hspace{0.9mm}Svenningsson}} \cite{Svenningsson2020}. 
\\
The mentioned different boundaries of the runaway region, for the magnitude of the relativistic momentum $p$ from \ref{part_screen_section}, are applied and the results of the resulting calculation rules are analyzed. In addition, one extends those calculation schemes by taking into account, that the runaway region is anisotropic in the two-dimensional momentum space from \ref{mom_space_coord_section}, which will be further discussed in the subsection \ref{pitch_RE_region_subsection}. As a result, computation rules for the \textit{hot-tail} runaway density, the mean velocity and the mean kinetic energy density will be stated for four different approximations of the momentum magnitude boundaries. Additionally, the four calculation schemes are extended, by means of the consideration of the pitch-angle dependent runaway region, while all results are computed for a deuterium plasma and for a case with a present neon impurity.

\subsection{Isotropic distribution function for the \textit{hot-tail} runaway electron generation in the \textit{Smith-Verwichte} approach}\label{ht_dist_func_subsection}

The isotropic electron distribution function for the \textit{hot-tail} runaway generation mechanism was derived by \textit{H.\hspace{0.8mm}M.\hspace{0.9mm}Smith} and \textit{E.\hspace{0.9mm}Verwichte} in the paper \cite{hottailREdistfunc}. However, its representation with respect to the relativistic momentum:\vspace*{-3.1mm}
\begin{equation}\label{SV_ht_dist_func}
f^{\mathrm{ht}}_{\mathrm{RE}}(p,\,t)= \dfrac{n_{\mathrm{e}}}{\pi^{\frac{3}{2}}p_{\mathrm{th},0}^3}\hspace{-0.6mm}\cdot\hspace{-0.3mm}\exp{\hspace{-0.6mm}\left(\hspace{-0.7mm}-\dfrac{\left(p^3+3\hspace{-0.5mm}\cdot\hspace{-0.3mm}\textup{I}_{\tau_{rel}}(t)\right)^{\frac{2}{3}}}{p_{\mathrm{th},0}^2}\hspace{-0.5mm}\right)}\;\;;\;\;\textup{I}_{\tau_{rel}}\hspace{-0.1mm}(t)\coloneqq\hspace{-0.75mm}\displaystyle{\int\limits_{\tilde{t}=t_{0}}^{t}} \hspace{-0.5mm}\dfrac{1}{\tau_{rel}(\tilde{t}\hspace{0.1mm})}\;\mathrm{d}\tilde{t} \,
\end{equation}
\vspace*{-7.5mm}\\originates from the work of \textit{I.\hspace{0.9mm}Svenningsson} \cite{Svenningsson2020}, which states as well that this distribution function neglects momentum diffusion. The occurring parameters are the free electron density $n_{e}$ and the normalized initial thermal momentum corresponding to the initial electron temperature \mbox{$T_{e,0}\coloneqq T_{e}(t_{0})$} in electron volts, in accordance with the relation $(\ref{v_th})$:\vspace*{-6.5mm}
\begin{equation}\label{p_th0_def}
p_{\mathrm{th},0}=\sqrt{\dfrac{2\hspace{0.25mm}e\hspace{0.25mm}k_{B}T_{e,0}}{m_{e0}\hspace{0.25mm}c^2}}\,.
\end{equation}
\vspace*{-8.0mm}\\Additionally, the relativistic collision time $\tau_{rel}$ as defined in $(\ref{tau_rel})$ appears in the integral $\textup{I}_{\tau_{rel}}\hspace{-0.1mm}(t)$, which ensures a decay of the distribution function in time, resulting from the electron interactions connected to the plasma cooling. The main proportionality is noticed to be a exponential decay in time and with respect to momentum, because the appearing integral and the magnitude of the momentum are positive, so that the argument of the exponential function in $(\ref{SV_ht_dist_func})$ is always negative. Here, it can be remarked, that this is also the relation, which is indirectly utilized in the simulation software DREAM, where it appears in the growth rate for the runaway electron density, similarly to equation $(\ref{growth_rate_hot_tail})$, due to the \textit{hot-tail} generation mechanism. 

In the definition $(\ref{SV_ht_dist_func})$, one notices the absence of an electric field dependency, which explains the characterization as an isotropic distribution function \cite{Svenningsson2020}. On this occasion, one remarks that the mentioned isotropy is related to the neglection of the pitch-angle dependent runaway region, which is further elaborated in the following subsection. Moreover, the distribution function is found to overestimate the true electron distribution in momentum space for the early phase of the plasma cooling, because it does not include momentum diffusion \cite{Svenningsson2020}. Therefore, the correction $\textup{I}_{\tau_{rel}}\hspace{-0.1mm}(t)=\tau_{ee,0}^{-1}(t-t_{TQ})$ \cite{hottailREdistfunc} can be applied, where \mbox{$\tau_{ee,0}=\tau_{ee}(t=t_{0})$} is the thermal collision time, as defined in $(\ref{tau_ee})$ at the starting time $t_{0}$, and $t_{TQ}$ is the cooling time, which is the timescale of the thermal quench for the case of a disruption. But in the context of this work, this relation is not applied, in order to calculate $\textup{I}_{\tau_{rel}}\hspace{-0.1mm}(t)$ in a similar manner to the DREAM-code \cite{Hoppe_2021}. A second reason is the fact, that the diffusion also influences the time evolution of the distribution function for times larger than the cooling time, if the final electron temperature of the thermal quench has an order of magnitude of \mbox{$100\,\mathrm{eV}$}. More precise, the mentioned relation should not be used for \mbox{$T_{e,fin}\gtrsim 10\,\mathrm{eV}$}, where one refers to the model $(\ref{T_e_func})$ for the time evolution of the electron temperature from subsection \ref{Hot_tail_subsection}.

Note, that within all \textsc{MATLAB}-implementations associated with this chapter the integral $\textup{I}_{\tau_{rel}}\hspace{-0.1mm}(t)$, which was defined in the equation $(\ref{SV_ht_dist_func})$, is evaluated with a trapezoidal rule. In detail, the \textsc{MATLAB}-routine \qq{\texttt{trapz}} \cite{trapz} is applied, which is able to numerically integrate the integrand with respect to the given time step spacing of the ITER-simulation data with the initial time \mbox{$t_{0}=0\,\mathrm{ms}$}.

\subsection{Pitch angle-dependent runaway region}\label{pitch_RE_region_subsection}

The pitch-angle resolved momentum of an electron leads to a parallel and orthogonal component in relation to a present electric field. This electric field parameter is commonly assumed to be solely parallel to the local magnetic field in runaway simulations \cite{Hoppe_2021}. Subsequently, one has to compare the parallel momentum component of an electron with the critical momentum, in order to decide, if it will become a runaway electron for \mbox{$p_{\|}>p_{c}$}. This leads to the insight, that for a certain pitch coordinate \textit{$\xi<1$} the critical momentum is not exceeded. Hence, an anisotropy of the runaway region appears in the two-dimensional $(p,\,\xi)$-momentum space, since electrons with a pitch coordinate within a distinct interval are favored to run away. The maximum interval of the pitch coordinate is \mbox{$\xi=\cos{(\theta)}\in[-1,\,1]$} for pitch-angles with \mbox{$\theta\in[-\pi,\,0]\,$}. At that, \textit{$\xi=1$} denotes an electron momentum, which is exclusively parallel to the magnetic field and consequently represents the upper boundary of the runaway region with respect to the pitch coordinate. The lower boundary is clear for the parallel direction with \textit{$\xi=1$}, where it is defined directly through the equality \textit{$p=p_{c}$}, where $p_{c}$ is the critical momentum. The momentum-dependent pitch coordinate \textit{$\xi_{sep}(p)$}, defining the anisotropic runaway region for the rest of the momentum space, is then found from the trajectory of electrons, where the acceleration force of the electric field and the slowing-down forces, like the dynamical friction force and reaction forces due to radiation, balance each other out, so that they are neither accelerated nor decelerated. Technically, those electrons are neither runaway nor thermal electrons. This described trajectory is called \textit{separatrix} and its pitch dependence was found from the analysis of particle trajectories in phase
space by \textit{G.\hspace{0.9mm}Fussmann} \cite{Fussmann1979}. According to the references \cite{Smith_2005,stahl,Svenningsson2020}, it is equivalently expressible as a pitch- or momentum-dependent function:\vspace*{-3.5mm}
\begin{equation}\label{xi_sep_def}
p_{\mathrm{sep}}\left(\xi,\,E_{\mathrm{c}}\right)=\left(\dfrac{\xi+1}{2}\cdot\dfrac{\vert E_{\|}\vert}{E_{\mathrm{c}}}-1\right)^{-\frac{1}{2}}\;\;\longleftrightarrow\;\;\xi_{\mathrm{sep}}\hspace{-0.7mm}\left(p,\,E_{\mathrm{c}}\right)= \dfrac{2\cdot E_{\mathrm{c}}}{p^2\cdot\vert E_{\|}\vert}-1\,.
\end{equation}
\vspace*{-7.5mm}\\This separatrix is the anisotropic, electric field dependent lower boundary of the two-dimensional runaway region. The \textit{Smith-Verwichte} \textit{hot-tail} runaway electron distribution function has no electric field dependency, which finally allows to understand, why it was classified as isotropic in the previous subsection \ref{ht_dist_func_subsection}.

\begin{figure}[H]
\begin{center}
\includegraphics[trim=45 24 89 44,width=0.8\textwidth,clip]{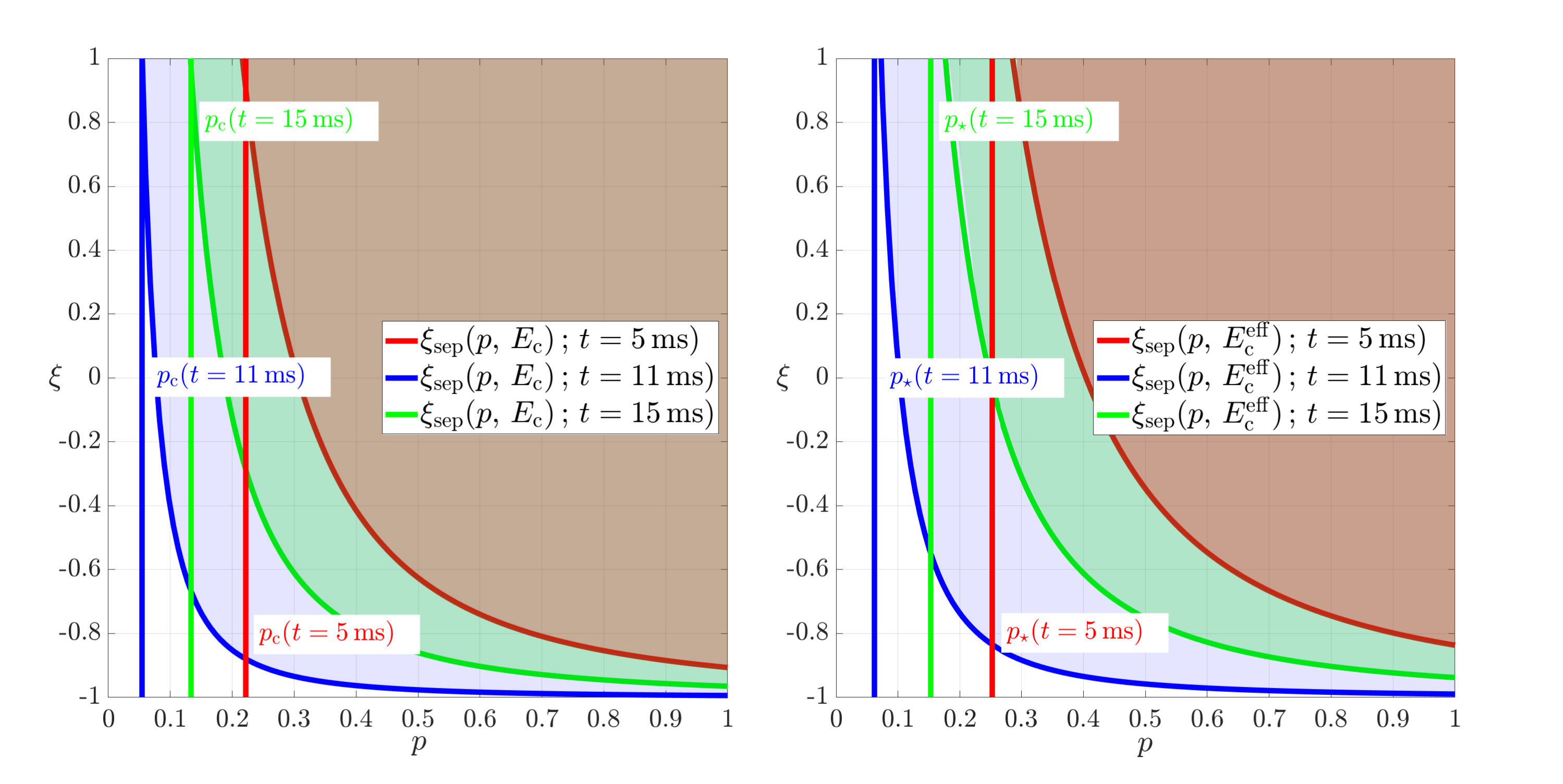}
\caption[Snapshots of the time evolution of the pitch-dependent runaway region, determined by the \textit{generalized separatrix} \mbox{$\xi_{\mathrm{sep}}(p,\,\tilde{E}_{\mathrm{c}})$} for the choices of the generalized critical electric field \mbox{$\tilde{E}_{\mathrm{c}}=E_{\mathrm{c}}^{\mathrm{eff}}$} and \mbox{$\tilde{E}_{\mathrm{c}}=E_{\mathrm{c}}$}, during the disruption from figure \ref{fig_RE_ITER_simu} for the ITER-scenario \cite{Smith_2009}.]{Snapshots\protect\footnotemark{} of the time evolution of the pitch-dependent runaway region, determined by the \textit{generalized separatrix} \mbox{$\xi_{\mathrm{sep}}(p,\,\tilde{E}_{\mathrm{c}})$} for the choices of the generalized critical electric field \mbox{$\tilde{E}_{\mathrm{c}}=E_{\mathrm{c}}^{\mathrm{eff}}$} and \mbox{$\tilde{E}_{\mathrm{c}}=E_{\mathrm{c}}$}, during the disruption from figure \ref{fig_RE_ITER_simu} for the ITER-scenario \cite{Smith_2009}.}
\label{fig_RE_region_pitch}
\end{center}
\end{figure}

Note, that the dependency of the separatrix on the critical electric field in the relations in $(\ref{xi_sep_def})$ was written explicitly. Subsequently, one can define four different representations for each the pitch-dependent and the isotropic runaway region, based on the results from the sections \ref{RE_phenom_section} and \ref{part_screen_section}. 
\\
First, the \textit{Connor-Hastie} critical momentum from $(\ref{p_crit})$ is used in combination with the associated critical electric field from $(\ref{E_crit})$. From this the isotropic and anisotropic runaway region is determined by:\vspace*{-3.5mm}
\begin{equation}\label{CH_RE_region}
\begin{split}
\begin{gathered}
\xi\in[-1,\,1]\;\wedge\;p\in\left[p_{\mathrm{c}},\,\infty\right)\;\wedge\;E_{\|}>E_{\mathrm{c}}
 \\
 \xi\in\left[\xi_{\mathrm{sep}}\hspace{-0.7mm}\left(p,\,E_{\mathrm{c}}\right),\,1\right]\;\wedge\;p\in\left[p_{\mathrm{c}},\,\infty\right)\;\wedge\;E_{\|}>E_{\mathrm{c}}\,.
\end{gathered}
\end{split}
\end{equation}
\vspace*{-7.0mm}\\Second, the effective critical electric field $E^{\mathrm{eff}}_{\mathrm{c}}$, which can be computed with the\linebreak\mbox{\textsc{MATLAB}-script} \qq{\texttt{calculate_E_c_eff.m}} from \textit{L. Hesslow} \cite{Hesslow_2018}, describes the runaway region together with the effective critical momentum $p_{\star}$, which can be found as the root of the function from $(\ref{func_p_c_eff_def})$. They also account for the effects of partial screening, as explained in section \ref{part_screen_section}, and should define the physically more accurate runaway region with and without the consideration of its pitch-dependency:\vspace*{-3.5mm}
\begin{equation}\label{H_RE_region}
\begin{split}
\begin{gathered}
\xi\in[-1,\,1]\;\wedge\;p\in\left[p_{\star},\,\infty\right)\;\wedge\;E_{\|}>E_{\mathrm{c}}^{\mathrm{eff}}
 \\
 \xi\in\left[\xi_{\mathrm{sep}}\hspace{-0.7mm}\left(p,\,E_{\mathrm{c}}^{\mathrm{eff}}\right),\,1\right]\;\wedge\;p\in\left[p_{\star},\,\infty\right)\;\wedge\;E_{\|}>E_{\mathrm{c}}^{\mathrm{eff}}\,.
\end{gathered}
\end{split}
\end{equation}
\vspace*{-7.0mm}\\Third, the runaway region momentum boundaries $p_{min}$ and $p_{max}$ can be calculated from the net acceleration force, as the roots of the function in $(\ref{F_acc})$, including the effects of partial screening and synchrotron radiation as well as \textit{Bremstrahlung}. This leads to the following representations of the isotropic and anisotropic runaway region:\vspace*{-3.5mm}
\begin{equation}\label{minmax_RE_region}
\begin{split}
\begin{gathered}
\xi\in[-1,\,1]\;\wedge\;p\in\left[p_{min},\,p_{max}\right]\;\wedge\;E_{\|}>E_{\mathrm{c}}^{\mathrm{eff}}
 \\
 \xi\in\left[\xi_{\mathrm{sep}}\left(p,\,E_{\mathrm{c}}^{\mathrm{eff}}\right),\,1\right]\;\wedge\;p\in\left[p_{min},\,p_{max}\right]\;\wedge\;E_{\|}>E_{\mathrm{c}}^{\mathrm{eff}}\,.
\end{gathered}
\end{split}
\end{equation}
\vspace*{-7.0mm}\\The fourth possibility of the definition of the runaway region with and without pitch-dependence is motivated by the result from section \ref{part_screen_section}, that the upper limit for the runaway momentum is determined through the largest change in the poloidal magnetic flux for tokamak reactors. This was discussed in section \ref{part_screen_section}, based on the simulation of an ITER-like disruption \cite{Smith_2009}. In particular it was estimated, that for such a disruption, as depicted in the figures \ref{fig_RE_ITER_simu} and \ref{fig_RE_mom_bound}, the maximum runaway electron momentum satisfies the inequalities \mbox{$p_{max}\lesssim 196.7$} for the highest possible energy of the electrons \mbox{$\mathcal{E}_{e,max}\approx 100\,\mathrm{MeV}$} \cite{Papp_2011}. Those estimations for the maximum energies were also obtained for a typical ITER-scenario and result from a self-consistent simulation including loss effects and electric field diffusion, as it can be read in the publication \cite{Papp_2011}. Hence, one introduces a simplifies the computation of $p_{max}$ and only computes it as the root of the function in $(\ref{F_acc})$, if it is below its estimation based on the maximum reachable energy. A possible expression for the considered ITER-scenario, could be given by:\vspace*{-3.5mm}
\begin{equation}\label{tilde_p_max}
\tilde{p}_{\mathrm{max}}\hspace{-0.4mm} =\hspace{-0.2mm} 196.7 \cdot \mathcal{H}_{196.7}(p_{\mathrm{max}}) +p_{\mathrm{max}}\hspace{-0.02mm}\cdot\hspace{-0.02mm}\left[1-\mathcal{H}_{196.7}(p_{\mathrm{max}})\right]\hspace{-0.4mm}=\hspace{-0.4mm}\biggl\lbrace\begin{aligned}
\hspace{0.2mm}p_{\mathrm{max}}\;;\; p_{\mathrm{max}}<196.7 \\[-4pt]
\hspace{0.2mm}196.7 \;;\;\hspace{-0.3mm} p_{\mathrm{max}}\geq 196.7
\end{aligned} \;,
\end{equation}
\vspace*{-7.5mm}\\where the \textit{Heaviside} function was used, which shall be defined as follows \cite{WolframHeaviside}:\vspace*{-4.0mm}
\begin{equation}\label{Heaviside}
\mathcal{H}_{b}(x)\,\coloneqq\, \mathcal{H}(x-b)\,=\, \biggl\lbrace\begin{aligned}
\;0\;\;;\;\; x<b \\[-7pt]
\;1 \;\;;\;\; x\geq b
\end{aligned}\;\,.
\end{equation}
\vspace*{-8.0mm}\\Consequently, one finds a last expression for the isotropic and anisotropic runaway region for the considered ITER-scenario, which reads:\vspace*{-3.5mm}
\begin{equation}\label{tilde_minmax_RE_region}
\begin{split}
\begin{gathered}
\xi\in[-1,\,1]\;\wedge\;p\in\left[p_{min},\,\tilde{p}_{max}\right]\;\wedge\;E_{\|}>E_{\mathrm{c}}^{\mathrm{eff}}
 \\
 \xi\in\left[\xi_{\mathrm{sep}}\hspace{-0.7mm}\left(p,\,E_{\mathrm{c}}^{\mathrm{eff}}\right),\,1\right]\;\wedge\;p\in\left[p_{min},\,\tilde{p}_{max}\right]\;\wedge\;E_{\|}>E_{\mathrm{c}}^{\mathrm{eff}}\,.
\end{gathered}
\end{split}
\end{equation}
\vspace*{-7.0mm}\\The pitch-dependent and thus anisotropic runaway region, determined by the \textit{generalized separatrix} \mbox{$\xi_{\mathrm{sep}}\left(p,\,\tilde{E}_{\mathrm{c}}\right)$}, can be viewed in figure \ref{fig_RE_ITER_simu}, for three different times during the disruption depicted in figure \ref{fig_RE_region_pitch}.  \vspace{-0.2mm}
\footnotetext{\label{fig_RE_region_pitch_footnote} The snapshots in the diagram in figure \ref{fig_RE_region_pitch} were produced with the \textsc{MATLAB}-file\\ \hspace*{8.7mm}\qq{\texttt{RE_ht_moments_SV.m}}. The script and its \qq{\texttt{output_RE_ht_moments_SV.txt}} can be\\ \hspace*{8.7mm}found in the digital appendix.}
\vspace{-9.5mm}Note, that the generalized electric field $\tilde{E}_{\mathrm{c}}$ in the defining equation $(\ref{xi_sep_def})$ of the separatix, was set to the \textit{Connor-Hastie} critical electric field $E_{\mathrm{c}}$ for the left diagram, whilst the effective critical electric field $E_{\mathrm{c}}^{\mathrm{eff}}$ was used for the computation of the right plot. At this, one can observe how the anisotropic runaway region grows with the increasing electric field from the start of the runaway generation at approximately \mbox{$4\,\mathrm{ms}$} until the electric field reaches its maximum at \mbox{$8.5\,\mathrm{ms}$}. After that, the runaway region shrinks, which is visible in figure \ref{fig_RE_region_pitch}, if the snapshots of the pitch-dependent runaway region at \mbox{$t=11\,\mathrm{ms}$} and \mbox{$t=15\,\mathrm{ms}$} are compared.

For the subsequent derivations, the following isotropic description of the runaway region shall be introduced, which is given by:\vspace*{-4.0mm} 
\begin{equation}\label{iso_region_RE_HT_def}
\xi\in[-1,\,1]\;\wedge\;p\in\left[p_{low},\,p_{high}\right]\;\wedge\;E_{\|}>\tilde{E}_{\mathrm{c}}\,,
\end{equation} 
\vspace*{-10.0mm}\\so that the mentioned representations $(\ref{CH_RE_region})$, $(\ref{H_RE_region})$, $(\ref{minmax_RE_region})$ and $(\ref{tilde_minmax_RE_region})$ of the pitch-independent runaway region are generalized by this definition. 
\\ 
A similar combination of conditions can be found for the generalized anisotropic or pitch-dependent runaway region:\vspace*{-3.0mm} 
\begin{equation}\label{aniso_region_RE_HT_def}
\xi\in\left[\xi_{\mathrm{sep}}\left(p,\,\tilde{E}_{\mathrm{c}} \right),\,1\right]\;\wedge\;p\in\left[p_{1},\,p_{2}\right]\;\wedge\;E_{\|}>\tilde{E}_{\mathrm{c}}\,.
\end{equation} 
\vspace*{-9.5mm}\\Consequently, one obtains the representations $(\ref{CH_RE_region})$, $(\ref{H_RE_region})$, $(\ref{minmax_RE_region})$ and $(\ref{tilde_minmax_RE_region})$ by inserting the according expressions into the generalized critical electric field $\tilde{E}_{\mathrm{c}}$ and the generalized lower and upper boundary of the momentum magnitude $p_{low}$ and $p_{high}$. At this, it should be noted, that $p_{1}$ and $p_{2}$ are used as shorthand notation throughout this chapter.

\clearpage

\section{Runaway electron density due to the \textit{hot-tail} generation mechanism in the \textit{Smith-Verwichte} approach}\label{ht_n_section}

\subsection{\textit{Hot-tail} runaway electron density for isotropic descriptions of the runaway region}\label{ht_n_iso_subsection}

The \textit{hot-tail} runaway electron density was defined in equation $(\ref{n_RE_HT_def})$ and can be calculated from the isotropic electron distribution function in the \textit{Smith-Verwichte} approach. In this subsection, an isotropic runaway region as defined in $(\ref{iso_region_RE_HT_def})$ is considered. Together with the two-dimensional volume or area element in momentum space from $(\ref{volelem_sphere_2D})$ this leads to the calculation rule:\vspace*{-4.3mm} 
\begin{equation}\label{n_RE_HT_iso_def}
\begin{split}
\begin{gathered}
n_{\mathrm{RE}}^{\mathrm{ht}}(t) = \hspace{-3.05mm}\displaystyle{ \int\limits_{p=p_{1}}^{p_{2}}\int\limits_{\xi=-1}^{1}}\hspace{-0.6mm} f_{RE}^{\textup{ht}}(p,\,t)\,2\pi\,p^2\,\mathrm{d}\xi\, \mathrm{d}p
= \hspace{-2.5mm}\displaystyle{  \int\limits_{p=p_{1}}^{p_{2}}}\hspace{-0.5mm} \dfrac{4\pi\hspace{0.35mm} n_{\mathrm{e}}\hspace{0.35mm}p^2}{\pi^{\frac{3}{2}}\hspace{0.25mm}p_{\mathrm{th},0}^3}\hspace{-0.6mm}\cdot\hspace{-0.3mm}\exp{\hspace{-0.7mm}\left(\hspace{-0.7mm}-\dfrac{\left(p^3+3\hspace{-0.5mm}\cdot\hspace{-0.3mm}\textup{I}_{\tau_{rel}}(t)\right)^{\frac{2}{3}}}{p_{\mathrm{th},0}^2}\hspace{-0.5mm}\right)} \mathrm{d}p
\\[2pt]
= \dfrac{4\,n_{\mathrm{e}} }{\sqrt{\pi}\hspace{0.35mm}p_{\mathrm{th},0}^3} \underbrace{\displaystyle{  \int\limits_{p=p_{1}}^{p_{2}}}\hspace{-0.5mm} p^2\hspace{-0.6mm}\cdot\hspace{-0.3mm}\exp{\hspace{-0.6mm}\left(\hspace{-0.7mm}-\dfrac{\left(p^3+3\hspace{-0.5mm}\cdot\hspace{-0.3mm}\textup{I}_{\tau_{rel}}(t)\right)^{\frac{2}{3}}}{p_{\mathrm{th},0}^2}\hspace{-0.5mm}\right)}  \mathrm{d}p}_{\eqqcolon\,\textup{I}_{n_{\mathrm{RE}}^{\mathrm{ht}}}}=\dfrac{4\,n_{\mathrm{e}} }{\sqrt{\pi}\hspace{0.35mm}p_{\mathrm{th},0}^3} \cdot\textup{I}_{n_{\mathrm{RE}}^{\mathrm{ht}}}\,.
\end{gathered}
\end{split} 
\end{equation} 
\vspace*{-7.3mm}\\The integral $\textup{I}_{n_{\mathrm{RE}}^{\mathrm{ht}}}$ has an analytical solution, which was calculated in subsection \ref{int_n_RE_ht_appendix_subsection} of the appendix and reads:\vspace*{-4mm} 
\begin{equation}\label{n_RE_HT_iso_analyt_sol}
n_{\mathrm{RE}}^{\mathrm{ht}}(t) = \dfrac{2\cdot n_{\mathrm{e}} }{\sqrt{\pi} } \cdot\left[\dfrac{\sqrt{\pi}}{2}\cdot\textup{erf}\left(\varrho(p)\right)-\varrho(p)\cdot\textup{e}^{-\left(\varrho(p)\right)^2}\right]_{\varrho(p_{1})}^{\varrho(p_{2})}\,,
\end{equation} 
\vspace*{-7.3mm}\\where the \textit{error function} $\textup{erf}(z)$ \cite{helander} and the substitution variable:\vspace*{-4.2mm} 
\begin{equation}\label{n_RE_HT_iso_analyt_variable}
\varrho(p)\coloneqq\dfrac{\left(p^3+3\hspace{-0.5mm}\cdot\hspace{-0.3mm}\textup{I}_{\tau_{rel}}(t)\right)^{\frac{1}{3}}}{p_{\mathrm{th},0}} 
\end{equation} 
\vspace*{-7.5mm}\\were introduced. With regard to the representations of the isotropic runaway region in $(\ref{CH_RE_region})$ and $(\ref{H_RE_region})$, one can simplify the result from $(\ref{n_RE_HT_iso_analyt_sol})$ in the limit \mbox{$p_{2}\rightarrow\infty$}. The analysis of the relation $(\ref{n_RE_HT_iso_analyt_variable})$ leads to the insight, that \mbox{$\varrho(p_{2}\rightarrow\infty)\rightarrow\infty$}. Hence, one can write:\vspace*{-5.0mm} 
\begin{equation}\label{n_RE_HT_iso_analyt_sol_inf}
n_{\mathrm{RE}}^{\mathrm{ht}}(t) \overset{{\scriptsize \underbrace{p_{2}\rightarrow\infty}{}}}{=} \dfrac{2\cdot n_{\mathrm{e}} }{\sqrt{\pi} } \cdot\left(\dfrac{\sqrt{\pi}}{2}\cdot\textup{erfc}\left(\varrho(p_{1})\right)+\varrho(p_{1})\cdot\textup{e}^{-\left(\varrho(p_{1})\right)^2}\right)\,,
\end{equation} 
\vspace*{-7.3mm}\\which is also derived in subsection \ref{int_n_RE_ht_appendix_subsection} of the appendix and contains the \textit{complementary error function} $\textup{erfc}(z)$ \cite{WolframERFC}. 

In addition, one can compute the integral $\textup{I}_{n_{\mathrm{RE}}^{\mathrm{ht}}}$ numerically as a control criterion for an numerical integration implementation or it might be used as computation rule for the \textit{hot-tail} runaway electron density by itself. A transformation of the integration domain to the interval \mbox{$w\in[0,\,1]$} is possible, with the help of the substitution:\vspace*{-4.0mm}
\begin{equation}\label{substitution_num_I_HT}
p=p_{1}+\dfrac{w}{1-w} \;;\;\dfrac{\mathrm{d}p }{\mathrm{d}w}= \dfrac{1}{(1-w)^2}\;;\; w\left(p =p_{1}\right)=0\;,\;w(p_{2} \rightarrow\infty)=1\,,
\end{equation} 
\vspace*{-7.0mm}\\for the isotropic runaway regions defined in $(\ref{CH_RE_region})$ and $(\ref{H_RE_region})$. Subsequently, the integral $\textup{I}_{n_{\mathrm{RE}}^{\mathrm{ht}}}$ can be evaluated numerically from:\vspace*{-5.0mm}
\begin{equation}\label{I_num_n_HT}
\textup{I}_{num}^{\hspace{0.25mm}n_{\mathrm{RE}}^{\mathrm{ht}}}=\hspace{-1.5mm} \displaystyle{  \int\limits_{w=0}^{1}}\hspace{-0.6mm} \dfrac{\left(p_{1}\cdot(1-w)+w\right)^2}{(1-w)^4}\hspace{-0.3mm}\cdot\hspace{-0.1mm}\exp{\hspace{-0.6mm}\left(\hspace{-0.7mm}-\dfrac{1}{p_{\mathrm{th},0}^2}\hspace{-0.5mm}\cdot\hspace{-0.3mm}\left(\left(p_{1}+\frac{w}{1-w}\right)^{\hspace{-0.7mm}3}\hspace{-0.5mm}+3\hspace{-0.5mm}\cdot\hspace{-0.3mm}\textup{I}_{\tau_{rel}}(t)\right)^{\hspace{-0.7mm}\frac{2}{3}} \hspace{-0.5mm}\right)}  \mathrm{d}w\,.
\end{equation} 
\vspace*{-7.0mm}\\For the determining conditions $(\ref{minmax_RE_region})$ and $(\ref{tilde_minmax_RE_region})$ of the isotropic runaway region with finite generalized momentum magnitude boundaries $p_{1}$ and $p_{2}$ the different substitution:\vspace*{-5.0mm}
\begin{equation}\label{substitution_num_I_HT_finite}
p=p_{1}+(p_{2}-p_{1})\cdot w \;;\;\dfrac{\mathrm{d}p }{\mathrm{d}w}= p_{2}-p_{1}\;;\; w\left(p =p_{1}\right)=0\;,\;w(p =p_{2} )=1 
\end{equation} 
\vspace*{-7.0mm}\\can be applied, which yields the computation rule:\vspace*{-4.5mm}
\begin{equation}\label{I_num_n_HT_finite}
\begin{split}
\begin{gathered}
\tilde{\textup{I}}_{num}^{\hspace{0.25mm}n_{\mathrm{RE}}^{\mathrm{ht}}}= \hspace{-1.6mm}\displaystyle{  \int\limits_{w=0}^{1}}\hspace{-0.5mm} (p_{2}-p_{1})\cdot(p_{1}+(p_{2}-p_{1})\cdot w)^2;\times
\\[1pt]
\times\,\exp{\left(-\frac{1}{p_{\mathrm{th},0}^2}\hspace{-0.5mm}\cdot\hspace{-0.3mm}\left((p_{1}+(p_{2}-p_{1})\cdot w)^3+3 \cdot \textup{I}_{\tau_{rel}}(t)\right)^{\frac{2}{3}} \right) }  \mathrm{d}w\,.
\end{gathered}
\end{split} 
\end{equation}

\subsection{\textit{Hot-tail} runaway electron density for anisotropic descriptions of the runaway region}\label{ht_n_aniso_subsection}

The calculation rule $(\ref{n_RE_HT_iso_def})$ from the previous subsection contains the integration of the \textit{Smith-Verwichte} electron distribution function over the maximum interval for the pitch coordinate \mbox{$\xi\in[-1,\,1]$}. This can be thought of as the assumption, that electrons with an arbitrary pitch angle will runaway, if the parallel component of electric field, with respect to the local magnetic field as depicted in figure \ref{fig_mom_coord}, is greater than the critical electric field and the parallel momentum component exceeds the critical momentum. Hence, also electrons of the considered isotropic distribution function, with a momentum in a mainly anti-parallel direction to the local magnetic field or with a momentum, which is solely orthogonal to the magnetic field vector, will be classified as runaway electrons. However, this is not physically accurate, due to the fact, that electrons with a pitch coordinate \mbox{$\xi<\xi_{sep}(p,\,\tilde{E}_{c})$} will not accelerate towards ultra-relativistic velocities, so that they stay in the thermal electron population. In the subsection \ref{pitch_RE_region_subsection}, this pitch-anisotropy of the runaway region was introduced together with the generalized separatrix \mbox{$\xi_{sep}(p,\,\tilde{E}_{c})$}, which depends on the momentum magnitude variable $p$ and the generalized critical electric field $\tilde{E}_{c}$ and is determined by the relation $(\ref{xi_sep_def})$. With this, one was able to state the general combination of conditions $(\ref{aniso_region_RE_HT_def})$ for the pitch-dependent runaway region, allowing oneself to derive a calculation rule for the \textit{hot-tail} runaway electron density, under consideration of the anisotropic two-dimensional runaway region. Here, one might comment, that the distribution function from $(\ref{SV_ht_dist_func})$ is strictly positive for its whole domain of definition \mbox{$p\in\mathbb{R}^{+}$}. Consequently, the anisotropic runaway region, which is smaller than the isotropic runaway region, will decrease the zeroth moment and hence the runaway density, due to the fact, that it is defined as the integral over a positive product of the momentum space volume element and the distribution function. Thus, the general definition $(\ref{n_RE_HT_def})$ of the \textit{hot-tail} runaway electron density, the two-dimensional volume element in momentum space from $(\ref{volelem_sphere_2D})$ and the distribution function in the \textit{Smith-Verwichte} approach, as written in the paper \cite{Svenningsson2020} by \textit{I.\hspace{0.9mm}Svenningsson}, determine the following formula for the seed runaway electron density for a pitch-dependent runaway region, satisfying the subsequent condition $(\ref{aniso_region_RE_HT_def})$:\vspace*{-7.5mm} 
\begin{equation}\label{n_RE_HT_aniso_def}
\begin{split}
\begin{gathered}
n_{\mathrm{RE}}^{\mathrm{ht},\xi}(t) = \hspace{-3.05mm}\displaystyle{\int\limits_{p=p_{1}}^{p_{2}}\int\limits_{\xi=\xi_{sep}(p,\,\tilde{E}_{c})}^{1} }\hspace{-0.6mm} f_{RE}^{\textup{ht}}(p,\,t)\,2\pi\,p^2\,\mathrm{d}\xi\, \mathrm{d}p 
\\[-5pt]
= \dfrac{ 2\pi\,n_{\mathrm{e}} }{ \pi^{\frac{2}{3}}\hspace{0.35mm}p_{\mathrm{th},0}^3}  \displaystyle{  \int\limits_{p=p_{1}}^{p_{2}}}\hspace{-1.5mm} \left(1-\xi_{sep}\xi_{sep}\hspace{-0.7mm}\left(p,\,\tilde{E}_{c}\right)\right)\hspace{-0.6mm}\cdot\hspace{-0.3mm}p^2\hspace{-0.6mm}\cdot\hspace{-0.3mm}\exp{\hspace{-0.6mm}\left(\hspace{-0.7mm}-\dfrac{\left(p^3+3\hspace{-0.5mm}\cdot\hspace{-0.3mm}\textup{I}_{\tau_{rel}}(t)\right)^{\frac{2}{3}}}{p_{\mathrm{th},0}^2}\hspace{-0.5mm}\right)}  \mathrm{d}p
\\[-1pt]
\overset{(\ref{xi_sep_def})}{=}\hspace{-0.9mm} \dfrac{ 4 \,n_{\mathrm{e}} }{ \sqrt{\pi}\hspace{0.35mm}p_{\mathrm{th},0}^3} \underbrace{\displaystyle{  \int\limits_{p=p_{1}}^{p_{2}}}\hspace{-1.75mm} \left(\hspace{-0.5mm}1\hspace{-0.3mm}-\hspace{-0.3mm}\dfrac{\tilde{E}_{c}}{p^2\hspace{0.25mm}E_{\|}}\hspace{-0.5mm}\right)\hspace{-0.8mm}\cdot\hspace{-0.3mm}p^2\hspace{-0.7mm}\cdot\hspace{-0.3mm}\exp{\hspace{-0.6mm}\left(\hspace{-0.9mm}-\dfrac{\left(p^3+3\hspace{-0.5mm}\cdot\hspace{-0.3mm}\textup{I}_{\tau_{rel}}(t)\right)^{\frac{2}{3}}}{p_{\mathrm{th},0}^2}\hspace{-0.5mm}\right)}  \mathrm{d}p}_{\eqqcolon\,\textup{I}_{n_{\mathrm{RE}}^{\mathrm{ht},\xi}}}
=\dfrac{4\,n_{\mathrm{e}} }{\sqrt{\pi}\hspace{0.35mm}p_{\mathrm{th},0}^3} \hspace{-0.6mm}\cdot\hspace{-0.4mm}\textup{I}_{n_{\mathrm{RE}}^{\mathrm{ht},\xi}}\,.
\end{gathered}
\end{split} 
\end{equation} 
\vspace*{-7.3mm}\\Note, that the absolute value of the electric field component in parallel to the local magnetic field \mbox{$E_{\|}\coloneqq \vert E_{\|}\vert$} was chosen as a determining parameter of the separatrix.

An analytic solution of the appearing integral $\textup{I}_{n_{\mathrm{RE}}^{\mathrm{ht},\xi}}$ has not been found, although a computation is nevertheless possible with a one-dimensional numerical integration. Hence, a preparation for the application of standard quadrature formulas is appropriate. For this purpose, the transformation of the integration domain, for the anisotropic runaway region with a half-open momentum interval \mbox{$p\in[p_{1},\,\infty)$}, to the interval \mbox{$w\in[0,\,1]$} should be applied, by means of the substitution $(\ref{substitution_num_I_HT})$. Subsequently, the integral $\textup{I}_{n_{\mathrm{RE}}^{\mathrm{ht},\xi}}$ can be evaluated numerically, for the different descriptions of the pitch-dependent runaway region defined in $(\ref{CH_RE_region})$ and $(\ref{H_RE_region})$, with the integral expression:\vspace*{-3.3mm}
\begin{equation}\label{I_num_n_HT_sep}
\begin{split}
\begin{gathered}
\textup{I}_{num}^{\hspace{0.25mm}n_{\mathrm{RE}}^{\mathrm{ht},\xi}}=\hspace{-1.2mm} \displaystyle{  \int\limits_{w=0}^{1}}\hspace{-0.5mm} \dfrac{\left(p_{1}\cdot(1-w)+w\right)^2-\frac{\tilde{E}_{c}}{ E_{\|}}\cdot(1-w)^2}{(1-w)^4} \;\times
\\[1pt]
\times\,\exp{\hspace{-0.6mm}\left( \hspace{-0.6mm}-\frac{1}{p_{\mathrm{th},0}^2}\cdot\left(\left(p_{1}+\frac{w}{1-w}\right)^3+3 \cdot\textup{I}_{\tau_{rel}}(t)\right)^{\frac{2}{3}}\right)  }  \mathrm{d}w\,.
\end{gathered}
\end{split}
\end{equation} 
\vspace*{-6.3mm}\\For the conditions $(\ref{minmax_RE_region})$ and $(\ref{tilde_minmax_RE_region})$ representing the anisotropic runaway region with finite generalized momentum magnitude boundaries $p_{1}$ and $p_{2}$ the different substitution $(\ref{substitution_num_I_HT_finite})$ can be utilized, which yields to the integral definiton:\vspace*{-4.5mm}
\begin{equation}\label{I_num_n_HT_finite_sep}
\begin{split}
\begin{gathered}
\tilde{\textup{I}}_{num}^{\hspace{0.25mm}n_{\mathrm{RE}}^{\mathrm{ht},\xi}}= \hspace{-1.6mm}\displaystyle{  \int\limits_{w=0}^{1}}\hspace{-0.5mm} (p_{2}-p_{1})\hspace{-0.6mm}\cdot\hspace{-0.6mm}\left(\hspace{-0.5mm}(p_{1}\hspace{-0.2mm}+\hspace{-0.2mm}(p_{2}\hspace{-0.2mm}-\hspace{-0.2mm}p_{1})\hspace{-0.4mm}\cdot\hspace{-0.3mm} w)^2 \hspace{-0.4mm}-\hspace{-0.2mm}\dfrac{\tilde{E}_{c}}{E_{\|}}\hspace{-0.4mm}\right)\;\times
\\[2pt]
\times\,\exp{\left(-\frac{\left((p_{1}+(p_{2}-p_{1})\cdot w)^3+3 \cdot\textup{I}_{\tau_{rel}}(t)\right)^{\frac{2}{3}}}{p_{\mathrm{th},0}^2} \right) } \; \mathrm{d}w\,.
\end{gathered}
\end{split}
\end{equation}

\subsection{Computation and evaluation of the \textit{hot-tail} runaway electron density}\label{n_RE_ht_eval_comp_section}

The formulas for the zeroth moment of the \textit{Smith-Verwichte} distribution function for the consideration of an isotropic or anisotropic runaway region from the subsections \ref{ht_n_iso_subsection} and \ref{ht_n_aniso_subsection} allow the computation of the hot-tail runaway electron density and shall now be validated on the basis of the simulation of a disruption for an ITER-scenario. As explained in the beginning of this \hyperref[hot_tail_chapter]{third chapter}, the simulation covers the thermal and current quench phase for a total plasma current of \mbox{$I_{p}=15\;\mathrm{MA}$}, a magnetic field of \mbox{$B=5.3\,\mathrm{T}$} and a time-independent electron density of \mbox{$n_{e}=1.06\cdot10^{20}\;\mathrm{m}^{-3}$} and is accompanied by the paper \cite{Smith_2009}. The time evolution of the main physical quantities was displayed in figure \ref{fig_RE_ITER_simu} and in addition the results for different models of the boundaries of the runaway electron generation region with respect to the momentum magnitude can be viewed in figure \ref{fig_RE_mom_bound}. 

In order to receive the data for the \textit{hot-tail} runaway electron density two \textsc{MATLAB}-\linebreak implementations$^{\ref{fig_n_ht_RE_footnote},\ref{fig_n_ht_RE_sep_footnote}}$ are used. The first script$^{\ref{fig_n_ht_RE_footnote}}$ corresponds to the isotropic description of the runaway region and produces results for the \textit{hot-tail} runaway electron density for the four isotropic representations of the runaway region $(\ref{CH_RE_region})$, $(\ref{H_RE_region})$, $(\ref{minmax_RE_region})$ and $(\ref{tilde_minmax_RE_region})$. At that, the simplified analytic formula $(\ref{n_RE_HT_iso_analyt_sol_inf})$ is used for the first two representations, which intrinsically contain an infinite upper momentum boundary, while the more general analytic expression $(\ref{n_RE_HT_iso_analyt_sol})$ was applied for the other two descriptions. The required function evaluations of the error and the complementary error function are provided by the \textsc{MATLAB}-routines \qq{\texttt{erf}} \cite{erfMATLAB} and \qq{\texttt{erfc}} \cite{erfcMATLAB}. The input data from the ITER-simulation are the time evolutions of the electron temperature and the parallel component of the electric field as displayed in figure \ref{fig_RE_ITER_simu}, which appear at the radius \mbox{$r_{\perp}=0.75\,\mathrm{m}$}, because the simulation is based on a one-dimensional cylindrical plasma model, so that the solutions are the time evolutions of radial profiles of the physical quantities \cite{Smith_2009}. The associated output of the \textsc{MATLAB}-\linebreak implementations$^{\ref{fig_n_ht_RE_footnote},\ref{fig_n_ht_RE_sep_footnote}}$ can be found in the listings \labelcref{outMATLABoutput_RE_ht_moments_SV,outMATLABoutput_RE_ht_moments_SV_imp,outMATLABoutput_RE_ht_moments_SV_sep,outMATLABoutput_RE_ht_moments_SV_sep_imp} in subsection \ref{output_matlab_appendix_subsection} of the\vspace{-17cm}\linebreak\newpage\noindent
\begin{figure}[H]
\centering
  \subfloat[][Time-dependent quantities at \mbox{$r_{\perp}=0.75\,\mathrm{m}$} for a constant electron density of \mbox{$n_{e}=n_{_{1}^{2}\mathrm{H}^{+}}=1.06\cdot 10^{20}\,\mathrm{m}^{-3}$} within a singly-ionized deuterium plasma with \mbox{$B=5.3\,\textup{T}$} and \mbox{$Z_{eff}=1.0$}.]{\label{n_ht_RE_main} 
   \includegraphics[trim=82 18 120 25,width=0.90\textwidth,clip]
    {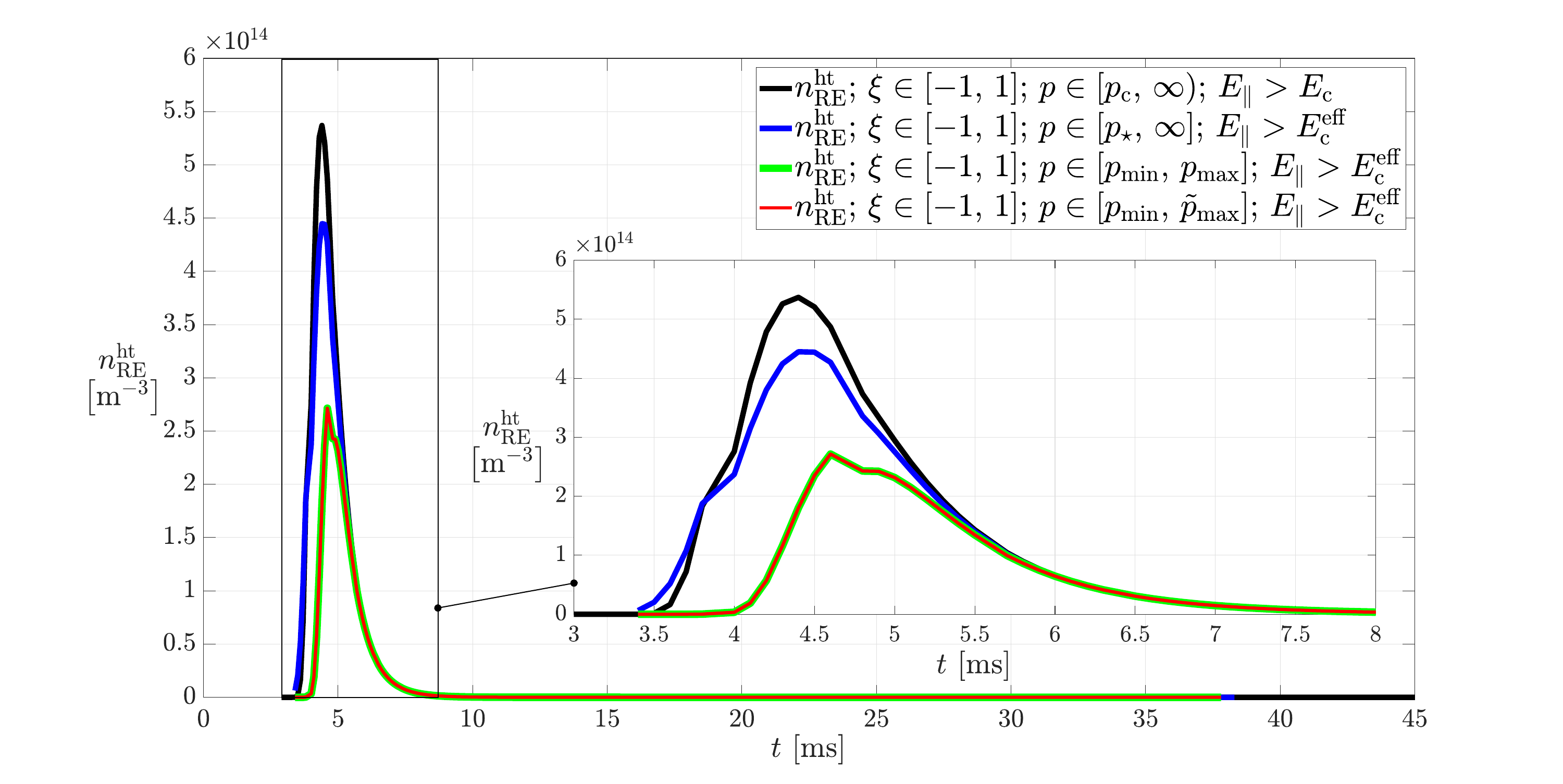}}\\[7pt]
  \subfloat[][Time-dependent quantities at \mbox{$r_{\perp}=0.75\,\mathrm{m}$} for a constant singly-ionized deuterium density of \mbox{$n_{_{1}^{2}\mathrm{H}^{+}}=1.06\cdot 10^{20}\,\mathrm{m}^{-3}$} within a plasma with \mbox{$B=5.3\,\textup{T}$} and \mbox{$Z_{eff}=1.0$}, in the presence of a singly-ionized neon impurity density of \mbox{$n_{_{10}^{20}\mathrm{Ne}^{+}}=2.40\cdot 10^{20}\,\mathrm{m}^{-3}$}.]{\label{n_ht_RE_imp_main}  
    \includegraphics[trim=95 16 120 25,width=0.90\textwidth,clip]
    {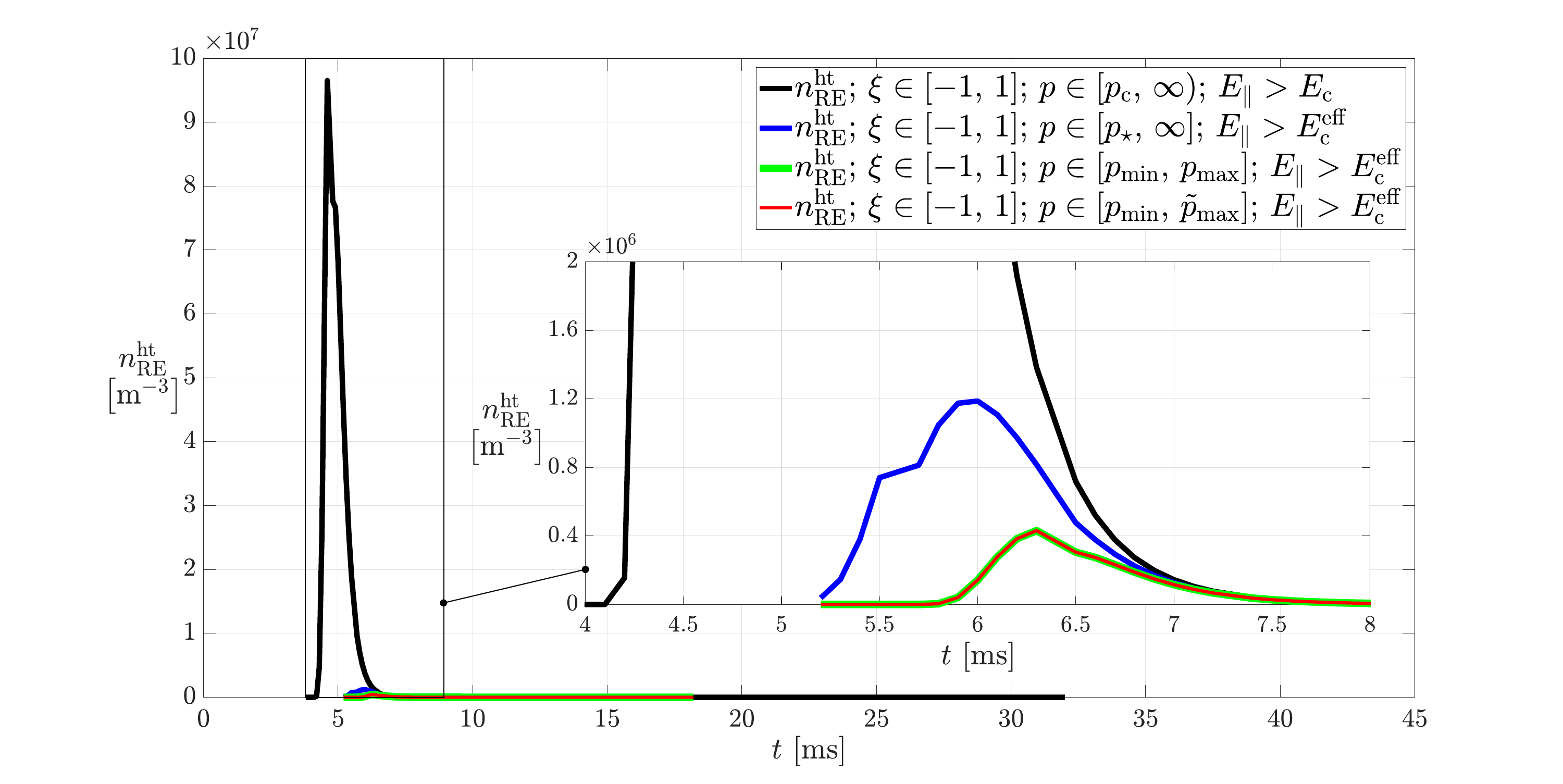}}
\caption[Time-evolution of the \textit{hot-tail} runaway electron density \mbox{$n^{\mathrm{ht}}_{\mathrm{RE}}$}, during an ITER-like disruption (see also figure \ref{fig_RE_ITER_simu}, \cite{Smith_2009}) in the \textit{Smith-Verwichte} model in combination with four different isotropic descriptions of the runaway region with respect to the relativistic momentum magnitude $p$.]{Time-evolution\protect\footnotemark{} of the \textit{hot-tail} runaway electron density \mbox{$n^{\mathrm{ht}}_{\mathrm{RE}}$}, during an ITER-like disruption (see also figure \ref{fig_RE_ITER_simu}, \cite{Smith_2009}) in the \textit{Smith-Verwichte} model in combination with four different isotropic descriptions of the runaway region with respect to the relativistic momentum magnitude $p$.}
\label{fig_n_ht_RE}
\end{figure} 
\footnotetext{\label{fig_n_ht_RE_footnote} The displayed results were generated by means of the \textsc{MATLAB}-implementations\\ \hspace*{8.7mm}\qq{\texttt{RE_ht_moments_SV.m}} and \qq{\texttt{RE_ht_moments_SV_imp.m}}, which utilize the script\\ \hspace*{8.7mm}\qq{\texttt{calculate_E_c_eff.m}} \cite{Hesslow_2018} and the data \qq{\texttt{ITER_data.txt}} in accordance with\\ \hspace*{8.7mm}reference \cite{Smith_2009}. The associated console output is stored in the files\\ \hspace*{8.7mm}\qq{\texttt{output_RE_ht_moments_SV.txt}} and \qq{\texttt{output_RE_ht_moments_SV_imp.txt}} in\\ \hspace*{8.7mm}the digital appendix.}
\noindent\newpage\noindent
appendix. With those, one can also verify the implementation on the basis of the analytical and numerical computation of the zeroth moment, due to the analytic expressions $(\ref{n_RE_HT_iso_analyt_sol})$ and $(\ref{n_RE_HT_iso_analyt_sol_inf})$, as well as from the numerical integral formulas $(\ref{I_num_n_HT})$ and $(\ref{I_num_n_HT_finite})$, for the isotropic models of the runaway region and with the computation rules $(\ref{I_num_n_HT_sep})$ and $(\ref{I_num_n_HT_finite_sep})$ for the pitch-angle dependent runaway region.

The results of the \textsc{MATLAB}-computations for the isotropic descriptions of the runaway region are visualized in the figure \ref{fig_n_ht_RE}. At this, the subfigure \ref{n_ht_RE_main} depicts the results for the ITER-simulation as presented in the reference \cite{Smith_2009} for a constant electron density of \mbox{$n_{e}=n_{_{1}^{2}\mathrm{H}^{+}}=1.06\cdot 10^{20}\,\mathrm{m}^{-3}$} within a singly-ionized deuterium plasma. In contrast, the presence of a singly-ionized neon impurity density of \mbox{$n_{_{10}^{20}\mathrm{Ne}^{+}}=2.26\cdot n_{e}$} leads to the time evolution of the \textit{hot-tail} runaway electron density plotted in the subfigure \ref{n_ht_RE_imp_main}.

An estimation for the order of magnitude for the \textit{hot-tail} runaway electron density can be found in the paper \cite{hottailREdistfunc} by \textit{H.\hspace{0.7mm}M.\hspace{0.8mm}Smith} and \textit{E.\hspace{0.8mm}Verwichte}, where the runaway density fractions with $10^{-8} \lesssim n_{RE}^{\mathrm{ht}}/n_{e}\lesssim 10^{-3}$ were calculated from the distribution function $(\ref{f_RE_hot_tail})$, which is similar to the momentum representation $(\ref{SV_ht_dist_func})$ used in \hyperref[hot_tail_chapter]{this chapter}. Hence, the developing \textit{hot-tail} runaway seed density seems to be plausible. Furthermore, one notices that the \textit{hot-tail} runaway electrons are produced between \mbox{$3.5\,\mathrm{ms}$} and \mbox{$7.5\,\mathrm{ms}$} for the case without the presence of an impurity density. This is in well correspondence to the end of the thermal and the start of the current quench, where the electron temperature drops, whilst the electric field simultaneously increases, which can be viewed in figure \ref{fig_RE_ITER_simu}. In addition, one observes that the production of the \textit{hot-tail} runaway electrons is predicted lower for the descriptions of the runaway region with a finite upper momentum, where it is apparent, that the choice $\tilde{p}_{max}$ gives equivalent results in comparison to the computational more expensive upper momentum boundary $p_{max}$. For the effective critical momentum $p_{\star}$ one notices also a lower runaway density, due to the \textit{hot-tail} generation mechanism, than for the application of the \textit{Connor-Hastie} critical momentum $p_{c}$ as the lower boundary of the isotropic runaway region, which was expected, since $p_{\star}$ is mostly larger, as it can be recognized in the figures \ref{fig_p_comparison} and \ref{fig_RE_mom_bound} from section \ref{part_screen_section}. Also it was expected, that the two descriptions $(\ref{minmax_RE_region})$ and $(\ref{tilde_minmax_RE_region})$ lead to a smaller density than the representations $(\ref{CH_RE_region})$ and $(\ref{H_RE_region})$ of the pitch-independent runaway region, because the further decrease the integration domain of the zeroth moment of a positive function and respectively its interpretation as the \textit{hot-tail} runaway electron density. Note, that the runaway generation starts later for the two models with the lower momentum $p_{min}$, because it is larger than the other approximations of the lower runaway momentum until $t\approx5\,\mathrm{ms}$, which can be verified as well in the enlarged area of the diagram of the figure \ref{fig_RE_mom_bound}.
\\
The presence of a neon impurity produces results for the runaway seed density, which\vspace{-17cm}\linebreak\newpage\noindent
\begin{figure}[H]
\centering
  \subfloat[][Time-dependent quantities at \mbox{$r_{\perp}=0.75\,\mathrm{m}$} for a constant electron density of \mbox{$n_{e}=n_{_{1}^{2}\mathrm{H}^{+}}=1.06\cdot 10^{20}\,\mathrm{m}^{-3}$} within a singly-ionized deuterium plasma with \mbox{$B=5.3\,\textup{T}$} and \mbox{$Z_{eff}=1.0$}.]{\label{n_ht_RE_sep_main} 
    \includegraphics[trim=82 17 119 23,width=0.9\textwidth,clip]
    {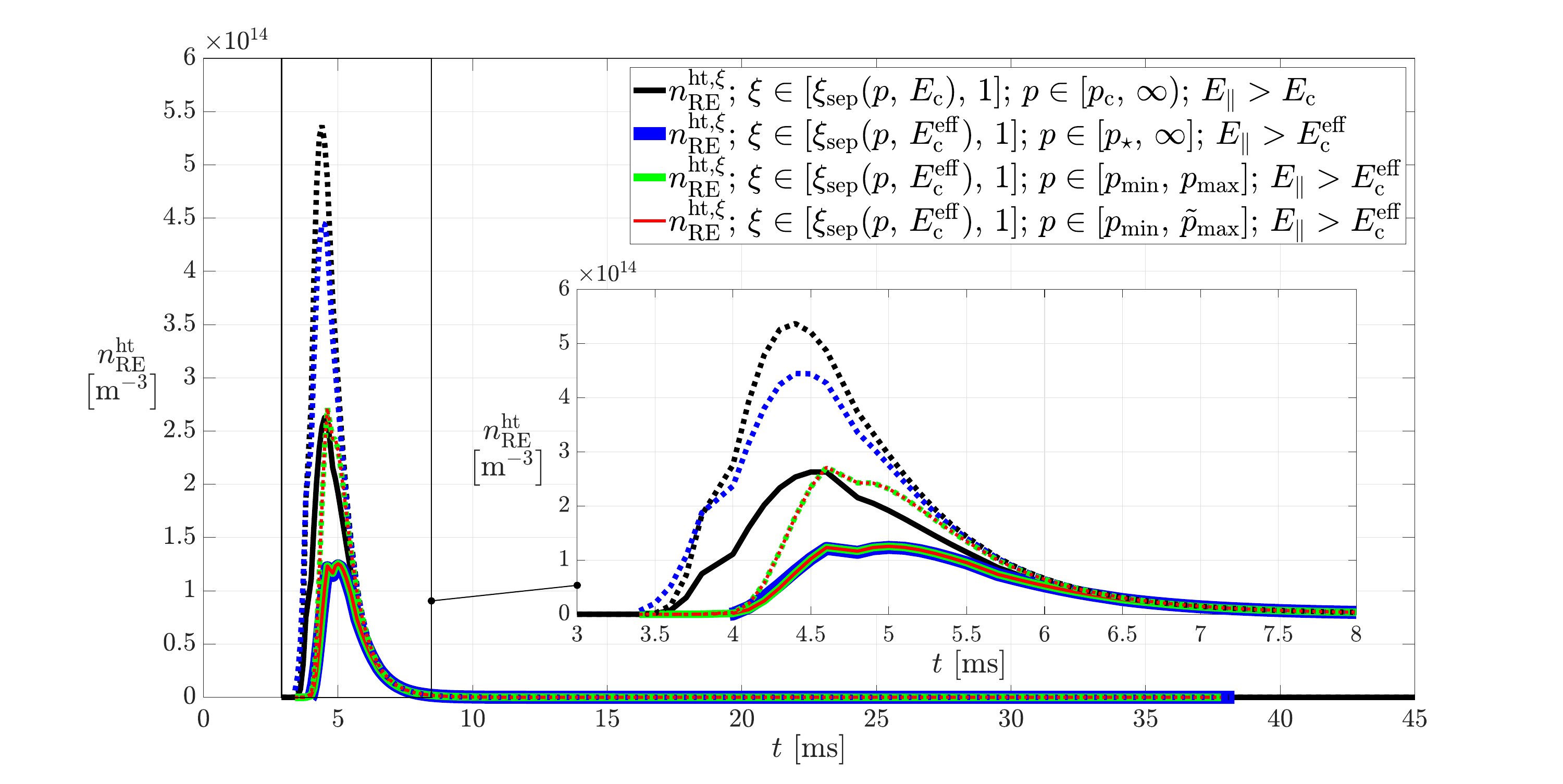}}\\[7pt]
  \subfloat[][Time-dependent quantities at \mbox{$r_{\perp}=0.75\,\mathrm{m}$} for a constant singly-ionized deuterium density of \mbox{$n_{_{1}^{2}\mathrm{H}^{+}}=1.06\cdot 10^{20}\,\mathrm{m}^{-3}$} within a plasma with \mbox{$B=5.3\,\textup{T}$} and \mbox{$Z_{eff}=1.0$}, in the presence of a singly-ionized neon impurity density of \mbox{$n_{_{10}^{20}\mathrm{Ne}^{+}}=2.40\cdot 10^{20}\,\mathrm{m}^{-3}$}.]{\label{n_ht_RE_imp_sep_main}  
    \includegraphics[trim=97 18 121 23,width=0.9\textwidth,clip]
    {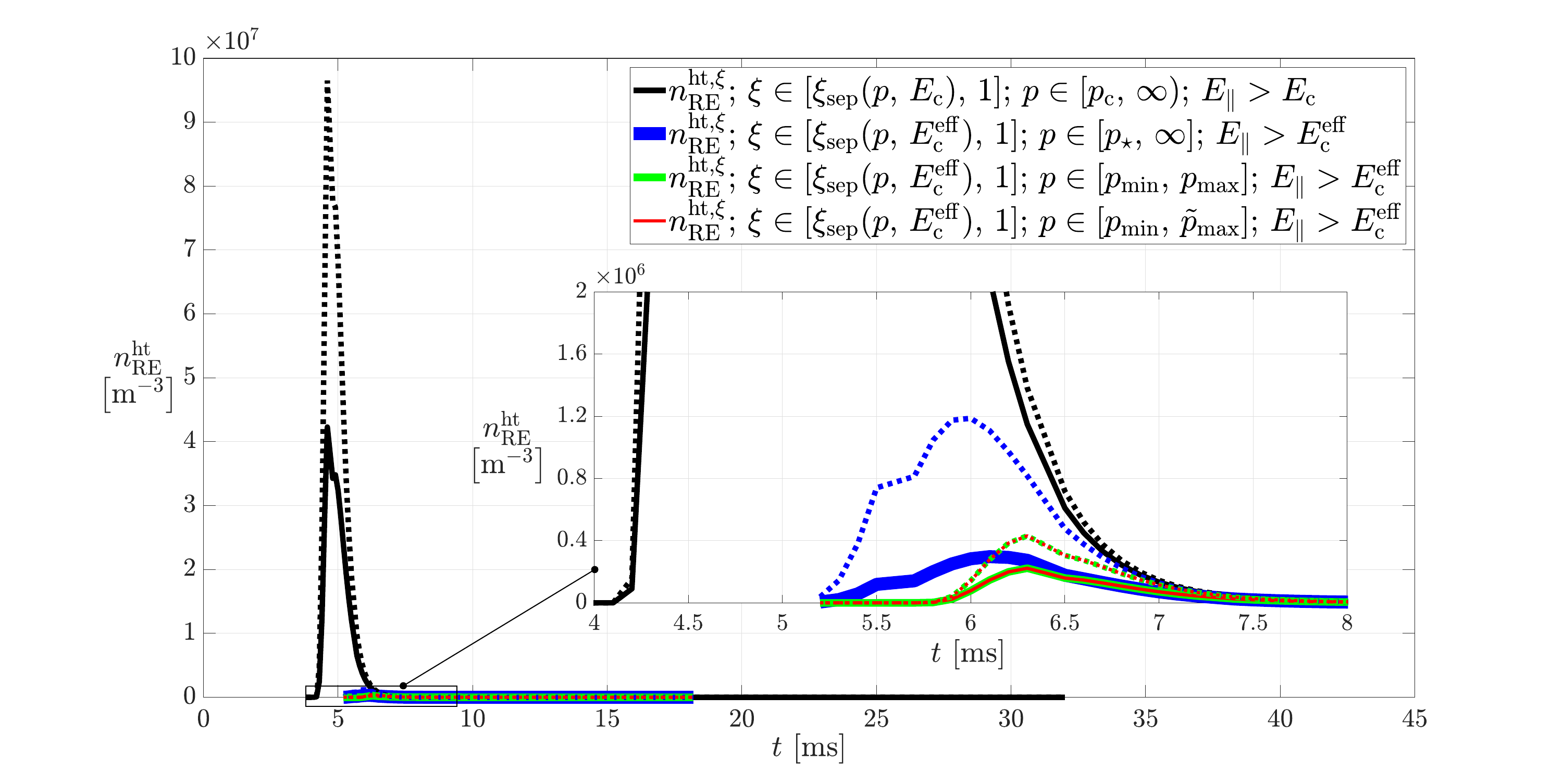}} 
    \caption[Time-evolution of the \textit{hot-tail} runaway electron density \mbox{$n^{\mathrm{ht},\xi}_{\mathrm{RE}}$} in the \textit{Smith-Verwichte} model for with four different pitch-dependent descriptions of the runaway region (solid lines) and for the corresponding isotropic representations (dotted lines, figure \ref{fig_n_ht_RE}), during an ITER-like disruption (see also figure \ref{fig_RE_ITER_simu}, \cite{Smith_2009}).]{Time-evolution\protect\footnotemark{} of the \textit{hot-tail} runaway electron density \mbox{$n^{\mathrm{ht},\xi}_{\mathrm{RE}}$} in the \textit{Smith-Verwichte} model for with four different pitch-dependent descriptions of the runaway region (solid lines) and for the corresponding isotropic representations (dotted lines, figure \ref{fig_n_ht_RE}), during an ITER-like disruption (see also figure \ref{fig_RE_ITER_simu}, \cite{Smith_2009}).}
\label{fig_n_ht_RE_sep}
\end{figure} \vspace{-6mm}
\footnotetext{\label{fig_n_ht_RE_sep_footnote} The displayed results were generated by means of the \textsc{MATLAB}-implementations\\ \hspace*{8.7mm}\qq{\texttt{RE_ht_moments_SV_sep.m}} and \qq{\texttt{RE_ht_moments_SV_sep_imp.m}}, which make use\\ \hspace*{8.7mm}of the script \qq{\texttt{calculate_E_c_eff.m}} \cite{Hesslow_2018} and the data \qq{\texttt{ITER_data.txt}} in accordance\\ \hspace*{8.7mm}with reference \cite{Smith_2009}. The associated console output is stored in the files\\ \hspace*{8.7mm}\qq{\texttt{output_RE_ht_moments_SV_sep.txt}} and\\ \hspace*{8.7mm}\qq{\texttt{output_RE_ht_moments_SV_sep_imp.txt}} in the digital appendix.}
\noindent\newpage\noindent
are smaller than the results for the pure deuterium case approximately by a factor of $10^7$. Furthermore, a distinct delay in the starting time of the \textit{hot-tail} runaway electron generation can be seen in the subfigure \ref{n_ht_RE_main} for the isotropic runaway region models $(\ref{H_RE_region})$, $(\ref{minmax_RE_region})$ and $(\ref{tilde_minmax_RE_region})$, which take into account the effects of partial screening. Their more accurate generated data for the \textit{hot-tail} runaway electron density is roughly smaller by a factor of ten. Hence, one can deduce, that those three models describe disruptions with impurity injection more accurate and might avoid an overestimation, as it often occurs for the \textit{Connor-Hastie} runaway region $(\ref{CH_RE_region})$.

The results of the \textsc{MATLAB}-computations for the anisotropic descriptions of the runaway region are shown in the figure \ref{fig_n_ht_RE_sep}, where the subfigure \ref{n_ht_RE_sep_main} depicts the results for the ITER-simulation as presented in the reference \cite{Smith_2009} for a constant electron density of \mbox{$n_{e}=n_{_{1}^{2}\mathrm{H}^{+}}=1.06\cdot 10^{20}\,\mathrm{m}^{-3}$} within a singly-ionized deuterium plasma. In contrast, the \textit{hot-tail} runaway electron density is plotted for the presence of a singly-ionized neon impurity density of \mbox{$n_{_{10}^{20}\mathrm{Ne}^{+}}=2.26\cdot n_{e}$} in the subfigure \ref{n_ht_RE_imp_main}. It has to be remarked, that non-physical negative values appear for the calculation of the \textit{hot-tail} runaway electron density, for the pure deuterium plasma scenario, with the anisotropic description $(\ref{H_RE_region})$ of the runaway region, according to the minimum of the according data, which is stated in the listing \ref{outMATLABoutput_RE_ht_moments_SV_sep}. The reason for this is, that the time evolution of the electric field from the considered ITER-simulation does not take into account the effect of partial screening and radiation \cite{Smith_2009}.

Generally, one can state, that the anisotropy of the runaway region leads to \textit{hot-tail} runaway electron densities, which are approximately smaller, compared to an isotropic description of the runaway region, by a factor of two. This can be specified to relative deviations of approximately $-57\,\%$ for the calculation scheme using $p_{\star}$ and to relative differences of roughly $-33\,\%$ for the other three calculation rules and is presented in detail in the listings \labelcref{outMATLABoutput_RE_ht_moments_SV_sep,outMATLABoutput_RE_ht_moments_SV_sep_imp} in subsection \ref{output_matlab_appendix_subsection} of the appendix. Otherwise, the same deductions as for figure \ref{fig_n_ht_RE} hold. However, it is remarkable, that the results for the pure deuterium plasma in subfigure \ref{n_ht_RE_imp_sep_main}, are similar for the three representations $(\ref{H_RE_region})$, $(\ref{minmax_RE_region})$ and $(\ref{tilde_minmax_RE_region})$ of the runaway region, which take the effects of partial screening into account. Only for the case with a neon impurity in subfigure \ref{n_ht_RE_imp_sep_main}, a large deviation occurs in between the models $(\ref{H_RE_region})$, $(\ref{minmax_RE_region})$ and $(\ref{tilde_minmax_RE_region})$. In the context of this scenario, one again ascertains the significant difference in the predictions of the magnitude and the time behaviour of the \textit{hot-tail} runaway electron density in particular in comparison of the mentioned three models to the \textit{Connor-Hastie} runaway region $(\ref{CH_RE_region})$.

\clearpage

\section{Current density of a \textit{hot-tail} runaway electron population in the \textit{Smith-Verwichte} approach}\label{ht_j_section}

\subsection{\textit{Hot-tail} runaway electron current density the isotropic descriptions of the runaway region}\label{ht_j_iso_subsection}

The \textit{hot-tail} runaway electron current density $j_{\mathrm{RE}}^{\mathrm{ht}}$ was defined in equation $(\ref{RE_HT_curr_dens_def})$ as the product of the density with the mean velocity of a \textit{hot-tail} runaway electron population scaled with the elementary charge. Thus, it can be calculated from the isotropic electron distribution function in the \textit{Smith-Verwichte} approach, due to the fact that the runaway electron density and the mean velocity can be expressed as moments of this function as shown in the equations $(\ref{n_RE_HT_def})$ and $(\ref{u_RE_HT_def})$. Calculation rules for the \textit{hot-tail} runaway electron density were derived, evaluated for a typical disruption in an ITER-setup and therefore physically validated based on this example in section \ref{ht_n_section}. As a consequence, the computation of the mean velocity has to be regarded, in order to be able to obtain results, which are easy to evaluate and concurrently allow the calculation of the current density of the \textit{hot-tail} runaway electrons. 
\\
In this subsection, an isotropic runaway region as defined in $(\ref{iso_region_RE_HT_def})$ is considered with the aim to determine a first formula for the mean velocity related to a \textit{hot-tail} runaway electron density. From this one starts with the definition of the mean velocity $(\ref{u_RE_HT_def})$ for the considered pitch-independent runaway region. Furthermore, the \textit{Smith-Verwichte} distribution function $(\ref{SV_ht_dist_func})$ and the two-dimensional volume or area element in momentum space from $(\ref{volelem_sphere_2D})$ are utilized together with the momentum representation of the velocity magnitude $v$ from the expression $(\ref{p_norm_gamma_def})$. By means of this preparatory work, a calculation rule for the mean velocity normalized to the speed of light can be deduced as follows:\vspace*{-6.0mm} 
\begin{equation}\label{u_RE_HT_iso_def}
\begin{split}
\begin{gathered}
\frac{u_{\mathrm{RE}}^{\mathrm{ht}}(t)}{c} =  \dfrac{1}{n_{\mathrm{RE}}^{\mathrm{ht}}(t)}\,\displaystyle{ \int\limits_{p=p_{1}}^{p_{2}}\int\limits_{\xi=-1}^{1}}\hspace{-0.6mm}\dfrac{v}{c}\hspace{-0.6mm}\cdot\hspace{-0.3mm} f_{RE}^{\textup{ht}}(p,\,t)\,2\pi\,p^2\,\mathrm{d}\xi\, \mathrm{d}p
\\[2pt]
=  \dfrac{4\pi}{n_{\mathrm{RE}}^{\mathrm{ht}}(t)}\,\displaystyle{ \int\limits_{p=p_{1}}^{p_{2}} }\hspace{-0.6mm}\dfrac{p^3}{\sqrt{1+p^2}}\hspace{-0.6mm}\cdot\hspace{-0.3mm} f_{RE}^{\textup{ht}}(p,\,t)\,   \mathrm{d}p
\\[-3pt]
=   \dfrac{4 \hspace{0.35mm} n_{\mathrm{e}} }{\sqrt{\pi}\hspace{0.35mm}p_{\mathrm{th},0}^3 \hspace{0.25mm}n_{\mathrm{RE}}^{\mathrm{ht}}(t)}\,\underbrace{\displaystyle{ \int\limits_{p=p_{1}}^{p_{2}} }\hspace{-0.8mm}\dfrac{p^3}{\sqrt{1+p^2}}\hspace{-0.6mm}\cdot\hspace{-0.3mm}\exp{\hspace{-0.7mm}\left(\hspace{-0.7mm}-\dfrac{\left(p^3+3\hspace{-0.5mm}\cdot\hspace{-0.3mm}\textup{I}_{\tau_{rel}}(t)\right)^{\frac{2}{3}}}{p_{\mathrm{th},0}^2}\hspace{-0.5mm}\right)} \mathrm{d}p}_{\eqqcolon\,\textup{I}_{u_{\mathrm{RE}}^{\mathrm{ht}}}} 
\\[-5pt]
= \dfrac{4 \hspace{0.35mm} n_{\mathrm{e}}}{\sqrt{\pi}\hspace{0.35mm}p_{\mathrm{th},0}^3 \hspace{0.25mm}n_{\mathrm{RE}}^{\mathrm{ht}}(t)}\cdot \textup{I}_{u_{\mathrm{RE}}^{\mathrm{ht}}}  \,.
\end{gathered}
\end{split} 
\end{equation} 
\vspace*{-6.5mm}\\The derived integral $\textup{I}_{u_{\mathrm{RE}}^{\mathrm{ht}}}$ can not be evaluated in an analytic manner and hence requires a numerical integration scheme. This can be supported by the substitution $(\ref{substitution_num_I_HT})$, which allows the application of standard quadrature formulas for the integration interval \mbox{$w\in[0,\,1]$} and expressions for the isotropic runaway regions, as defined in $(\ref{CH_RE_region})$ and $(\ref{H_RE_region})$. The mean velocity to a given \textit{hot-tail} runaway electron density is than computable as the numerical solution of the definite integral:\vspace*{-4.5mm}
\begin{equation}\label{I_num_u_HT_iso}
\textup{I}_{num}^{\hspace{0.25mm}u_{\mathrm{RE}}^{\mathrm{ht}}} \,=\hspace{-1.3mm} \displaystyle{  \int\limits_{w=0}^{1}}\hspace{-0.5mm} \dfrac{\left(p_{1}\cdot(1-w)+w\right)^3 }{(1-w)^5\cdot\sqrt{1+\left(p_{1}+\frac{w}{1-w}\right)^2}} \hspace{-0.7mm}\cdot\hspace{-0.3mm}\textup{e}^{\, -\frac{1}{p_{\mathrm{th},0}^2}\cdot\left(\left(p_{1}+\frac{w}{1-w}\right)^3+3 \cdot\textup{I}_{\tau_{rel}}(t)\right)^{\frac{2}{3}}  }  \mathrm{d}w\,.
\end{equation} 
\vspace*{-6.5mm}\\For the conditions $(\ref{minmax_RE_region})$ and $(\ref{tilde_minmax_RE_region})$ of the pitch-independent runaway region with finite generalized momentum magnitude boundaries $p_{1}$ and $p_{2}$ the substitution $(\ref{substitution_num_I_HT_finite})$ allows the transformation of the integral $\textup{I}_{u_{\mathrm{RE}}^{\mathrm{ht}}}$, so that a numerical integration can take place in the interval \mbox{$w\in[0,\,1]$}. In this case the mean velocity of a \textit{hot-tail} runaway electron population can be received from the evaluation of the rewritten integral $\textup{I}_{u_{\mathrm{RE}}^{\mathrm{ht}}}$, which reads:\vspace*{-4.0mm}
\begin{equation}\label{I_num_u_HT_iso_finite}
\tilde{\textup{I}}_{num}^{\hspace{0.25mm}u_{\mathrm{RE}}^{\mathrm{ht}}}= \hspace{-1.7mm}\displaystyle{  \int\limits_{w=0}^{1}}\hspace{-0.5mm} \frac{(p_{2}-p_{1})\cdot(p_{1}+(p_{2}-p_{1})\cdot w)^3}{\sqrt{1+(p_{1}+(p_{2}-p_{1})\cdot w)^2}} \hspace{-0.2mm}\cdot\hspace{-0.1mm}\textup{e}^{\,-\frac{1}{p_{\mathrm{th},0}^2}\cdot \left((p_{1}+(p_{2}-p_{1})\cdot w)^3+3 \cdot \textup{I}_{\tau_{rel}}(t)\right)^{\frac{2}{3}}   } \; \mathrm{d}w\,.
\end{equation}

\subsection{\textit{Hot-tail} runaway electron current density for anisotropic descriptions of the runaway region}\label{ht_j_aniso_subsection}

The calculation rule $(\ref{u_RE_HT_iso_def})$ for the mean velocity of \textit{hot-tail} runaway electrons from the previous subsection does not consider the pitch-angle or pitch-dependency of the runaway region, which leads to the smaller generalized pitch coordinate interval\linebreak\mbox{$\xi\in\bigl[\xi_{sep}(p,\,\tilde{E}_{c}) ,\,1\bigr]$} for the anisotropic description of the runaway region. Here, the generalized separatrix \mbox{$\xi_{sep}(p,\,\tilde{E}_{c})$} with a dependency on the momentum magnitude variable $p$ and the generalized critical electric field $\tilde{E}_{c}$, as defined in the relation $(\ref{xi_sep_def})$, determines the pitch-anisotropy of the runaway generation, which was introduced in detail in the subsection \ref{pitch_RE_region_subsection}. There, a general combination of conditions $(\ref{aniso_region_RE_HT_def})$ for the pitch-dependent runaway region was presented, allowing oneself to derive a calculation rule for the \textit{hot-tail} runaway electron mean velocity, taking the anisotropic two-dimensional runaway region into account. This approach, together with the general definition $(\ref{u_RE_HT_def})$ of the mean velocity of \textit{hot-tail} runaway electrons, the two-dimensional volume element in momentum space from $(\ref{volelem_sphere_2D})$ and the \textit{Smith-Verwichte} distribution function $(\ref{SV_ht_dist_func})$ \cite{Svenningsson2020}, determines a formula for the \textit{hot-tail} runaway electron mean velocity for a pitch-dependent runaway region:\vspace*{-4.0mm} 
\begin{equation}\label{u_RE_HT_aniso_def}
\begin{split}
\begin{gathered}
\frac{u_{\mathrm{RE}}^{\mathrm{ht},\xi}(t)}{c} = \hspace{-0.35mm}\dfrac{1}{n_{\mathrm{RE}}^{\mathrm{ht},\xi}(t)}\,\displaystyle{ \int\limits_{p=p_{1}}^{p_{2}}\int\limits_{\xi=\xi_{sep}(p,\,\tilde{E}_{c})}^{1}}\hspace{-0.6mm}\dfrac{v}{c}\hspace{-0.6mm}\cdot\hspace{-0.3mm} f_{RE}^{\textup{ht}}(p,\,t)\,2\pi\,p^2\,\mathrm{d}\xi\, \mathrm{d}p
\\[0pt]
= \hspace{-0.35mm}\dfrac{2\pi}{n_{\mathrm{RE}}^{\mathrm{ht},\xi}(t)}\,\displaystyle{ \int\limits_{p=p_{1}}^{p_{2}} }\hspace{-0.6mm}\dfrac{p^3}{\sqrt{1+p^2}}\hspace{-0.6mm}\cdot\hspace{-0.3mm}\left(1-\xi_{sep}\hspace{-0.7mm}\left(p,\,\tilde{E}_{c}\right)\right)\hspace{-0.6mm}\cdot\hspace{-0.3mm} f_{RE}^{\textup{ht}}(p,\,t)\,   \mathrm{d}p  
\end{gathered}
\end{split} 
\end{equation}
\noindent\newpage \noindent
\begin{equation*}
\begin{split}
\begin{gathered}
\overset{(\ref{xi_sep_def})}{=} \hspace{-1.05mm} \dfrac{4 \hspace{0.35mm} n_{\mathrm{e}} }{\sqrt{\pi}\hspace{0.35mm}p_{\mathrm{th},0}^3 \hspace{0.25mm}n_{\mathrm{RE}}^{\mathrm{ht},\xi}(t)}\underbrace{\displaystyle{ \int\limits_{p=p_{1}}^{p_{2}} }\hspace{-0.8mm}\dfrac{p^3}{\sqrt{1+p^2}}\hspace{-0.6mm}\cdot\hspace{-0.7mm}\left(\hspace{-0.5mm}1\hspace{-0.3mm}-\hspace{-0.3mm}\dfrac{\tilde{E}_{c}}{p^2\hspace{0.25mm}E_{\|}}\hspace{-0.5mm}\right)\hspace{-0.8mm}\cdot\hspace{-0.3mm}\exp{\hspace{-0.7mm}\left(\hspace{-0.7mm}-\dfrac{\left(p^3+3\hspace{-0.5mm}\cdot\hspace{-0.3mm}\textup{I}_{\tau_{rel}}(t)\right)^{\frac{2}{3}}}{p_{\mathrm{th},0}^2}\hspace{-0.5mm}\right)} \mathrm{d}p}_{\eqqcolon\,\textup{I}_{u_{\mathrm{RE}}^{\mathrm{ht},\xi}}} 
\\[-5.5pt]
= \dfrac{4 \hspace{0.35mm} n_{\mathrm{e}}}{\sqrt{\pi}\hspace{0.35mm}p_{\mathrm{th},0}^3 \hspace{0.25mm}n_{\mathrm{RE}}^{\mathrm{ht},\xi}(t)}\cdot \textup{I}_{u_{\mathrm{RE}}^{\mathrm{ht},\xi}}  \,.
\end{gathered}
\end{split} 
\end{equation*} 
\vspace*{-6.5mm}\\At this, the electric field parameter of the separatrix was chosen to be the absolute value of the electric field component in parallel to the local magnetic field \mbox{$E_{\|}\coloneqq \vert E_{\|}\vert$}.

An analytic solution of the appearing integral $\textup{I}_{u_{\mathrm{RE}}^{\mathrm{ht},\xi}}$ could not be determined. Nevertheless, a computation is possible with a one-dimensional numerical integration routine. For the application of for instance a standard quadrature formula, a transformation of the integration domain for the anisotropic runaway region with a half-open momentum interval \mbox{$p\in[p_{1},\,\infty)$}, to the interval \mbox{$w\in[0,\,1]$} should be applied, with the help of the substitution $(\ref{substitution_num_I_HT})$. Consequently, the integral $\textup{I}_{u_{\mathrm{RE}}^{\mathrm{ht},\xi}}$ can be calculated with an implementation, for the different descriptions of the pitch-dependent runaway region defined in $(\ref{CH_RE_region})$ and $(\ref{H_RE_region})$, by means of the integral expression:\vspace*{-3.5mm}
\begin{equation}\label{I_num_u_HT_sep}
\textup{I}_{num}^{\hspace{0.25mm}u_{\mathrm{RE}}^{\mathrm{ht},\xi}}=\hspace{-1.2mm} \displaystyle{  \int\limits_{w=0}^{1}}\hspace{-0.5mm} \dfrac{ \left(p_{1}+\frac{w}{1-w}\right)\hspace{-0.6mm}\cdot\hspace{-0.7mm}\left(\hspace{-0.5mm}\left(p_{1}+\frac{w}{1-w}\right)^2\hspace{-0.3mm}-\hspace{-0.3mm}\frac{\tilde{E}_{c}}{ \hspace{0.25mm}E_{\|}}\hspace{-0.5mm}\right)  }{(1-w)^2\cdot\sqrt{1+\left(p_{1}+\frac{w}{1-w}\right)^2}}\hspace{-0.6mm}\cdot\hspace{-0.7mm} \textup{e}^{\, -\frac{1}{p_{\mathrm{th},0}^2}\cdot\left(\left(p_{1}+\frac{w}{1-w}\right)^3+3 \cdot\textup{I}_{\tau_{rel}}(t)\right)^{\frac{2}{3}}  }  \mathrm{d}w\,.
\end{equation} 
\vspace*{-6.5mm}\\The conditions $(\ref{minmax_RE_region})$ and $(\ref{tilde_minmax_RE_region})$ represent the anisotropic runaway region with finite generalized momentum magnitude boundaries $p_{1}$ and $p_{2}$. For this approach, the mean velocity moment of a \textit{hot-tail} runaway electron population as deduced in $(\ref{u_RE_HT_aniso_def})$ should be transformed with the different substitution $(\ref{substitution_num_I_HT_finite})$, so that the subsequent integral definition:\vspace*{-2.8mm}
\begin{equation}\label{I_num_u_HT_finite_sep}
\begin{split}
\begin{gathered}
\tilde{\textup{I}}_{num}^{\hspace{0.25mm}u_{\mathrm{RE}}^{\mathrm{ht},\xi}}= \hspace{-1.6mm}\displaystyle{  \int\limits_{w=0}^{1}}\hspace{-0.5mm} (p_{2}-p_{1})\hspace{-0.6mm}\cdot\hspace{-0.6mm}(p_{1}\hspace{-0.2mm}+\hspace{-0.2mm}(p_{2}\hspace{-0.2mm}-\hspace{-0.2mm}p_{1})\hspace{-0.4mm}\cdot\hspace{-0.3mm} w)\hspace{-0.35mm}\cdot\hspace{-0.4mm}\dfrac{ \hspace{-0.5mm}(p_{1}\hspace{-0.2mm}+\hspace{-0.2mm}(p_{2}\hspace{-0.2mm}-\hspace{-0.2mm}p_{1})\hspace{-0.4mm}\cdot\hspace{-0.3mm} w)^2 \hspace{-0.4mm}-\hspace{-0.2mm}\frac{\tilde{E}_{c}}{E_{\|}}\hspace{-0.4mm} }{\sqrt{1+(p_{1}\hspace{-0.2mm}+\hspace{-0.2mm}(p_{2}\hspace{-0.2mm}-\hspace{-0.2mm}p_{1})\hspace{-0.4mm}\cdot\hspace{-0.3mm} w)^2}}\; \times
\\
\times \;\exp{\hspace{-0.6mm}\left(\hspace{-0.75mm}-\frac{\left((p_{1}+(p_{2}-p_{1})\cdot w)^3+3 \cdot\textup{I}_{\tau_{rel}}(t)\right)^{\frac{2}{3}}}{p_{\mathrm{th},0}^2}\hspace{-0.6mm}\right)  } \; \mathrm{d}w\,,
\end{gathered}
\end{split} 
\end{equation}\vspace*{-6.0mm}\\defines a conveniently implementable computation rule.

\clearpage

\section{Mean rest mass-related kinetic energy density of a \textit{hot-tail} runaway electron population in the \textit{Smith-Verwichte} approach}\label{ht_k_section}

\subsection{Mean kinetic energy density of \textit{hot-tail} runaway electrons for isotropic descriptions of the runaway region}\label{ht_k_iso_subsection}

The mean rest mass-related kinetic energy density of \textit{hot-tail} runaway electrons $k_{\mathrm{RE}}^{\mathrm{ht}}$ was defined in equation $(\ref{k_RE_HT_def})$ as a moment of the isotropic electron distribution function in the \textit{Smith-Verwichte} approach normalized to the square of the speed of light. Thus, this moment can be evaluated as well, if the calculation rules from section \ref{ht_n_section} are used for the generation of results for the \textit{hot-tail} runaway electron density. 
\\
In this subsection, an isotropic runaway region, as defined in $(\ref{iso_region_RE_HT_def})$, is considered for the purpose of a derivation of a formula for the mean rest-mass related kinetic energy density of a \textit{hot-tail} runaway electron population. In order to achieve this, one starts with the definition for the kinetic energy density $(\ref{k_RE_HT_def})$ for the considered pitch-independent runaway region and recapitulates the \textit{Smith-Verwichte} distribution function $(\ref{SV_ht_dist_func})$. From this, the relations $(\ref{volelem_sphere_2D})$ and $(\ref{p_norm_gamma_def})$, for the two-dimensional momentum space volume element and the momentum representation of the \textit{Lorentz} factor $\gamma(p)$, a calculation rule for the mean rest-mass related kinetic energy density of a \textit{hot-tail} runaway electron population normalized to the square of the speed of light can be deduced:\vspace*{-3.5mm} 
\begin{equation}\label{k_RE_HT_iso_def}
\begin{split}
\begin{gathered}
\frac{k_{\mathrm{RE}}^{\mathrm{ht}}(t)}{c^2} = \hspace{-1.05mm}\dfrac{1}{n_{\mathrm{RE}}^{\mathrm{ht}}(t)}\,\displaystyle{ \int\limits_{p=p_{1}}^{p_{2}}\int\limits_{\xi=-1}^{1}}\hspace{-0.6mm}\gamma(p)\hspace{-0.6mm}\cdot\hspace{-0.3mm} f_{RE}^{\textup{ht}}(p,\,t)\,2\pi\,p^2\,\mathrm{d}\xi\, \mathrm{d}p\,-\,1
\\[0pt]
= \hspace{-1.05mm}\dfrac{4\pi}{n_{\mathrm{RE}}^{\mathrm{ht}}(t)}\,\displaystyle{ \int\limits_{p=p_{1}}^{p_{2}} }\hspace{-0.6mm} p^2\hspace{-0.6mm}\cdot\hspace{-0.3mm} \sqrt{1+p^2} \hspace{-0.6mm}\cdot\hspace{-0.3mm} f_{RE}^{\textup{ht}}(p,\,t)\,   \mathrm{d}p\,-\,1
\\[-5.5pt]
= \hspace{-1.05mm} \dfrac{4 \hspace{0.35mm} n_{\mathrm{e}} }{\sqrt{\pi}\hspace{0.35mm}p_{\mathrm{th},0}^3 \hspace{0.25mm}n_{\mathrm{RE}}^{\mathrm{ht}}(t)}\,\underbrace{\displaystyle{ \int\limits_{p=p_{1}}^{p_{2}} }\hspace{-0.8mm}p^2\hspace{-0.6mm}\cdot\hspace{-0.3mm} \sqrt{1+p^2}\hspace{-0.6mm}\cdot\hspace{-0.3mm}\exp{\hspace{-0.7mm}\left(\hspace{-0.7mm}-\dfrac{\left(p^3+3\hspace{-0.5mm}\cdot\hspace{-0.3mm}\textup{I}_{\tau_{rel}}(t)\right)^{\frac{2}{3}}}{p_{\mathrm{th},0}^2}\hspace{-0.5mm}\right)} \mathrm{d}p}_{\eqqcolon\,\textup{I}_{k_{\mathrm{RE}}^{\mathrm{ht}}}} \,-\,1
\\[-5.5pt]
= \dfrac{4 \hspace{0.35mm} n_{\mathrm{e}}}{\sqrt{\pi}\hspace{0.35mm}p_{\mathrm{th},0}^3 \hspace{0.25mm}n_{\mathrm{RE}}^{\mathrm{ht}}(t)}\cdot \textup{I}_{k_{\mathrm{RE}}^{\mathrm{ht}}}  \,-\,1 \,.
\end{gathered}
\end{split} 
\end{equation} 
\vspace*{-6.0mm}\\For the appearing integral $\textup{I}_{k_{\mathrm{RE}}^{\mathrm{ht}}}$ no analytic solution could be found. Thus, it is again appropriate to prepare a numerical integration by means of the application of the substitution $(\ref{substitution_num_I_HT})$, so that a standard quadrature formula can be used for the integration interval \mbox{$w\in[0,\,1]$} and for expressions of the isotropic runaway regions as defined in $(\ref{CH_RE_region})$ and $(\ref{H_RE_region})$. The normalized mean mass-related kinetic energy density, corresponding to a given \textit{hot-tail} runaway electron density, can subsequently be computed, with an implementation in a common programming language, from the result of the definite integral:\vspace*{-7.5mm}
\begin{equation}\label{I_num_k_HT_iso}
\begin{split}
\begin{gathered}
\textup{I}_{num}^{\hspace{0.25mm}k_{\mathrm{RE}}^{\mathrm{ht}}}=\hspace{-1.2mm} \displaystyle{  \int\limits_{w=0}^{1}}\hspace{-0.5mm} \dfrac{\left(p_{1}\cdot(1-w)+w\right)^2 }{(1-w)^4}\hspace{-0.7mm}\cdot\hspace{-0.3mm}\sqrt{1+\left(p_{1}+\frac{w}{1-w}\right)^2}\;\times
\\[2pt]
\times\;\exp{\hspace{-0.7mm}\left(\hspace{-0.8mm}-\frac{1}{p_{\mathrm{th},0}^2}\cdot\left(\left(p_{1}+\frac{w}{1-w}\right)^3+3 \cdot\textup{I}_{\tau_{rel}}(t)\right)^{\frac{2}{3}} \hspace{-0.75mm}\right) }  \mathrm{d}w\,.
\end{gathered}
\end{split}
\end{equation} 
\vspace*{-6.5mm}\\The representations $(\ref{minmax_RE_region})$ and $(\ref{tilde_minmax_RE_region})$ of the pitch-independent runaway region with finite generalized momentum magnitude boundaries $p_{1}$ and $p_{2}$ require the substitution $(\ref{substitution_num_I_HT_finite})$, in order to allow a transformation of the integral $\textup{I}_{k_{\mathrm{RE}}^{\mathrm{ht}}}$, so that a numerical integration can take place in the interval \mbox{$w\in[0,\,1]$}. Thus, the mean kinetic energy of a \textit{hot-tail} runaway electron population normalized with the electron rest mass follows from the evaluation of the transformed integral $\textup{I}_{k_{\mathrm{RE}}^{\mathrm{ht}}}$, which reads:\vspace*{-3.5mm}
\begin{equation}\label{I_num_k_HT_iso_finite}
\begin{split}
\begin{gathered}
\tilde{\textup{I}}_{num}^{\hspace{0.25mm}k_{\mathrm{RE}}^{\mathrm{ht}}}= \hspace{-1.6mm}\displaystyle{  \int\limits_{w=0}^{1}}\hspace{-0.5mm} (p_{2}-p_{1})\hspace{-0.4mm}\cdot\hspace{-0.4mm}(p_{1}+(p_{2}-p_{1})\cdot w)^2\hspace{-0.4mm}\cdot\hspace{-0.4mm}\sqrt{1+(p_{1}+(p_{2}-p_{1})\cdot w)^2}  \;\times
\\[2pt]
\times\; \exp{\hspace{0.8mm}\left(\hspace{0.8mm}-\frac{1}{p_{\mathrm{th},0}^2}\cdot \left((p_{1}+(p_{2}-p_{1})\cdot w)^3+3 \cdot \textup{I}_{\tau_{rel}}(t)\right)^{\frac{2}{3}}\hspace{0.75mm} \right)  } \; \mathrm{d}w\,.
\end{gathered}
\end{split}
\end{equation}

\subsection{Mean kinetic energy density of \textit{hot-tail} runaway electrons for anisotropic descriptions of the runaway region}\label{ht_k_aniso_subsection}

The calculation rule $(\ref{k_RE_HT_iso_def})$ for the mean kinetic energy of \textit{hot-tail} runaway electrons from the previous subsection can be improved, if the pitch-dependent runaway region is considered, which introduces the generalized smaller pitch coordinate interval\linebreak\mbox{$\xi\in\bigl[\xi_{sep}(p,\,\tilde{E}_{c}) ,\,1\bigr]$} for the anisotropic description of the runaway region. This lower momentum boundary is the so-called separatrix \mbox{$\xi_{sep}(p,\,\tilde{E}_{c})$} with a dependency on the momentum magnitude variable $p$ and the generalized critical electric field $\tilde{E}_{c}$, as defined in the relation $(\ref{xi_sep_def})$. The separatrix was introduced in detail in the subsection \ref{pitch_RE_region_subsection} together with a general combination of conditions $(\ref{aniso_region_RE_HT_def})$, describing the pitch-dependent runaway region. This allows oneself to derive a calculation rule for the mean kinetic energy density of \textit{hot-tail} runaway electrons, whilst taking the anisotropic two-dimensional runaway region into account. Therefore the general definition $(\ref{k_RE_HT_def})$ of the moment related to the mean kinetic energy of \textit{hot-tail} runaway electrons normalized to the square of the speed of light, the two-dimensional volume element in momentum space from $(\ref{volelem_sphere_2D})$ and the \textit{Smith-Verwichte} distribution function $(\ref{SV_ht_dist_func})$ \cite{Svenningsson2020} are used to derive an expression for the normalized mean kinetic energy of \textit{hot-tail} runaway electrons for a pitch-dependent runaway region:\vspace*{-3.7mm} 
\begin{equation}\label{k_RE_HT_aniso_def}
\begin{split}
\begin{gathered}
\frac{k_{\mathrm{RE}}^{\mathrm{ht},\xi}(t)}{c^2} = \hspace{-1.05mm}\dfrac{1}{n_{\mathrm{RE}}^{\mathrm{ht},\xi}(t)}\,\displaystyle{ \int\limits_{p=p_{1}}^{p_{2}}\int\limits_{\xi=\xi_{sep}(p,\,\tilde{E}_{c})}^{1}}\hspace{-5.0mm}\gamma(p)\hspace{-0.6mm}\cdot\hspace{-0.3mm} f_{RE}^{\textup{ht}}(p,\,t)\,2\pi\,p^2\,\mathrm{d}\xi\, \mathrm{d}p
\\[-1pt]
= \hspace{-1.05mm}\dfrac{2\pi}{n_{\mathrm{RE}}^{\mathrm{ht},\xi}(t)}\,\displaystyle{ \int\limits_{p=p_{1}}^{p_{2}} }\hspace{-0.6mm} p^2\hspace{-0.6mm}\cdot\hspace{-0.3mm}\sqrt{1+p^2} \hspace{-0.6mm}\cdot\hspace{-0.3mm}\left(1-\xi_{sep}\hspace{-0.7mm}\left(p,\,\tilde{E}_{c}\right)\right)\hspace{-0.6mm}\cdot\hspace{-0.3mm} f_{RE}^{\textup{ht}}(p,\,t)\,   \mathrm{d}p  
\\[-5.5pt]
\overset{(\ref{xi_sep_def})}{=} \hspace{-1.05mm} \dfrac{4 \hspace{0.35mm} n_{\mathrm{e}} }{\sqrt{\pi}\hspace{0.35mm}p_{\mathrm{th},0}^3 \hspace{0.25mm}n_{\mathrm{RE}}^{\mathrm{ht},\xi}(t)}\,\underbrace{\displaystyle{ \int\limits_{p=p_{1}}^{p_{2}} }\hspace{-0.8mm} \sqrt{1+p^2} \hspace{-0.4mm}\cdot\hspace{-0.8mm}\left(\hspace{-0.5mm}p^2\hspace{-0.3mm}-\hspace{-0.3mm}\dfrac{\tilde{E}_{c}}{ E_{\|}}\hspace{-0.5mm}\right)\hspace{-0.9mm}\cdot\hspace{-0.3mm}\exp{\hspace{-0.7mm}\left(\hspace{-0.7mm}-\dfrac{\left(p^3+3\hspace{-0.5mm}\cdot\hspace{-0.3mm}\textup{I}_{\tau_{rel}}(t)\right)^{\frac{2}{3}}}{p_{\mathrm{th},0}^2}\hspace{-0.5mm}\right)} \mathrm{d}p}_{\eqqcolon\,\textup{I}_{k_{\mathrm{RE}}^{\mathrm{ht},\xi}}} 
\\[-6.0pt]
= \dfrac{4 \hspace{0.35mm} n_{\mathrm{e}}}{\sqrt{\pi}\hspace{0.35mm}p_{\mathrm{th},0}^3 \hspace{0.25mm}n_{\mathrm{RE}}^{\mathrm{ht},\xi}(t)}\cdot \textup{I}_{k_{\mathrm{RE}}^{\mathrm{ht},\xi}}  \,.
\end{gathered}
\end{split} 
\end{equation} 
\vspace*{-6.2mm}\\Here, the absolute value of the electric field component in parallel to the local magnetic field \mbox{$E_{\|}\coloneqq \vert E_{\|}\vert$} was used as the descriptive parameter for the present electric field, which influences the behaviour of the separatrix in the momentum space.

Since, no analytic solution of the appearing integral $\textup{I}_{k_{\mathrm{RE}}^{\mathrm{ht},\xi}}$ has been found, one prepares the application of a standard quadrature for the purpose of a numerical evaluation of the integral. For the anisotropic runaway region with a half-open momentum interval \mbox{$p\in[p_{1},\,\infty)$} this can be achieved by transforming the integration domain to the closed interval \mbox{$w\in[0,\,1]$}, with the help of the substitution $(\ref{substitution_num_I_HT})$. Consequently, the integral $\textup{I}_{k_{\mathrm{RE}}^{\mathrm{ht},\xi}}$ for the corresponding descriptions of the pitch-dependent runaway region, defined in $(\ref{CH_RE_region})$ and $(\ref{H_RE_region})$, can also be evaluated in the following form:\vspace*{-3.7mm}
\begin{equation}\label{I_num_k_HT_sep}
\begin{split}
\begin{gathered}
\textup{I}_{num}^{\hspace{0.25mm}k_{\mathrm{RE}}^{\mathrm{ht},\xi}}=\hspace{-1.2mm} \displaystyle{  \int\limits_{w=0}^{1}}\hspace{-0.5mm}   \dfrac{ \left(\hspace{-0.5mm}\left(p_{1}+\frac{w}{1-w}\right)^2\hspace{-0.3mm}-\hspace{-0.3mm}\frac{\tilde{E}_{c}}{ \hspace{0.25mm}E_{\|}}\hspace{-0.5mm}\right) }{(1-w)^2} \cdot  \sqrt{1+\left(p_{1}+\frac{w}{1-w}\right)^2} \; \times
\\
\times \;
 \exp{\hspace{-0.6mm}\left(\hspace{-0.75mm}-\frac{1}{p_{\mathrm{th},0}^2}\cdot\left(\left(p_{1}+\frac{w}{1-w}\right)^3+3 \cdot\textup{I}_{\tau_{rel}}(t)\right)^{\frac{2}{3}} \hspace{-0.6mm}\right)  } \; \mathrm{d}w\,.
\end{gathered}
\end{split} 
\end{equation} 
\vspace*{-7.0mm}\\For the anisotropic runaway region with the finite generalized momentum magnitude boundaries $p_{1}$ and $p_{2}$, one has defined the conditions $(\ref{minmax_RE_region})$ and $(\ref{tilde_minmax_RE_region})$ in subsection \ref{pitch_RE_region_subsection}. For those representations of the runaway region, the different substitution $(\ref{substitution_num_I_HT_finite})$ should be used for the transformation of the moment, related to the normalized mean rest mass-related kinetic energy density of a \textit{hot-tail} runaway electron population as stated in $(\ref{k_RE_HT_aniso_def})$. Hence, the integral $\textup{I}_{u_{\mathrm{RE}}^{\mathrm{ht}}}$ is rewritten in the following form:\vspace*{-4.5mm}
\begin{equation}\label{I_num_k_HT_finite_sep}
\begin{split}
\begin{gathered}
\tilde{\textup{I}}_{num}^{\hspace{0.25mm}k_{\mathrm{RE}}^{\mathrm{ht},\xi}}= \hspace{-1.6mm}\displaystyle{  \int\limits_{w=0}^{1}}\hspace{-0.5mm} (p_{2}-p_{1})\hspace{-0.6mm}\cdot\hspace{-0.6mm}(p_{1}\hspace{-0.2mm}+\hspace{-0.2mm}(p_{2}\hspace{-0.2mm}-\hspace{-0.2mm}p_{1})\hspace{-0.4mm}\cdot\hspace{-0.3mm} w)^2\hspace{-0.35mm}\cdot\hspace{-0.4mm} \sqrt{1+(p_{1}\hspace{-0.2mm}+\hspace{-0.2mm}(p_{2}\hspace{-0.2mm}-\hspace{-0.2mm}p_{1})\hspace{-0.4mm}\cdot\hspace{-0.3mm} w)^2} \; \times
\\[-1pt]
\times \;\exp{\hspace{-0.6mm}\left(\hspace{-0.75mm}-\frac{1}{p_{\mathrm{th},0}^2}\cdot\left((p_{1}+(p_{2}-p_{1})\cdot w)^3+3 \cdot\textup{I}_{\tau_{rel}}(t)\right)^{\frac{2}{3}}  \hspace{-0.6mm}\right)  } \; \mathrm{d}w\,.
\end{gathered}
\end{split} 
\end{equation}


\clearpage

\section{Computation and evaluation of the mean velocity and the mean kinetic energy density of a \textit{hot-tail} runaway electron population in the \textit{Smith-Verwichte} approach}\label{ht_eval_comp_u_and_k_section}

The mean velocity and the mean kinetic energy density of a \textit{hot-tail} runaway electron population were related to calculation rules of certain moments of the distribution function by \textit{I.\hspace{0.8mm}Svenningsson} \cite{Svenningsson2020} in the sections \ref{ht_j_section} and \ref{ht_k_section}. At that, the calculation schemes for the moments are related to the approach of \textit{H.\hspace{0.7mm}M.\hspace{0.8mm}Smith} and \textit{E.\hspace{0.8mm}Verwichte}, presented in their publication \cite{hottailREdistfunc} from 2008, for the modeling of the \textit{hot-tail} generation of runaway electrons. Furthermore, this applied approach, which interprets the runaway region as isotropic in momentum space, was extended with the consideration of a pitch-dependent runaway region as explained in subsection \ref{pitch_RE_region_subsection}.
\\
In general, the expressions for the computation of the moments from the subsections \ref{ht_j_iso_subsection} and \ref{ht_k_iso_subsection} were derived for an isotropic representation as defined $(\ref{iso_region_RE_HT_def})$. In contrast, the integrals in the subsections \ref{ht_j_aniso_subsection} and \ref{ht_k_aniso_subsection} are connected to the generalized definition $(\ref{aniso_region_RE_HT_def})$ of the runaway region. Within these different frameworks a further distinction is made based on different approximations of the momentum boundaries of the runaway region, which were discussed in section \ref{part_screen_section}. Those possible approximations of the lower and upper momentum of the runaway region can be recapitulated in figure \ref{fig_RE_mom_bound} for an ITER-like disruption. The simulation of this disruption for an ITER-scenario, covering the thermal and current quench phase, from the paper \cite{Smith_2009} with a plasma current of \mbox{$I_{p}=15\;\mathrm{MA}$}, a magnetic field of \mbox{$B=5.3\,\mathrm{T}$} and a time-independent electron density of \mbox{$n_{e}=1.06\cdot10^{20}\;\mathrm{m}^{-3}$}, is reused from section \ref{part_screen_section}. The associated simulation results are the time evolution of the electron temperature and the parallel component of the electric field for the duration of the disruption, which are displayed in figure \ref{fig_RE_ITER_simu}. In particular, the depicted data set corresponds to the solution of the simulation at the radius \mbox{$r_{\perp}=0.75\,\mathrm{m}$}, because the simulation is based on a one-dimensional cylindrical plasma model, so that the results are time- and cylindrical radius dependent fields of the physical quantities \cite{Smith_2009}. 
\\
The input data from the ITER-simulation at \mbox{$r_{\perp}=0.75\,\mathrm{m}$} was then utilized in two \mbox{\textsc{MATLAB}-implementations$^{\ref{fig_n_ht_RE_footnote},\ref{fig_n_ht_RE_sep_footnote}}$}, which separately cover the calculation of the moments for a pitch-dependent and a pitch-independent modeling of the runaway region. Their console output can be found in the listings \labelcref{outMATLABoutput_RE_ht_moments_SV,outMATLABoutput_RE_ht_moments_SV_imp,outMATLABoutput_RE_ht_moments_SV_sep,outMATLABoutput_RE_ht_moments_SV_sep_imp} in subsection \ref{output_matlab_appendix_subsection} of the appendix. On this occasion it shall be remarked, that the integrals for the computation of the mean velocity and the mean kinetic energy density are one-dimensional and hence allow an numerical evaluation with the \textsc{MATLAB}-routine \qq{\texttt{integral}} \cite{integral}. In the subsequent paragraphs, the computed data shall be analysed on the basis of the graphs of the time evolution of the different results for the mean velocity and the kinetic energy density of the hot-tail runaway electron population, that is generated during the considered disruption scenario.

\subsection{Evaluation of the computational results for the isotropic or pitch-independent descriptions of the runaway region}\label{ht_eval_iso_subsection}

\begin{figure}[H]
\centering
  \subfloat[][Time-dependent quantities at \mbox{$r_{\perp}=0.75\,\mathrm{m}$} for a constant electron density of \mbox{$n_{e}=n_{_{1}^{2}\mathrm{H}^{+}}=1.06\cdot 10^{20}\,\mathrm{m}^{-3}$} within a singly-ionized deuterium plasma with \mbox{$B=5.3\,\textup{T}$} and \mbox{$Z_{eff}=1.0$}.]{\label{u_k_ht_RE_main} 
  \includegraphics[trim=10 15 9 28,width=0.97\textwidth,clip]
    {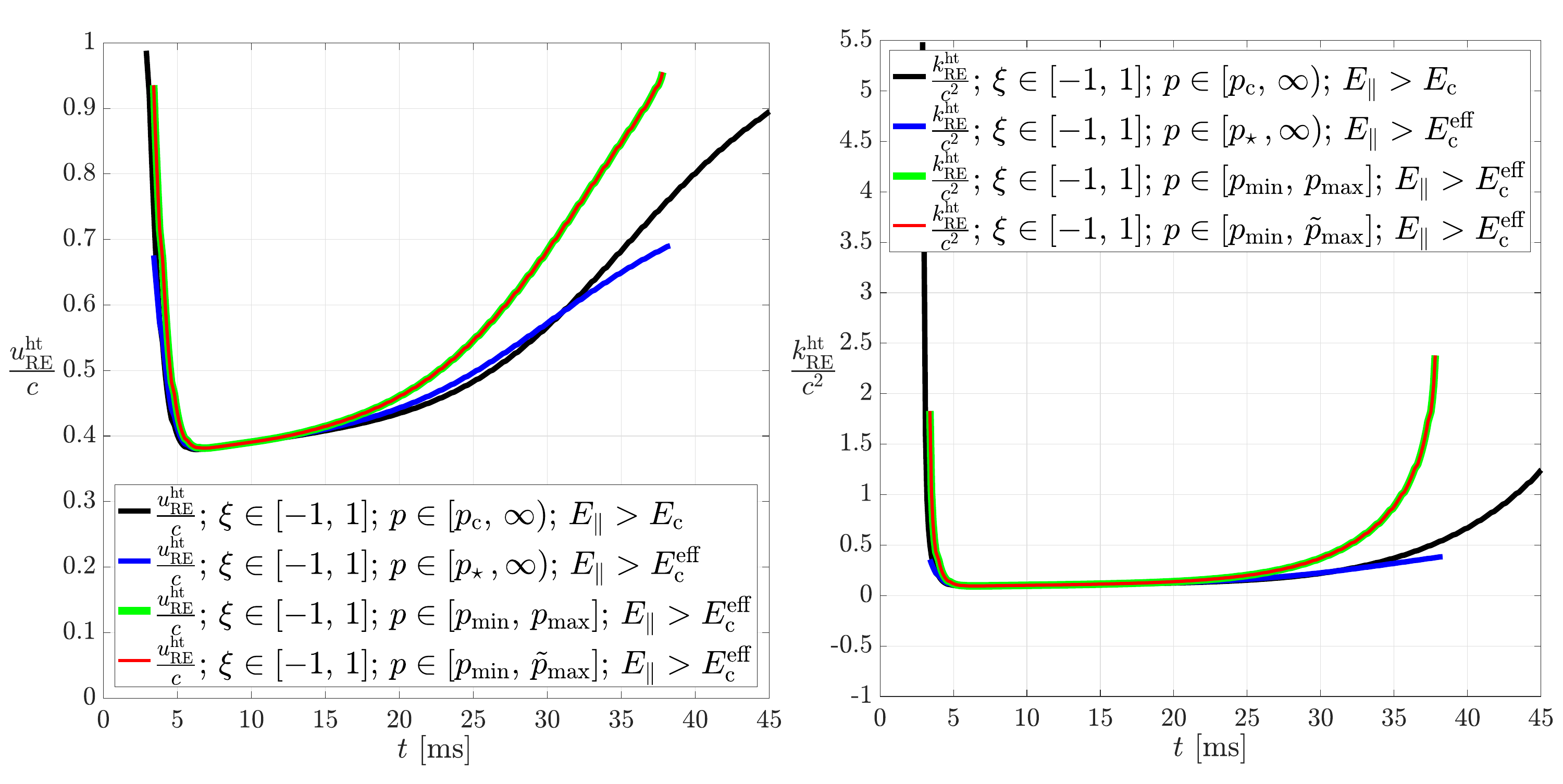}}\\[7pt]
  \subfloat[][Time-dependent quantities at \mbox{$r_{\perp}=0.75\,\mathrm{m}$} for a constant singly-ionized deuterium density of \mbox{$n_{_{1}^{2}\mathrm{H}^{+}}=1.06\cdot 10^{20}\,\mathrm{m}^{-3}$} within a plasma with \mbox{$B=5.3\,\textup{T}$} and \mbox{$Z_{eff}=1.0$}, in the presence of a singly-ionized neon impurity density of \mbox{$n_{_{10}^{20}\mathrm{Ne}^{+}}=2.40\cdot 10^{20}\,\mathrm{m}^{-3}$}.]{\label{u_k_ht_RE_imp_main}  
  \includegraphics[trim=10 17 9 29,width=0.97\textwidth,clip]
    {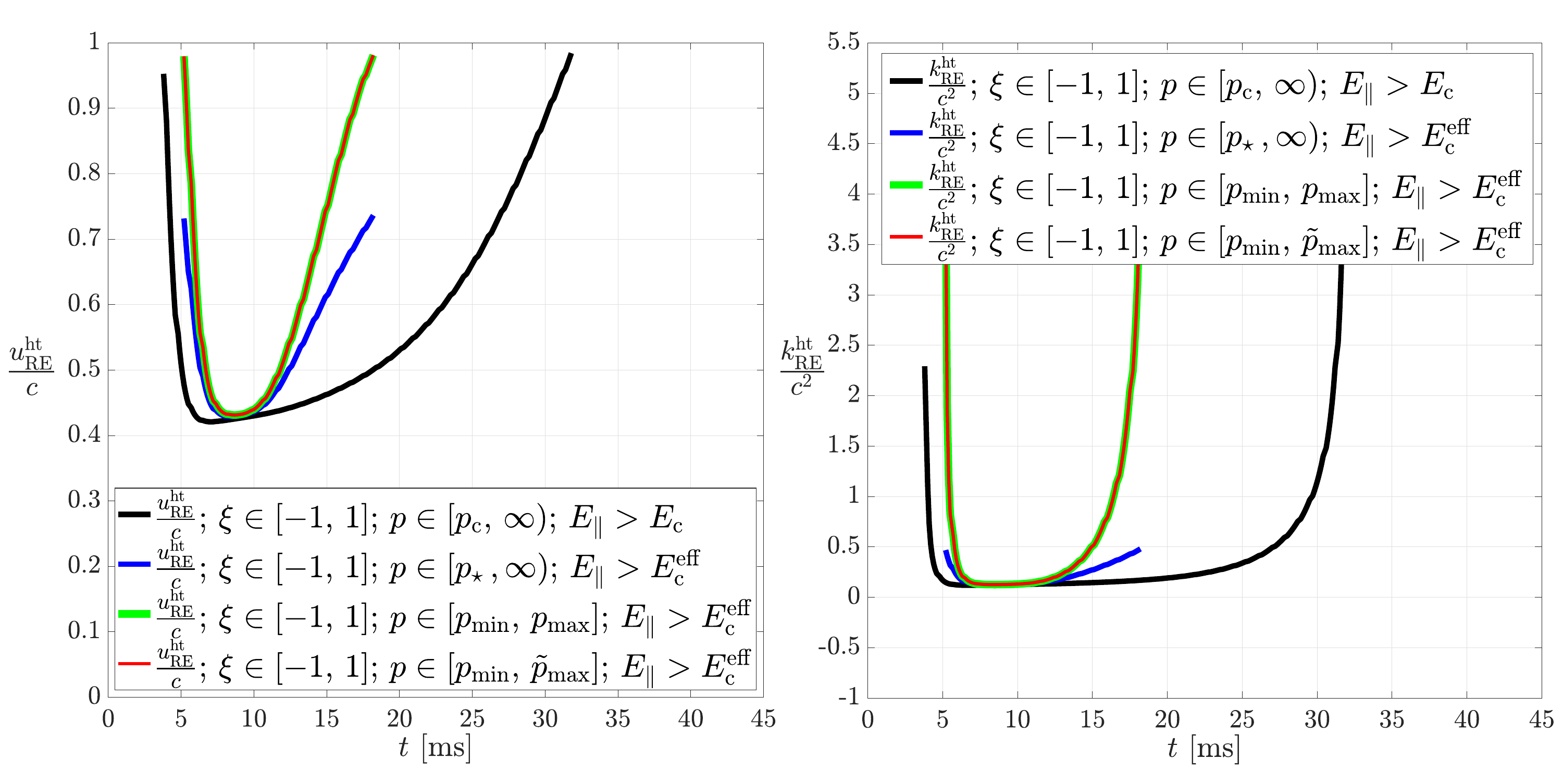}}
\caption[Time-dependent behaviour of the normalized mean velocity \mbox{$u^{\mathrm{ht}}_{\mathrm{RE}}/c$} and the normalized mean rest mass-related kinetic energy density \mbox{$k^{\mathrm{ht}}_{\mathrm{RE}}/c^2$} of a \textit{hot-tail} electron population, during an ITER-like disruption (see also figure \ref{fig_RE_ITER_simu}, \cite{Smith_2009}). Here, the \textit{Smith-Verwichte} model was used in combination with four different isotropic descriptions of the runaway region with respect to the relativistic momentum magnitude $p$.]{Time-dependent behaviour$^{\ref{fig_n_ht_RE_footnote}}$ of the normalized mean velocity \mbox{$u^{\mathrm{ht}}_{\mathrm{RE}}/c$} and the normalized mean rest mass-related kinetic energy density \mbox{$k^{\mathrm{ht}}_{\mathrm{RE}}/c^2$} of a \textit{hot-tail} electron population, during an ITER-like disruption (see also figure \ref{fig_RE_ITER_simu}, \cite{Smith_2009}). Here, the \textit{Smith-Verwichte} model was used in combination with four different isotropic descriptions of the runaway region with respect to the relativistic momentum magnitude $p$.}
\label{fig_u_k_ht_RE}
\end{figure}

First, the results for the mean velocity and the mean kinetic energy of \textit{hot-tail} electrons, according to the integral definitions $(\ref{I_num_u_HT_iso})$, $(\ref{I_num_u_HT_iso_finite})$, $(\ref{I_num_k_HT_iso})$ and $(\ref{I_num_k_HT_iso_finite})$ for the isotropic representations of the runaway region in $(\ref{CH_RE_region})$, $(\ref{H_RE_region})$, $(\ref{minmax_RE_region})$ and $(\ref{tilde_minmax_RE_region})$, are going to be analyzed. For this purpose, the visualized results from the computation of the mentioned rules in the figure \ref{fig_u_k_ht_RE} are presented. In detail, the subfigure \ref{u_k_ht_RE_main} depicts the results for the ITER-simulation, as presented in the reference \cite{Smith_2009}, for a constant electron density of \mbox{$n_{e}=n_{_{1}^{2}\mathrm{H}^{+}}=1.06\cdot 10^{20}\,\mathrm{m}^{-3}$} within a singly-ionized deuterium plasma. In addition, the influences of the presence of a singly-ionized neon impurity density of \mbox{$n_{_{10}^{20}\mathrm{Ne}^{+}}=2.26\cdot n_{e}$} on the time evolution of the normalized mean velocity and the mean rest mass-related kinetic energy density of the developing \textit{hot-tail} runaway electron population can be seen in the subfigure \ref{u_k_ht_RE_imp_main}. 

A general behaviour, concerning the time evolution of the moments, is apparent for all of the applied calculation rules, because the mean velocity and the kinetic energy density decrease during the thermal quench until \textit{$t\approx 5\,\mathrm{ms}$}. This is correlated to the temperature drop, which can be seen in figure \ref{fig_RE_ITER_simu}. Moreover, one recognizes a correlation to the increasing \textit{hot-tail} electron density, which reaches its maximum until the end of the thermal quench, as it can be verified in figure \ref{fig_n_ht_RE} in subsection \ref{n_RE_ht_eval_comp_section}, where this was explained in detail. This described trend in the self-consistent time evolution seems to be physically accurate, due to the following explanation. If the first runaway electron is generated through the \textit{hot-tail} mechanism, it will accelerate rapidly to a velocity close to the speed of light. At that point in time, at \textit{$t\approx 3.55\,\mathrm{ms}$}, the runaway generation starts and the corresponding density is low. Therefore the mean velocity is high and also close to the speed of light, since only a few runaway electrons with \mbox{$u^{\mathrm{ht}}_{\mathrm{RE}}\approx c$} exist, which experience less collisions and thus a small friction force. Until the end of the thermal quench, the runaway density increases and within the runaway electron population more interactions lead to lower velocities, which deviate stronger from the speed of light. Consequently, the mean velocity and the kinetic energy density become smaller. However, the maximum of the runaway seed density is reached before the end of the thermal quench\vspace{-11cm}\linebreak\newpage\noindent 
\vspace{-5.0mm}at \mbox{$t\approx 5\,\mathrm{ms}$}, since the electric field is increasing for \mbox{$t>4\,\mathrm{ms}$} as the transition towards the phase of the current quench happens. In this transition phase of the disruption, one still notices a decreasing mean velocity and kinetic energy density, which is a delay in the reaction of the runaway population to the change in the electric field parameter, due to new characteristics of the interactions or respectively collisions of the runaway electrons with the thermal particles within the plasma. However, the accuracy of the modeling of this delay can not be validated with the computed data and requires the inclusion of the calculation rules into fully self-consistent disruption simulations. Another aspect of the noticed general time behaviour of the mean velocity and the kinetic energy density is their slower growth until the end of the disruption. This is also explained and correlated to the \textit{hot-tail} runaway density, which decays slower than it had grown until the end of the simulation time at \mbox{$t\approx 45\,\mathrm{ms}$}. Subsequently, fewer electrons are present in the runaway region and interact less frequently, so that the mean velocity and hence the kinetic energy density ramps up. However, one has to consider, that the current quench phase ends at \mbox{$t\approx 15\,\mathrm{ms}$}, when the electric field has reached a maximum and decayed back to lower values with \mbox{$E_{\|}<2\,\mathrm{V/m}$}. Within this time more runaway electrons are produced by the \textit{hot-tail} mechanism, as it was seen in the density time evolution, due to the high electric field. But the mean velocity and the kinetic energy density only slowly increase for \mbox{$6\,\mathrm{ms}<t< 15\,\mathrm{ms}$}, because the distribution describes a cooling down electron population. This is also represented by the computed moments, although an exact verification is not possible here either. Furthermore, it has to be admitted, that the influences of other runaway generation and decay mechanisms, and especially the dominant avalanche generation during this simulated disruption on the moments of the \textit{hot-tail} runaway electrons, can not be evaluated with the calculated data. Therefore again a further elaboration of all derived calculation rules in simulation software like the DREAM-code is required, in order to receive a complete validation and verification as well as a determination of parameter ranges, in which the moment-based computations will improve existing calculation schemes without requiring too much additional runtime.

The four different calculation rules for the mean velocity and the mean kinetic energy density $(\ref{I_num_u_HT_iso})$, $(\ref{I_num_u_HT_iso_finite})$, $(\ref{I_num_k_HT_iso})$ and $(\ref{I_num_k_HT_iso_finite})$ connected to the different approaches for the modeling of the momentum magnitude boundaries are now evaluated. They use the isotropic representations of the runaway region stated in $(\ref{CH_RE_region})$, $(\ref{H_RE_region})$, $(\ref{minmax_RE_region})$ and $(\ref{tilde_minmax_RE_region})$ and are in good agreement for the early phase of the disruption and tend to deviate more towards the end of the simulation time. 
\\
The computed results are found to satisfy the inequalities \mbox{$0.38<u^{\mathrm{ht}}_{\mathrm{RE}}/c<0.99$} for the normalized mean velocity and \mbox{$0.09<k^{\mathrm{ht}}_{\mathrm{RE}}/c^2<5.48$} for the normalized mean kinetic energy density. Those ranges seem physically plausible for the considered input data in terms of the time evolution of the electron temperature and the component of the electric field parallel to the magnetic field from the simulation of a typical disruption in ITER \cite{Smith_2009}. Furthermore, one observes, that the calculation rules $(\ref{I_num_u_HT_iso})$ and $(\ref{I_num_k_HT_iso})$ with the lower momentum $p_{\star}$ in the description $(\ref{H_RE_region})$ model the behaviour of the results for the moments, calculated with the \textit{Connor-Hastie} critical momentum $p_{\mathrm{c}}$ in the runaway region condition $(\ref{CH_RE_region})$, nearly over the entire time of the disruption. On the contrary, it is apparent, that the conditions $(\ref{minmax_RE_region})$ and $(\ref{tilde_minmax_RE_region})$ lead to increasingly higher mean velocities and kinetic energies in the second half of the simulation time. Furthermore, an oscillation in the data has to be avoided for both moments, if the finite momentum boundaries $p_{\mathrm{min}}$ and $p_{\mathrm{max}}$ are used, as defined in the representation $(\ref{minmax_RE_region})$ of the runaway region. For this purpose, one has to increase the absolute default precision of the \textsc{MATLAB}-routine \qq{\texttt{integral}} \cite{integral} from $10^{-6}$ to $10^{-18}$ at the minimum, implying a negative impact on the computational effort. However, the physical accuracy of those calculations schemes can again only be validated by self-consistent simulations. On that occasion, the introduced modified upper momentum $\tilde{p}_{\mathrm{max}}$, which appears in the isotropic description of $(\ref{tilde_minmax_RE_region})$, is found to be helpful in suppressing the numerical oscillatory effects, if it is used in the computation rules $(\ref{I_num_u_HT_iso_finite})$ and $(\ref{I_num_k_HT_iso_finite})$. This might be helpful for the usage of the calculation rules for the moments in disruption simulation software, because it allows the usage of a lower absolute error tolerance for the numerical integration, so that the runtime of the integration does not increases. In addition, the computation of the upper runaway momentum $p_{\mathrm{max}}$ is only carried out, if the electric field is close to the critical electric field and thus $p_{\mathrm{max}}$ is smaller and has an enhanced influence on the result of the integration of the moments. It should be remarked, that both upper runaway momenta $\tilde{p}_{\mathrm{max}}$ and $p_{\mathrm{max}}$ should be applied in self-consistent simulations, in order to further evaluate the needed precision for the computations and a possible propagation of the numerical oscillations into the results. Based on the comparison of the results with fully kinetic calculations, one can subsequently evaluate a conceivable loss in physical accuracy as well, which might be caused by the introduced relation $(\ref{tilde_p_max})$ for $\tilde{p}_{\mathrm{max}}$. 
\\
In addition, the lowest, highest and the mean value of the magnitude of the current density \mbox{$j_{RE}^{\mathrm{ht}}=e\hspace{-0.3mm}\cdot\hspace{-0.3mm} n_{RE}^{\mathrm{ht}}\hspace{-0.3mm}\cdot\hspace{-0.3mm} u_{RE}^{\mathrm{ht}}$} at \mbox{$r_{\perp}=0.75\,\mathrm{m}$} is shown in the listing \ref{outMATLABoutput_RE_ht_moments_SV}. This allows oneself to state, that the order of magnitude is approximately between \mbox{$0.22\,\mathrm{kA/m}^2$} and \mbox{$0.40\,\mathrm{kA/m}^2$}, whereat the maximum values are found to be between \mbox{$6.20\,\mathrm{kA/m}^2$} and \mbox{$11.88\,\mathrm{kA/m}^2$}. These values are in the expected range for the \textit{hot-tail} runaway current density for ITER disruptions \cite{Hollmann2014}.

In the subfigure \ref{u_k_ht_RE_imp_main}, the influences of the presence of a singly-ionized neon impurity density of \mbox{$n_{_{10}^{20}\mathrm{Ne}^{+}}=2.26\cdot n_{e}$} on the time evolution of the normalized mean velocity and the mean rest mass-related kinetic energy density of the developing \textit{hot-tail} runaway electron population is depicted for the four calculation schemes including the different descriptions of the isotropic runaway region. 
\\
The analysis of the computed data reveals two general differences to the pure deuterium case in subfigure \ref{u_k_ht_RE_main}. The first is, that time evolution happens faster, which is correlated to the behaviour of the \textit{hot-tail} runaway density, as visualized in subfigure \ref{n_ht_RE_imp_main}. The second observation concerns the enhanced deviations between the four calculation approaches and the higher absolute error tolerance of $10^{-28}$, which is necessary for the calculation schemes, using the maximum runaway momentum $p_{\mathrm{max}}$. Nevertheless, it is apparent, that all results are positive for all times and might therefore be classified as physically possible. However, one intentionally avoids any deliberations about physical processes, which might explain the observed behavior for the presence of an impurity density, due to the fact that the utilized input data for the electron temperature the parallel component of the electric field was originally simulated for a pure deuterium plasma. The ranges of the computed results are in the approximate ranges \mbox{$0.38<u^{\mathrm{ht}}_{\mathrm{RE}}/c<0.99$} and \mbox{$0.09<k^{\mathrm{ht}}_{\mathrm{RE}}/c^2<5.48$}, so that they are less extreme than in the scenario with a pure deuterium plasma. Thus, one is able to again verify, that the derived calculation schemes produce plausible results and are additionally able to model influences of impurities within the plasma. Note, that this is further validated with regard to the negligibly small maximum values for the current density, displayed in the listing \ref{outMATLABoutput_RE_ht_moments_SV_imp}.

Finally, it should be remarked, that for instance the lower runaway momentum approximation $p_{\star}$ and the critical electric field $E_{\mathrm{c}}^{\mathrm{eff}}$, which appear in the conditions for the isotropic and the anisotropic runaway region from $(\ref{H_RE_region})$, have an intrinsic applicability threshold in the limit \mbox{$E_{\|}\rightarrow E_{\mathrm{c}}^{\mathrm{eff}}$}, as explained in section \ref{part_screen_section}. Hence, the calculation rules $(\ref{I_num_u_HT_iso_finite})$ and $(\ref{I_num_k_HT_iso_finite})$ connected to the representations of the runaway region stated in $(\ref{minmax_RE_region})$ and $(\ref{tilde_minmax_RE_region})$ should be more accurate in this limit. For the considered disruption, this would mainly apply for \mbox{$t>20\,\mathrm{ms}$} for the pure deuterium case, regarded in subfigure \ref{u_k_ht_RE_main}. Nevertheless, a final suggestion concerning the superiority of certain calculation rules for electric fields close to the critical electric field and the time periods of the opening and closing of the runaway region in momentum space, as it can be recapitulated by means of figure \ref{fig_RE_mom_bound}, can not be made without an analysis of a fully self-consistent simulation, which utilizes the presented calculation schemes.

\subsection{Evaluation of the computational results for the anisotropic or pitch-dependent descriptions of the runaway region}\label{ht_eval_aniso_subsection}

\begin{figure}[H]
\centering
  \subfloat[][Time-dependent quantities at \mbox{$r_{\perp}=0.75\,\mathrm{m}$} for a constant electron density of \mbox{$n_{e}=n_{_{1}^{2}\mathrm{H}^{+}}=1.06\cdot 10^{20}\,\mathrm{m}^{-3}$} within a singly-ionized deuterium plasma with \mbox{$B=5.3\,\textup{T}$} and \mbox{$Z_{eff}=1.0$}.]{\label{u_k_ht_RE_sep_main} 
   \includegraphics[trim=12 12 5 30,width=0.97\textwidth,clip]
    {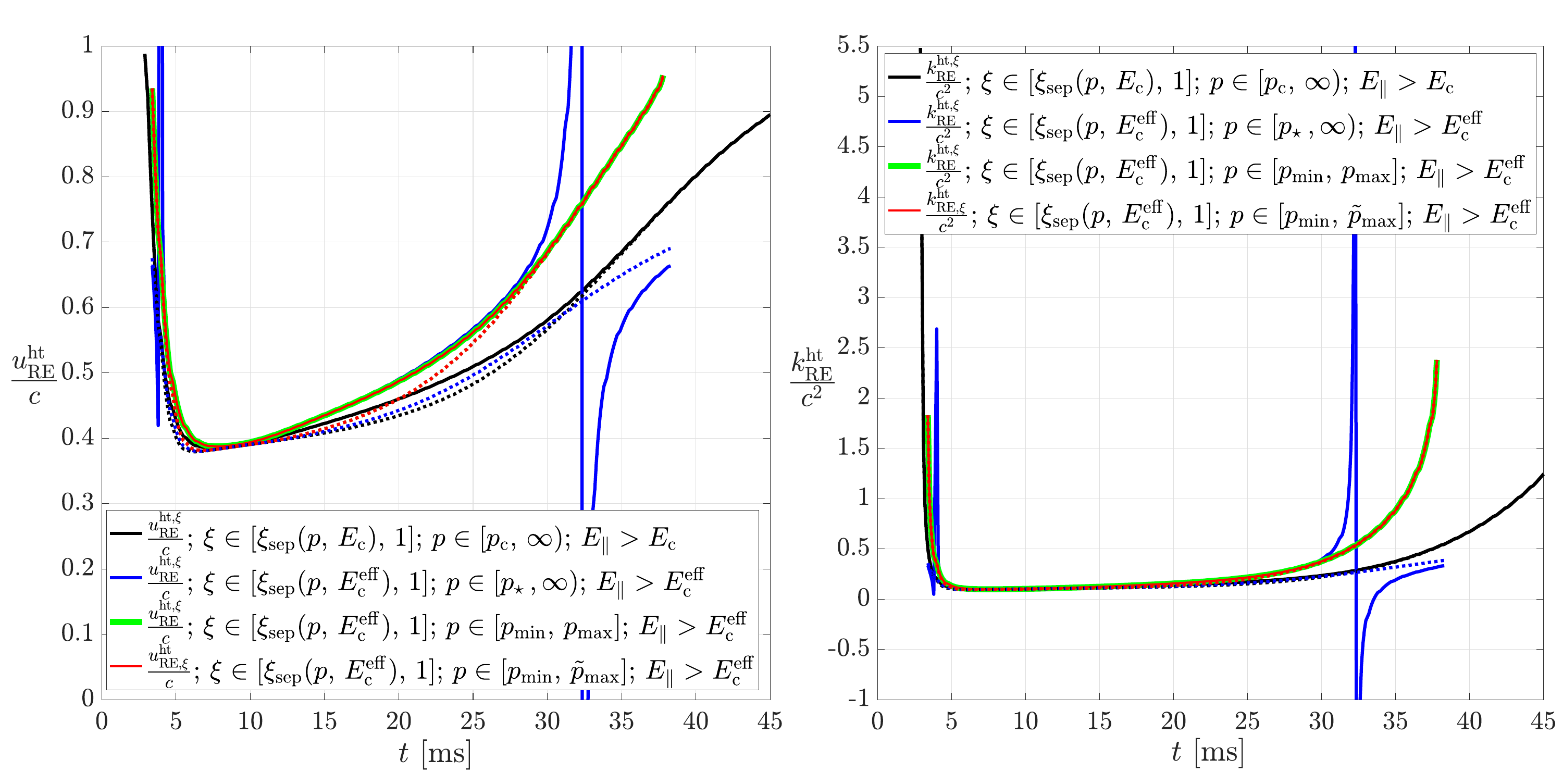}} \\[7pt]
  \subfloat[][Time-dependent quantities at \mbox{$r_{\perp}=0.75\,\mathrm{m}$} for a constant singly-ionized deuterium density of \mbox{$n_{_{1}^{2}\mathrm{H}^{+}}=1.06\cdot 10^{20}\,\mathrm{m}^{-3}$} within a plasma with \mbox{$B=5.3\,\textup{T}$} and \mbox{$Z_{eff}=1.0$}, in the presence of a singly-ionized neon impurity density of \mbox{$n_{_{10}^{20}\mathrm{Ne}^{+}}=2.40\cdot 10^{20}\,\mathrm{m}^{-3}$}.]{\label{u_k_ht_RE_imp_sep_main}  
  \includegraphics[trim=9 12 6 33,width=0.97\textwidth,clip]
    {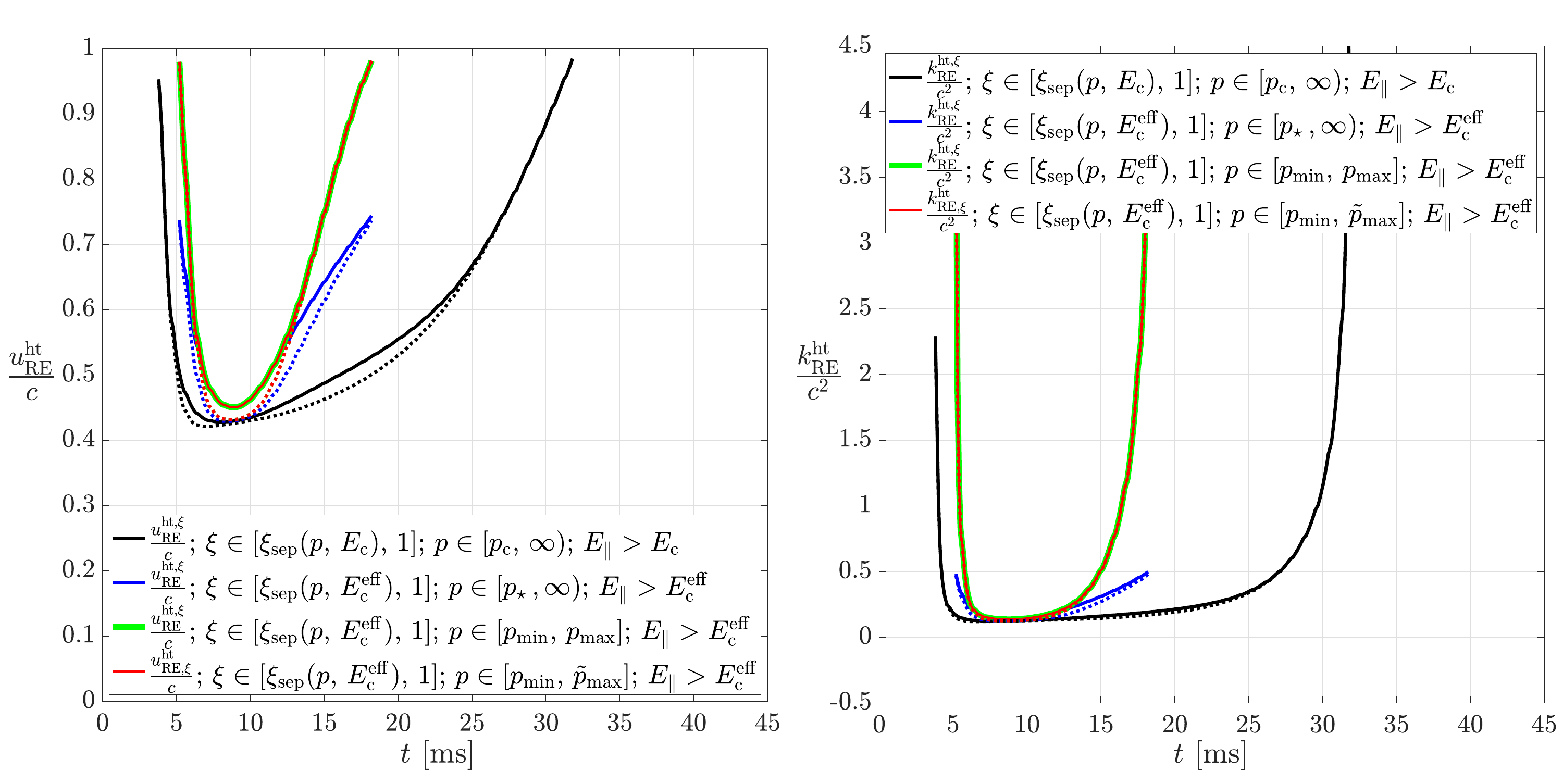}}  
\caption[Time-dependent behaviour of the normalized mean velocity \mbox{$u^{\mathrm{ht},\xi}_{\mathrm{RE}}/c$} and the normalized mean rest mass-related kinetic energy density \mbox{$k^{\mathrm{ht},\xi}_{\mathrm{RE}}/c^2$} of a \textit{hot-tail} electron population, during an ITER-like disruption (see also figure \ref{fig_RE_ITER_simu}, \cite{Smith_2009}). Here, the results of the \textit{Smith-Verwichte} model for the four different pitch-dependent descriptions of the runaway region (solid lines) are compared with the results for the corresponding isotropic representations (dotted lines, figure \ref{fig_u_k_ht_RE}).]{Time-dependent behaviour$^{\ref{fig_n_ht_RE_sep_footnote}}$ of the normalized mean velocity \mbox{$u^{\mathrm{ht},\xi}_{\mathrm{RE}}/c$} and the normalized mean rest mass-related kinetic energy density \mbox{$k^{\mathrm{ht},\xi}_{\mathrm{RE}}/c^2$} of a \textit{hot-tail} electron population, during an ITER-like disruption (see also figure \ref{fig_RE_ITER_simu}, \cite{Smith_2009}). Here, the results of the \textit{Smith-Verwichte} model for the four different pitch-dependent descriptions of the runaway region (solid lines) are compared with the results for the corresponding isotropic representations (dotted lines, figure \ref{fig_u_k_ht_RE}).}
\label{fig_u_k_ht_RE_sep}
\end{figure}

Second, the results for the mean velocity and the mean kinetic energy of \textit{hot-tail} electrons, according to the integral-based calculation rules $(\ref{I_num_u_HT_sep})$, $(\ref{I_num_u_HT_finite_sep})$, $(\ref{I_num_k_HT_sep})$ and $(\ref{I_num_k_HT_finite_sep})$ for the anisotropic representations of the runaway region in $(\ref{CH_RE_region})$, $(\ref{H_RE_region})$, $(\ref{minmax_RE_region})$ and $(\ref{tilde_minmax_RE_region})$, are going to be analyzed. The results of the carried out computations with the mentioned rules are presented in the figure \ref{fig_u_k_ht_RE_sep} for the above mentioned purpose. More specific, the subfigure \ref{u_k_ht_RE_sep_main} depicts the results for the ITER-simulation, as presented in\vspace{-11cm}\linebreak\newpage\noindent 
\vspace{-5.0mm}the reference \cite{Smith_2009} for a constant electron density of \mbox{$n_{e}=n_{_{1}^{2}\mathrm{H}^{+}}=1.06\cdot 10^{20}\,\mathrm{m}^{-3}$} within a singly-ionized deuterium plasma. Moreover, the influences of the presence of a singly-ionized neon impurity density of \mbox{$n_{_{10}^{20}\mathrm{Ne}^{+}}=2.26\cdot n_{e}$} on the time evolution of the normalized mean velocity and the mean rest mass-related kinetic energy density of the developing \textit{hot-tail} runaway electron population are shown in the subfigure \ref{u_k_ht_RE_imp_sep_main}. Additionally, the results from the computations for the isotropic descriptions of the runaway region from the figure \ref{fig_u_k_ht_RE} were plotted. This intended to show the deviation, which originates from the neglection of the pitch-dependency of the runaway region as it was discussed in subsection \ref{pitch_RE_region_subsection} and displayed in figure \ref{fig_RE_region_pitch}. The observation and analysis of the figure \ref{fig_u_k_ht_RE_sep} indicates, that the consideration of a pitch-dependent runaway region leads to minor increases in the mean velocity and the mean kinetic energy density. This can be understood, because the distribution function $(\ref{SV_ht_dist_func})$ from the work of \mbox{\textit{I.\hspace{0.9mm}Svenningsson}} \cite{Svenningsson2020} has no intrinsic pitch-dependency and in addition the runaway region is only marginally decreased by its lower boundary. This small effect of the separatrix $(\ref{xi_sep_def})$ can be validated by means of figure \ref{fig_RE_region_pitch} and is responsible for mean deviations between the isotropic and anisotropic calculation schemes below \mbox{$7.2\,\%$} for the mean velocity and below \mbox{$16.0\,\%$} for the mean kinetic energy density. At this point, one refers to the listings \labelcref{outMATLABoutput_RE_ht_moments_SV_sep,outMATLABoutput_RE_ht_moments_SV_sep_imp} in subsection \ref{output_matlab_appendix_subsection} of the appendix, where the relative deviations between the pitch-dependent and the pitch-independent computations are stated.
\\
Remarkable is, that the pitch-dependent runaway region leads to nonphysical artifacts at \mbox{$t\approx 4\,\mathrm{ms}$} and at \mbox{$t\approx 32.5\,\mathrm{ms}$} for the calculation rules $(\ref{I_num_u_HT_sep})$ and $(\ref{I_num_k_HT_sep})$ for the anisotropic representation of the runaway region in $(\ref{H_RE_region})$, which applies the lower runaway momentum $p_{\star}$. The artifact is not explainable physically and might be reasoned by the limit of the applicability of the approximation $p_{\star}$ of the effective critical momentum, since at the mentioned points in time the electric field is close to the effective critical electric field. Besides, one emphasizes, that the graphs for the mean velocity and the kinetic energy have different asymptotic functions before and after the appearance of the artifact. For \mbox{$t < 32.5\,\mathrm{ms}$} the graph of the results using $p_{\star}$ seems to approximate the data from the calculation rules $(\ref{I_num_u_HT_finite_sep})$ and $(\ref{I_num_k_HT_finite_sep})$ for the anisotropic representations of the runaway region in $(\ref{minmax_RE_region})$ and $(\ref{tilde_minmax_RE_region})$, i.a.\ with the effective critical momentum $p_{min}$, whereas the moments calculated with the integrals $(\ref{I_num_u_HT_sep})$ and $(\ref{I_num_k_HT_sep})$ for the anisotropic description of the runaway region in $(\ref{CH_RE_region})$ with the \textit{Connor-Hastie} critical momentum $p_{c}$ might by the asymptotic function for \mbox{$t > 32.5\,\mathrm{ms}$}. Hence, one can imagine that the calculation scheme including $p_{\star}$, as the choice for the effective critical momentum, leads to results in between the data from the calculation rules with $p_{c}$ and $p_{min}$, which is as well noticed as a general tendency for the regarded disruption simulation.
\\
Moreover, the minimum, maximum and mean value of the magnitude of the current density \mbox{$j_{RE}^{\mathrm{ht},\xi}=e\hspace{-0.3mm}\cdot\hspace{-0.3mm} n_{RE}^{\mathrm{ht},\xi}\hspace{-0.3mm}\cdot\hspace{-0.3mm} u_{RE}^{\mathrm{ht},\xi}$} at \mbox{$r_{\perp}=0.75\,\mathrm{m}$} is presented in the listing \ref{outMATLABoutput_RE_ht_moments_SV_sep}. From this, one can approximate the order of magnitude to be between \mbox{$0.13\,\mathrm{kA/m}^2$} and \mbox{$0.2\,\mathrm{kA/m}^2$}, where the maximum values satisfy \mbox{$2.96\,\mathrm{kA/m}^2<j_{RE}^{\mathrm{ht},\xi}<5.86\,\mathrm{kA/m}^2$}. These values reproduce the expected range for the \textit{hot-tail} runaway current density for ITER disruptions \cite{Hollmann2014}.

In the subfigure \ref{u_k_ht_RE_imp_sep_main}, the influences of the presence of a singly-ionized neon impurity density of \mbox{$n_{_{10}^{20}\mathrm{Ne}^{+}}=2.26\cdot n_{e}$} on the time evolution of the normalized mean velocity and the mean rest mass-related kinetic energy density of the developing \textit{hot-tail} runaway electron population is depicted for the four calculation schemes including the different descriptions of the pitch-dependent runaway region. 
\\
The analysis of the computed data shows the same general difference to the utilization of an isotropic representation of the runaway region, because the results of the computations are higher on average by less than \mbox{$6.0\,\%$} for the mean velocity and by up to \mbox{$14.5\,\%$} for the mean kinetic energy density, according to the listing \labelcref{outMATLABoutput_RE_ht_moments_SV_sep_imp} in subsection \ref{output_matlab_appendix_subsection} of the appendix.
\\
Besides, no artifacts appear for the calculation rules $(\ref{I_num_u_HT_sep})$ and $(\ref{I_num_k_HT_sep})$ for the anisotropic representation of the runaway region in $(\ref{H_RE_region})$, utilizing the lower runaway momentum $p_{\star}$. The choice of $p_{\star}$ as the effective critical momentum is motivated, by the goal to take the effects of partial screening into account, which dominate particularity, if impurities with high nuclear charge are present. Since, the artifacts seem to vanish in this most likely application of the mentioned calculation scheme, one should consider this approach for simulations.

Finally, it can be stated, that the consideration of a pitch-dependent runaway region leads to minor corrections. However, the computational effort should only increase negligibly, because the calculation rules are only modified by a multiplicative function in their integrands. This is why, one should preferably use the calculation rules $(\ref{I_num_u_HT_sep})$, $(\ref{I_num_u_HT_finite_sep})$, $(\ref{I_num_k_HT_sep})$ and $(\ref{I_num_k_HT_finite_sep})$ for the anisotropic representations of the runaway region in $(\ref{CH_RE_region})$, $(\ref{minmax_RE_region})$ and $(\ref{tilde_minmax_RE_region})$. Only in case of the anisotropic description of the runaway region in $(\ref{H_RE_region})$, the calculation rules $(\ref{I_num_u_HT_iso})$ and $(\ref{I_num_k_HT_iso})$ together with the isotropic runaway region in $(\ref{H_RE_region})$ might be better applicable. 
\\
Moreover, the calculation rules $(\ref{I_num_u_HT_finite_sep})$ and $(\ref{I_num_k_HT_finite_sep})$ should be used together with the anisotropic representations of the runaway region in $(\ref{tilde_minmax_RE_region})$, which applies the modified maximum runaway momentum $\tilde{p}_{max}$, instead of the conditions $(\ref{minmax_RE_region})$ in connection with $p_{max}$. This statement holds, if the required higher precision for the computations with $p_{max}$ leads to an intolerable computational effort for the received gain in physical accuracy. For this, the analysis of the application of the mentioned calculation schemes in self-consistent simulations is essential and hence emphasized again.

\clearpage

%% file: Calculation_of_the_moments_of_the_avalanche_runaway_electron_distribution_function.tex
\chapter{Calculation of the moments of \textit{avalanche} runaway electron distribution functions}\label{avalanche_chapter}

The avalanche generation mechanism was found to be the dominant source of runaway electrons during a disruption in a large tokamak device \cite{Hender_2007,Rosenbluth_1997}. This was discussed in detail in section \ref{Avalanche_subsection}, revealing that in particular larger fusion reactors like ITER are vulnerable to damages, resulting from impacts of a runaway beam, which mainly consists of avalanche runaway electrons \cite{HesslowPHD}. 
\\
In consequence, physically accurate and preferably efficient simulation tools are needed to increase the understanding of the appearance, generation and characteristics of avalanche runaway electron populations, in order to develop strategies, controlling algorithms and further tools for the prevention, prediction and mitigation of disruptions and in particular runaway electron beams with a non-negligible damage potential.

This leads to the finding from section \ref{kin_equa_section}, that in order to develop efficient and accurate simulation codes, it is of interest to calculate the moments of a distribution function. In the case of the avalanche runaway electrons, the calculation of such quantities, could utilize analytically or numerically given distribution functions, which might even be based on experimental data. Furthermore, one can imagine the computation of certain moments like the mean velocity or the mean  kinetic energy density for a wide parameter space. Those results could then enhance the efficiency and applicability of existing simulation codes or might find their usage as training sets for neural networks with the goal of improved simulations on the basis of machine learning. Eventually, it is also thinkable, that well-chosen moments are used as a criterion, in order to decide, when certain assumptions are useful or to what extent physical phenomena have to be simulated. At that, they should improve the understanding of the behaviour of physical quantities, if certain parameters are changed.  

Hereinafter, two models will be evaluated, which provide analytic distribution functions based on the growth rates from section \ref{Avalanche_subsection}. Namely, the \textit{Rosenbluth-Putvinski} model with the growth rate $\Gamma_{ava}$ from (\ref{growth_rate_Avalanche}) and the \textit{Hesslow} model with the growth rate $\Gamma^{\hspace{0.3mm}\mathrm{scr}}_{ava}$ from (\ref{growth_rate_Avalanche_Hesslow}) are considered. On that point, one first defines and analyses the distribution functions derived in those models. Second, an approach for the computation of the mean velocity and the mean kinetic energy density, which are connected to the first and second moment of a distribution function, is stated for each model. With those approaches, calculations are carried out and discussed, in order to comment the applicability and efficiency of the models, so that the derived calculation rules can be used in existing simulations.  

\section{Distribution functions for the \textit{avalanche} runaway electron generation}\label{avalanche_dist_section}

Physical quantities like the density, the mean velocity or the mean mass-related kinetic energy density of a runway electron population are related to certain moments of a distribution function. Those determining functions shall be analysed in the following subsection. At this, the \textit{Rosenbluth-Putvinski} model with its completely analytically representable avalanche runaway electron distribution function $f_{RE}^{\textup{ava}}$, derived by \mbox{\textit{T.\hspace{0.9mm}Fülöp et \hspace{-0.4mm}al.}} in reference \cite{REdistfuncderivation}, is evaluated first. Based on this, the avalanche runaway electron distribution function $\tilde{f}_{0,RE}^{\textup{ava,scr}}$, proposed by \textit{P.\hspace{0.9mm}Svensson} in the paper \cite{Svensson_2021}, is investigated, since its associated growth rate $(\ref{growth_rate_Avalanche_Hesslow})$, as defined in \cite{Hesslow_2019} by \textit{L.\hspace{0.9mm}Hesslow} and discussed in the subsection \ref{Avalanche_subsection}, extends the \textit{Rosenbluth-Putvinski} growth rate $(\ref{growth_rate_Avalanche})$. Hence, the distribution function $\tilde{f}_{0,RE}^{\textup{ava,scr}}$ in the \textit{Hesslow} model can also be seen as a modification or improvement of the \textit{Rosenbluth-Putvinski} model, due to the fact that it includes the effects of partial screening, which were introduced in section \ref{part_screen_section}.

\subsection{Distribution function for the \textit{avalanche} runaway electron generation in the \textit{Rosenbluth-Putvinski} model}\label{RP_avalanche_dist_subsection}

\begin{figure}[H]
\begin{center}
\includegraphics[trim=36 165 39 94,width=0.99\textwidth,clip]{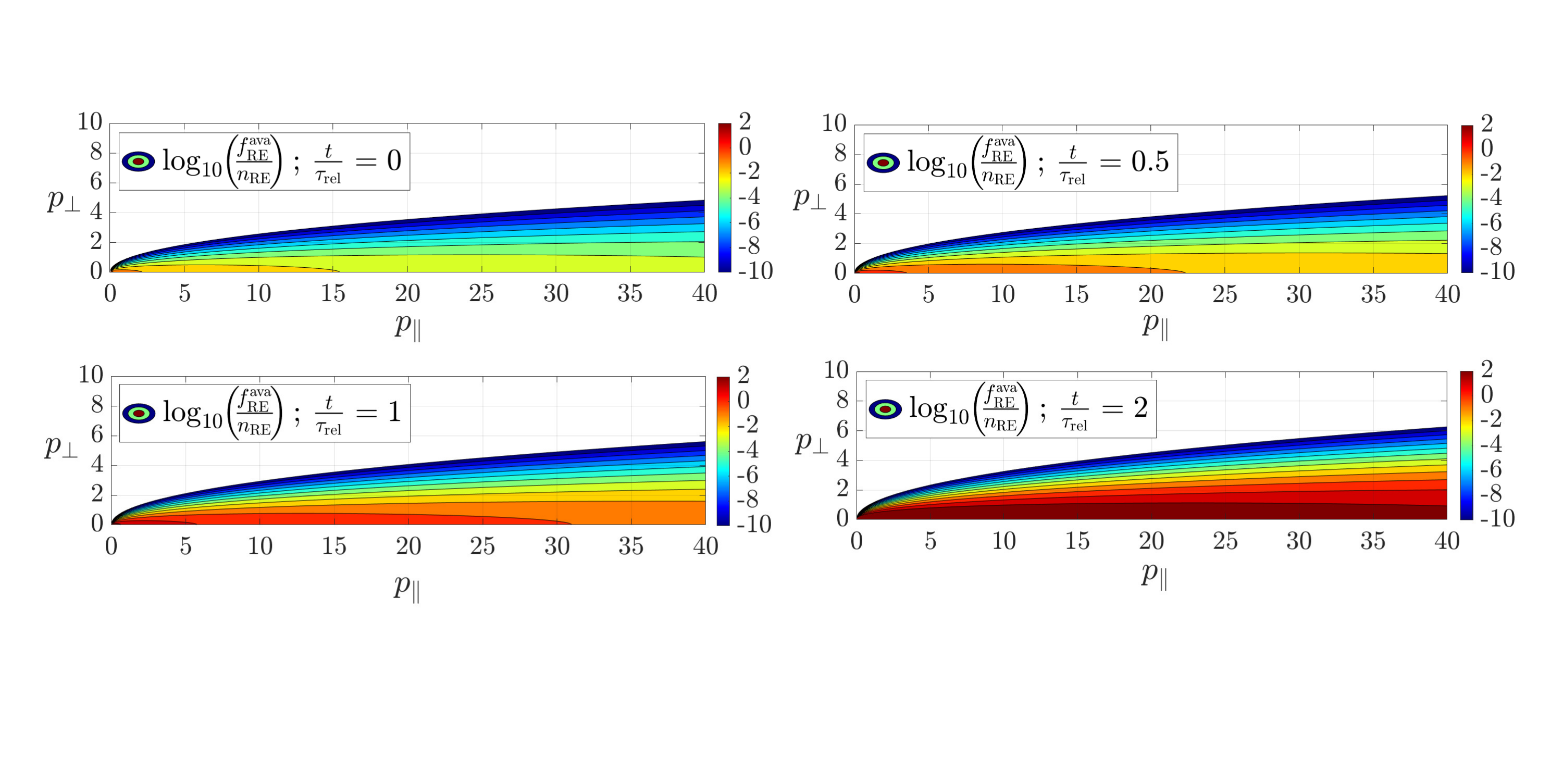}
\caption[Snapshots of the time-evolving contour plot of the analytical \textit{Rosenbluth-Putvinski} avalanche runaway electron distribution function $f_{RE}^{\mathrm{ava}}$ for\linebreak\mbox{$k_{B}\,T_{e}=100 \,\textup{eV}$}, \mbox{$n_{e}=10^{20}\,\textup{m}^{-3}$}, \mbox{$Z_{eff}=1.0$} and \mbox{$E_{\|}=10\,\textup{V/m}$}.]{Snapshots of the time-evolving contour plot\protect\footnotemark{} of the analytic \textit{Rosenbluth-Putvinski} avalanche runaway electron distribution function $f_{RE}^{\mathrm{ava}}$ for \mbox{$k_{B}\,T_{e}=100 \,\textup{eV}$}, \mbox{$n_{e}=10^{20}\,\textup{m}^{-3}$}, \mbox{$Z_{eff}=1.0$} and \mbox{$E_{\|}=10\,\textup{V/m}$}.}
\label{fig_ava_dist_func_RP}
\end{center}
\end{figure}

The \textit{Rosenbluth-Putvinski} avalanche runaway electron distribution function was introduced by \mbox{\textit{T.\hspace{0.9mm}Fülöp et \hspace{-0.4mm}al.}} in 2006, as a time-dependent momentum space distribution function with respect to two momentum coordinates. At this, the implied two-dimensional momentum space was described in the section \ref{mom_space_coord_section}, where the two coordinates are the component of the momentum vector in parallel and perpendicular direction to the local magnetic field vector, as depicted in figure \ref{fig_mom_coord}. According to the publication \cite{REdistfuncderivation} by \mbox{\textit{T.\hspace{0.9mm}Fülöp et \hspace{-0.4mm}al.}}, one can define the \textit{Rosenbluth-Putvinski} distribution function by the following expression:\vspace*{-3.5mm}
\begin{equation}\label{dist_func_RP_ava_def}
f_{RE}^{\textup{ava}}(p_{\|},\,p_{\perp},\,t)=\frac{n_{\mathrm{RE}} \,\tilde{E}}{\pi\,c_{\mathrm{Z}_{\mathrm{eff}}}\ln{\hspace{-0.45mm}\Lambda_{rel}}}\hspace{-0.2mm}\cdot\hspace{-0.35mm}\dfrac{1}{p_{\|}}\hspace{-0.35mm}\cdot\hspace{-0.2mm}\exp{\left(\dfrac{2\,(\hat{E}-1)}{c_{\mathrm{Z}_{\mathrm{eff}}}\ln{\hspace{-0.45mm}\Lambda_{rel}}}\hspace{-0.45mm}\cdot\hspace{-0.45mm}\dfrac{t}{\tau_{rel}}-\dfrac{p_{\|}}{c_{\mathrm{Z}_{\mathrm{eff}}}\ln{\hspace{-0.45mm}\Lambda_{rel}}}-\tilde{E}\hspace{-0.35mm}\cdot\hspace{-0.35mm}\dfrac{p_{\perp}^{2}}{p_{\|}}\right)}\,,
\end{equation} 
\vspace*{-7.0mm}\\where the absolute value of the parallel component of the electric field with respect to the magnetic field $E_{\|}\coloneqq \vert E_{\|}\vert$ is used as the representing parameter for the accelerating electric field strength. Moreover, the abbreviations:\vspace*{-3.5mm}
\begin{equation}\label{abrev_dist_func_RP_ava_def}
\hat{E}\coloneqq\dfrac{  E_{\|} }{E_{\mathrm{c}}}\;\;;\;\;\tilde{E}\coloneqq\frac{\hat{E}-1}{2\,(Z_{eff}+1)}\;\;;\;\;c_{\mathrm{Z}_{\mathrm{eff}}}\coloneqq\sqrt{\dfrac{3\,(Z_{\mathrm{eff}}+5)}{\pi}} 
\end{equation}
\vspace*{-8.0mm}\\are introduced, while the physical parameters $Z_{\mathrm{eff}}$, $\ln{\hspace{-0.45mm}\Lambda_{rel}}$ and $\tau_{rel}$ are defined in the expressions $(\ref{Z_eff_def})$, $(\ref{CoulombLogrel})$ and $(\ref{tau_rel})$. 
\\
The general proportionalities $f_{RE}^{\textup{ava}}\propto\textup{e}^{t}$ and $f_{RE}^{\textup{ava}}\propto\textup{e}^{-p_{\perp}^2}$ in $(\ref{dist_func_RP_ava_def})$ for a constant critical electric field mean, that with linearly progressing time the distribution function increases exponentially, while it decays exponentially for large squares of the orthogonal momentum. This indicates a concentration of the runaway electrons around smaller perpendicular momenta in momentum space. However, the exponential growth in time in reality is impaired by the critical electric field and the accelerating field, which will eventually close the runaway region for \mbox{$\hat{E}-1=E_{\|}-E_{c} \rightarrow 0$}, during a disruption as it was shown previously in the figure \ref{fig_RE_mom_bound}. The $p_{\|}$-dependency and the verification of the deduced predictions, concerning the time and orthogonal momentum behaviour, can be seen, with the help of a visualization of the distribution function. Hence, a \mbox{\textsc{MATLAB}-script$^{\ref{RP_ava_dist_func_fig_footnote}}$} resolves the momentum space dependency by means of contour plots, while the time evolution is expressed through snapshots. The results are computed for typical tokamak plasma parameters and can be viewed in figure \ref{fig_ava_dist_func_RP}. 
\vspace{0.5mm}
\footnotetext{\label{RP_ava_dist_func_fig_footnote} The depiction in figure \ref{fig_ava_dist_func_RP} was generated by means of the \textsc{MATLAB}-script\\ \hspace*{8.7mm}\qq{\texttt{RE_ava_dist_func_RP.m}}, which can be viewed in the digital appendix together with its\\ \hspace*{8.7mm}output \qq{\texttt{output_RE_ava_dist_func_RP.txt}}.}
\vspace*{-8.5mm}
Note, that in figure \ref{fig_ava_dist_func_RP} all quantities are dimensionless and hence no units are given. The corresponding console output is shown in subsection \ref{output_matlab_appendix_subsection} of the appendix and states the critical momentum, the critical electric field, the \textit{Dreicer} field and the slide-away field, calculated from the definitions of those characteristic quantities from section \ref{RE_phenom_section}. In conclusion, one can confirm the estimated concentration of the runaway electrons at lower perpendicular momenta, where \mbox{$p_{\|}\gg p_{\perp}$}. Furthermore, one notices the maximum plateau of the distribution, which grows in time and parallel momentum, representing the non-trivial parallel momentum behaviour in combination with the expected exponentially behaving time evolution. In addition, it should be remarked, that the distribution function solely models runaway electrons moving in the same direction as the local magnetic field vector and the accelerating electric field, because \mbox{$p_{\|}\in [0,\,\infty)$}, although in reality the number of anti-parallel moving runaway electrons is not identically zero \cite{Landreman2014}.  
\\
Finally, it is also important to mention, that the distribution function was originally derived for a homogeneous magnetic field. Consequently, the distribution function should be applied for parameter scenarios, as present near the magnetic axis of the tokamak, where the inhomogeneity in the magnetic field over a flux surface is less distinctive and the assumption of a constant magnetic field holds. Due to the fact, that runaway electron populations typically appear in beam-like structures, mainly move along the magnetic field lines, because of their parallel momentum concentration, and are primarily generated in the core of the plasma, one can generally use the distribution function from \mbox{\textit{T.\hspace{0.9mm}Fülöp et \hspace{-0.4mm}al.}} for the simulation of avalanche runaway electrons.

The normalization of the distribution function in the \textit{Rosenbluth-Putvinski} model can be derived from the zeroth moment, which is related to the runaway electron density through the definition $(\ref{zeroth_moment})$. Consequently, one has to evaluate the following two-dimensional integration and obtains:\vspace*{-3.0mm}
\begin{equation}\label{RE_density_integral_RP}
n_{RE}(t)\hspace{-0.4mm}=\hspace{-6.4mm}\displaystyle{\int\limits_{p_{\|}=p_{\|,\mathrm{min}}}^{\infty}\hspace{-0.8mm}\int\limits_{p_{\perp}=0}^{\infty}}\hspace{-2.5mm} f_{RE}^{\textup{ava}}(p_{\|},\,p_{\perp},\,t)\,2\pi\,p_{\perp}\mathrm{d}p_{\perp}\mathrm{d}p_{\|}
\hspace{-0.6mm}\overset{\text{\hspace{-0.35mm}\cite{study_thesis}}}{=}\hspace{-0.6mm} n_{\mathrm{RE}}\hspace{-0.13mm}\cdot\hspace{-0.01mm}  \exp{\hspace{-1mm}\left(\hspace{-0.9mm}\dfrac{2\hspace{0.3mm}(\hat{E}-1)\,t\,\tau_{rel}^{-1}\hspace{-0.3mm}-\hspace{-0.15mm}p_{\|,\mathrm{min}}}{c_{\mathrm{Z}_{\mathrm{eff}}}\ln{\hspace{-0.45mm}\Lambda_{rel}}}  \hspace{-0.9mm}\right)} .
\end{equation}
\vspace*{-6.0mm}\\The analytic derivation was carried out in detail in previous work \cite{study_thesis}, where a finite lower integration bound \mbox{$p_{\|,\mathrm{min}}>-\infty$} had to be introduced, in order to ensure a finite result. In consequence, this leads to a normalization of $ f_{RE}^{\textup{ava}}$ to $n_{RE}(t)$, if the condition, which follows from $(\ref{RE_density_integral_RP})$ and includes the lower parallel momentum integration boundary $p_{\|,\mathrm{min}}$, is satisfied:\vspace*{-5.0mm}
\begin{equation}\label{normalization_conditon_RP}
p_{\|,\mathrm{min}}\,=\,2\hspace{-0.2mm}\cdot\hspace{-0.4mm}\left(\hspace{-0.4mm}\hat{E}-1\right)\hspace{-0.6mm} \cdot\hspace{-0.7mm}\dfrac{t}{\tau_{rel}}\,\underset{(\ref{abrev_dist_func_RP_ava_def})}{\overset{(\ref{E_crit})}{=}}\,2\hspace{-0.2mm}\cdot\hspace{-0.4mm}\dfrac{e\left(\hspace{-0.4mm}E_{\|}-E_{c}\right)t}{m_{e0}\,c}\,\eqqcolon\, 2\hspace{-0.2mm}\cdot\hspace{-0.4mm}\dfrac{\delta p(t)}{p_{norm}}  \,.
\end{equation} 
\vspace*{-7.0mm}\\This condition can be interpreted as the time-dependent momentum difference $\delta p(t)$, normalized to the momentum $p_{norm}$, gained by the runaway electrons. Moreover, one notices, that the condition is important for the modeling of the time-dependent moments of the distribution function, since it changes the integration domain.

Although the zeroth moment in equation $(\ref{RE_density_integral_RP})$ can be evaluated analytically, one can use it as a numerical control criterion within an implementation. It can be formulated as follows \cite{study_thesis}:\vspace*{-3.0mm}
\begin{equation}\label{check_n_RE_ava_def_general}
1\underset{}{\overset{(\ref{normalization_conditon_RP})}{\equiv}}\hspace{-0.3mm}\exp{\hspace{-1mm}\left(\hspace{-0.8mm}\dfrac{2\hspace{0.3mm}(\hat{E}-1)\,t\,\tau_{rel}^{-1}\hspace{-0.3mm}-\hspace{-0.15mm}p_{\|,\mathrm{min}}}{c_{\mathrm{Z}_{\mathrm{eff}}}\ln{\hspace{-0.45mm}\Lambda_{rel}}}  \hspace{-0.8mm}\right)}\hspace{-1.0mm} \underset{(\ref{RE_density_integral_RP})}{\overset{!}{=}}\hspace{-0.0mm}\displaystyle{\int\limits_{p_{\|}=p_{\|,\mathrm{min}}}^{\infty}\int\limits_{p_{\perp}=0}^{\infty}}\hspace{-0.8mm} \dfrac{2\pi\hspace{0.25mm}p_{\perp}}{n_{\mathrm{RE}} }\hspace{-0.3mm}\cdot\hspace{-0.3mm}f_{RE}^{\textup{ava}}(p_{\|},\,p_{\perp},\,t)\,\mathrm{d}p_{\perp}\mathrm{d}p_{\|}
\,,
\end{equation}
\vspace*{-6.5mm}\\by means of the expressions in $(\ref{RE_density_integral_RP})$, and is applicable for all times $t$, if the condition for from $(\ref{normalization_conditon_RP})$ is satisfied. 
\\
In the subsequent analysis, one focuses on the steady-state, implying \mbox{$t=0\,\mathrm{s}$} in $(\ref{dist_func_RP_ava_def})$, in order to ensure the comparability to the non-time-dependent distribution function in the \textit{Hesslow} model, as it is introduced in the next subsection \ref{Hesslow_avalanche_dist_subsection}. Therefore, the criterion from $(\ref{check_n_RE_ava_def_general})$ is modified, so that the simplified version reads:\vspace*{-3.5mm}
\begin{equation}\label{check_n_RE_ava_def}
1 \underset{\nobarfrac{(\ref{dist_func_RP_ava_def}),(\ref{normalization_conditon_RP}),}{(\ref{check_n_RE_ava_def_general})}}{\overset{!}{=}}  \dfrac{2\,\tilde{E}}{c_{\mathrm{Z}_{\mathrm{eff}}} \ln{\hspace{-0.45mm}\Lambda_{rel}}}\,\underbrace{\displaystyle{\int\limits_{p_{\|}=p_{\|,min}}^{\infty}\int\limits_{p_{\perp}=0}^{\infty}} \dfrac{p_{\perp}}{p_{\|}}\hspace{-0.45mm}\cdot\hspace{-0.3mm}\exp{\hspace{-0.9mm}\left(\hspace{-0.95mm}-\dfrac{p_{\|}}{c_{\mathrm{Z}_{\mathrm{eff}}} \ln{\hspace{-0.45mm}\Lambda_{rel}}}\hspace{-0.25mm}-\hspace{-0.25mm}\tilde{E}\hspace{-0.4mm}\cdot\hspace{-0.4mm}\dfrac{p_{\perp}^{2}}{p_{\|}}\hspace{-0.7mm}\right)}\mathrm{d}p_{\perp}\mathrm{d}p_{\|}}_{\coloneqq\,\textup{I}^{\textit{n}^{\hspace{0.25mm}\mathrm{ava}}_{\mathrm{RE}}}_{\,\mathrm{num}}}
\end{equation}
\vspace*{-6.5mm}\\Note, that \mbox{$p_{\|,min}=0$} is required to fulfill the condition $(\ref{normalization_conditon_RP})$ for the steady-state with \mbox{$t=0\,\mathrm{s}$}, although the lower parallel momentum boundary in $\ref{check_n_RE_ava_def}$ is kept in a more general from, so that the subsequently stated computation rules also hold for \mbox{$p_{\|,min}(t)\neq 0$} from $(\ref{normalization_conditon_RP})$ in time-dependent simulations.
\\
By application of the two substitutions:\vspace*{-3.0mm}
\begin{equation}\label{substitutions_num_k_again}
\begin{split}
\begin{gathered}
p_{\|}=p_{\|,\mathrm{min}}+\dfrac{w}{1-w} \;;\;\dfrac{\mathrm{d}p_{\|}}{\mathrm{d}w}= \dfrac{1}{(1-w)^2}\;;\; w(p_{\|}=p_{\|,\mathrm{min}})=0\;,\;w(p_{\|}\rightarrow\infty)=1\,;
\\ 
p_{\perp}=\dfrac{z}{1-z} \;;\;\dfrac{\mathrm{d}p_{\perp}}{\mathrm{d}z}= \dfrac{1}{(1-z)^2}\;;\; z(p_{\perp}=0)=0\;,\;z(p_{\perp}\rightarrow\infty)=1\,,
\end{gathered}
\end{split}
\end{equation} 
\vspace*{-6.0mm}\\a conveniently computable expression for the occurring integral $\textup{I}^{\,n^{\hspace{0.25mm}\mathrm{ava}}_{\mathrm{RE}}}_{\,\mathrm{num}}$ can be obtained:\vspace*{-3.2mm}
\begin{equation}\label{check_n_RE_ava_Integral}
\textup{I}^{\,n^{\hspace{0.25mm}\mathrm{ava}}_{\mathrm{RE}}}_{\,\mathrm{num}}\,=\displaystyle{\int\limits_{w=0}^{1}\int\limits_{z=0}^{1}} \,\frac{z\hspace{-0.4mm}\cdot\hspace{-0.4mm}\exp{\left(-\frac{p_{\|,min}+\frac{w}{1-w}}{c_{\mathrm{Z}_{\mathrm{eff}}}\cdot\;\ln{\hspace{-0.45mm}\Lambda_{rel}}}-\frac{\tilde{E} \,\cdot \,\left(\frac{z}{1-z}\right)^2}{p_{\|,min}+\frac{w}{1-w}}\right)}}{(p_{\|,min}\hspace{-0.4mm}\cdot\hspace{-0.4mm}(1-w)+w)\hspace{-0.4mm}\cdot\hspace{-0.4mm}(1-w)\hspace{-0.4mm}\cdot\hspace{-0.4mm}(1-z)^3} \;\mathrm{d}z\,\mathrm{d}w
\end{equation}  
\vspace*{-6.0mm}\\The integral $\textup{I}^{\,n^{\hspace{0.25mm}\mathrm{ava}}_{\mathrm{RE}}}_{\,\mathrm{num}}$ is then evaluable by means of a programming language like\linebreak\mbox{\textsc{MATLAB}}, with the help of a numerical two-dimensional integration routine or an appropriate nested quadrature formula with reference to \textit{Fubini's} theorem. Hereinafter, the \textsc{MATLAB}-routine \qq{\texttt{integral2}} \cite{integral2} is used for the analysed steady state with $p_{\|,min}=0$. This routine has a default precision of $10^{-6}$ and either performs an iterated integration with the \textsc{MATLAB}-function \qq{\texttt{integral}} \cite{integral}, based on a global adaptive quadrature and said default error tolerance, or transforms the integration domain to a rectangular shape and subdivides into smaller rectangles. Multiple examples of the application of this test criterion during the computation of the moments of distribution functions, are shown in the outputs displayed in the listings \cref{MATLABoutput_plot_p_scr_E3,MATLABoutput_plot_p_scr_E10,MATLABoutput_plot_p_scr_E30,MATLABoutput_plot_p_scr_E100,MATLABoutput_plot_p_star_E3,MATLABoutput_plot_p_star_E10,MATLABoutput_plot_p_star_E30,MATLABoutput_plot_p_star_E100} of the utilized \textsc{MATLAB}-scripts in subsection \ref{output_matlab_appendix_subsection} of the appendix.

\subsection{Distribution function for the \textit{avalanche} runaway electron generation in the \textit{Hesslow} model}\label{Hesslow_avalanche_dist_subsection}

Since the avalanche generation of runaway electrons is a secondary production mechanism, one can assume, that the electric field exceeds the \textit{Connor-Hastie} critical electric field $E_{c}$ and more precisely the effective critical electric field $E_{c}^{\mathrm{eff}}$, if one includes the effect of partial screening. In consequence, the effective critical momentum $p_{c}^{\mathrm{eff}}$ can be approximated by \mbox{$p_{c}^{\mathrm{eff}}\approx p_{\star}$}, as the root of the function denoted in $(\ref{func_p_c_eff_def})$, because this determining function was derived for \mbox{$E_{\|} \gtrsim E_{\mathrm{c}}^{\mathrm{eff}}>E_{\mathrm{c}}$} by \textit{L.\hspace{0.9mm}Hesslow} \cite{Hesslow_2019}. Moreover, the validity of this approximation was shown in figure \ref{fig_RE_mom_bound} of section \ref{part_screen_section}. This motivates the application of an avalanche runaway electron distribution function, which is based on the \textit{Rosenbluth-Putvinski} model and allows to consider the phenomena, which occur in not fully ionized plasma, as explained in section \ref{part_screen_section}. This request is fulfilled by the distribution function from \textit{P.\hspace{0.9mm}Svensson}, since it is derived from the improved growth rate $\Gamma^{\hspace{0.3mm}\mathrm{scr}}_{ava}$ from \textit{L.\hspace{0.9mm}Hesslow}, considering partial screening effects. 

The mentioned effective one-dimensional avalanche runaway distribution function can be found in zeroth order approximation in the publication \cite{Svensson_2021} from 2021. It ignores radial transport influences to the momentum distribution of the runaway electrons and does not resolve the time evolution or a second momentum dimension. Thus, it represents the integral over the maximum pitch coordinate interval \mbox{$\xi\in[-1,\,1]$} of a two-dimensional distribution function, which would be comparable to the function $(\ref{dist_func_RP_ava_def})$ from \mbox{\textit{T.\hspace{0.9mm}Fülöp et \hspace{-0.4mm}al.}} in the \textit{Rosenbluth-Putvinski} model. Hence, one has \cite{Svensson_2021}:\vspace*{-3.5mm}
\begin{equation}\label{dist_func_H}
\begin{split}
\begin{gathered}
\tilde{f}_{0,RE}^{\hspace{0.25mm}\mathrm{ava,scr}}(p) = \hspace{-0.5mm}\displaystyle{\int\limits_{\xi=-1}^{1} }\hspace{-0.5mm} f_{0,RE}^{\hspace{0.25mm}\textup{ava,scr}}(p,\,\xi)\,2\pi\,p^2\,\mathrm{d}\xi 
\\[-5pt]
= \dfrac{n_{\mathrm{e}}^{\mathrm{tot}}\hspace{-0.25mm}\cdot\hspace{-0.25mm} n_{RE} }{n_{\mathrm{e}}\cdot\ln{\hspace{-0.45mm}\Lambda_{rel}}\hspace{-0.25mm}\cdot\hspace{-0.25mm}\sqrt{4+\tilde{\nu}_{\mathrm{s}}(p_{c}^{\mathrm{eff}})\hspace{-0.25mm}\cdot\hspace{-0.25mm}\tilde{\nu}_{\mathrm{d}}(p_{c}^{\mathrm{eff}})}} \hspace{-0.25mm}\cdot\hspace{-0.25mm}\exp{\hspace{-0.5mm}\left(\hspace{-0.45mm}-\dfrac{n_{\mathrm{e}}^{\mathrm{tot}}\hspace{-0.25mm}\cdot\hspace{-0.25mm}\left(p-p_{c}^{\mathrm{eff}} \right)}{n_{\mathrm{e}}\hspace{-0.25mm}\cdot\hspace{-0.25mm}\ln{\hspace{-0.45mm}\Lambda_{rel}}\hspace{-0.25mm}\cdot\hspace{-0.25mm}\sqrt{4+\tilde{\nu}_{\mathrm{s}}(p_{c}^{\mathrm{eff}})\hspace{-0.25mm}\cdot\hspace{-0.25mm}\tilde{\nu}_{\mathrm{d}}(p_{c}^{\mathrm{eff}})}}\hspace{-0.45mm} \right)}\,,
\end{gathered}
\end{split} 
\end{equation} 
\vspace*{-5.0mm}\\where one calculates the relativistic Coulomb logarithm $\ln{\hspace{-0.45mm}\Lambda_{rel}}$ from $(\ref{CoulombLogrel})$, the relativistic collision time $\tau_{rel}$ from $(\ref{tau_rel})$ and the ultra-relativistic limits \mbox{$\tilde{\nu}_{\mathrm{d}}(p_{c}^{\mathrm{eff}})$} and \mbox{$\tilde{\nu}_{\mathrm{s}}(p_{c}^{\mathrm{eff}})$} of the deflection and the slowing-down frequency, evaluated at the effective critical momentum from $(\ref{nue_s_nue_d_def})$. In addition, one might set \mbox{$p_{c}^{\mathrm{eff}}\approx p_{\star}$} and utilizes the \textsc{MATLAB}-script\footnote{\label{Matlab_Hesslow_script} \qq{\texttt{calculate_E_c_eff.m}}} from \textit{L.\hspace{0.9mm}Hesslow} \cite{Hesslow_2018}.

In general, the one-dimensional steady-state distribution function \mbox{$\tilde{f}_{0,RE}^{\hspace{0.25mm}\mathrm{ava,scr}}(p)$} in the \textit{Hesslow} model is comparable to the \textit{Rosenbluth-Putvinski} model with its one-dimensional distribution function \mbox{$f_{RE}^{\mathrm{ava}}(p_{\|},\,p_{\perp},\,t)$}. This is possible, because although \mbox{$f_{RE}^{\mathrm{ava}}(p_{\|},\,p_{\perp},\,t)$} resolves two momentum dimensions, it does not account for the effects of partial screening. Furthermore, one can model the time evolution of \mbox{$\tilde{f}_{0,RE}^{\hspace{0.25mm}\mathrm{ava,scr}}$} similarly to the distribution function $f_{RE}^{\mathrm{ava}}$ from \mbox{\textit{T.\hspace{0.9mm}Fülöp et \hspace{-0.4mm}al.}}, by inserting the time-dependent factor $(\ref{dist_func_RP_ava_def})$ into the expression $(\ref{dist_func_H})$ for the distribution function, so that:\vspace*{-4.0mm}
\begin{equation}\label{dist_func_H_time}
\tilde{f}_{0,RE}^{\hspace{0.25mm}\mathrm{ava,scr}}(p,\,t) = \tilde{f}_{0,RE}^{\hspace{0.25mm}\mathrm{ava,scr}}(p)\cdot\exp{\left(\dfrac{2\,(\hat{E}_{eff}-1)}{c_{\mathrm{Z}_{\mathrm{eff}}}\ln{\hspace{-0.45mm}\Lambda_{rel}}}\hspace{-0.45mm}\cdot\hspace{-0.45mm}\dfrac{t}{\tau_{rel}}\right)}\;\;;\;\;\hat{E}_{eff}\coloneqq \frac{E}{E_{\mathrm{c}}^{\mathrm{eff}}}\,.
\end{equation} 
\vspace*{-7.5mm}\\In consequence, the two models are physically comparable. Additionally, the comparability of the two distribution functions is given in terms of the computation efficiency of their moments. The reason therefore is, that the moments are integrals over the runaway region in momentum space, whose numerical calculation is more efficient for a one-dimensional integrand function, because the number of function evaluations, and thus the runtime of numerical integration routines, strongly increases with the dimensionality of the integral. In consequence, the distribution function $\tilde{f}_{0,RE}^{\hspace{0.25mm}\mathrm{ava,scr}}$ from $(\ref{dist_func_H_time})$ should lead to less time consuming computations. However, it requires the additional computation of an appropriate lower momentum boundary $p_{c}^{\mathrm{eff}}$ as well as the deflection and slowing-down frequencies at this momentum, requiring iterative calculations as implemented in the mentioned \textsc{MATLAB}-script$^{\ref{Matlab_Hesslow_script}}$.

\begin{figure}[H]
\begin{center}
\includegraphics[trim=20 15 80 40,width=\textwidth,clip]{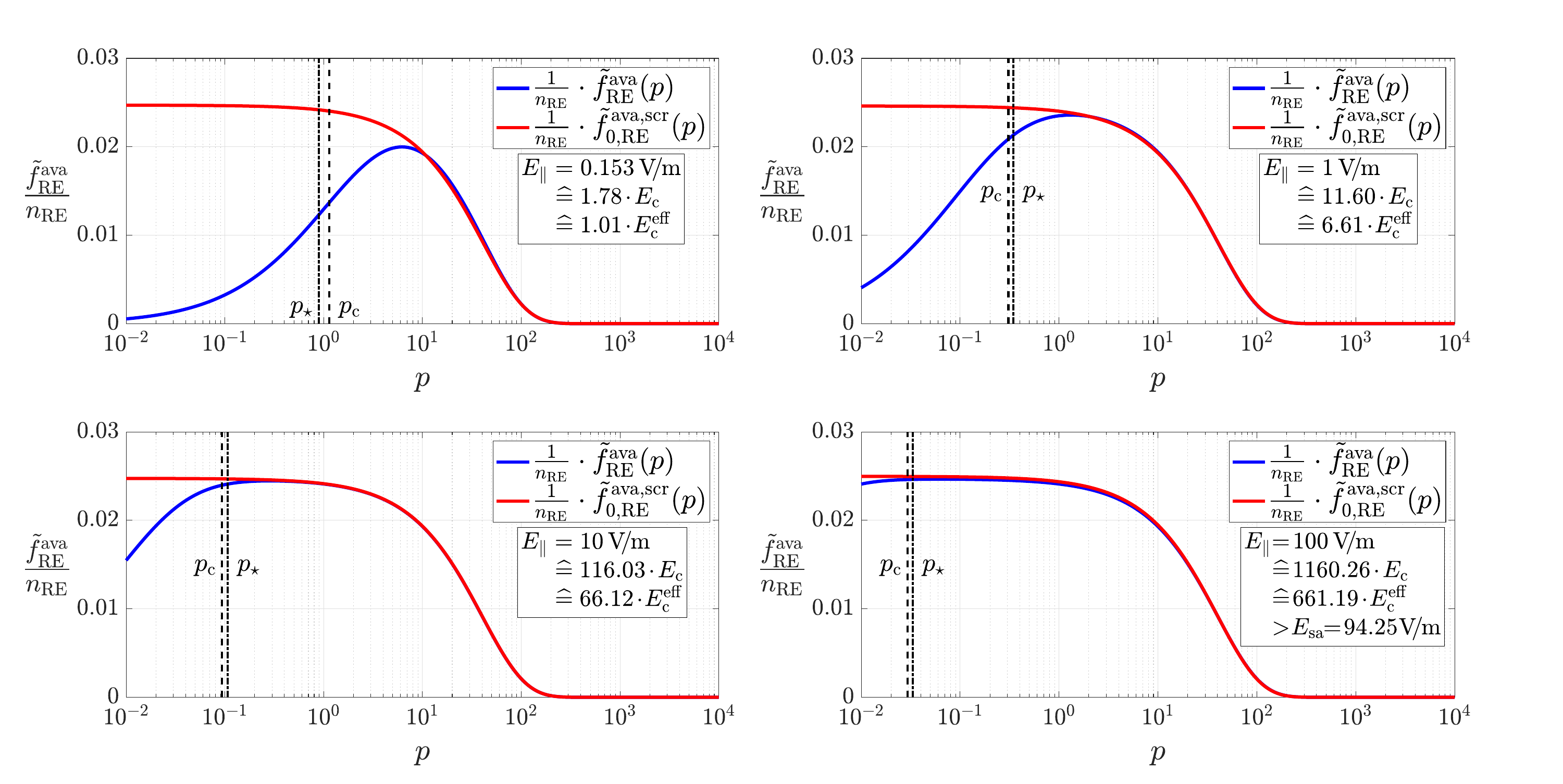} 
\caption[Comparison of the graphs of the avalanche runaway electron distribution function $\tilde{f}_{\mathrm{RE}}^{\mathrm{ava}}(p)$ in the \textit{Rosenbluth-Putvinski} and the \textit{Hesslow} model for different values of the electric field $E_{\|}$, considering a pure deuterium plasma with \mbox{$k_{B}\,T_{e}=100 \,\textup{eV}$}, \mbox{$B=5.25 \,\textup{T}$}, \mbox{$n_{e}=10^{20}\,\textup{m}^{-3}$} and \mbox{$Z_{eff}=1.0$}.]{Comparison of the graphs\protect\footnotemark{} of the avalanche runaway electron distribution function in the \textit{Rosenbluth-Putvinski} and the \textit{Hesslow} model for different values of the electric field $E_{\|}$, considering a pure deuterium plasma with \mbox{$k_{B}\,T_{e}=100 \,\textup{eV}$}, \mbox{$B=5.25 \,\textup{T}$}, \mbox{$n_{e}=10^{20}\,\textup{m}^{-3}$} and \mbox{$Z_{eff}=1.0$}.}
\label{fig_ava_dist_func_H}
\end{center}
\end{figure}

This comparability can be elucidated by a depiction of the one-dimensional steady-state distribution functions \mbox{$\tilde{f}_{RE}^{\hspace{0.25mm}\mathrm{ava}}(p)$} and \mbox{$\tilde{f}_{0,RE}^{\hspace{0.25mm}\mathrm{ava,scr}}(p)$} for different values of the electric field, because the lower momentum boundary and the distribution function $\tilde{f}_{RE}^{\hspace{0.25mm}\mathrm{ava}}$ from the \textit{Rosenbluth-Putvinski} model depends on this parameter. For that purpose, one can plot $\tilde{f}_{0,RE}^{\hspace{0.25mm}\mathrm{ava,scr}}$ directly from its defining equation $(\ref{dist_func_H})$, while the function $\tilde{f}_{RE}^{\hspace{0.25mm}\mathrm{ava}}$ requires the integration of its determining expression from $(\ref{dist_func_RP_ava_def})$ over the pitch coordinate. This can be written as:\vspace*{-6.5mm}
\begin{equation}\label{dist_func_RP_onedimensional}
\tilde{f}_{RE}^{\hspace{0.25mm}\mathrm{ava}}(p) = \hspace{-0.5mm}\displaystyle{\int\limits_{\xi=0}^{1} }\hspace{-0.5mm} f_{RE}^{\mathrm{ava}}(p,\,\xi)\,2\,\pi\,p^2\,\mathrm{d}\xi\,.
\end{equation} 
\vspace*{-6.5mm}\\At that, the steady-state distribution function \mbox{$f_{RE}^{\textup{ava}}(p_{\|},\,p_{\perp},\,t=0\,\mathrm{s})$}, according to $(\ref{dist_func_RP_ava_def})$, is expressed in the coordinates $p$ and $\xi$, which were described alongside their connection to the coordinates $p_{\|}$ and $p_{\perp}$ in section \ref{mom_space_coord_section}. In addition, it was made use of the appropriate momentum space area element from $(\ref{volelem_sphere_2D})$ and the integration interval for $\xi$, which was elaborated in the previous subsection \ref{RP_avalanche_dist_subsection}. The integration was then carried out, by means of the one-dimensional numerical integration \mbox{\textsc{MATLAB}-routine} \qq{\texttt{integral}}, which is based on a global adaptive quadrature and a default error tolerance of $10^{-6}$ \cite{integral}. Thus, the figure \ref{fig_ava_dist_func_H} can be produced with the \mbox{\textsc{MATLAB}-scripts$^{\ref{Matlab_Hesslow_script},\ref{H_ava_dist_func_fig_footnote}}$} for an increasing electric field component parallel to the magnetic field, whereat the console output can be found in the listing \ref{outMATLABoutput_RE_ava_dist_func_H} of subsection \ref{output_matlab_appendix_subsection} in the appendix. In figure \ref{fig_ava_dist_func_H} all quantities are dimensionless and therefore no units are viewable in the graphics. Note, that the last subplot in figure \ref{fig_ava_dist_func_H} corresponds to the slide-away phenomenon, explained in section \ref{RE_phenom_section}, because in this case \mbox{$E_{\|}>E_{sa}$} holds, as one can verify\vspace{-7cm}\linebreak\newpage\noindent
\footnotetext{\label{H_ava_dist_func_fig_footnote} The diagrams in figure \ref{fig_ava_dist_func_H} were computed with the help of the \textsc{MATLAB}-script\\ \hspace*{8.7mm}\qq{\texttt{RE_ava_dist_func_H.m}}, which is stored in the digital appendix together with its output\\ \hspace*{8.7mm}\qq{\texttt{output_RE_ava_dist_func_H.txt}}.}
\vspace{-9.0mm}
in the listing \ref{outMATLABoutput_RE_ava_dist_func_H}. As well, it should be remarked, that the electric field dependency is weak for the distribution function $\tilde{f}_{0,RE}^{\hspace{0.25mm}\mathrm{ava,scr}}$ from \textit{P.\hspace{0.9mm}Svensson}, because it is only implicitly given through the governing equation of $p_{c}^{\mathrm{eff}}\approx p_{\star}$. Hereinafter, the figure \ref{fig_ava_dist_func_H} is analysed and discussed with regard to the appropriate choice of the momentum integration bound for the runaway region.

First, one observes that both distribution functions are exponentially suppressed for $p \rightarrow \infty$, as it can be seen as well in their definitions $(\ref{dist_func_RP_ava_def})$ and $(\ref{dist_func_H})$. Therefore the moments are not sensitive to a sufficiently high upper integration bound $p_{max}$. Hence, it is more efficient to set the upper bound to infinity than to carry out an additional computation for $p_{max}$ as the upper boundary of the runaway region in momentum space, as it would be possible on the basis of the equation $(\ref{F_acc})$ in section \ref{part_screen_section}. 
\\
In contrast, the results of the integration are strongly dependent on the lower integration bound, because in the limit $p\rightarrow 0$ the distribution function in the \textit{Hesslow} model reaches its maximum, contrary to the decreasing distribution function $\tilde{f}_{RE}^{\mathrm{ava}}$ from \mbox{\textit{T.\hspace{0.9mm}Fülöp et \hspace{-0.4mm}al.}}. This can be seen in the figure \ref{fig_ava_dist_func_RP} and \ref{fig_ava_dist_func_H} and motivates a thoughtful choice of the critical momentum as the lower integration bound. 
\\
Possible choices for $p_{c}^{\mathrm{eff}}$ are $p_{c}$ from (\ref{p_crit}), $p_{c}^{\mathrm{scr}}$ from (\ref{p_c_scr_def}), $p_{\star}$ defined by (\ref{func_p_c_eff_def}) and $p_{min}$ determined by (\ref{F_acc}), which shall be discussed successively. One should exclude \textit{Connor-Hastie} critical momentum $p_{c}$, due to the fact, that it does not account for partial screening effects and hence models less physical phenomena. In addition, one has to admit, that the computation of $p_{\star}$ and $p_{min}$, as the roots of the non-linear functions $(\ref{func_p_c_eff_def})$ and $(\ref{F_acc})$, requires more runtime than for the evaluation of analytic expressions. Additionally, the non-trivial choice of a starting value is necessary, so that both possibilities do not serve the purpose of an efficient calculation. However, the governing function for $p_{\star}$ is simpler than for $p_{min}$. Moreover, the starting value \mbox{$p_{\star,0}=1$} was suggested by \textit{O.\hspace{0.9mm}Linder} \cite{LinderPHD} and \mbox{$p_{\star,0}=p_{c}^{\mathrm{scr}}$} was used in this thesis. Thus, a computation of $p_{\star}$ with a finite number of iterations should be possible for a wide range of parameters with one of the mentioned starting values. For the utilization of $p_{min}$ a further analysis is required, which explains why hereinafter \mbox{$p_{c}^{\mathrm{eff}}=p_{min}$} is treated as an efficient calculation option. Therefore, the most promising choices are $p_{\star}$ or an approximated analytic expression like $p_{c}^{\mathrm{scr}}$, which might provide a more efficient way to evaluate the lower momentum threshold. 

\begin{figure}[H]
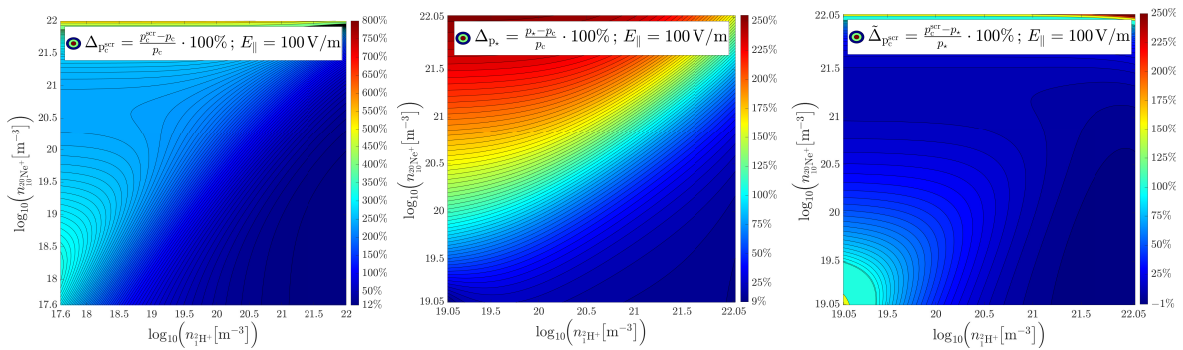

  \centering
  \subfloat{\label{fig_rel_p_c_scr_ava_E100_main0} 
   \includegraphics[trim=310 23 360 21,width=0.322\textwidth,clip]
    {rel_p_c_scr_ava_p_c_scr_E100_2.pdf}}\hfill
  \subfloat{\label{fig_rel_p_star_ava_E100_main0}
    \includegraphics[trim=321 24 355 18,width=0.322\textwidth,clip]
    {rel_p_star_ava_p_star_E100_2.pdf}}\hfill
  \subfloat{\label{fig_tilde_rel_p_c_scr_ava_E100_main0} 
    \includegraphics[trim=310 23 360 21,width=0.329\textwidth,clip]
    {tilde_rel_p_c_scr_ava_p_star_E100_2.pdf}} 
 \caption[Relative deviations $\Delta_{p^{\mathrm{scr}}_{c}}$ and $\Delta_{p_{\star}}$ of the approximations for the effective critical momentum $p^{\mathrm{scr}}_{c}$ and $p_{\star}$ from the \textit{Connor-Hastie} critical momentum $p_{c}$, as well as the relative difference $\tilde{\Delta}_{p^{\mathrm{scr}}_{c}}$ between them, displayed through contour plots over a density parameter space for a singly-ionized deuterium-neon plasma with \mbox{$k_{\mathrm{B}}T_{\mathrm{e}}=10\,\textup{eV}$}, \mbox{$B=5.25\,\textup{T}$}, \mbox{$E_{\|}=100\,\textup{V/m}$} and \mbox{$Z_{\mathrm{eff}}=1$} (larger view in figures \labelcref{fig_rel_p_c_scr_ava,fig_rel_p_star_ava,fig_tilde_rel_p_c_scr_ava} of the appendix).]{Relative deviations $\Delta_{p^{\mathrm{scr}}_{c}}$ and $\Delta_{p_{\star}}$ of the approximations for the effective critical momentum $p^{\mathrm{scr}}_{\mathrm{c}}$ and $p_{\star}$ from the \textit{Connor-Hastie} critical momentum $p_{\mathrm{c}}$, as well as the relative difference $\tilde{\Delta}_{p^{\mathrm{scr}}_{\mathrm{c}}}$ between them, displayed through contour plots\protect\footnotemark{} over a density parameter space for a singly-ionized deuterium-neon plasma with \mbox{$k_{\mathrm{B}}T_{\mathrm{e}}=10\,\textup{eV}$}, \mbox{$B=5.25\,\textup{T}$}, \mbox{$E_{\|}=100\,\textup{V/m}$} and \mbox{$Z_{\mathrm{eff}}=1$} (larger view in figures \labelcref{fig_rel_p_c_scr_ava,fig_rel_p_star_ava,fig_tilde_rel_p_c_scr_ava} of the appendix).}
\label{fig_rel_p_ava}
\end{figure}

Admittedly, this is connected to a deviation between $p_{c}^{\mathrm{scr}}$ and $p_{\star}$, which has to be considered, because it reduces the accuracy of the calculated moments. That is the reason why an evaluation, as displayed e.g.\ in figure \ref{fig_rel_p_ava}, is useful to determine the influence of a certain choice as an approximation of the effective critical momentum. 
\footnotetext{\label{fig_rel_p_ava_footnote} The contour plots were computed with the help of the \textsc{MATLAB}-scripts\\ \hspace*{8.7mm}\qq{\texttt{plot_num_data_densities_p_c_scr_E100.m}} and\\ \hspace*{8.7mm}\qq{\texttt{plot_num_data_densities_p_star_E100.m}}, under utilization of the data from the\\ \hspace*{8.7mm}implementations \qq{\texttt{generate_num_data_densities_p_c_scr_E100.m}} and\\ \hspace*{8.7mm}\qq{\texttt{generate_num_data_densities_p_star_E100.m}}, which are all stored in the digital\\ \hspace*{8.7mm}appendix.}
\vspace{-5.0mm}Based on the analysis of figure \ref{fig_rel_p_ava} one can generalize, that both choices $p^{\mathrm{scr}}_{\mathrm{c}}$ and $p_{\star}$ yield to a greater lower momentum boundary than the \textit{Connor-Hastie} critical momentum $p_{\mathrm{c}}$. Moreover, one might consider the analytical relation for $p^{\mathrm{scr}}_{\mathrm{c}}$ for efficient computations, because it overestimates the physically more accurate values of $p_{\star}$ only up to a factor of two. Since, this deviation or error does not have to propagate uninfluenced into the results of the calculation of the moments, one could save runtime by choosing $p^{\mathrm{scr}}_{\mathrm{c}}$. Nevertheless, the propagation of said error should be investigated, as it is shown in the following sections, in order to relate the deviations in the approximations of the lower momentum integration boundary of the runaway region to relative errors in the calculated moments.

The normalization of the distribution function in the \textit{Hesslow} model is, equivalently to the procedure from section \ref{RP_avalanche_dist_subsection}, deduced from the zeroth moment, which is related to the runaway electron density through the definition $(\ref{zeroth_moment})$. Thus the runaway electron density results from the one-dimensional integral:\vspace*{-3.0mm}
\begin{equation}\label{RE_density_integral_H}
n_{RE}(t)\,= \hspace{-0.75mm}\displaystyle{\int\limits_{p =p_{\mathrm{c}}^{\mathrm{eff}}}^{\infty} } \hspace{-0.35mm} \tilde{f}_{0,RE}^{\hspace{0.25mm}\mathrm{ava,scr}}(p,\,t) \; \mathrm{d}p .
\end{equation}
\vspace*{-6.0mm}\\As mentioned previously, one considers the steady-state for the comparison of the two distribution functions modeling the avalanche runaway electron phenomenon. Hence, the time dependency in $(\ref{dist_func_H})$ is dropped, allowing oneself to prove that \mbox{$\tilde{f}_{0,RE}^{\hspace{0.25mm}\mathrm{ava,scr}}(p)$} can indeed be normalized to the runaway electron density $n_{RE}$:\vspace*{-3.2mm}
\begin{equation}\label{RE_density_integral_H_calc1}
\begin{split}
\begin{gathered}
 n_{RE} = \hspace{-0.75mm}\displaystyle{\int\limits_{p =p_{\mathrm{c}}^{\mathrm{eff}}}^{\infty} } \hspace{-0.35mm} \tilde{f}_{0,RE}^{\hspace{0.25mm}\mathrm{ava,scr}}(p) \; \mathrm{d}p 
 \\[-2pt]
=\hspace{-0.75mm}\displaystyle{\int\limits_{p =p_{\mathrm{c}}^{\mathrm{eff}}}^{\infty} } \hspace{-0.35mm}  \dfrac{n_{\mathrm{e}}^{\mathrm{tot}}\cdot n_{RE}  }{n_{\mathrm{e}}\ln{\hspace{-0.45mm}\Lambda_{rel}}\sqrt{4+\tilde{\nu}_{\mathrm{s}}(p_{c}^{\mathrm{eff}})\tilde{\nu}_{\mathrm{d}}(p_{c}^{\mathrm{eff}})}} \hspace{-0.25mm}\cdot\hspace{-0.25mm}\exp{\hspace{-0.5mm}\left(\hspace{-0.45mm}-\dfrac{n_{\mathrm{e}}^{\mathrm{tot}}\hspace{-0.25mm}\cdot\hspace{-0.25mm}\left(p-p_{c}^{\mathrm{eff}} \right)}{n_{\mathrm{e}}\ln{\hspace{-0.45mm}\Lambda_{rel}}\sqrt{4+\tilde{\nu}_{\mathrm{s}}(p_{c}^{\mathrm{eff}})\tilde{\nu}_{\mathrm{d}}(p_{c}^{\mathrm{eff}})}}\hspace{-0.45mm} \right)} \, \mathrm{d}p .
\end{gathered}
\end{split}
\end{equation}
\vspace*{-5.9mm}\\Note, that the ultra-relativistic limits of the deflection and the slowing-down frequency $\tilde{\nu}_{\mathrm{d}}(p_{c}^{\mathrm{eff}})$ and $\tilde{\nu}_{\mathrm{s}}(p_{c}^{\mathrm{eff}})$ are constants with respect to the integration variable $p$, if they are evaluated for a known and fixed effective critical momentum, before the integration has been conducted. Therefore, one can rewrite the integral from $(\ref{RE_density_integral_H_calc1})$, with the help of a numerical constant, implicitly defined in $(\ref{RE_density_integral_H_calc1})$ and evaluate the occurring standard integral:\vspace*{-6.0mm}
\begin{equation}\label{RE_density_integral_H_calc2}
\begin{split}
\begin{gathered}
 n_{RE} = \zeta\hspace{-0.25mm}\cdot\hspace{-0.25mm} n_{RE} \hspace{-1mm}\displaystyle{\int\limits_{p =p_{\mathrm{c}}^{\mathrm{eff}}}^{\infty} } \hspace{-1.4mm}  \textup{e}^{ -\zeta \cdot \left(p-p_{c}^{\mathrm{eff}} \right)   } \; \mathrm{d}p =\zeta\hspace{-0.25mm}\cdot\hspace{-0.25mm} n_{RE} \hspace{-0.25mm}\cdot\hspace{-0.25mm}\left[-\dfrac{1}{\zeta}\hspace{-0.25mm}\cdot\hspace{-0.25mm}\textup{e}^{-\zeta \cdot \left(p-p_{c}^{\mathrm{eff}} \right)   }\right]^{\infty}_{p=p_{\mathrm{c}}^{\mathrm{eff}}}
 \\
 \hspace{0.4cm} =n_{RE}\cdot\left(\textup{e}^{-\zeta \cdot \left(p_{c}^{\mathrm{eff}}-p_{c}^{\mathrm{eff}}\right)}- \lim\limits_{p\rightarrow\infty} \left\lbrace \textup{e}^{-\zeta \cdot \left(p-p_{c}^{\mathrm{eff}}\right)}  \right\rbrace\right)=n_{RE}\cdot\left(1-0\right)\equiv n_{RE}\,.  \hspace{0.6cm}\blacksquare 
\end{gathered}
\end{split}
\end{equation}
\vspace*{-6.0mm}\\Again, a control condition associated with the zeroth moment from $(\ref{RE_density_integral_H_calc1})$ can be defined:\vspace*{-4.0mm}
\begin{equation}\label{check_n_RE_ava_def_H}
1 \underset{ }{\overset{!}{=}}\textup{I}^{\,n^{\hspace{0.25mm}\mathrm{ava}}_{\mathrm{RE}}}_{\,\mathrm{num,scr}} \coloneqq \dfrac{n_{RE}}{n_{RE}} \underset{ }{\overset{(\ref{RE_density_integral_H})}{=}}\hspace{-0.75mm}\displaystyle{\int\limits_{p =p_{\mathrm{c}}^{\mathrm{eff}}}^{\infty} } \hspace{-0.35mm} \dfrac{1}{n_{RE}}\cdot\tilde{f}_{0,RE}^{\hspace{0.25mm}\mathrm{ava,scr}}(p,\,t) \; \mathrm{d}p 
\end{equation}
\begin{equation*}
\begin{split}
\begin{gathered} 
=\hspace{-2.65mm}\displaystyle{\int\limits_{p =p_{\mathrm{c}}^{\mathrm{eff}}}^{\infty} } \hspace{-0.35mm}  \dfrac{n_{\mathrm{e}}^{\mathrm{tot}}  }{n_{\mathrm{e}}\ln{\hspace{-0.45mm}\Lambda_{rel}}\sqrt{4+\tilde{\nu}_{\mathrm{s}}(p_{c}^{\mathrm{eff}})\tilde{\nu}_{\mathrm{d}}(p_{c}^{\mathrm{eff}})}} \cdot\exp{\hspace{-0.5mm}\left(\hspace{-0.95mm}-\dfrac{n_{\mathrm{e}}^{\mathrm{tot}}\cdot\left(p-p_{c}^{\mathrm{eff}} \right)}{n_{\mathrm{e}}\ln{\hspace{-0.45mm}\Lambda_{rel}}\sqrt{4+\tilde{\nu}_{\mathrm{s}}(p_{c}^{\mathrm{eff}})\tilde{\nu}_{\mathrm{d}}\left(p_{c}^{\mathrm{eff}} \right)}} \hspace{-0.95mm}\right)} \, \mathrm{d}p
\\[3pt]
=\hspace{-1.6mm}\displaystyle{\int\limits_{w =0}^{1} } \hspace{-0.15mm}  \dfrac{n_{\mathrm{e}}^{\mathrm{tot}}\hspace{-0.25mm}\cdot\hspace{-0.25mm}(1-w)^{-2} }{n_{\mathrm{e}}\ln{\hspace{-0.45mm}\Lambda_{rel}}\sqrt{4+\tilde{\nu}_{\mathrm{s}}(p_{c}^{\mathrm{eff}})\tilde{\nu}_{\mathrm{d}}(p_{c}^{\mathrm{eff}})}} \cdot\exp{\hspace{-0.5mm}\left(\hspace{-0.95mm}-\dfrac{n_{\mathrm{e}}^{\mathrm{tot}}\hspace{-0.25mm}\cdot\hspace{-0.25mm} w\hspace{-0.25mm}\cdot\hspace{-0.25mm}\left(1-w\right)^{-1}}{n_{\mathrm{e}}\ln{\hspace{-0.45mm}\Lambda_{rel}}\sqrt{4+\tilde{\nu}_{\mathrm{s}}(p_{c}^{\mathrm{eff}})\tilde{\nu}_{\mathrm{d}}\left(p_{c}^{\mathrm{eff}} \right)}} \hspace{-0.95mm}\right)} \, \mathrm{d}w \,,
\end{gathered}
\end{split}
\end{equation*}
\vspace*{-5.5mm}\\implying a computation rule for the integral $\textup{I}^{\,n^{\hspace{0.25mm}\mathrm{ava}}_{\mathrm{RE}}}_{\,\mathrm{num,scr}}$. In the last equality a substitution was inserted, which is form-invariant to the substitutions presented in $(\ref{substitutions_num_k_again})$, and shall be defined as follows:\vspace*{-2.0mm}
\begin{equation}\label{substitutions_num_k_again_H}
p=p_{c}^{\mathrm{eff}}+\dfrac{w}{1-w} \;;\;\dfrac{\mathrm{d}p }{\mathrm{d}w}= \dfrac{1}{(1-w)^2}\;;\; w\left(p =p_{c}^{\mathrm{eff}}\right)=0\;,\;w(p \rightarrow\infty)=1\,.
\end{equation} 
\vspace*{-7.0mm}\\The integral $\textup{I}^{\,n^{\hspace{0.25mm}\mathrm{ava}}_{\mathrm{RE}}}_{\,\mathrm{num,scr}}$ is thus a computable criterion, which can for instance be calculated with the already introduced \textsc{MATLAB}-routine \qq{\texttt{integral}} \cite{integral} and might be used to ensure, that a certain accuracy is reached. Moreover, it is suitable to determine the order of magnitude of the mean runtime per calculation of a moment. Exemplary results of the computation of the condition $(\ref{check_n_RE_ava_def_H})$ are stated in the listings \cref{MATLABoutput_plot_p_scr_E3,MATLABoutput_plot_p_scr_E10,MATLABoutput_plot_p_scr_E30,MATLABoutput_plot_p_scr_E100,MATLABoutput_plot_p_star_E3,MATLABoutput_plot_p_star_E10,MATLABoutput_plot_p_star_E30,MATLABoutput_plot_p_star_E100} in subsection \ref{output_matlab_appendix_subsection} of the appendix.

\clearpage

\section{Current density of an \textit{avalanche} runaway electron population}\label{ava_j_section}

The current density describes the current strength per unit area and was defined in equation $(\ref{RE_curr_dens_def})$ from section \ref{kin_equa_section}. It is i.a.\ connected to the mean velocity of the particles, which generate the current, in this thesis runaway electrons, and plays an important role in governing equations of runaway electron simulations. In particular, the component of the runaway current density of an avalanche runaway electron population parallel to the magnetic field is of interest, since simulation tools typically model the runaway current to solely move in direction parallel to the magnetic field. On this occasion, the parallel component of the mean velocity is usually set to the speed of light in vacuum, leading to the following approximation of the parallel current density:\vspace{-3.9mm}
\begin{equation}\label{RE_ava_curr_dens_def_general}
j_{\|,\mathrm{RE}}^{\hspace{0.25mm}\mathrm{ava}}= -\,e\cdot n_{\mathrm{RE}}\cdot u^{\mathrm{ava}}_{\|,\mathrm{RE}}\approx -\,e\cdot n_{\mathrm{RE}}\cdot c\;.
\end{equation}  
\vspace{-10.4mm}\\This approximation can be improved, by a calculation of the mean velocity, as the first moment of a distribution function, according to the definition $(\ref{first_moment})$ from section \ref{kin_equa_section}. This is only possible for distribution functions with at least two momentum dimensions, which leads to the fact, that only the magnitude of the mean velocity is defined as a moment in case of a distribution function with at one momentum dimension. However, the behaviour and contribution of the perpendicular component of the current density might provide further insight in the physics of runaway electrons. Furthermore, it should be remarked, that the mean velocity has additional applications, for instance in the description of the radial diffusion of runaway electrons, where the test particle diffusion coefficient is estimated to be linear proportional to the parallel component of the mean runaway electron velocity \cite{Entrop_1998,Rechester_1978} and also often approximated as \mbox{$u_{\|,\mathrm{RE}}\approx c$} \cite{pappPHD}.
\\
Therefore, calculation rules for the parallel and orthogonal component as well as the magnitude of the mean velocity will be derived in the subsequent subsection \ref{RP_avalanche_j_subsection} and \ref{Hesslow_avalanche_j_subsection}. At that, those quantities are expressed as moments of the distribution functions in the \textit{Rosenbluth-Putvinski} and the \textit{Hesslow} model, as introduced in the section \ref{avalanche_dist_section}. Note, that in the \textit{Hesslow} model only the magnitude of the mean avalanche runaway electron velocity is defined, due to the fact that the distribution function from \textit{P.\hspace{0.9mm}Svensson} already contains the pitch integration, neglects radial transport influences and is thus one-dimensional with respect to momentum, in accordance with the analysis in subsection \ref{Hesslow_avalanche_dist_subsection}. Consequently, the parallel and the perpendicular mean velocity moment for avalanche runaway electrons can only be obtained from the \textit{Rosenbluth-Putvinski} distribution function by \mbox{\textit{T.\hspace{0.9mm}Fülöp et \hspace{-0.4mm}al.}} from subsection \ref{RP_avalanche_dist_subsection}, which was shown in the study thesis \cite{study_thesis}, stored in the digital appendix. 
\\
Hereinafter, the current density magnitude and hence the mean velocity magnitude is evaluated based on the results of computations, which apply the derived calculation rules for the two considered models of the avalanche runaway electron phenomenon. Thus the comparability of the approaches is ensured and the discussion and evaluation in subsection \ref{comparison_avalanche_j_subsection} is possible, although for the most simulation codes, the parallel component of the current density, would be more interesting. But since the magnitude of the momentum of a runaway population receives its main contribution from its parallel component, as shown in figure \ref{fig_ava_dist_func_RP}, one can expect the current density magnitude to differ only physically non-significantly from its parallel component. This can be understood in detail in the study thesis \cite{study_thesis}.

\subsection{Current density of an \textit{avalanche} runaway electron population in the \textit{Rosenbluth-Putvinski} model}\label{RP_avalanche_j_subsection}

The magnitude of the two-dimensional current density vector of an avalanche runaway electron population in the cylindrical gyro-averaged momentum space coordinate system, as described in section \ref{mom_space_coord_section}, can be calculated from the first moment of a distribution function. According to the definition $(\ref{first_moment})$ of the first moment from section \ref{kin_equa_section}, the relation:\vspace*{-6.2mm}
\begin{equation}\label{u_ava_moment_def1}
 u_{\,\mathrm{RE}}^{\,\textup{ava}}  (\mathbf{r},\,t)= \dfrac{1}{n_{\mathrm{RE}} (\mathbf{r},\,t)}\, \displaystyle{\iiint\limits_{\mathbb{R}^3}} \vert\mathbf{v}\vert \,f_{RE}^{\textup{ava}}(\mathbf{r},\,\mathbf{p},\,t)\,\mathrm{d}^3p 
\end{equation}
\vspace*{-6.8mm}\\represents a calculation rule for the magnitude of the mean velocity of an avalanche runaway electron population. In the \textit{Rosenbluth-Putvinski} model, the distribution function to be used, can be recapitulated in equation $(\ref{dist_func_RP_ava_def})$ of section \ref{RP_avalanche_dist_subsection}. Thus, by means of the definition of the current density from $(\ref{RE_curr_dens_def})$, the expression for the mean velocity $(\ref{u_ava_moment_def1})$, the relation $(\ref{p_norm_gamma_def})$, the notation \mbox{$v\coloneqq\vert\mathbf{v}\vert$} and the momentum space volume element from $(\ref{volelem_cyl_2D})$, one can write:\vspace*{-3.9mm}
\begin{equation}\label{j_ava_RE_mag_def}
\begin{split}
\begin{gathered}
j_{RE}^{\hspace{0.25mm}\textup{ava}}\underset{(\ref{volelem_cyl_2D}),(\ref{u_ava_moment_def1})}{\overset{(\ref{RE_curr_dens_def})}{=}}-\,e\,\displaystyle{\int\limits_{p_{\|}=-\infty}^{\infty}\int\limits_{p_{\perp}=0}^{\infty}} v \cdot f_{RE}^{\textup{ava}}(p_{\|},\,p_{\perp},\,t)\,2\pi\,p_{\perp} \mathrm{d}p_{\perp} \mathrm{d}p_{\|}
\\[0pt]
\overset{(\ref{p_norm_gamma_def})}{=}-\,2\pi\,c\,e\,\displaystyle{\int\limits_{p_{\|}=p_{\|,min}}^{\infty}\int\limits_{p_{\perp}=0}^{\infty}} p_{\perp}\sqrt{\frac{ p_{\|}^{2}+p_{\perp}^2}{1+p_{\|}^{2}+p_{\perp}^2}}\cdot f_{RE}^{\textup{ava}}(p_{\|},\,p_{\perp},\,t) \,\mathrm{d}p_{\perp} \mathrm{d}p_{\|}
\\[0pt]
\underset{(\ref{abrev_dist_func_RP_ava_def})}{\overset{(\ref{dist_func_RP_ava_def})}{=}}\hspace{-1.75mm}-\dfrac{2\hspace{0.3mm}c\hspace{0.3mm}e\hspace{0.3mm}n_{RE}\hspace{0.3mm}\tilde{E}}{c_{\mathrm{Z}_{\mathrm{eff}}} \ln{\hspace{-0.45mm}\Lambda_{rel}}}\,\textup{e}^{\,\frac{2\,(\hat{E}-1)}{c_{\mathrm{Z}_{\mathrm{eff}}} \ln{\hspace{-0.45mm}\Lambda_{rel}}} \frac{t}{\tau_{rel}}}\hspace{-6.55mm}\underbrace{\displaystyle{\int\limits_{p_{\|}=p_{\|,min}}^{\infty}\int\limits_{p_{\perp}=0}^{\infty}} \hspace{-0.65mm}\frac{p_{\perp}}{ p_{\|}}\sqrt{\hspace{-0.25mm}\frac{ p_{\|}^{2}\hspace{-0.3mm}+\hspace{-0.3mm}p_{\perp}^2}{1\hspace{-0.3mm}+\hspace{-0.3mm}p_{\|}^{2}\hspace{-0.3mm}+\hspace{-0.3mm}p_{\perp}^2}}\,\textup{e}^{\,-\frac{p_{\|}}{c_{\mathrm{Z}_{\mathrm{eff}}} \ln{\hspace{-0.45mm}\Lambda_{rel}}}-\tilde{E}\cdot\frac{p_{\perp}^2}{p_{\|}}}\hspace{0.35mm}\mathrm{d}p_{\perp}\mathrm{d}p_{\|} }_{\eqqcolon \,\textup{I}_{\,\textup{num}}^{\,\textit{j}_{RE}^{\hspace{0.25mm}\textup{ava}}}}.
\end{gathered}
\end{split}
\end{equation}
\vspace*{-6.2mm}\\Here, it should be remarked, as already discussed in section \ref{RP_avalanche_dist_subsection}, that a finite lower parallel momentum boundary \mbox{$p_{\|,min}>-\infty$} has to be introduced necessarily, which has to satisfy the condition $(\ref{normalization_conditon_RP})$, in order to receive finite results. In addition, it was discussed in the reference \cite{study_thesis}, that the appearing integral has to be solved with a two-dimensional numerical integration method, due to the fact, that no full analytic solution was found.
\\
A computation of the mean avalanche runaway electron velocity normalized to the speed of light $c$, from the distribution function by \mbox{\textit{T.\hspace{0.9mm}Fülöp et \hspace{-0.4mm}al.}} in the \textit{Rosenbluth-Putvinski} model, is then possible, with the help of the definition of the current density from $(\ref{RE_ava_curr_dens_def_general})$ and the evaluation of the integral $\textup{I}_{\,\textup{num}}^{\,j_{RE}^{\hspace{0.25mm}\textup{ava}}}$:
\vspace*{-3.4mm}
\begin{equation}\label{RE_ava_u_mag_def}
\dfrac{u_{\mathrm{RE}}^{\mathrm{ava}}}{c}= \dfrac{j^{\hspace{0.25mm}\mathrm{ava}}_{\mathrm{RE}}}{-\,e\,c\,n_{\mathrm{RE}}}  =\dfrac{2\,\tilde{E}}{c_{\mathrm{Z}_{\mathrm{eff}}} \ln{\hspace{-0.45mm}\Lambda_{rel}}}\cdot\textup{exp}\left(\dfrac{2\,(\hat{E}-1)}{c_{\mathrm{Z}_{\mathrm{eff}}} \ln{\hspace{-0.45mm}\Lambda_{rel}}}\cdot \frac{t}{\tau_{rel}}\right) \cdot\textup{I}_{\,\textup{num}}^{\,\textit{j}_{RE}^{\hspace{0.25mm}\textup{ava}}} \,.
\end{equation}  
\vspace*{-7.5mm}\\
Although, it was suggested in the study thesis \cite{study_thesis} to transform the integration domain of the integral $\textup{I}_{\,\textup{num}}^{\,j_{RE}^{\hspace{0.25mm}\textup{ava}}}$ to the unit square \mbox{$[0,\,1]\times[0,\,1]$}, by applying the substitutions from $(\ref{substitutions_num_k_again})$, with the purpose of an enhanced runtime efficiency and the easier application of standard quadrature schemes. This consequently results in the following expression for the integral from $(\ref{j_ava_RE_mag_def})$:\vspace*{-5.0mm}
\begin{equation}\label{I_j_ava_num_RE_def}
\textup{I}_{\,\textup{num}}^{\,j_{RE}^{\hspace{0.25mm}\textup{ava}}}\;\coloneqq \displaystyle{\int\limits_{w=0}^{1}\int\limits_{z=0}^{1}} \;\frac{z\cdot\sqrt{\frac{\left(p_{\|,min}+\frac{w}{1-w}\right)^{2}+\left(\frac{z}{1-z}\right)^2}{1+\left(p_{\|,min}+\frac{w}{1-w}\right)^{2}+\left(\frac{z}{1-z}\right)^2}} \cdot\textup{e}^{\,-\frac{p_{\|,min}+\frac{w}{1-w}}{c_{\mathrm{Z}_{\mathrm{eff}}} \ln{\hspace{-0.45mm}\Lambda_{rel}}}-\frac{\tilde{E}\cdot\left(\frac{z}{1-z}\right)^2}{p_{\|,min}+\frac{w}{1-w}}}}{\left(p_{\|,min}+\frac{w}{1-w}\right)\cdot(1-w)^2 \cdot(1-z)^3}\;\,\mathrm{d}z\,\mathrm{d}w\,
\end{equation}
\vspace*{-5.8mm}\\which can be computed with a \textsc{MATLAB}-implementation, using the previously mentioned \textsc{MATLAB}-routine \qq{\texttt{integral2}} \cite{integral2}.

However in this work, one obtains the current density and also, under utilization of the relation $(\ref{RE_ava_u_mag_def})$, the mean avalanche runaway electron velocity from the \textit{Euclidean} norm of the current density vector in the cylindrical two-dimensional coordinate system:\vspace*{-3.0mm}
\begin{equation}\label{RE_ava_j_norm_def}
j_{\mathrm{RE}}^{\hspace{0.25mm}\textup{ava}}=\sqrt{\left(j_{\|,RE}^{\hspace{0.25mm}\textup{ava}}\right)^2+\left(j_{\perp,RE}^{\hspace{0.25mm}\textup{ava}}\right)^2}\,. 
\end{equation}\vspace*{-10.0mm}\\Hence, the components $j_{\|,RE}^{\hspace{0.25mm}\textup{ava}}$ and $j_{\perp,RE}^{\hspace{0.25mm}\textup{ava}}$ of the current density have to be calculated from their corresponding moments.
\\
For that purpose, one states a calculation rule for the parallel component of the avalanche runaway current density $j_{\|,RE}^{\hspace{0.25mm}\textup{ava}}$, with the help of the definition $(\ref{first_moment})$ and the relations for the parallel velocity component $(\ref{p_norm_gamma_def})$ and $(\ref{p_relations})$:\vspace*{-3.9mm}
\begin{equation}\label{j_ava_RE_parallel_def}
\begin{split}
\begin{gathered}
j_{\|,RE}^{\hspace{0.25mm}\textup{ava}}\underset{(\ref{volelem_cyl_2D}),(\ref{first_moment})}{\overset{(\ref{RE_curr_dens_def})}{=}}-\,e\,\displaystyle{\int\limits_{p_{\|}=-\infty}^{\infty}\int\limits_{p_{\perp}=0}^{\infty}} v_{\|}\cdot f_{RE}^{\textup{ava}}(p_{\|},\,p_{\perp},\,t)\,2\pi\,p_{\perp} \mathrm{d}p_{\perp} \mathrm{d}p_{\|}
\\[-2pt]
\underset{(\ref{p_relations})}{\overset{(\ref{p_norm_gamma_def})}{=}}-\,2\pi\,c\,e\,\displaystyle{\int\limits_{p_{\|}=p_{\|,min}}^{\infty}\int\limits_{p_{\perp}=0}^{\infty}} \frac{p_{\perp}\,p_{\|}}{\sqrt{1+p_{\|}^{2}+p_{\perp}^2}}\cdot f_{RE}^{\textup{ava}}(p_{\|},\,p_{\perp},\,t)\;\mathrm{d}p_{\perp}\mathrm{d}p_{\|}
\\[-6pt]
\underset{(\ref{abrev_dist_func_RP_ava_def})}{\overset{(\ref{dist_func_RP_ava_def})}{=}}\hspace{-0.7mm}-\dfrac{2\,c\,e\,n_{RE}\,\tilde{E}}{c_{\mathrm{Z}_{\mathrm{eff}}} \ln{\hspace{-0.45mm}\Lambda_{rel}}}\,\textup{e}^{\,\frac{2\,(\hat{E}-1)}{c_{\mathrm{Z}_{\mathrm{eff}}} \ln{\hspace{-0.45mm}\Lambda_{rel}}} \frac{t}{\tau_{rel}}}\hspace{-4.4mm}\underbrace{\displaystyle{\int\limits_{p_{\|}=p_{\|,min}}^{\infty}\int\limits_{p_{\perp}=0}^{\infty}} \hspace{-0.6mm}\frac{p_{\perp}\cdot\textup{e}^{\,-\frac{p_{\|}}{c_{\mathrm{Z}_{\mathrm{eff}}} \ln{\hspace{-0.45mm}\Lambda_{rel}}}-\tilde{E}\cdot\frac{p_{\perp}^2}{p_{\|}}} }{\sqrt{1+p_{\|}^{2}+p_{\perp}^2}}\;\mathrm{d}p_{\perp}\mathrm{d}p_{\|} }_{=:\,\textup{I}_{\,\textup{num}}^{\,\textit{j}_{\|,RE}^{\hspace{0.25mm}\textup{ava}}}}\,.
\end{gathered}
\end{split}
\end{equation}\vspace*{-7.5mm}\\
Equivalently, one proceeds with the derivation of a relation for the calculation of the orthogonal component of the avalanche runaway current density $j_{\perp,RE}^{\hspace{0.25mm}\textup{ava}}$:\vspace*{-3.3mm}
\begin{equation}\label{j_ava_RE_perp_def}
\begin{split}
\begin{gathered}
j_{\perp,RE}^{\hspace{0.25mm}\textup{ava}}\underset{(\ref{volelem_cyl_2D}),(\ref{u_ava_moment_def1})}{\overset{(\ref{RE_curr_dens_def})}{=}}-\,e\,\displaystyle{\int\limits_{p_{\|}=-\infty}^{\infty}\int\limits_{p_{\perp}=0}^{\infty}} v_{\perp}\cdot f_{RE}^{\textup{ava}}(p_{\|},\,p_{\perp},\,t)\,2\pi\,p_{\perp}\,\mathrm{d}p_{\perp}\,\mathrm{d}p_{\|}
\\[-2pt]
\underset{(\ref{p_relations})}{\overset{(\ref{p_norm_gamma_def})}{=}}-\,2\pi\,c\,e\,\displaystyle{\int\limits_{p_{\|}=p_{\|,min}}^{\infty}\int\limits_{p_{\perp}=0}^{\infty}} \frac{p_{\perp}^{2} }{\sqrt{1+p_{\|}^2+p_{\perp}^{2}}}\cdot f_{RE}^{\textup{ava}}(p_{\|},\,p_{\perp},\,t)\;\mathrm{d}p_{\perp}\,\mathrm{d}p_{\|}
\\[-6pt]
\underset{(\ref{abrev_dist_func_RP_ava_def})}{\overset{(\ref{dist_func_RP_ava_def})}{=}}\hspace{-1.8mm}- \frac{2 \, c\,e\,n_{\mathrm{RE}}\,\tilde{E} }{ c_{\mathrm{Z}_{\mathrm{eff}}} \ln{\hspace{-0.45mm}\Lambda_{rel}}}\cdot\textup{e}^{\,\frac{2\,(\hat{E}-1)}{c_{\mathrm{Z}_{\mathrm{eff}}}\ln{\hspace{-0.45mm}\Lambda_{rel}}}\frac{t}{\tau_{rel}}}\hspace{-4.5mm}\underbrace{\displaystyle{\int\limits_{p_{\|}=p_{\|,min}}^{\infty}\int\limits_{p_{\perp}=0}^{\infty}}  \frac{p_{\perp}^{2}\cdot\textup{e}^{\,-\frac{p_{\|}}{c_{\mathrm{Z}_{\mathrm{eff}}} \ln{\hspace{-0.45mm}\Lambda_{rel}}}-\tilde{E}\cdot\frac{p_{\perp}^2}{p_{\|}}}}{p_{\|}\sqrt{1\hspace{-0.4mm}+\hspace{-0.4mm}p_{\|}^2\hspace{-0.4mm}+\hspace{-0.4mm}p_{\perp}^{2}}}\;\mathrm{d}p_{\perp}\mathrm{d}p_{\|} }_{=:\,\textup{I}_{\,\textup{num}}^{\,\textit{j}_{\perp,RE}^{\hspace{0.25mm}\textup{ava}}}}
\end{gathered}
\end{split}
\end{equation}\vspace*{-7.0mm}\\Note, that the parallel momentum integration for the moments in $(\ref{j_ava_RE_parallel_def})$ and $(\ref{j_ava_RE_perp_def})$ was defined, similarly to the moment related to the magnitude of the current density from $(\ref{j_ava_RE_mag_def})$, for the lower momentum boundary \mbox{$p_{\|,min}>-\infty$}, which is required to fulfill the condition $(\ref{normalization_conditon_RP})$.
\\
With the calculation rules for the components of the avalanche runaway current density, one implicitly obtains the mean velocity components parallel and orthogonal to the local magnetic field. Results for the steady-state for different values of the electric field can be generated with a \textsc{MATLAB}-implementation. Depictions of those results are shown, alongside a detailed discussion and evaluation, in the study thesis \cite{study_thesis}.

The normalized magnitude of the mean velocity of a runaway electron population $u_{\mathrm{RE}}^{\hspace{0.25mm}\mathrm{ava}}/c$ can then be received from the current density, as written in $(\ref{RE_ava_curr_dens_def_general})$, and its representation through their components, stated in $(\ref{RE_ava_j_norm_def})$:\vspace*{-2.5mm}
\begin{equation}\label{u_ava_def_from_norm}
\dfrac{u_{\mathrm{RE}}^{\hspace{0.25mm}\textup{ava}}}{c}=\dfrac{j_{\mathrm{RE}}^{\hspace{0.25mm}\textup{ava}}}{-\,e\,c\,n_{\mathrm{RE}}}=\dfrac{2\,\tilde{E}}{c_{\mathrm{Z}_{\mathrm{eff}}} \ln{\hspace{-0.45mm}\Lambda_{rel}}}\cdot\textup{exp}\left(\dfrac{2\,(\hat{E}-1)}{c_{\mathrm{Z}_{\mathrm{eff}}} \ln{\hspace{-0.45mm}\Lambda_{rel}}}\cdot \frac{t}{\tau_{rel}}\right)\cdot\sqrt{\left( \textup{I}_{\,\textup{num}}^{\,\textit{j}_{\|,RE}^{\hspace{0.25mm}\textup{ava}}}\right)^2+ \left( \textup{I}_{\,\textup{num}}^{\,\textit{j}_{\perp,RE}^{\hspace{0.25mm}\textup{ava}}}\right)^2}\,.
\end{equation}
\vspace*{-6.5mm}\\
The runtime for the computation of the mean velocity magnitude as defined in $(\ref{u_ava_def_from_norm})$ requires two numerical two-dimensional integrations. This is, because only the $p_{\perp}$-integration of the integral $\textup{I}_{\,\textup{num}}^{\,\textit{j}_{\|,RE}^{\hspace{0.25mm}\textup{ava}}}$ is analytically possible and the resulting numerical one-dimensional integration is less runtime efficient than the numerical two-dimensional integration, which was shown based on \textsc{MATLAB}-implementations in the study thesis \cite{study_thesis}. In consequence, it is appropriate to again apply the substitutions from $(\ref{substitutions_num_k_again})$, so that the integral, related to the parallel component of the current density:\vspace{-3.2mm}
\begin{equation}\label{u_ava_def_Integrals_par}
\textup{I}_{\,\textup{num}}^{\,j_{\|,RE}^{\hspace{0.25mm}\textup{ava}}}\;\coloneqq \displaystyle{\int\limits_{w=0}^{1}\int\limits_{z=0}^{1}} \,\frac{\frac{z}{(1-w)^2(1-z)^3}\hspace{-0.55mm}\cdot\hspace{-0.45mm}\exp{\left(-\frac{ p_{\|,min}\hspace{-0.1mm}+\hspace{-0.1mm}\frac{w}{1-w} }{c_{\mathrm{Z}_{\mathrm{eff}}} \ln{\hspace{-0.45mm}\Lambda_{rel}}}-\frac{\tilde{E}\cdot\left(\frac{z}{1-z}\right)^2}{p_{\|,min}\hspace{-0.1mm}+\hspace{-0.1mm}\frac{w}{1-w}}\right)}}{\sqrt{1+\left(p_{\|,min}+\frac{w}{1-w}\right)^{2}+\left(\frac{z}{1-z}\right)^2}} \;\mathrm{d}z\,\mathrm{d}w
\end{equation}
\vspace*{-6.2mm}\\and the integral, which determines the orthogonal component of the current density:\vspace{-3.2mm}
\begin{equation}\label{u_ava_def_Integrals_perp}
\textup{I}_{\,\textup{num}}^{\,j_{\perp,RE}^{\hspace{0.25mm}\textup{ava}}} \,\hspace{-0.1mm}\coloneqq\hspace{-1.2mm} \displaystyle{\int\limits_{w=0}^{1}\int\limits_{z=0}^{1}} \,  \frac{ z^2 \cdot(1-w)^2 \hspace{-0.65mm}\cdot\hspace{-0.3mm}(1-z)^4 \hspace{-0.65mm}\cdot\hspace{-0.3mm}\exp{\left(-\frac{ p_{\|,min}\hspace{-0.1mm}+\hspace{-0.1mm}\frac{w}{1-w} }{c_{\mathrm{Z}_{\mathrm{eff}}} \ln{\hspace{-0.45mm}\Lambda_{rel}}}-\frac{\tilde{E}\cdot\left(\frac{z}{1-z}\right)^2}{p_{\|,min}\hspace{-0.1mm}+\hspace{-0.1mm}\frac{w}{1-w}}\right)}}{\left(p_{\|,min}\hspace{-0.3mm}+\hspace{-0.3mm}\frac{w}{1-w}\right)\hspace{-0.65mm}\cdot\hspace{-0.3mm}\sqrt{1\hspace{-0.3mm}+\hspace{-0.3mm}\left(p_{\|,min}\hspace{-0.3mm}+\hspace{-0.3mm}\frac{w}{1-w}\right)^{2}\hspace{-0.3mm}+\hspace{-0.3mm}\left(\frac{z}{1-z}\right)^2}}\;\mathrm{d}z\,\mathrm{d}w\hspace{0.3mm},
\end{equation}\vspace{-6.2mm}\\can be computed efficiently. 

\begin{figure}[H]
  \centering
  \subfloat{\label{fig_u_ava_p_c_scr_E3} 
    \includegraphics[trim=316 19 347 20,width=0.41\textwidth,clip]
    {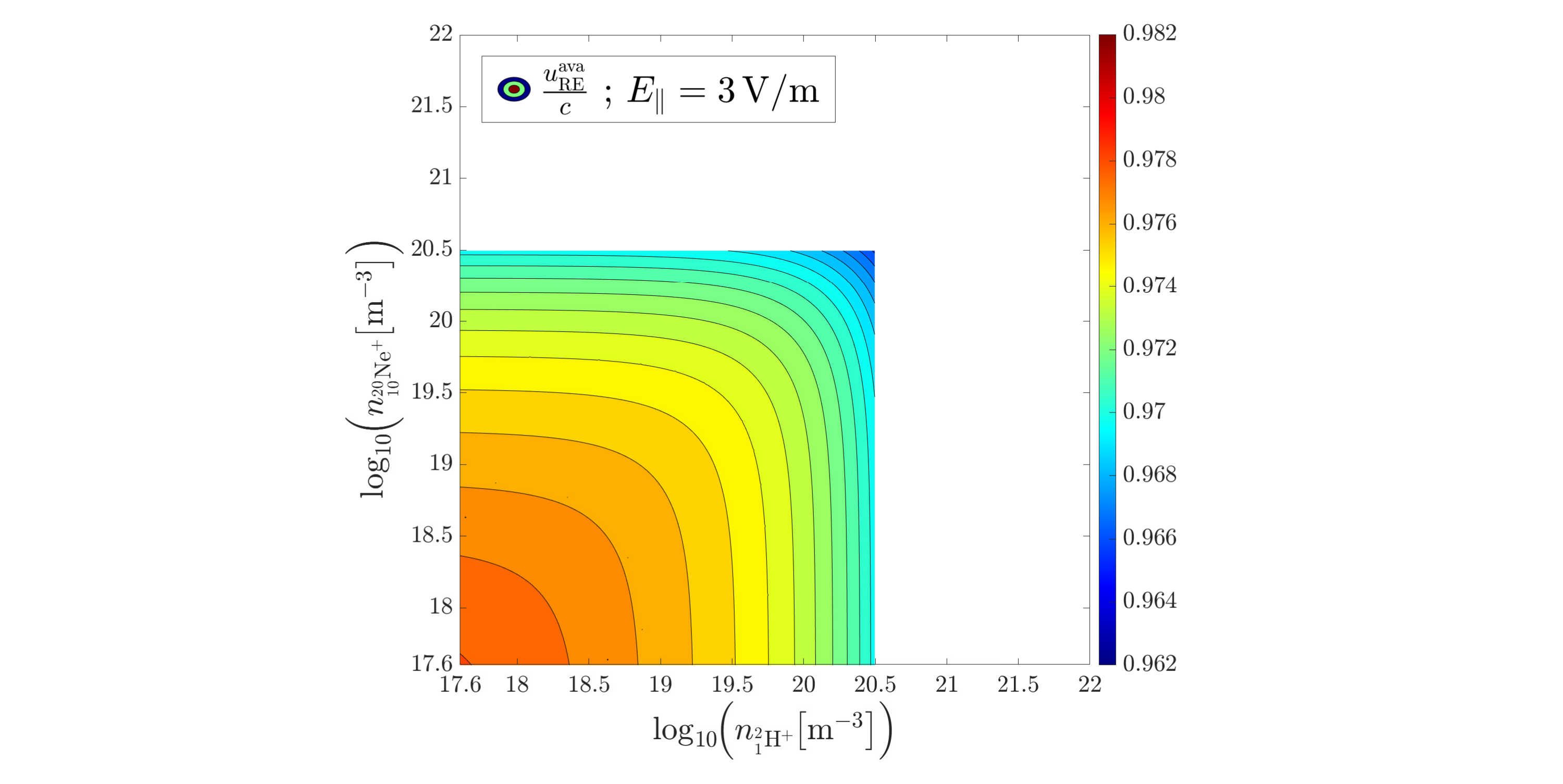}}\quad
  \subfloat{\label{fig_u_ava_p_c_scr_E10}
    \includegraphics[trim=317 23 348 22,width=0.41\textwidth,clip]
    {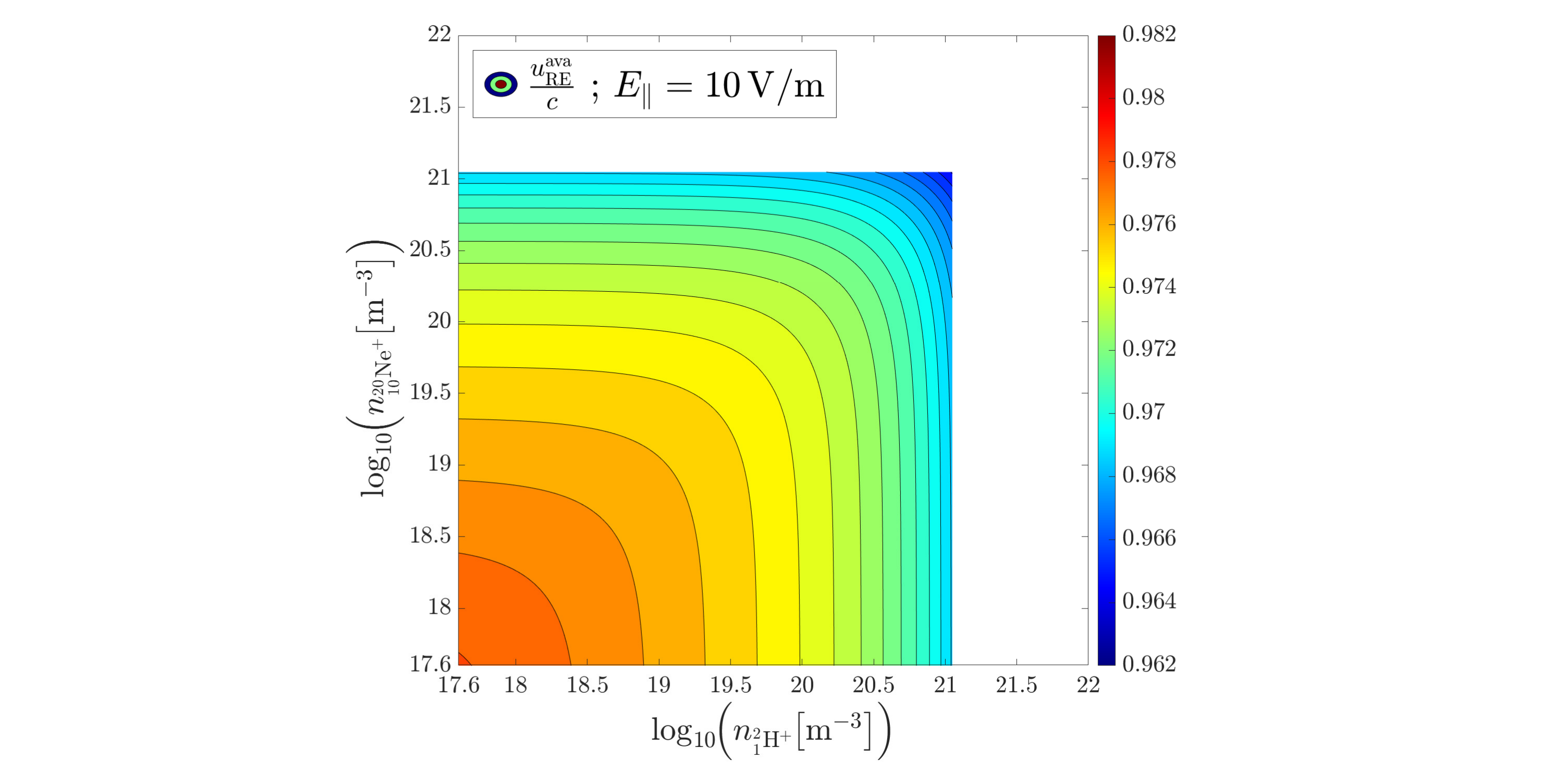}}\\[4pt]
  \subfloat{\label{fig_u_ava_p_c_scr_E30} 
    \includegraphics[trim=314 19 354 20,width=0.41\textwidth,clip]
    {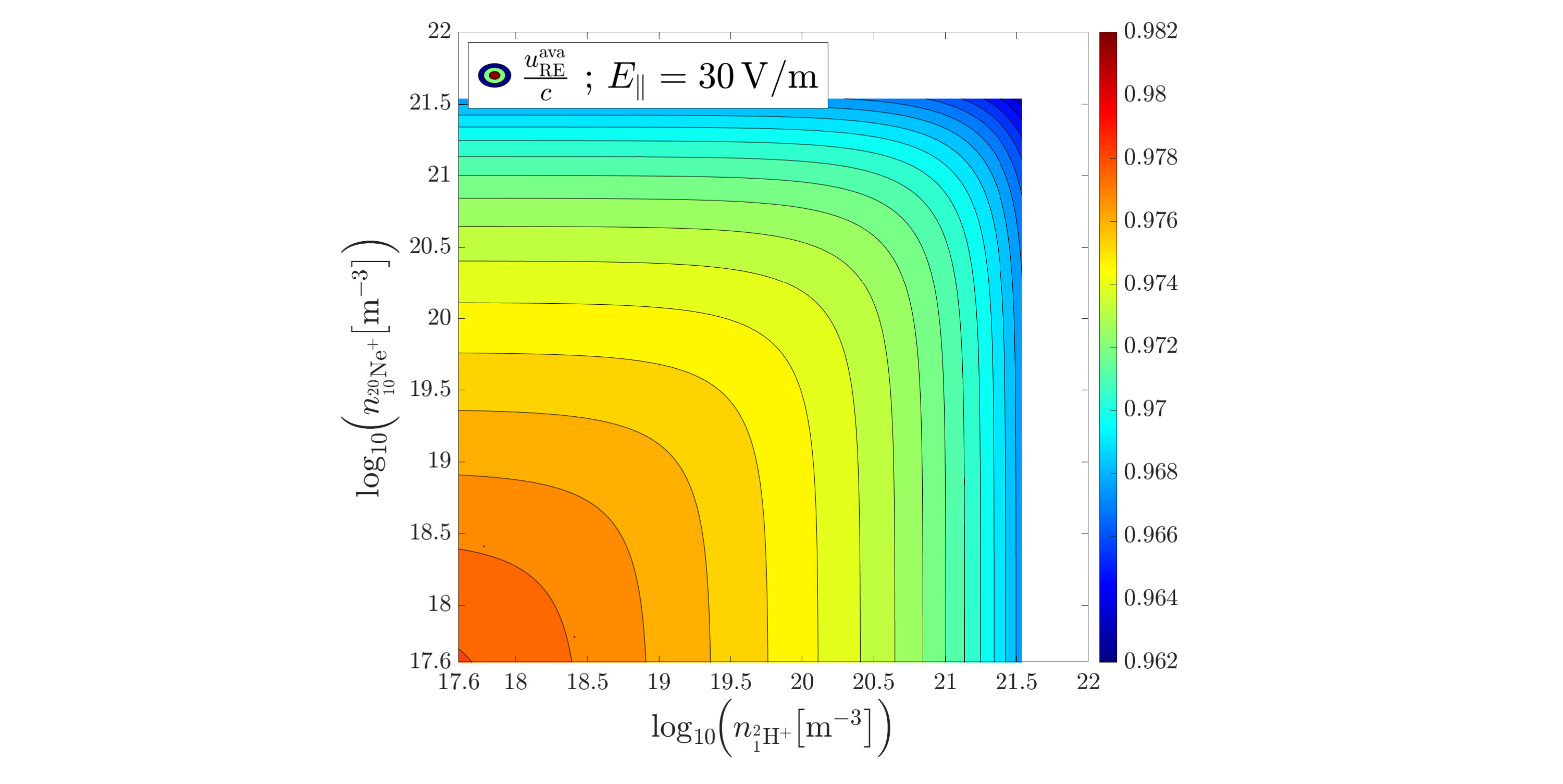}}\quad
  \subfloat{\label{fig_u_ava_p_c_scr_E100}
    \includegraphics[trim=314 19 348 24,width=0.41\textwidth,clip]
    {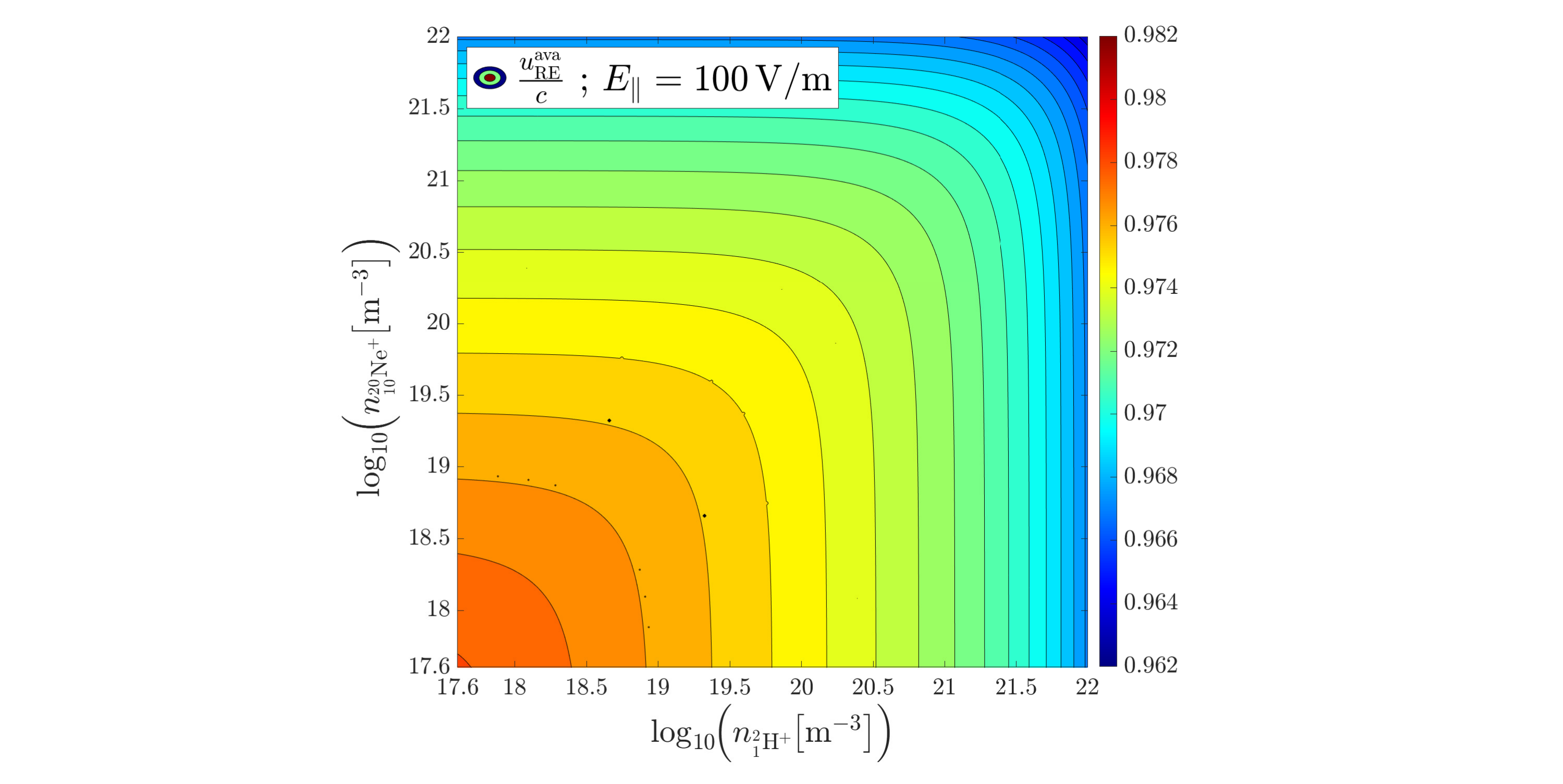}}
  \caption[Contour plots of the normalized mean velocity \mbox{$u_{\mathrm{RE}}^{\mathrm{ava}}/c$} as a function of the deuterium and neon ion density in the \textit{Rosenbluth-Putvinski} model for an avalanche runaway electron population with \mbox{$k_{\mathrm{B}}T_{\mathrm{e}}=10\,\textup{eV}$}, \mbox{$B=5.25\,\textup{T}$} and \mbox{$Z_{\mathrm{eff}}=1$} for different values of the electric field strength \mbox{$E_{\|}\coloneqq\vert E_{\|}\vert$} (larger view in figure \ref{fig_u_ava_p_c_scr} of the appendix).]{Contour plots\protect\footnotemark{} of the normalized mean velocity \mbox{$u_{\mathrm{RE}}^{\mathrm{ava}}/c$} as a function of the deuterium and neon ion density in the \textit{Rosenbluth-Putvinski} model for an avalanche runaway electron population with \mbox{$k_{\mathrm{B}}T_{\mathrm{e}}=10\,\textup{eV}$}, \mbox{$B=5.25\,\textup{T}$} and \mbox{$Z_{\mathrm{eff}}=1$} for different values of the electric field strength \mbox{$E_{\|}\coloneqq\vert E_{\|}\vert$} (larger view in figure \ref{fig_u_ava_p_c_scr} of the appendix).}
\label{fig_u_ava_p_c_scr_main}
\end{figure}

The mean avalanche runaway electron velocity magnitude was evaluated with the help of equation $(\ref{u_ava_def_from_norm})$ and the integrals defined in $(\ref{u_ava_def_Integrals_par})$ and $(\ref{u_ava_def_Integrals_perp})$. At that, the computation was carried out in \textsc{MATLAB} for the steady-state with \mbox{$t=0\,\mathrm{s}$}, corresponding to \mbox{$p_{\|,min}=0$}. In particular, the implementation made use of the routine \qq{\texttt{integral2}} and calculated results for different densities of singly-ionized deuterium and neon atoms for the research plasma, which was introduced in section \ref{nuclear_fusion_section} and different values of the electric field.
\\
The produced plots for the different values of the electric field, which at first approximation increase logarithmically, result from different \textsc{MATLAB}-scripts$^{\ref{fig_plot_u_RP_footnote}}$ and are arranged in figure \ref{fig_u_ava_p_c_scr_main}. 
\footnotetext{\label{fig_plot_u_RP_footnote} The contour plots were computed, with the help of the \textsc{MATLAB}-scripts\\ \hspace*{8.7mm}\qq{\texttt{generate_num_data_densities_p_c_scr_E3.m}},\\ \hspace*{8.7mm}\qq{\texttt{plot_num_data_densities_p_c_scr_E3.m}},\\ \hspace*{8.7mm}\qq{\texttt{generate_num_data_densities_p_c_scr_E10.m}},\\\hspace*{9mm} \qq{\texttt{plot_num_data_densities_p_c_scr_E10.m}},\\ \hspace*{8.7mm}\qq{\texttt{generate_num_data_densities_p_c_scr_E30.m}},\\ \hspace*{8.7mm}\qq{\texttt{plot_num_data_densities_p_c_scr_E30.m}},\\ \hspace*{8.7mm}\qq{\texttt{generate_num_data_densities_p_c_scr_E100.m}} and\\ \hspace*{8.7mm}\qq{\texttt{plot_num_data_densities_p_c_scr_E100.m}}, which can be viewed in the digital\\ \hspace*{8.7mm}appendix.}
\newpage\noindent At this, the relation $(\ref{CoulombLogrel})$ was used for the computation of the relativistic \textit{Coulomb} logarithm. The corresponding outputs, however, can be found in the listings \cref{MATLABoutput_plot_p_scr_E3,MATLABoutput_plot_p_scr_E10,MATLABoutput_plot_p_scr_E30,MATLABoutput_plot_p_scr_E100} in subsection \ref{output_matlab_appendix_subsection} of the appendix. There it displays the values of the minima, maxima and mean values of each contour plot, defines the general parameter settings and allows the verification of the compliance with the control criterion $(\ref{check_n_RE_ava_def})$ for all computations. 

An analysis of the figure \ref{fig_u_ava_p_c_scr_main} yields, that the mean velocity of runaway electrons, generated from the avalanche mechanism, deviates more significantly from the speed of light in vacuum for high electron densities, according to the analytic distribution function based on the avalanche runaway electron growth rate proposed by \mbox{\textit{M.\hspace{0.9mm}Rosenbluth}} and \mbox{\textit{S.\hspace{0.9mm}Putvinski}}. Furthermore, one notices the symmetry of the contour lines to the diagonal \mbox{$n_{_{10}^{20}\mathrm{Ne}^{+}}(n_{_{1}^{2}\mathrm{H}^{+}})=n_{_{1}^{2}\mathrm{H}^{+}}$}. From this, one can deduce, that this distribution function does not resolve a different behaviour for a change in either the neon- or the deuterium ion density and therefore ignores all effects of partial screening as discussed in the section \ref{part_screen_section}. In addition, one can confirm the expectation, that the avalanche runaway electron generation region grows within the density parameter space as a consequence of an increase in the electric field strength, by the observation of the densities associated with for instance the yellow area for increasing values of the electric field strength. It shall be remarked, that this expectation can for example be seen from the definition of \textit{Connor-Hastie} critical electric field from $(\ref{E_crit})$, which explicitly shows its growth with greater values of the free electron density.
\\
Finally, one prediction, resulting from the discussion of the data, produced with the \textit{Rosenbluth-Putvinski} model and displayed in the figure \ref{fig_u_ava_p_c_scr_main}, would be, that the mean avalanche runaway electron velocity would monotonically increase, if more material is injected. However, this does not model the influences, achieved with the injection of impurities with higher nuclear charge than the main plasma particles, like for example neon or argon.

\subsection{Current density of an \textit{avalanche} runaway electron population in the \textit{Hesslow} model}\label{Hesslow_avalanche_j_subsection}

The mean velocity of an avalanche runaway electron population $u_{\mathrm{RE}}^{\mathrm{ava,scr}}$, under consideration of the effects of partial screening, is connected to the first moment of a distribution function, which models the influences of a not fully ionized plasma as discussed in section \ref{part_screen_section}. It is determined, in accordance with the definition of the first moment from section \ref{kin_equa_section} and the notation \mbox{$v\coloneqq\vert\mathbf{v}\vert$}, by the following relation:\vspace*{-3.1mm}
\begin{equation}\label{u_ava_scr_def1}
 u_{\mathrm{RE}}^{\hspace{0.25mm}\textup{ava,\,scr}} \hspace{-0.5mm}= \hspace{-0.5mm}\dfrac{1}{n_{\mathrm{RE}} }\, \displaystyle{\iiint\limits_{\mathbb{R}^3}} \vert\mathbf{v}\vert\hspace{-0.3mm} \cdot \hspace{-0.3mm}f_{RE}^{\hspace{0.25mm}\textup{ava,scr}}(\mathbf{p})\,\mathrm{d}^3p\hspace{-0.3mm}=\hspace{-0.3mm}\dfrac{1}{n_{\mathrm{RE}} }\, \displaystyle{\int\limits_{p=p_{\mathrm{c}}^{\mathrm{eff}}}^{\infty}}\underbrace{\displaystyle{\int\limits_{\xi=-1}^{1}  }\hspace{-0.8mm}v\hspace{-0.35mm}\cdot\hspace{-0.35mm} f_{RE}^{\hspace{0.25mm}\textup{ava,scr}}(p,\,\xi)\,2 \pi\,p^2\,\mathrm{d}\xi}_{=\,\tilde{f}_{0,RE}^{\hspace{0.25mm}\mathrm{ava,scr}}(p)}\mathrm{d}p\,,
\end{equation}
\vspace*{-6.9mm}\\which makes use of the \mbox{$(p,\,\xi)$}-momentum space coordinate system from \ref{mom_space_coord_section} and the corresponding volume element, given in $(\ref{volelem_sphere_2D})$. Further, the effective one-dimensional avalanche runaway distribution function in zeroth order approximation \mbox{$\tilde{f}_{0,RE}^{\hspace{0.25mm}\mathrm{ava,scr}}(p)$}, as written in equation $(\ref{dist_func_H})$ is identified. Note, that this is possible as a consequence of the discussion from section \ref{Hesslow_avalanche_dist_subsection}. There, it was reasoned that the moments of the distribution function in the \textit{Hesslow} model are sensitive to the lowest momentum for runaway electrons, originating from the avalanche mechanism, while the upper momentum bound has only an exponentially suppressed influence and should be set to infinity, in order to save runtime for an accurate calculation of maximum momentum of the runaway region. Hence, the integration in the momentum magnitude coordinate $p$ has to take place between the effective critical momentum $p_{\mathrm{c}}^{\mathrm{eff}}$ and infinity. Therewith, a calculation rule for the current density of an avalanche runaway electron population in the \textit{Hesslow} model, based on the distribution function \mbox{$\tilde{f}_{0,RE}^{\hspace{0.25mm}\mathrm{ava,scr}}(p)$} by \textit{P.\hspace{0.9mm}Svensson} results from the velocity magnitude expressed in $p$, as written in $(\ref{p_norm_gamma_def})$, the definition of the current density from $(\ref{RE_curr_dens_def})$ and the distribution function from $(\ref{dist_func_H})$:
\vspace*{-3.0mm}
\begin{equation}\label{u_ava_scr_def2}
\begin{split}
\begin{gathered}
j_{\mathrm{RE}}^{\hspace{0.25mm}\textup{ava,\,scr}}  \underset{(\ref{u_ava_scr_def1})}{\overset{(\ref{RE_curr_dens_def})}{=}} -\,e \displaystyle{\int\limits_{p=p_{c}^{\mathrm{eff}}}^{\infty} }v\cdot \tilde{f}_{0,RE}^{\hspace{0.25mm}\mathrm{ava,scr}}(p)\;\mathrm{d}p \overset{(\ref{p_norm_gamma_def})}{=}  -\,e\,c \displaystyle{\int\limits_{p=p_{c}^{\mathrm{eff}}}^{\infty} }\dfrac{p }{\sqrt{1+p^2}}\cdot \tilde{f}_{0,RE}^{\hspace{0.25mm}\mathrm{ava,scr}}(p)\;\mathrm{d}p
\\[-1pt]
\overset{(\ref{dist_func_H})}{=}  -\,e\,c\,\hspace{-1.5mm}\displaystyle{\int\limits_{p=p_{c}^{\mathrm{eff}}}^{\infty} }\dfrac{n_{\mathrm{e}}^{\mathrm{tot}}\cdot p\cdot n_{RE} }{n_{\mathrm{e}}\cdot\sqrt{1+p^2}}\cdot\dfrac{\textup{e}^{\, -\frac{n_{\mathrm{e}}^{\mathrm{tot}} \cdot (p-p_{c}^{\mathrm{eff}})}{n_{\mathrm{e}} \cdot \ln{\hspace{-0.45mm}\Lambda_{rel}} \cdot \sqrt{4+\tilde{\nu}_{\mathrm{s}}(p_{c}^{\mathrm{eff}}) \cdot \tilde{\nu}_{\mathrm{d}}(p_{c}^{\mathrm{eff}})}}  }   }{ \ln{\hspace{-0.45mm}\Lambda_{rel}}\cdot\sqrt{4+\tilde{\nu}_{\mathrm{s}}(p_{c}^{\mathrm{eff}}) \cdot\tilde{\nu}_{\mathrm{d}}(p_{c}^{\mathrm{eff}})}}   \;\mathrm{d}p 
\\[-1pt]
\overset{(\ref{substitutions_num_k_again_H})}{=}\hspace{-1.45mm} -\,e\,c\,n_{RE}\,\hspace{-1.75mm}\underbrace{\displaystyle{\int\limits_{w=0}^{1} }\dfrac{n_{\mathrm{e}}^{\mathrm{tot}}\hspace{-0.45mm}\cdot\hspace{-0.35mm}\left(p_{c}^{\mathrm{eff}}+\frac{w}{1-w}\right)  }{n_{\mathrm{e}}\hspace{-0.7mm}\cdot\hspace{-0.5mm}\sqrt{1\hspace{-0.25mm}+\hspace{-0.25mm}\left(p_{c}^{\mathrm{eff}}\hspace{-0.15mm}+\hspace{-0.15mm}\frac{w}{1-w}\right)^2}}\hspace{-0.75mm}\cdot\hspace{-0.75mm} \dfrac{\textup{e}^{ -\frac{n_{\mathrm{e}}^{\mathrm{tot}}\cdot w}{n_{\mathrm{e}}\cdot\ln{\hspace{-0.45mm}\Lambda_{rel}}\cdot(1-w)\cdot\sqrt{4+\tilde{\nu}_{\mathrm{s}}(p_{c}^{\mathrm{eff}})\tilde{\nu}_{\mathrm{d}}(p_{c}^{\mathrm{eff}})}}  }}{\ln{\hspace{-0.45mm}\Lambda_{rel}}(1-w)^{2}\sqrt{4+\tilde{\nu}_{\mathrm{s}}(p_{c}^{\mathrm{eff}}) \tilde{\nu}_{\mathrm{d}}(p_{c}^{\mathrm{eff}})}}\,\mathrm{d}w}_{\eqqcolon \,\textup{I}_{\,\textup{num}}^{\,\textit{j}_{RE}^{\hspace{0.25mm}\textup{ava,scr}}}}\hspace{0.35mm},
\end{gathered}
\end{split}
\end{equation}
\vspace*{-5.7mm}\\where in the last equality the substitution from $(\ref{substitutions_num_k_again_H})$ was used again. Note, that the integration in the momentum coordinate $p$ has to take place between the effective critical momentum $p_{\mathrm{c}}^{\mathrm{eff}}$ and infinity, due to the discussion from section \ref{Hesslow_avalanche_dist_subsection}. There, it was found, that the results of the moments of the distribution function in the \textit{Hesslow} model might be more sensitive to the lower momentum boundary, requiring an appropriate choice of the approximation of $p_{\mathrm{c}}^{\mathrm{eff}}$, while the upper momentum boundary has a minor influence on the integration results, because the integration contributions of the distribution function are exponentially suppressed for large momenta. Consequently, it is justified to set the maximum momentum to infinity, in order to save runtime by avoiding its calculation as discussed in section \ref{part_screen_section}, whilst one might set \mbox{$p_{c}^{\mathrm{eff}}\approx p^{\mathrm{scr}}_{c}$}, in accordance with the expression $(\ref{p_c_scr_def})$ or \mbox{$p_{c}^{\mathrm{eff}}\approx p_{\star}$} with $p_{\star}$, as the root of the function defined in $(\ref{func_p_c_eff_def})$. The relativistic \textit{Coulomb} logarithm $\ln{\hspace{-0.45mm}\Lambda_{rel}}$ follows for instance from the relation $(\ref{CoulombLogrel})$ with the corresponding relativistic collision time $\tau_{rel}$ as stated in $(\ref{tau_rel})$ and the ultra-relativistic limits \mbox{$\tilde{\nu}_{\mathrm{d}}(p_{c}^{\mathrm{eff}})$} and \mbox{$\tilde{\nu}_{\mathrm{s}}(p_{c}^{\mathrm{eff}})$} of the deflection and the slowing-down frequency, evaluated at the effective critical momentum from $(\ref{nue_s_nue_d_def})$. In addition, the \textsc{MATLAB}-script$^{\ref{Matlab_Hesslow_script}}$ from \textit{L.\hspace{0.9mm}Hesslow} \cite{Hesslow_2018} has to be used, for the calculation of the effective critical electric field $E^{\mathrm{eff}}_{c}$, which is used in $p^{\mathrm{scr}}_{c}$ and for the constants needed in the equation $(\ref{nue_s_nue_d_def})$ for the computation of \mbox{$\tilde{\nu}_{\mathrm{d}}(p_{c}^{\mathrm{eff}})$} and \mbox{$\tilde{\nu}_{\mathrm{s}}(p_{c}^{\mathrm{eff}})$}. Finally, it should be remarked, that all of the mentioned quantities are calculated before the integration of $\textup{I}_{\,\textup{num}}^{\,j_{RE}^{\hspace{0.25mm}\textup{ava,scr}}}$, which means that they can be treated as constants, since they do not depend on the integration variable.

\begin{figure}[H]
  \centering
  \subfloat{\label{fig_u_ava_screen_p_c_scr_E3_main} 
   \includegraphics[trim=319 19 366 24,width=0.41\textwidth,clip]
    {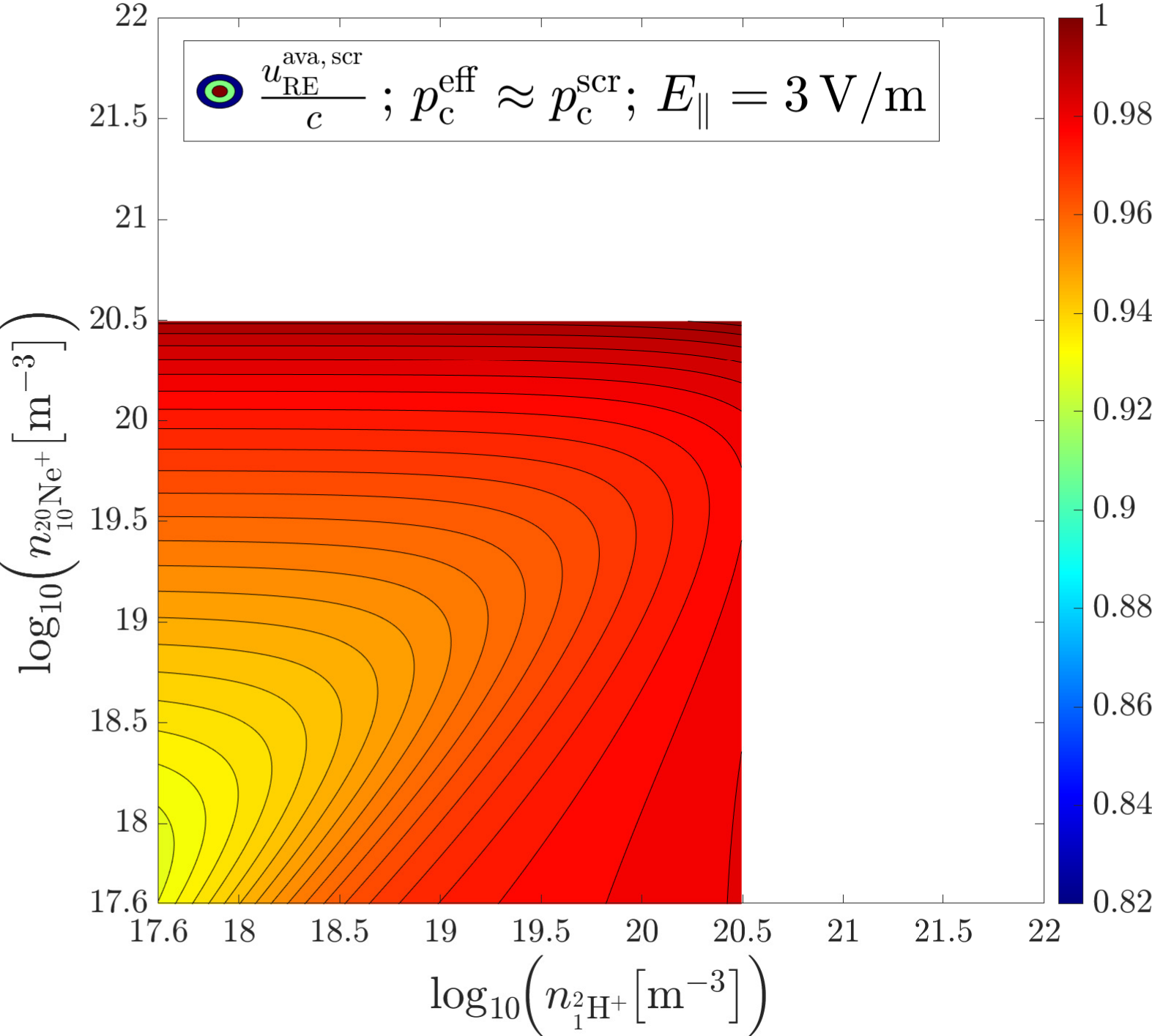}}\quad
  \subfloat{\label{fig_u_ava_screen_p_c_scr_E10_main}
    \includegraphics[trim=321 25 368 18,width=0.41\textwidth,clip]
    {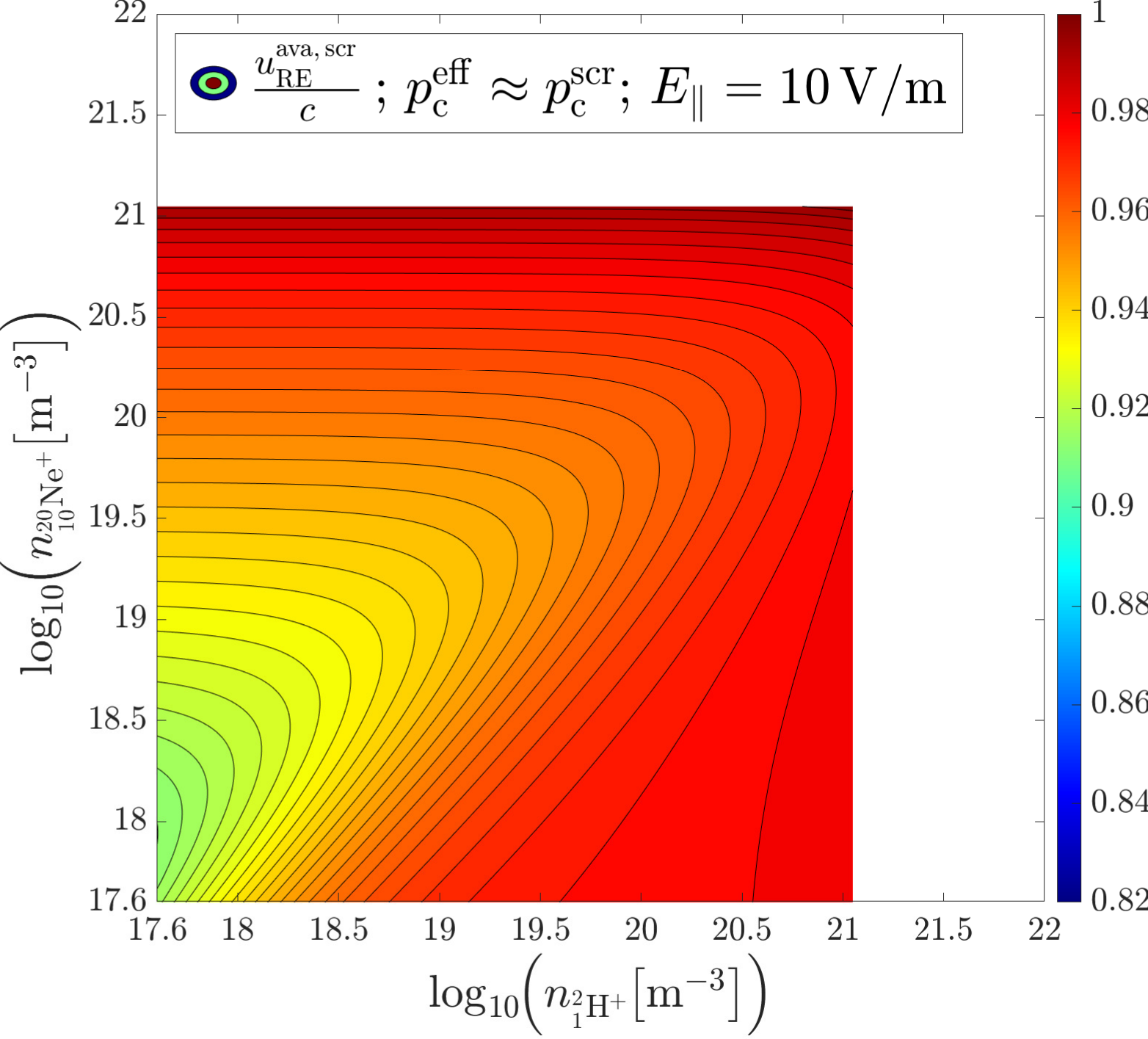}}\\[4pt]
  \subfloat{\label{fig_u_ava_screen_p_c_scr_E30_main} 
   \includegraphics[trim=319 19 364 21,width=0.41\textwidth,clip]
    {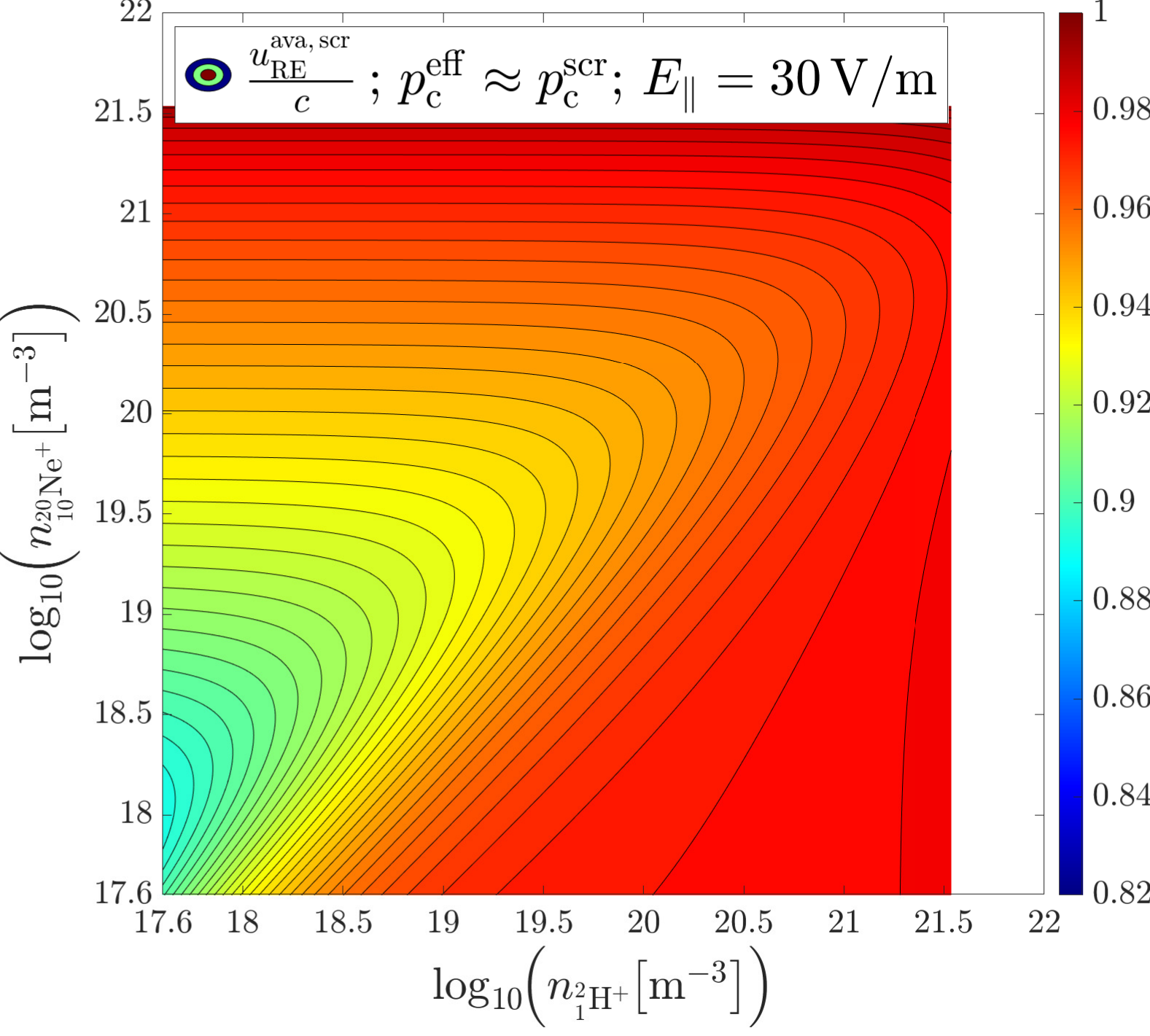}}\quad
  \subfloat{\label{fig_u_ava_screen_p_c_scr_E100_main}
    \includegraphics[trim=314 22 367 21,width=0.41\textwidth,clip]
    {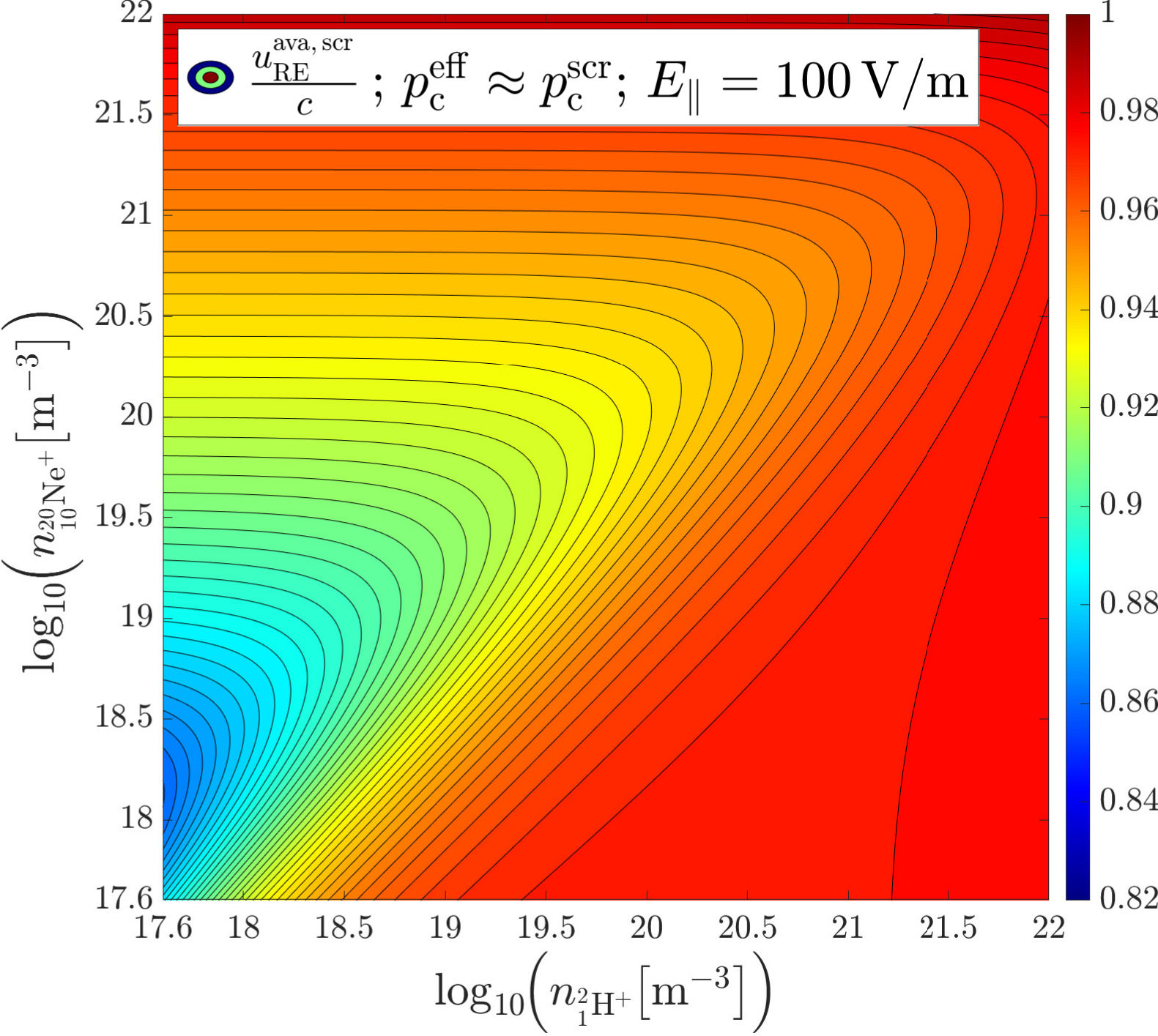}}
  \caption[Contour plots of the normalized mean velocity \mbox{$u_{\mathrm{RE}}^{\hspace{0.25mm}\mathrm{ava,scr}}/c$}, in the \textit{Hesslow} model with the effective critical momentum \mbox{$p_{\mathrm{c}}^{\mathrm{eff}}\approx p_{\mathrm{c}}^{\mathrm{scr}}$}, of an avalanche runaway electron population with \mbox{$k_{\mathrm{B}}T_{\mathrm{e}}=10\,\textup{eV}$}, \mbox{$B=5.25\,\textup{T}$} and \mbox{$Z_{\mathrm{eff}}=1$} for approximately logarithmically increasing values of the electric field strength \mbox{$E_{\|}\coloneqq\vert E_{\|}\vert$} (larger view in figure \ref{fig_u_ava_screen_p_c_scr} of the appendix).]{Contour plots$^{\ref{fig_plot_u_RP_footnote}}$ of the normalized mean velocity \mbox{$u_{\mathrm{RE}}^{\hspace{0.25mm}\mathrm{ava,scr}}/c$}, in the \textit{Hesslow} model with the effective critical momentum \mbox{$p_{\mathrm{c}}^{\mathrm{eff}}\approx p_{\mathrm{c}}^{\mathrm{scr}}$}, of an avalanche runaway electron population with \mbox{$k_{\mathrm{B}}T_{\mathrm{e}}=10\,\textup{eV}$}, \mbox{$B=5.25\,\textup{T}$} and \mbox{$Z_{\mathrm{eff}}=1$} for approximately logarithmically increasing values of the electric field strength \mbox{$E_{\|}\coloneqq\vert E_{\|}\vert$} (larger view in figure \ref{fig_u_ava_screen_p_c_scr} of the appendix).}
\label{fig_u_ava_screen_p_c_scr_main}
\end{figure}

\begin{figure}[H]
  \centering
  \subfloat{\label{fig_u_ava_screen_p_star_E3_main} 
   \includegraphics[trim=321 25 361 18,width=0.41\textwidth,clip]
    {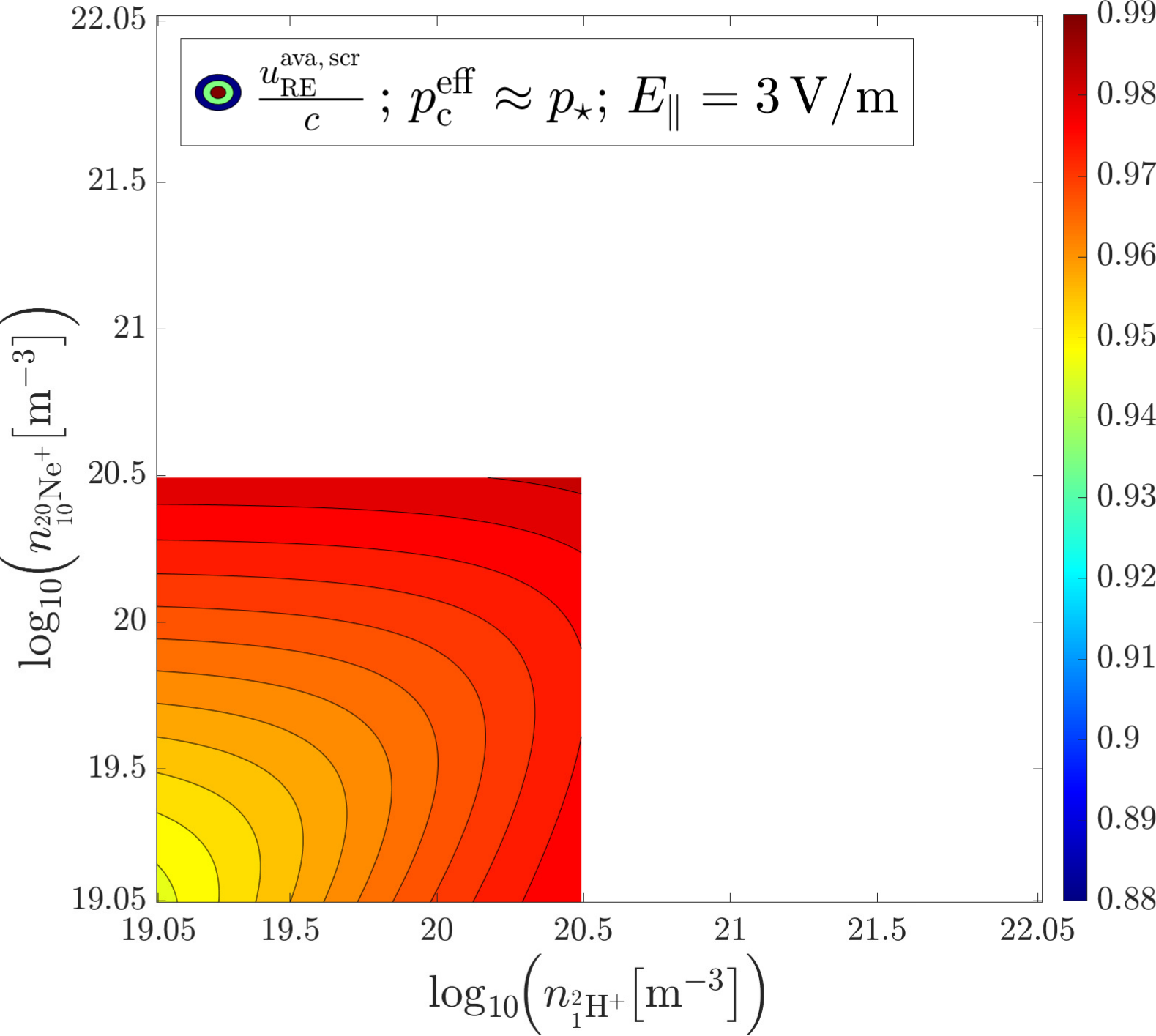}}\quad
  \subfloat{\label{fig_u_ava_screen_p_star_E10_main}
    \includegraphics[trim=322 25 367 18,width=0.41\textwidth,clip]
    {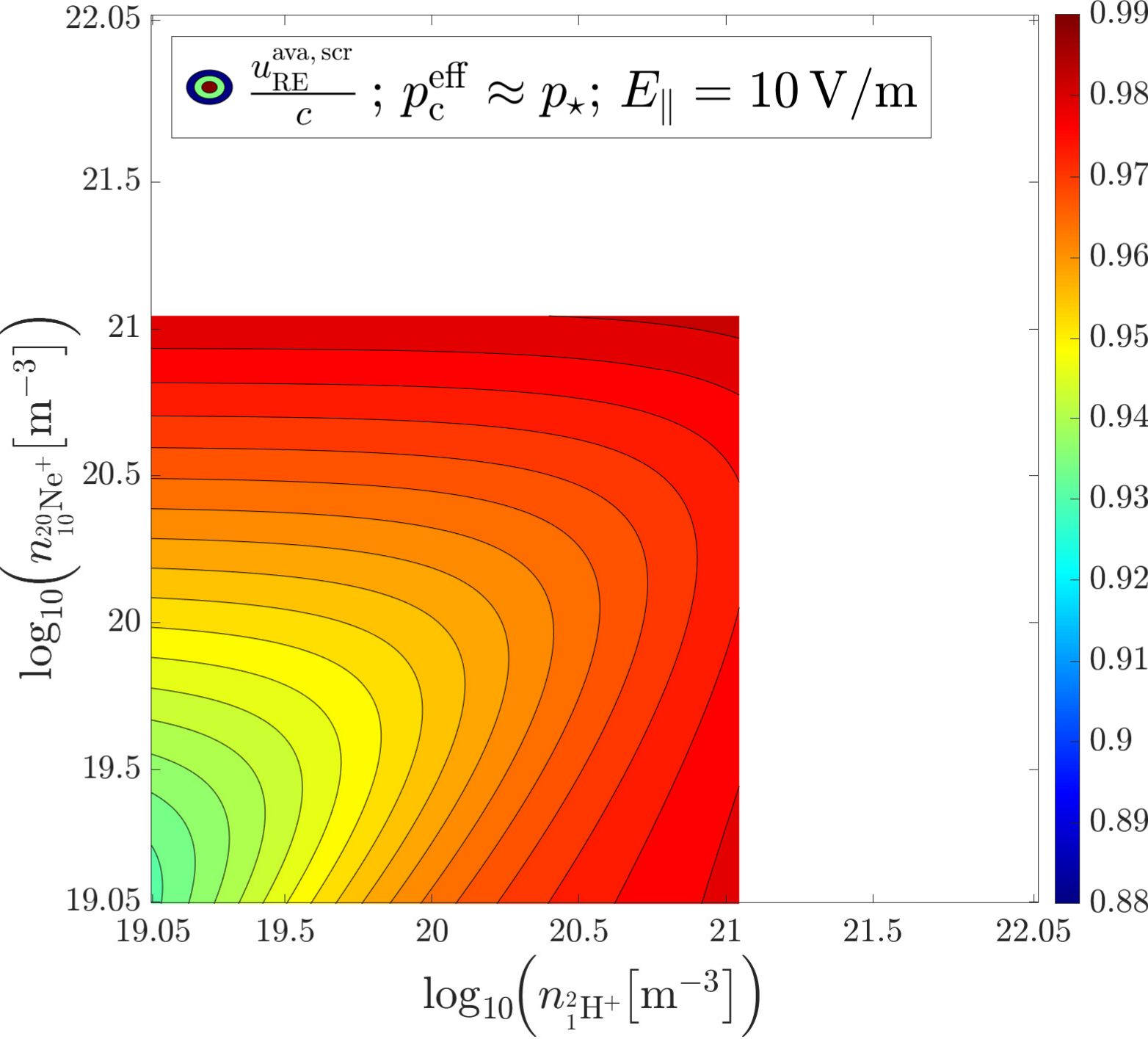}}\\[4pt]
  \subfloat{\label{fig_u_ava_screen_p_star_E30_main} 
    \includegraphics[trim=323 21 368 21,width=0.41\textwidth,clip]
    {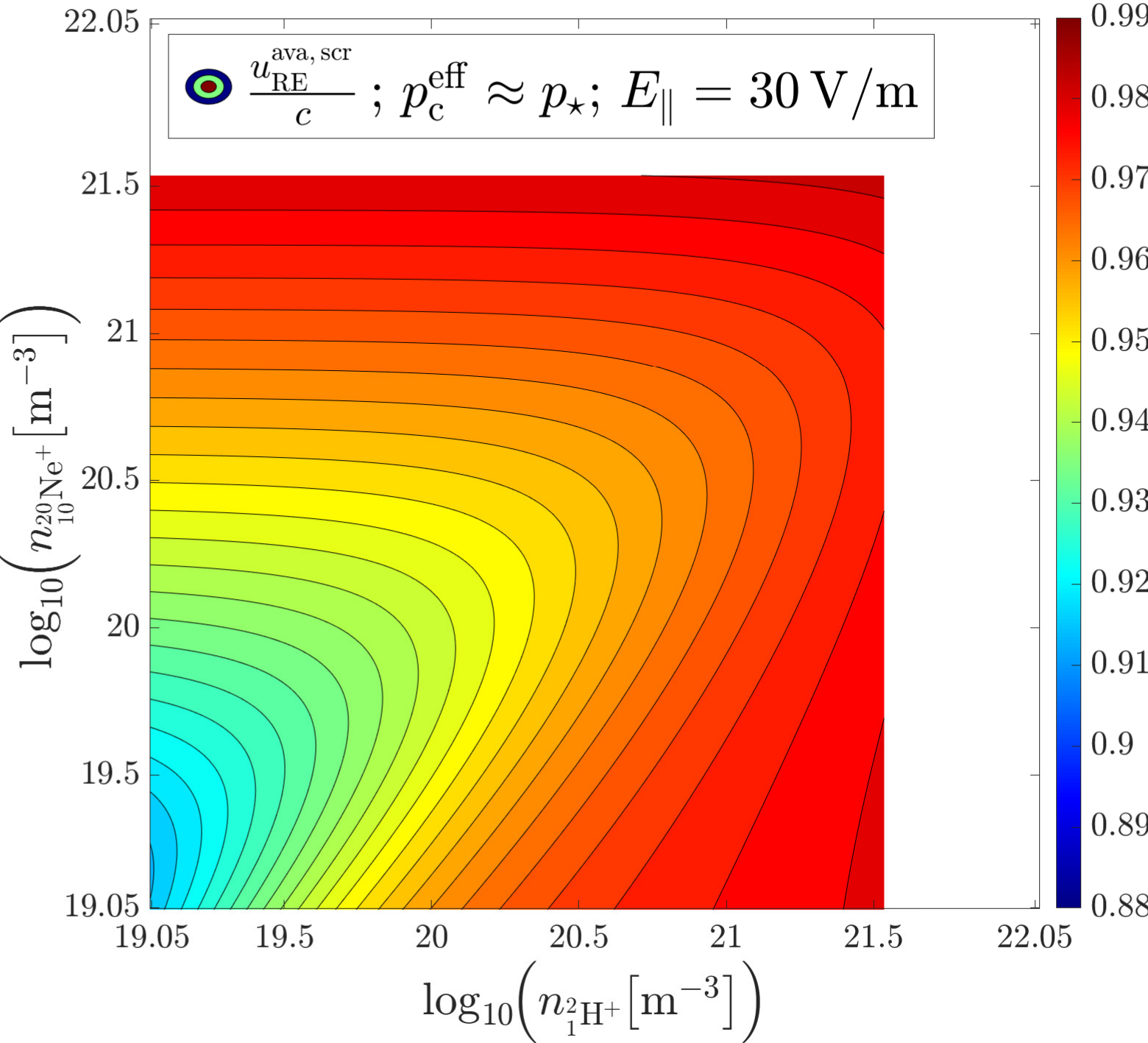}}\quad
  \subfloat{\label{fig_u_ava_screen_p_star_E100_main}
     \includegraphics[trim=315 21 366 20,width=0.41\textwidth,clip]
    {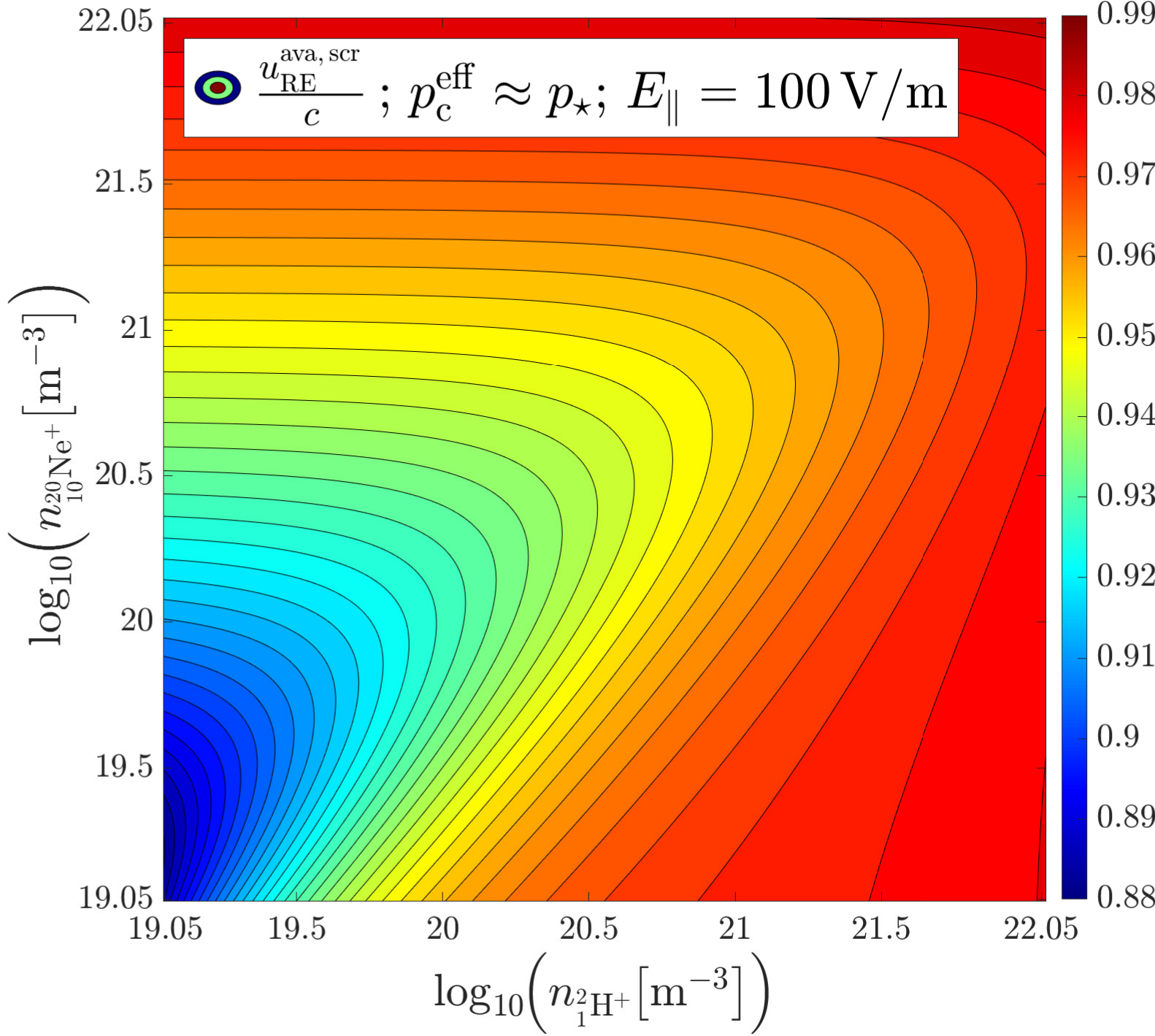}}
  \caption[Contour plots of the normalized mean velocity \mbox{$u_{\mathrm{RE}}^{\hspace{0.25mm}\mathrm{ava,scr}}/c$}, in the \textit{Hesslow} model with the effective critical momentum \mbox{$p_{\mathrm{c}}^{\mathrm{eff}}\approx p_{\star}$}, of an avalanche runaway electron population with \mbox{$k_{\mathrm{B}}T_{\mathrm{e}}=10\,\textup{eV}$}, \mbox{$B=5.25\,\textup{T}$} and \mbox{$Z_{\mathrm{eff}}=1$} for approximately logarithmically increasing values of the electric field strength \mbox{$E_{\|}\coloneqq\vert E_{\|}\vert$} (larger view in figure \ref{fig_u_ava_screen_p_star} of the appendix).]{Contour plots\protect\footnotemark{} of the normalized mean velocity \mbox{$u_{\mathrm{RE}}^{\hspace{0.25mm}\mathrm{ava,scr}}/c$}, in the \textit{Hesslow} model with the effective critical momentum \mbox{$p_{\mathrm{c}}^{\mathrm{eff}}\approx p_{\star}$}, of an avalanche runaway electron population with \mbox{$k_{\mathrm{B}}T_{\mathrm{e}}=10\,\textup{eV}$}, \mbox{$B=5.25\,\textup{T}$} and \mbox{$Z_{\mathrm{eff}}=1$} for approximately logarithmically increasing values of the electric field strength \mbox{$E_{\|}\coloneqq\vert E_{\|}\vert$} (larger view in figure \ref{fig_u_ava_screen_p_star} of the appendix).}
\label{fig_u_ava_screen_p_star_main}
\end{figure}

The magnitude of the normalized mean velocity \mbox{$u_{\mathrm{RE}}^{\mathrm{ava,\,scr}}/c$} of an avalanche runaway electron population can be obtained from a computation of the integral $\textup{I}_{\,\textup{num}}^{\,j_{RE}^{\hspace{0.25mm}\textup{ava,scr}}}$, because the definition $(\ref{RE_curr_dens_def})$ of the current density and the computation rule $(\ref{u_ava_scr_def2})$ lead to the following statement:
\vspace{-4.5mm}
\begin{equation}\label{u_ava_scr_over_c_def}
 \dfrac{ u_{\mathrm{RE}}^{\hspace{0.25mm}\textup{ava,\,scr}}}{c} \underset{ }{\overset{(\ref{RE_curr_dens_def})}{=}} \dfrac{j_{\mathrm{RE}}^{\hspace{0.25mm}\textup{ava,\,scr}}  }{-\,e\,c\,n_{RE}} \underset{}{\overset{(\ref{u_ava_scr_def2})}{=}}\textup{I}_{\,\textup{num}}^{\,j_{RE}^{\hspace{0.25mm}\textup{ava,scr}}}
\end{equation}
\vspace{-8.0mm}\\Hence, the one-dimensional numerical integration of the integral $\textup{I}_{\,\textup{num}}^{\,j_{RE}^{\hspace{0.25mm}\textup{ava,scr}}}$ with the\linebreak\mbox{\textsc{MATLAB}-}routine \qq{\texttt{integral}} produces results for the mean velocity magnitude of avalanche runaway electrons, according to the steady-state distribution function in the \textit{Hesslow} model. At this, the \textsc{MATLAB}-scripts$^{\ref{fig_plot_u_RP_footnote},\ref{fig_plot_footnote_2_main}}$ made use of the electric field values and nearly the same density intervals in the parameter space of the densities of singly-ionized deuterium and neon atoms, as for the calculations in the \textit{Rosenbluth-Putvinski} model. Their produced outputs are viewable in the listings \cref{MATLABoutput_plot_p_scr_E3,MATLABoutput_plot_p_scr_E10,MATLABoutput_plot_p_scr_E30,MATLABoutput_plot_p_scr_E100,MATLABoutput_plot_p_star_E3,MATLABoutput_plot_p_star_E10,MATLABoutput_plot_p_star_E30,MATLABoutput_plot_p_star_E100} in subsection \ref{output_matlab_appendix_subsection} of the appendix. They provide the minima, maxima and mean values of the results, related to each contour plot, state the general parameter settings and show, that the control criterion $(\ref{check_n_RE_ava_def_H})$ was satisfied for all computations. In consequence of the similar calculation settings, a comparability of the contour plots of \mbox{$u_{\mathrm{RE}}^{\hspace{0.25mm}\mathrm{ava,scr}}/c$} in figure \ref{fig_u_ava_screen_p_c_scr_main} and figure \ref{fig_u_ava_screen_p_star_main}, which account for the effects of partial screening, with the data for \mbox{$u_{\mathrm{RE}}^{\hspace{0.25mm}\mathrm{ava}}/c$} from figure \ref{fig_u_ava_p_c_scr_main} of section \ref{RP_avalanche_j_subsection}, is ensured, since also the same relation $(\ref{CoulombLogrel})$ was used to obtain values for the relativistic \textit{Coulomb} logarithm. On this occasion, it has to be remarked, that the computation of the mean velocity of the avalanche runaway electrons in the \textit{Hesslow} model was carried out for two approximations of the effective critical momentum $p_{\mathrm{c}}^{\mathrm{eff}}$, in order to evaluate their influence on the first moment of the one-dimensional distribution function from \textit{P.\hspace{0.9mm}Svensson}. Therefore, two figures were produced and their data was additionally compared to the results of \mbox{$u_{\mathrm{RE}}^{\hspace{0.25mm}\mathrm{ava}}/c$}, displayed in figure \ref{fig_u_ava_p_c_scr_main}, by means of the relative deviation between the computed values for each point in the density parameter space. Although, the associated figures for the relative deviations are shown in the subsequent section, the first figure to analyse is the figure \ref{fig_u_ava_screen_p_c_scr_main} were the quantity \mbox{$u_{\mathrm{RE}}^{\hspace{0.25mm}\mathrm{ava}}/c$} originates from the integration of the integral $\textup{I}_{\,\textup{num}}^{\,j_{RE}^{\hspace{0.25mm}\textup{ava,scr}}}$ with the lower momentum boundary \mbox{$p_{\mathrm{c}}^{\mathrm{eff}}\approx p_{\mathrm{c}}^{\mathrm{scr}}$}.
\vspace*{-4.0mm}An evaluation of the figure \ref{fig_u_ava_screen_p_c_scr_main} again depicts the growth of the avalanche runaway electron generation region within the density parameter space as a consequence of an increase in the electric field strength. Moreover, one observes a more distinct deviation of the mean velocity from the speed of light than in figure \ref{fig_u_ava_p_c_scr_main} of the previous section \ref{RP_avalanche_j_subsection}. Furthermore, an asymmetry of the contour lines to the diagonal \mbox{$n_{_{10}^{20}\mathrm{Ne}^{+}}(n_{_{1}^{2}\mathrm{H}^{+}})=n_{_{1}^{2}\mathrm{H}^{+}}$} is apparent, which is contrary to the results in the \textit{Rosenbluth-Putvinski} model. This means, that the distribution function from the \textit{Hesslow} model resolves the different behaviour for a change in the neon- or the deuterium ion density and therefore includes influences of partial screening on the mean velocity of the avalanche runaway electrons. For instance, a non-negligible minimum of the normalized mean velocity magnitude is visible, if one considers a fixed deuterium density and looks at the plotted data for an increasing neon density. This illustrates one of the reasons why impurity injection is studied for runaway mitigation in tokamak disruptions. Interestingly, one notices a decrease of the velocity for lower density parameter points, if the prevalent electric field is enhanced logarithmically for the four contour plots. Nevertheless, the discovered velocity minimum is only distinct, if the electron density is not too close to the maximum electron density, which corresponds to the point of a closing runaway region, where the critical field exceeds the present accelerating electric field. In the scenarios with high densities, one finds the velocity to be close to the speed of light. However, it has to be remarked, that this is also the region, where the condition \mbox{$E_{\|}\gg E_{c}$} does not hold and the applicability and accuracy of the \textit{Hesslow} model is not validated. 

Next the figure \ref{fig_u_ava_screen_p_star_main} shall be discussed, which displays the normalized mean avalanche runaway electron velocity \mbox{$u_{\mathrm{RE}}^{\hspace{0.25mm}\mathrm{ava,scr}}/c$}, computed from the integral from equation $(\ref{u_ava_scr_def2})$, but for the more accurate approximation \mbox{$p_{\mathrm{c}}^{\mathrm{eff}}\approx p_{\star}$} for the effective critical momentum as the lower integration boundary.
\\
First, it needs to be mentioned, that the lower interval boundaries for the deuterium and neon ion densities are larger than in figure \ref{fig_u_ava_screen_p_c_scr_main}, because the calculation of $p_{\star}$, as the root of the defining function $f_{p_{\star}}(p)$ from $(\ref{func_p_c_eff_def})$, did not converge for lower densities. Hence, the depicted density parameter space also represents the density region in which this approximation of the lower momentum boundary of the runaway region is applicable.
\\
By comparison of the values of the velocity for certain density data points in figure \ref{fig_u_ava_screen_p_c_scr_main} and figure \ref{fig_u_ava_screen_p_star_main} only minor deviations are recognizable. Although, it seems to be the case, that the calculation of $p_{\star}$ is indeed physically more accurate, due to the fact that it reveals, that the region, where the velocity is approximately equal to the speed of light, does not extend as far in the direction of lower densities, as predicted from the approximation \mbox{$p_{\mathrm{c}}^{\mathrm{eff}}\approx p_{\mathrm{c}}^{\mathrm{scr}}$}, as utilized for the calculation of the data shown in figure \ref{fig_u_ava_screen_p_c_scr_main}. Finally, one should remark, that the visible minimum of the mean velocitiy within the parameter space, is correlated with the maximum deviation between the effective critical electric field and the \textit{Connor-Hastie} critical electric field, as depicted in\vspace{-7cm}\linebreak \newpage\noindent
\footnotetext{\label{fig_plot_footnote_2_main} The contour plots were produced, by means of the \textsc{MATLAB}-scripts\\ \hspace*{8.7mm}\qq{\texttt{generate_num_data_densities_p_star_E3.m}},\\ \hspace*{8.7mm}\qq{\texttt{plot_num_data_densities_p_star_E3.m}},\\ \hspace*{8.7mm}\qq{\texttt{generate_num_data_densities_p_star_E10.m}},\\ \hspace*{8.7mm}\qq{\texttt{plot_num_data_densities_p_star_E10.m}},\\ \hspace*{8.7mm}\qq{\texttt{generate_num_data_densities_p_star_E30.m}},\\ \hspace*{8.7mm}\qq{\texttt{plot_num_data_densities_p_star_E30.m}},\\ \hspace*{8.7mm}\qq{\texttt{generate_num_data_densities_p_star_E100.m}} and\\ \hspace*{8.7mm}\qq{\texttt{plot_num_data_densities_p_star_E100.m}}, which are stored in the digital appendix.}
\vspace*{-4.0mm}
figure \ref{fig_E_C_ava_p_c_scr_main}. This explains the lower velocities in this density region, due to the fact that for a fixed electric field a larger critical electric field, as proposed by the \textit{Hesslow} model, leads to a slower net acceleration of the runaway electrons. 

\subsection{Comparison of the models by means of the current density of an \textit{avalanche} runaway electron population}\label{comparison_avalanche_j_subsection}

\begin{figure}[H]
  \centering
  \subfloat{\label{fig_rel_u_ava_p_c_scr_E3_main} 
   \includegraphics[trim=312 19 345 23,width=0.41\textwidth,clip]
    {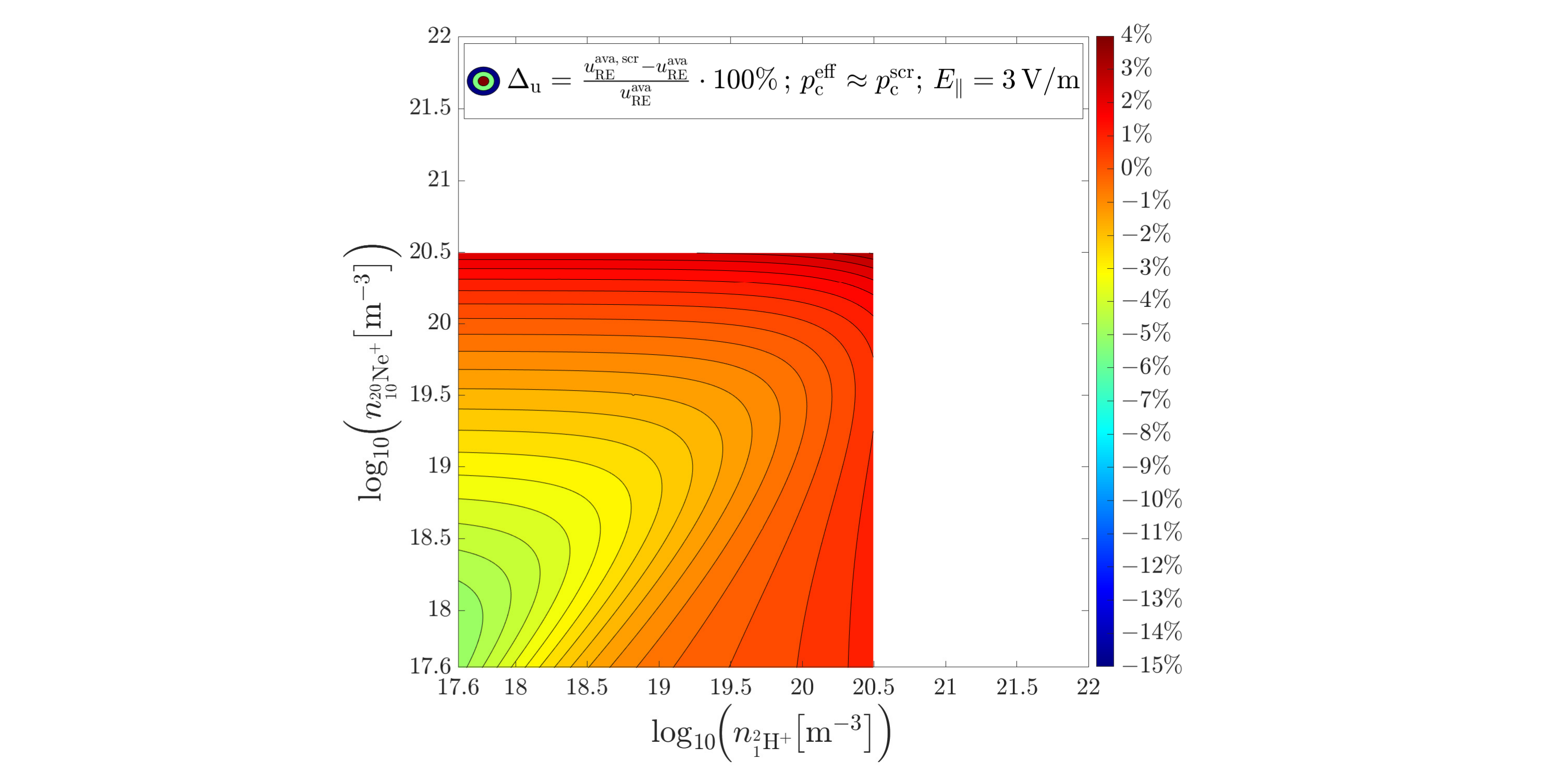}}\quad
  \subfloat{\label{fig_rel_u_ava_p_c_scr_E10_main}
    \includegraphics[trim=311 27 353 13,width=0.41\textwidth,clip]
    {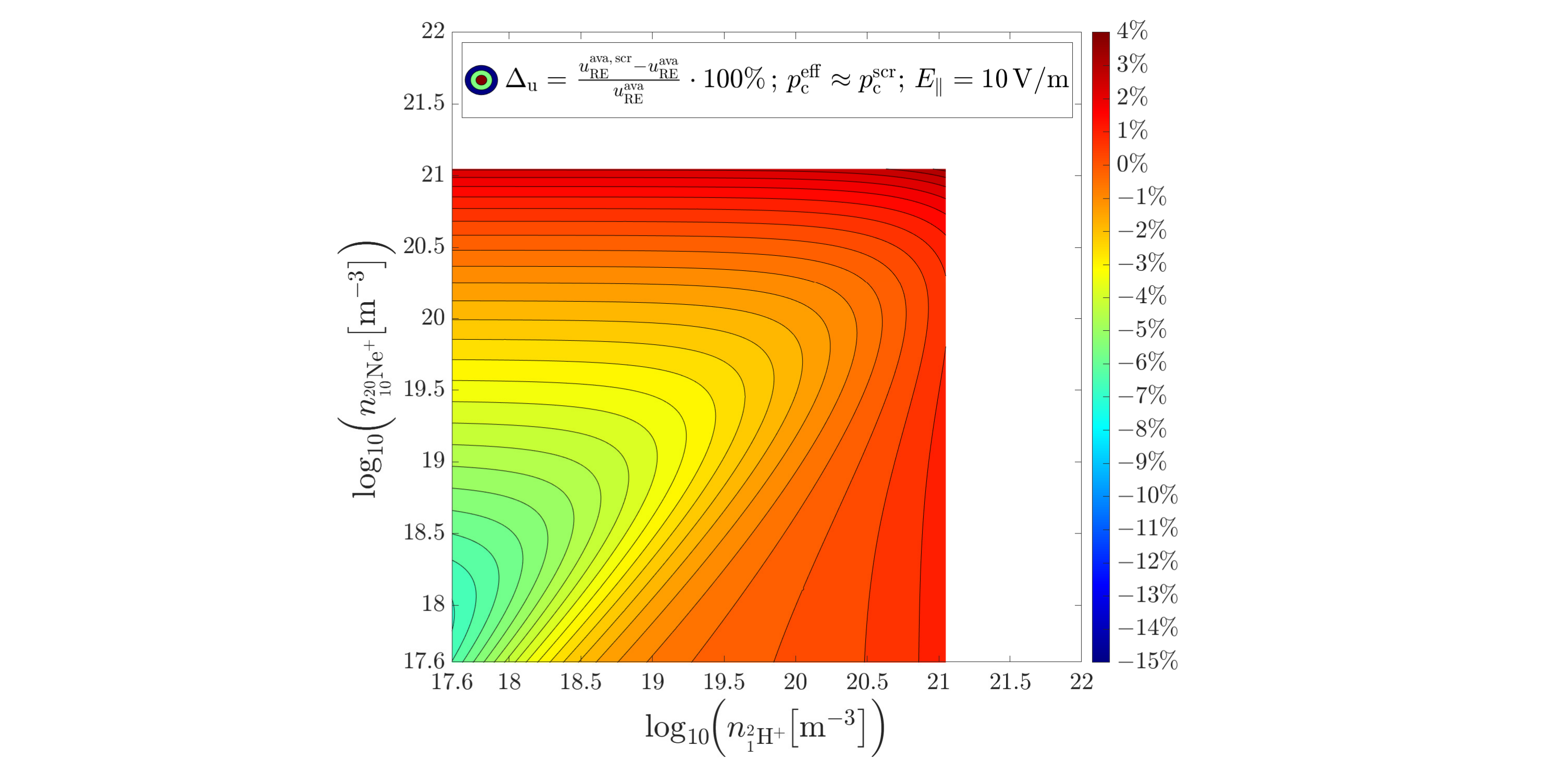}}\\[4pt]
  \subfloat{\label{fig_rel_u_ava_p_c_scr_E30_main} 
    \includegraphics[trim=312 21 346 20,width=0.41\textwidth,clip]
    {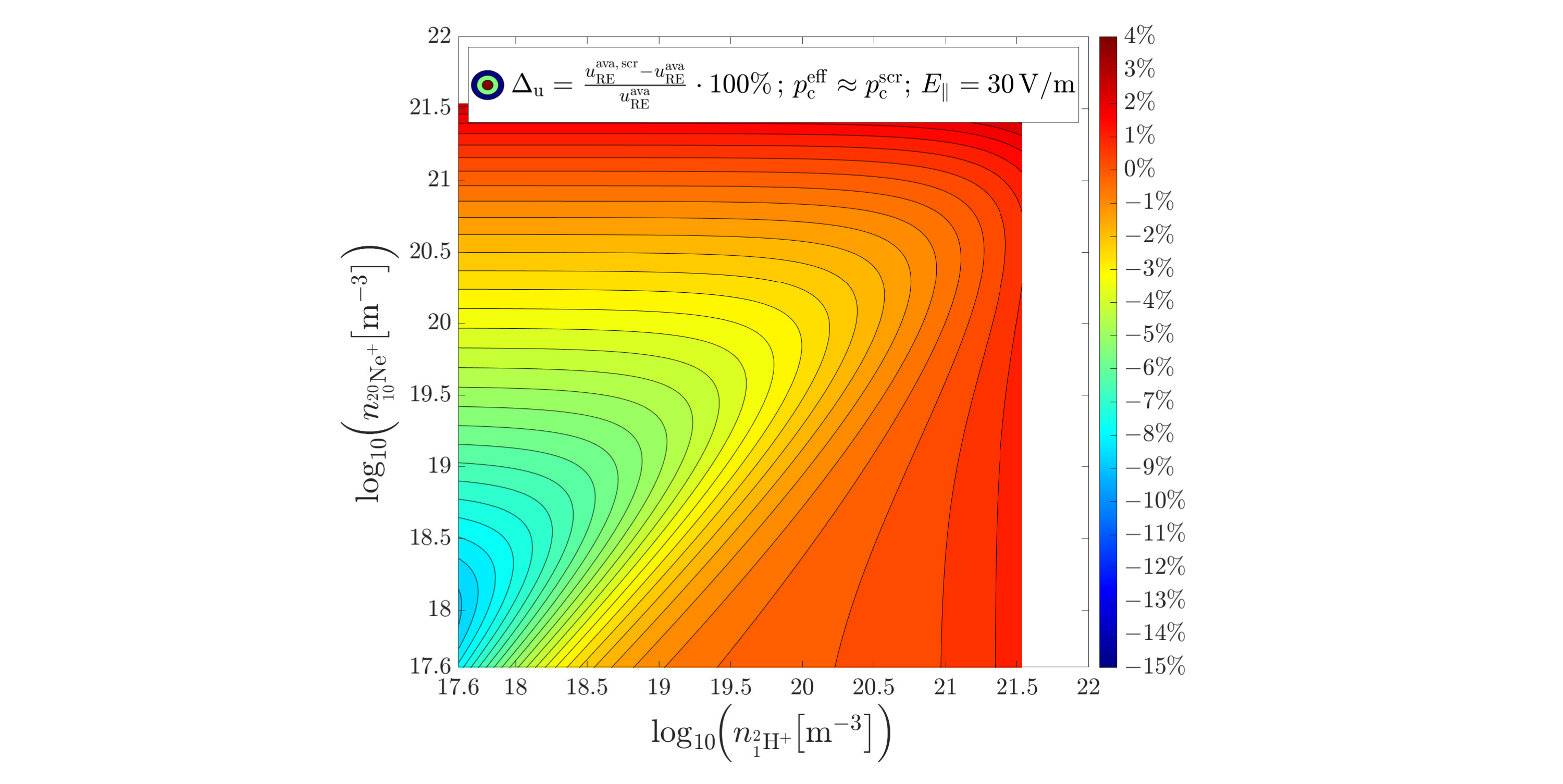}}\quad
  \subfloat{\label{fig_rel_u_ava_p_c_scr_E100_main}
    \includegraphics[trim=314 27 350 14,width=0.41\textwidth,clip]
    {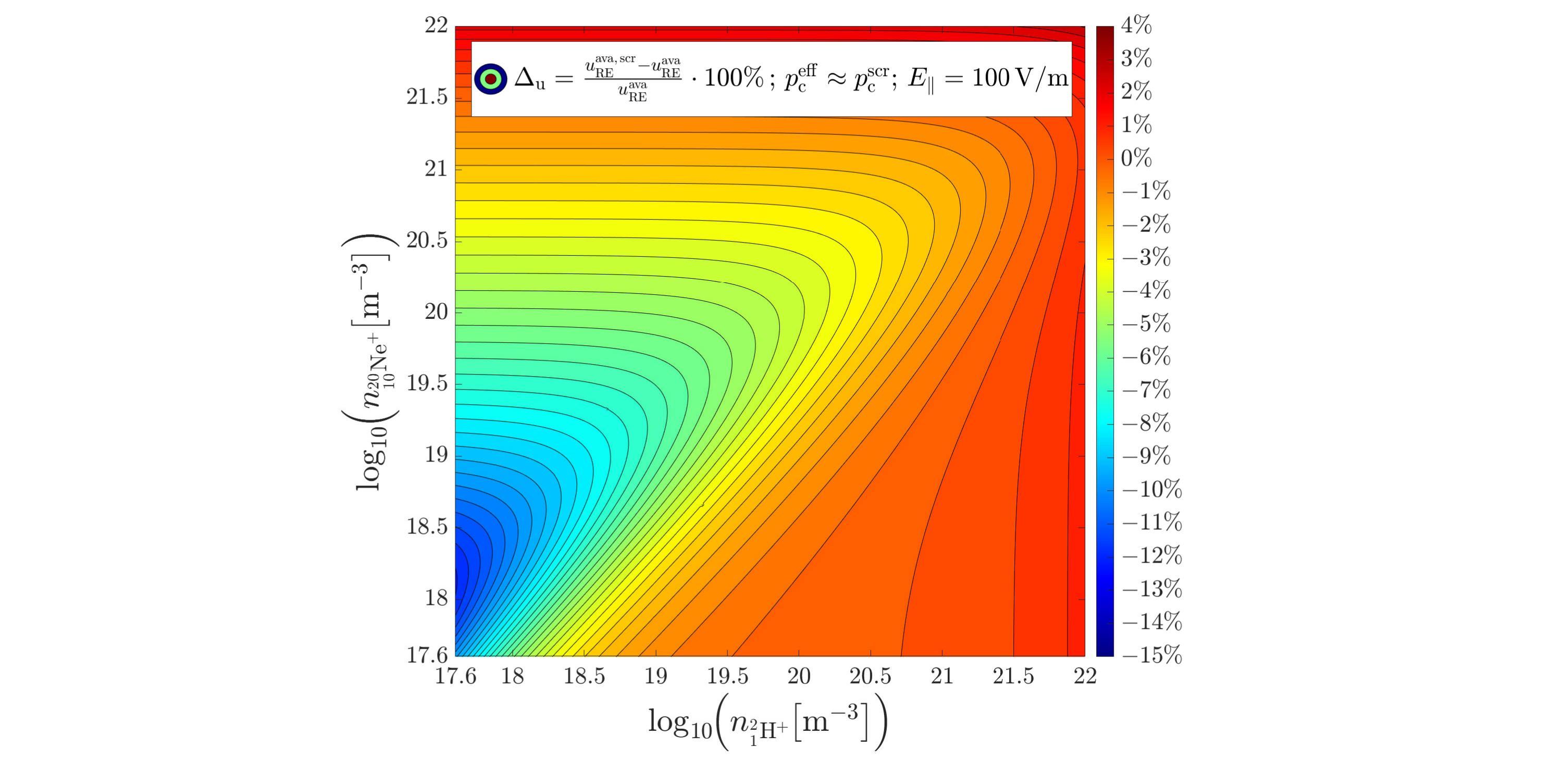}}
  \caption[Contour plots of the relative deviation $\Delta_{u}$ for the mean velocity of an avalanche runaway electron population with \mbox{$k_{\mathrm{B}}T_{\mathrm{e}}=10\,\textup{eV}$}, \mbox{$B=5.25\,\textup{T}$} and \mbox{$Z_{\mathrm{eff}}=1$}, due to the effect of partial screening with the effective critical momentum \mbox{$p_{\mathrm{c}}^{\mathrm{eff}}\approx p_{\mathrm{c}}^{\mathrm{scr}}$}, displayed for different values of the electric field strength \mbox{$E_{\|}\coloneqq\vert E_{\|}\vert$} (larger view in figure \ref{fig_rel_u_ava_p_c_scr} of the appendix).]{Contour plots$^{\ref{fig_plot_u_RP_footnote}}$ of the relative deviation $\Delta_{u}$ for the mean velocity of an avalanche runaway electron population with \mbox{$k_{\mathrm{B}}T_{\mathrm{e}}=10\,\textup{eV}$}, \mbox{$B=5.25\,\textup{T}$} and \mbox{$Z_{\mathrm{eff}}=1$}, due to the effect of partial screening with the effective critical momentum \mbox{$p_{\mathrm{c}}^{\mathrm{eff}}\approx p_{\mathrm{c}}^{\mathrm{scr}}$}, displayed for approximately logarithmically increasing values of the electric field strength \mbox{$E_{\|}\coloneqq\vert E_{\|}\vert$} (larger view in figure \ref{fig_rel_u_ava_p_c_scr} of the appendix).}
\label{fig_rel_u_ava_p_c_scr_main}
\end{figure}

As mentioned in the previous section, one is able to compare the data for \mbox{$u_{\mathrm{RE}}^{\hspace{0.25mm}\mathrm{ava,scr}}/c$} from the figure \ref{fig_u_ava_screen_p_c_scr_main} and \ref{fig_u_ava_screen_p_star_main}, originating from the \textit{Hesslow} model for the different lower momentum boundary approximations \mbox{$p_{\mathrm{c}}^{\mathrm{eff}}\approx p_{\mathrm{c}}^{\mathrm{scr}}$} and \mbox{$p_{\mathrm{c}}^{\mathrm{eff}}\approx p_{\star}$}, by means of their relative deviation from the results for \mbox{$u_{\mathrm{RE}}^{\hspace{0.25mm}\mathrm{ava}}/c$} in the \textit{Rosenbluth-Putvinski} model from figure \ref{fig_u_ava_p_c_scr_main}. For this purpose, the relative deviation between the normalized mean velocity in the \textit{Hesslow} model \mbox{$u_{\mathrm{RE}}^{\hspace{0.25mm}\mathrm{ava,scr}}/c$}, related to the approximated effective critical momentum $p_{\mathrm{c}}^{\mathrm{scr}}$, as defined in $\ref{p_c_scr_def}$, is presented in figure \ref{fig_rel_u_ava_p_c_scr_main}.\vspace*{-0.5mm}
\vspace*{-5.0mm}An observation leads to the insight, that the \textit{Hesslow} model in combination with \mbox{$p_{\mathrm{c}}^{\mathrm{eff}}\approx p_{\mathrm{c}}^{\mathrm{scr}}$} corrects the \textit{Rosenbluth-Putvinski} approach to higher mean velocity magnitudes, especially for higher electric field strength values and large densities. However, this underestimation of \mbox{$u_{\mathrm{RE}}^{\hspace{0.25mm}\mathrm{ava,scr}}/c$} by \mbox{$u_{\mathrm{RE}}^{\hspace{0.25mm}\mathrm{ava}}/c$} in the high density limit, is smaller than the overestimation of \mbox{$u_{\mathrm{RE}}^{\hspace{0.25mm}\mathrm{ava,scr}}/c$} by \mbox{$u_{\mathrm{RE}}^{\hspace{0.25mm}\mathrm{ava}}/c$} for lower densities. Additionally, the asymmetric contour lines and the minimum of the relative deviation are similar to figure \ref{fig_u_ava_screen_p_c_scr_main}. This means, that the reasons for the deviations between the two models are indeed the effects of partial screening, which leads to an enhanced gradient in the mean velocity \mbox{$u_{\mathrm{RE}}^{\hspace{0.25mm}\mathrm{ava,scr}}/c$} along the neon density direction in comparison to the change in the mean velocity in the deuterium density direction. 
 
\begin{figure}[H]
  \centering
  \subfloat{\label{fig_rel_u_ava_p_star_E3_main} 
    \includegraphics[trim=325 20 345 21,width=0.41\textwidth,clip]
    {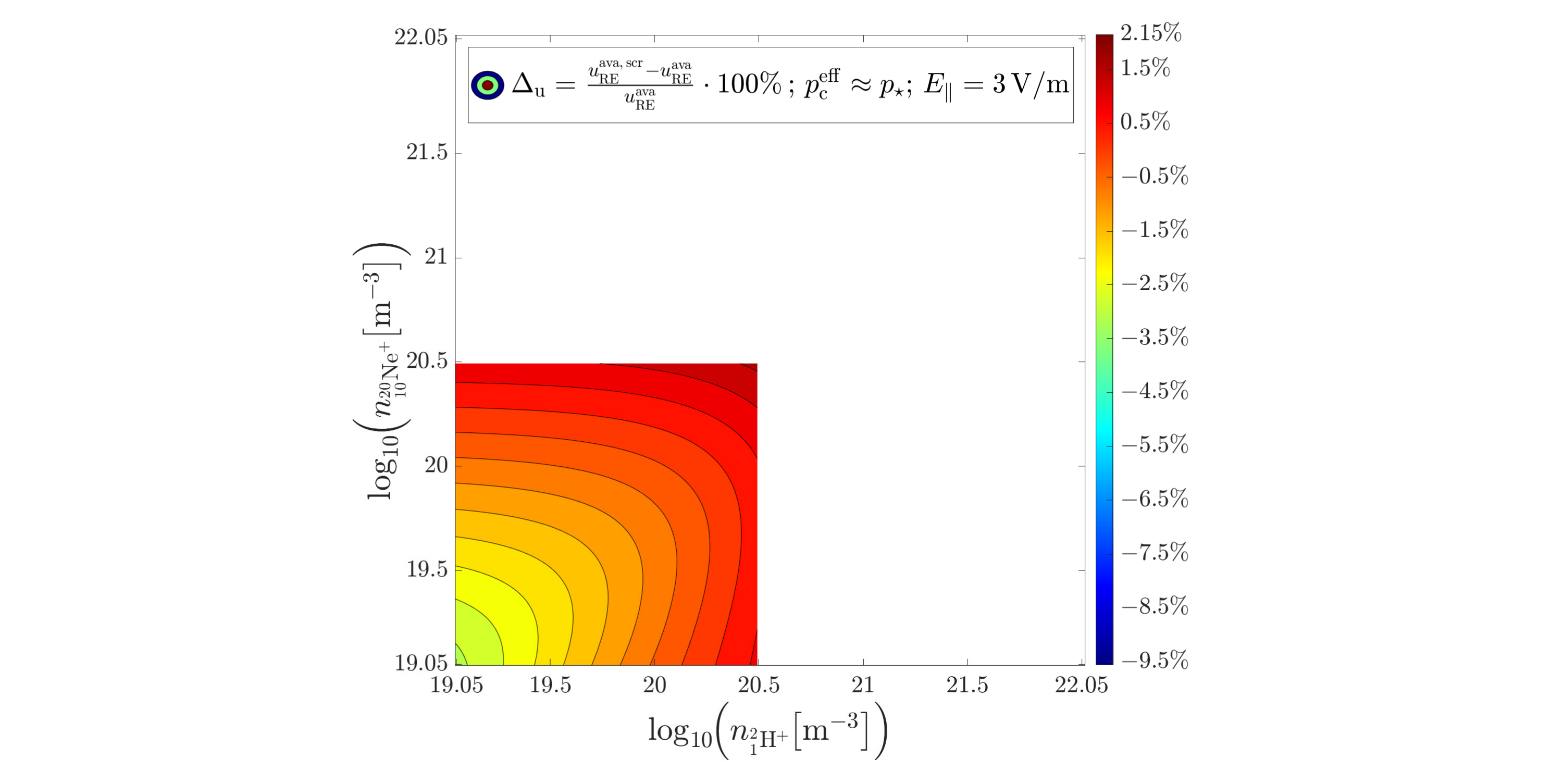}}\quad
  \subfloat{\label{fig_rel_u_ava_p_star_E10_main}
     \includegraphics[trim=329 25 341 17,width=0.41\textwidth,clip]
    {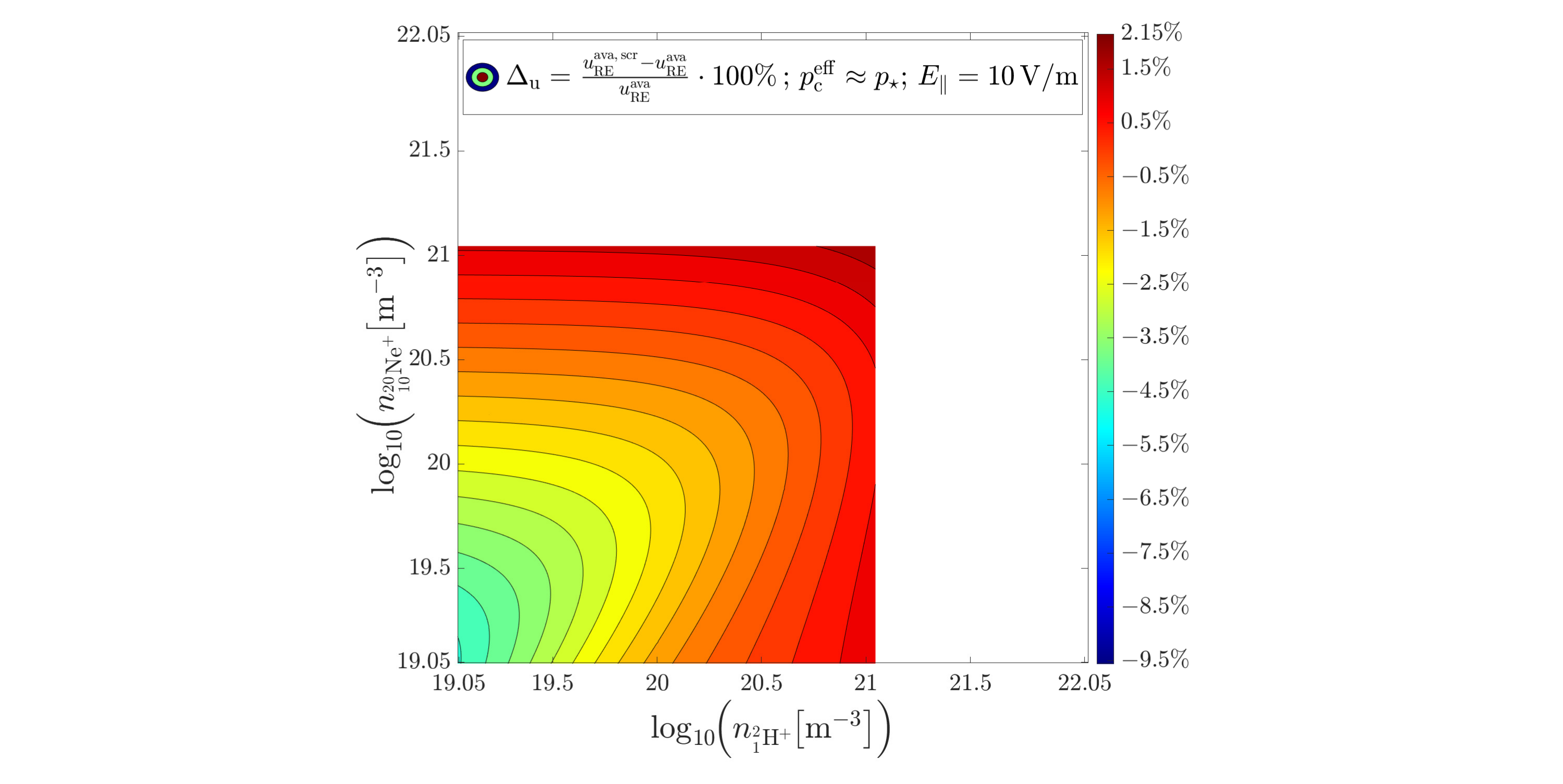}}\\[4pt]
  \subfloat{\label{fig_rel_u_ava_p_star_E30_main} 
    \includegraphics[trim=325 20 346 20,width=0.41\textwidth,clip]
    {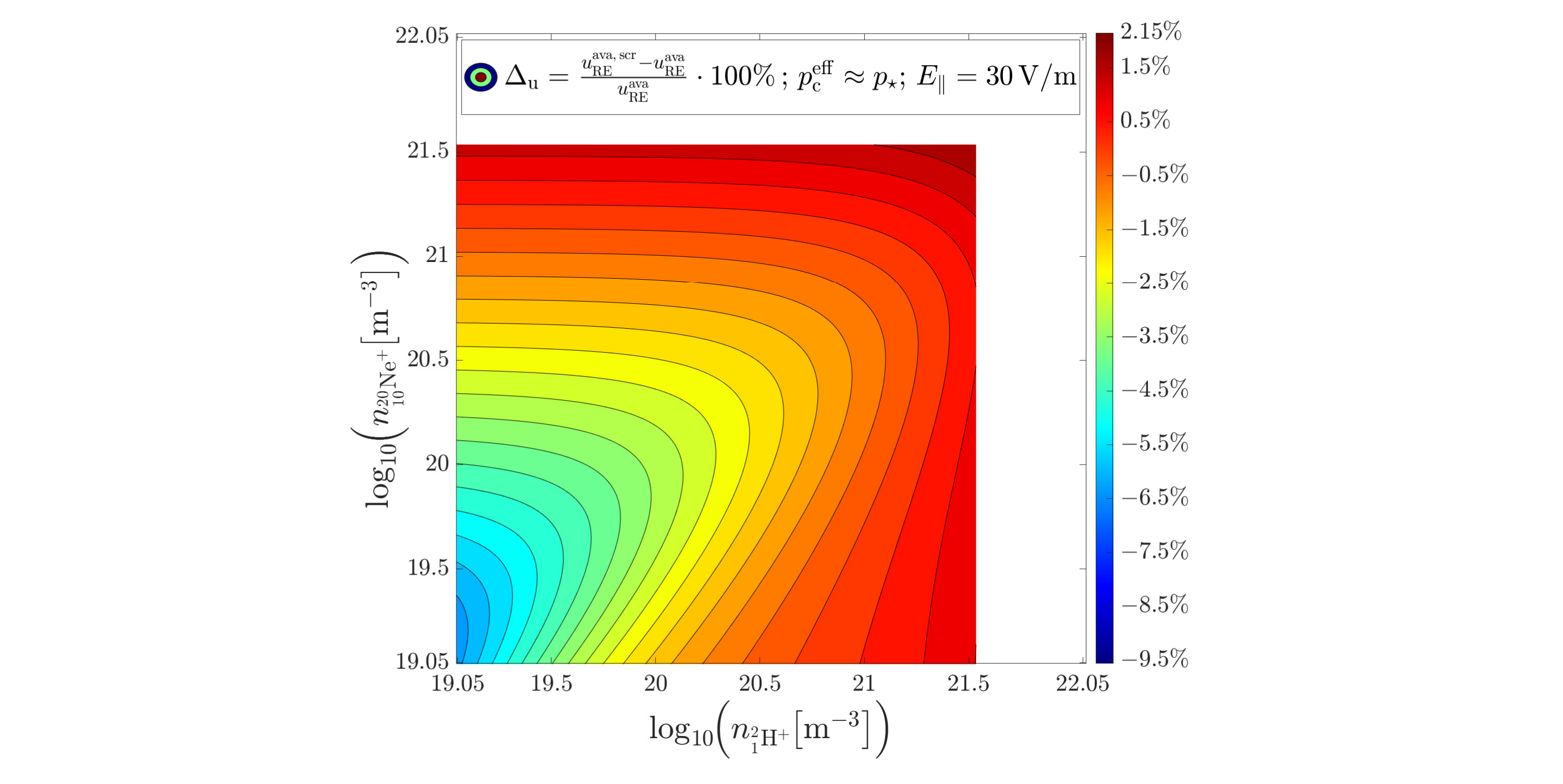}}\quad
  \subfloat{\label{fig_rel_u_ava_p_star_E100_main}
    \includegraphics[trim=319 21 346 20,width=0.41\textwidth,clip]
    {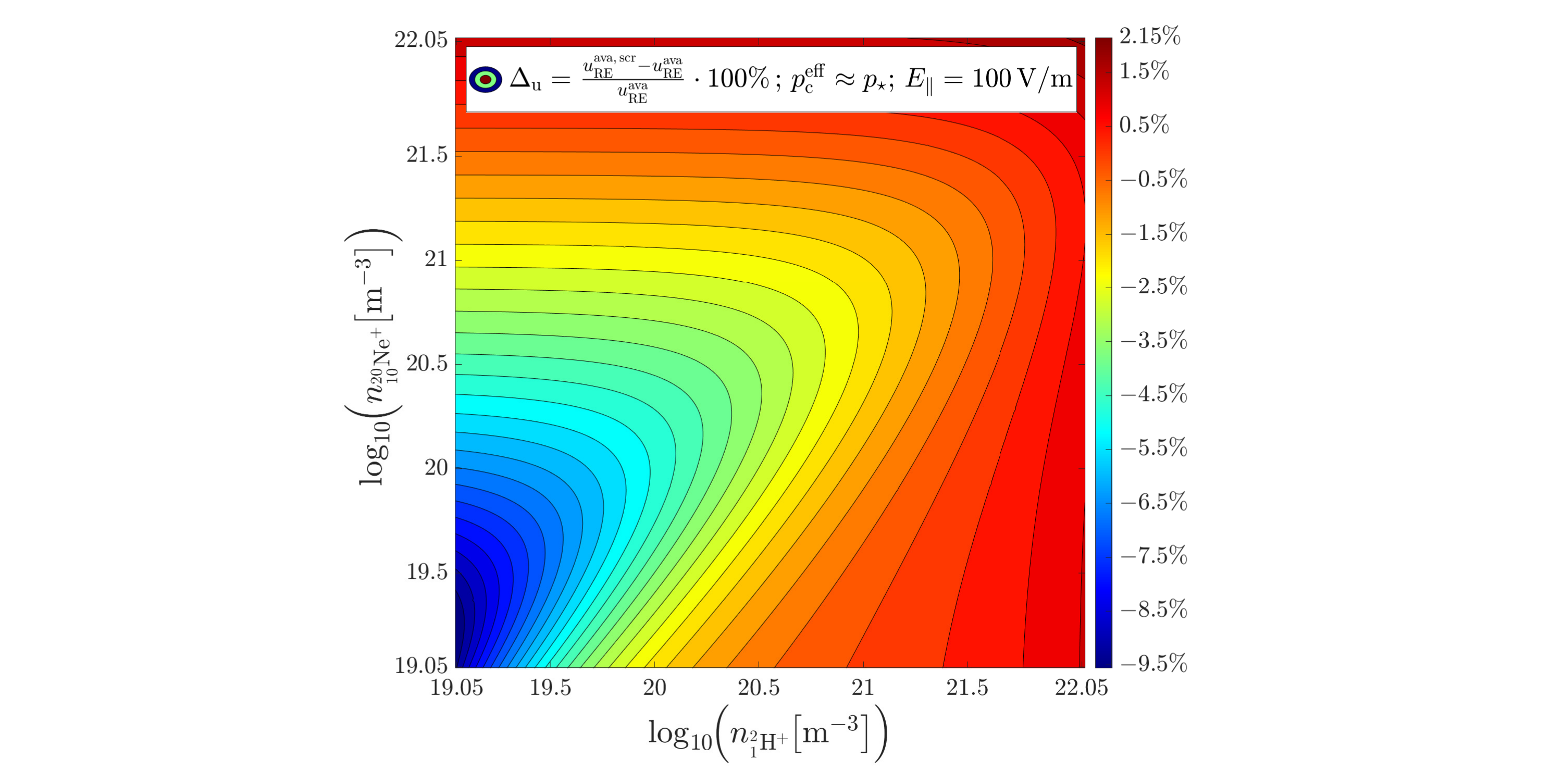}}
  \caption[Contour plots of the relative deviation $\Delta_{u}$ for the mean velocity of an avalanche runaway electron population with \mbox{$k_{\mathrm{B}}T_{\mathrm{e}}=10\,\textup{eV}$}, \mbox{$B=5.25\,\textup{T}$} and \mbox{$Z_{\mathrm{eff}}=1$}, due to the effect of partial screening with the effective critical momentum \mbox{$p_{\mathrm{c}}^{\mathrm{eff}}\approx p_{\star}$}, displayed for approximately logarithmically increasing values of the electric field strength \mbox{$E_{\|}\coloneqq\vert E_{\|}\vert$} (larger view in figure \ref{fig_rel_u_ava_p_star} of the appendix).]{Contour plots$^{\ref{fig_plot_footnote_2_main}}$ of the relative deviation $\Delta_{u}$ for the mean velocity of an avalanche runaway electron population with \mbox{$k_{\mathrm{B}}T_{\mathrm{e}}=10\,\textup{eV}$}, \mbox{$B=5.25\,\textup{T}$} and \mbox{$Z_{\mathrm{eff}}=1$}, due to the effect of partial screening with the effective critical momentum \mbox{$p_{\mathrm{c}}^{\mathrm{eff}}\approx p_{\star}$}, displayed for approximately logarithmically increasing values of the electric field strength \mbox{$E_{\|}\coloneqq\vert E_{\|}\vert$} (larger view in figure \ref{fig_rel_u_ava_p_star} of the appendix).}
\label{fig_rel_u_ava_p_star_main}
\end{figure}

\begin{figure}[H]
  \centering
  \subfloat{\label{fig_tilde_rel_u_ava_p_star_E3_main} 
    \includegraphics[trim=319 35 335 8,width=0.41\textwidth,clip]
    {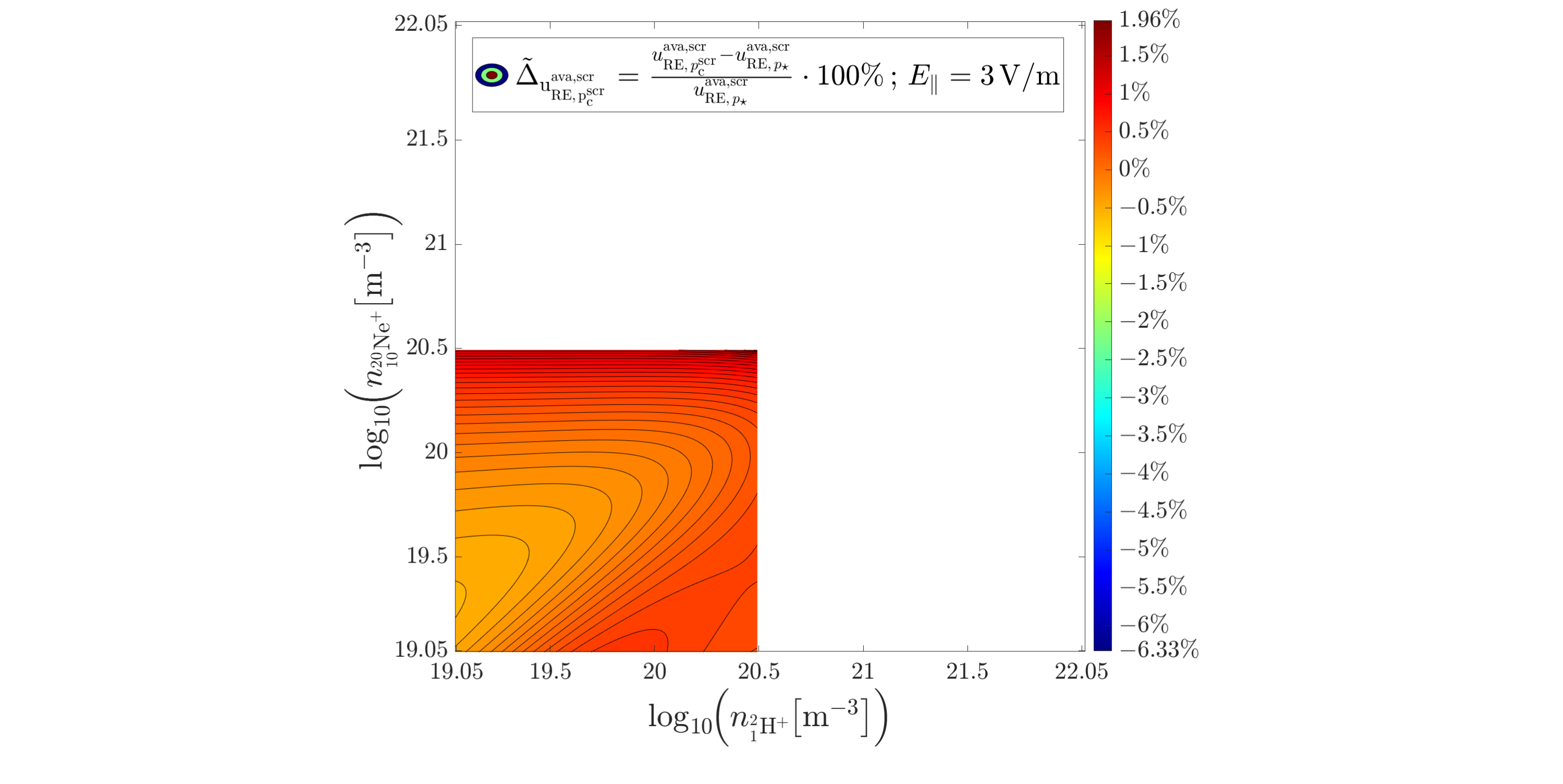}}\quad
  \subfloat{\label{fig_tilde_rel_u_ava_p_star_E10_main}
    \includegraphics[trim=323 37 334 8,width=0.41\textwidth,clip]
    {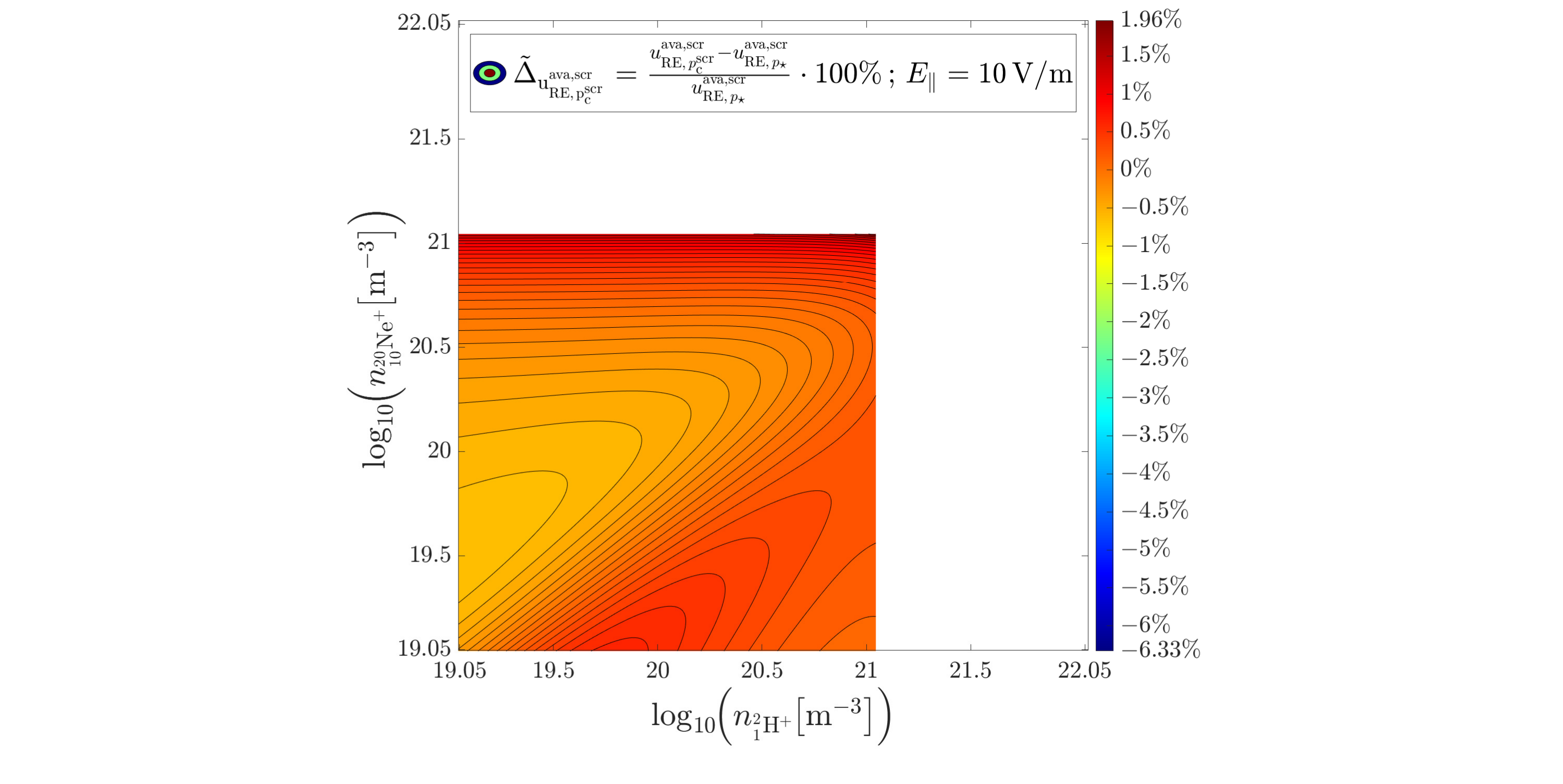}}\\[4pt]
  \subfloat{\label{fig_tilde_rel_u_ava_p_star_E30_main} 
    \includegraphics[trim=322 33 333 10,width=0.41\textwidth,clip]
    {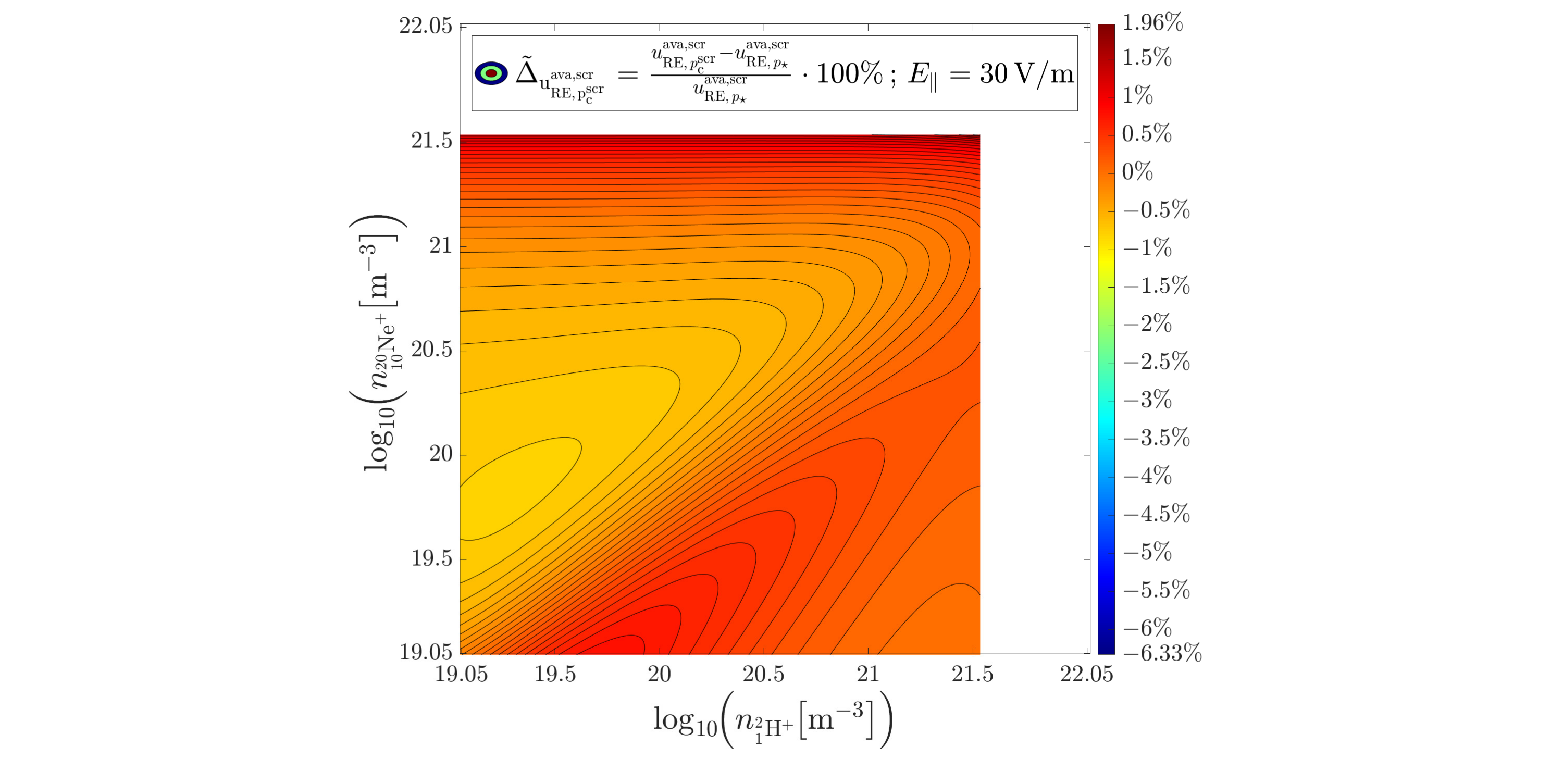}}\quad
  \subfloat{\label{fig_tilde_rel_u_ava_p_star_E100_main}
  \includegraphics[trim=320 34 335 8,width=0.41\textwidth,clip]
    {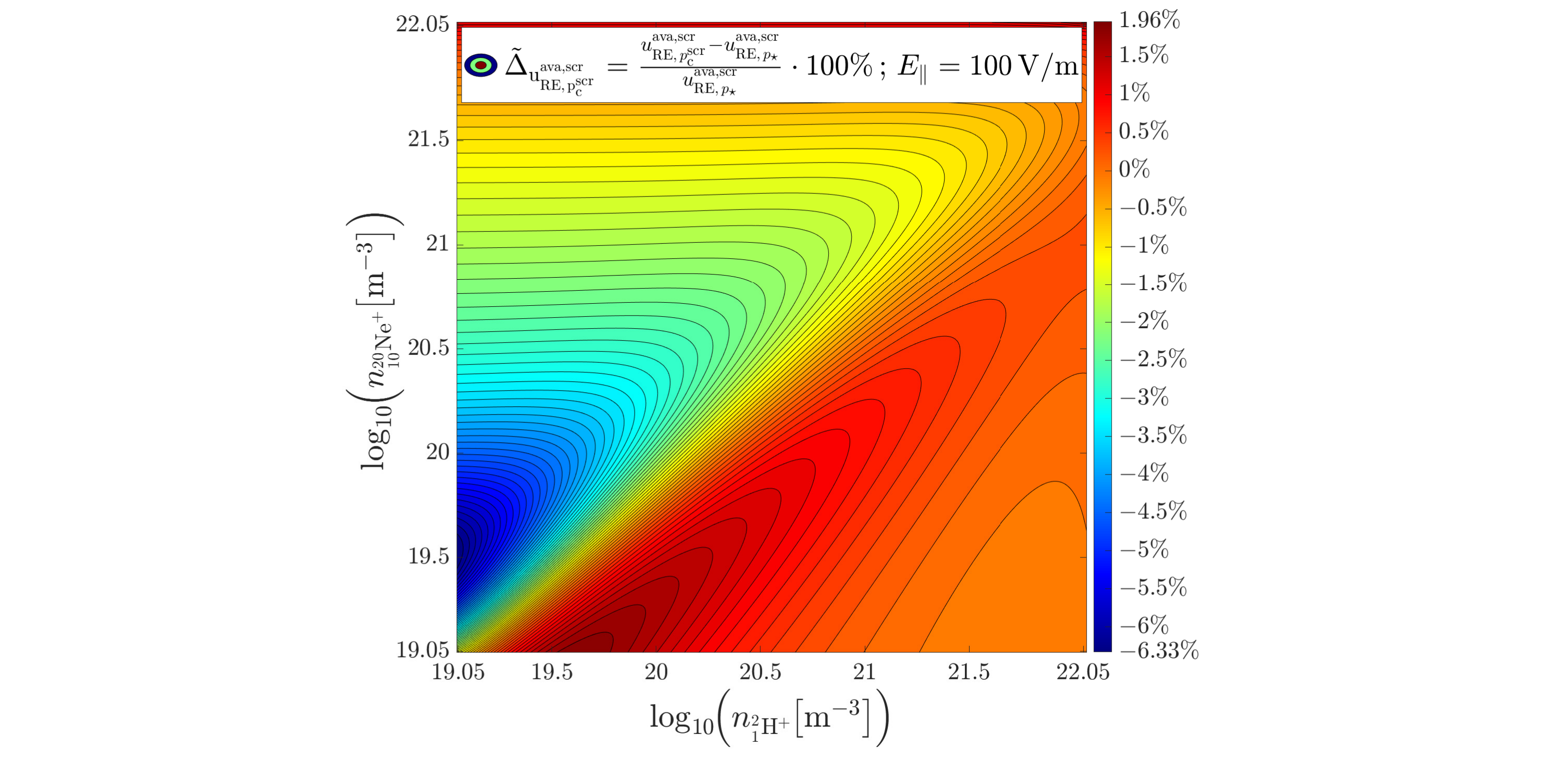}}
  \caption[Contour plots of the relative deviation $\tilde{\Delta}_{u_{\mathrm{RE},\,p^{\mathrm{scr}}_{\mathrm{c}}}^{\mathrm{ava,scr}}}$ for the mean velocity of an avalanche runaway electron population with \mbox{$k_{\mathrm{B}}T_{\mathrm{e}}=10\,\textup{eV}$}, \mbox{$B=5.25\,\textup{T}$} and \mbox{$Z_{\mathrm{eff}}=1$} in the \textit{Hesslow} model, due to the different approximations of the effective critical momentum \mbox{$p_{\mathrm{c}}^{\mathrm{eff}}\approx p_{\mathrm{c}}^{\mathrm{scr}}$} and \mbox{$p_{\mathrm{c}}^{\mathrm{eff}}\approx p_{\star}$}, displayed for approximately logarithmically increasing values of the electric field strength \mbox{$E_{\|}\coloneqq\vert E_{\|}\vert$} (larger view in figure \ref{fig_tilde_rel_u_ava_p_star} of the appen-\linebreak dix).]{Contour plots$^{\ref{fig_plot_footnote_2_main}}$ of the relative deviation $\tilde{\Delta}_{u_{\mathrm{RE},\,p^{\mathrm{scr}}_{\mathrm{c}}}^{\mathrm{ava,scr}}}$ for the mean velocity of an avalanche runaway electron population with \mbox{$k_{\mathrm{B}}T_{\mathrm{e}}=10\,\textup{eV}$}, \mbox{$B=5.25\,\textup{T}$} and \mbox{$Z_{\mathrm{eff}}=1$} in the \textit{Hesslow} model, due to the different approximations of the effective critical momentum \mbox{$p_{\mathrm{c}}^{\mathrm{eff}}\approx p_{\mathrm{c}}^{\mathrm{scr}}$} and \mbox{$p_{\mathrm{c}}^{\mathrm{eff}}\approx p_{\star}$}, displayed for approximately logarithmically increasing values of the electric field strength \mbox{$E_{\|}\coloneqq\vert E_{\|}\vert$} (larger view in figure \ref{fig_tilde_rel_u_ava_p_star} of the appendix).}
\label{fig_tilde_rel_u_ava_p_star_main}
\end{figure}

Similar, deductions follow from the discussion of the figure \ref{fig_rel_u_ava_p_star_main}, which shows the relative deviation between the normalized mean velocity in the \textit{Hesslow} model \mbox{$u_{\mathrm{RE}}^{\hspace{0.25mm}\mathrm{ava,scr}}/c$}, computed with approximated effective critical momentum $p_{\star}$ and the normalized mean velocity \mbox{$u_{\mathrm{RE}}^{\hspace{0.25mm}\mathrm{ava}}/c$} from the \textit{Rosenbluth-Putvinski} approach.
\vspace*{-4.0mm}In comparison to the relative deviations displayed in the above figure \ref{fig_rel_u_ava_p_c_scr_main}, it might be remarked, that the calculation based on $p_{\star}$ leads to smaller positive deviations in the high density limit and smaller negative relative differences for lower densities. Since the analytic approximation $p_{\mathrm{c}}^{\mathrm{scr}}$ is less accurate than $p_{\star}$, one can deduce, that the approximative formula for $p_{\mathrm{c}}^{\mathrm{scr}}$ leads to results, which differ more distinctly from the \textit{Rosenbluth-Putvinski} approach. Hence, one could say, that the utilization of \mbox{$p_{\mathrm{c}}^{\mathrm{eff}}\approx p_{\mathrm{c}}^{\mathrm{scr}}$} generally overestimates the magnitude of the mean velocity of an avalanche runaway electron population. In this context, a recapitulation of the relative deviations $\Delta_{p^{\mathrm{scr}}_{c}}$ and $\Delta_{p_{\star}}$, as plotted in figure \ref{fig_p_comparison}, is recommended. The analysis of those contour plots reveals a higher range of the results for the deviation $\Delta_{p^{\mathrm{scr}}_{c}}$ in comparison to $\Delta_{p^{\star}}$, so that the higher deviations of \mbox{$u_{\mathrm{RE}}^{\hspace{0.25mm}\mathrm{ava,scr}}/c$} with \mbox{$p_{\mathrm{c}}^{\mathrm{eff}}\approx p_{\mathrm{c}}^{\mathrm{scr}}$} from the \textit{Rosenbluth-Putvinski} model compared to the lower relative differences of \mbox{$u_{\mathrm{RE}}^{\hspace{0.25mm}\mathrm{ava,scr}}/c$} with \mbox{$p_{\mathrm{c}}^{\mathrm{eff}}\approx p_{\star}$} from \mbox{$u_{\mathrm{RE}}^{\hspace{0.25mm}\mathrm{ava}}/c$} can be related to the influence of the approximation of the effective critical momentum. In addition, it is remarkable that the high order of magnitude of the relative deviations between the approximations of the effective critical momentum and the \textit{Connor-Hastie} critical momentum, used in the \textit{Rosenbluth-Putvinski} approach, as it can be seen in figure \ref{fig_p_comparison}, do not inherit directly in the deviation in the mean velocity moment. Much more correct would be the statement, that the lower momentum boundary of the runaway region influences the first moment of the avalanche runaway distribution functions of both considered models, the propagated influence is strongly suppressed by the multiplication of the distribution with the velocity, or similarly its relativistic momentum expression, as it can be seen from the definition of the first moment from $(\ref{first_moment})$. This means, that the sensitivity of the lower momentum boundary is reduced, because the contributions of the integration results for lower momenta are multiplied with the lower velocities, while the function values of the distribution function are enhanced for larger velocities respectively momenta. Nevertheless, it could be verified, that both replacements for the lower momentum boundary deliver physically possible results, which account for the effects of partial screening. Moreover, the analytic relation $p_{\mathrm{c}}^{\mathrm{scr}}$ leads to tolerable deviations below $10\%$, so that it might be applied in simulations instead of $p_{\star}$, if the saved runtime is found to be sufficient for the loss in accuracy. This can be verified, by reference to the contour plots in figure \ref{fig_tilde_rel_u_ava_p_star_main}, which depict the relative difference between the calculations of the mean velocity \mbox{$u_{\mathrm{RE}}^{\hspace{0.25mm}\mathrm{ava,scr}}/c$} from $p_{\mathrm{c}}^{\mathrm{scr}}$ and $p_{\star}$. Additionally, the minimum, maximum and mean values of this displayed relative deviation $\tilde{\Delta}_{u_{\mathrm{RE},\,p^{\mathrm{scr}}_{\mathrm{c}}}^{\mathrm{ava,scr}}}$ support the previous hypothesis, which can be found in the listings \cref{MATLABoutput_plot_p_star_E3,MATLABoutput_plot_p_star_E10,MATLABoutput_plot_p_star_E30,MATLABoutput_plot_p_star_E100} of the appendix. As well, it was found, that a large deviation between the lower momentum expressions of the runaway region, does not necessarily propagate directly into the mean velocity moment, much more one can expect a suppressed sensitivity to the choice of a relation for the effective critical momentum.
\\
Finally, it shall be mentioned, that both models reach their boundary of applicability and validity in the high density limit, where the \textit{Rosenbluth-Putvinski} model predicts lower avalanche runaway electron velocities than the \textit{Hesslow} approach. In this limit,\vspace{-7cm}\linebreak \newpage\noindent 
\vspace*{-4.0mm}it is not possible to decide which model is more accurate, since the relative deviations for both models are marginal. However, the best agreement between both models, with a zero deviation, is found for slightly lower densities than their possible maximum values, which is approximately between $10^{-20}\,\mathrm{m}^{-3}$ and $10^{-21}\,\mathrm{m}^{-3}$ for electric fields between $10\,\mathrm{V/m}$ and $30\,\mathrm{V/m}$. These are typical parameters for tokamak fusion plasmas, which means, that the difference in the application of the mean velocity in simulations of the runaway current in tokamak disruptions, might not be significant between the two presented analytic models of an avalanche runaway electron distribution function. In contrast, it is however obvious, that for lower densities the \textit{Hesslow} model predicts different results than the \textit{Rosenbluth-Putvinski} distribution function, while being more accurate, due to the inclusion of the effects of a non-fully ionized plasma. Conclusively, the results \mbox{$u_{\mathrm{RE}}^{\hspace{0.25mm}\mathrm{ava,scr}}/c$} should be used for further simulations, instead of the normalized mean velocity \mbox{$u_{\mathrm{RE}}^{\hspace{0.25mm}\mathrm{ava}}/c$} in the \textit{Rosenbluth-Putvinski} model, in order to resolve the correct value of the mean velocity magnitude more accurately, especially for lower densities, high electric fields and the presence of impurities with high nuclear charge.

\clearpage

\section{Mean rest mass-related kinetic energy density of an \textit{avalanche} runaway electron population}\label{ava_k_moment_section}

The most important characteristic of an avalanche runaway electron is its ultra-relati-\linebreak vistic velocity, which reaches values of more than $90\%$ of the speed of light. This was also a deduction of the analysis of the data for the mean velocity, produced for a large region of density combinations of a singly-ionized deuterium-neon research plasma, from the previous section \ref{ava_j_section}. Besides, it was derived in subsection \ref{Avalanche_subsection}, that an avalanche runaway population can arise from small seed runaway electron densities and could therefore be responsible for the major fraction of a runaway plateau, as it can occur in tokamak disruptions as described in section \ref{tokamak_disruptions_section}. In addition, it is empirically known, that such a runaway beam can potentially damage plasma-facing components \cite{Bazylev_2011}. Furthermore, it was mentioned in reference \cite{Breizman_2019}, that the conversion of magnetic to kinetic energy of the runaway electrons, is expected to be the dominant source of wall-damaging runaway-wall strikes in ITER. This motivates i.a.\ the calculation and analysis of the kinetic energy density of an avalanche runaway electron population. At that, one can expect interesting insights, in particular from the \textit{Hesslow} model, which depends on the magnetic field through the effective critical electric field calculation, based on the \textsc{MATLAB}-script \mbox{\qq{\texttt{calculate_E_c_eff.m}}}, because it requires the magnetic field as an input parameter. However, the avalanche runaway electron kinetic energy density is also inherently interesting, due to the fact that is a characteristic quantity especially for the rapidly moving runaway electrons. 

A suitable quantity for the purpose of an analysis and comparison of the kinetic energy of an avalanche runaway electron population in the \textit{Rosenbluth-Putvinski} and the \textit{Hesslow} model might be chosen to be the mean rest mass-related kinetic energy density $k_{RE}$ normalized to the square of the speed of light in vacuum $c^2$. This is equivalent to a normalization of the kinetic energy to the electron rest mass energy \mbox{$m_{e0}\hspace{0.25mm}c^2$}. Its definition with respect to a moment of a distribution function was given in section \ref{kin_equa_section} and reads:\vspace{-4.8mm} 
\begin{equation}\label{kin_dens_ava_def_recap}
\dfrac{k_{\mathrm{RE}}}{c^2} =\dfrac{\langle K_{\mathrm{RE}}\rangle }{m_{e0}\,c^2}= \langle\gamma-1\rangle = \dfrac{1}{n_{\mathrm{RE}}}\,\iiint\limits_{\mathbb{R}^3}  \gamma \,f_{\mathrm{RE}}(\mathbf{r},\,\mathbf{p},\,t)\,\mathrm{d}^3 p -1\,.
\end{equation}
\vspace*{-7.1mm}\\Thus, one is able to derive calculation rules for this kinetic energy density in the following sections, on the basis of the two distribution functions for avalanche electrons, associated with the two models, introduced in the beginning of chapter \ref{avalanche_chapter}.
\\
Beforehand, a first estimation of the results for \mbox{$k_{\mathrm{RE}}/c^2$} shall be carried out, in order to get an idea of the order of magnitude to be expected. For this, one considers post-disruption plasmas, as apparent in large tokamaks, where the energy of the runaway electrons is typically assumed to be between $10$ and \mbox{$20\,\mathrm{MeV}$} \cite{RunawayPositrons}. If those energies are multiplied by the elementary charge and divided by the electron rest mass, a first estimated interval for the mean rest mass energy-related kinetic energy of runaway electron populations can be obtained. Consequently, one supposes, that the results for \mbox{$k_{\mathrm{RE}}/c^2$} in the following subsections will be of the order of magnitude \mbox{$19.57\lesssim k_{\mathrm{RE}}/c^2 \lesssim 39.14$}. However, since the energy of runaway electrons can also reach up to \mbox{$100\,\mathrm{MeV}$} \cite{Papp_2011} or be as low as \mbox{$100\,\mathrm{keV}$} \cite{Stahl_2013_Syn}, one might additionally consider the coarser estimation \mbox{$0.20\lesssim k_{\mathrm{RE}}/c^2 \lesssim 195.70$}.

\subsection{Mean rest mass-related kinetic energy density of an \textit{avalanche} runaway electron population in the \textit{Rosenbluth-Putvinski} model}\label{RP_avalanche_k_subsection}

\begin{figure}[H]
  \centering
  \subfloat{\label{fig_k_ava_p_c_scr_E3_main} 
  \includegraphics[trim=317 22 376 23,width=0.41\textwidth,clip]
    {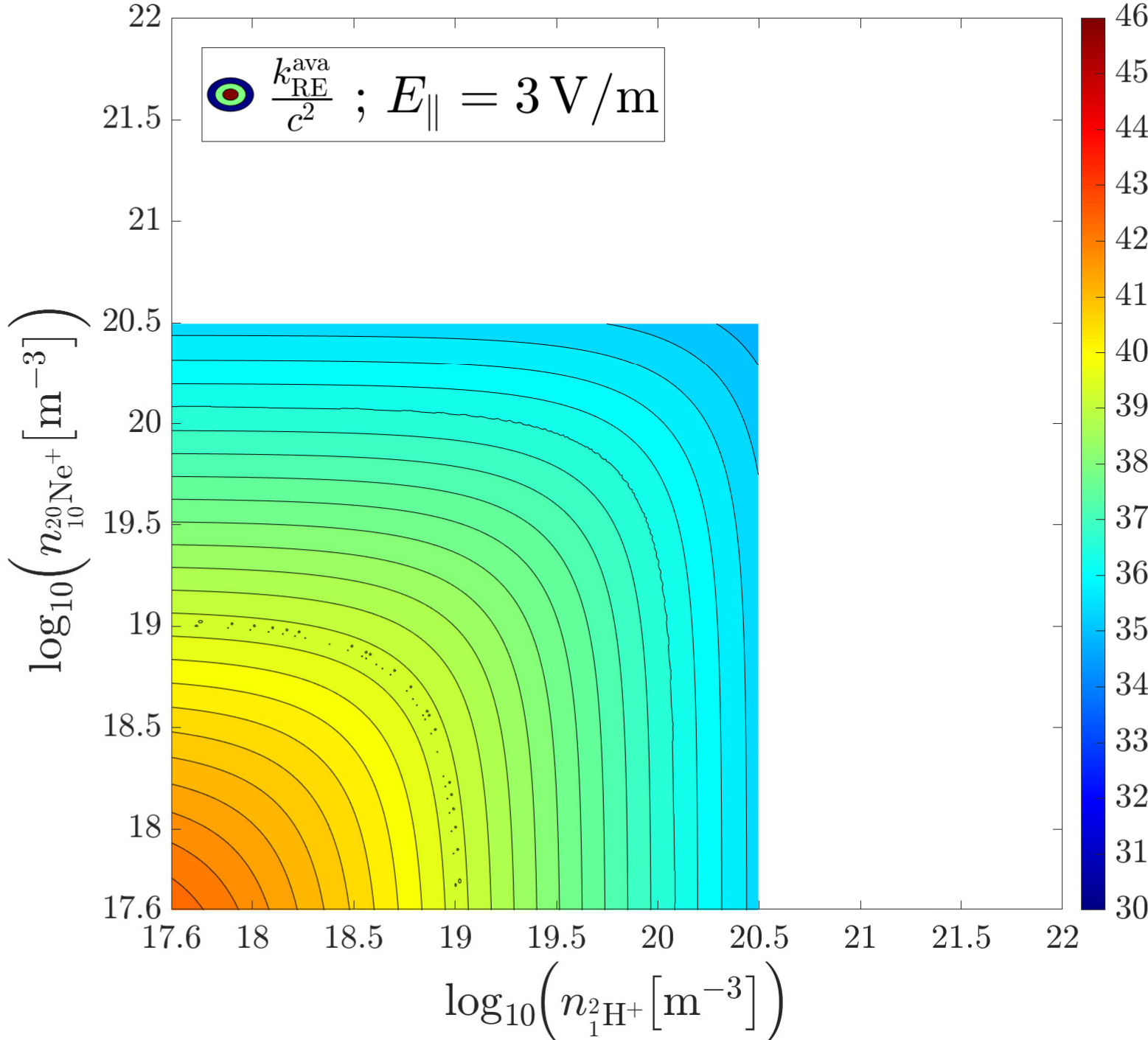}}\quad
  \subfloat{\label{fig_k_ava_p_c_scr_E10_main}
  \includegraphics[trim=317 17 379 25,width=0.41\textwidth,clip]
    {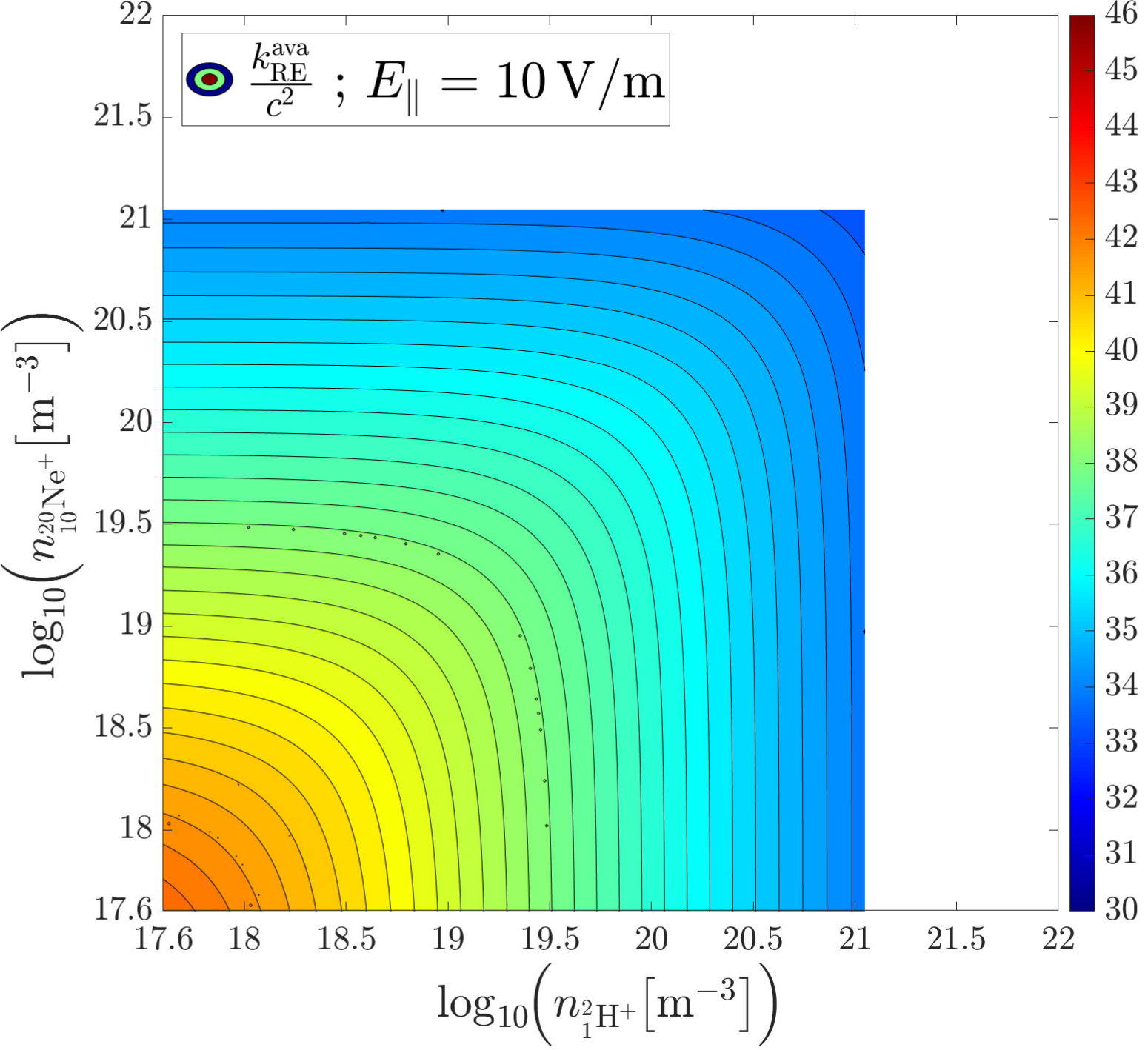}}\\[4pt]
  \subfloat{\label{fig_k_ava_p_c_scr_E30_main} 
    \includegraphics[trim=315 14 379 24,width=0.41\textwidth,clip]
    {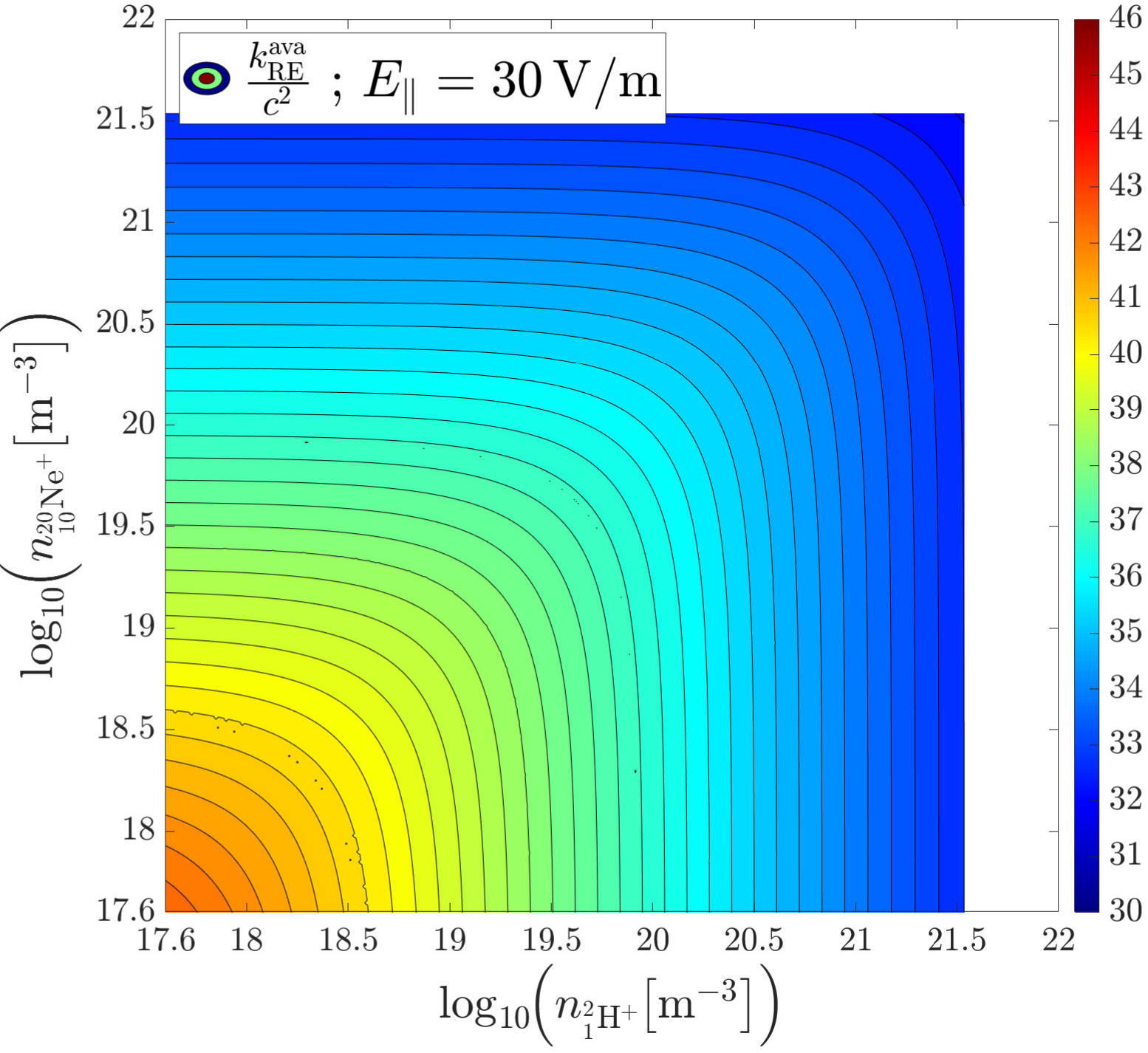}}\quad
  \subfloat{\label{fig_k_ava_p_c_scr_E100_main}
    \includegraphics[trim=313 18 373 18,width=0.41\textwidth,clip]
    {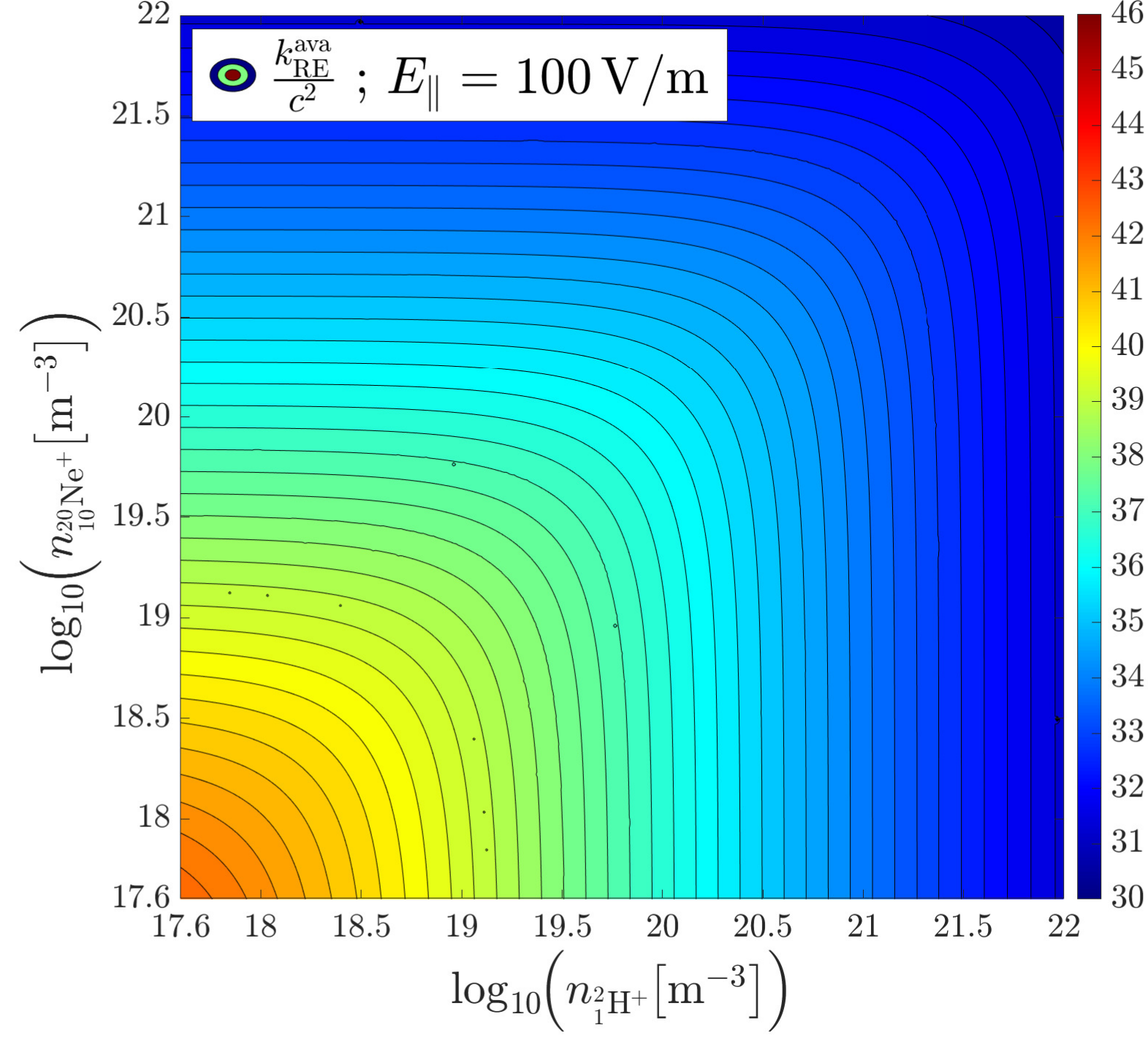}}
  \caption[Contour plots of the normalized mean rest mass-related kinetic energy density \mbox{$k_{\mathrm{RE}}^{\hspace{0.25mm}\mathrm{ava}}/c^2$}, in the \textit{Rosenbluth-Putvinski} model, of an avalanche runaway electron population with \mbox{$k_{\mathrm{B}}T_{\mathrm{e}}=10\,\textup{eV}$}, \mbox{$B=5.25\,\textup{T}$} and \mbox{$Z_{\mathrm{eff}}=1$} for approximately logarithmically increasing values of the electric field strength \mbox{$E_{\|}\coloneqq\vert E_{\|}\vert$} (larger view in figure \ref{fig_k_ava_p_c_scr} of the appendix).]{Contour plots$^{\ref{fig_plot_u_RP_footnote}}$ of the normalized mean rest mass-related kinetic energy density \mbox{$k_{\mathrm{RE}}^{\hspace{0.25mm}\mathrm{ava}}/c^2$}, in the \textit{Rosenbluth-Putvinski} model, of an avalanche runaway electron population with \mbox{$k_{\mathrm{B}}T_{\mathrm{e}}=10\,\textup{eV}$}, \mbox{$B=5.25\,\textup{T}$} and \mbox{$Z_{\mathrm{eff}}=1$} for approximately logarithmically increasing values of the electric field strength \mbox{$E_{\|}\coloneqq\vert E_{\|}\vert$} (larger view in figure \ref{fig_k_ava_p_c_scr} of the appendix).}
\label{fig_k_ava_p_c_scr_main}
\end{figure}

The mean rest mass-related kinetic energy density of an avalanche runaway electron population in the \textit{Rosenbluth-Putvinski} model may be denoted as $k_{\mathrm{RE}}^{\hspace{0.25mm}\mathrm{ava}}$ and shall be normalized to the square of the speed of light in vacuum $c^2$. Its connection to a particular moment of the distribution function $f_{\mathrm{RE}}^{\textup{ava}}$, proposed by \mbox{\textit{T.\hspace{0.9mm}Fülöp et \hspace{-0.4mm}al.}}, was recapitulated in the equation $(\ref{kin_dens_ava_def_recap})$. If now the expression $(\ref{dist_func_RP_ava_def})$ for the distribution function and the momentum representation of the gamma factor \mbox{$\gamma=\sqrt{1+p_{\|}^2+p_{\perp}^2}$}, in accordance with the relation $(\ref{p_norm_gamma_def})$, as well as the volume element $\mathrm{d}^3p =2\pi\,p_{\perp}\mathrm{d}p_{\perp}\mathrm{d}p_{\|}$ for the gyro-radius-averaged momentum space coordinates from section \ref{mom_space_coord_section} are inserted in the calculation rule $(\ref{kin_dens_ava_def_recap})$, one receives:\vspace{-2.5mm}
\begin{equation}\label{kin_dens_V1}
\begin{split}
\begin{gathered}
\dfrac{k_{\mathrm{RE}}^{\hspace{0.25mm}\mathrm{ava}}}{c^2}\underset{}{\overset{(\ref{kin_dens_ava_def_recap})}{=}}\dfrac{1}{n_{\mathrm{RE}}}\, \displaystyle{\int\limits_{p_{\|}=-\infty}^{\infty}\int\limits_{p_{\perp}=0}^{\infty}}  \sqrt{1+p_{\|}^2+p_{\perp}^2} \cdot f_{\mathrm{RE}}^{\textup{ava}}(p_{\|},\,p_{\perp},\,t)\;2\pi\,p_{\perp}\mathrm{d}p_{\perp}\mathrm{d}p_{\|}\,-\,1
\\[0pt]
\overset{(\ref{dist_func_RP_ava_def})}{=}\hspace{-0.5mm}\dfrac{2\,\tilde{E}\,\textup{e}^{\,\frac{2\,(\hat{E}-1)}{c_{\mathrm{Z}_{\mathrm{eff}}} \ln{\hspace{-0.45mm}\Lambda_{rel}}} \frac{t}{\tau_{rel}}}  }{ c_{\mathrm{Z}_{\mathrm{eff}}}\ln{\hspace{-0.45mm}\Lambda_{rel}}} \hspace{-3.55mm}\underbrace{\displaystyle{\int\limits_{p_{\|}=p_{\|,min}}^{\infty}\int\limits_{p_{\perp}=0}^{\infty}}  \hspace{-0.75mm}\dfrac{p_{\perp}}{p_{\|}}\,\sqrt{1+p_{\|}^2+p_{\perp}^2}  \,\hspace{-0.5mm}\textup{e}^{\,-\frac{p_{\|}}{c_{\mathrm{Z}_{\mathrm{eff}}} \ln{\hspace{-0.45mm}\Lambda_{rel}}}-\tilde{E}\cdot\frac{p_{\perp}^2}{p_{\|}}}\hspace{0.35mm}\mathrm{d}p_{\perp}\mathrm{d}p_{\|}}_{\coloneqq\,\textup{I}_{ \,\mathrm{num}}^{ \,\textit{k}_{\mathrm{RE}}^{\hspace{0.25mm}\mathrm{ava}}}}\,-\,1
\\[0pt]
=\dfrac{2\cdot\tilde{E}  }{ c_{\mathrm{Z}_{\mathrm{eff}}}\cdot\ln{\hspace{-0.45mm}\Lambda_{rel}}}\cdot\exp{\hspace{-0.45mm}\left(\hspace{-0.45mm} \dfrac{2\cdot\left(\hat{E}-1\right)}{c_{\mathrm{Z}_{\mathrm{eff}}} \cdot\ln{\hspace{-0.45mm}\Lambda_{rel}}}\cdot \frac{t}{\tau_{rel}}\hspace{-0.45mm}\right)} \cdot\textup{I}_{\,\mathrm{num}}^{\,\textit{k}_{\mathrm{RE}}^{\hspace{0.25mm}\mathrm{ava}}}\;-\;1\;. 
\end{gathered}
\end{split}
\end{equation}
\vspace{-6.0mm}\\Note, that the integral $\textup{I}_{\,\mathrm{num}}^{\,k_{\mathrm{RE}}^{\hspace{0.25mm}\mathrm{ava}}}$ is not solvable analytically and furthermore one has to ensure the convergence of a numerical integration with a finite lower integration bound \mbox{$p_{\|,\mathrm{min}}>-\infty$}, which was discussed in detail in the document \cite{study_thesis}. In addition, a reference has to be made to the normalization condition $(\ref{normalization_conditon_RP})$, which defines the time-dependent lower integration boundary for the parallel momentum. For the preparation of the integral $\textup{I}_{\,\mathrm{num}}^{\,k_{\mathrm{RE}}^{\hspace{0.25mm}\mathrm{ava}}}$ for a numerical integration, it is possible to apply the substitutions from $(\ref{substitutions_num_k_again})$, so that the subsequent expression of the integral allows the utilization of standard quadrature schemes:\vspace{-2.5mm}
\begin{equation}\label{kin_dens_integral_num}
\begin{split}
\begin{gathered}
\textup{I}_{\,\mathrm{num}}^{\,k_{\mathrm{RE}}^{\hspace{0.25mm}\mathrm{ava}}}\hspace{-0.3mm}=\hspace{-2.7mm}\displaystyle{\int\limits_{w=0}^{1}\int\limits_{z=0}^{1}}  \hspace{-0.6mm}\frac{ z \hspace{-0.6mm}\cdot\hspace{-0.5mm}\sqrt{1\hspace{-0.5mm}+\hspace{-0.5mm}\left(p_{\|,\mathrm{min}}\hspace{-0.45mm}+\hspace{-0.35mm}\frac{w}{1-w}\right)^{2}\hspace{-0.7mm}+\hspace{-0.5mm}\left(\frac{z}{1-z}\right)^2}}{(1-w) (1-z)^3 (p_{\|\mathrm{min}}(1\hspace{-0.5mm}-\hspace{-0.5mm}w)\hspace{-0.5mm}+\hspace{-0.5mm}w)}
\hspace{-0.78mm}\cdot\hspace{-0.4mm}\textup{e}^{\,-\frac{p_{\|,min}+\frac{w}{1-w}}{c_{\mathrm{Z}_{\mathrm{eff}}} \ln{\hspace{-0.45mm}\Lambda_{rel}}}-\frac{\tilde{E}\cdot\left(\frac{z}{1-z}\right)^2}{p_{\|,min}+\frac{w}{1-w}}}\hspace{0.7mm}\mathrm{d}z\,\mathrm{d}w\,.
\end{gathered}
\end{split}
\end{equation}
\vspace{-6.0mm}\\
The previously derived computation rules $(\ref{kin_dens_V1})$ and $(\ref{kin_dens_integral_num})$ allow a calculation of the mean kinetic energy density normalized with the electron rest mass of an avalanche runaway electron population in the \textit{Rosenbluth-Putvinski} model, for different densities of singly-ionized deuterium and neon atoms for the research plasma, presented in section \ref{nuclear_fusion_section} and logarithmically increasing values of the electric field. At that, the steady-state with \mbox{$t = 0\,\mathrm{s}$} is considered in a \textsc{MATLAB}-implementation. This state corresponds to \mbox{$p_{\|,\mathrm{min}}=0$}, satisfying the condition $(\ref{normalization_conditon_RP})$. In detail, the \textsc{MATLAB}-scripts$^{\ref{fig_plot_u_RP_footnote}}$ apply the routine \qq{\texttt{integral2}} and present the calculated data in four contour plots for four different values of the electric field, which at roughly increase logarithmically. Those plots are subsequently arranged in figure \ref{fig_k_ava_p_c_scr_main}. It is remarked that, the relation $(\ref{CoulombLogrel})$ was used for the computation of the relativistic \textit{Coulomb} logarithm. The corresponding outputs, is shown in the listings \cref{MATLABoutput_plot_p_scr_E3,MATLABoutput_plot_p_scr_E10,MATLABoutput_plot_p_scr_E30,MATLABoutput_plot_p_scr_E100} in subsection \ref{output_matlab_appendix_subsection} of the appendix. It provides the minima, maxima and mean values of each contour plot, shows the general parameter settings and allows oneself to verify, that the control criterion $(\ref{check_n_RE_ava_def})$ was satisfied for all computations. 

Based on this preparatory work, an analysis of the figure \ref{fig_k_ava_p_c_scr_main} is possible. In doing so, it is apparent, that the results for the normalized kinetic energy density of avalanche runaway electrons indeed reproduce the predicated estimations about the order of magnitude from the beginning of this section \ref{ava_k_moment_section}. Moreover, the same symmetry of the contour lines to the diagonal \mbox{$n_{_{10}^{20}\mathrm{Ne}^{+}}(n_{_{1}^{2}\mathrm{H}^{+}})=n_{_{1}^{2}\mathrm{H}^{+}}$}, can be seen, which was also present for the velocity magnitude in the \textit{Rosenbluth-Putvinski} model. Hence, one can state again, that the distribution function proposed by \mbox{\textit{T.\hspace{0.9mm}Fülöp et \hspace{-0.4mm}al.}} does not resolve a different behaviour for a change in either the neon- or the deuterium ion density and therefore neglects the effects of a not fully ionized plasma as explained in the section \ref{part_screen_section}. In addition, one can also observe the growth of avalanche runaway electron generation region within the density parameter space, due to an increase in the electric field strength. 
\\
In conclusion, one can imagine, that the \textit{Rosenbluth-Putvinski} approach predicts a too low critical electric field particularly for lower densities, because it assumes less collisions and therefore a decreased friction force, so that the runaway electrons experience a larger net acceleration. However, if one looks at the deviation between the critical electric fields in the \textit{Hesslow} and in the \textit{Rosenbluth-Putvinski} model, displayed in figure \ref{fig_E_C_ava_p_c_scr_main}, it is obvious, that exactly for the mentioned lower density region,\vspace{-11cm}\linebreak\newpage\noindent 
\vspace{-6.0mm}the phenomenon of partial screening plays a significant role. This is reasoned by the fact, that the consideration of this effect leads to enhanced collision rates and thus stronger deflection and pitch-angle scattering. Consequently, the effective critical electric field is higher than the \textit{Connor-Hastie} critical electric field, which is used by the \textit{Rosenbluth-Putvinski} approach. This entails, that for a fixed external electric field the net acceleration is smaller than the prediction from the \textit{Rosenbluth-Putvinski} model. Here, one could recapitulate the runaway region from figure \ref{fig_RE_region}, where a larger critical electric field decreases the height of the runaway region for each momentum between the critical and the maximum momentum. But since this height is a measure for the net acceleration of the runaway electrons, one can understand, why in reality the runaway electrons reach lower velocities and thus, lower kinetic energy densities. This can only be resolved by the \textit{Hesslow} model, as it will be shown in the next subsection, while the \textit{Rosenbluth-Putvinski} approach overestimates those quantities.

\subsection{Mean rest mass-related kinetic energy density of an \textit{avalanche} runaway electron population in the \textit{Hesslow} model}\label{Hesslow_avalanche_k_subsection}

For the \textit{Hesslow} model, one reproduces the procedure for the derivation of a convenient computation rule for the mean rest mass energy-related kinetic energy of avalanche runaway electrons from the previous subsection. In consequence, one again uses the moment defined in equation $(\ref{kin_dens_ava_def_recap})$, but inserts the time-dependent modification $(\ref{dist_func_H_time})$ of the distribution function $\tilde{f}_{\mathrm{RE}}^{\hspace{0.25mm}\textup{ava,scr}}(p)$, proposed by \mbox{\textit{P.\hspace{0.9mm}Svensson}} and extended with the modified time evolution factor from the \textit{Rosenbluth-Putvinski} model. Furthermore, the notation $k_{\mathrm{RE}}^{\hspace{0.25mm}\mathrm{ava,\,scr}}$ is chosen for the mean rest mass-related kinetic energy density of an avalanche runaway electron population under consideration of the effects of partial screening in the \textit{Hesslow} model. Also the previously used normalization to the square of the speed of light $c^2$ is applied. Now the relation $(\ref{dist_func_H})$ for $\tilde{f}_{\mathrm{RE}}^{\hspace{0.25mm}\textup{ava,scr}}$ together with the time-dependent extension $(\ref{dist_func_H_time})$ is identified, after the momentum representation of the gamma factor from $(\ref{p_norm_gamma_def})$ and the expression $(\ref{volelem_sphere_2D})$ for the volume element for the gyro-radius-averaged momentum space coordinates were combined in the calculation rule $(\ref{kin_dens_ava_def_recap})$. Thus, one has:\vspace{-2.5mm}
\begin{equation}\label{kin_dens_ava_scr_def}
\begin{split}
\begin{gathered}
\dfrac{k_{\mathrm{RE}}^{\hspace{0.25mm}\mathrm{ava,scr}}}{c^2}  \underset{(\ref{dist_func_H})}{\overset{(\ref{kin_dens_ava_def_recap})}{=}} \dfrac{1}{n_{\mathrm{RE}} }\, \displaystyle{\int\limits_{p=p_{\mathrm{c}}^{\mathrm{eff}}}^{\infty}}\underbrace{\displaystyle{\int\limits_{\xi=-1}^{1}  }\hspace{-0.8mm}\gamma\hspace{-0.35mm}\cdot\hspace{-0.35mm} f_{RE}^{\hspace{0.25mm}\textup{ava,scr}}(p,\,\xi,\,t)\;2 \pi\,p^2\,\mathrm{d}\xi}_{=\,\tilde{\textit{f}}_{ \,0,RE}^{\;\mathrm{ava,scr}}(p,\,t)}\mathrm{d}p\; - \; 1
\\[0pt]
\underset{(\ref{dist_func_H_time})}{\overset{(\ref{p_norm_gamma_def})}{=}}\dfrac{1}{n_{\mathrm{RE}}}\,\displaystyle{\int\limits_{p=p_{\mathrm{c}}^{\mathrm{eff}}}^{\infty} }\sqrt{1+p^2}\cdot\exp{\left(\dfrac{2\,(\hat{E}-1)}{c_{\mathrm{Z}_{\mathrm{eff}}}\ln{\hspace{-0.45mm}\Lambda_{rel}}}\hspace{-0.45mm}\cdot\hspace{-0.45mm}\dfrac{t}{\tau_{rel}}\right)}\cdot \tilde{f}_{\mathrm{RE}}^{\hspace{0.25mm}\textup{ava,scr}}(p)\;\mathrm{d}p \; - \; 1
\\[-2pt]
\overset{(\ref{dist_func_H})}{=} \textup{e}^{\,\frac{2\,(\hat{E}-1)}{c_{\mathrm{Z}_{\mathrm{eff}}} \ln{\hspace{-0.45mm}\Lambda_{rel}}} \frac{t}{\tau_{rel}}}\hspace{-0.5mm}\underbrace{\displaystyle{\int\limits_{p=p_{c}^{\mathrm{eff}}}^{\infty} } \hspace{-0.4mm}\dfrac{n_{\mathrm{e}}^{\mathrm{tot}}\cdot\sqrt{1+p^2} \cdot\textup{e}^{\, -\frac{n_{\mathrm{e}}^{\mathrm{tot}} \cdot (p-p_{c}^{\mathrm{eff}})}{n_{\mathrm{e}} \cdot \ln{\hspace{-0.45mm}\Lambda_{rel}} \cdot \sqrt{4+\tilde{\nu}_{\mathrm{s}}(p_{c}^{\mathrm{eff}}) \cdot \tilde{\nu}_{\mathrm{d}}(p_{c}^{\mathrm{eff}})}}  }  }{n_{\mathrm{e}} \cdot \ln{\hspace{-0.45mm}\Lambda_{rel}} \cdot\sqrt{4+\tilde{\nu}_{\mathrm{s}}(p_{c}^{\mathrm{eff}}) \cdot\tilde{\nu}_{\mathrm{d}}(p_{c}^{\mathrm{eff}})}} \;\mathrm{d}p }_{\eqqcolon \,\textup{I}_{\,\textup{num}}^{\,\textit{k}_{RE}^{\hspace{0.25mm}\textup{ava,scr}}} }\, - \, 1
\\[0pt]
= \exp{\left(\dfrac{2\,(\hat{E}-1)}{c_{\mathrm{Z}_{\mathrm{eff}}}\ln{\hspace{-0.45mm}\Lambda_{rel}}}\hspace{-0.45mm}\cdot\hspace{-0.45mm}\dfrac{t}{\tau_{rel}}\right)}\cdot \textup{I}_{\,\textup{num}}^{\,\textit{k}_{RE}^{\hspace{0.25mm}\textup{ava,scr}}}\, - \; 1\;.
\end{gathered}
\end{split}
\end{equation}
\vspace{-6.0mm}\\Note, that the integration in the momentum magnitude coordinate $p$ has to take place between the effective critical momentum $p_{\mathrm{c}}^{\mathrm{eff}}$ and infinity, due to the discussion from section \ref{Hesslow_avalanche_dist_subsection}. There, it was argued, that the moments of the distribution function in the \textit{Hesslow} model might be more sensitive to the lowest momentum for the avalanche generation of runaway electrons, while the upper momentum bound has a minor influence on the integration results, because the integration contributions of the distribution function are exponentially suppressed for large momenta. Consequently, it is justified to set the maximum momentum to infinity, in order to decrease the total runtime of a possible computation. 

The calculation rule $(\ref{kin_dens_ava_scr_def})$ for the mean kinetic energy divided by the electron rest mass can now be rehashed, in order to make a numerical integration, based on quadrature formulae, more convenient. Therefore, one inserts the previously used substitution $(\ref{substitutions_num_k_again_H})$ into the integral $\textup{I}_{\,\textup{num}}^{\,\textit{k}_{RE}^{\hspace{0.25mm}\textup{ava,scr}}}$, which yields:\vspace{-2.5mm}
\begin{equation}\label{kin_dens_ava_scr_interagl_num_def}
 \textup{I}_{\,\textup{num}}^{\,\textit{k}_{RE}^{\hspace{0.25mm}\textup{ava,scr}}}= \displaystyle{\int\limits_{w=0}^{1} } \hspace{-0.4mm}\dfrac{n_{\mathrm{e}}^{\mathrm{tot}} \hspace{-0.5mm}\cdot\hspace{-0.3mm}\sqrt{1+\left(p_{c}^{\mathrm{eff}}+\frac{w}{1-w}\right)^2}  \hspace{-0.4mm}\cdot\hspace{-0.3mm}\textup{e}^{\, -\frac{n_{\mathrm{e}}^{\mathrm{tot}} \cdot w}{n_{\mathrm{e}} \cdot \ln{\hspace{-0.45mm}\Lambda_{rel}}\cdot(1-w) \cdot \sqrt{4+\tilde{\nu}_{\mathrm{s}}(p_{c}^{\mathrm{eff}}) \cdot \tilde{\nu}_{\mathrm{d}}(p_{c}^{\mathrm{eff}})}}  }  }{n_{\mathrm{e}} \cdot \ln{\hspace{-0.45mm}\Lambda_{rel}}\cdot(1-w)^2 \cdot\sqrt{4+\tilde{\nu}_{\mathrm{s}}(p_{c}^{\mathrm{eff}}) \cdot\tilde{\nu}_{\mathrm{d}}(p_{c}^{\mathrm{eff}})}} \;\mathrm{d}w\,   
\end{equation}
\vspace{-6.0mm}\\where one might use the relativistic \textit{Coulomb} logarithm $\ln{\hspace{-0.45mm}\Lambda_{rel}}$, as determined in the relation $(\ref{CoulombLogrel})$, the relativistic collision time $\tau_{rel}$ from $(\ref{tau_rel})$ and the ultra-relativistic limits \mbox{$\tilde{\nu}_{\mathrm{d}}(p_{c}^{\mathrm{eff}})$} and \mbox{$\tilde{\nu}_{\mathrm{s}}(p_{c}^{\mathrm{eff}})$} of the deflection and the slowing-down frequency evaluated at the effective critical momentum from $(\ref{nue_s_nue_d_def})$. In addition, one might set \mbox{$p_{c}^{\mathrm{eff}}\approx p^{\mathrm{scr}}_{c}$}, in accordance with the expression $(\ref{p_c_scr_def})$ or \mbox{$p_{c}^{\mathrm{eff}}\approx p_{\star}$} with $p_{\star}$ as the root of the function defined in $(\ref{func_p_c_eff_def})$. This also requires the utilization of the \textsc{MATLAB}-script$^{\ref{Matlab_Hesslow_script}}$ from \textit{L.\hspace{0.9mm}Hesslow} \cite{Hesslow_2018}, for the calculation of the effective critical electric field $E^{\mathrm{eff}}_{c}$, which is used in $p^{\mathrm{scr}}_{c}$ and for the constants needed in the equation $(\ref{nue_s_nue_d_def})$ for the computation of \mbox{$\tilde{\nu}_{\mathrm{d}}(p_{c}^{\mathrm{eff}})$} and \mbox{$\tilde{\nu}_{\mathrm{s}}(p_{c}^{\mathrm{eff}})$}. Here it should be remarked, that the approximation of the effective critical momentum $p_{c}^{\mathrm{eff}}$, the \textit{Coulomb} logarithm, the relativistic collision time and the ultra-relativistic limits \mbox{$\tilde{\nu}_{\mathrm{d}}(p_{c}^{\mathrm{eff}})$} and \mbox{$\tilde{\nu}_{\mathrm{s}}(p_{c}^{\mathrm{eff}})$} of the deflection and the slowing-down frequency evaluated at this effective critical momentum, are calculated before the integration of $\textup{I}_{\,\textup{num}}^{\,\textit{k}_{RE}^{\hspace{0.25mm}\textup{ava,scr}}}$, so that those quantities appear as constants and do not depend on the integration variable.

The calculation rule $(\ref{kin_dens_ava_scr_def})$ together with the integral $(\ref{kin_dens_ava_scr_interagl_num_def})$ define an implementable computation scheme for the mean kinetic energy density, normalized with the electron rest mass, of an avalanche runaway electron population in the \textit{Hesslow} model. This is now proven, by means of an implementation$^{\ref{fig_plot_u_RP_footnote},\ref{fig_plot_footnote_2_main}}$ in \textsc{MATLAB}. In order to allow a comparison, with the results from the previous section and especially with figure \ref{fig_k_ava_p_c_scr_main}, a singly-ionized deuterium and neon plasma with different density combinations is used as a set of test cases in the density parameter space. Further, four logarithmically increasing values of a prevalent accelerating electric field are considered, whilst \mbox{$t = 0\,\mathrm{s}$} is kept fixed, so that all results correspond to a steady-state. Note, that the last equality in $(\ref{kin_dens_ava_scr_def})$ and the steady-state entail, that the computed data \mbox{$k_{\mathrm{RE}}^{\hspace{0.25mm}\mathrm{ava,scr}}/c^2$} directly follows from the definite integral $\textup{I}_{\,\textup{num}}^{\,\textit{k}_{RE}^{\hspace{0.25mm}\textup{ava,scr}}}$ reduced by one. 
 
The implementation evaluates the integral numerically with the help of the \mbox{\textsc{MATLAB}-} routine \qq{\texttt{integral}} and produces results for the mean rest mass-related kinetic energy density normalized to the square of the speed of light of avalanche runaway electrons, based on the steady-state distribution function in the \textit{Hesslow} model. At that, two approximations of the effective critical momentum $p_{\mathrm{c}}^{\mathrm{eff}}$ are used, in order to evaluate their influence on the first moment of the one-dimensional distribution function from \textit{P.\hspace{0.9mm}Svensson}. It should be recalled, that the effective critical momentum represents the lower integration boundary in the derivation of the calculation rule $(\ref{kin_dens_ava_scr_def})$ for \mbox{$k_{\mathrm{RE}}^{\hspace{0.25mm}\mathrm{ava,scr}}/c^2$} and that it appears as a parameter in the rewritten form $(\ref{kin_dens_ava_scr_interagl_num_def})$ of the integral $\textup{I}_{\,\textup{num}}^{\,\textit{k}_{RE}^{\hspace{0.25mm}\textup{ava,scr}}}$. Hence, the computed data contains influences due to partial screening, in contrast to the data produced with the distribution function by \mbox{\textit{T.\hspace{0.9mm}Fülöp et \hspace{-0.4mm}al.}} in the \textit{Rosenbluth-Putvinski} model, which led to the figure \ref{fig_k_ava_p_c_scr_main}.

\begin{figure}[H]
  \centering
  \subfloat{\label{fig_k_ava_screen_p_c_scr_E3_main} 
  \includegraphics[trim=314 22 377 23,width=0.41\textwidth,clip]
    {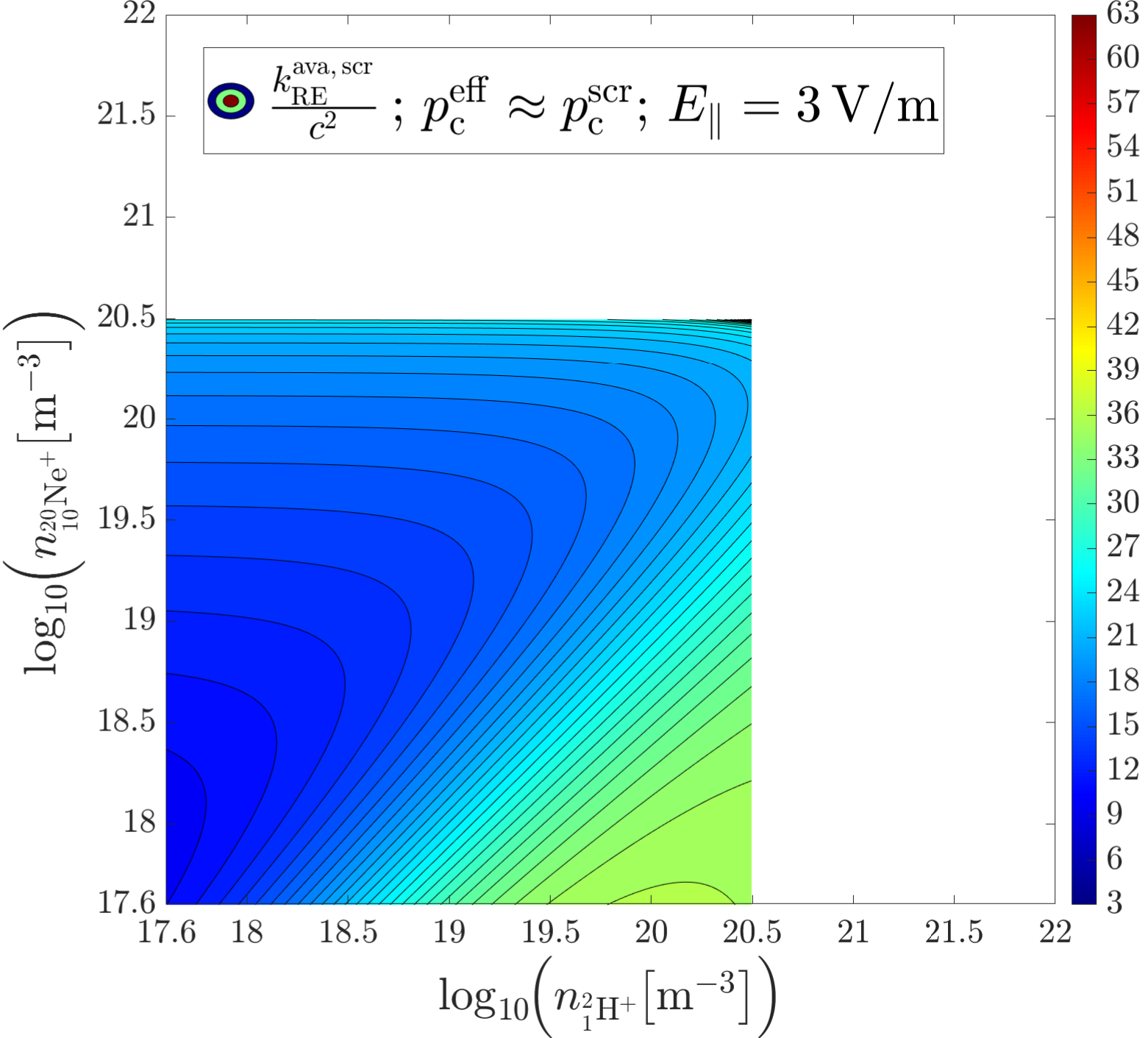}}\quad
  \subfloat{\label{fig_k_ava_screen_p_c_scr_E10_main}
   \includegraphics[trim=316 21 381 22,width=0.41\textwidth,clip]
    {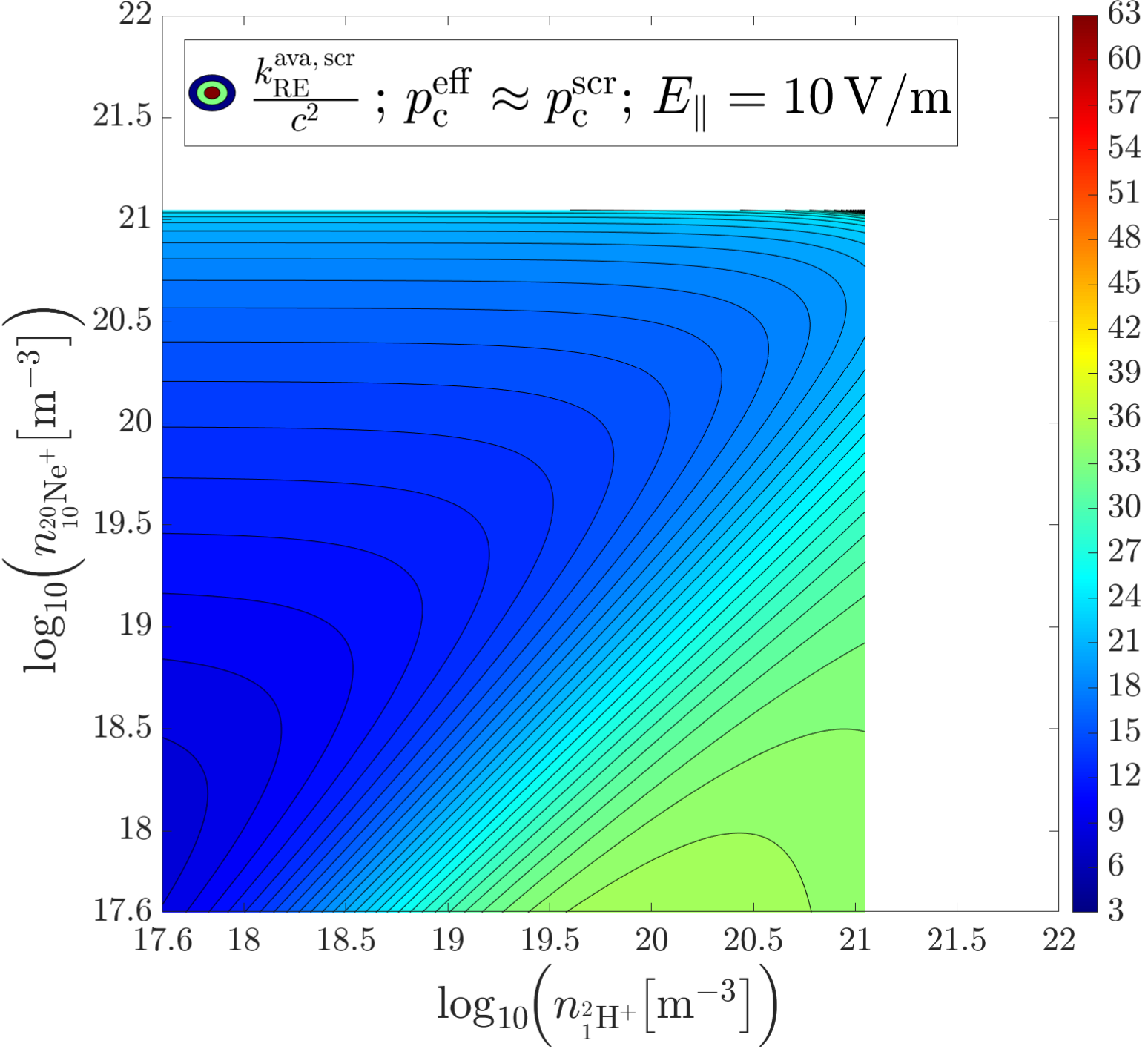}}\\[4pt]
  \subfloat{\label{fig_k_ava_screen_p_c_scr_E30_main} 
  \includegraphics[trim=317 21 378 21,width=0.41\textwidth,clip]
    {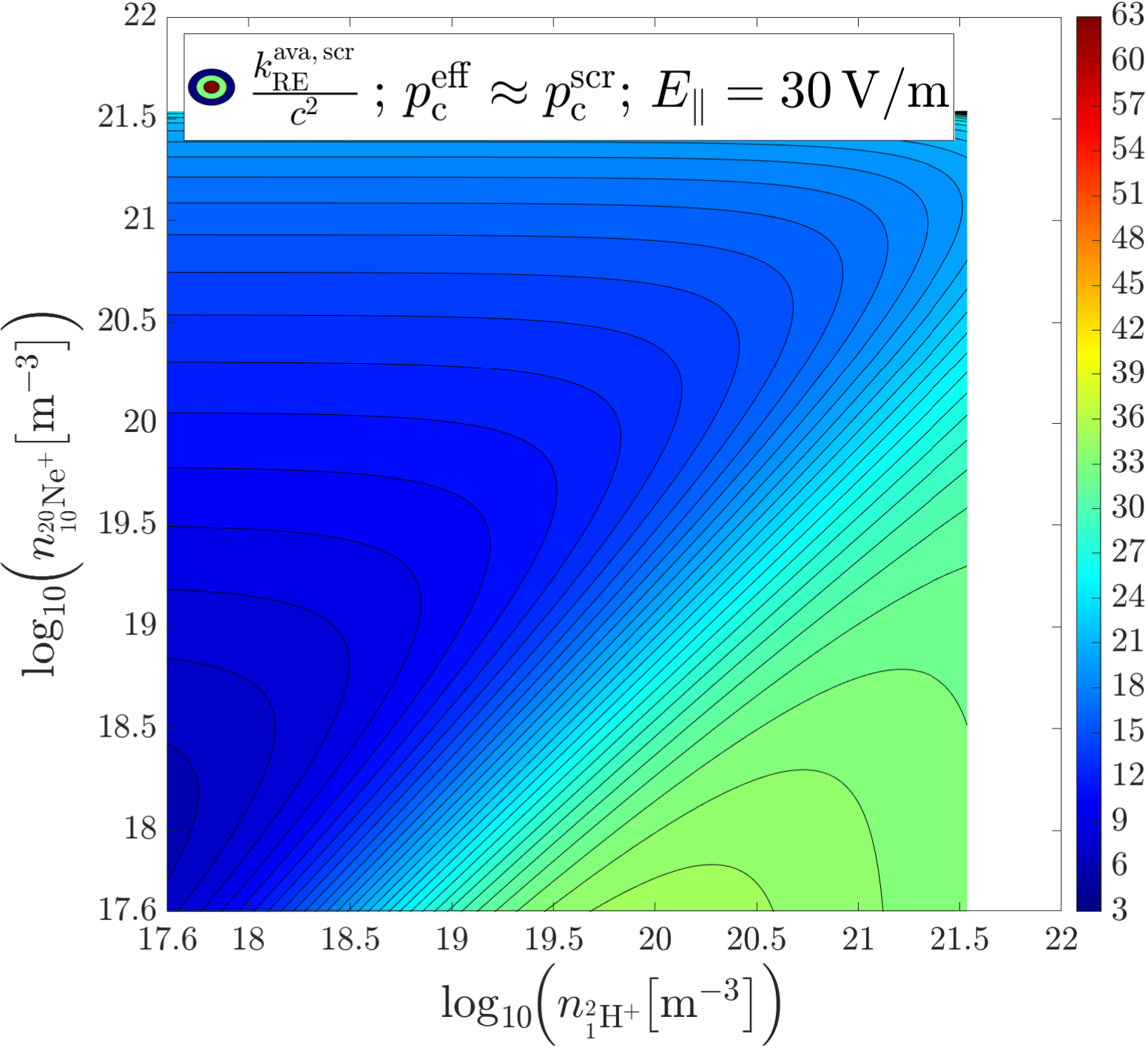}}\quad
  \subfloat{\label{fig_k_ava_screen_p_c_scr_E100_main}
  \includegraphics[trim=317 22 379 21,width=0.41\textwidth,clip]
    {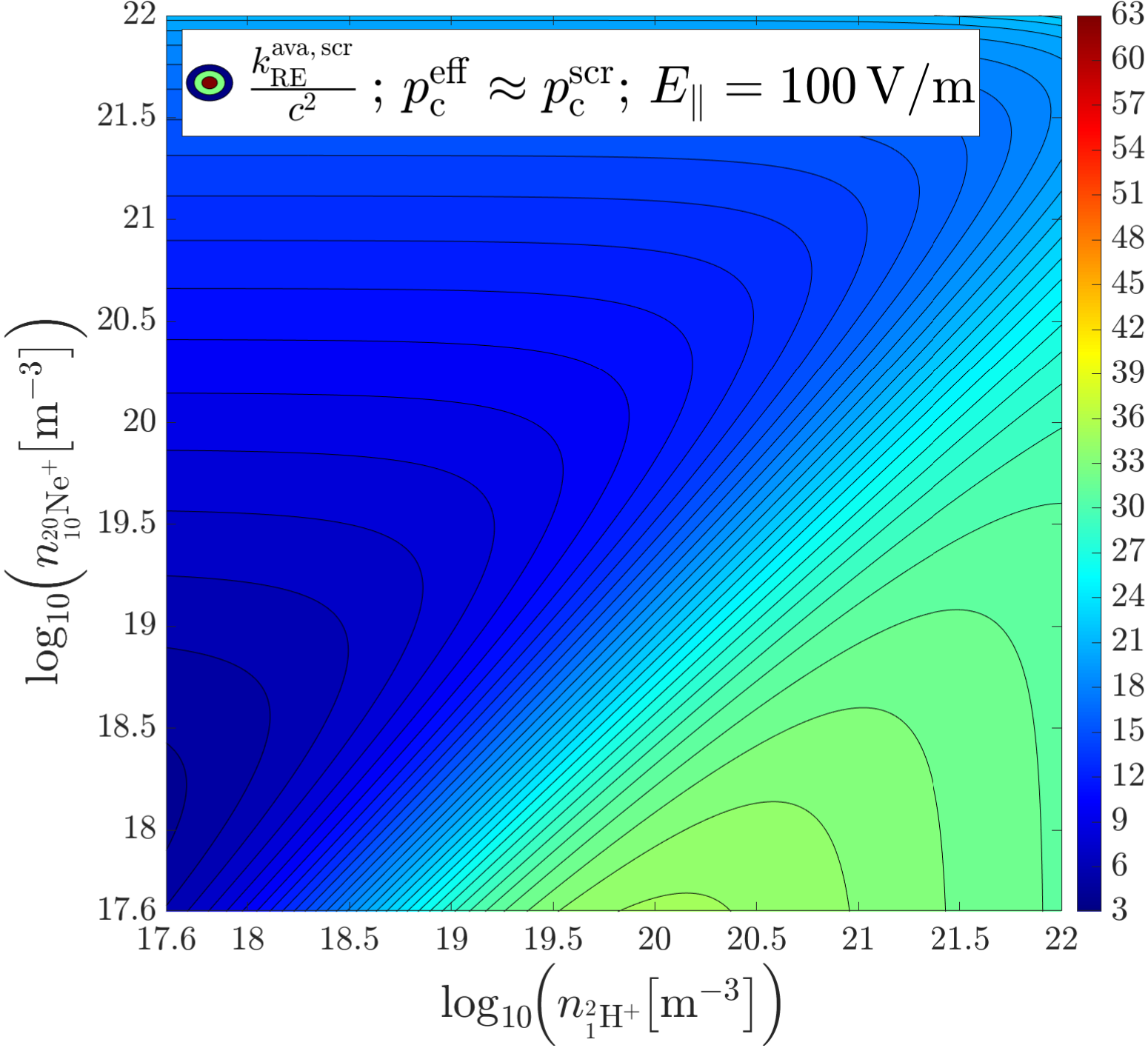}}
  \caption[Contour plots of the normalized mean rest mass-related kinetic energy density \mbox{$k_{\mathrm{RE}}^{\hspace{0.25mm}\mathrm{ava,scr}}/c^2$}, in the \textit{Hesslow} model with the effective critical momentum \mbox{$p_{\mathrm{c}}^{\mathrm{eff}}\approx p_{\mathrm{c}}^{\mathrm{scr}}$}, of an avalanche runaway electron population with \mbox{$k_{\mathrm{B}}T_{\mathrm{e}}=10\,\textup{eV}$}, \mbox{$B=5.25\,\textup{T}$} and \mbox{$Z_{\mathrm{eff}}=1$} for approximately logarithmically increasing values of the electric field strength \mbox{$E_{\|}\coloneqq\vert E_{\|}\vert$} (larger view in figure \ref{fig_k_ava_screen_p_c_scr} of the appendix).]{Contour plots$^{\ref{fig_plot_u_RP_footnote}}$ of the normalized mean rest mass-related kinetic energy density \mbox{$k_{\,\mathrm{RE}}^{\,\mathrm{ava,scr}}/c^2$}, in the \textit{Hesslow} model with the effective critical momentum \mbox{$p_{\mathrm{c}}^{\mathrm{eff}}\approx p_{\mathrm{c}}^{\mathrm{scr}}$}, of an avalanche runaway electron population with \mbox{$k_{\mathrm{B}}T_{\mathrm{e}}=10\,\textup{eV}$}, \mbox{$B=5.25\,\textup{T}$} and \mbox{$Z_{\mathrm{eff}}=1$} for approximately logarithmically increasing values of the electric field strength \mbox{$E_{\|}\coloneqq\vert E_{\|}\vert$} (larger view in figure \ref{fig_k_ava_screen_p_c_scr} of the appendix).}
\label{fig_k_ava_screen_p_c_scr_main}
\end{figure}

First, the choice of \mbox{$p_{\mathrm{c}}^{\mathrm{eff}}\approx p_{\mathrm{c}}^{\mathrm{scr}}$} shall be considered, based on the computed results of the \textsc{MATLAB}-scripts$^{\ref{fig_plot_u_RP_footnote}}$, which also provide visualizations of the produced data in the form of the plots in figure \ref{fig_k_ava_screen_p_c_scr_main}.\vspace{0.5mm} 
 \vspace*{-4.0mm}
The console outputs of said scripts are shown in the listings \cref{MATLABoutput_plot_p_scr_E3,MATLABoutput_plot_p_scr_E10,MATLABoutput_plot_p_scr_E30,MATLABoutput_plot_p_scr_E100} in subsection \ref{output_matlab_appendix_subsection} of the appendix. They provide the minima, maxima and mean values of the data connected to the contour plots, state the general parameter settings and show the compliance with the control criterion $(\ref{check_n_RE_ava_def_H})$ for all computations. 
\\
The analysis of the figure \ref{fig_k_ava_screen_p_c_scr_main}, first and foremost confirms the predictions about the order of magnitude of the normalized kinetic energy density of avalanche runaway electrons from the beginning of this section \ref{ava_k_moment_section}. As well, it can be suspected, that the computation is more accurate than the made estimations. This is, because for typical deuterium densities of \mbox{$n_{{}_{2}^{1}\mathrm{H}^{+}}\approx 10^{20}\,\mathrm{m}^{-3}$} and neon impurity densities with \mbox{$n_{{}_{20}^{10}\mathrm{Ne}^{+}}< 10^{19.2}\,\mathrm{m}^{-3}$} the estimation \mbox{$19.57\lesssim k_{\mathrm{RE}}/c^2 \lesssim 39.14$} for larger tokamaks, is not violated. Furthermore, the contour lines of the results are asymmetric to the diagonal \mbox{$n_{_{10}^{20}\mathrm{Ne}^{+}}(n_{_{1}^{2}\mathrm{H}^{+}})=n_{_{1}^{2}\mathrm{H}^{+}}$}, which shows that the \textit{Hesslow} model describes the effects of partial screening and thus the different variation of the kinetic energy density for changes in the neon ion density contrary to the deuterium density. Note, that this insight is similar to the findings concerning the mean velocity from subsection \ref{Hesslow_avalanche_j_subsection}. Hence, one ascertains, that the \textit{Hesslow} model is generally superior over the \textit{Rosenbluth-Putvinski} model in terms of the physical accuracy. In accordance with the previous analysis, the growth of the runaway electron generation region within the density parameter space, due to an increase in the electric field strength, is visible in figure \ref{fig_k_ava_screen_p_c_scr_main}. Remarkable is, that the values for \mbox{$k_{\mathrm{RE}}^{\hspace{0.25mm}\mathrm{ava,scr}}/c^2$} are smaller than in the \textit{Rosenbluth-Putvinski}. In addition, the empirically known influence of the presence of impurities of high nuclear charge, like neon or argon can be seen in the figure \ref{fig_k_ava_screen_p_c_scr_main}. Moreover, one notices, that the results also reproduce, that for high deuterium densities, larger impurity densities need to be added, in order to reduce the runaway electron energy. 

\begin{figure}[H]
  \centering
  \subfloat{\label{fig_k_ava_screen_p_star_E3_main} 
   \includegraphics[trim=313 23 370 23,width=0.41\textwidth,clip]
    {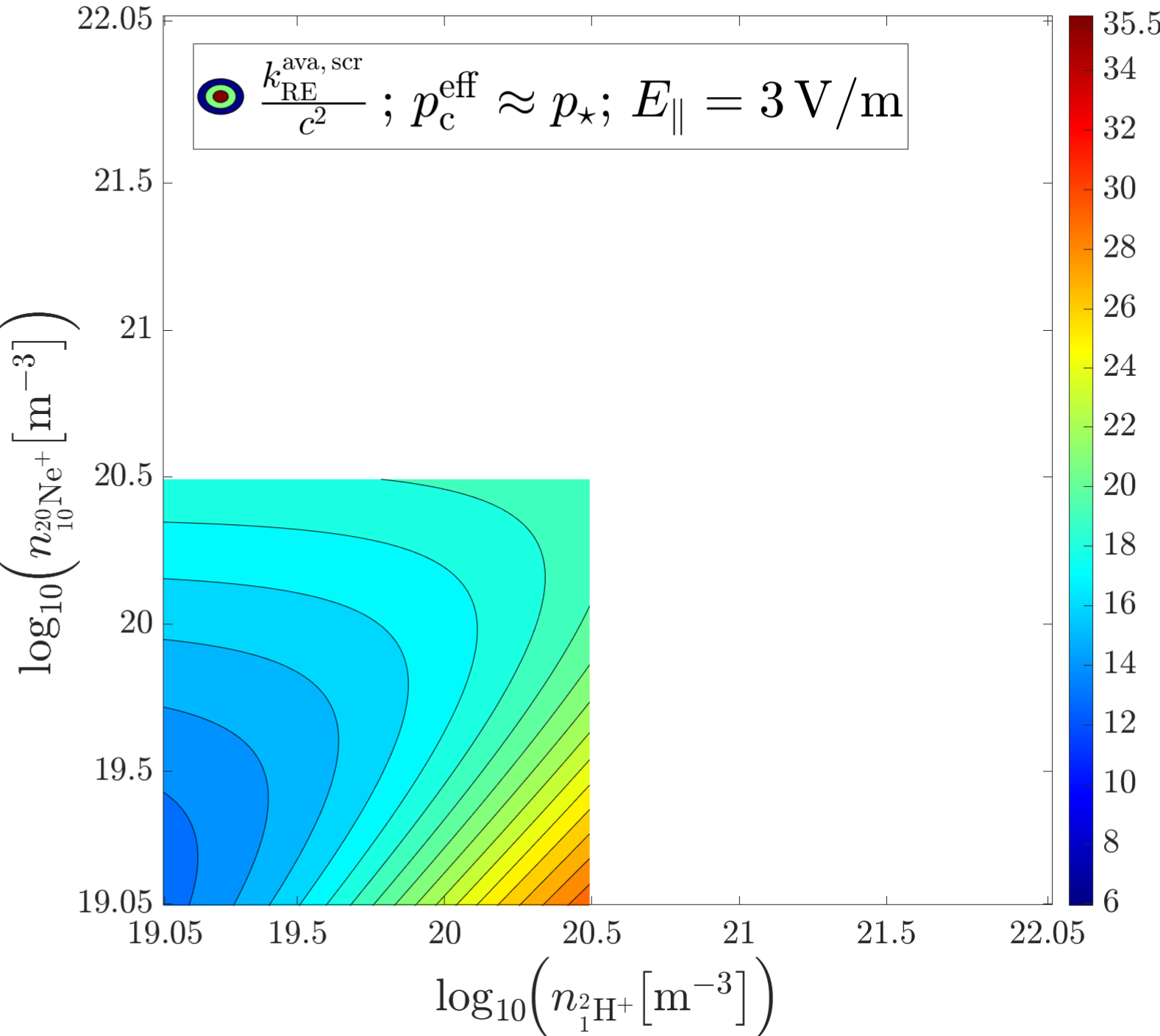}}\quad
  \subfloat{\label{fig_k_ava_screen_p_star_E10_main}
   \includegraphics[trim=324 25 369 23,width=0.41\textwidth,clip]
    {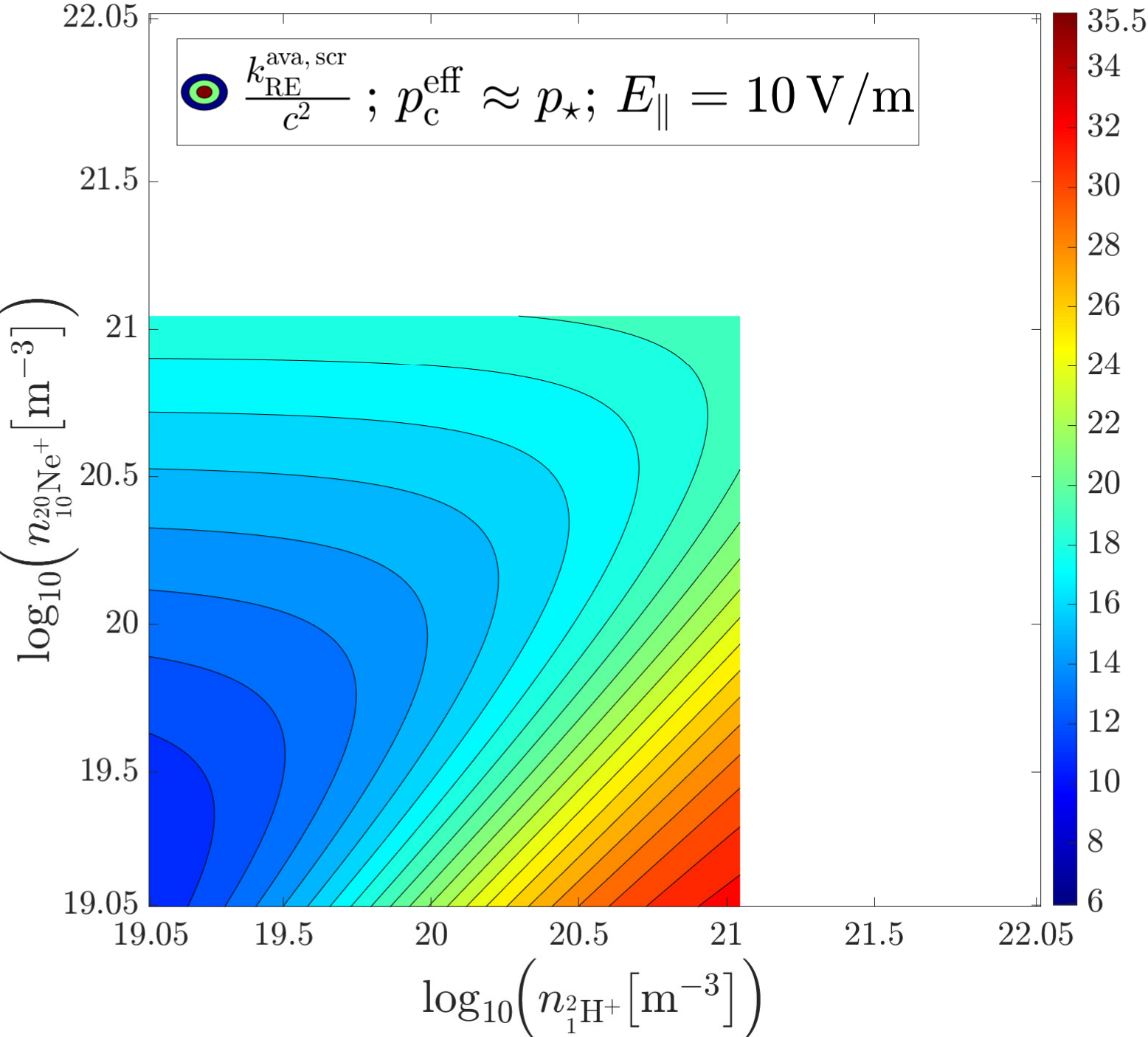}}\\[4pt]
  \subfloat{\label{fig_k_ava_screen_p_star_E30_main} 
   \includegraphics[trim=320 21 370 22,width=0.41\textwidth,clip]
    {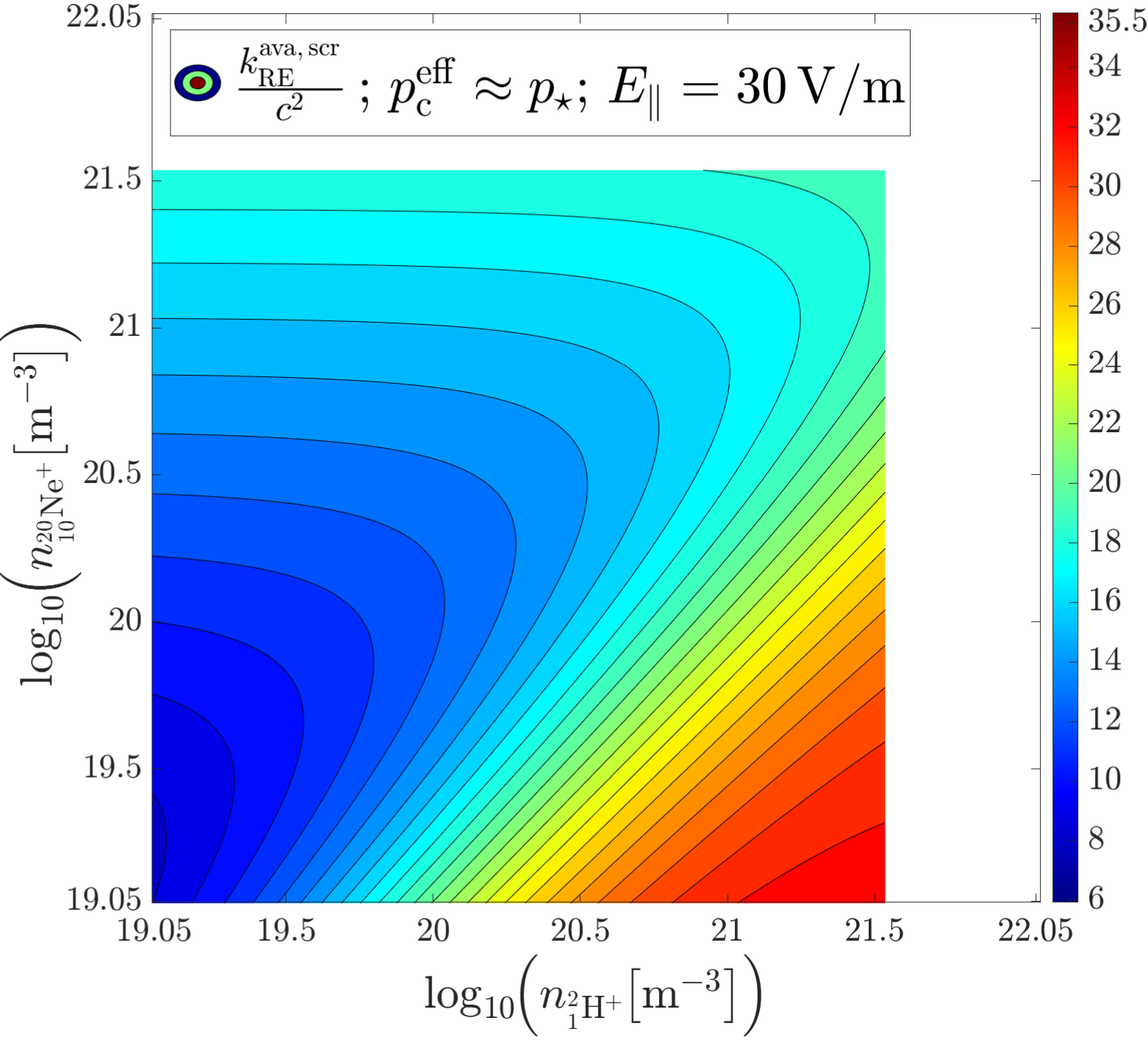}}\quad
  \subfloat{\label{fig_k_ava_screen_p_star_E100_main}
   \includegraphics[trim=321 21 368 22,width=0.41\textwidth,clip]
    {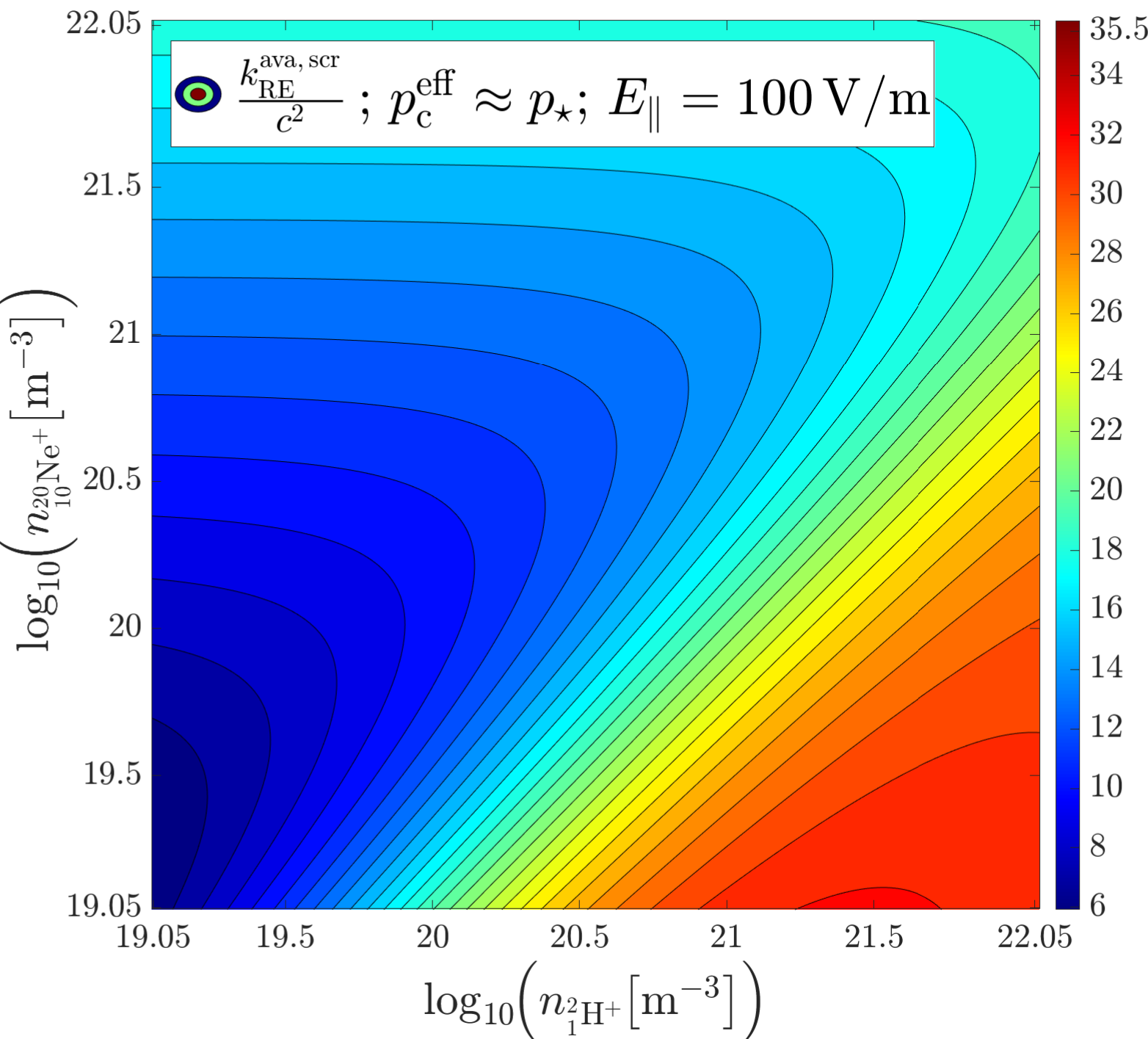}}
  \caption[Contour plots of the normalized mean rest mass-related kinetic energy density \mbox{$k_{\mathrm{RE}}^{\hspace{0.25mm}\mathrm{ava,scr}}/c^2$}, in the \textit{Hesslow} model with the effective critical momentum \mbox{$p_{\mathrm{c}}^{\mathrm{eff}}\approx p_{\star}$}, of an avalanche runaway electron population with \mbox{$k_{\mathrm{B}}T_{\mathrm{e}}=10\,\textup{eV}$}, \mbox{$B=5.25\,\textup{T}$} and \mbox{$Z_{\mathrm{eff}}=1$} for approximately logarithmically increasing values of the electric field strength \mbox{$E_{\|}\coloneqq\vert E_{\|}\vert$} (larger view in figure \ref{fig_k_ava_screen_p_star} of the appendix).]{Contour plots$^{\ref{fig_plot_footnote_2_main}}$ of the normalized mean rest mass-related kinetic energy density \mbox{$k_{\mathrm{RE}}^{\hspace{0.25mm}\mathrm{ava,scr}}/c^2$}, in the \textit{Hesslow} model with the effective critical momentum \mbox{$p_{\mathrm{c}}^{\mathrm{eff}}\approx p_{\star}$}, of an avalanche runaway electron population with \mbox{$k_{\mathrm{B}}T_{\mathrm{e}}=10\,\textup{eV}$}, \mbox{$B=5.25\,\textup{T}$} and \mbox{$Z_{\mathrm{eff}}=1$} for approximately logarithmically increasing values of the electric field strength \mbox{$E_{\|}\coloneqq\vert E_{\|}\vert$} (larger view in figure \ref{fig_k_ava_screen_p_star} of the appendix).}
\label{fig_k_ava_screen_p_star_main}
\end{figure}

Second, the approximation \mbox{$p_{\mathrm{c}}^{\mathrm{eff}}\approx p_{\mathrm{c}}^{\mathrm{scr}}$} of the effective critical momentum is regarded, on the basis of the results of the \textsc{MATLAB}-scripts$^{\ref{fig_plot_footnote_2_main}}$. They also present the data in contour plots, which are arranged in figure \ref{fig_k_ava_screen_p_star_main}. In addition to the graphical representation of the results, a console output is obtained for each of the mentioned \mbox{\textsc{MATLAB}-scripts}. It provides the minima, maxima and mean values of the data related to a contour plot, states the general parameter settings and also proves that the control criterion $(\ref{check_n_RE_ava_def_H})$ was satisfied for all computations. The outputs can be found in the listings \cref{MATLABoutput_plot_p_star_E3,MATLABoutput_plot_p_star_E10,MATLABoutput_plot_p_star_E30,MATLABoutput_plot_p_star_E100} in subsection \ref{output_matlab_appendix_subsection} of the appendix.
\\
Basically, the same deduction as for figure \ref{fig_k_ava_screen_p_c_scr_main} follows from a discussion of figure \ref{fig_k_ava_screen_p_star_main}. Nevertheless, one observes less extreme values for the kinetic energy density than in figure \ref{fig_k_ava_screen_p_star_main} for the lower and upper density boundaries. Presumably, this is caused by the analytic expression $p_{\mathrm{c}}^{\mathrm{scr}}$, which was used as the approximation for the effective critical momentum $p_{\mathrm{c}}^{\mathrm{eff}}$. Apparently, it has to be admitted, that this formula is not accurate for extremely low densities and for high densities, which are also connected to critical electric field values, which are close to the present electric field. However,\vspace{-7cm}\linebreak\newpage\noindent 
\vspace*{-4.0mm}this was expected, since it only approximates the more accurate representation of the effective critical momentum $p_{\star}$ and further it has to be remembered, that in particular the high density limit or respectively the limit \mbox{$E_{c}\rightarrow E_{\|}$}, is connected to a decreasing accuracy and applicability of the \textit{Hesslow} model. For the purpose of a further evaluation of the influences of the choice of the effective critical momentum on the mean kinetic energy of avalanche runaway electrons, one is referred to the next section.

\subsection{Comparison of the models by means of the mean rest mass-related kinetic energy density of an \textit{avalanche} runaway electron population}\label{comparison_avalanche_k_subsection}

In the following, a successive comparison of different calculation schemes for the mean rest mass-related kinetic energy density of an \textit{avalanche} runaway electron population shall be carried out. This is done, by means of the results \mbox{$k_{\mathrm{RE}}^{\hspace{0.25mm}\mathrm{ava}}/c$} from the \textit{Rosenbluth-Putvinski} model and the data \mbox{$k_{\mathrm{RE}}^{\hspace{0.25mm}\mathrm{ava,scr}}/c$} from the \textit{Hesslow} model. In case of the latter model an additional distinction is made for the choice of the approximation for the effective critical momentum, which is either the physically more accurate relation $p_{\star}$, computed as the root of the function $(\ref{func_p_c_eff_def})$ or the analytic expression $p_{\mathrm{c}}^{\mathrm{scr}}$. More precise, one utilizes the relative deviations between the results, computed by means of the mentioned calculation approaches. 

\begin{figure}[H]
  \centering
  \subfloat{\label{fig_rel_k_ava_p_c_scr_E3_main} 
   \includegraphics[trim=318 19 345 25,width=0.41\textwidth,clip]
    {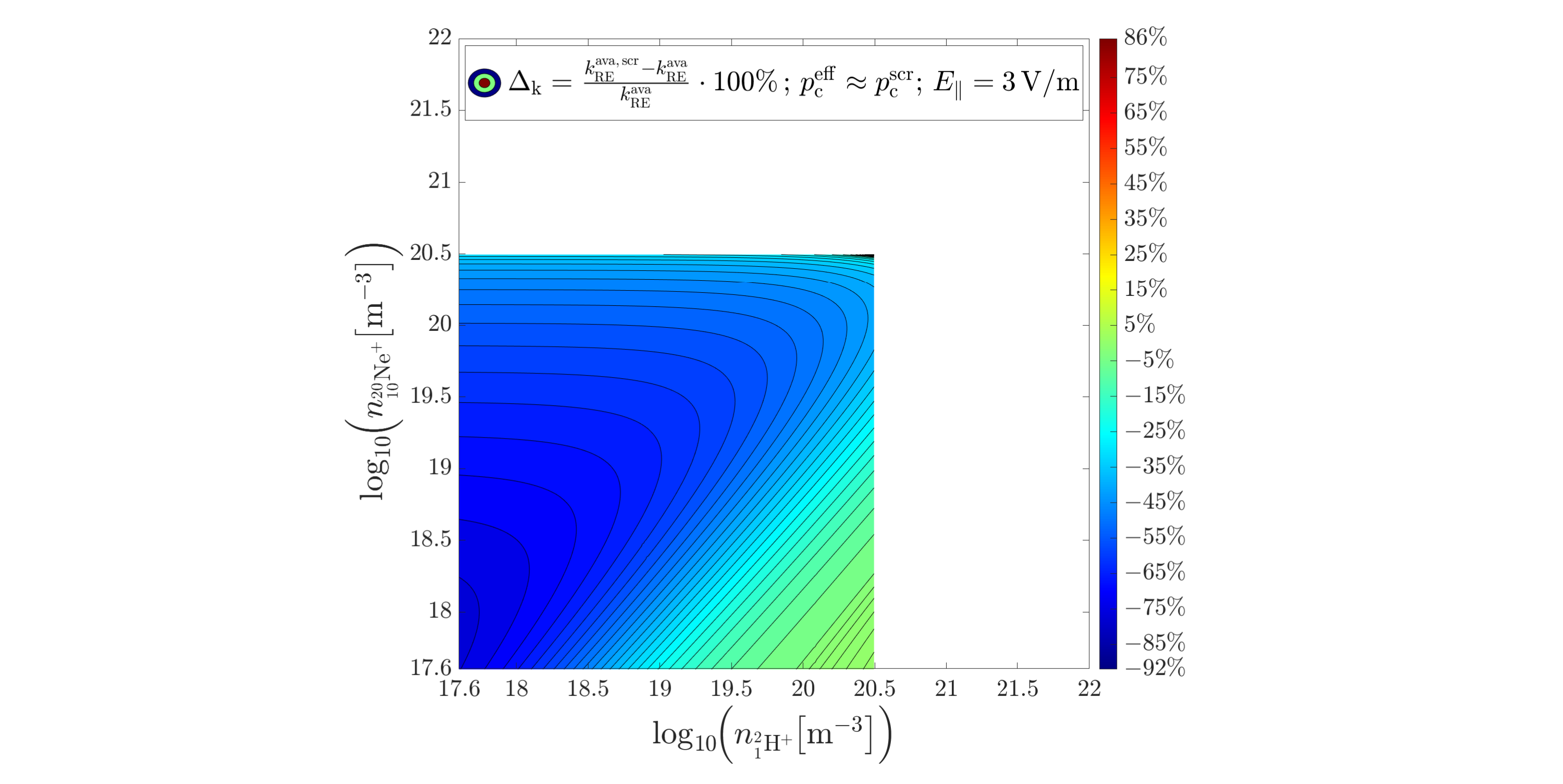}}\quad
  \subfloat{\label{fig_rel_k_ava_p_c_scr_E10_main}
    \includegraphics[trim=316 25 348 19,width=0.41\textwidth,clip]
    {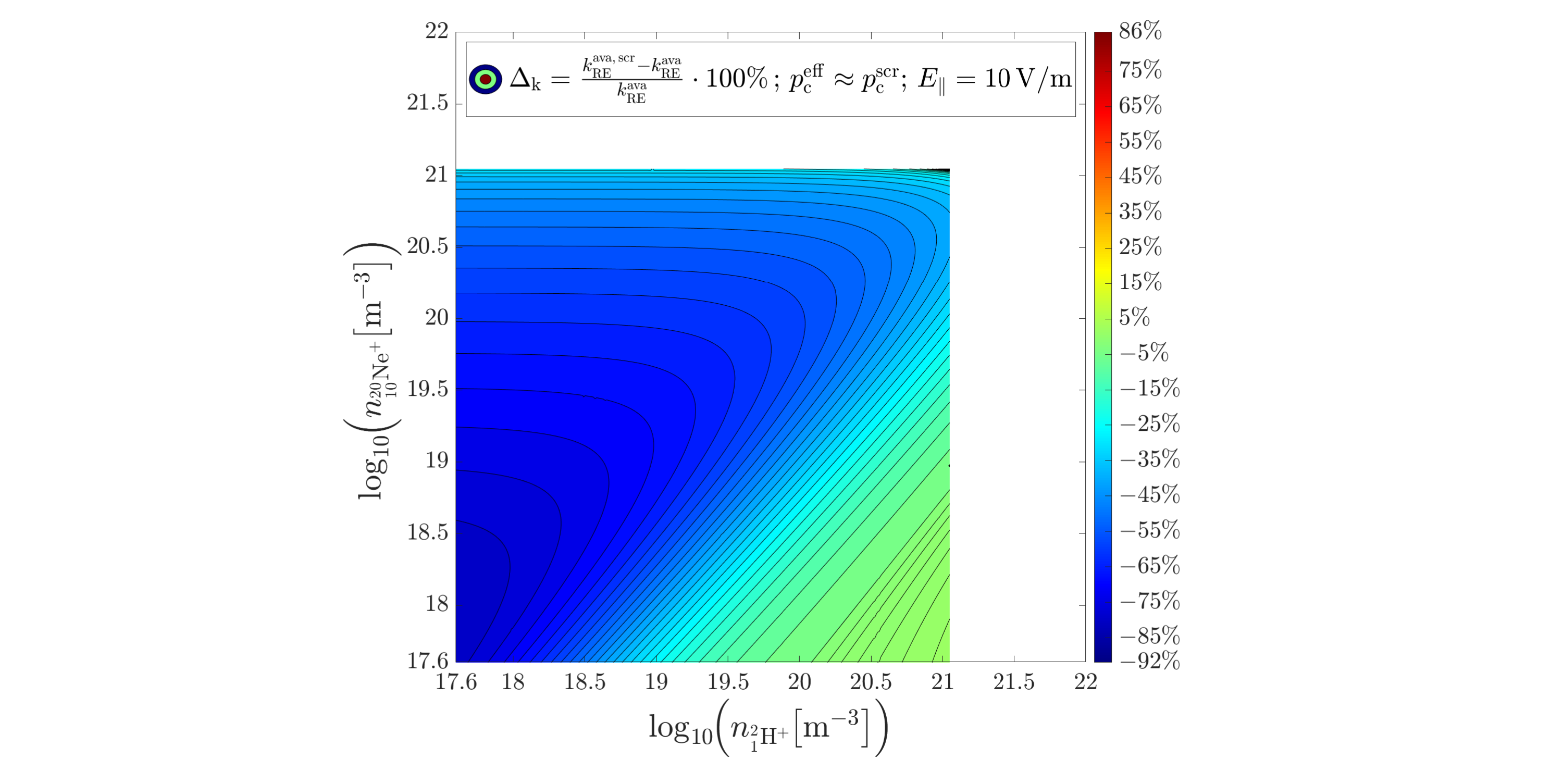}}\\[4pt]
  \subfloat{\label{fig_rel_k_ava_p_c_scr_E30_main} 
    \includegraphics[trim=315 29 343 17,width=0.41\textwidth,clip]
    {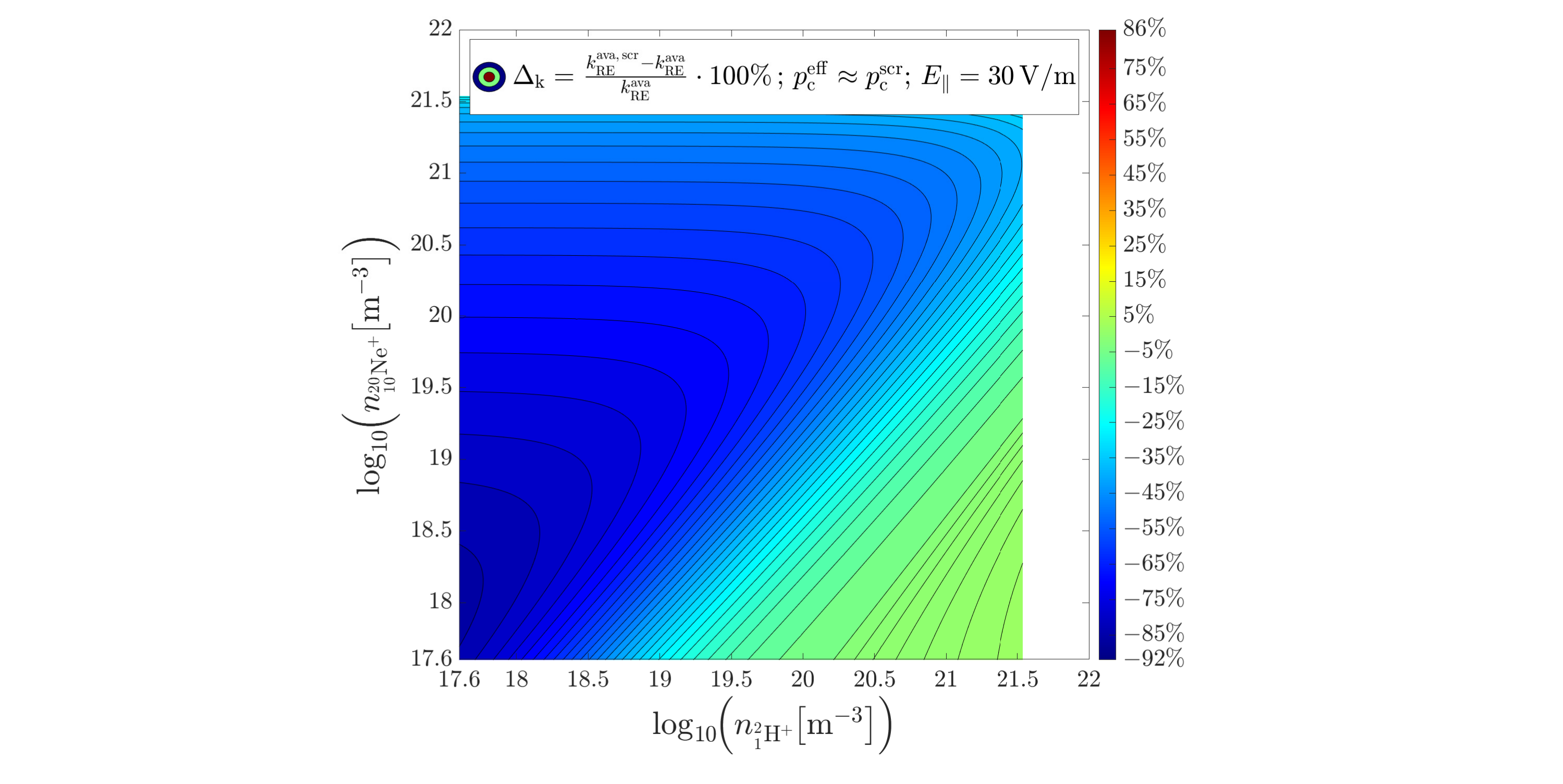}}\quad
  \subfloat{\label{fig_rel_k_ava_p_c_scr_E100_main}
     \includegraphics[trim=319 25 341 17,width=0.41\textwidth,clip]
    {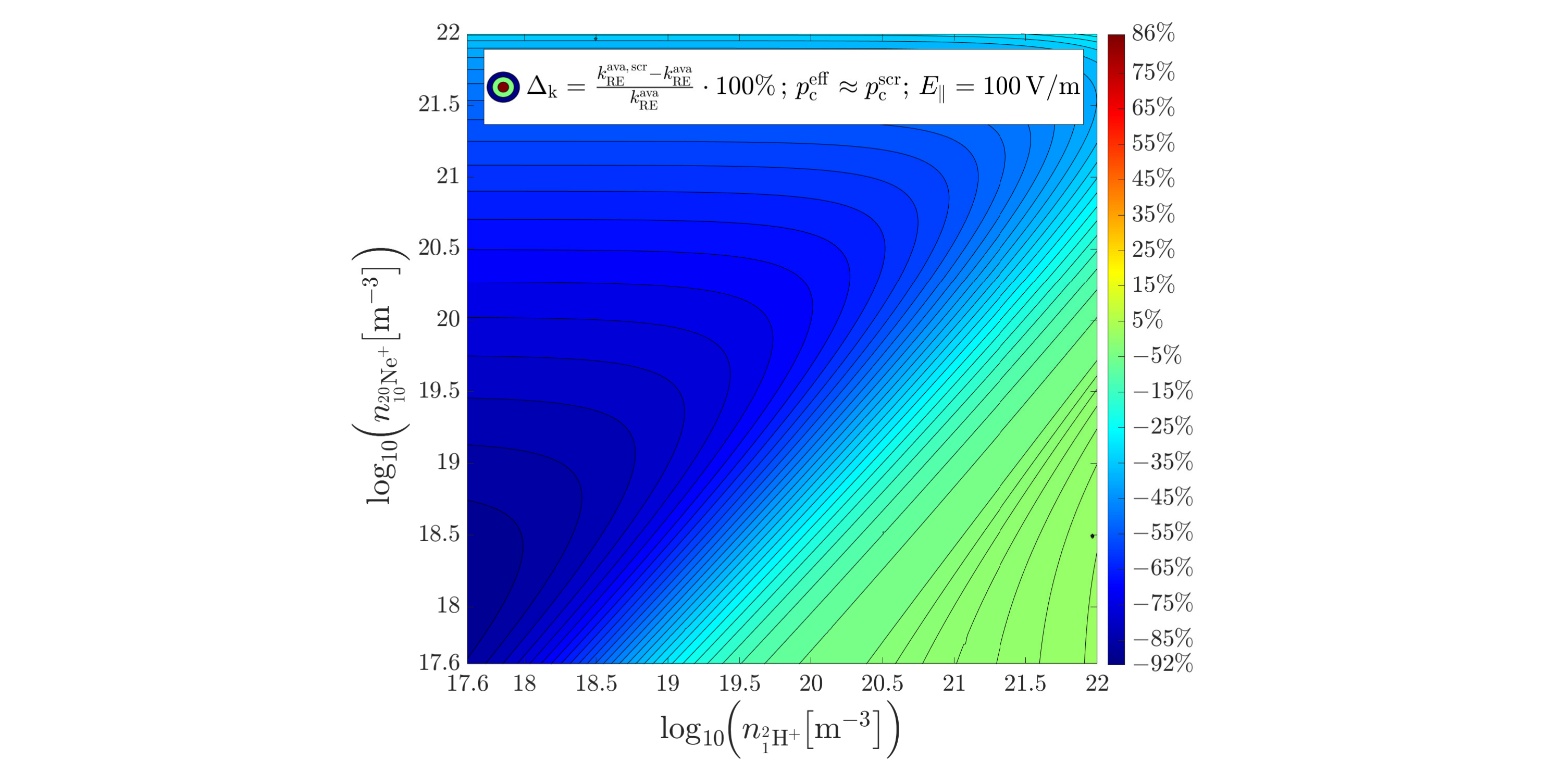}}
  \caption[Contour plots of the relative deviation $\Delta_{k}$ for the mean rest mass-related kinetic energy density of an avalanche runaway electron population with \mbox{$k_{\mathrm{B}}T_{\mathrm{e}}=10\,\textup{eV}$}, \mbox{$B=5.25\,\textup{T}$} and \mbox{$Z_{\mathrm{eff}}=1$}, due to the effect of partial screening with the effective critical momentum \mbox{$p_{\mathrm{c}}^{\mathrm{eff}}\approx p_{\mathrm{c}}^{\mathrm{scr}}$}, displayed for approximately logarithmically increasing values of the electric field strength \mbox{$E_{\|}\coloneqq\vert E_{\|}\vert$} (larger view in figure \ref{fig_rel_k_ava_p_c_scr} of the appendix).]{Contour plots$^{\ref{fig_plot_u_RP_footnote}}$ of the relative deviation $\Delta_{k}$ for the mean rest mass-related kinetic energy density of an avalanche runaway electron population with \mbox{$k_{\mathrm{B}}T_{\mathrm{e}}=10\,\textup{eV}$}, \mbox{$B=5.25\,\textup{T}$} and \mbox{$Z_{\mathrm{eff}}=1$}, due to the effect of partial screening with the effective critical momentum \mbox{$p_{\mathrm{c}}^{\mathrm{eff}}\approx p_{\mathrm{c}}^{\mathrm{scr}}$}, displayed for approximately logarithmically increasing values of the electric field strength \mbox{$E_{\|}\coloneqq\vert E_{\|}\vert$} (larger view in figure \ref{fig_rel_k_ava_p_c_scr} of the appendix).}
\label{fig_rel_k_ava_p_c_scr_main}
\end{figure}

First, the figure \ref{fig_rel_k_ava_p_c_scr_main} shall be discussed, which shows the relative difference between the kinetic energy of avalanche electrons in the \textit{Rosenbluth-Putvinski} model \mbox{$k_{\mathrm{RE}}^{\hspace{0.25mm}\mathrm{ava}}/c$} and its equivalent \mbox{$k_{\mathrm{RE}}^{\hspace{0.25mm}\mathrm{ava,scr}}/c$}, calculated in the \textit{Hesslow} model with \mbox{$p_{\mathrm{c}}^{\mathrm{eff}}\approx p_{\mathrm{c}}^{\mathrm{scr}}$}. 
\vspace{1.0mm}
\vspace*{-4.0mm}Note, that this figure also represents the relative deviation between the figure \ref{fig_k_ava_p_c_scr_main} and \ref{fig_k_ava_screen_p_c_scr_main} from the previous two subsections. 
\\
In general, one recognizes deviations below \mbox{$\pm 100\,\%$} and more detailed an underestimation of the \textit{Hesslow} model by the \textit{Rosenbluth-Putvinski} model for lower densities as well as a tendential overestimation for high deuterium and lower neon ion densities. In particular the area of underestimation correlates with the immense underestimation of the effective critical electric field by the \textit{Connor-Hastie} critical electric field, which is used in the \textit{Rosenbluth-Putvinski} approach. This can be recapitulated with the help of figure \ref{fig_E_C_ava_p_c_scr_main} from section \ref{part_screen_section}. Furthermore, a correlation to the difference between the approximation of the effective critical momentum $p_{\mathrm{c}}^{\mathrm{scr}}$ and the \textit{Connor-Hastie} critical momentum $p_{\mathrm{c}}$ is noticeable, if figure \ref{fig_p_comparison} is analysed again. At this, $p_{\mathrm{c}}^{\mathrm{scr}}\gg p_{\mathrm{c}}$ holds in the density region of the minimum relative deviation in figure \ref{fig_rel_k_ava_p_c_scr_main}. Therefore, the results \mbox{$k_{\mathrm{RE}}^{\hspace{0.25mm}\mathrm{ava,scr}}/c$} have to be smaller than the data for \mbox{$k_{\mathrm{RE}}^{\hspace{0.25mm}\mathrm{ava}}/c$}, because all moments are integrals over the always positive distribution functions and a larger lower integration boundary leads to a smaller integration interval for a fixed upper boundary. Since the maximum momentum boundary is set to infinity, stays unchanged, and for the lowest momentum of the avalanche runaway electron region one has $p_{\mathrm{c}}^{\mathrm{scr}}\gg p_{\mathrm{c}}$, the smaller value, compared to \mbox{$k_{\mathrm{RE}}^{\hspace{0.25mm}\mathrm{ava}}/c$}, of the definite integral related to \mbox{$k_{\mathrm{RE}}^{\hspace{0.25mm}\mathrm{ava,scr}}/c$}, is inevitable. However, one can conclude, that the relative deviation in the kinetic energy density between the two avalanche calculation schemes has a larger order of magnitude than the relative difference in the mean velocity, which can be looked up in subsection \ref{comparison_avalanche_j_subsection} and particularly in figure \ref{fig_rel_u_ava_p_c_scr_main}. This forces the comment, that the choice between the \textit{Rosenbluth-Putvinski} and the \textit{Hesslow} approach is more significant for the kinetic energy moment than for the mean velocity moment. Moreover and due to the fact that solely the \textit{Hesslow} model accounts for the influences of a non-full ionized plasma, one should use the \textit{Rosenbluth-Putvinski} model carefully, in order to avoid the usage of data for the kinetic energy density, which might be wrong by a factor of two. 

\begin{figure}[H]
  \centering
  \subfloat{\label{fig_rel_k_ava_p_star_E3_main} 
   \includegraphics[trim=319 22 332 17,width=0.41\textwidth,clip]
    {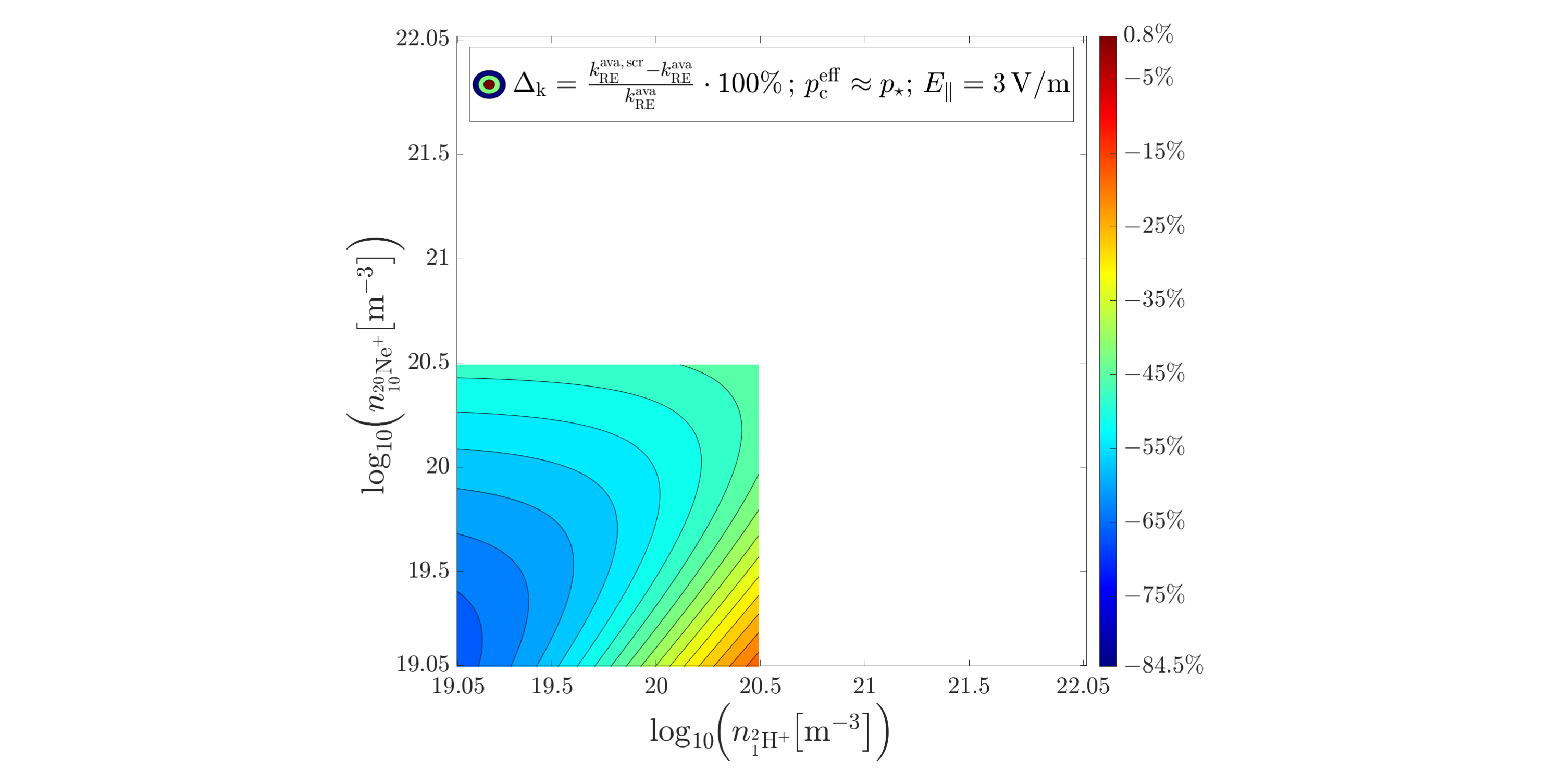}}\quad
  \subfloat{\label{fig_rel_k_ava_p_star_E10_main}
    \includegraphics[trim=325 26 329 19,width=0.41\textwidth,clip]
    {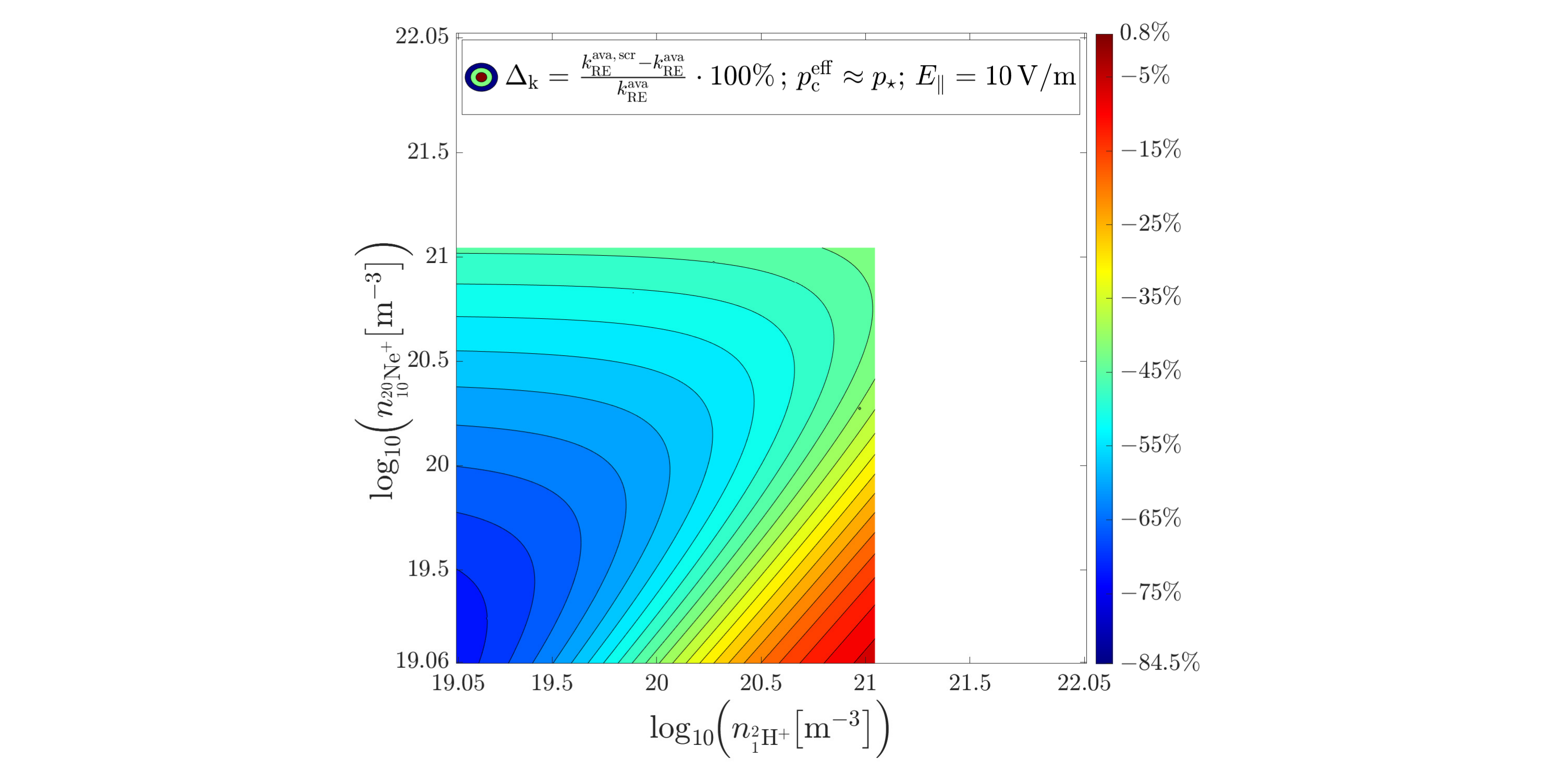}}\\[4pt]
  \subfloat{\label{fig_rel_k_ava_p_star_E30_main} 
    \includegraphics[trim=321 28 340 14,width=0.41\textwidth,clip]
    {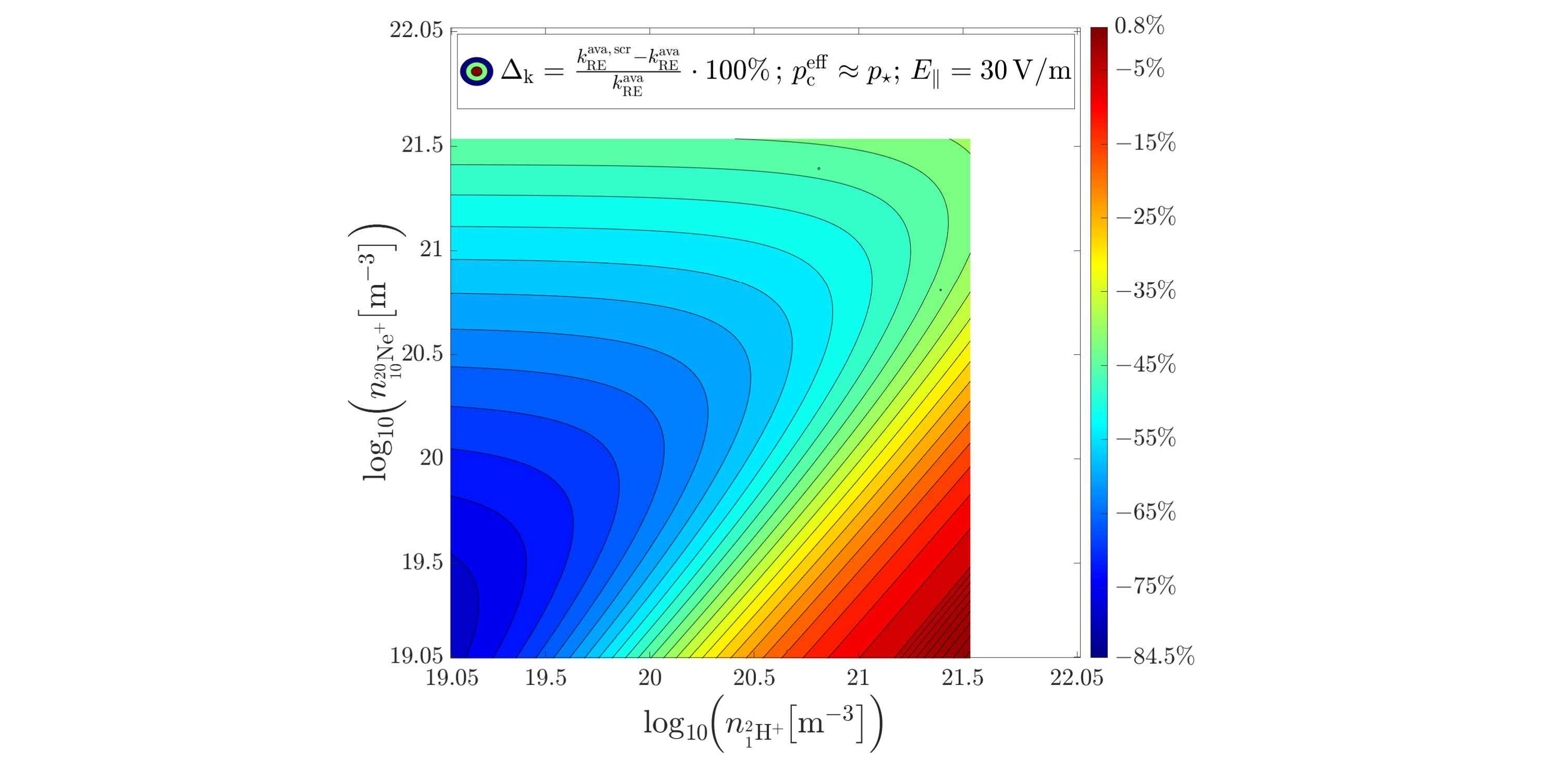}}\quad
  \subfloat{\label{fig_rel_k_ava_p_star_E100_main}
   \includegraphics[trim=322 21 334 18,width=0.41\textwidth,clip]
    {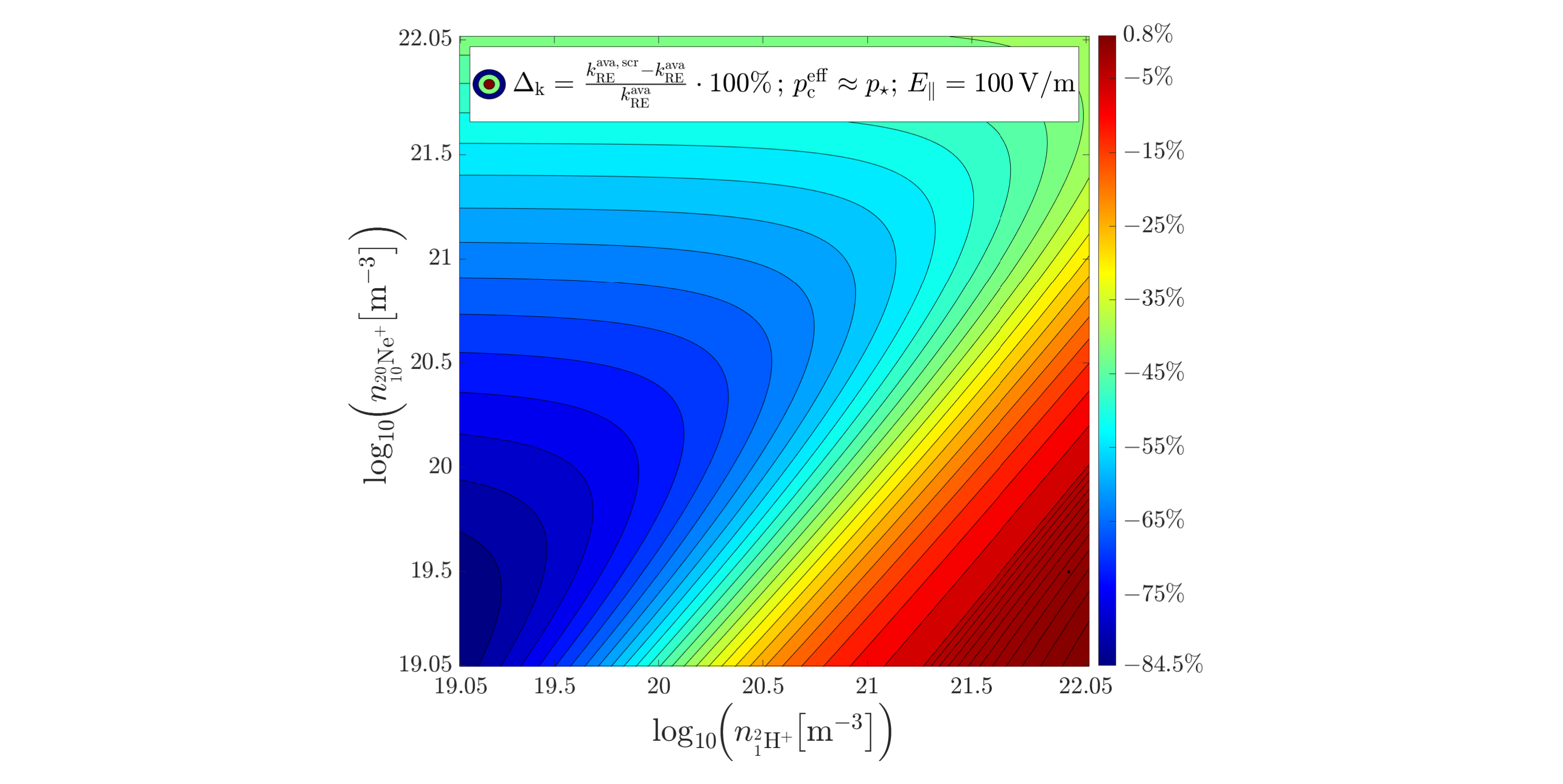}}
  \caption[Contour plots of the relative deviation $\Delta_{k}$ for the mean rest mass-related kinetic energy density of an avalanche runaway electron population with \mbox{$k_{\mathrm{B}}T_{\mathrm{e}}=10\,\textup{eV}$}, \mbox{$B=5.25\,\textup{T}$} and \mbox{$Z_{\mathrm{eff}}=1$}, due to the effect of partial screening with the effective critical momentum \mbox{$p_{\mathrm{c}}^{\mathrm{eff}}\approx p_{\star}$}, displayed for approximately logarithmically increasing values of the electric field strength \mbox{$E_{\|}\coloneqq\vert E_{\|}\vert$} (larger view in figure \ref{fig_rel_k_ava_p_star} of the appendix).]{Contour plots$^{\ref{fig_plot_footnote_2_main}}$ of the relative deviation $\Delta_{k}$ for the mean rest mass-related kinetic energy density of an avalanche runaway electron population with \mbox{$k_{\mathrm{B}}T_{\mathrm{e}}=10\,\textup{eV}$}, \mbox{$B=5.25\,\textup{T}$} and \mbox{$Z_{\mathrm{eff}}=1$}, due to the effect of partial screening with the effective critical momentum \mbox{$p_{\mathrm{c}}^{\mathrm{eff}}\approx p_{\star}$}, displayed for approximately logarithmically increasing values of the electric field strength \mbox{$E_{\|}\coloneqq\vert E_{\|}\vert$} (larger view in figure \ref{fig_rel_k_ava_p_star} of the appendix).}
\label{fig_rel_k_ava_p_star_main}
\end{figure}

Second, the figure \ref{fig_rel_k_ava_p_star_main} is going to be analysed, which depicts the relative error in the kinetic energy of avalanche electrons between the \textit{Rosenbluth-Putvinski} model and the \textit{Hesslow} model. For the latter model, the values of \mbox{$k_{\mathrm{RE}}^{\hspace{0.25mm}\mathrm{ava,scr}}/c$} were calculated with \mbox{$p_{\mathrm{c}}^{\mathrm{eff}}\approx p_{\star}$}, which is supposed to be the most accurate of the three presented calculation schemes.
\\
The evaluation of figure \ref{fig_rel_k_ava_p_star_main} reveals, that the \textit{Hesslow} model together with the choice \mbox{$p_{\mathrm{c}}^{\mathrm{eff}}\approx p_{\star}$} leads categorically to smaller values for the kinetic energy density in comparison to the \textit{Rosenbluth-Putvinski} approach. This means, that one can expect up to \mbox{$100\,\%$} smaller vales for the \textit{Hesslow} model with \mbox{$p_{\mathrm{c}}^{\mathrm{eff}}\approx p_{\star}$}, whilst one might receive relative differences of up to \mbox{$\pm 100\,\%$} for \mbox{$p_{\mathrm{c}}^{\mathrm{eff}}\approx p_{\mathrm{c}}^{\mathrm{scr}}$}, as it was seen in the above figure \ref{fig_rel_k_ava_p_c_scr_main}. Therefore, one has to relativize oneselfs comment on the applicability of the \textit{Rosenbluth-Putvinski} model, because the deviation from the more accurate \mbox{\textit{Hesslow}} model is negligible for low neon and high deuterium densities. In this region, both models yield to comparable results, although for the whole parameter space a larger deviation of the different calculation schemes is apparent. Further, one basically observes\vspace{-7cm}\linebreak\newpage\noindent 
\vspace*{-4.0mm}similar characteristics as in figure \ref{fig_rel_k_ava_p_c_scr_main}, like for example the minimum of the deviation for low densities, which correlates with the deviation in the critical electric field from the \textit{Connor-Hastie} critical electric field. 

\begin{figure}[H]
  \centering
  \subfloat{\label{fig_tilde_rel_k_ava_p_star_E3_main} 
 \includegraphics[trim=322 33 327 9,width=0.41\textwidth,clip]
    {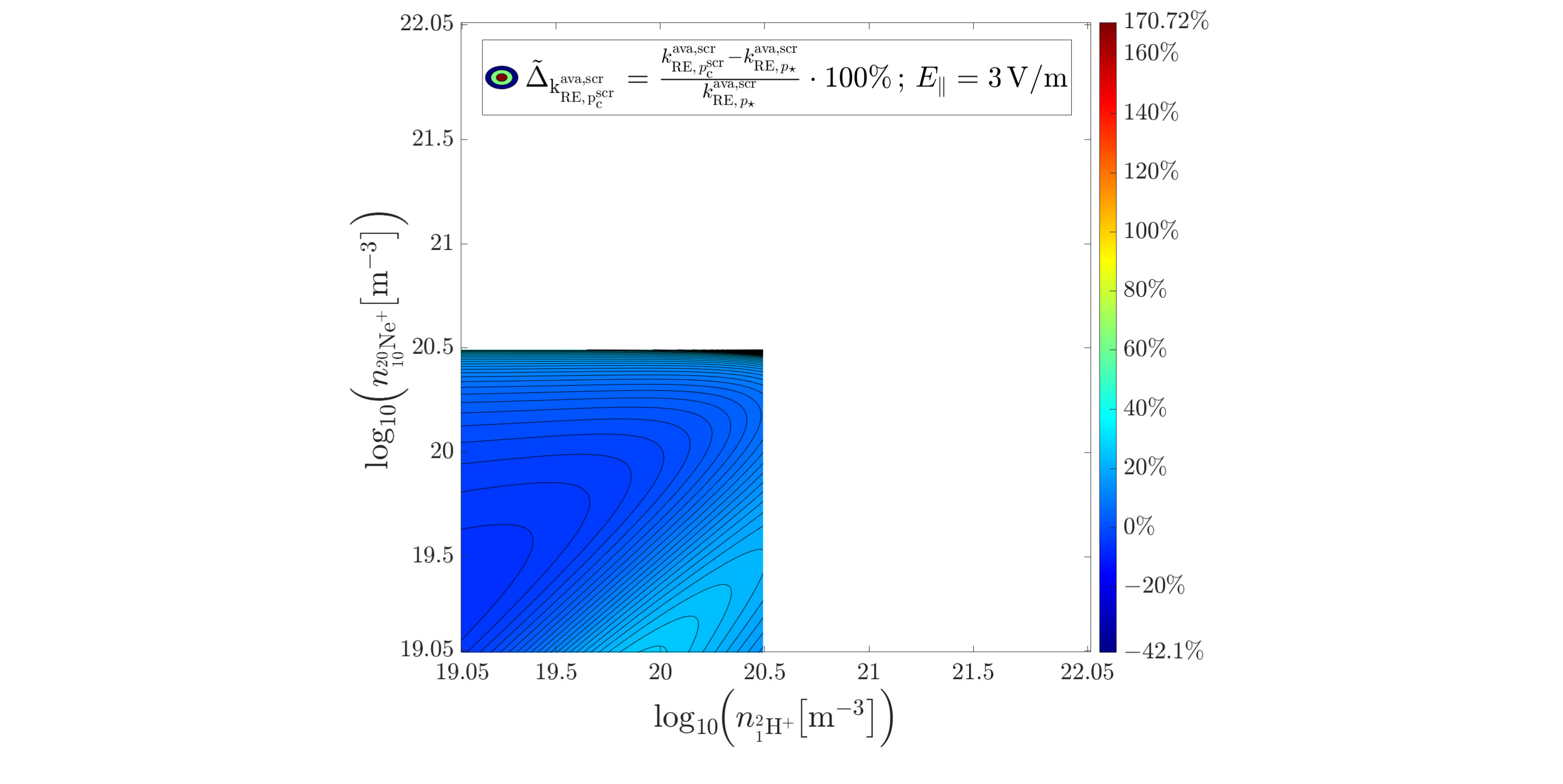}}\quad
  \subfloat{\label{fig_tilde_rel_k_ava_p_star_E10_main}
   \includegraphics[trim=317 33 331 8,width=0.41\textwidth,clip]
    {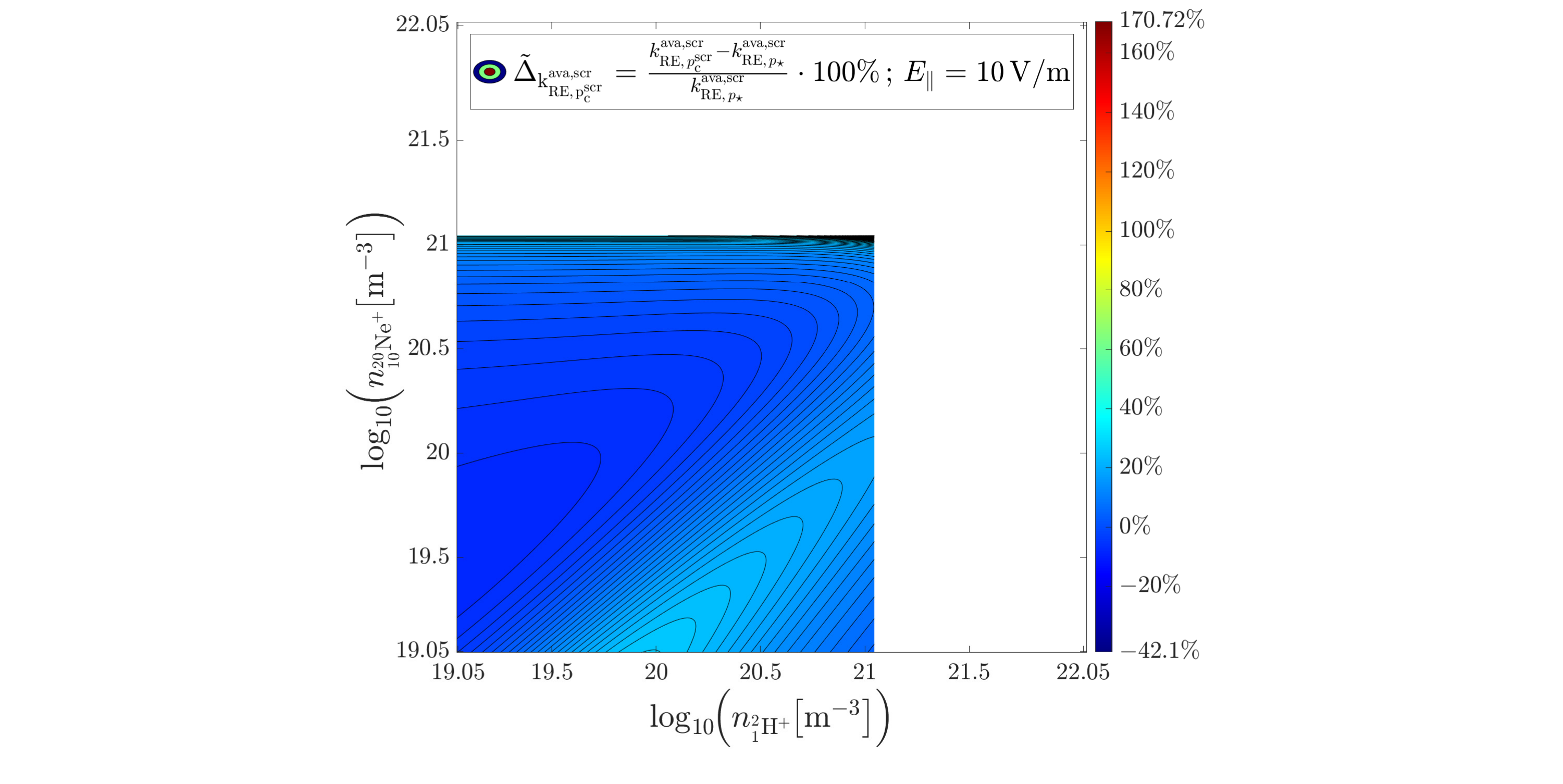}}\\[4pt]
  \subfloat{\label{fig_tilde_rel_k_ava_p_star_E30_main} 
  \includegraphics[trim=317 32 335 7,width=0.41\textwidth,clip]
    {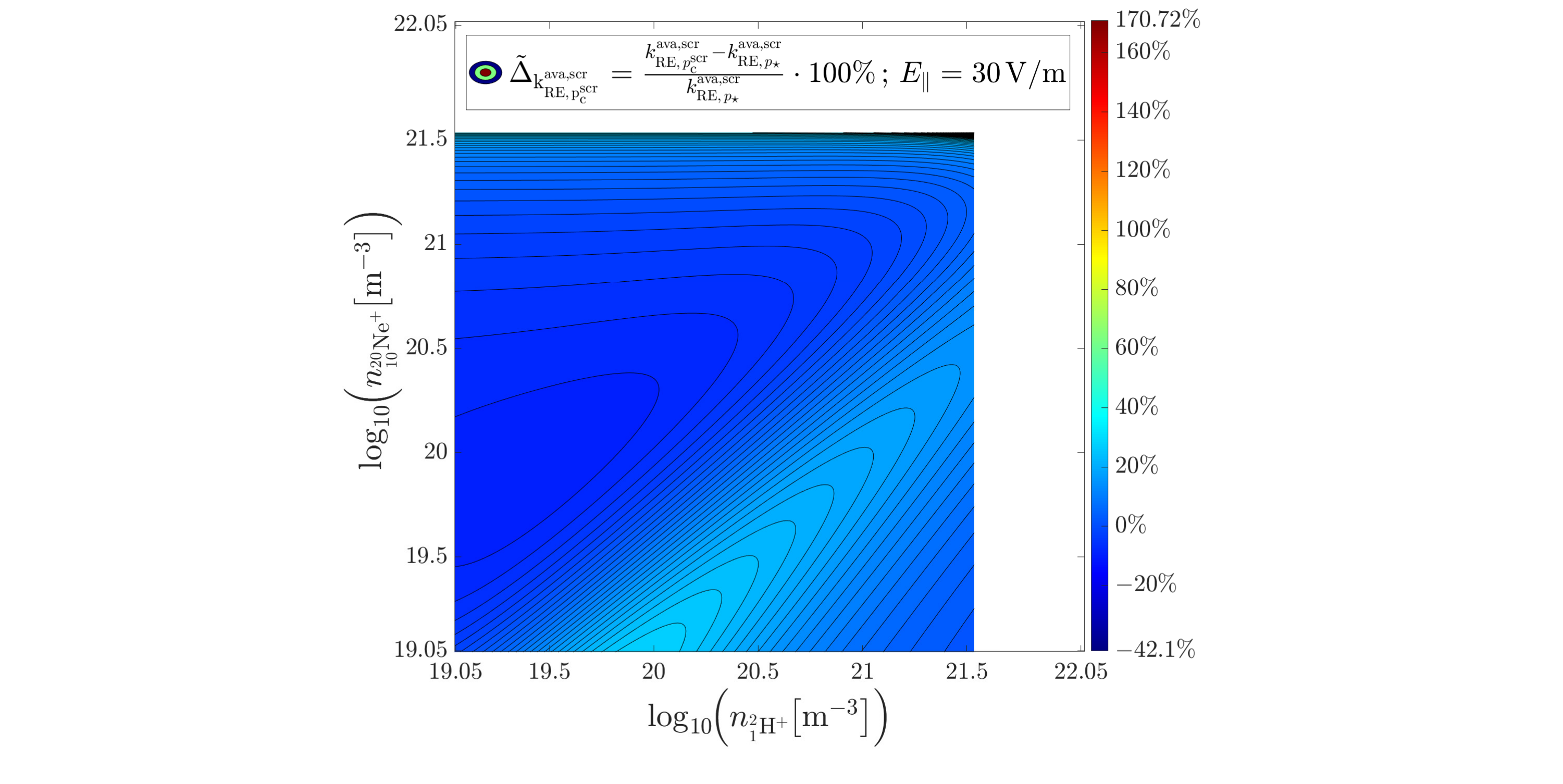}}\quad
  \subfloat{\label{fig_tilde_rel_k_ava_p_star_E100_main}
   \includegraphics[trim=320 33 332 6,width=0.41\textwidth,clip]
    {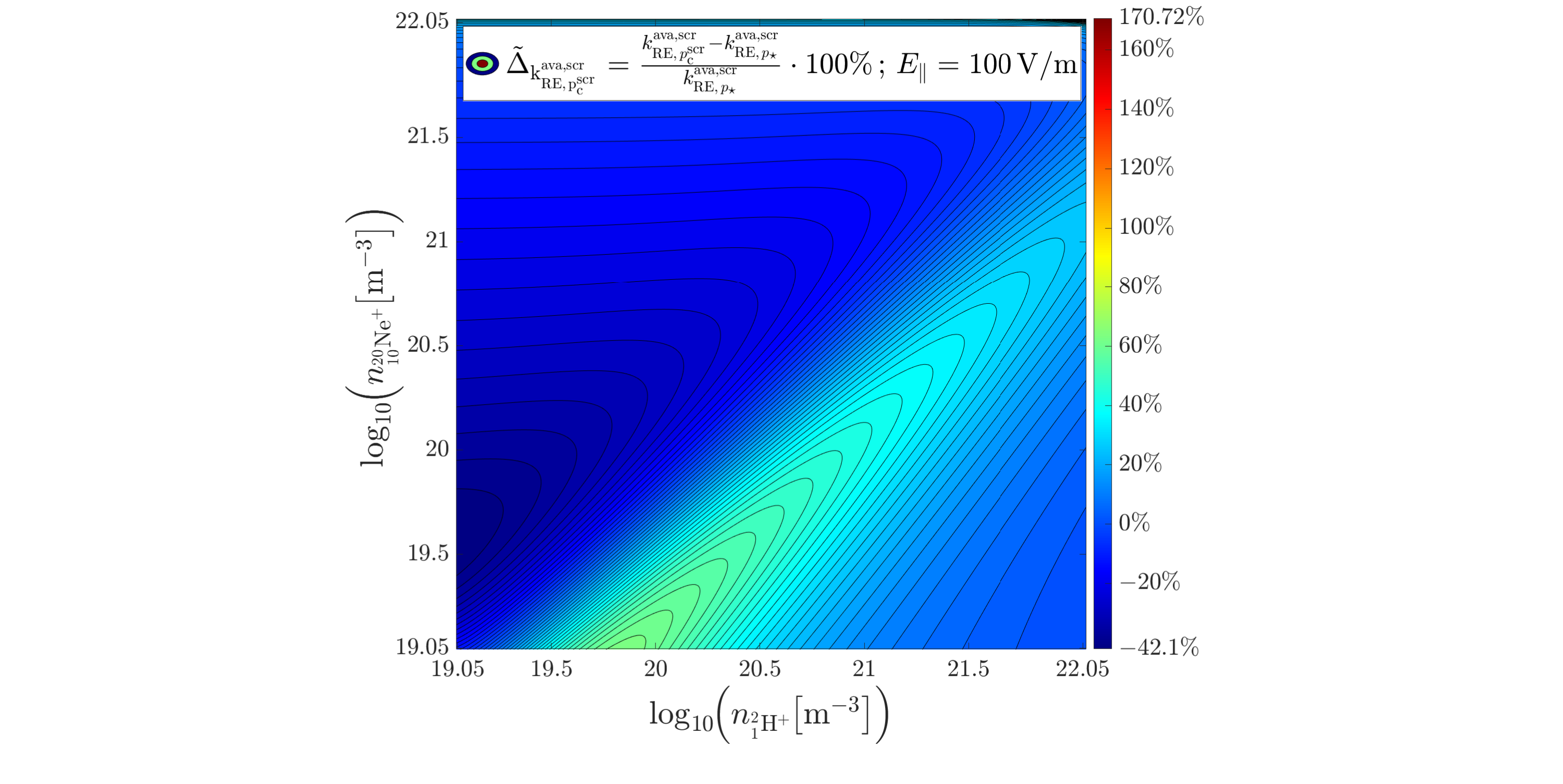}}
  \caption[Contour plots of the relative deviation $\tilde{\Delta}_{k_{\mathrm{RE},\,p^{\mathrm{scr}}_{\mathrm{c}}}^{\mathrm{ava,scr}}}$ for the mean rest mass-related kinetic energy density of an avalanche runaway electron population with \mbox{$k_{\mathrm{B}}T_{\mathrm{e}}=10\,\textup{eV}$}, \mbox{$B=5.25\,\textup{T}$} and \mbox{$Z_{\mathrm{eff}}=1$} in the \textit{Hesslow} model, due to the different approximations of the effective critical momentum \mbox{$p_{\mathrm{c}}^{\mathrm{eff}}\approx p_{\mathrm{c}}^{\mathrm{scr}}$} and \mbox{$p_{\mathrm{c}}^{\mathrm{eff}}\approx p_{\star}$}, displayed for approximately logarithmically increasing values of the electric field strength \mbox{$E_{\|}\coloneqq\vert E_{\|}\vert$} (larger view in figure \ref{fig_tilde_rel_k_ava_p_star} of the appendix).]{Contour plots$^{\ref{fig_plot_footnote_2_main}}$ of the relative deviation $\tilde{\Delta}_{k_{\mathrm{RE},\,p^{\mathrm{scr}}_{\mathrm{c}}}^{\mathrm{ava,scr}}}$ for the mean rest mass-related kinetic energy density of an avalanche runaway electron population with \mbox{$k_{\mathrm{B}}T_{\mathrm{e}}=10\,\textup{eV}$}, \mbox{$B=5.25\,\textup{T}$} and \mbox{$Z_{\mathrm{eff}}=1$} in the \textit{Hesslow} model, due to the different approximations of the effective critical momentum \mbox{$p_{\mathrm{c}}^{\mathrm{eff}}\approx p_{\mathrm{c}}^{\mathrm{scr}}$} and \mbox{$p_{\mathrm{c}}^{\mathrm{eff}}\approx p_{\star}$}, displayed for approximately logarithmically increasing values of the electric field strength \mbox{$E_{\|}\coloneqq\vert E_{\|}\vert$} (larger view in figure \ref{fig_tilde_rel_k_ava_p_star} of the appendix).}
\label{fig_tilde_rel_k_ava_p_star_main}
\end{figure}

Third, the influence of the approximation of the effective critical momentum is discussed on the basis of figure \ref{fig_tilde_rel_k_ava_p_star_main}. It arranges the contour plots of the relative deviation in \mbox{$k_{\mathrm{RE}}^{\hspace{0.25mm}\mathrm{ava,scr}}/c$} between the choices \mbox{$p_{\mathrm{c}}^{\mathrm{eff}}\approx p_{\mathrm{c}}^{\mathrm{scr}}$} and \mbox{$p_{\mathrm{c}}^{\mathrm{eff}}\approx p_{\star}$} in the \textit{Hesslow} model for the four logarithmically increasing electric field strength values. 
\\
By means of the contour plots in figure \ref{fig_tilde_rel_k_ava_p_star_main}, one identifies deviations of approximately up to \mbox{$\pm 50\,\%$} between the two calculation rules, which are associated with $p_{\mathrm{c}}^{\mathrm{scr}}$ and $p_{\star}$. Note, that the range of the relative deviation increases with larger electric fields. In addition, significantly greater results for the kinetic energy density can be seen in the high neon ion density limit, where \mbox{$p_{\mathrm{c}}^{\mathrm{scr}}>p_{\star}$} holds, which can be verified with the relative deviation $\tilde{\Delta}_{p^{\mathrm{scr}}_{\mathrm{c}}}$ from figure \ref{fig_p_comparison}. Additionally, the minimum, maximum and mean values of this displayed relative deviation $\tilde{\Delta}_{k_{\mathrm{RE},\,p^{\mathrm{scr}}_{\mathrm{c}}}^{\mathrm{ava,scr}}}$ can be found in the listings\vspace{-7cm}\linebreak\newpage\noindent  
\vspace*{-4.0mm}\cref{MATLABoutput_plot_p_star_E3,MATLABoutput_plot_p_star_E10,MATLABoutput_plot_p_star_E30,MATLABoutput_plot_p_star_E100} of the appendix. Hence, one can claim, that the two possibilities for the effective critical momentum lead to results for the kinetic energy density, which lead to a mean relative difference below approximately \mbox{$10\,\%$}. This implies, that a calculation based on \mbox{$p_{\mathrm{c}}^{\mathrm{eff}}\approx p_{\mathrm{c}}^{\mathrm{scr}}$} overestimates the more accurate computation with \mbox{$p_{\mathrm{c}}^{\mathrm{eff}}\approx p_{\star}$}, where larger errors are expected for high electric fields and low densities.  

For a comparison of the influence of a certain calculation of $p_{\mathrm{c}}^{\mathrm{eff}}$ on different moments, one regards the relative deviations $\tilde{\Delta}_{k_{\mathrm{RE},\,p^{\mathrm{scr}}_{\mathrm{c}}}^{\mathrm{ava,scr}}}$ and $\tilde{\Delta}_{u_{\mathrm{RE},\,p^{\mathrm{scr}}_{\mathrm{c}}}^{\mathrm{ava,scr}}}$ displayed in the figure \ref{fig_tilde_rel_k_ava_p_star_main} and \ref{fig_tilde_rel_u_ava_p_star_main}. In doing so, it can first be deduced, that the order of magnitude of the deviation between $p_{\mathrm{c}}^{\mathrm{scr}}$ and $p_{\star}$ is not inherited unchanged into the relative difference of the moments, calculated with the two different computation schemes. At that, the relative deviation $\tilde{\Delta}_{p^{\mathrm{scr}}_{\mathrm{c}}}$ from figure \ref{fig_p_comparison} is suggested for verification purposes. Secondly, one finds that the moment itself determines the propagation of the deviation between the approximations of the effective critical momentum into the final result. To this, one remarks, that the calculated values for the kinetic energy density differ by an order of magnitude of the second power of ten, whilst for the mean velocity the deviation $\tilde{\Delta}_{u_{\mathrm{RE},\,p^{\mathrm{scr}}_{\mathrm{c}}}^{\mathrm{ava,scr}}}$ shows only a magnitude of the first power of ten. Consequently, it can be summarized, that the choice of the lower momentum boundary has to be tailored to the specific moments, which shall be calculated, and as well to the density parameter region, which is expected for the case under consideration. Apart from that, neither the \textit{Rosenbluth-Putvinski} nor the \textit{Hesslow} model can be generally classified as unsuitable. This holds as well for the approximation of the effective critical momentum in the \textit{Hesslow} model, because for different parameter domains, error thresholds with varying degrees of rigor, and different moments to be calculated, one could receive the outcome, to choose between all, none or only one or two of the presented calculation schemes. However, one should always prefer the utilization of the \textit{Hesslow} calculation approach with \mbox{$p_{\mathrm{c}}^{\mathrm{eff}}\approx p_{\star}$}, if possible, due to the fact, that it contains the most physical knowledge and should therefore yield to more accurate results, in comparison to the utilization of the approximation \mbox{$p_{\mathrm{c}}^{\mathrm{eff}}\approx p_{\mathrm{c}}^{\mathrm{scr}}$} or the \textit{Rosenbluth-Putvinski} model with the \textit{Connor-Hastie} critical electric field $p_{c}$, representing the two other calculation schemes.

\clearpage

%% file: outlook.tex
\chapter{Summary and outlook}

At the present day, the magnetic plasma confinement in a toroidally symmetric \textit{tokamak} reactor is the most advanced fusion reactor concept \cite{wesson}. This is due to the extensive underlying research, which is motivated by the aim to provide a future environmentally friendly and secure alternative to coal, gas or nuclear fission power plants with a similar base load capability in energy supply, in order to cope with the overall increase in energy demand \cite{IEA_report,EPSpostitionpaper}. 
\\
As explained in section \ref{tokamak_disruptions_section}, an abrupt loss of the energy and the magnetic confinement in a plasma discharge can for instance be caused by a large plasma current, high plasma densities or a significant presence of impurity atoms with high nuclear charge number \cite{Kowslowski,Hender_2007}. Such a \textit{disruption} can be accompanied by the generation of a runaway electron beam with the potential to damage plasma-facing reactor components \cite{Bazylev_2011,Breizman_2019,Matthews2016,Reux2015}, the avoidance of which is essential for future reactors such as ITER \cite{Hoppe_2021,REsimulation,Hoppe_2022}.
\\
As a consequence, computationally efficient simulations of the runaway current, which provide sufficient physical accuracy are required in research. At this, the \mbox{\textit{reduced}} \mbox{\textit{kinetic}}\linebreak\mbox{\textit{modeling}} approach, as implemented i.a.\ in the DREAM-code \cite{Hoppe_2022}, is a compromise between the computation-intensive and highly accurate calculations based on the complete solution of the kinetic equation and the simplified simulation on the bases of the so-called \textit{fluid description} of a plasma, as it was elaborated in section \ref{kin_equa_section}. Its governing equations rely on the modeling of the runaway electron current density, which itself contains the product of the density and the mean velocity of the runaway electrons. Those quantities can be expressed as certain moments of analytic runaway electron distribution functions. This is also possible for the mean runaway electron kinetic energy density, which is used to understand the influences of a runaway electron population on e.g.\ the equilibrium confinement \cite{Ficker2019,Ficker2021}, the evolution of atomic physics processes \cite{Breizman_2019} or the electron impact ablation of mitigation pellet injections \cite{James_2011,Hollmann2016}.
\\
For the purpose of establishing calculation rules for said moments of analytic distribution functions, the cylindrical coordinates \mbox{$(p_{\|},\,p_{\perp})$} and the spherical coordinates \mbox{$(p,\,\xi)$} of a gyro-averaged two-dimensional momentum space coordinate system were introduced in section \ref{mom_space_coord_section}. However, the description of the runaway region was carried out by means of the momentum magnitude $p$. At that, a classical representation on the basis of the critical momentum $p_{c}$ and electric field $E_{c}$ proposed by \mbox{\textit{J.\hspace{0.7mm}W.\hspace{0.9mm}Connor} }and \mbox{\textit{R.\hspace{0.7mm}J.\hspace{0.9mm}Hastie}} \cite{Connor_1975} from section \ref{RE_phenom_section} was compared to an approach, using the effective critical quantities $p_{c}^{\mathrm{eff}}$ and $E_{c}^{\mathrm{eff}}$, on the basis of the work of \mbox{\textit{L.\hspace{0.9mm}Hesslow}} \cite{Hesslow_2018}, which includes the effects of the partial screening of nuclear charges in not fully ionized plasmas. A detailed discussion by means of an ITER-like disruption simulation \cite{Smith_2009} and with the help of contour plots for the parameter space of singly-ionized 
deuterium and neon ions has been conducted in section \ref{part_screen_section}. This allows to chose a lower runaway momentum boundary \mbox{$p_{low}\in\lbrace p_{c},\, p_{\star},\,p_{c}^{\mathrm{scr}},\,p_{min}\rbrace$} and a maximum momentum magnitude \mbox{$p_{high}\in\lbrace p_{max},\,\infty\rbrace$} for the runaway region. In particular, one should consider the physically more accurate approximations \mbox{$p_{low}=p_{c}^{\mathrm{eff}}\in\lbrace p_{\star},\,p_{c}^{\mathrm{scr}},\,p_{min}\rbrace$} together with the effective critical electric field $E_{c}^{\mathrm{eff}}$, in order to account for a partially ionized plasma. This is reasoned, by the higher sensitivity of the moment calculation rules to $p_{low}$, so that for $p_{high}$ a fixed value in correspondence to the maximum runaway electron energy or an infinite momentum are suitable, in order to increase the efficiency of the calculation by for instance avoiding a highly accurate computation of $p_{max}$ as the larger root of the parallel, pitch-averaged net acceleration force balance $(\ref{F_acc})$. Note, that this thought also applies to the physically more precise, but computational expensive approximations $p_{\star}$ and $p_{min}$ of the effective critical momentum, in contrast to the analytic relations $p_{c}$ and $p_{c}^{\mathrm{scr}}$.
\\
Thereupon, the generation and loss mechanisms of runaway electrons were elaborated in section \ref{mechanisms_section}. This led to the decision, that moment-based calculation rules should be derived and evaluated for the primary and the secondary generation mechanisms, since without a primary seed density of runaway electrons the \textit{avalanche} mechanism would not be triggered and a runaway beam would not be explainable, since it forms due to the multiplication of the seed runaway electron density with an avalanche multiplication factor of for instance \mbox{$\textup{e}^{\hspace{0.35mm}\mathcal{M}_{ava}}\approx 10^{35}$} as predicted in the paper \cite{Hesslow_2019} for an ITER-like deuterium density and impurity densities near \mbox{$10^{20}\,\textup{m}^{-3}$}. Hence, the focus of this work was placed on the elucidation, analysis and evaluation of calculation schemes for the density, the mean velocity and the kinetic energy density of was expressed as moments of distribution functions for primary \textit{hot-tail} and secondary \textit{avalanche} runaway electrons.  

First, the \textit{hot-tail} generation mechanism was modeled on the basis of the time- and momentum magnitude-dependent isotropic electron distribution function by \mbox{\textit{H.\hspace{0.7mm}M.\hspace{0.9mm}Smith}} and \textit{E.\hspace{0.9mm}Verwichte} \cite{hottailREdistfunc}, expressed as stated in the work of \textit{I.\hspace{0.9mm}Svenningsson} \cite{Svenningsson2020}. Further, a distinction has been made between an \textit{isotropic} description of the runaway region, where the full pitch coordinate interval \mbox{$\xi\in[-1,\,1]$} is considered, and an \textit{anisotropic} representation of the runaway region \mbox{$\xi\in[\xi_{sep}(p,\,\tilde{E}),\,1]$} with a pitch-dependent lower boundary, referred to as \textit{separatrix}. Beyond that, the mentioned relations for the momentum magnitude boundaries of the runaway region from section \ref{part_screen_section} were applied together with the corresponding choices for the generalized electric field \mbox{$\tilde{E}_{c}\in\lbrace E_{c},\,E_{c}^{\mathrm{eff}}\rbrace$}. On this occasion, the modified upper runaway momentum $\tilde{p}_{max}$ was introduced, in order to make use of its physically more accurate values below a fixed momentum threshold, related to the highest possible runaway energy during a disruption, while saving runtime above this threshold.
\\
Based on that, one was able to establish numerical calculation rules for the \textit{hot-tail} runaway electron density, the mean velocity and the kinetic energy density in the sections \ref{ht_n_section}, \ref{ht_j_section} and \ref{ht_k_section}, which allow a computation by means of standard quadrature formulas and are stated for the isotropic and anisotropic pitch interval in combination with four different representations of the momentum magnitude description of the runaway region. Thus, three of the four calculation rules for the pitch-dependent and the isotropic consideration of the runaway region account for the influences of partial screening. For the case of the \textit{hot-tail} runaway electron density and an isotropic respectively pitch-independent interpretation of the runaway region a known analytical calculation rule was derived rigorously and in a more general manner, so that it can be used as a control criterion for the numerical one-dimensional integration. 
\\  
The evaluation of the deduced calculation schemes was then carried out with the help of a \textsc{MATLAB}-implementation, under utilization of the results of an ITER-disruption simulation \cite{Smith_2009} for the evolution of the electric field and the electron temperature in time, which was previously introduced in section \ref{part_screen_section}. Hence, one was able to analyse and validate the computation rules of the \textit{Smith-Verwichte} approach for a deuterium plasma with and without the presence of a time-independent neon impurity in section \ref{ht_eval_comp_u_and_k_section}, with the aid of the order of magnitude of the of the three moments and the current density. In the process, minor deviations were discovered between the different descriptions of the runaway region, which are are marginally enhanced for electric fields close to the critical electric field and if an impurity density with an order of magnitude of the deuterium density is considered. Apart from that a suggestion for a specific calculation rule could not be presented terminally without an analysis of results from self-consistent disruption simulations, which might also include a time-dependent presence of impurities. Nevertheless, it was deduced, that for anisotropic representations of the runaway region one should avoid the improvident use of $p_{\star}$ and $p_{max}$, although they are applicable in a modified form, which was shown in the example of $\tilde{p}_{max}$.

Second, two approaches for the modeling of \textit{avalanche} generation of runaway electrons were compared in chapter \ref{avalanche_chapter}, which provide analytic distribution functions based on the growth rates introduced in section \ref{Avalanche_subsection}. At that, the \textit{Rosenbluth-Putvinski} model with its two-dimensional distribution function as stated by \mbox{\textit{T.\hspace{0.9mm}Fülöp et \hspace{-0.4mm}al.}} \cite{REdistfuncderivation}, in contrast to the one-dimensional distribution function proposed by \textit{P.\hspace{0.9mm}Svensson} \cite{Svensson_2021} in the \textit{Hesslow} model, does not restore the effect of partial screening. Moreover, the runaway region was modeled with an infinite upper momentum boundary, while for the lower momentum boundary the possibilities $p_{c}$, $p_{c}^{\mathrm{scr}}$ and $p_{\star}$ were used, in order to ascertain how their relative deviations with respect to the singly-ionized deuterium and neon ion density combinations propagates into the final results of the calculated moments.
\\
Again, numerical calculation schemes, which allow the direct application of quadrature schemes, for the moments related to the density, the mean velocity and the kinetic energy density of runaway electrons were determined, in this case for \textit{avalanche} runaway electron distribution functions. For the \textit{Hesslow} model, they require a one-dimensional integration, whilst for \textit{Rosenbluth-Putvinski} model a two-dimensional integration is necessary \cite{study_thesis}. Furthermore, control criteria were defined for both of the models on the basis of the runaway electron density, which might be used to verify the accuracy of an implementation. This was presented in the \textsc{MATLAB}-scripts, that computed the moments for the derived calculation rules over a wide singly-ionized deuterium and neon ion density parameter space and four approximately logarithmically increasing electric field strengths. 
\\
By means of contour plots of the computed results and their relative deviation, a discussion and evaluation of the two models and the different lower boundaries for the runaway region in the one-dimensional momentum space was possible in the subsections \ref{comparison_avalanche_j_subsection} and \ref{comparison_avalanche_k_subsection}. A first understanding is, that relative deviations between the considered representations for the critical momentum are inherited to varying degrees to the final results. In detail, it was found, that the deviations in the lower boundary of the runaway region momentum are stronger suppressed for the mean velocity than for the kinetic energy density. In addition, the contour plots revealed, that the \textit{Rosenbluth-Putvinski} model, which is computationally more expensive due to its requirement of two-dimensional integration methods, is not able to resolve the effects of partial screening. Thus, this reconfirms, that the \textit{Hesslow} model is superior in terms of physical accuracy and runtime efficiency. Furthermore, a distinction can be made for this model concerning the utilization of the approximations of the effective critical momentum $p_{c}^{\mathrm{eff}}$. Regarding this, it was noticed, that the analytic relation $p_{c}^{\mathrm{scr}}$ leads to tolerable relative deviations below \mbox{$10\,\%$} from the more accurate calculations, which apply $p_{\star}$. Therefore, it might be applied in simulations instead of $p_{\star}$, if the saved runtime, from not calculating $p_{\star}$ as the root of the function $(\ref{func_p_c_eff_def})$, is found to be sufficient for the loss in accuracy. A final remark, deduced with the help of the study thesis \cite{study_thesis}, concerns the dependence of the three analysed moments on the electric field, which can be assessed as weaker than the influence of a changing electron temperature.

In summary, it can be asserted, that the presented calculation schemes were physically validated and compared against each other on the basis of relative deviations, so that it is possible to assess their physical precision and their level of computational efficiency. However, it has been repeatedly stated throughout the thesis, that an application in a self-consistent disruption simulation software like the DREAM-code is vital, in order to be able to fully evaluate the applicability of the computation rules and their total influence on the final results like for instance the runaway current strength or the time evolution of the electric field. 
\\
Due to this consideration, the implementation of the deduced calculation rules in the programming language \texttt{C} has been started by the DREAM-project group. At that, their goal is to achieve a global examination of the application and utility of moment-based calculations in the \textit{reduced kinetic modeling} approach. 
\\
Nevertheless, the presented computational methods can be ascribed an independent importance, since they allow the rapid investigation of physical quantities even over large parameter spaces. This has been evinced first in the evaluation of the moments for the \textit{avalanche} generation mechanism for different electric field strengths and various plasma density compositions and second in the analysis of the time- and electric field-dependent computed moment-related quantities for the \textit{hot-tail} generation of runaway electrons with and without an present impurity density. Ultimately, this allows to deduce, that the analysis of the moments of distribution functions yields valuable insight into the behaviour and validity of the applied assumptions.

\clearpage

%% file: Appendix.tex
\chapter{Appendix}

\section{Analytical calculations}

\subsection{Calculation of the integral $\textup{\normalfont{I}}_{n_{\mathrm{RE}}^{\mathrm{ht}}}$}\label{int_n_RE_ht_appendix_subsection}

The integral $\textup{I}_{n_{\mathrm{RE}}^{\mathrm{ht}}}$ is recapitulated from the equation $(\ref{n_RE_HT_iso_def})$ from subsection \ref{ht_n_iso_subsection}:\vspace*{-4mm} 
\begin{equation}\label{n_RE_HT_iso_RECAP}
\textup{I}_{n_{\mathrm{RE}}^{\mathrm{ht}}} =  \displaystyle{  \int\limits_{p=p_{1}}^{p_{2}}}\hspace{-0.5mm} p^2\hspace{-0.6mm}\cdot\hspace{-0.3mm}\exp{\hspace{-0.6mm}\left(\hspace{-0.7mm}-\dfrac{\left(p^3+3\hspace{-0.5mm}\cdot\hspace{-0.3mm}\textup{I}_{\tau_{rel}}(t)\right)^{\frac{2}{3}}}{p_{\mathrm{th},0}^2}\hspace{-0.5mm}\right)}  \mathrm{d}p \,.
\end{equation} 
\vspace*{-6.5mm}\\The subsequently defined \textit{substitution}:\vspace*{-3.5mm}
\begin{equation}\label{substitution_num_I_n_HT_app}
\varrho =\dfrac{\left(p^3+3\hspace{-0.5mm}\cdot\hspace{-0.3mm}\textup{I}_{\tau_{rel}}(t)\right)^{\frac{1}{3}}}{p_{\mathrm{th},0}}  \;\;;\;\;\dfrac{\mathrm{d}\varrho }{\mathrm{d}p}= \dfrac{p^2}{p_{\mathrm{th},0}}\cdot\left(p^3+3\hspace{-0.5mm}\cdot\hspace{-0.3mm}\textup{I}_{\tau_{rel}}(t)\right)^{-\frac{2}{3}} =\dfrac{p^2}{\varrho^2\cdot p_{\mathrm{th},0}^3}
\end{equation} 
\vspace*{-7.5mm}\\can now be used to rewrite the integral. Thus, one inserts the expressions from $(\ref{substitution_num_I_n_HT_app})$ into the definition $(\ref{n_RE_HT_iso_RECAP})$ of the integral, applies the method of the \textit{integration by parts} and receives:
\vspace*{-4.0mm} 
\begin{equation}\label{n_RE_HT_iso_1}
\begin{split}
\begin{gathered}
\textup{I}_{n_{\mathrm{RE}}^{\mathrm{ht}}} =   p_{\mathrm{th},0}^3\cdot\hspace{-3mm} \displaystyle{  \int\limits_{\varrho=\varrho(p_{1})}^{\varrho(p_{2})}}\hspace{-0.4mm} \varrho^2\hspace{-0.6mm}\cdot\hspace{-0.3mm}\textup{e}^{\,-\varrho^2}  \mathrm{d}\varrho\overset{{\scriptsize \underbrace{\textup{I.b.P.}}{}}}{=}  p_{\mathrm{th},0}^3\cdot\left(\left[-\dfrac{\varrho}{2}\cdot\textup{e}^{\,-\varrho^2}\right]_{\varrho(p_{1})}^{\varrho(p_{2})}+\frac{1}{2}\displaystyle{  \int\limits_{\varrho=\varrho(p_{1})}^{\varrho(p_{2})}}\hspace{-0.4mm} \textup{e}^{\,-\varrho^2}  \mathrm{d}\varrho\right)
\\[2pt]
= \frac{p_{\mathrm{th},0}^3}{2}\hspace{-0.3mm}\cdot\hspace{-0.3mm}\left(\hspace{-0.6mm}-\left[ \varrho \cdot\textup{e}^{\,-\varrho^2}\right]_{\varrho(p_{1})}^{\varrho(p_{2})}+\hspace{-0.3mm} \displaystyle{  \int\limits_{\varrho=0}^{\varrho(p_{2})}}\hspace{-0.5mm} \textup{e}^{\,-\varrho^2}  \mathrm{d}\varrho-\hspace{-0.3mm}\displaystyle{  \int\limits_{\varrho=0}^{\varrho(p_{1})}}\hspace{-0.5mm} \textup{e}^{\,-\varrho^2}  \mathrm{d}\varrho\right)  
\\
= \frac{p_{\mathrm{th},0}^3}{2}\hspace{-0.3mm}\cdot\hspace{-0.3mm}\left(-\left[ \varrho \cdot\textup{e}^{\,-\varrho^2}\right]_{\varrho(p_{1})}^{\varrho(p_{2})}+ \textup{erf}\left(\varrho(p_{2}) \right)-\textup{erf}\left(\varrho(p_{1}) \right)\right)
\\
= \frac{p_{\mathrm{th},0}^3}{2}\hspace{-0.3mm}\cdot\hspace{-0.3mm}\left(-\left[ \varrho \cdot\textup{e}^{\,-\varrho^2}\right]_{\varrho(p_{1})}^{\varrho(p_{2})}+ \left[ \textup{erf}\left(\varrho \right)\right]_{\varrho(p_{1})}^{\varrho(p_{2})}\right)=\frac{p_{\mathrm{th},0}^3}{2}\cdot \left[\textup{erf}\left(\varrho \right)-\varrho \cdot\textup{e}^{\,-\varrho^2}\right]_{\varrho(p_{1})}^{\varrho(p_{2})}\,.
\end{gathered}
\end{split} 
\end{equation} 
\vspace*{-4.0mm}\\Note, that the definition of the \textit{error function} \cite{helander}, from page \pageref{erf_label}, was used. 

The \textit{hot-tail} runaway electron density was defined in equation $(\ref{n_RE_HT_iso_def})$ and can now be written with the analytic result $(\ref{n_RE_HT_iso_1})$ of the integral $\textup{I}_{n_{\mathrm{RE}}^{\mathrm{ht}}}$, so that one obtains:\vspace*{-3.0mm} 
\begin{equation}\label{n_RE_HT_iso_app}
\begin{split}
\begin{gathered}
n_{\mathrm{RE}}^{\mathrm{ht}}(t) =  \dfrac{4\cdot n_{\mathrm{e}} }{\sqrt{\pi}\cdot p_{\mathrm{th},0}^3} \cdot\textup{I}_{n_{\mathrm{RE}}^{\mathrm{ht}}}=\dfrac{2\cdot n_{\mathrm{e}} }{\sqrt{\pi}  }  \cdot \left[\textup{erf}\left(\varrho \right)-\varrho \cdot\textup{e}^{\,-\varrho^2}\right]_{\varrho(p_{1})}^{\varrho(p_{2})}\,.
\end{gathered}
\end{split} 
\end{equation} 
\vspace*{-8.0mm}\\With regard to the representations of the isotropic runaway region in $(\ref{CH_RE_region})$ and $(\ref{H_RE_region})$, one can simplify the result from $(\ref{n_RE_HT_iso_analyt_sol})$ in the limit \mbox{$p_{2}\rightarrow\infty$}. The analysis of the relation $(\ref{n_RE_HT_iso_analyt_variable})$ leads to the insight, that \mbox{$\varrho(p_{2}\rightarrow\infty)\rightarrow\infty$}. Furthermore, it holds, that \mbox{$\textup{erf}(x\rightarrow\infty)\rightarrow 1$} \cite{WolframERFC} and that \mbox{$\textup{erfc}(x)=1-\textup{erf}(x)$} is the \textit{complementary error function} \cite{WolframERFC}, which was also introduced on page \pageref{erf_label}. Therefore, one can state a rewritten version of the analytic expression from $(\ref{n_RE_HT_iso_app})$:\vspace*{-3.0mm} 
\begin{equation}\label{n_RE_HT_iso_analyt_sol_inf_app}
\begin{split}
n_{\mathrm{RE}}^{\mathrm{ht}}(t) \overset{{\scriptsize \underbrace{p_{2}\rightarrow\infty}{}}}{=}\;&\dfrac{2\cdot n_{\mathrm{e}} }{\sqrt{\pi} } \cdot\left(\dfrac{\sqrt{\pi}}{2}\cdot\left(\lim \limits_{p_{2} \to \infty}\left\lbrace\textup{erf}\left(p_{2}\right)\right\rbrace-\textup{erf}\left(\varrho(p_{1})\right)\right)\right.
\\[-1pt]
\;&-\left.\left(\lim \limits_{p_{2} \to \infty}\left\lbrace\varrho(p_{2})\cdot\textup{e}^{-\left(\varrho(p_{2})\right)^2}\right\rbrace-\varrho(p_{1})\cdot\textup{e}^{-\left(\varrho(p_{1})\right)^2}\right)\right)
\\[-1pt]
= \hspace{4.3mm}&\dfrac{2\cdot n_{\mathrm{e}} }{\sqrt{\pi} } \cdot\left(\dfrac{\sqrt{\pi}}{2}\cdot\left(1-\textup{erf}\left(\varrho(p_{1})\right)\right)- \left(0-\varrho(p_{1})\cdot\textup{e}^{-\left(\varrho(p_{1})\right)^2}\right)\right)
\\[3pt]
= \hspace{4.3mm}&\dfrac{2\cdot n_{\mathrm{e}} }{\sqrt{\pi} } \cdot\left(\dfrac{\sqrt{\pi}}{2}\cdot\textup{erfc}\left(\varrho(p_{1})\right)+\varrho(p_{1})\cdot\textup{e}^{-\left(\varrho(p_{1})\right)^2}\right)\,.
\end{split} 
\end{equation}

\clearpage

\section{Numerical calculations}

\subsection{Console outputs of the utilized \textsc{MATLAB}-scripts}\label{output_matlab_appendix_subsection}

\vspace*{2mm}
\begin{lstlisting}[language=Matlab,keywordstyle=\empty,frame=single, caption={Output\protect\footnotemark{} of the \textsc{MATLAB}-script \qq{\texttt{RE_ht_moments_SV.m}} \vspace{1mm}},mathescape=true,label={outMATLABoutput_RE_ht_moments_SV} ]
set of parameters:

B = 5.30 T; Z_eff = 1.00;
n_e = 1.06e+20 m^-3; n_D2 = 1.06e+20 m^-3; n_Ne = 0.00e+00 m^-3;

Smith-Verwichte-model [-1<xi<1; p_c<p<inf; E>E_c]:

      n_RE_ht [m^-3]  n_RE_ht_num [m^-3]  rel_n_RE_ht [$\%$]    u_RE_ht/c    k_RE_ht/c^2
min    2.6025e-256       2.6025e-256       -1.7001e-11       0.37935       0.094117    
max     5.3702e+14        5.3702e+14          2.85e-12       0.98803         5.4814    
mean    1.6473e+13        1.6473e+13        2.0445e-14       0.55524         0.3183    

                              min                 max          mean
j_RE_ht [A/m^2]          1.2351e-266             11876        350.23
  
Smith-Verwichte-model [-1<xi<1; p_star<p<inf; E>E_c_eff]:

      n_RE_ht [m^-3]  n_RE_ht_num [m^-3]  rel_n_RE_ht [$\%$]    u_RE_ht/c    k_RE_ht/c^2
min      0.0030472         0.0030472       -1.4651e-12        0.3804       0.094575    
max     4.4482e+14        4.4482e+14        1.6557e-12       0.69032        0.38464    
mean     1.815e+13         1.815e+13         6.287e-14        0.4921         0.1749    

                              min                 max          mean
j_RE_ht [A/m^2]          1.0104e-13              10038        395.78     

Smith-Verwichte-model [-1<xi<1; p_min<p<p_max; E>E_c_eff]:

      n_RE_ht [m^-3]  n_RE_ht_num [m^-3]  rel_n_RE_ht [$\%$]    u_RE_ht/c    k_RE_ht/c^2
min              0         3.103e-57           -9.0153       0.38151       0.095189     
max     2.7144e+14        2.7144e+14           0.68421       0.95521         2.3792     
mean     1.035e+13         1.035e+13         -0.028201       0.55363        0.32783     

                              min                max           mean
j_RE_ht [A/m^2]                0                6291.7        220.86     

Smith-Verwichte-model [-1<xi<1; p_min<p<tilde_p_max; E>E_c_eff]:

      n_RE_ht [m^-3]  n_RE_ht_num [m^-3]  rel_n_RE_ht [$\%$]    u_RE_ht/c    k_RE_ht/c^2
min              0         3.103e-57           -9.0153       0.38151       0.095189    
max     2.7144e+14        2.7144e+14           0.68421       0.95521         2.3792    
mean     1.035e+13         1.035e+13         -0.028201       0.55363        0.32783    

                              min                max           mean
j_RE_ht [A/m^2]                0                6291.7        220.86     
\end{lstlisting}
\footnotetext{\label{footnote_output_RE_ht_moments_SV} Stored in the file \qq{\texttt{output_RE_ht_moments_SV.txt}} in the digital appendix.}
\noindent\newpage\noindent
\begin{lstlisting}[language=Matlab,keywordstyle=\empty,frame=single, caption={Output\protect\footnotemark{} of the \textsc{MATLAB}-script \qq{\texttt{RE_ht_moments_SV_imp.m}} \vspace{1mm}},mathescape=true,label={outMATLABoutput_RE_ht_moments_SV_imp} ]
set of parameters:

B = 5.30 T; Z_eff = 1.00;
n_e = 3.46e+20 m^-3; n_D2 = 1.06e+20 m^-3; n_Ne = 2.40e+20 m^-3;

Smith-Verwichte-model [-1<xi<1; p_c<p<inf; E>E_c]:

      n_RE_ht [m^-3]  n_RE_ht_num [m^-3]  rel_n_RE_ht [$\%$]    u_RE_ht/c    k_RE_ht/c^2
min              0                 0       -7.8409e-12       0.42093        0.11986  
max     9.6527e+07        9.6527e+07        6.3506e-12       0.98429         4.6644  
mean     2.506e+06         2.506e+06        1.4979e-13       0.57679        0.39709  

     j_RE_ht [A/m^2]
min     8.359e-211    
max      0.0027081 
mean    6.4743e-05 

Smith-Verwichte-model [-1<xi<1; p_star<p<inf; E>E_c_eff]:

      n_RE_ht [m^-3]  n_RE_ht_num [m^-3]  rel_n_RE_ht [$\%$]    u_RE_ht/c    k_RE_ht/c^2
min     1.5193e-10        1.5193e-10       -3.0944e-12       0.42902         0.1246  
max     1.1873e+06        1.1873e+06        2.8583e-12       0.73651        0.48036  
mean         89740             89740        8.8606e-14        0.5423        0.22325  

     j_RE_ht [A/m^2]
min     5.3748e-21    
max     3.2528e-05 
mean    2.3938e-06 

Smith-Verwichte-model [-1<xi<1; p_min<p<p_max; E>E_c_eff]:

      n_RE_ht [m^-3]  n_RE_ht_num [m^-3]  rel_n_RE_ht [$\%$]    u_RE_ht/c    k_RE_ht/c^2
min              0       3.8078e-161          -0.22219       0.43136        0.12598 
max     4.3108e+05        4.3108e+05            17.238       0.98123         4.1857 
mean         24526             24526           0.16999       0.62712        0.58401 
   
     j_RE_ht [A/m^2]
min              0    
max     1.1535e-05 
mean    6.3388e-07  

Smith-Verwichte-model [-1<xi<1; p_min<p<tilde_p_max; E>E_c_eff]:

      n_RE_ht [m^-3]  n_RE_ht_num [m^-3]  rel_n_RE_ht [$\%$]    u_RE_ht/c    k_RE_ht/c^2
min              0       3.8078e-161          -0.22219       0.43136        0.12598  
max     4.3108e+05        4.3108e+05            17.238       0.98123         4.1857  
mean         24526             24526           0.16999       0.62712        0.58401  

     j_RE_ht [A/m^2]
min              0    
max     1.1535e-05 
mean    6.3388e-07 
\end{lstlisting}
\footnotetext{\label{footnote_output_RE_ht_moments_SV_imp} Stored in the file \qq{\texttt{output_RE_ht_moments_SV_imp.txt}} in the digital appendix.}
\noindent\newpage\noindent
\begin{lstlisting}[language=Matlab,keywordstyle=\empty,frame=single, caption={Output\protect\footnotemark{} of the \textsc{MATLAB}-script \qq{\texttt{RE_ht_moments_SV_sep.m}} \vspace{1mm}},mathescape=true,label={outMATLABoutput_RE_ht_moments_SV_sep} ]
set of parameters:

B = 5.30 T; Z_eff = 1.00;
n_e = 1.06e+20 m^-3; n_D2 = 1.06e+20 m^-3; n_Ne = 0.00e+00 m^-3;

Smith-Verwichte-model [xi_sep(p,E_c)<xi<1; p_c<p<inf; E>E_c]:

      n_RE_ht_num [m^-3]  rel_n_RE_ht_sep [$\%$]       u_RE_ht/c    rel_u_RE_ht_sep [$\%$] 
min        2.5406e-256           -59.768             0.38523         7.1053e-08              
max         2.6282e+14          -0.96409             0.98803             6.7239                   
mean        9.5051e+12            -31.07             0.56668             2.4932             

           k_RE_ht/c^2    rel_k_RE_ht_sep [$\%$]   j_RE_ht [A/m^2]
min           0.097368        3.4525e-06         1.2057e-266
max             5.4814            14.903              5861.1
mean           0.32666            5.5171              205.68
 
Smith-Verwichte-model [xi_sep(p,E_c_eff)<xi<1; p_star<p<inf; E>E_c_eff]:

      n_RE_ht_num [m^-3]  rel_n_RE_ht_sep [$\%$]       u_RE_ht/c    rel_u_RE_ht_sep [$\%$] 
min        -6.4508e+12           -112.01             -5.2478            -959.81  
max         1.2344e+14           -1.6929              4.7117             674.23                
mean        6.0429e+12           -57.187              0.5244             7.1853               

           k_RE_ht/c^2    rel_k_RE_ht_sep [$\%$]   j_RE_ht [A/m^2]
min            -7.6182           -2954.9             -166.53
max              5.744            2069.8              2957.1
mean            0.1916            16.029              132.57

Smith-Verwichte-model [xi_sep(p,E_c_eff)<xi<1; p_min<p<p_max; E>E_c_eff]:

      n_RE_ht_num [m^-3]  rel_n_RE_ht_sep [$\%$]       u_RE_ht/c    rel_u_RE_ht_sep [$\%$] 
min         2.8156e-57           -57.811             0.38757         1.8499e-05             
max         1.2541e+14           -1.6691             0.95521             6.2907           
mean        6.1689e+12           -32.615             0.56531             2.5629

           k_RE_ht/c^2    rel_k_RE_ht_sep [$\%$]   j_RE_ht [A/m^2]
min           0.098561        0.00027619          1.2918e-67   
max             2.3792            13.832              2967.2  
mean           0.33628            5.6842              133.77

Smith-Verwichte-model [xi_sep(p,E_c_eff)<xi<1; p_min<p<tilde_p_max; E>E_c_eff]:

      n_RE_ht_num [m^-3]  rel_n_RE_ht_sep [$\%$]        u_RE_ht/c    rel_u_RE_ht_sep [$\%$] 
min         2.8156e-57           -57.811             0.38757         1.8499e-05          
max         1.2541e+14           -1.6691             0.95521             6.2907            
mean        6.1689e+12           -32.615             0.56581             2.5629 

           k_RE_ht/c^2    rel_k_RE_ht_sep [$\%$]   j_RE_ht [A/m^2]
min           0.098561        0.00027619          1.2918e-67
max             2.3792            13.832              2967.2
mean           0.33628            5.6842              133.77
\end{lstlisting}
\footnotetext{\label{footnote_output_RE_ht_moments_SV_sep} Stored in the file \qq{\texttt{output_RE_ht_moments_SV_sep.txt}} in the digital appendix.}
\noindent\newpage\noindent
\begin{lstlisting}[language=Matlab,keywordstyle=\empty,frame=single, caption={Output\protect\footnotemark{} of the \textsc{MATLAB}-script \qq{\texttt{RE_ht_moments_SV_sep_imp.m}} \vspace{1mm}},mathescape=true,label={outMATLABoutput_RE_ht_moments_SV_sep_imp} ]
set of parameters:

B = 5.30 T; Z_eff = 1.00;
n_e = 3.46e+20 m^-3; n_D2 = 1.06e+20 m^-3; n_Ne = 2.40e+20 m^-3;

Smith-Verwichte-model [xi_sep(p,E_c)<xi<1; p_c<p<inf; E>E_c]:

      n_RE_ht_num [m^-3]  rel_n_RE_ht_sep [$\%$]        u_RE_ht/c    rel_u_RE_ht_sep [$\%$] 
min                  0           -56.174             0.42785         2.2506e-07                
max         4.2304e+07           -2.2692             0.98429              6.179                 
mean        1.2435e+06           -29.614             0.58947             2.5849                

           k_RE_ht/c^2    rel_k_RE_ht_sep [$\%$]   j_RE_ht [A/m^2]
min            0.12425        8.5219e-06         8.0985e-211
max             4.6644            13.969           0.0012053
mean           0.40744            5.8641          3.2773e-05

Smith-Verwichte-model [xi_sep(p,E_c_eff)<xi<1; p_star<p<inf; E>E_c_eff]:

      n_RE_ht_num [m^-3]  rel_n_RE_ht_sep [$\%$]        u_RE_ht/c    rel_u_RE_ht_sep [$\%$] 
min         2.4638e-11            -85.52              0.4514            0.74739            
max         2.9719e+05             -16.3             0.74393             10.128                  
mean             24818           -56.831             0.57268             6.0169                   

           k_RE_ht/c^2    rel_k_RE_ht_sep [$\%$]   j_RE_ht [A/m^2]
min            0.13664            2.8675          8.8039e-22 
max            0.49932            24.035          8.2304e-06
mean           0.25168            14.528          6.7867e-07

Smith-Verwichte-model [xi_sep(p,E_c_eff)<xi<1; p_min<p<p_max; E>E_c_eff]:

      n_RE_ht_num [m^-3]  rel_n_RE_ht_sep [$\%$]        u_RE_ht/c    rel_u_RE_ht_sep [$\%$] 
min        3.6622e-161           -48.283             0.45044         4.7221e-07                 
max         2.2294e+05           -6.0108             0.98123             6.0243              
mean             13632           -32.869             0.63902             2.5275             

           k_RE_ht/c^2    rel_k_RE_ht_sep [$\%$]   j_RE_ht [A/m^2]
min            0.13626        1.5203e-05          1.726e-171
max             4.1857            13.106          6.0804e-06
mean            0.5928            5.5458          3.5986e-07

Smith-Verwichte-model [xi_sep(p,E_c_eff)<xi<1; p_min<p<tilde_p_max; E>E_c_eff]:

      n_RE_ht_num [m^-3]  rel_n_RE_ht_sep [$\%$]        u_RE_ht/c    rel_u_RE_ht_sep [$\%$] 
min        3.6622e-161           -48.283             0.45044         4.7221e-07           
max         2.2294e+05           -6.0108             0.98123             6.0243            
mean             13632           -32.869             0.63902             2.5275           

           k_RE_ht/c^2    rel_k_RE_ht_sep [$\%$]   j_RE_ht [A/m^2]
min            0.13626        1.5203e-05          1.726e-171  
max             4.1857            13.106          6.0804e-06  
mean            0.5928            5.5458          3.5986e-07
\end{lstlisting}
\footnotetext{\label{footnote_output_RE_ht_moments_SV_sep_imp} Stored in the file \qq{\texttt{output_RE_ht_moments_SV_sep_imp.txt}} in the digital appendix.}
\noindent\newpage\noindent
\begin{lstlisting}[language=Matlab,keywordstyle=\empty,frame=single, caption={Output\protect\footnotemark{} of the \textsc{MATLAB}-script \qq{\texttt{RE_ava_dist_func_RP.m}} \vspace{1mm}},mathescape=true,label={outMATLABoutput_RE_ava_dist_func_RP} ]
set of parameters:

E = 10.00 V/m; n_e = 1.00e+20 m^-3;
k_B*T_e = 100.00 eV; Z_eff = 1.00;

calculated quantities:

lnLambda_rel = 16.903; tau_rel = 0.020 s;
E_c = 0.086 V/m; EoverEc = 116.025; p_c = 0.093;
E_D = 440.420 V/m; E_sa = 94.250 V/m 
\end{lstlisting}
\footnotetext{\label{footnote_output_RE_ava_dist_func_RP} Stored in the file \qq{\texttt{output_RE_ava_dist_func_RP.txt}} in the digital appendix.}
\vspace*{5mm}
\begin{lstlisting}[language=Matlab,keywordstyle=\empty,frame=single, caption={Output\protect\footnotemark{} of the \textsc{MATLAB}-script \qq{\texttt{RE_ava_dist_func_H.m}} \vspace{1mm}},mathescape=true,label={outMATLABoutput_RE_ava_dist_func_H} ]
set of parameters:

B = 5.25 T; n_e = 1.00e+20 m^-3;
k_B*T_e = 100 eV; Z_eff = 1.0;

calculated quantities:

lnLambda_rel = 16.903;
E = 0.153 V/m; E_c = 0.086 V/m; E_c_eff = 0.151 V/m;
EoverEc = 1.775; EoverEceff = 1.012; 
E_D = 440.420 V/m; E_sa = 94.250 V/m 
p_c = 1.136; p_star = 0.891;

lnLambda_rel = 16.903;
E = 1.000 V/m; E_c = 0.086 V/m; E_c_eff = 0.151 V/m;
EoverEc = 11.603; EoverEceff = 6.612; 
E_D = 440.420 V/m; E_sa = 94.250 V/m 
p_c = 0.307; p_star = 0.344;

lnLambda_rel = 16.903;
E = 10.000 V/m; E_c = 0.086 V/m; E_c_eff = 0.151 V/m;
EoverEc = 116.025; EoverEceff = 66.119; 
E_D = 440.420 V/m; E_sa = 94.250 V/m 
p_c = 0.093; p_star = 0.107;

lnLambda_rel = 16.903;
E = 100.000 V/m; E_c = 0.086 V/m; E_c_eff = 0.151 V/m;
EoverEc = 1160.255; EoverEceff = 661.193; 
E_D = 440.420 V/m; E_sa = 94.250 V/m 
p_c = 0.029; p_star = 0.033;
\end{lstlisting}
\footnotetext{\label{footnote_output_RE_ava_dist_func_H} Stored in the file \qq{\texttt{output_RE_ava_dist_func_H.txt}} in the digital appendix.}
\noindent\newpage\noindent
\begin{lstlisting}[language=Matlab,keywordstyle=\empty,frame=single, caption={Output\protect\footnotemark{} of the \textsc{MATLAB}-script\\\qq{\texttt{plot_num_data_densities_p_c_scr_E3.m}} \vspace{1mm}},mathescape=true,label={MATLABoutput_plot_p_scr_E3} ]
set of parameters:  

B = 5.25 T; E = 3.0 V/m; k_BT_e = 10 eV; Z_eff = 1.00;

numerical data:

min_n_RE_ava_screen_check = 1.00; max_n_RE_ava_screen_check = 1.00; 
mean_n_RE_ava_screen_check = 1.00;

min_n_RE_ava_check = 1.00; max_n_RE_ava_check = 1.00; 
mean_n_RE_ava_check = 1.00;

min_u_RE_ava_over_c_screen = 0.903361; max_u_RE_ava_over_c_screen = 0.999804;

min_u_RE_ava_over_c = 0.966059; max_u_RE_ava_over_c = 0.979540;

min_rel_u = -7.75$\,\%$; max_rel_u = 3.49$\,\%$; mean_rel_u = -1.38$\,\%$;

min_k_RE_ava_over_csq_screen = 7.486815; max_k_RE_ava_over_csq_screen = 62.796057;

min_k_RE_ava_over_csq = 34.922911; max_k_RE_ava_over_csq = 45.579139;

min_rel_k = -83.29$\,\%$; max_rel_k = 79.81$\,\%$; mean_rel_k = -46.46$\,\%$;

min_rel_E_c_tot = 12.28$\,\%$; max_rel_E_c_tot = 367.55$\,\%$; mean_rel_E_c_tot = 97.40$\,\%$;

min_rel_E_c_eff = 54.55$\,\%$; max_rel_E_c_eff = 2824.89$\,\%$; mean_rel_E_c_eff = 882.48$\,\%$;

min_rel_p_c_scr = 27.41$\,\%$; max_rel_p_c_scr = 7358.51$\,\%$;
mean_rel_p_c_scr = 227.62$\,\%$;
\end{lstlisting}
\footnotetext{\label{footnote_output_plot_p_scr_E3} Stored in the file \qq{\texttt{output_plot_num_data_densities_p_c_scr_E3.txt}} in the digital\\ \hspace*{8.7mm}appendix.}
\vspace*{5mm}
\begin{lstlisting}[language=Matlab,keywordstyle=\empty,frame=single, caption={Output\protect\footnotemark{} of the \textsc{MATLAB}-script\\\qq{\texttt{plot_num_data_densities_p_c_scr_E10.m}} \vspace{1mm}},mathescape=true,label={MATLABoutput_plot_p_scr_E10} ]
set of parameters:  

B = 5.25 T; E = 10.0 V/m; k_BT_e = 10 eV; Z_eff = 1.00;

numerical data:

min_n_RE_ava_screen_check = 1.00; max_n_RE_ava_screen_check = 1.00; 
mean_n_RE_ava_screen_check = 1.00;

min_n_RE_ava_check = 1.00; max_n_RE_ava_check = 1.00; 
mean_n_RE_ava_check = 1.00;

min_u_RE_ava_over_c_screen = 0.897237; max_u_RE_ava_over_c_screen = 0.999707;

min_u_RE_ava_over_c = 0.964701; max_u_RE_ava_over_c = 0.978876;

min_rel_u = -8.31$\,\%$; max_rel_u = 3.63$\,\%$; mean_rel_u = -1.47$\,\%$;

min_k_RE_ava_over_csq_screen = 6.953834; max_k_RE_ava_over_csq_screen = 54.221278;

min_k_RE_ava_over_csq = 33.411872; max_k_RE_ava_over_csq = 44.057929;

min_rel_k = -83.94$\,\%$; max_rel_k = 62.28$\,\%$; mean_rel_k = -46.25$\,\%$;

min_rel_E_c_tot = 9.13$\,\%$; max_rel_E_c_tot = 262.83$\,\%$; mean_rel_E_c_tot = 64.58$\,\%$;

min_rel_E_c_eff = 37.60$\,\%$; max_rel_E_c_eff = 2188.64$\,\%$; mean_rel_E_c_eff = 728.70$\,\%$;

min_rel_p_c_scr = 19.35$\,\%$; max_rel_p_c_scr = 5554.99$\,\%$; 
mean_rel_p_c_scr = 201.21$\,\%$;
\end{lstlisting}
\footnotetext{\label{footnote_output_plot_p_scr_E10} Stored in the file \qq{\texttt{output_plot_num_data_densities_p_c_scr_E10.txt}} in the digital\\ \hspace*{8.7mm}appendix.}
\vspace{5mm}
\begin{lstlisting}[language=Matlab,keywordstyle=\empty,frame=single, caption={Output\protect\footnotemark{} of the \textsc{MATLAB}-script\\\qq{\texttt{plot_num_data_densities_p_c_scr_E30.m}} \vspace{1mm}},mathescape=true,label={MATLABoutput_plot_p_scr_E30} ]
set of parameters:  

B = 5.25 T; E = 30.0 V/m; k_BT_e = 10 eV; Z_eff = 1.00;

numerical data:

min_n_RE_ava_screen_check = 1.00; max_n_RE_ava_screen_check = 1.00; 
mean_n_RE_ava_screen_check = 1.00;

min_n_RE_ava_check = 1.00; max_n_RE_ava_check = 1.00; 
mean_n_RE_ava_check = 1.00;

min_u_RE_ava_over_c_screen = 0.890703; max_u_RE_ava_over_c_screen = 0.999774;

min_u_RE_ava_over_c = 0.963638; max_u_RE_ava_over_c = 0.978249;

min_rel_u = -8.92$\,\%$; max_rel_u = 3.75$\,\%$; mean_rel_u = -1.56$\,\%$;

min_k_RE_ava_over_csq_screen = 6.447293; max_k_RE_ava_over_csq_screen = 59.340833;

min_k_RE_ava_over_csq = 32.057708; max_k_RE_ava_over_csq = 42.708040;

min_rel_k = -84.63$\,\%$; max_rel_k = 85.11$\,\%$; mean_rel_k = -46.00$\,\%$;

min_rel_E_c_tot = 9.11$\,\%$; max_rel_E_c_tot = 192.08$\,\%$; mean_rel_E_c_tot = 44.65$\,\%$;

min_rel_E_c_eff = 28.58$\,\%$; max_rel_E_c_eff = 1758.45$\,\%$; mean_rel_E_c_eff = 638.77$\,\%$;

min_rel_p_c_scr = 14.87$\,\%$; max_rel_p_c_scr = 6670.54$\,\%$; 
mean_rel_p_c_scr = 185.39$\,\%$;
\end{lstlisting}
\footnotetext{\label{footnote_output_plot_p_scr_E30} Stored in the file \qq{\texttt{output_plot_num_data_densities_p_c_scr_E30.txt}} in the digital\\ \hspace*{8.7mm}appendix.}
\noindent\newpage\noindent
\begin{lstlisting}[language=Matlab,keywordstyle=\empty,frame=single, caption={Output\protect\footnotemark{} of the \textsc{MATLAB}-script\\\qq{\texttt{plot_num_data_densities_p_c_scr_E100.m}} \vspace{1mm}},mathescape=true,label={MATLABoutput_plot_p_scr_E100} ]
set of parameters:  

B = 5.25 T; E = 100.0 V/m; k_BT_e = 10 eV; Z_eff = 1.00;

numerical data:

min_n_RE_ava_screen_check = 1.00; max_n_RE_ava_screen_check = 1.00; 
mean_n_RE_ava_screen_check = 1.00;

min_n_RE_ava_check = 1.00; max_n_RE_ava_check = 1.00; 
mean_n_RE_ava_check = 1.00;

min_u_RE_ava_over_c_screen = 0.832187; max_u_RE_ava_over_c_screen = 0.999684;

min_u_RE_ava_over_c = 0.962520; max_u_RE_ava_over_c = 0.980985;

min_rel_u = -14.93$\,\%$; max_rel_u = 3.86$\,\%$; mean_rel_u = -2.30$\,\%$;

min_k_RE_ava_over_csq_screen = 3.652350; max_k_RE_ava_over_csq_screen = 52.586091;

min_k_RE_ava_over_csq = 30.586835; max_k_RE_ava_over_csq = 44.002389;

min_rel_k = -91.47$\,\%$; max_rel_k = 71.92$\,\%$; mean_rel_k = -46.05$\,\%$;

min_rel_E_c_tot = 9.11$\,\%$; max_rel_E_c_tot = 259.55$\,\%$; mean_rel_E_c_tot = 47.22$\,\%$;

min_rel_E_c_eff = 23.23$\,\%$; max_rel_E_c_eff = 2168.73$\,\%$; mean_rel_E_c_eff = 650.79$\,\%$;

min_rel_p_c_scr = 12.14$\,\%$; max_rel_p_c_scr = 5440.71$\,\%$; 
mean_rel_p_c_scr = 179.13$\,\%$;
\end{lstlisting}
\footnotetext{\label{footnote_output_plot_p_scr_E100} Stored in the file \qq{\texttt{output_plot_num_data_densities_p_c_scr_E100.txt}} in the\\ \hspace*{8.7mm}digital appendix.}
\vspace*{5mm}
\begin{lstlisting}[language=Matlab,keywordstyle=\empty,frame=single, caption={Output\protect\footnotemark{} of the \textsc{MATLAB}-script\\\qq{\texttt{plot_num_data_densities_p_star_E3.m}} \vspace{1mm}},mathescape=true,label={MATLABoutput_plot_p_star_E3} ]
set of parameters:  

B = 5.25 T; E = 3.0 V/m; k_BT_e = 10 eV; Z_eff = 1.00;

numerical data:

min_n_RE_ava_screen_check = 1.00; max_n_RE_ava_screen_check = 1.00; 
mean_n_RE_ava_screen_check = 1.00;

min_n_RE_ava_check = 1.00; max_n_RE_ava_check = 1.00; 
mean_n_RE_ava_check = 1.00;

min_u_RE_ava_over_c_screen = 0.906527; max_u_RE_ava_over_c_screen = 0.982909;

min_u_RE_ava_over_c = 0.966119; max_u_RE_ava_over_c = 0.977964;

min_rel_u = -7.29$\,\%$; max_rel_u = 1.74$\,\%$; mean_rel_u = -1.40$\,\%$;

min_k_RE_ava_over_csq_screen = 7.810221; max_k_RE_ava_over_csq_screen = 35.836320;

min_k_RE_ava_over_csq = 34.929245; max_k_RE_ava_over_csq = 42.152145;

min_rel_k = -81.31$\,\%$; max_rel_k = 0.34$\,\%$; mean_rel_k = -50.37$\,\%$;

min_rel_E_c_tot = 12.33$\,\%$; max_rel_E_c_tot = 167.71$\,\%$; mean_rel_E_c_tot = 59.31$\,\%$;

min_rel_E_c_eff = 56.57$\,\%$; max_rel_E_c_eff = 1610.87$\,\%$; mean_rel_E_c_eff = 714.79$\,\%$;

min_rel_p_star = 12.62$\,\%$; max_rel_p_star = 233.37$\,\%$; mean_rel_p_star = 109.33$\,\%$;

min_tilde_rel_p_c_scr = 13.96$\,\%$; max_tilde_rel_p_c_scr = 942.61$\,\%$; 
mean_tilde_rel_p_c_scr = 50.68$\,\%$;

min_tilde_rel_u_p_c_scr = -0.61$\,\%$; max_tilde_rel_u_p_c_scr = 1.72$\,\%$; 
mean_tilde_rel_u_p_c_scr = 0.13$\,\%$;

min_tilde_rel_k_p_c_scr = -7.13$\,\%$; max_tilde_rel_k_p_c_scr = 215.93$\,\%$; 
mean_tilde_rel_k_p_c_scr = 8.10$\,\%$;
\end{lstlisting}
\footnotetext{\label{footnote_output_plot_p_star_E3} Stored in the file \qq{\texttt{output_plot_num_data_densities_p_star_E3.txt}} in the digital\\ \hspace*{8.7mm}appendix.}
\vspace*{5mm}
\begin{lstlisting}[language=Matlab,keywordstyle=\empty,frame=single, caption={Output\protect\footnotemark{} of the \textsc{MATLAB}-script\\\qq{\texttt{plot_num_data_densities_p_star_E10.m}} \vspace{1mm}},mathescape=true,label={MATLABoutput_plot_p_star_E10} ]
set of parameters:  

B = 5.25 T; E = 10.0 V/m; k_BT_e = 10 eV; Z_eff = 1.00;

numerical data:

min_n_RE_ava_screen_check = 1.00; max_n_RE_ava_screen_check = 1.00; 
mean_n_RE_ava_screen_check = 1.00;

min_n_RE_ava_check = 1.00; max_n_RE_ava_check = 1.00; 
mean_n_RE_ava_check = 1.00;

min_u_RE_ava_over_c_screen = 0.897057; max_u_RE_ava_over_c_screen = 0.983271;

min_u_RE_ava_over_c = 0.964750; max_u_RE_ava_over_c = 0.977415;

min_rel_u = -8.21$\,\%$; max_rel_u = 1.92$\,\%$; mean_rel_u = -1.44$\,\%$;

min_k_RE_ava_over_csq_screen = 6.968900; max_k_RE_ava_over_csq_screen = 34.633435;

min_k_RE_ava_over_csq = 33.416860; max_k_RE_ava_over_csq = 41.050472;

min_rel_k = -82.88$\,\%$; max_rel_k = 0.69$\,\%$; mean_rel_k = -49.30$\,\%$;

min_rel_E_c_tot = 9.14$\,\%$; max_rel_E_c_tot = 126.47$\,\%$; mean_rel_E_c_tot = 40.12$\,\%$;

min_rel_E_c_eff = 38.85$\,\%$; max_rel_E_c_eff = 1363.82$\,\%$; mean_rel_E_c_eff = 622.97$\,\%$;

min_rel_p_star = 12.25$\,\%$; max_rel_p_star = 240.57$\,\%$; mean_rel_p_star = 110.31$\,\%$;

min_tilde_rel_p_c_scr = 6.77$\,\%$; max_tilde_rel_p_c_scr = 945.45$\,\%$; 
mean_tilde_rel_p_c_scr = 39.53$\,\%$;

min_tilde_rel_u_p_c_scr = -0.62$\,\%$; max_tilde_rel_u_p_c_scr = 1.67$\,\%$; 
mean_tilde_rel_u_p_c_scr = 0.06$\,\%$;

min_tilde_rel_k_p_c_scr = -7.33$\,\%$; max_tilde_rel_k_p_c_scr = 174.01$\,\%$; 
mean_tilde_rel_k_p_c_scr = 6.24$\,\%$;
\end{lstlisting}
\footnotetext{\label{footnote_output_plot_p_star_E10} Stored in the file \qq{\texttt{output_plot_num_data_densities_p_star_E10.txt}} in the digital\\ \hspace*{8.7mm}appendix.}
\vspace{5mm}
\begin{lstlisting}[language=Matlab,keywordstyle=\empty,frame=single, caption={Output\protect\footnotemark{} of the \textsc{MATLAB}-script\\\qq{\texttt{plot_num_data_densities_p_star_E30.m}} \vspace{1mm}},mathescape=true,label={MATLABoutput_plot_p_star_E30} ]
set of parameters:  

B = 5.25 T; E = 30.0 V/m; k_BT_e = 10 eV; Z_eff = 1.00;

numerical data:

min_n_RE_ava_screen_check = 1.00; max_n_RE_ava_screen_check = 1.00; 
mean_n_RE_ava_screen_check = 1.00;

min_n_RE_ava_check = 1.00; max_n_RE_ava_check = 1.00; 
mean_n_RE_ava_check = 1.00;

min_u_RE_ava_over_c_screen = 0.889168; max_u_RE_ava_over_c_screen = 0.983308;

min_u_RE_ava_over_c = 0.963688; max_u_RE_ava_over_c = 0.981224;

min_rel_u = -8.96$\,\%$; max_rel_u = 2.04$\,\%$; mean_rel_u = -1.47$\,\%$;

min_k_RE_ava_over_csq_screen = 6.368502; max_k_RE_ava_over_csq_screen = 33.455638;

min_k_RE_ava_over_csq = 32.062867; max_k_RE_ava_over_csq = 39.927134;

min_rel_k = -83.92$\,\%$; max_rel_k = 0.81$\,\%$; mean_rel_k = -48.49$\,\%$;

min_rel_E_c_tot = 9.12$\,\%$; max_rel_E_c_tot = 92.49$\,\%$; mean_rel_E_c_tot = 28.33$\,\%$;

min_rel_E_c_eff = 29.56$\,\%$; max_rel_E_c_eff = 1167.91$\,\%$; mean_rel_E_c_eff = 570.47$\,\%$;

min_rel_p_star = 12.13$\,\%$; max_rel_p_star = 247.20$\,\%$; mean_rel_p_star = 111.72$\,\%$;

min_tilde_rel_p_c_scr = 2.50$\,\%$; max_tilde_rel_p_c_scr = 946.96$\,\%$; 
mean_tilde_rel_p_c_scr = 32.47$\,\%$;

min_tilde_rel_u_p_c_scr = -0.75$\,\%$; max_tilde_rel_u_p_c_scr = 1.67$\,\%$; 
mean_tilde_rel_u_p_c_scr = -0.00$\,\%$;

min_tilde_rel_k_p_c_scr = -8.60$\,\%$; max_tilde_rel_k_p_c_scr = 202.25$\,\%$; 
mean_tilde_rel_k_p_c_scr = 4.96$\,\%$;
\end{lstlisting}
\footnotetext{\label{footnote_output_plot_p_star_E30} Stored in the file \qq{\texttt{output_plot_num_data_densities_p_star_E30.txt}} in the digital\\ \hspace*{8.7mm}appendix.}
\vspace*{5mm}
\begin{lstlisting}[language=Matlab,keywordstyle=\empty,frame=single, caption={Output\protect\footnotemark{} of the \textsc{MATLAB}-script\\\qq{\texttt{plot_num_data_densities_p_star_E100.m}} \vspace{1mm}},mathescape=true,label={MATLABoutput_plot_p_star_E100} ]
set of parameters:  

B = 5.25 T; E = 100.0 V/m; k_BT_e = 10 eV; Z_eff = 1.00;

numerical data:

min_n_RE_ava_screen_check = 1.00; max_n_RE_ava_screen_check = 1.00;
mean_n_RE_ava_screen_check = 1.00;

min_n_RE_ava_check = 1.00; max_n_RE_ava_check = 1.00; 
mean_n_RE_ava_check = 1.00;

min_u_RE_ava_over_c_screen = 0.882621; max_u_RE_ava_over_c_screen = 0.983192;

min_u_RE_ava_over_c = 0.962576; max_u_RE_ava_over_c = 0.976058;
min_rel_u = -9.56$\,\%$; max_rel_u = 2.14$\,\%$; mean_rel_u = -1.48$\,\%$;

min_k_RE_ava_over_csq_screen = 5.927737; max_k_RE_ava_over_csq_screen = 32.069514;

min_k_RE_ava_over_csq = 30.592786; max_k_RE_ava_over_csq = 38.560246;

min_rel_k = -84.5$\,\%$; max_rel_k = 0.81$\,\%$; mean_rel_k = -47.64$\,\%$;

min_rel_E_c_tot = 9.15$\,\%$; max_rel_E_c_tot = 60.58$\,\%$; mean_rel_E_c_tot = 20.68$\,\%$;

min_rel_E_c_eff = 24.23$\,\%$; max_rel_E_c_eff = 1035.67$\,\%$; mean_rel_E_c_eff = 540.60$\,\%$;

min_rel_p_star = 8.92$\,\%$; max_rel_p_star = 254.74$\,\%$; mean_rel_p_star = 114.09$\,\%$;

min_tilde_rel_p_c_scr = -0.37$\,\%$; max_tilde_rel_p_c_scr = 852.20$\,\%$; 
mean_tilde_rel_p_c_scr = 27.22$\,\%$;

min_tilde_rel_u_p_c_scr = -6.33$\,\%$; max_tilde_rel_u_p_c_scr = 1.96$\,\%$; 
mean_tilde_rel_u_p_c_scr = -0.68$\,\%$;

min_tilde_rel_k_p_c_scr = -42.10$\,\%$; max_tilde_rel_k_p_c_scr = 170.72$\,\%$; 
mean_tilde_rel_k_p_c_scr = 2.31$\,\%$;
\end{lstlisting}
\footnotetext{\label{footnote_output_plot_p_star_E100} Stored in the file \qq{\texttt{output_plot_num_data_densities_p_star_E100.txt}} in the digital\\ \hspace*{8.7mm}appendix.} 

\clearpage

\subsection{Contour plots of the critical electric field strength for the generation of runaway electrons}\label{contour_plots_E_c_ava_appendix_subsection}

\vspace*{-0.9mm}

\begin{figure}[H]
  \centering
  \subfloat{\label{fig_E_c_ava_p_c_scr_E100} 
    \includegraphics[trim=320 27 381 17,width=0.405\textwidth,clip]
    {E_c_ava_p_c_scr_E100_1.pdf}}\quad
  \subfloat{\label{fig_E_c_tot_ava_p_c_scr_E10}
    \includegraphics[trim=320 27 381 17,width=0.411\textwidth,clip]
    {E_c_tot_ava_p_c_scr_E100_1.pdf}}\\[3.5pt]
  \subfloat{\label{fig_E_c_eff_ava_p_c_scr_E100} 
    \includegraphics[trim=320 22 366 23,width=0.405\textwidth,clip]
    {E_c_eff_ava_p_c_scr_E100_1.pdf}}\quad
  \subfloat{\label{fig_rel_E_c_tot_ava_p_c_scr_E100}
    \includegraphics[trim=320 28 348 14,width=0.411\textwidth,clip]
    {rel_E_c_tot_ava_p_c_scr_E100_1.pdf}}\\[7pt]
  \subfloat{\label{fig_rel_E_c_eff_ava_p_c_scr_E100} 
    \includegraphics[trim=319 22 342 16,width=0.425\textwidth,clip]
    {rel_E_c_eff_ava_p_c_scr_E100_1.pdf}}
  \caption[Contour plots of the \textit{Connor-Hastie} critical electric field from the free electron density $E_{\mathrm{c}}$, from the total electron density $E_{\mathrm{c}}^{\mathrm{tot}}$, the effective critical electric field $E_{\mathrm{c}}^{\mathrm{eff}}$ and the relative deviations $\Delta_{E_{\mathrm{c}}^{\mathrm{tot}}}$ and $\Delta_{E_{\mathrm{c}}^{\mathrm{eff}}}$ for \mbox{$k_{\mathrm{B}}T_{\mathrm{e}}=10\,\textup{eV}$}, \mbox{$B=5.25\,\textup{T}$}, \mbox{$Z_{\mathrm{eff}}=1$} and \mbox{$E_{\|}=100\,\mathrm{V/m}$}.]{Contour plots\protect\footnotemark{} of the \textit{Connor-Hastie} critical electric field from the free electron density $E_{\mathrm{c}}$, from the total electron density $E_{\mathrm{c}}^{\mathrm{tot}}$, the effective critical electric field $E_{\mathrm{c}}^{\mathrm{eff}}$ and the relative deviations $\Delta_{E_{\mathrm{c}}^{\mathrm{tot}}}$ and $\Delta_{E_{\mathrm{c}}^{\mathrm{eff}}}$ for \mbox{$k_{\mathrm{B}}T_{\mathrm{e}}=10\,\textup{eV}$}, \mbox{$B=5.25\,\textup{T}$}, \mbox{$Z_{\mathrm{eff}}=1$} and \mbox{$E_{\|}=100\,\mathrm{V/m}$}.}
\label{fig_E_C_ava_p_c_scr}
\end{figure} 
\footnotetext{\label{fig_plot_footnote_0} The contour plots were generated, with the help of the \textsc{MATLAB}-scripts\\ \hspace*{8.7mm}\qq{\texttt{generate_num_data_densities_p_c_scr_E100.m}} and\\ \hspace*{8.7mm}\qq{\texttt{plot_num_data_densities_p_c_scr_E100.m}}, which can be found in the digital\\ \hspace*{8.7mm}appendix.}

\clearpage

\subsection{Contour plots of the effective critical momentum for the generation of runaway electrons}\label{contour_plots_p_c_eff_ava_appendix_subsection}

\vspace*{0.3cm}

\begin{figure}[H]
  \centering
  \subfloat{\label{fig_p_c_ava_p_c_scr_E3} 
   \includegraphics[trim=318 20 366 21,width=0.49\textwidth,clip]
    {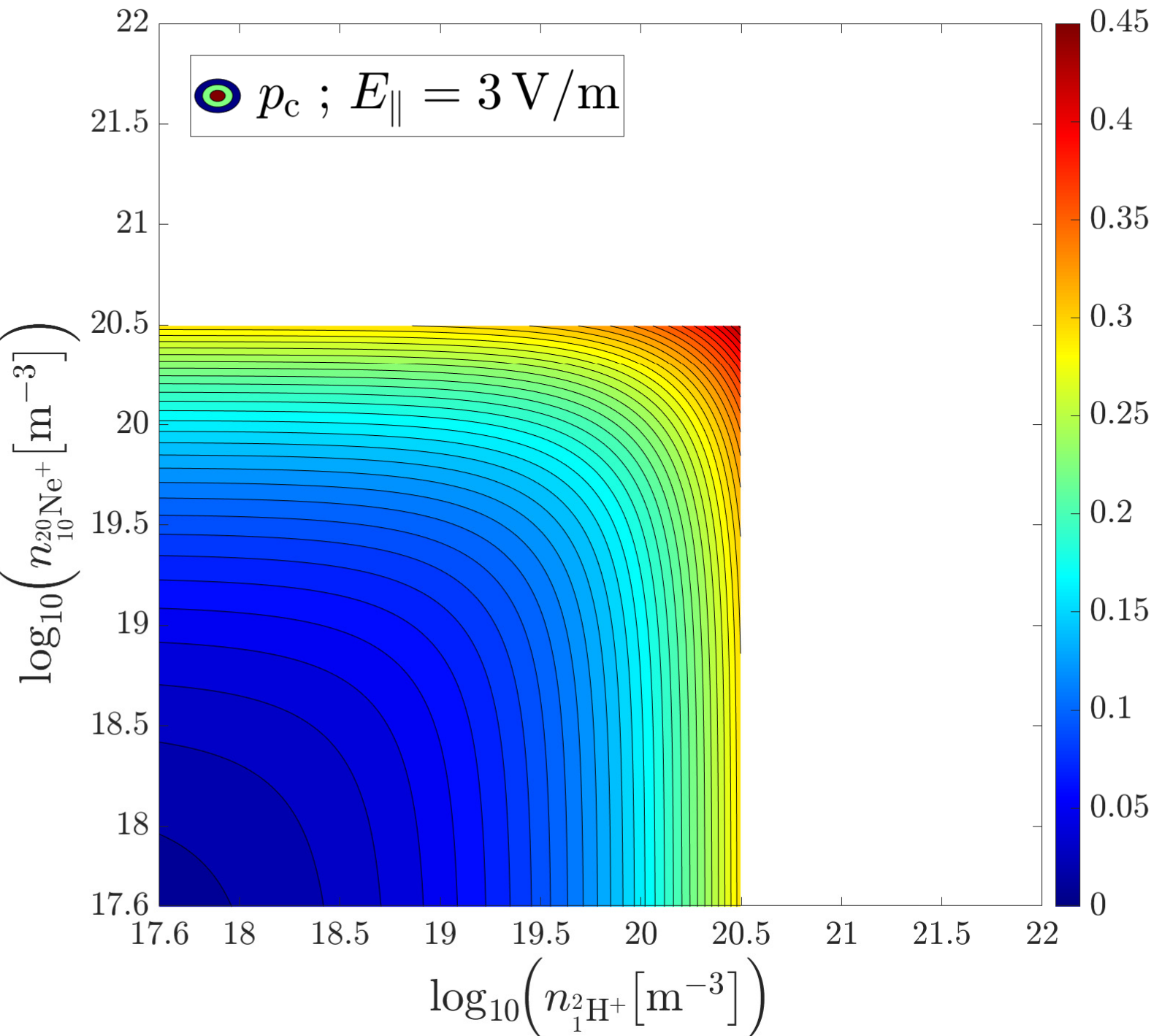}}\hfill
  \subfloat{\label{fig_p_c_ava_p_c_scr_E10}
  \includegraphics[trim=318 21 366 20,width=0.49\textwidth,clip]
    {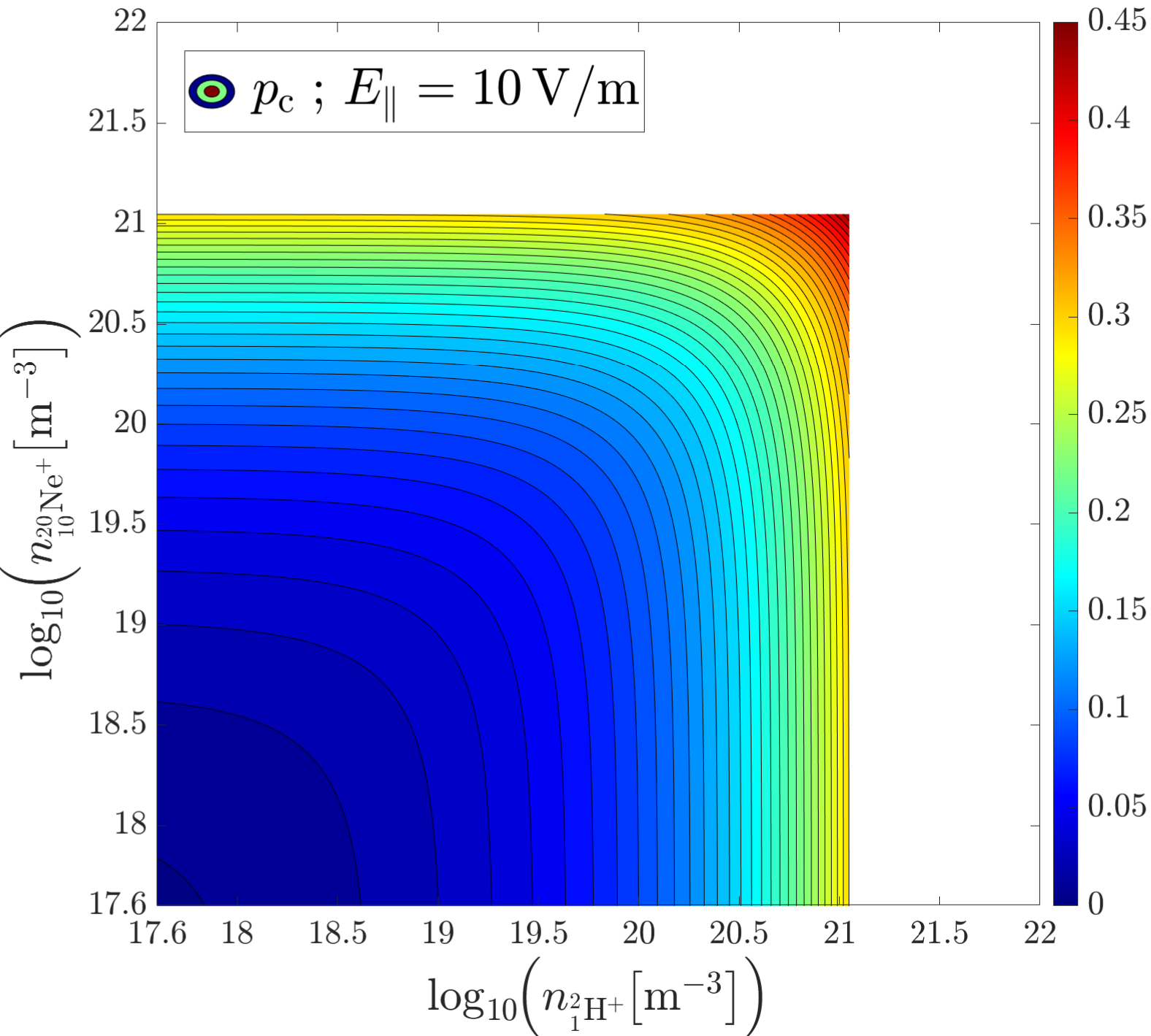}}\\[11pt]
  \subfloat{\label{fig_p_c_ava_p_c_scr_E30} 
    \includegraphics[trim=324 15 362 24,width=0.49\textwidth,clip]
    {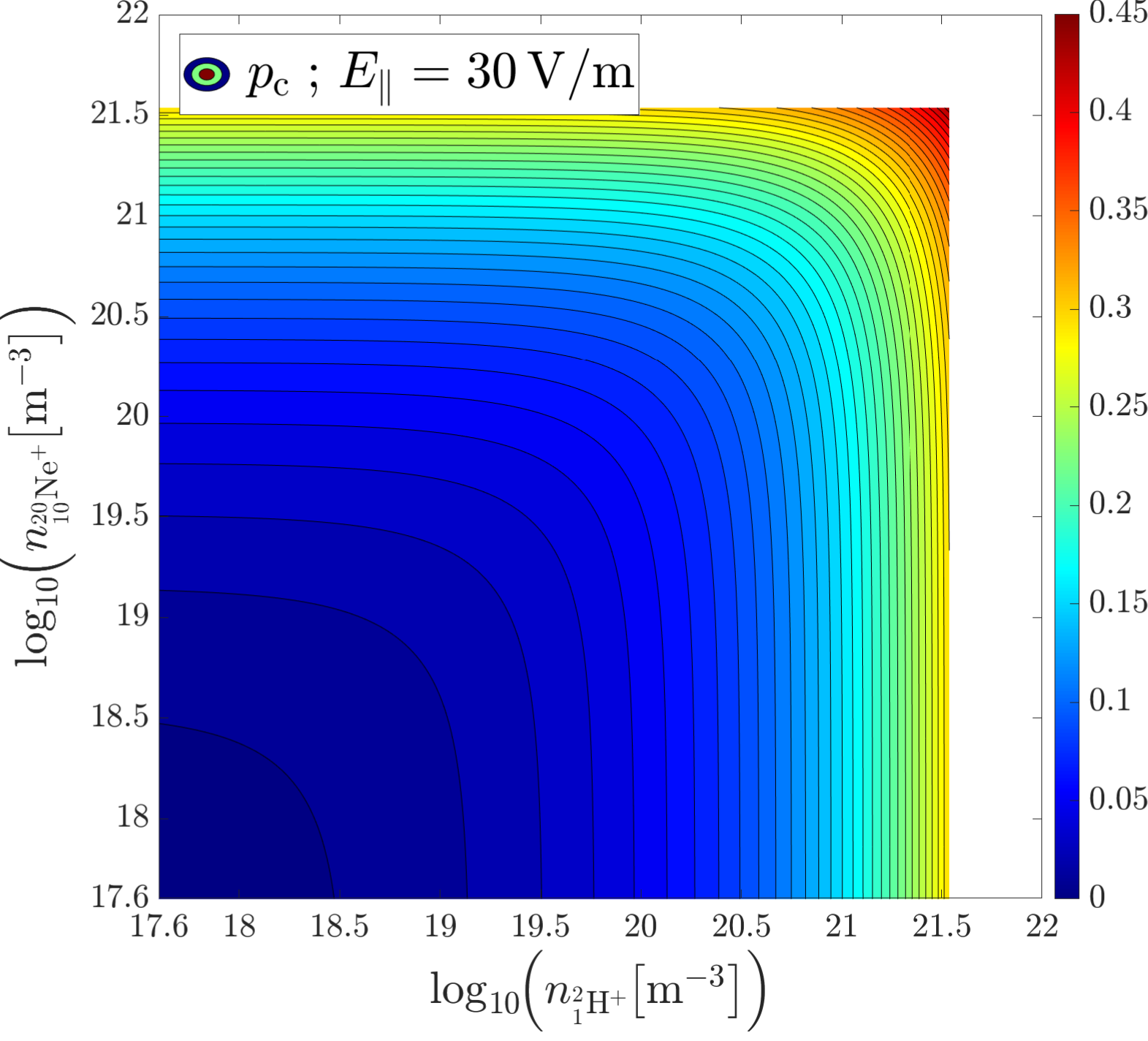}}\hfill
  \subfloat{\label{fig_p_c_ava_p_c_scr_E100}
    \includegraphics[trim=318 23 363 23,width=0.49\textwidth,clip]
    {p_c_ava_p_c_scr_E100_2.pdf}}
  \caption[Contour plots of the normalized \textit{Connor-Hastie} critical momentum $p_{\mathrm{c}}$ of an avalanche runaway electron population with \mbox{$k_{\mathrm{B}}T_{\mathrm{e}}=10\,\textup{eV}$}, \mbox{$B=5.25\,\textup{T}$} and \mbox{$Z_{\mathrm{eff}}=1$} for approximately logarithmically increasing values of the electric field strength \mbox{$E_{\|}\coloneqq\vert E_{\|}\vert$}.]{Contour plots\protect\footnotemark{} of the normalized \textit{Connor-Hastie} critical momentum $p_{\mathrm{c}}$ of an avalanche runaway electron population with \mbox{$k_{\mathrm{B}}T_{\mathrm{e}}=10\,\textup{eV}$}, \mbox{$B=5.25\,\textup{T}$} and \mbox{$Z_{\mathrm{eff}}=1$} for approximately logarithmically increasing values of the electric field strength \mbox{$E_{\|}\coloneqq\vert E_{\|}\vert$}.}
\label{fig_p_c_ava_p_c_scr}
\end{figure}
\footnotetext{\label{fig_plot_footnote_1} The contour plots were computed, with the help of the \textsc{MATLAB}-scripts\\ \hspace*{8.7mm}\qq{\texttt{generate_num_data_densities_p_c_scr_E3.m}},\\ \hspace*{8.7mm}\qq{\texttt{plot_num_data_densities_p_c_scr_E3.m}},\\ \hspace*{8.7mm}\qq{\texttt{generate_num_data_densities_p_c_scr_E10.m}},\\ \hspace*{8.7mm}\qq{\texttt{plot_num_data_densities_p_c_scr_E10.m}},\\ \hspace*{8.7mm}\qq{\texttt{generate_num_data_densities_p_c_scr_E30.m}},\\ \hspace*{8.7mm}\qq{\texttt{plot_num_data_densities_p_c_scr_E30.m}},\\ \hspace*{8.7mm}\qq{\texttt{generate_num_data_densities_p_c_scr_E100.m}} and\\ \hspace*{8.7mm}\qq{\texttt{plot_num_data_densities_p_c_scr_E100.m}}, which can be viewed in the digital\\ \hspace*{8.7mm}appendix.}
 
\noindent\newpage\noindent

\vspace*{2.8cm}

\begin{figure}[H]
  \centering
  \subfloat{\label{fig_p_c_ava_screen_p_c_scr_E3} 
   \includegraphics[trim=322 18 379 24,width=0.49\textwidth,clip]
    {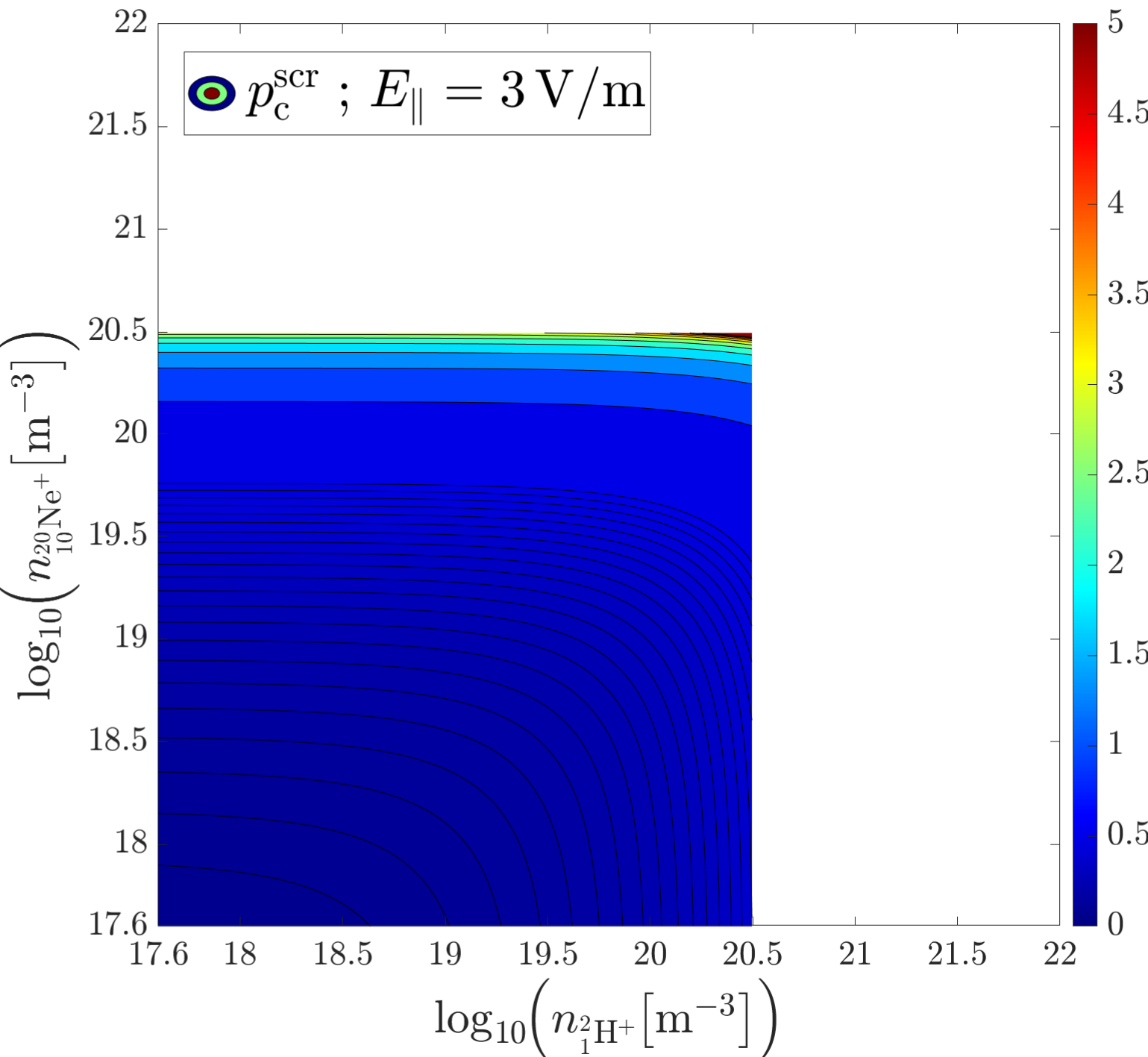}}\hfill
  \subfloat{\label{fig_p_c_ava_screen_p_c_scr_E10}
    \includegraphics[trim=318 16 378 24,width=0.49\textwidth,clip]
    {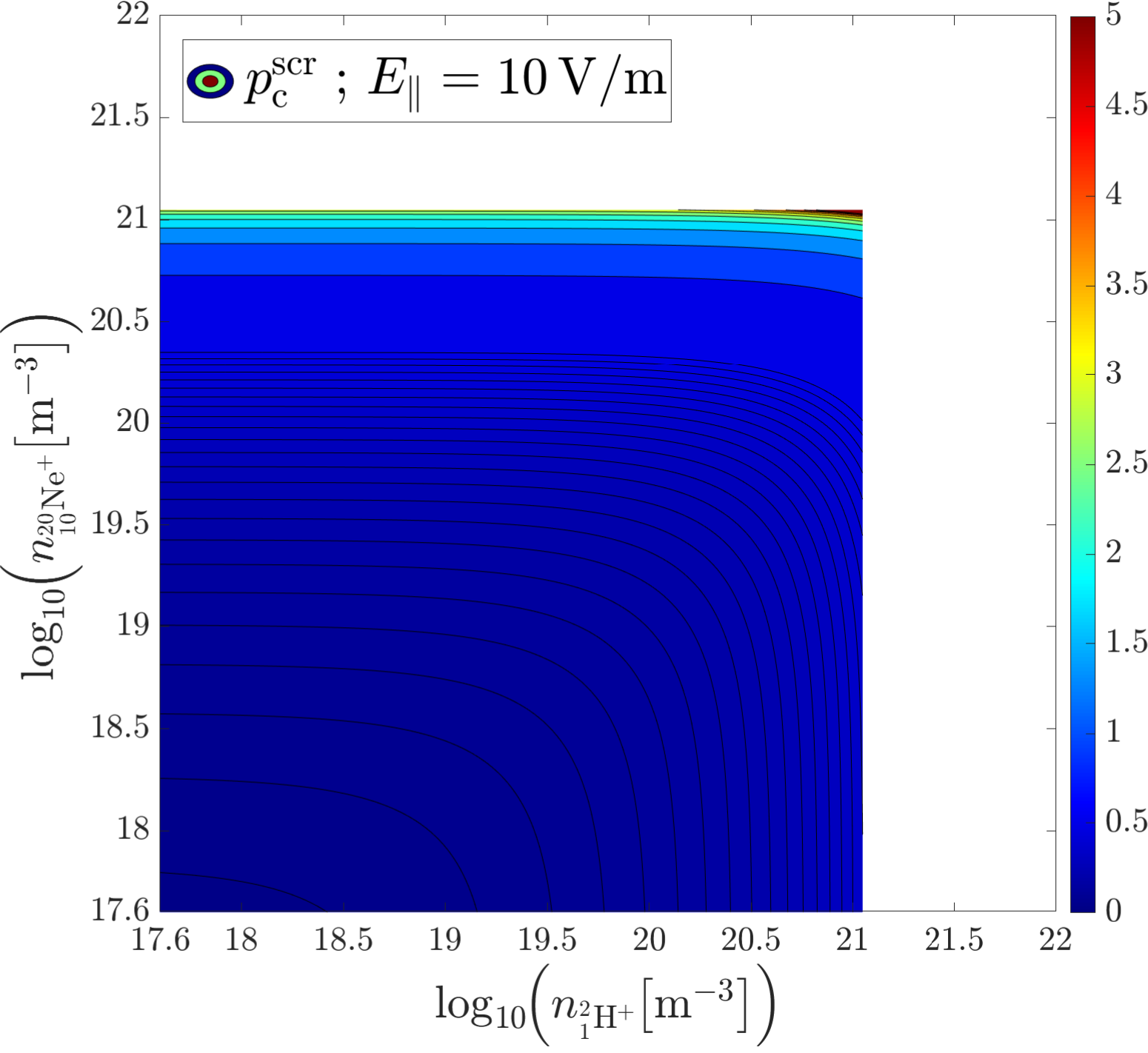}}\\[11pt]
  \subfloat{\label{fig_p_c_ava_screen_p_c_scr_E30} 
  \includegraphics[trim=319 17 375 23,width=0.49\textwidth,clip]
    {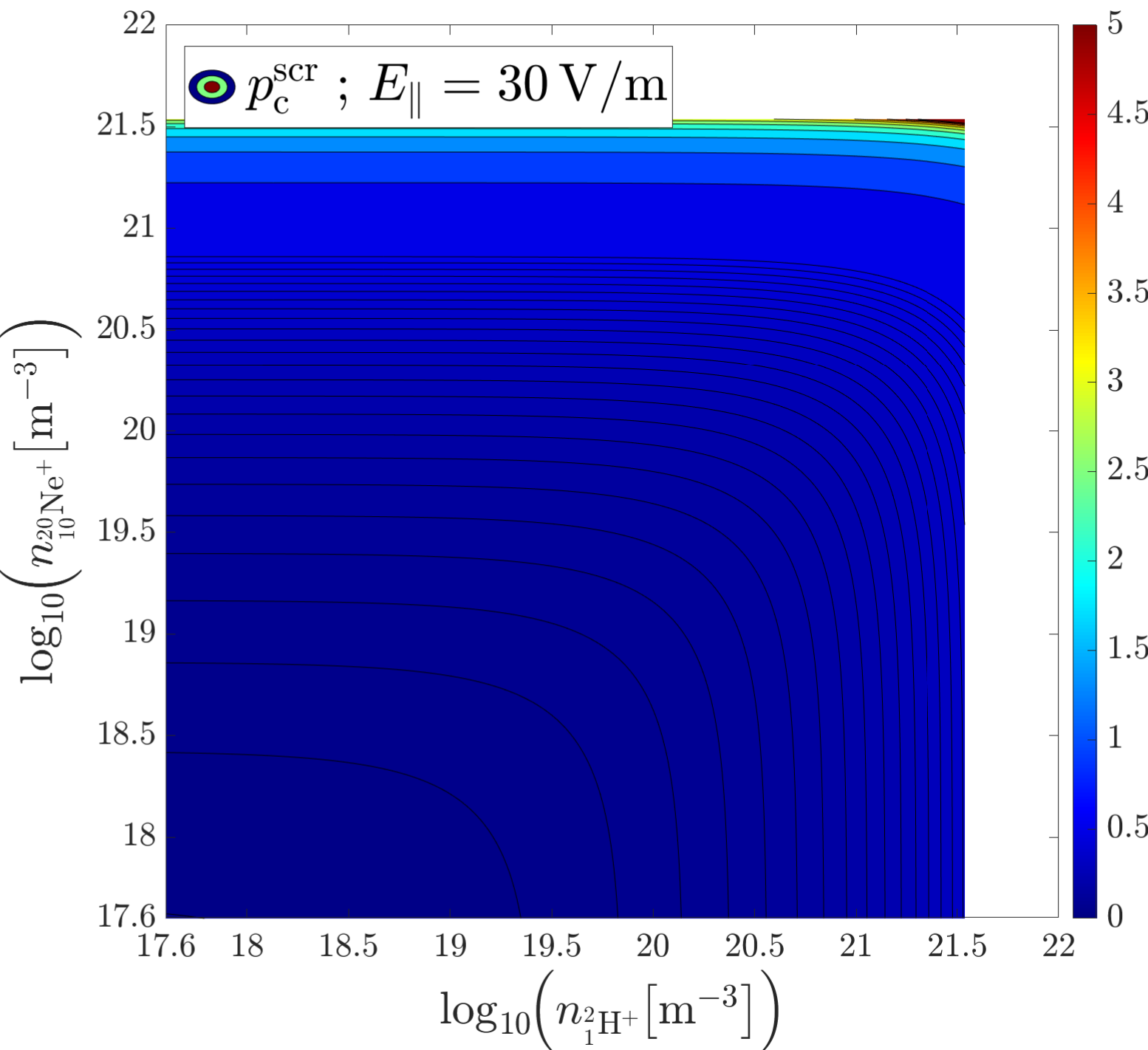}}\hfill
  \subfloat{\label{fig_p_c_ava_screen_p_c_scr_E100}
    \includegraphics[trim=318 24 371 20,width=0.49\textwidth,clip]
    {p_c_ava_screen_p_c_scr_E100_2.pdf}}
  \caption[Contour plots of the normalized approximated effective critical momentum $p_{\mathrm{c}}^{\mathrm{scr}}$ of an avalanche runaway electron population with \mbox{$k_{\mathrm{B}}T_{\mathrm{e}}=10\,\textup{eV}$}, \mbox{$B=5.25\,\textup{T}$} and \mbox{$Z_{\mathrm{eff}}=1$} for approximately logarithmically increasing values of the electric field strength \mbox{$E_{\|}\coloneqq\vert E_{\|}\vert$}, under consideration of the effects of partial screening.]{Contour plots$^{\ref{fig_plot_footnote_1}}$ of the normalized approximated effective critical momentum $p_{\mathrm{c}}^{\mathrm{scr}}$ of an avalanche runaway electron population with \mbox{$k_{\mathrm{B}}T_{\mathrm{e}}=10\,\textup{eV}$}, \mbox{$B=5.25\,\textup{T}$} and \mbox{$Z_{\mathrm{eff}}=1$} for approximately logarithmically increasing values of the electric field strength \mbox{$E_{\|}\coloneqq\vert E_{\|}\vert$}, under consideration of the effects of partial screening.}
\label{fig_p_c_ava_screen_p_c_scr}
\end{figure}
 
\noindent\newpage\noindent

\vspace*{0.28cm}

\begin{figure}[H]
  \centering
  \subfloat{\label{fig_p_star_ava_p_star_E3} 
   \includegraphics[trim=320 22 377 24,width=0.49\textwidth,clip]
    {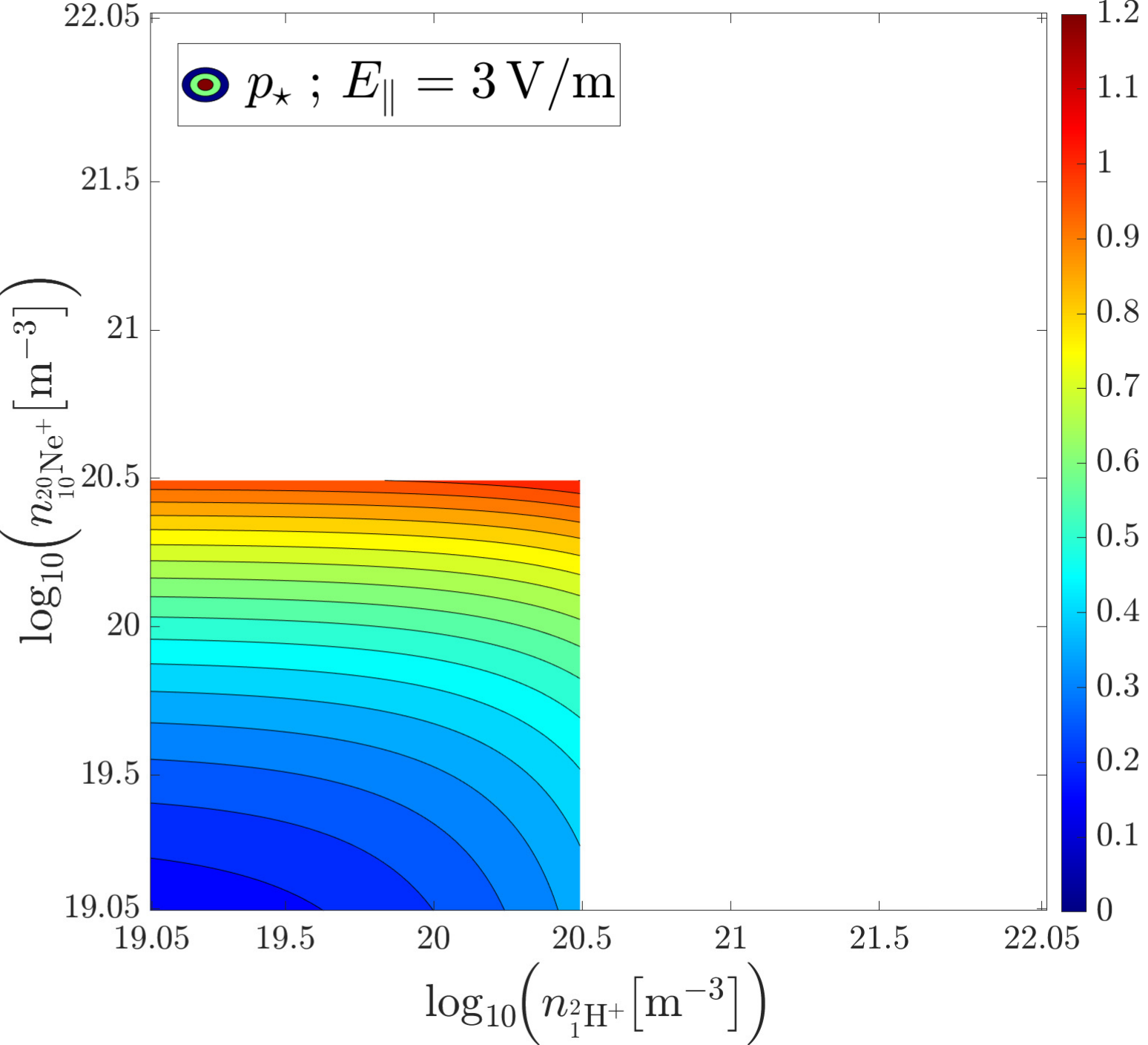}}\hfill
  \subfloat{\label{fig_p_star_ava_p_star_E10}
   \includegraphics[trim=333 25 374 18,width=0.49\textwidth,clip]
    {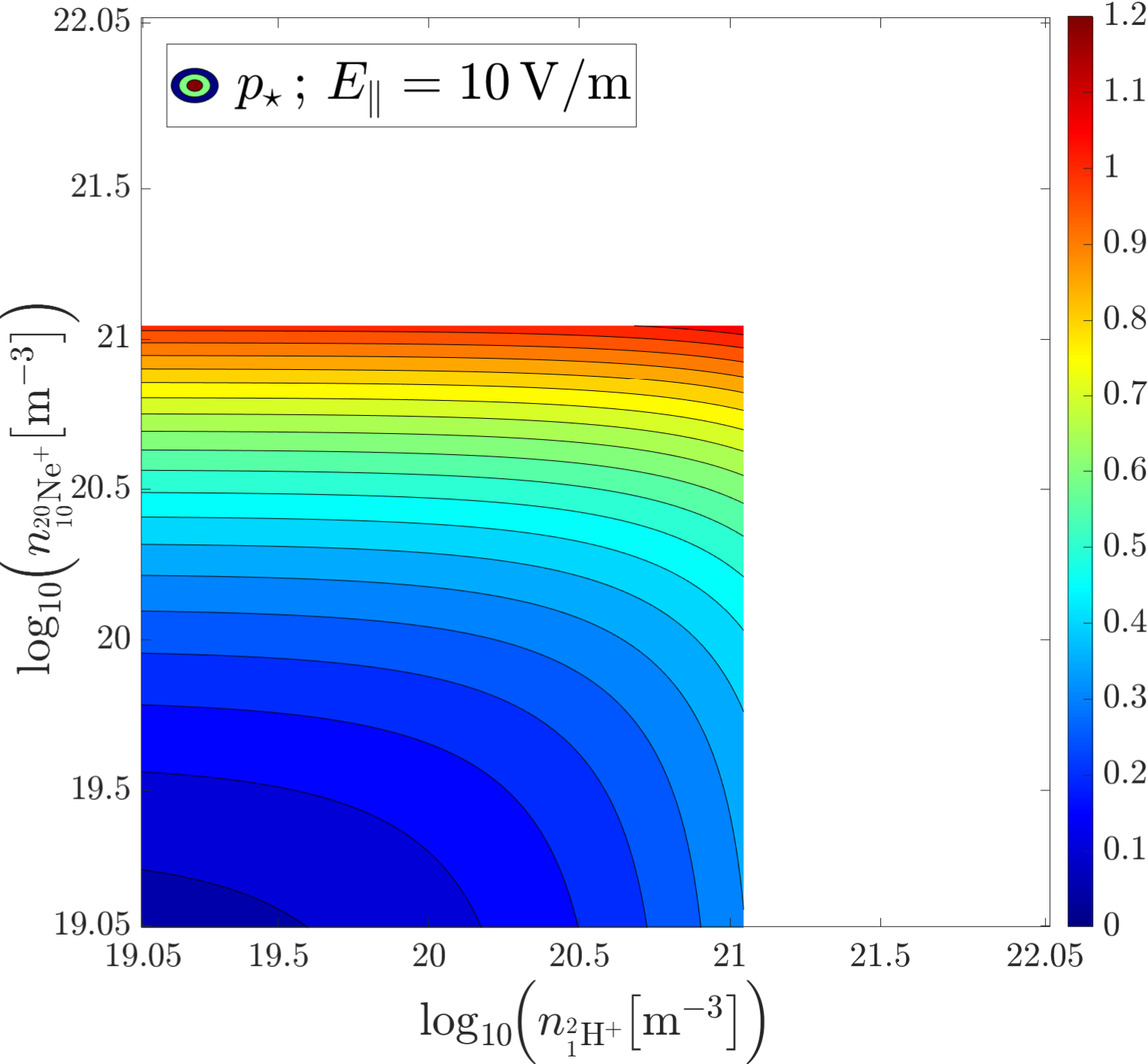}}\\[11pt]
  \subfloat{\label{fig_p_star_ava_p_star_E30} 
    \includegraphics[trim=330 21 372 19,width=0.49\textwidth,clip]
    {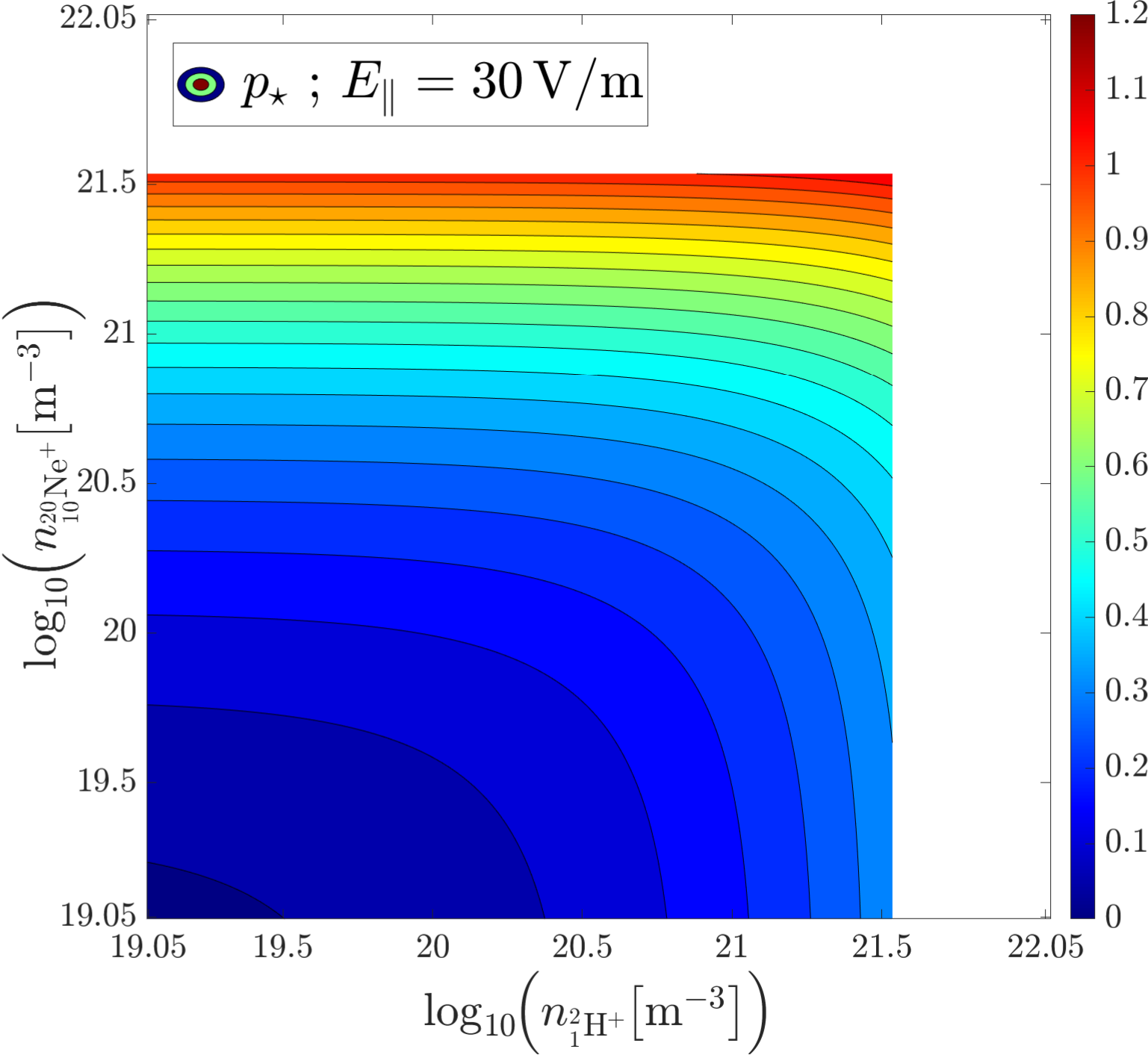}}\hfill
  \subfloat{\label{fig_p_star_ava_p_star_E100}
   \includegraphics[trim=325 23 367 24,width=0.49\textwidth,clip]
    {p_star_ava_p_star_E100_2.pdf}}
  \caption[Contour plots of the normalized effective critical momentum $p_{\star}$\linebreak\mbox{(\textit{L.\hspace{0.7mm}Hesslow})} of an avalanche runaway electron population with\linebreak\mbox{$k_{\mathrm{B}}T_{\mathrm{e}}=10\,\textup{eV}$}, \mbox{$B=5.25\,\textup{T}$} and \mbox{$Z_{\mathrm{eff}}=1$} for approximately logarithmically increasing values of the electric field strength \mbox{$E_{\|}\coloneqq\vert E_{\|}\vert$}, under consideration
of the effects of partial screening.]{Contour plots\protect\footnotemark{} of the normalized effective critical momentum $p_{\star}$ (\textit{L.\hspace{0.7mm}Hesslow}) of an avalanche runaway electron population with \mbox{$k_{\mathrm{B}}T_{\mathrm{e}}=10\,\textup{eV}$}, \mbox{$B=5.25\,\textup{T}$} and \mbox{$Z_{\mathrm{eff}}=1$} for approximately logarithmically increasing values of the electric field strength \mbox{$E_{\|}\coloneqq\vert E_{\|}\vert$}, under consideration
of the effects of partial screening.}
\label{fig_p_star_ava}
\end{figure}
\footnotetext{\label{fig_plot_footnote_2} The contour plots were produced, by means of the \textsc{MATLAB}-scripts\\ \hspace*{8.7mm}\qq{\texttt{generate_num_data_densities_p_star_E3.m}},\\ \hspace*{8.7mm}\qq{\texttt{plot_num_data_densities_p_star_E3.m}},\\ \hspace*{8.7mm}\qq{\texttt{generate_num_data_densities_p_star_E10.m}},\\ \hspace*{8.7mm}\qq{\texttt{plot_num_data_densities_p_star_E10.m}},\\ \hspace*{8.7mm}\qq{\texttt{generate_num_data_densities_p_star_E30.m}},\\ \hspace*{8.7mm}\qq{\texttt{plot_num_data_densities_p_star_E30.m}},\\ \hspace*{8.7mm}\qq{\texttt{generate_num_data_densities_p_star_E100.m}} and\\ \hspace*{8.7mm}\qq{\texttt{plot_num_data_densities_p_star_E100.m}}, which are stored in the digital appendix.}
 
\noindent\newpage\noindent

\vspace*{2.9cm}

\begin{figure}[H]
  \centering
  \subfloat{\label{fig_rel_p_c_scr_ava_p_c_scr_E3} 
  \includegraphics[trim=317 18 354 24,width=0.49\textwidth,clip]
    {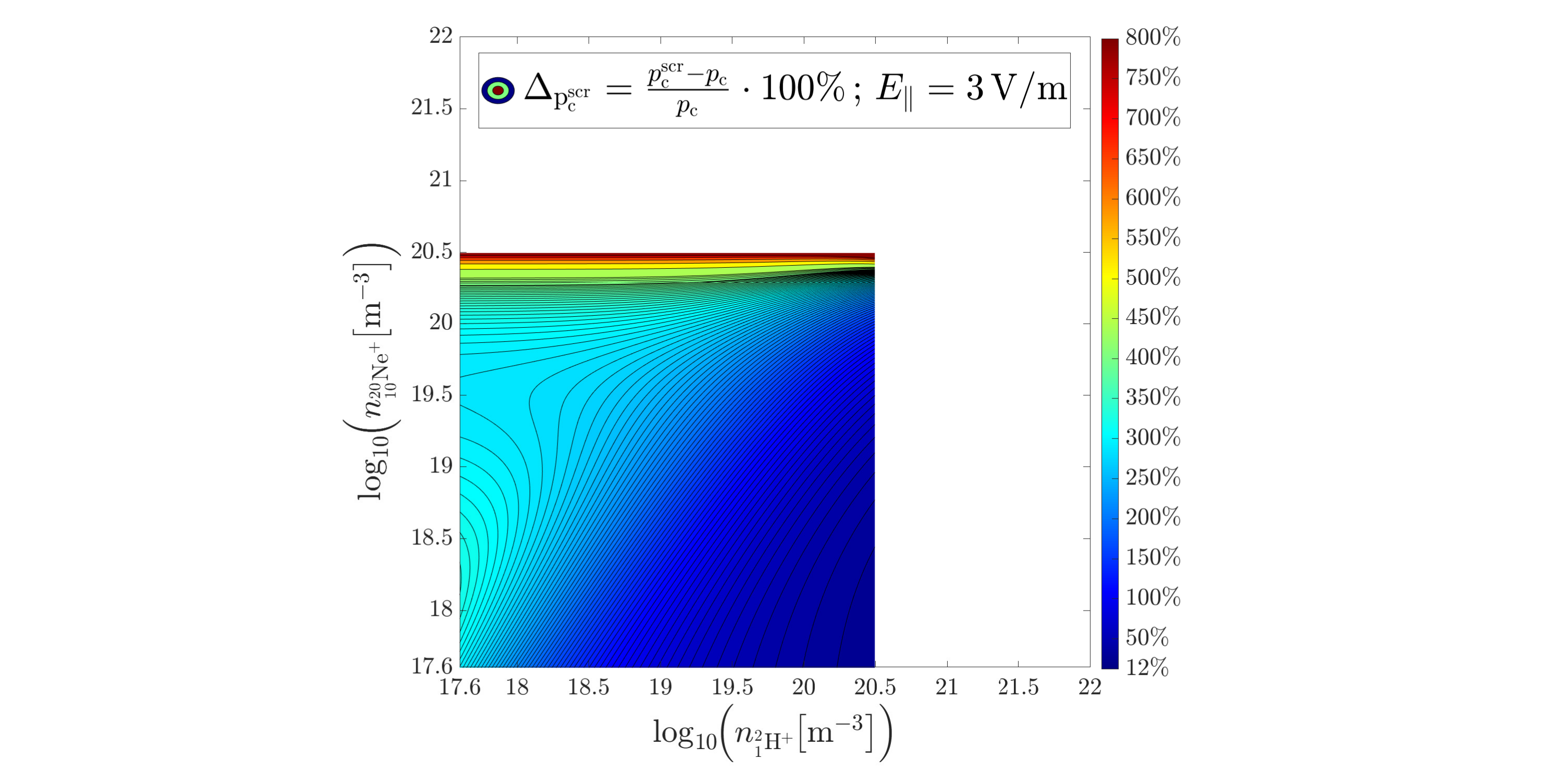}}\hfill
  \subfloat{\label{fig_rel_p_c_scr_ava_p_c_scr_E10}
   \includegraphics[trim=318 19 355 22,width=0.49\textwidth,clip]
    {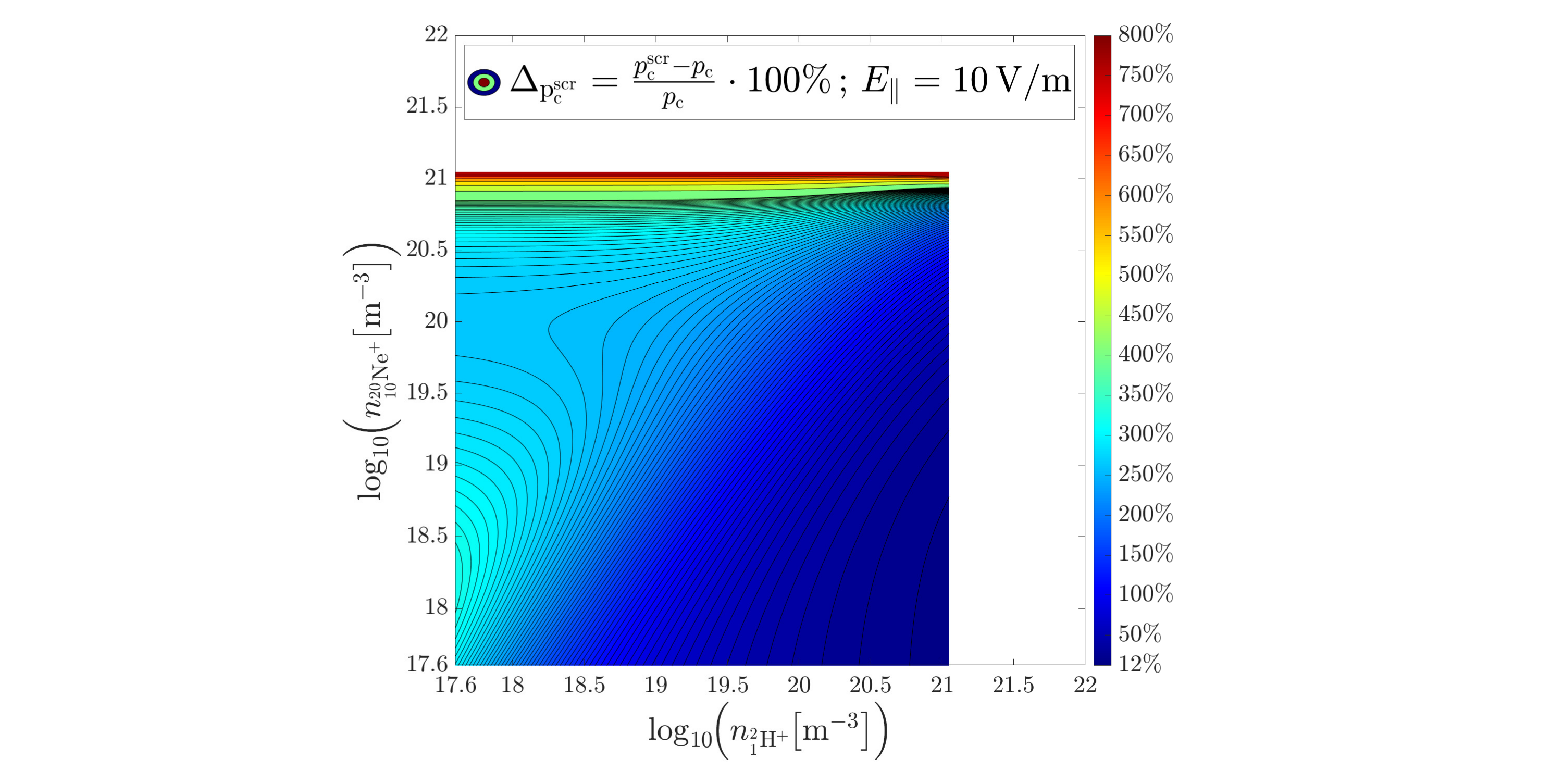}}\\[11pt]
  \subfloat{\label{fig_rel_p_c_scr_ava_p_c_scr_E30} 
    \includegraphics[trim=313 18 355 21,width=0.49\textwidth,clip]
    {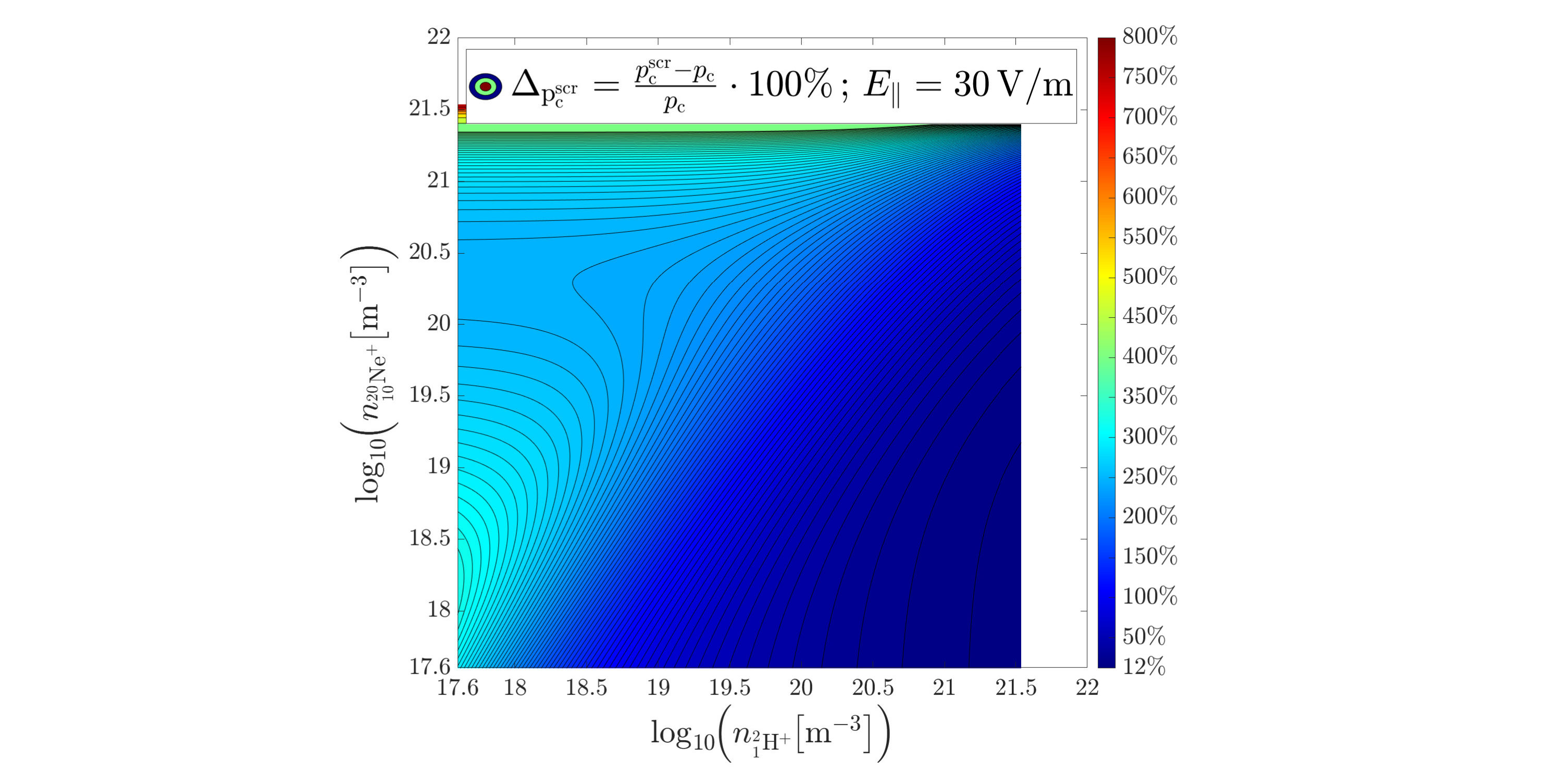}}\hfill
  \subfloat{\label{fig_rel_p_c_scr_ava_p_c_scr_E100}
    \includegraphics[trim=310 23 360 21,width=0.49\textwidth,clip]
    {rel_p_c_scr_ava_p_c_scr_E100_2.pdf}}
  \caption[Contour plots of the relative deviation $\Delta_{p_{\mathrm{c}}^{\mathrm{scr}}}$ between the \textit{Connor-Hastie} critical momentum $p_{\mathrm{c}}$ and the approximated effective critical momentum $p_{\mathrm{c}}^{\mathrm{scr}}$ for approximately logarithmically increasing values of the electric field strength \mbox{$E_{\|}\coloneqq\vert E_{\|}\vert$}.]{Contour plots$^{\ref{fig_plot_footnote_1}}$ of the relative deviation $\Delta_{p_{\mathrm{c}}^{\mathrm{scr}}}$ between the \textit{Connor-Hastie} critical momentum $p_{\mathrm{c}}$ and the approximated effective critical momentum $p_{\mathrm{c}}^{\mathrm{scr}}$ for approximately logarithmically increasing values of the electric field strength \mbox{$E_{\|}\coloneqq\vert E_{\|}\vert$}.}
\label{fig_rel_p_c_scr_ava}
\end{figure}
 
\noindent\newpage\noindent

\vspace*{2.9cm}

\begin{figure}[H]
  \centering
  \subfloat{\label{fig_rel_p_star_ava_p_star_E3} 
 \includegraphics[trim=317 27 356 18,width=0.49\textwidth,clip]
    {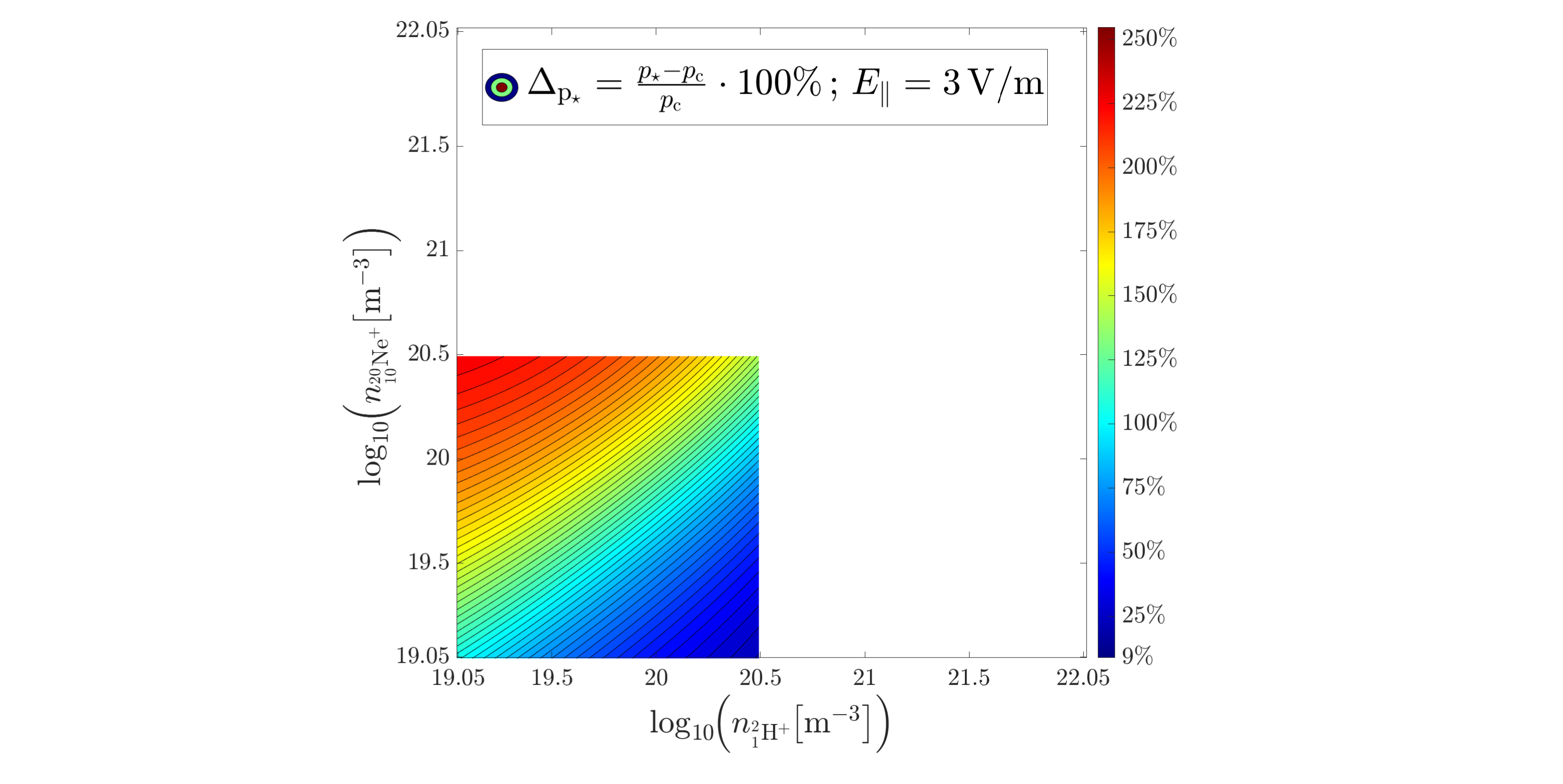}}\hfill
  \subfloat{\label{fig_rel_p_star_ava_p_star_E10}
   \includegraphics[trim=323 25 356 20,width=0.49\textwidth,clip]
    {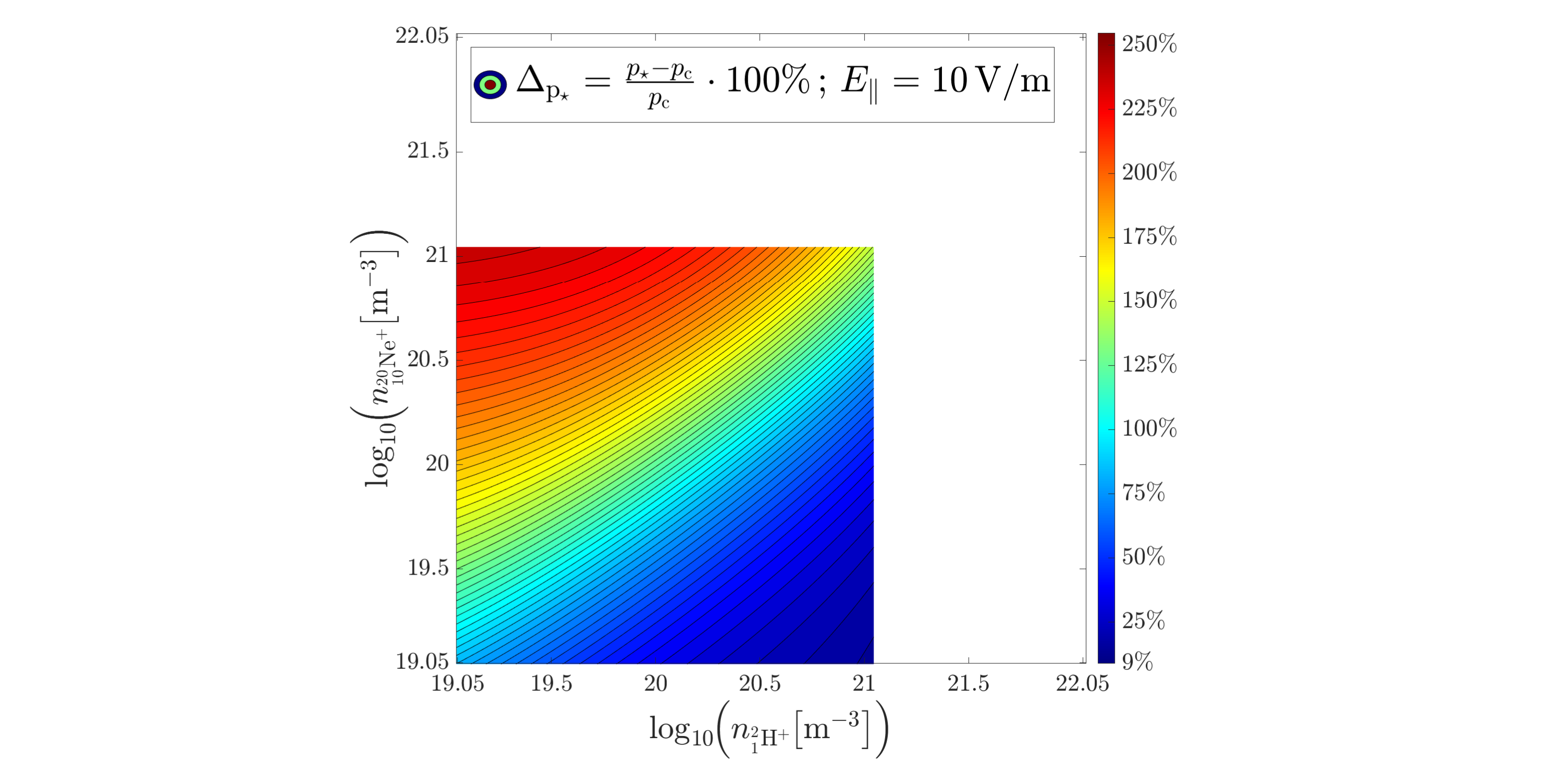}}\\[11pt]
  \subfloat{\label{fig_rel_p_star_ava_p_star_E30} 
   \includegraphics[trim=322 20 356 21,width=0.49\textwidth,clip]
    {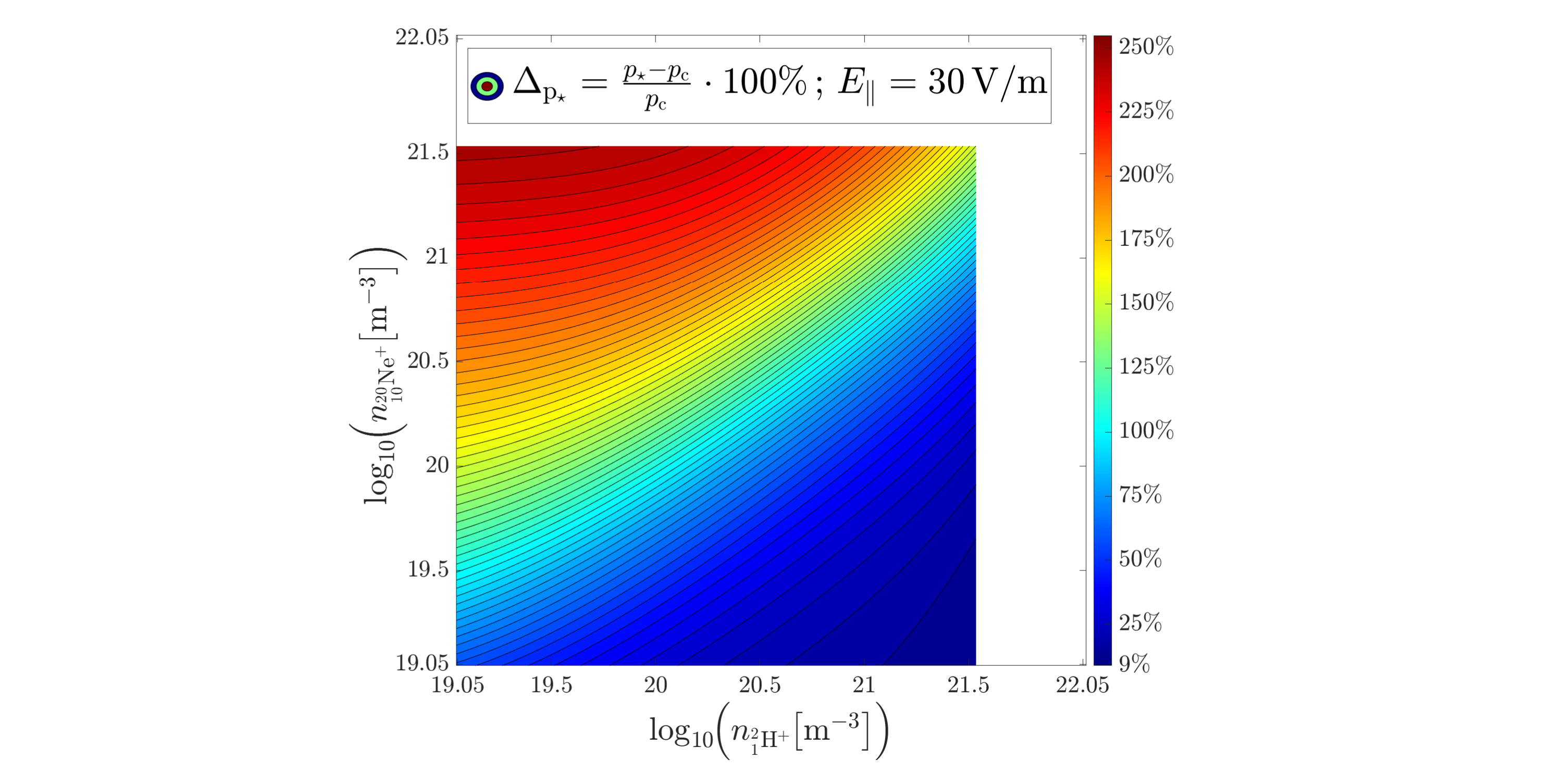}}\hfill
  \subfloat{\label{fig_rel_p_star_ava_p_star_E100}
    \includegraphics[trim=321 24 355 18,width=0.49\textwidth,clip]
    {rel_p_star_ava_p_star_E100_2.pdf}}
  \caption[Contour plots of the relative deviation $\Delta_{p_{\star}}$ between the \textit{Connor-Hastie} critical momentum $p_{\mathrm{c}}$ and the normalized effective critical momentum $p_{\star}$ (\textit{L.\hspace{0.7mm}Hesslow}) of an avalanche runaway electron population with\linebreak\mbox{$k_{\mathrm{B}}T_{\mathrm{e}}=10\,\textup{eV}$}, \mbox{$B=5.25\,\textup{T}$} and \mbox{$Z_{\mathrm{eff}}=1$} for approximately logarithmically increasing values of the electric field strength \mbox{$E_{\|}\coloneqq\vert E_{\|}\vert$}.]{Contour plots$^{\ref{fig_plot_footnote_2}}$ of the relative deviation $\Delta_{p_{\star}}$ between the \textit{Connor-Hastie} critical momentum $p_{\mathrm{c}}$ and the normalized effective critical momentum $p_{\star}$ (\textit{L.\hspace{0.7mm}Hesslow}) for an avalanche runaway electron population with \mbox{$k_{\mathrm{B}}T_{\mathrm{e}}=10\,\textup{eV}$}, \mbox{$B=5.25\,\textup{T}$} and \mbox{$Z_{\mathrm{eff}}=1$} for approximately logarithmically increasing values of the electric field strength \mbox{$E_{\|}\coloneqq\vert E_{\|}\vert$}.}
\label{fig_rel_p_star_ava}
\end{figure}
 
\noindent\newpage\noindent

\vspace*{2.9cm}

\begin{figure}[H]
  \centering
  \subfloat{\label{fig_tilde_rel_p_c_scr_ava_p_star_E3} 
   \includegraphics[trim=319 23 354 16,width=0.49\textwidth,clip]
    {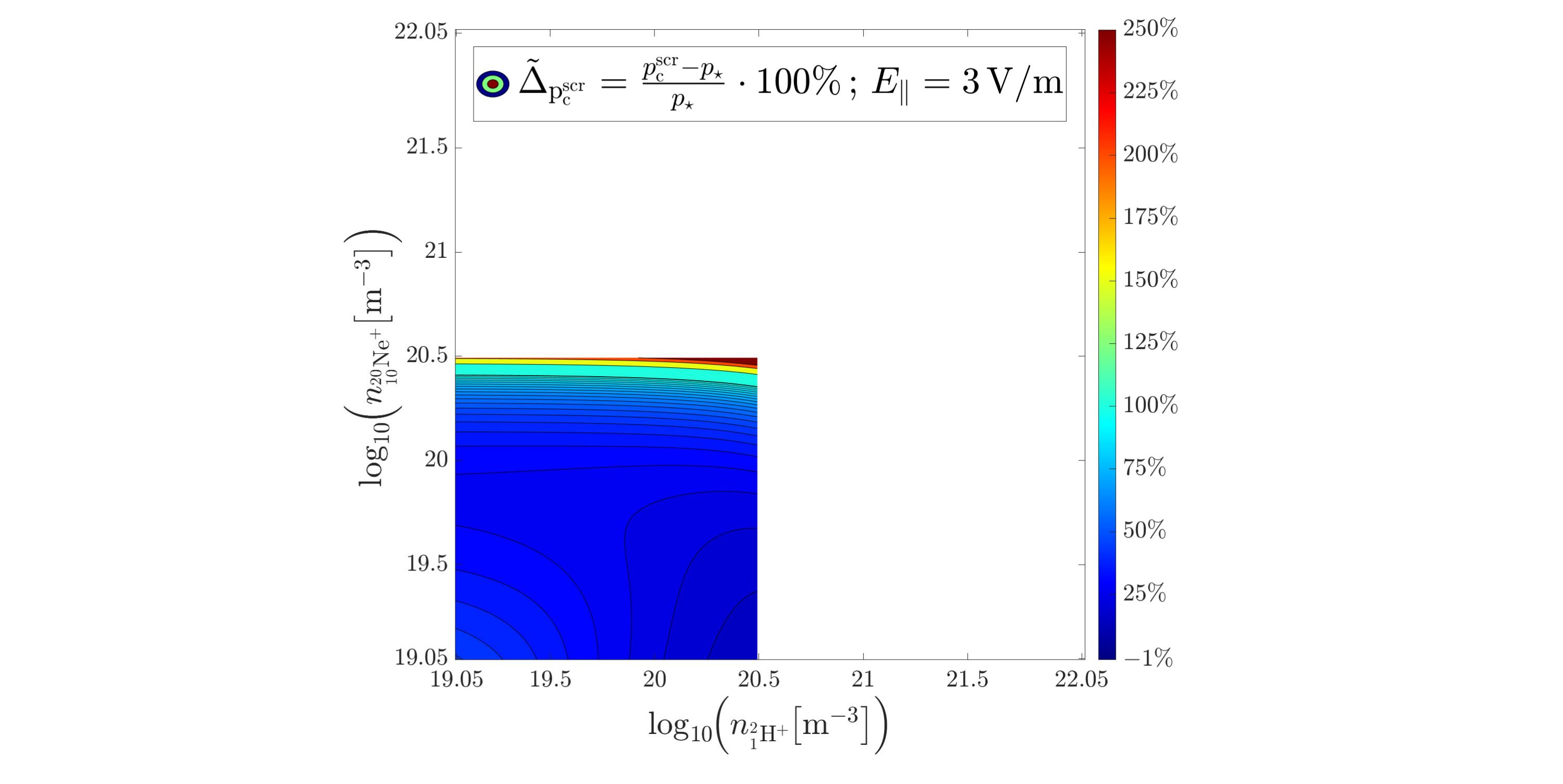}}\hfill
  \subfloat{\label{fig_tilde_rel_p_c_scr_ava_p_star_E10}
  \includegraphics[trim=327 25 356 18,width=0.49\textwidth,clip]
    {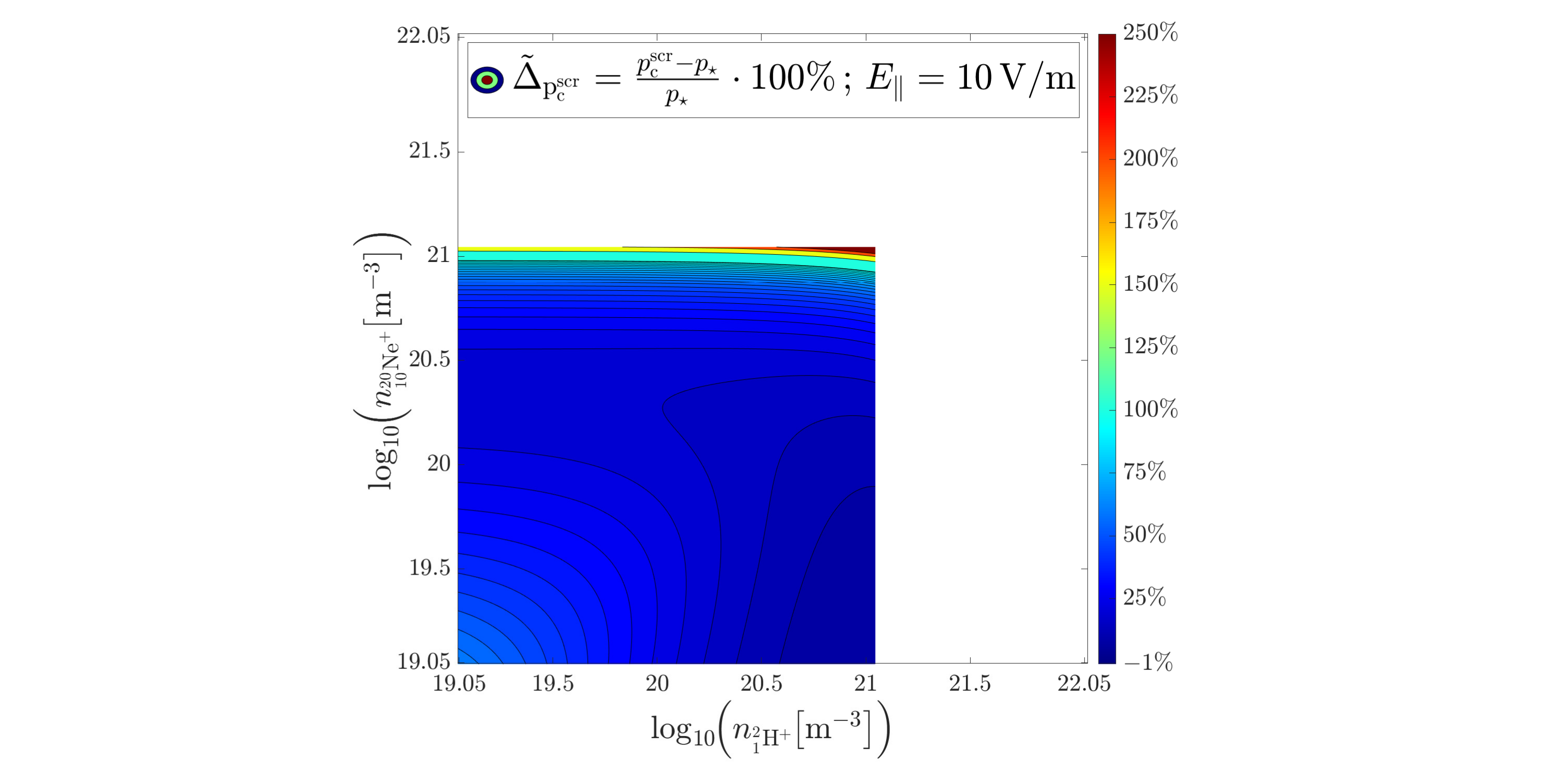}}\\[11pt]
  \subfloat{\label{fig_tilde_rel_p_c_scr_ava_p_star_E30} 
   \includegraphics[trim=322 21 355 19,width=0.49\textwidth,clip]
    {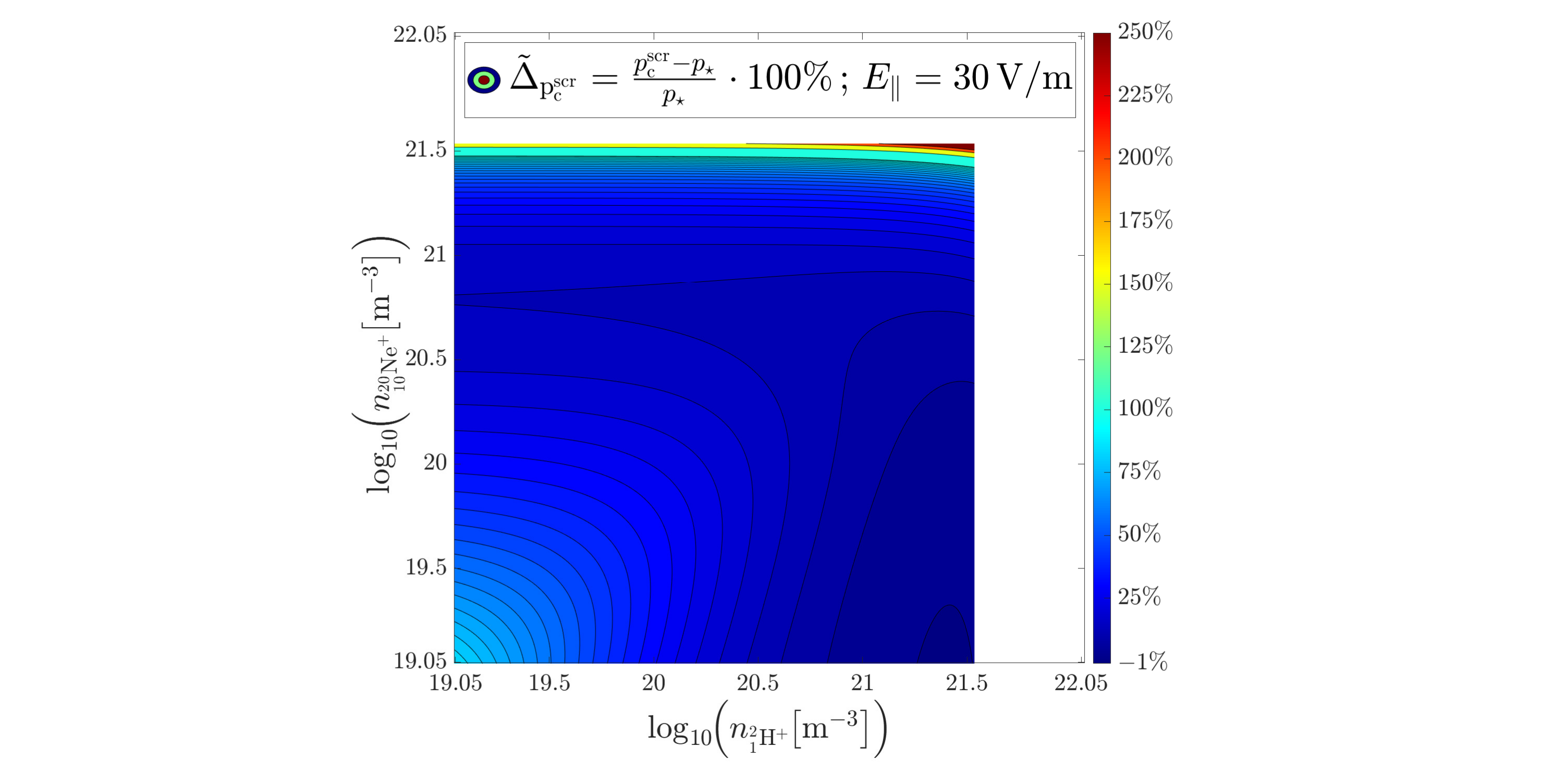}}\hfill
  \subfloat{\label{fig_tilde_rel_p_c_scr_ava_p_star_E100}
    \includegraphics[trim=320 24 355 19,width=0.49\textwidth,clip]
    {tilde_rel_p_c_scr_ava_p_star_E100_2.pdf}}
  \caption[Contour plots$^{\ref{fig_plot_footnote_2}}$ of the relative deviation $\tilde{\Delta}_{p_{\mathrm{c}}^{\mathrm{scr}}}$ between the normalized effective critical momentum $p_{\star}$ (\textit{L.\hspace{0.7mm}Hesslow}) and the approximation $p_{\mathrm{c}}^{\mathrm{scr}}$ for an avalanche runaway electron population with \mbox{$k_{\mathrm{B}}T_{\mathrm{e}}=10\,\textup{eV}$}, \mbox{$B=5.25\,\textup{T}$} and \mbox{$Z_{\mathrm{eff}}=1$} for different values of the electric field strength \mbox{$E_{\|}\coloneqq\vert E_{\|}\vert$}.]{Contour plots$^{\ref{fig_plot_footnote_2}}$ of the relative deviation $\tilde{\Delta}_{p_{\mathrm{c}}^{\mathrm{scr}}}$ between the normalized effective critical momentum $p_{\star}$ (\textit{L.\hspace{0.7mm}Hesslow}) and the approximation $p_{\mathrm{c}}^{\mathrm{scr}}$ for an avalanche runaway electron population with \mbox{$k_{\mathrm{B}}T_{\mathrm{e}}=10\,\textup{eV}$}, \mbox{$B=5.25\,\textup{T}$} and \mbox{$Z_{\mathrm{eff}}=1$} for different values of the electric field strength \mbox{$E_{\|}\coloneqq\vert E_{\|}\vert$}.}
\label{fig_tilde_rel_p_c_scr_ava}
\end{figure}

\clearpage

\subsection{Contour plots of the mean velocity of an avalanche runaway electron population}\label{contour_plots_u_ava_appendix_subsection}

\vspace*{2.8cm}

\begin{figure}[H]
  \centering
  \subfloat{\label{fig_u_ava_p_c_scr_E3} 
    \includegraphics[trim=318 19 349 20,width=0.49\textwidth,clip]
    {u_ava_p_c_scr_E3_2.pdf}}\hfill
  \subfloat{\label{fig_u_ava_p_c_scr_E10}
    \includegraphics[trim=319 23 350 22,width=0.49\textwidth,clip]
    {u_ava_p_c_scr_E10_2.pdf}}\\[11pt]
  \subfloat{\label{fig_u_ava_p_c_scr_E30} 
   \includegraphics[trim=316 19 355 20,width=0.49\textwidth,clip]
    {u_ava_p_c_scr_E30_2.pdf}}\hfill
  \subfloat{\label{fig_u_ava_p_c_scr_E100}
    \includegraphics[trim=316 19 355 24,width=0.49\textwidth,clip]
    {u_ava_p_c_scr_E100_2.pdf}}
  \caption[Contour plots of the normalized mean velocity \mbox{$u_{\mathrm{RE}}^{\hspace{0.25mm}\mathrm{ava}}/c$}, in the \textit{Rosenbluth-Putvinski} model, of an avalanche runaway electron population with \mbox{$k_{\mathrm{B}}T_{\mathrm{e}}=10\,\textup{eV}$}, \mbox{$B=5.25\,\textup{T}$} and \mbox{$Z_{\mathrm{eff}}=1$} for different values of the electric field strength \mbox{$E_{\|}\coloneqq\vert E_{\|}\vert$}.]{Contour plots$^{\ref{fig_plot_footnote_1}}$ of the normalized mean velocity \mbox{$u_{\mathrm{RE}}^{\hspace{0.25mm}\mathrm{ava}}/c$}, in the \textit{Rosenbluth-Putvinski} model, of an avalanche runaway electron population with \mbox{$k_{\mathrm{B}}T_{\mathrm{e}}=10\,\textup{eV}$}, \mbox{$B=5.25\,\textup{T}$} and \mbox{$Z_{\mathrm{eff}}=1$} for different values of the electric field strength \mbox{$E_{\|}\coloneqq\vert E_{\|}\vert$}.}
\label{fig_u_ava_p_c_scr}
\end{figure}

\noindent\newpage\noindent

\vspace*{2.9cm}

\begin{figure}[H]
  \centering
  \subfloat{\label{fig_u_ava_screen_p_c_scr_E3} 
   \includegraphics[trim=319 19 366 24,width=0.49\textwidth,clip]
    {u_ava_screen_p_c_scr_E3_2.pdf}}\hfill
  \subfloat{\label{fig_u_ava_screen_p_c_scr_E10}
    \includegraphics[trim=321 25 368 18,width=0.49\textwidth,clip]
    {u_ava_screen_p_c_scr_E10_2.pdf}}\\[11pt]
  \subfloat{\label{fig_u_ava_screen_p_c_scr_E30} 
   \includegraphics[trim=319 19 364 21,width=0.49\textwidth,clip]
    {u_ava_screen_p_c_scr_E30_2.pdf}}\hfill
  \subfloat{\label{fig_u_ava_screen_p_c_scr_E100}
    \includegraphics[trim=314 22 367 21,width=0.49\textwidth,clip]
    {u_ava_screen_p_c_scr_E100_2.pdf}}
  \caption[Contour plots of the normalized mean velocity \mbox{$u_{\mathrm{RE}}^{\hspace{0.25mm}\mathrm{ava}}/c$}, in the \textit{Hesslow} model with the effective critical momentum \mbox{$p_{\mathrm{c}}^{\mathrm{eff}}\approx p_{\mathrm{c}}^{\mathrm{scr}}$}, of an avalanche runaway electron population with \mbox{$k_{\mathrm{B}}T_{\mathrm{e}}=10\,\textup{eV}$}, \mbox{$B=5.25\,\textup{T}$} and \mbox{$Z_{\mathrm{eff}}=1$} for different values of the electric field strength \mbox{$E_{\|}\coloneqq\vert E_{\|}\vert$}.]{Contour plots$^{\ref{fig_plot_footnote_1}}$ of the normalized mean velocity \mbox{$u_{\mathrm{RE}}^{\hspace{0.25mm}\mathrm{ava}}/c$}, in the \textit{Hesslow} model with the effective critical momentum \mbox{$p_{\mathrm{c}}^{\mathrm{eff}}\approx p_{\mathrm{c}}^{\mathrm{scr}}$}, of an avalanche runaway electron population with \mbox{$k_{\mathrm{B}}T_{\mathrm{e}}=10\,\textup{eV}$}, \mbox{$B=5.25\,\textup{T}$} and \mbox{$Z_{\mathrm{eff}}=1$} for different values of the electric field strength \mbox{$E_{\|}\coloneqq\vert E_{\|}\vert$}.}
\label{fig_u_ava_screen_p_c_scr}
\end{figure}
  
\noindent\newpage\noindent

\vspace*{2.9cm}

\begin{figure}[H]
  \centering
  \subfloat{\label{fig_u_ava_screen_p_star_E3} 
   \includegraphics[trim=321 25 361 18,width=0.49\textwidth,clip]
    {u_ava_screen_p_star_E3_2.pdf}}\hfill
  \subfloat{\label{fig_u_ava_screen_p_star_E10}
    \includegraphics[trim=322 25 367 18,width=0.49\textwidth,clip]
    {u_ava_screen_p_star_E10_2.pdf}}\\[11pt]
  \subfloat{\label{fig_u_ava_screen_p_star_E30} 
    \includegraphics[trim=323 21 368 21,width=0.49\textwidth,clip]
    {u_ava_screen_p_star_E30_2.pdf}}\hfill
  \subfloat{\label{fig_u_ava_screen_p_star_E100}
     \includegraphics[trim=315 21 366 20,width=0.49\textwidth,clip]
    {u_ava_screen_p_star_E100_2.pdf}}
  \caption[Contour plots of the normalized mean velocity \mbox{$u_{\mathrm{RE}}^{\hspace{0.25mm}\mathrm{ava}}/c$}, in the \textit{Hesslow} model with the effective critical momentum \mbox{$p_{\mathrm{c}}^{\mathrm{eff}}\approx p_{\star}$}, of an avalanche runaway electron population with \mbox{$k_{\mathrm{B}}T_{\mathrm{e}}=10\,\textup{eV}$}, \mbox{$B=5.25\,\textup{T}$} and \mbox{$Z_{\mathrm{eff}}=1$} for different values of the electric field strength \mbox{$E_{\|}\coloneqq\vert E_{\|}\vert$}.]{Contour plots$^{\ref{fig_plot_footnote_2}}$ of the normalized mean velocity \mbox{$u_{\mathrm{RE}}^{\hspace{0.25mm}\mathrm{ava}}/c$}, in the \textit{Hesslow} model with the effective critical momentum \mbox{$p_{\mathrm{c}}^{\mathrm{eff}}\approx p_{\star}$}, of an avalanche runaway electron population with \mbox{$k_{\mathrm{B}}T_{\mathrm{e}}=10\,\textup{eV}$}, \mbox{$B=5.25\,\textup{T}$} and \mbox{$Z_{\mathrm{eff}}=1$} for different values of the electric field strength \mbox{$E_{\|}\coloneqq\vert E_{\|}\vert$}.}
\label{fig_u_ava_screen_p_star}
\end{figure}
 
\noindent\newpage\noindent

\vspace*{2.9cm}

\begin{figure}[H]
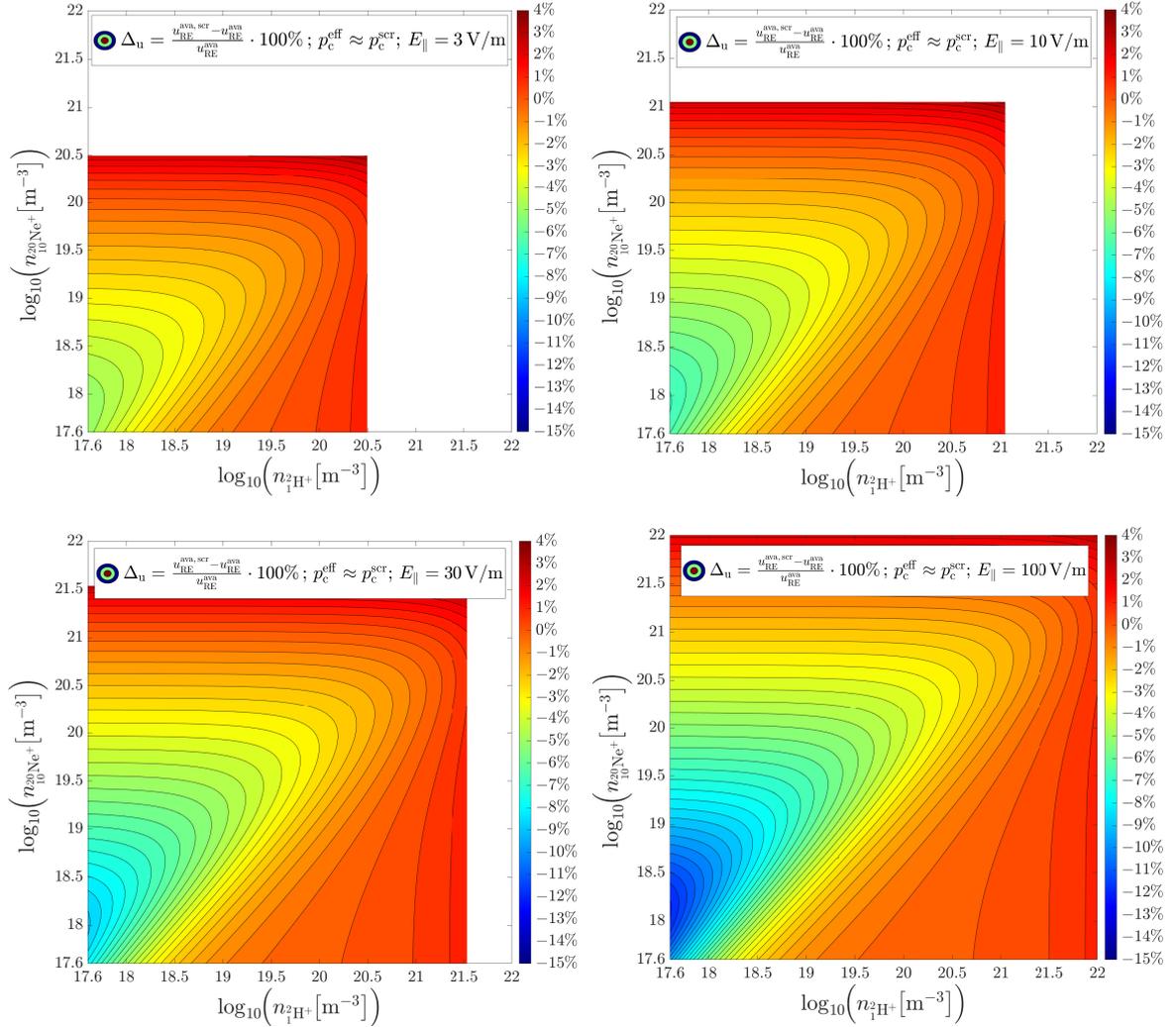

  \centering
  \subfloat{\label{fig_rel_u_ava_p_c_scr_E3} 
   \includegraphics[trim=312 19 345 22,width=0.49\textwidth,clip]
    {rel_u_ava_p_c_scr_E3_2.pdf}}\hfill
  \subfloat{\label{fig_rel_u_ava_p_c_scr_E10}
    \includegraphics[trim=311 27 353 13,width=0.49\textwidth,clip]
    {rel_u_ava_p_c_scr_E10_2.pdf}}\\[11pt]
  \subfloat{\label{fig_rel_u_ava_p_c_scr_E30} 
    \includegraphics[trim=312 21 346 20,width=0.49\textwidth,clip]
    {rel_u_ava_p_c_scr_E30_2.pdf}}\hfill
  \subfloat{\label{fig_rel_u_ava_p_c_scr_E100}
    \includegraphics[trim=314 27 350 14,width=0.49\textwidth,clip]
    {rel_u_ava_p_c_scr_E100_2.pdf}}
  \caption[Contour plots of the relative deviation $\Delta_{u}$ for the mean velocity of an avalanche runaway electron population with \mbox{$k_{\mathrm{B}}T_{\mathrm{e}}=10\,\textup{eV}$}, \mbox{$B=5.25\,\textup{T}$} and \mbox{$Z_{\mathrm{eff}}=1$}, due to the effect of partial screening with the effective critical momentum \mbox{$p_{\mathrm{c}}^{\mathrm{eff}}\approx p_{\mathrm{c}}^{\mathrm{scr}}$}, displayed for different values of the electric field strength \mbox{$E_{\|}\coloneqq\vert E_{\|}\vert$}.]{Contour plots$^{\ref{fig_plot_footnote_1}}$ of the relative deviation $\Delta_{u}$ for the mean velocity of an avalanche runaway electron population with \mbox{$k_{\mathrm{B}}T_{\mathrm{e}}=10\,\textup{eV}$}, \mbox{$B=5.25\,\textup{T}$} and \mbox{$Z_{\mathrm{eff}}=1$}, due to the effect of partial screening with the effective critical momentum \mbox{$p_{\mathrm{c}}^{\mathrm{eff}}\approx p_{\mathrm{c}}^{\mathrm{scr}}$}, displayed for different values of the electric field strength \mbox{$E_{\|}\coloneqq\vert E_{\|}\vert$}.}
\label{fig_rel_u_ava_p_c_scr}
\end{figure}
 
\noindent\newpage\noindent

\vspace*{2.9cm}

\begin{figure}[H]
  \centering
  \subfloat{\label{fig_rel_u_ava_p_star_E3} 
    \includegraphics[trim=325 20 345 21,width=0.49\textwidth,clip]
    {rel_u_ava_p_star_E3_2.pdf}}\hfill
  \subfloat{\label{fig_rel_u_ava_p_star_E10}
     \includegraphics[trim=329 25 341 17,width=0.49\textwidth,clip]
    {rel_u_ava_p_star_E10_2.pdf}}\\[11pt]
  \subfloat{\label{fig_rel_u_ava_p_star_E30} 
    \includegraphics[trim=325 20 346 20,width=0.49\textwidth,clip]
    {rel_u_ava_p_star_E30_2.pdf}}\hfill
  \subfloat{\label{fig_rel_u_ava_p_star_E100}
    \includegraphics[trim=319 21 346 20,width=0.49\textwidth,clip]
    {rel_u_ava_p_star_E100_2.pdf}}
  \caption[Contour plots of the relative deviation $\Delta_{u}$ for the mean velocity of an avalanche runaway electron population with \mbox{$k_{\mathrm{B}}T_{\mathrm{e}}=10\,\textup{eV}$}, \mbox{$B=5.25\,\textup{T}$} and \mbox{$Z_{\mathrm{eff}}=1$}, due to the effect of partial screening with the effective critical momentum \mbox{$p_{\mathrm{c}}^{\mathrm{eff}}\approx p_{\star}$}, displayed for approximately logarithmically increasing values of the electric field strength \mbox{$E_{\|}\coloneqq\vert E_{\|}\vert$}.]{Contour plots$^{\ref{fig_plot_footnote_2}}$ of the relative deviation $\Delta_{u}$ for the mean velocity of an avalanche runaway electron population with \mbox{$k_{\mathrm{B}}T_{\mathrm{e}}=10\,\textup{eV}$}, \mbox{$B=5.25\,\textup{T}$} and \mbox{$Z_{\mathrm{eff}}=1$}, due to the effect of partial screening with the effective critical momentum \mbox{$p_{\mathrm{c}}^{\mathrm{eff}}\approx p_{\star}$}, displayed for approximately logarithmically increasing values of the electric field strength \mbox{$E_{\|}\coloneqq\vert E_{\|}\vert$}.}
\label{fig_rel_u_ava_p_star}
\end{figure}

\noindent\newpage\noindent

\vspace*{2.9cm}

\begin{figure}[H]
  \centering
  \subfloat{\label{fig_tilde_rel_u_ava_p_star_E3} 
    \includegraphics[trim=319 35 335 8,width=0.49\textwidth,clip]
    {tilde_rel_u_p_c_scr_ava_p_star_E3_2.pdf}}\hfill
  \subfloat{\label{fig_tilde_rel_u_ava_p_star_E10}
    \includegraphics[trim=323 37 334 8,width=0.49\textwidth,clip]
    {tilde_rel_u_p_c_scr_ava_p_star_E10_2.pdf}}\\[11pt]
  \subfloat{\label{fig_tilde_rel_u_ava_p_star_E30} 
    \includegraphics[trim=322 33 333 10,width=0.49\textwidth,clip]
    {tilde_rel_u_p_c_scr_ava_p_star_E30_2.pdf}}\hfill
  \subfloat{\label{fig_tilde_rel_u_ava_p_star_E100}
   \includegraphics[trim=320 34 335 8,width=0.49\textwidth,clip]
    {tilde_rel_u_p_c_scr_ava_p_star_E100_2.pdf}}
  \caption[Contour plots of the relative deviation $\tilde{\Delta}_{u_{\mathrm{RE},\,p^{\mathrm{scr}}_{\mathrm{c}}}^{\mathrm{ava,scr}}}$ for the mean velocity of an avalanche runaway electron population with \mbox{$k_{\mathrm{B}}T_{\mathrm{e}}=10\,\textup{eV}$}, \mbox{$B=5.25\,\textup{T}$} and \mbox{$Z_{\mathrm{eff}}=1$} in the \textit{Hesslow} model, due to the different approximations of the effective critical momentum \mbox{$p_{\mathrm{c}}^{\mathrm{eff}}\approx p_{\mathrm{c}}^{\mathrm{scr}}$} and \mbox{$p_{\mathrm{c}}^{\mathrm{eff}}\approx p_{\star}$}, displayed for approximately logarithmically increasing values of the electric field strength \mbox{$E_{\|}\coloneqq\vert E_{\|}\vert$}.]{Contour plots$^{\ref{fig_plot_footnote_2}}$ of the relative deviation $\tilde{\Delta}_{u_{\mathrm{RE},\,p^{\mathrm{scr}}_{\mathrm{c}}}^{\mathrm{ava,scr}}}$ for the mean velocity of an avalanche runaway electron population with \mbox{$k_{\mathrm{B}}T_{\mathrm{e}}=10\,\textup{eV}$}, \mbox{$B=5.25\,\textup{T}$} and \mbox{$Z_{\mathrm{eff}}=1$} in the \textit{Hesslow} model, due to the different approximations of the effective critical momentum \mbox{$p_{\mathrm{c}}^{\mathrm{eff}}\approx p_{\mathrm{c}}^{\mathrm{scr}}$} and \mbox{$p_{\mathrm{c}}^{\mathrm{eff}}\approx p_{\star}$}, displayed for approximately logarithmically increasing values of the electric field strength \mbox{$E_{\|}\coloneqq\vert E_{\|}\vert$}.}
\label{fig_tilde_rel_u_ava_p_star}
\end{figure}

\clearpage

\subsection{Contour plots of the mean rest mass-related kinetic energy density of an avalanche runaway electron population}\label{contour_plots_k_ava_appendix_subsection}

\vspace*{2.4cm}

\begin{figure}[H]
  \centering
  \subfloat{\label{fig_k_ava_p_c_scr_E3} 
  \includegraphics[trim=317 22 376 23,width=0.49\textwidth,clip]
    {k_ava_p_c_scr_E3_2.pdf}}\hfill
  \subfloat{\label{fig_k_ava_p_c_scr_E10}
  \includegraphics[trim=317 17 379 25,width=0.49\textwidth,clip]
    {k_ava_p_c_scr_E10_2.pdf}}\\[11pt]
  \subfloat{\label{fig_k_ava_p_c_scr_E30} 
    \includegraphics[trim=315 14 379 24,width=0.49\textwidth,clip]
    {k_ava_p_c_scr_E30_2.pdf}}\hfill
  \subfloat{\label{fig_k_ava_p_c_scr_E100}
    \includegraphics[trim=313 18 373 18,width=0.49\textwidth,clip]
    {k_ava_p_c_scr_E100_2.pdf}}
  \caption[Contour plots of the normalized mean rest mass-related kinetic energy density \mbox{$k_{\mathrm{RE}}^{\hspace{0.25mm}\mathrm{ava}}/c^2$}, in the \textit{Rosenbluth-Putvinski} model, of an avalanche runaway electron population with \mbox{$k_{\mathrm{B}}T_{\mathrm{e}}=10\,\textup{eV}$}, \mbox{$B=5.25\,\textup{T}$} and \mbox{$Z_{\mathrm{eff}}=1$} for approximately logarithmically increasing values of the electric field strength \mbox{$E_{\|}\coloneqq\vert E_{\|}\vert$}.]{Contour plots$^{\ref{fig_plot_footnote_1}}$ of the normalized mean rest mass-related kinetic energy density \mbox{$k_{\mathrm{RE}}^{\hspace{0.25mm}\mathrm{ava}}/c^2$}, in the \textit{Rosenbluth-Putvinski} model, of an avalanche runaway electron population with \mbox{$k_{\mathrm{B}}T_{\mathrm{e}}=10\,\textup{eV}$}, \mbox{$B=5.25\,\textup{T}$} and \mbox{$Z_{\mathrm{eff}}=1$} for approximately logarithmically increasing values of the electric field strength \mbox{$E_{\|}\coloneqq\vert E_{\|}\vert$}.}
\label{fig_k_ava_p_c_scr}
\end{figure}
 
\noindent\newpage\noindent

\vspace*{2.9cm}

\begin{figure}[H]
  \centering
  \subfloat{\label{fig_k_ava_screen_p_c_scr_E3} 
  \includegraphics[trim=314 22 377 23,width=0.49\textwidth,clip]
    {k_ava_screen_p_c_scr_E3_2.pdf}}\hfill
  \subfloat{\label{fig_k_ava_screen_p_c_scr_E10}
   \includegraphics[trim=316 21 381 22,width=0.49\textwidth,clip]
    {k_ava_screen_p_c_scr_E10_2.pdf}}\\[11pt]
  \subfloat{\label{fig_k_ava_screen_p_c_scr_E30} 
  \includegraphics[trim=317 21 378 21,width=0.49\textwidth,clip]
    {k_ava_screen_p_c_scr_E30_2.pdf}}\hfill
  \subfloat{\label{fig_k_ava_screen_p_c_scr_E100}
  \includegraphics[trim=317 22 379 21,width=0.49\textwidth,clip]
    {k_ava_screen_p_c_scr_E100_2.pdf}}
  \caption[Contour plots of the normalized mean rest mass-related kinetic energy density \mbox{$k_{\mathrm{RE}}^{\hspace{0.25mm}\mathrm{ava,scr}}/c^2$}, in the \textit{Hesslow} model with the effective critical momentum \mbox{$p_{\mathrm{c}}^{\mathrm{eff}}\approx p_{\mathrm{c}}^{\mathrm{scr}}$}, of an avalanche runaway electron population with \mbox{$k_{\mathrm{B}}T_{\mathrm{e}}=10\,\textup{eV}$}, \mbox{$B=5.25\,\textup{T}$} and \mbox{$Z_{\mathrm{eff}}=1$} for approximately logarithmically increasing values of the electric field strength \mbox{$E_{\|}\coloneqq\vert E_{\|}\vert$}.]{Contour plots$^{\ref{fig_plot_footnote_1}}$ of the normalized mean rest mass-related kinetic energy density \mbox{$k_{\mathrm{RE}}^{\hspace{0.25mm}\mathrm{ava,scr}}/c^2$}, in the \textit{Hesslow} model with the effective critical momentum \mbox{$p_{\mathrm{c}}^{\mathrm{eff}}\approx p_{\mathrm{c}}^{\mathrm{scr}}$}, of an avalanche runaway electron population with \mbox{$k_{\mathrm{B}}T_{\mathrm{e}}=10\,\textup{eV}$}, \mbox{$B=5.25\,\textup{T}$} and \mbox{$Z_{\mathrm{eff}}=1$} for approximately logarithmically increasing values of the electric field strength \mbox{$E_{\|}\coloneqq\vert E_{\|}\vert$}.}
\label{fig_k_ava_screen_p_c_scr}
\end{figure}
  
\noindent\newpage\noindent

\vspace*{2.9cm}

\begin{figure}[H]
  \centering
  \subfloat{\label{fig_k_ava_screen_p_star_E3} 
   \includegraphics[trim=313 23 370 23,width=0.49\textwidth,clip]
    {k_ava_screen_p_star_E3_2.pdf}}\hfill
  \subfloat{\label{fig_k_ava_screen_p_star_E10}
   \includegraphics[trim=324 25 369 23,width=0.49\textwidth,clip]
    {k_ava_screen_p_star_E10_2.pdf}}\\[11pt]
  \subfloat{\label{fig_k_ava_screen_p_star_E30} 
   \includegraphics[trim=320 21 370 22,width=0.49\textwidth,clip]
    {k_ava_screen_p_star_E30_2.pdf}}\hfill
  \subfloat{\label{fig_k_ava_screen_p_star_E100}
   \includegraphics[trim=321 21 368 22,width=0.49\textwidth,clip]
    {k_ava_screen_p_star_E100_2.pdf}}
  \caption[Contour plots of the normalized mean rest mass-related kinetic energy density \mbox{$k_{\mathrm{RE}}^{\hspace{0.25mm}\mathrm{ava,scr}}/c^2$}, in the \textit{Hesslow} model with the effective critical momentum \mbox{$p_{\mathrm{c}}^{\mathrm{eff}}\approx p_{\star}$}, of an avalanche runaway electron population with \mbox{$k_{\mathrm{B}}T_{\mathrm{e}}=10\,\textup{eV}$}, \mbox{$B=5.25\,\textup{T}$} and \mbox{$Z_{\mathrm{eff}}=1$} for approximately logarithmically increasing values of the electric field strength \mbox{$E_{\|}\coloneqq\vert E_{\|}\vert$}.]{Contour plots$^{\ref{fig_plot_footnote_2}}$ of the normalized mean rest mass-related kinetic energy density \mbox{$k_{\mathrm{RE}}^{\hspace{0.25mm}\mathrm{ava,scr}}/c^2$}, in the \textit{Hesslow} model with the effective critical momentum \mbox{$p_{\mathrm{c}}^{\mathrm{eff}}\approx p_{\star}$}, of an avalanche runaway electron population with \mbox{$k_{\mathrm{B}}T_{\mathrm{e}}=10\,\textup{eV}$}, \mbox{$B=5.25\,\textup{T}$} and \mbox{$Z_{\mathrm{eff}}=1$} for approximately logarithmically increasing values of the electric field strength \mbox{$E_{\|}\coloneqq\vert E_{\|}\vert$}.}
\label{fig_k_ava_screen_p_star}
\end{figure}

\noindent\newpage\noindent

\vspace*{2.9cm}

\begin{figure}[H]
  \centering
  \subfloat{\label{fig_rel_k_ava_p_c_scr_E3} 
   \includegraphics[trim=318 19 345 25,width=0.49\textwidth,clip]
    {rel_k_ava_p_c_scr_E3_2.pdf}}\hfill
  \subfloat{\label{fig_rel_k_ava_p_c_scr_E10}
    \includegraphics[trim=316 25 348 19,width=0.49\textwidth,clip]
    {rel_k_ava_p_c_scr_E10_2.pdf}}\\[11pt]
  \subfloat{\label{fig_rel_k_ava_p_c_scr_E30} 
    \includegraphics[trim=315 29 343 17,width=0.49\textwidth,clip]
    {rel_k_ava_p_c_scr_E30_2.pdf}}\hfill
  \subfloat{\label{fig_rel_k_ava_p_c_scr_E100}
     \includegraphics[trim=314 25 347 17,width=0.49\textwidth,clip]
    {rel_k_ava_p_c_scr_E100_2.pdf}}
  \caption[Contour plots of the relative deviation $\Delta_{k}$ for the mean mass-related kinetic energy density of an avalanche runaway electron population with \mbox{$k_{\mathrm{B}}T_{\mathrm{e}}=10\,\textup{eV}$}, \mbox{$B=5.25\,\textup{T}$} and \mbox{$Z_{\mathrm{eff}}=1$}, due to the effect of partial screening with the effective critical momentum \mbox{$p_{\mathrm{c}}^{\mathrm{eff}}\approx p_{\mathrm{c}}^{\mathrm{scr}}$}, displayed for approximately logarithmically increasing values of the electric field strength \mbox{$E_{\|}\coloneqq\vert E_{\|}\vert$}.]{Contour plots$^{\ref{fig_plot_footnote_1}}$ of the relative deviation $\Delta_{k}$ for the mean rest mass-related kinetic energy density of an avalanche runaway electron population with \mbox{$k_{\mathrm{B}}T_{\mathrm{e}}=10\,\textup{eV}$}, \mbox{$B=5.25\,\textup{T}$} and \mbox{$Z_{\mathrm{eff}}=1$}, due to the effect of partial screening with the effective critical momentum \mbox{$p_{\mathrm{c}}^{\mathrm{eff}}\approx p_{\mathrm{c}}^{\mathrm{scr}}$}, displayed for approximately logarithmically increasing values of the electric field strength \mbox{$E_{\|}\coloneqq\vert E_{\|}\vert$}.}
\label{fig_rel_k_ava_p_c_scr}
\end{figure}
 
\noindent\newpage\noindent

\vspace*{2.9cm}

\begin{figure}[H]
  \centering
  \subfloat{\label{fig_rel_k_ava_p_star_E3} 
   \includegraphics[trim=319 22 332 17,width=0.49\textwidth,clip]
    {rel_k_ava_p_star_E3_2.pdf}}\hfill
  \subfloat{\label{fig_rel_k_ava_p_star_E10}
    \includegraphics[trim=325 26 329 19,width=0.49\textwidth,clip]
    {rel_k_ava_p_star_E10_2.pdf}}\\[11pt]
  \subfloat{\label{fig_rel_k_ava_p_star_E30} 
    \includegraphics[trim=321 28 340 14,width=0.49\textwidth,clip]
    {rel_k_ava_p_star_E30_2.pdf}}\hfill
  \subfloat{\label{fig_rel_k_ava_p_star_E100}
   \includegraphics[trim=322 21 334 18,width=0.49\textwidth,clip]
    {rel_k_ava_p_star_E100_2.pdf}}
  \caption[Contour plots of the relative deviation $\Delta_{k}$ for the mean rest mass-related kinetic energy density of an avalanche runaway electron population with \mbox{$k_{\mathrm{B}}T_{\mathrm{e}}=10\,\textup{eV}$}, \mbox{$B=5.25\,\textup{T}$} and \mbox{$Z_{\mathrm{eff}}=1$}, due to the effect of partial screening with the effective critical momentum \mbox{$p_{\mathrm{c}}^{\mathrm{eff}}\approx p_{\star}$}, displayed for approximately logarithmically increasing values of the electric field strength \mbox{$E_{\|}\coloneqq\vert E_{\|}\vert$}.]{Contour plots$^{\ref{fig_plot_footnote_2}}$ of the relative deviation $\Delta_{k}$ for the mean rest mass-related kinetic energy density of an avalanche runaway electron population with \mbox{$k_{\mathrm{B}}T_{\mathrm{e}}=10\,\textup{eV}$}, \mbox{$B=5.25\,\textup{T}$} and \mbox{$Z_{\mathrm{eff}}=1$}, due to the effect of partial screening with the effective critical momentum \mbox{$p_{\mathrm{c}}^{\mathrm{eff}}\approx p_{\star}$}, displayed for approximately logarithmically increasing values of the electric field strength \mbox{$E_{\|}\coloneqq\vert E_{\|}\vert$}.}
\label{fig_rel_k_ava_p_star}
\end{figure}

\noindent\newpage\noindent

\vspace*{2.9cm}

\begin{figure}[H]
  \centering
  \subfloat{\label{fig_tilde_rel_k_ava_p_star_E3} 
  \includegraphics[trim=322 33 327 9,width=0.49\textwidth,clip]
    {tilde_rel_k_p_c_scr_ava_p_star_E3_2.pdf}}\hfill
  \subfloat{\label{fig_tilde_rel_k_ava_p_star_E10}
   \includegraphics[trim=317 33 331 8,width=0.49\textwidth,clip]
    {tilde_rel_k_p_c_scr_ava_p_star_E10_2.pdf}}\\[11pt]
  \subfloat{\label{fig_tilde_rel_k_ava_p_star_E30} 
  \includegraphics[trim=317 32 335 7,width=0.49\textwidth,clip]
    {tilde_rel_k_p_c_scr_ava_p_star_E30_2.pdf}}\hfill
  \subfloat{\label{fig_tilde_rel_k_ava_p_star_E100}
   \includegraphics[trim=320 33 332 6,width=0.49\textwidth,clip]
    {tilde_rel_k_p_c_scr_ava_p_star_E100_2.pdf}}
  \caption[Contour plots of the relative deviation $\tilde{\Delta}_{k_{\mathrm{RE},\,p^{\mathrm{scr}}_{\mathrm{c}}}^{\mathrm{ava,scr}}}$ for the mean rest mass-related kinetic energy density of an avalanche runaway electron population with \mbox{$k_{\mathrm{B}}T_{\mathrm{e}}=10\,\textup{eV}$}, \mbox{$B=5.25\,\textup{T}$} and \mbox{$Z_{\mathrm{eff}}=1$} in the \textit{Hesslow} model, due to the different approximations of the effective critical momentum \mbox{$p_{\mathrm{c}}^{\mathrm{eff}}\approx p_{\mathrm{c}}^{\mathrm{scr}}$} and \mbox{$p_{\mathrm{c}}^{\mathrm{eff}}\approx p_{\star}$}, displayed for approximately logarithmically increasing values of the electric field strength \mbox{$E_{\|}\coloneqq\vert E_{\|}\vert$}.]{Contour plots$^{\ref{fig_plot_footnote_2}}$ of the relative deviation $\tilde{\Delta}_{k_{\mathrm{RE},\,p^{\mathrm{scr}}_{\mathrm{c}}}^{\mathrm{ava,scr}}}$ for the mean rest mass-related kinetic energy density of an avalanche runaway electron population with \mbox{$k_{\mathrm{B}}T_{\mathrm{e}}=10\,\textup{eV}$}, \mbox{$B=5.25\,\textup{T}$} and \mbox{$Z_{\mathrm{eff}}=1$} in the \textit{Hesslow} model, due to the different approximations of the effective critical momentum \mbox{$p_{\mathrm{c}}^{\mathrm{eff}}\approx p_{\mathrm{c}}^{\mathrm{scr}}$} and \mbox{$p_{\mathrm{c}}^{\mathrm{eff}}\approx p_{\star}$}, displayed for approximately logarithmically increasing values of the electric field strength \mbox{$E_{\|}\coloneqq\vert E_{\|}\vert$}.}
\label{fig_tilde_rel_k_ava_p_star}
\end{figure}

\clearpage